\documentclass[a4paper,12pt,twoside,openright,sort&compress,tbtags]{book}
\usepackage{color}
\usepackage[square,comma,numbers,sort&compress]{natbib}

\usepackage[T1]{fontenc} 
\usepackage{lmodern}

\usepackage[hang,flushmargin]{footmisc} 

\usepackage{bm}
\usepackage[skip=10pt,font=footnotesize,bf,labelsep=period]{caption}
\usepackage{url}
\usepackage{amsmath}
\usepackage{amssymb}
\usepackage{mathtools}
\usepackage[breaklinks,hidelinks,colorlinks=false,urlcolor=black,citecolor=black,linkcolor=black]{hyperref}
\usepackage{epsfig,graphicx,verbatim,xspace,multirow,mathtools}
\usepackage{booktabs}
\usepackage{xcolor}
\usepackage{caption}
\usepackage{subcaption}
\usepackage{microtype}
\usepackage[nottoc,numbib]{tocbibind}

\usepackage{tabularx}
\usepackage{afterpage}
\usepackage{adforn}
\usepackage{bookmark}
\usepackage{lscape}
\usepackage{rotating}
\usepackage{wrapfig}
\usepackage{pdflscape}
\usepackage{pdfpages}

\usepackage[english]{babel}
\usepackage{epigraph}
\usepackage{fancyhdr}
\usepackage{indentfirst}
\usepackage{cleveref}

\usepackage{etoolbox}                % <---
\AtBeginEnvironment{tabular}{\footnotesize\centering}


\usepackage{tabstackengine}
\setstackEOL{\cr}

\usepackage{etoolbox}
\patchcmd{\part}{\thispagestyle{plain}}{\thispagestyle{empty}}{}{}

\usepackage{pgf,tikz}
\usepackage{pgfplots}
\usetikzlibrary{%
    calc,
    decorations.pathreplacing,
    decorations.markings,
    decorations.pathmorphing,
    shapes.geometric,
    external,
    decorations.text%
}
\usetikzlibrary{intersections}
\usepgfplotslibrary{fillbetween}
\usetikzlibrary{decorations.text}

\pgfplotsset{compat=1.17}   % To remove some warnings
\newcommand{\arrowscale}{1.2}
\newcommand{\diagramxscale}{2}
\newcommand{\diagramyscale}{1.4}

\pgfmathsetlengthmacro\MajorTickLength{
      \pgfkeysvalueof{/pgfplots/major tick length} * 0.4
    }
\pgfmathsetlengthmacro\MinorTickLength{
      \pgfkeysvalueof{/pgfplots/minor tick length} * 0.3
    }

\tikzset{
    on each segment/.style={
        decorate,
        decoration={
              show path construction,
              moveto code={},
              lineto code={
                \path [#1]
                (\tikzinputsegmentfirst) -- (\tikzinputsegmentlast);
              },
              curveto code={
                    \path [#1] (\tikzinputsegmentfirst)
                    .. controls
                    (\tikzinputsegmentsupporta) and (\tikzinputsegmentsupportb)
                    ..
                    (\tikzinputsegmentlast);
              },
              closepath code={
                    \path [#1]
                    (\tikzinputsegmentfirst) -- (\tikzinputsegmentlast);
              },
        },
    },
    mid arrow/.style={postaction={decorate,decoration={
        markings,
        mark=at position #1 with {\arrow[scale=\arrowscale]{stealth}}
      }}},
    mid arrow/.default=0.55,
    mid barrow/.style={postaction={decorate,decoration={
        markings,
        mark=at position #1 with {\arrowreversed[scale=\arrowscale]{stealth}}
      }}},
    mid barrow/.default=0.45,
    prop/.style={thick,join=round},
    dprop/.style={prop,postaction={on each segment={mid arrow=#1}}},
    bdprop/.style={prop,postaction={on each segment={mid barrow=#1}}},
    sketch onshell prop/.style={very thick},
    sketch offshell prop/.style={sketch onshell prop, densely dotted},
    sketch blob/.style={draw=black,thick, fill=#1, anchor=center},
    sketch onshell blob/.style={sketch blob=#1,
                                shape=regular polygon, regular polygon sides=4},
    sketch offshell blob/.style={sketch blob=#1,
                                shape=ellipse}
}
\newcommand{\makeexternallegcoordinates}{%
    \coordinate (k1) at (+1,+1);%
    \coordinate (k2) at (+1, 0);%
    \coordinate (k3) at (+1,-1);%
    \coordinate (p1) at (-1,+1);%
    \coordinate (p2) at (-1, 0);%
    \coordinate (p3) at (-1,-1);%
    }
\newcommand{\makeexternallegs}{%
    \makeexternallegcoordinates
    \draw (k1) node [right] {$k_1$};%
    \draw (k2) node [right] {$k_2$};%
    \draw (k3) node [right] {$k_3$};%
    \draw (p1) node [left] {$p_1$};%
    \draw (p2) node [left] {$p_2$};%
    \draw (p3) node [left] {$p_3$};%
    }
\newcommand{\makeexternallegsshifted}{%
    \makeexternallegcoordinates
    \draw (k1) node [right] {$k_3$};%
    \draw (k2) node [right] {$k_1$};%
    \draw (k3) node [right] {$k_2$};%
    \draw (p1) node [left] {$p_1$};%
    \draw (p2) node [left] {$p_2$};%
    \draw (p3) node [left] {$p_3$};%
    }

\newcommand{\tikzineq}[2][]{%
	\tikz[%
		anchor=base,%
		baseline={([yshift=-1.0ex]current bounding box.center)},%
		#1]{#2}%
	}
	
\definecolor{plotI}{HTML}{0077BB}
\definecolor{plotII}{HTML}{33BBEE}
\definecolor{plotIII}{HTML}{009988}
\definecolor{plotIV}{HTML}{EE7733}
\definecolor{plotV}{HTML}{CC3311}
\definecolor{plotVI}{HTML}{DDAA33}
\definecolor{plotVII}{HTML}{AA3377}

\newcommand{\sketchoperatorscale}{1.6}
\definecolor{LOcolor_}{HTML}{6699CC}
\definecolor{NLOcolor_}{HTML}{EE99AA}
\colorlet{LOcolor}{LOcolor_!70}
\colorlet{NLOcolor}{NLOcolor_!70}

\definecolor{fitorange}{HTML}{FFA500}
\definecolor{fitblue}{HTML}{0000FF}
\definecolor{fitgray}{HTML}{808080}

\tikzset{
    LOdash/.style={dash pattern=on 3pt off 2.5pt},
    fitdash/.style={dash pattern=on 1pt off 1.3pt},
    zeroline/.style={mark=none, black, ultra thin},   
    verticalline/.style={thin, black, dash pattern=on 3pt off 3.5pt},
    KXline/.style={thick},
    DXline/.style={ultra thick},
    LOline/.style={KXline, black,   LOdash},
    NLOline/.style={KXline, fitgray},
    NLOupperline/.style={line width = 0pt, fitgray},
    NLOlowerline/.style={line width = 0pt, fitgray},
    Fitline/.style={KXline, fitorange, fitdash},
    Fitupperline/.style={line width = 0pt, fitorange},
    Fitlowerline/.style={line width = 0pt, fitorange},
    NLO2line/.style={NLOline},
    NLO2upperline/.style={NLOupperline},
    NLO2lowerline/.style={NLOlowerline},
    NLOAline/.style={NLOline, LOdash},
    NLOAupperline/.style={NLOupperline},
    NLOAlowerline/.style={NLOlowerline},
    numericline/.style={KXline, red},
    threshline/.style={KXline, blue, dash pattern=on 3pt off 2.5pt},
    D0line/.style={DXline, plotI,   solid},
    D1line/.style={DXline, plotII,  dash pattern=on 8pt off 2pt on 2pt off 2pt},
    D2line/.style={DXline, plotIII, dash pattern=on 6pt off 2pt on 2pt off 2pt on 2pt off 2pt},
    DAline/.style={DXline, plotIV,  dash pattern=on 4.4pt off 4.4pt},
    DBline/.style={DXline, plotV,   dash pattern=on 2pt off 2pt},
    allwaveline/.style={DXline, black},
    swaveline/.style={D1line},
    dwaveline/.style={DAline},
    gwaveline/.style={DBline},
}
\pgfkeys{/pgf/number format/1000 sep={}}
\pgfplotsset{
    general plot/.style={
        set layers=axis on top,
        width=1.0\textwidth,
        height=0.75\textwidth,
        minor tick num=4,
        every tick/.style={
                semithick,
            },
        major tick length=\MajorTickLength,
        minor tick length=\MinorTickLength,
        y tick label style={
                /pgf/number format/.cd,
                fixed,
             fixed zerofill,
                precision=1,
                /tikz/.cd
                },
            every tick label/.append style={font=\footnotesize},
            axis line style={black, semithick},
        xticklabel style={inner sep=1.5pt},
        yticklabel style={inner sep=1.5pt},
            },
    fit plot/.style={
        general plot,
        ylabel style = {rotate=-90}, ylabel shift=-1.1ex,
        legend style={font=\scriptsize}
    },
    legend image code/.code={
        \draw [#1] (0pt,0pt) -- (15pt,0pt);
    },
    filled legend/.style={legend image code/.code={
        \draw [#1] (0pt,0pt) -- (15pt,0pt);
        \path [draw=none, fill=#1, opacity=0.3] (0pt,-4pt) rectangle (15pt,4pt);
    }},
    double filled legend/.style={legend image code/.code={
        \draw [#1] (0pt,5pt) -- (15pt,5pt);
        \path [draw=none, fill=#1, opacity=0.3] (0pt,1pt) rectangle (15pt,9pt);
        \draw [#1, solid] (0pt,-5pt) -- (15pt,-5pt);
        \path [draw=none, fill=#1, opacity=0.3] (0pt,-1pt) rectangle (15pt,-9pt);
    }},
    legend with mark/.style={legend image code/.code={
        \draw [#1, thick, solid] (7.5pt,4pt) -- (7.5pt,-4pt);
        \draw [#1, thick, solid, fill=white] (7.5pt,0pt) circle[radius=1.5pt];
    }},
    filled legend with mark/.style={legend image code/.code={
        \draw [#1] (0pt,0pt) -- (15pt,0pt);
        \path [draw=none, fill=#1, opacity=0.3] (0pt,-4pt) rectangle (15pt,4pt);
        \draw [#1, thick, solid] (7.5pt,4pt) -- (7.5pt,-4pt);
        \draw [#1, thick, solid, fill=white] (7.5pt,0pt) circle[radius=1.5pt];
    }},
    wide legend/.style={legend image code/.code={
        \draw [#1] (0pt,0pt) -- (22pt,0pt);
    }},
    lattice data/.style={
        thick,
        scatter, scatter/use mapped color={draw=#1,fill=white},
        only marks, mark=*,
        mark options={scale=0.7},
        error bars/.cd, y dir = both, x dir = both, y explicit, y explicit, error bar style={color=#1, thick, mark size = 0pt},
        /pgfplots/.cd
    },
    text along plot/.style 2 args={
        mark=none, draw=none, 
        decoration={text along path, text align=center, raise=1ex, 
            text={#1}, 
            text align={left indent=#2} },
        postaction={decorate}}
}

\newcommand{\Nf}[0]{N_\text{f}}
\newcommand{\Nc}[0]{N_\text{c}}

\newcommand{\alphas}[0]{\alpha_{\text{s}}}
\newcommand{\LambdaQCD}[0]{\Lambda_{\text{QCD}}}
\newcommand{\LambdaUV}[0]{\Lambda_{\text{UV}}}

\newcommand{\cC}[0]{\mathcal{C}}
\newcommand{\cD}[0]{\mathcal{D}}
\newcommand{\cF}[0]{\mathcal{F}}
\newcommand{\cG}[0]{\mathcal{G}}
\newcommand{\cH}[0]{\mathcal{H}}
\newcommand{\cI}[0]{\mathcal{I}}
\newcommand{\cJ}[0]{\mathcal{J}}
\newcommand{\cK}[0]{\mathcal{K}}
\newcommand{\cL}[0]{\mathcal{L}}
\newcommand{\cM}[0]{\mathcal{M}}

\newcommand{\cO}[0]{\mathcal{O}}
\newcommand{\cP}[0]{\mathcal{P}}
\newcommand{\cQ}[0]{\mathcal{Q}}
\newcommand{\cR}[0]{\mathcal{R}}
\newcommand{\cS}[0]{\mathcal{S}}
\newcommand{\cT}[0]{\mathcal{T}}
\newcommand{\cZ}[0]{\mathcal{Z}}

\renewcommand{\d}{\mathrm{d}}
\newcommand{\tmin}{t_\mathrm{min}}
\newcommand{\tmax}{t_\mathrm{max}}

\renewcommand{\Re}{\mathrm{Re}}
\renewcommand{\Im}{\mathrm{Im}}
\newcommand{\Tr}{\mathrm{Tr}}

\newcommand{\Kdf}[0]{\cK_\text{df,3}}
\newcommand{\Mdf}[0]{\cM_\text{df,3}}

\newcommand{\uu}[0]{{(\text{u},\text{u})}}
\newcommand{\rr}[0]{\text{r}}
\newcommand{\LO}[0]{\text{LO}}
\newcommand{\NLO}[0]{\text{NLO}}
\newcommand{\OPE}[0]{\text{OPE}}
\newcommand{\sOPE}[0]{s\text{-OPE}}
\newcommand{\BH}[0]{\text{BH}}
\newcommand{\nOPE}[0]{\text{non-OPE}}
\newcommand{\off}[0]{\text{off}}
\newcommand{\on}[0]{\text{on}}
\newcommand{\phys}[0]{\text{phys}}
\newcommand{\free}[0]{\text{free}}
\newcommand{\latt}[0]{\text{latt}}
\newcommand{\inter}[0]{\text{int}}
\newcommand{\dec}[0]{\text{dec}}
\newcommand{\PV}[0]{\text{PV}}
\newcommand{\dof}[0]{\text{dof}}

\newcommand{\td}[0]{t_{\text{d}}}

\newcommand{\A}{\mathrm{A}}

\renewcommand{\SS}{\mathrm{SS}}
\newcommand{\SD}{\mathrm{SD}}
\newcommand{\DS}{\mathrm{DS}}
\newcommand{\DD}{\mathrm{DD}}
\newcommand{\AS}{\mathrm{AS}}

\newcommand{\Kiso}[0]{\cK_0}
\newcommand{\Kisoone}[0]{\cK_1}
\newcommand{\Kisotwo}[0]{\cK_2}
\newcommand{\KA}[0]{\cK_\text{A}}
\newcommand{\KB}[0]{\cK_\text{B}}
\newcommand{\KX}[2][]{\cK^{#2}_{#1}}
\newcommand{\KT}[1][0]{\KX[#1]{\text{T}}}
\newcommand{\KSS}[1][0]{\KX[#1]{\SS}}
\newcommand{\KSD}[1][0]{\KX[#1]{\SD}}
\newcommand{\KDS}[1][0]{\KX[#1]{\DS}}
\newcommand{\KDD}[1][0]{\KX[#1]{\DD}}
\newcommand{\KSSA}{\KSS[\mathrm A]}
\newcommand{\KSSB}{\KSS[\mathrm B]}
\newcommand{\KAS}[1][0]{\KX[#1]{\AS}}

\newcommand{\Diso}[0]{\cD_0}
\newcommand{\Disoone}[0]{\cD_1}
\newcommand{\Disotwo}[0]{\cD_2}
\newcommand{\DisoA}[0]{\cD_\text{A}}
\newcommand{\DisoB}[0]{\cD_\text{B}}
\newcommand{\DA}[0]{\Delta_\text{A}}
\newcommand{\DB}[0]{\Delta_\text{B}}
\newcommand{\DeltaAS}[1]{\Delta_\AS^{(#1)}}

\newcommand{\m}{\mathbf}
\newcommand{\mD}{\m D}
\newcommand{\mG}{\m G}
\newcommand{\mK}{\m K}
\newcommand{\mF}{\m F}
\newcommand{\mM}{\m M}
\newcommand{\mT}{\m T}
\newcommand{\mKdf}{\mK_{\text{df},3}}
\newcommand{\mMdf}{\mM_{\text{df},3}}

\newcommand{\Ipp}{I_{\pi\pi}}
\newcommand{\Ippp}{I_{\pi\pi\pi}}
\newcommand{\Iss}{I_{\sigma\sigma}}
\newcommand{\Isss}{I_{\sigma\sigma\sigma}}
\newcommand{\sign}{\text{sign}}
\newcommand{\Ncluster}{{N_\text{cl}}}

\newcommand{\Mpi}[0]{M_\pi}
\newcommand{\Mpiphys}[0]{M_{\pi,\phys}}

\newcommand{\Mrho}[0]{M_\rho}
\newcommand{\Metap}[0]{M_{\eta'}}

\newcommand{\Fpi}[0]{F_\pi}
\newcommand{\Fpiphys}[0]{F_{\pi,\phys}}

\newcommand{\MF}[0]{\left(\frac{\Mpi}{\Fpi}\right)}

\newcommand{\tij}[1]{\tilde t_{#1}}

\newcommand{\lrI}{\ell_1^\rr}
\newcommand{\lrII}{\ell_2^\rr}
\newcommand{\lrIII}{\ell_3^\rr}
\newcommand{\lrIV}{\ell_4^\rr}
\newcommand{\elliso}{{\ell_{(0)}^\rr}}
\newcommand{\ellisoone}{{\ell_{(1)}^\rr}}
\newcommand{\ellisotwo}{{\ell_{(2)}^\rr}}
\newcommand{\ellA}{{\ell_{(\mathrm{A})}^\rr}}
\newcommand{\ellB}{{\ell_{(\mathrm{B})}^\rr}}
\newcommand{\logthree}{{\log3}}

\newcommand{\lr}[1]{\ell_{#1}^\mathrm{r}}

\newcommand{\Lfrac}[2][1]{$\tfrac{#1}{#2}L$}
\newcommand{\lrfrac}[3][1]{$\tfrac{#1}{#2}\lr{#3}$}

\newcommand{\ket}[1]{|#1\rangle}
\newcommand{\qtwos}{q_2^*}
\newcommand{\pip}{\pi^+}
\newcommand{\pim}{\pi^-}
\newcommand{\pio}{\pi^0}
\newcommand{\pipm}{\pi^\pm}
\newcommand{\roo}{\rho^0}
\newcommand{\rop}{\rho^+}
\newcommand{\rom}{\rho^-}
\newcommand{\ropm}{\rho^\pm}
\newcommand{\Pio}{\Pi^0}
\newcommand{\Pip}{\Pi^+}
\newcommand{\Pim}{\Pi^-}

\newcommand{\Pipm}{\Pi^\pm}

\newcommand{\mKdfI}[1]{\mKdf^{\Ippp=#1}}
\newcommand{\mXi}[1]{\boldsymbol{\Xi}_{#1}}
\newcommand{\vxi}[1][]{{\boldsymbol{\xi}}^{\,#1}}

\newcommand{\vxip}[1][]{\boldsymbol{\xi}^{\,\prime #1}}
\newcommand{\vxiS}[1][]{\boldsymbol{\xi}(S)^{#1}}

\newcommand{\vxiSb}[1][]{\boldsymbol{\xi}(\bar S)^{#1}}
\newcommand{\vxipS}[1][]{\boldsymbol{\xi}^{\,\prime}(S)^{#1}}

\newcommand{\vxipSb}[1][]{\boldsymbol{\xi}^{\,\prime}(\bar S)^{#1}}

\newcommand{\n}[1]{n_{#1}}

\makeatletter
\newcommand{\presentxi}{\@ifstar{\align@presentxi}{\@presentxi}}
\newcommand{\@presentxi}[4][]{%
    \vxi[\if!#2!\else(#2)\fi #1] = \big(\ensuremath{\xi}^{\if!#2!\else(#2)\fi #1}_1,\ensuremath{\xi}^{\if!#2!\else(#2)\fi #1}_2\big)\,,%
    \qquad\text{where}%
    \quad\ensuremath{\xi}^{\if!#2!\else(#2)\fi}_1\equiv #3\,,%
    \quad\ensuremath{\xi}^{\if!#2!\else(#2)\fi}_2\equiv #4}
\newcommand{\align@presentxi}[4][]{%
    \quad\ensuremath{\xi}^{\if!#2!\else(#2)\fi}_1&\equiv #3\,,%
    &\quad\ensuremath{\xi}^{\if!#2!\else(#2)\fi}_2&\equiv #4}
\makeatother

\newcommand{\mathbbm}[1]{\text{\usefont{U}{bbm}{m}{n}#1}} % from mathbbm.sty

\newcommand{\CosmoLattice}{$\cC$osmo$\cL$attice }
\newcommand{\mpl}{m_\mathrm{p}}

\newcommand{\TT}{\text{TT}}
\newcommand{\GW}{\text{GW}}
\newcommand{\eff}{\text{eff}}
\newcommand{\NO}{\text{NO}}
\renewcommand{\NG}{\text{NG}}

\newcommand{\HL}{\text{HL}}
\newcommand{\str}{\text{str}}

\newcommand{\ef}{\text{ef}}
\newcommand{\per}{\text{per}}

\newcommand{\rhok}{\rho_\text{K}}
\newcommand{\rhog}{\rho_\text{G}}

\newcommand{\kUV}{k_\text{UV}}
\newcommand{\kIR}{k_\text{IR}}

\newcommand{\hbmk}{\hat{\bm{k}}}

\newcommand{\bmk}{\bm{k}}
\newcommand{\tn}{\tilde{\bm{n}}}
\newcommand{\hbmn}{\hat{\bm{n}}}
\newcommand{\tbmn}{\tilde{\bm{n}}}
\newcommand{\bmn}{\bm{n}}
\newcommand{\bmx}{\bm{x}}
\newcommand{\LambdaR}{\Lambda_\text{R}}

\newcommand{\fstar}{f_\star}
\newcommand{\omegastar}{\omega_\star}
\newcommand{\tphi}{\tilde{\phi}}
\newcommand{\tvarphi}{\tilde{\varphi}}
\newcommand{\tA}{\tilde{A}}
\newcommand{\tx}{\tilde{x}}
\newcommand{\teta}{\tilde{\eta}}

\newcommand{\PGW}{P_\text{GW}}
\newcommand{\Pphi}{P_{\varphi}}
\newcommand{\rc}{r_\text{c}}
\newcommand{\mh}{m_h}
\newcommand{\Lonefourth}{L_{1/4}}

\crefname{equation}{eq.}{eqs.}
\crefname{section}{sec.}{secs.}
\Crefname{equation}{Eq.}{Eqs.}
\Crefname{section}{Sec.}{Secs.}
\newcommand{\rcite}[1]{ref.~\cite{#1}}
\newcommand{\rrcite}[1]{refs.~\cite{#1}}

\newcommand{\RNum}[1]{\text{\uppercase\expandafter{\romannumeral #1\relax}}}

\newcommand{\lattice}{\textit{lattice} }

\newcommand{\scattering}{\textit{scattering} }
\newcommand{\loops}{\textit{loops} }
\newcommand{\network}{\textit{network} }

\def\eqref#1{{(\ref{#1})}}

\usepackage[hyperpageref]{backref}

\renewcommand*{\backref}[1]{}  

\usepackage{etoolbox} % for '\AtBeginEnvironment' directive
\AtBeginEnvironment{array}{\arraycolsep=1pt}

\makeatletter
\def \cleardoublepage {\clearpage \if@twoside
\ifodd \c@page
\else
\null\thispagestyle{empty}\clearpage
\fi
\fi}
\makeatother

\usepackage{fancyhdr}
\renewcommand{\chaptermark}[1]{%
\markboth{#1}{}}

\renewcommand{\headrulewidth}{0.0pt}

\fancypagestyle{plain}{%
\fancyhead{} % get rid of headers on plain pages
\fancyfoot[C]{\thepage}
\renewcommand{\headrulewidth}{0pt} % and the line
}

\definecolor{lightgray}{RGB}{192,192,192}

\usepackage{lipsum}
\usepackage[explicit]{titlesec}
\usepackage{blindtext}

\colorlet{chapnumcolor}{lightgray}
\newcommand*{\chapnumfont}{%
  \usefont{T1}{jkp}{b}{n}%
  \fontsize{100}{120}%
  \selectfont%
}
\newcommand*{\chaptitlefont}{%
  \fontsize{22}{26}%
  \selectfont%
}
\newcommand*{\partitlefont}{%
  \bfseries
  \fontsize{27.5}{31.5}%
  \selectfont%
}

\titleformat{\chapter}
{\thispagestyle{plain}\vspace{50pt}\normalfont\huge\bfseries}
{\filleft{\parbox[b][][b]{.17\textwidth}{\chapnumfont\color{chapnumcolor}\thechapter}}}
{0pt}
{\Huge{\parbox[b][][t]{.83\textwidth}{\flushleft\hyphenpenalty=10000\chaptitlefont #1\vspace{0.0cm}}}}[\vspace{0.5ex}{\titlerule[0.7pt]}\vspace{-0.7cm}]

\fancypagestyle{frontmatter}{%
  \pagestyle{plain}% Just like plain page style
  \fancyfoot[C]{\roman{page}}% Roman page number in footer centre
}

\let\oldfrontmatter\frontmatter
\renewcommand{\frontmatter}{
  \oldfrontmatter
  \pagestyle{frontmatter}
}

\usepackage{pdfpages}
\newlength{\imagewidth}

\makeatletter
\pretocmd{\part}{\addtocontents{toc}{\protect\addvspace{50\p@}}}{}{}
\makeatother

\begin{document}

\hypersetup{linkcolor = black}
\includepdf{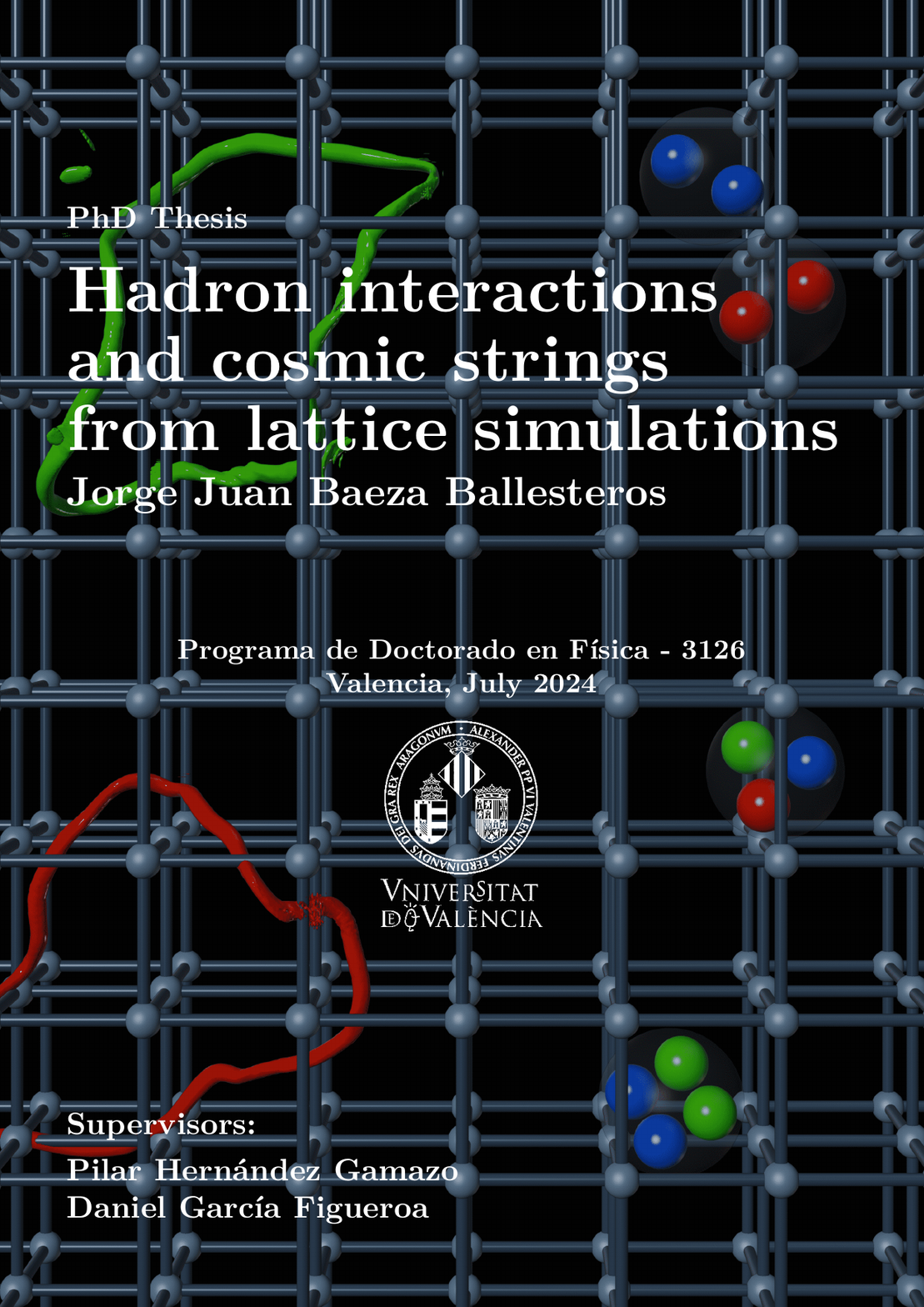}

%%%%%%%%%%%%%%%%% Portada interior %%%%%%%%%%%%%%%%%%%%%%%%%%

\setlength{\unitlength}{1cm} %Especificar unidad de trabajo
\thispagestyle{empty}
\begin{center}
\cleardoublepage

\begin{center}
\includegraphics[scale=0.07]{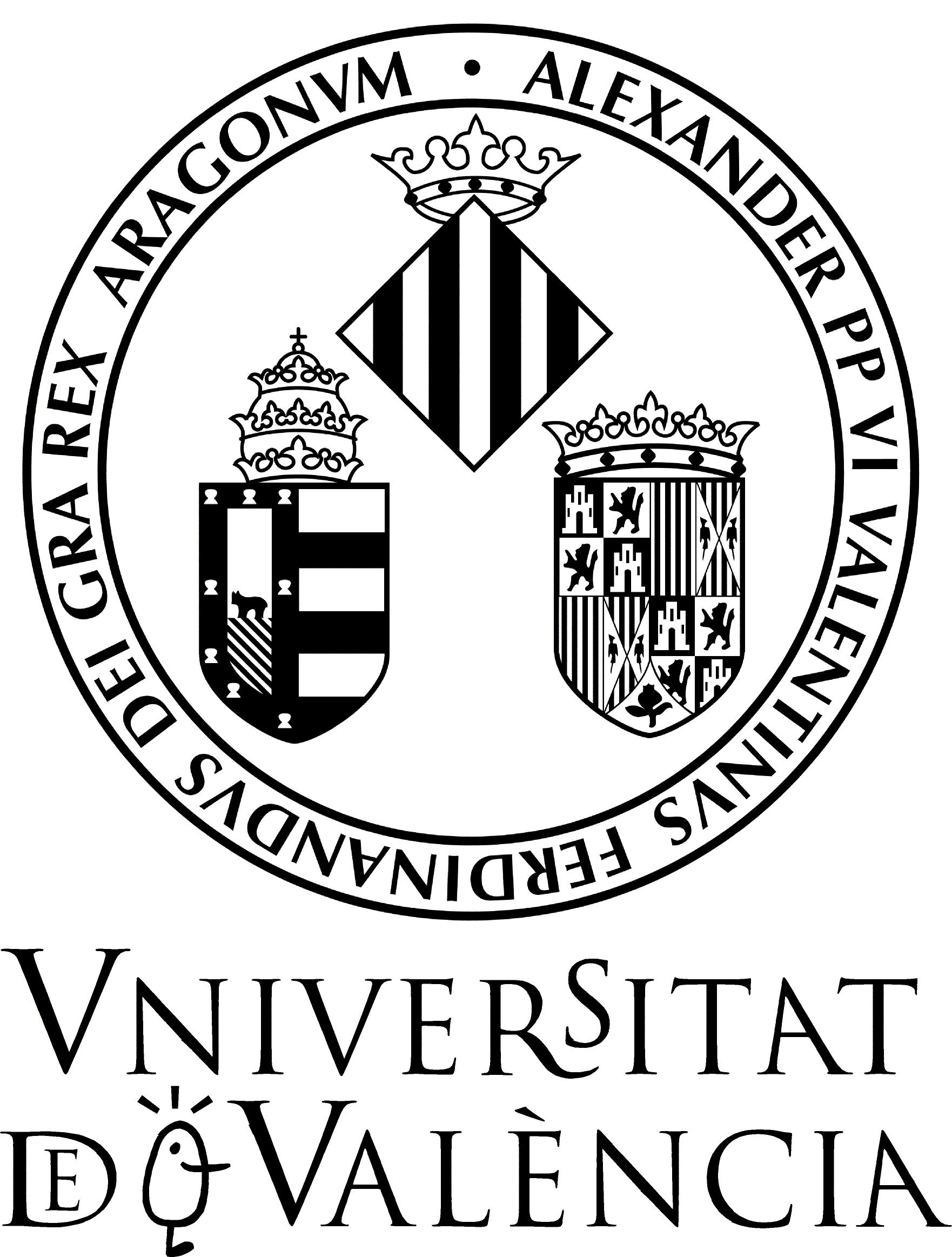}
\end{center}

\vspace{1cm}

\textbf{\huge{Hadron interactions\\[0.3ex] and cosmic strings\\[0.3ex] from lattice simulations}\\ \vspace{1.6cm}
{\large PhD Thesis by}}

\vspace{0.12cm}

{\Large \textbf{Jorge Juan Baeza Ballesteros}} \\[6ex]
{\large IFIC - Universitat de Val\`{e}ncia - CSIC}\\[0.5ex]
{\large Departament de F\'isica Te\`orica}\\[0.5ex]
{\large Programa de Doctorado en F\'isica - 3126 }\\[6ex]

{ \textbf{Under the supervision of}}\\[2ex]

 {\large \textbf{M Pilar Hernández Gamazo}} \\[0.7ex]
 {\large \textbf{Daniel García Figueroa}}

\vspace{1.5cm}

{\large \bf{Valencia, July 2024}}

\end{center}

%%%%%%%%%%%%%%%%%%%%%%%%%%%%%%%%%%%%%%%%%%%%%%%%%%%%%%%%%%%%

%%%%%%%%%%%%%%%%%%%%%%%%%%%%%%%%%%%%%%%%%%%%%%%%%%%%%%%%%%%%
\begin{titlepage}
\cleardoublepage
\thispagestyle{plain}

%%%%%%%%%%%%%%%%%%%%%%%%%%%%%%%%%%%%%%%%%%%%%%%%%%%%%%%%%%%%
% CERTIFICATION PAGE
%%%%%%%%%%%%%%%%%%%%%%%%%%%%%%%%%%%%%%%%%%%%%%%%%%%%%%%%%%%%
\thispagestyle{empty}
%%%%%%%%%%%%%%%%%%%%%%%%%%%%%%%%%%%%%%%%%%%%%%%%%%%%%%%%%%%%

\newpage
\thispagestyle{empty}

\newpage
\thispagestyle{empty}

\end{titlepage}
%%%%%%%%%%%%%%%%%%%%%%%%%%%%%%%%%%%%%%%%%%%%%%%%%%%%%%%%%%%%

\frontmatter
\cleardoublepage
%%%%%%%%%%%%%%%%%%%%%%%%%%%%%%%%%%%%%%%%%%%%%%%%%%%%%%%%%%%%
% Acknowledgements
%%%%%%%%%%%%%%%%%%%%%%%%%%%%%%%%%%%%%%%%%%%%%%%%%%%%%%%%%%%%
\hypersetup{urlcolor=black}
\chapter*{Agradecimientos}
\addcontentsline{toc}{chapter}{Agradecimientos}  

Cuando pienso que ya han pasado cuatro años desde que comenzó esta aventura por el mundo de la física, siento una mezcla de sorpresa y melancolía. Parece que fue ayer cuando empezó y, sin embargo, he vivido un sinnúmero de experiencias y adquirido una infinidad de conocimientos. A lo largo de este camino, muchas personas me han acompañado, sin las cuales no me habría sido posible llegar hasta aquí. %Os estoy enormemente agradecido.

A Pilar y Dani, mis directores, quienes me han guiado en este viaje y me han enseñado a pensar como un verdadero científico, quienes han estado ahí para aconsejarme cuando necesitaba ayuda y para mostrarme lo divertida que puede ser la investigación, quienes me han dado libertad para estudiar nuevas ideas y me han apoyado siempre. Pilar, me has transmitido esa visión crítica que es tan importante para un científico y siempre me has animado a explorar nuevos horizontes, sin dejar de preocuparte por mi progreso.   
Dani, de ti he aprendido la importancia de fijarme en los detalles y, al mismo tiempo, no perder de vista la relevancia global de cada resultado.
A los dos os estoy enormemente agradecido, y echaré realmente de menos ser vuestro estudiante de doctorado.

A Fernando, que ha sido un mentor para mí todos estos años y sin quien posiblemente este viaje nunca habría tenido lugar. Aún recuerdo cuando me contaste de qué iba la investigación en lattice. Ese fue, posiblemente, el punto de partida de mi camino como científico. Desde que nos conocemos siempre has estado ahí para darme consejos, no solo en el plano científico, sino también en el personal. Muchas gracias.

A Max, Steve, Raúl, André y Mark, por hacerme sentir a gusto en todos mis viajes, por proponerme nuevas ideas y enseñarme a plantear los problemas desde únicas perspectivas. Jamás olvidaré todo lo que he aprendido con vosotros y  los buenos momentos que he vivido alrededor del mundo.

Al resto de mis colaboradores: Hans, Mattias, Tomá{\v s}, Joanes, Ed y Andrea, con quienes he descubierto un sinfín de nuevos enfoques desde los que enfrentarme a los diferentes obstáculos que aparecen en el mundo de la ciencia.

A mis compañeros de SOM, especialmente a David y Nico, por todos los buenos momentos en el IFIC y fuera de él.

A mis padres, mis hermanos y mi abuela, que siempre han estado ahí para apoyarme y aconsejarme. Gracias a vosotros he crecido hasta ser quien soy.

Por último, a Lucía, que me ha acompañado durante todo este viaje, que ha estado a mi lado desde mucho antes que comenzara y que seguirá ahí en todas las aventuras venideras. Me has dado ánimo cuando lo necesitaba, distracción cuando no era capaz relajarme y cariño cuando me sentía perdido. Has escuchado atentamente mil historias sobre mi trabajo y me has aconsejado sobre cómo afrontar cada nuevo día. Sin ti no habría llegado hasta donde estoy hoy.

Muchas gracias a todos por estar a mi lado y por haberme acompañado en esta divertida aventura.

{\hfill Valencia, julio 2024}
\hypersetup{urlcolor=black}

%%%%%%%%%%%%%%%%%%%%%%%%%%%%%%%%%%%%%%%%%%%%%%%%%%%%%%%%%%%%
% Abbreviations
%%%%%%%%%%%%%%%%%%%%%%%%%%%%%%%%%%%%%%%%%%%%%%%%%%%%%%%%%%%%

\chapter*{List of publications}
\thispagestyle{plain} 
\addcontentsline{toc}{chapter}{List of publications}
\markboth{List of publications}{}

This doctoral thesis is based on the following publications and works in preparation, listed in the order they appear in the dissertation:
\begin{itemize}
\item[\cite{Baeza-Ballesteros:2022azb}] J. Baeza-Ballesteros, P. Hernández and F. Romero-López, \textit{A lattice study of $\pi\pi$ scattering at large $\Nc$}, JHEP 06 (2022) 049.
\item[\cite{Baeza-Ballesteros:largeNinprep}] J. Baeza-Ballesteros, P. Hernández and F. Romero-López, \textit{Meson-meson scattering at large $\Nc$}, 2024 (in preparation).
\item[\cite{Baeza-Ballesteros:2023ljl}] J. Baeza-Ballesteros, J. Bijnens, T. Husek, F. Romero-López, S. R. Sharpe and M. Sjö, \textit{The isospin-3 three-particle $K$-matrix at NLO in ChPT}, JHEP 05 (2023) 187.
\item[\cite{Baeza-Ballesteros:2024mii}] J. Baeza-Ballesteros, J. Bijnens, T. Husek, F. Romero-López, S. R. Sharpe and M. Sjö, \textit{The three-pion $K$-matrix at NLO in ChPT}, JHEP 03 (2024) 048.
\item[\cite{Baeza-Ballesteros:O3inprep}] J. Baeza-Ballesteros and M. T. Hansen, \textit{Two- and three-particle scattering in the (1+1)-dimensional O(3) non-linear sigma model}, 2024 (in preparation).
\item[\cite{Baeza-Ballesteros:2023say}] J. Baeza-Ballesteros, E. J. Copeland, D. G. Figueroa and J. Lizarraga, \textit{Gravitational wave emission from a cosmic string loop: Global case}, Phys. Rev. D 110 (2024) 043522.
\item[\cite{Baeza-Ballesteros:stringsinprep}] J. Baeza-Ballesteros, E. J. Copeland, D. G. Figueroa and J. Lizarraga, \textit{Gravitational Wave Emission from a Cosmic String Loop, II: Local Case}, 2024 (submitted  to Phys. Rev. Lett.).
\end{itemize}\newpage

The research presented also appeared in conference proceedings:
\begin{itemize}
\item[\cite{Baeza-Ballesteros:2021nxu}] J. Baeza-Ballesteros, P. Hernández and F. Romero-López, \textit{$\pi\pi$ scattering at Large $\Nc$}, PoS { LATTICE2021} (2022) 309.
\item[\cite{Baeza-Ballesteros:2022bsn}] J. Baeza-Ballesteros and M. T. Hansen, \textit{Two- and three-particle scattering in the (1+1)-dimensional O(3) non-linear sigma model}, PoS {LATTICE2022} (2023) 050.
\item[\cite{Baeza-Ballesteros:2024ogp}] J. Baeza-Ballesteros, P. Hernández and F. Romero-López, \textit{Progress in meson-meson scattering at large $\Nc$}, PoS { LATTICE2023} (2024) 059.
\end{itemize}

The author has also participated in the development of the \CosmoLattice package:
\begin{itemize}
\item[\cite{GWmodule:2022}] J.Baeza-Ballesteros, D. G. Figueroa, A. Florio and N. Loayza Romero, \textit{CosmoLattice Technical Note II: Gravitational Waves}, 2022.
\item[\cite{GWmodule:2023}] J.Baeza-Ballesteros, D. G. Figueroa and N. Loayza Romero, \textit{CosmoLattice Technical Note III: Gravitational Waves from U(1) gauge theories}, 2023.
\end{itemize}

Finally, other published works to which the author contributed, that are not presented in this dissertation, are:
\begin{itemize}
\item[\cite{Baeza-Ballesteros:2021tha}] J. Baeza-Ballesteros, A. Donini and S. Nadal-Gisbert, \textit{Dynamical measurements of deviations from Newton's $1/r^2$ law}, Eur. Phys. J. C 82 (2022) 154.
\item[\cite{Baeza-Ballesteros:2023par}] J. Baeza-Ballesteros, A. Donini, G. Molina-Terriza, F. Monrabal and A. Simón, \textit{Towards a realistic setup for a dynamical measurement of deviations from Newton's $1/r^2$ law: the impact of air viscosity}, Eur. Phys. J. C 84 (2024) 596
\end{itemize}
\chapter*{Abbreviations and conventions}
\markboth{Abbreviations}{Abbreviations and conventions}
\addcontentsline{toc}{chapter}{Abbreviations and conventions}  
\def\arraystretch{1.7}

\vspace{0.1cm} Throughout this dissertation, we made use of a number of abbreviations. They are summarized below, separated in two groups depending on whether they appear in the first or the second part of this doctoral thesis:\vspace{0.3cm}

\begin{tabular}{llll}
\multicolumn{4}{l}{\normalsize\textbf{Part I - Hadron interactions from lattice QCD}} \\
BH		& Bull's head & 
NLO 		& Next-to-leading order \\
ChPT		& Chiral perturbation theory &
NNLO 	& Next-to-next-to-leading order \\ 
CMF		& Center-of-mass frame & 
OPE		& One-particle exchange \\
df		& Divergence free		&
pNGB 	& Pseudo-Nambu-Goldstone bosons \\
DD		& Doublet-doublet &
PV		& Principal value \\
DS		& Doublet-singlet &
RChT		& Resonant chiral theory \\
dof 		& Degrees of freedom &
QC		& Quantization condition \\ 
EFT		& Effective field theory &
QCD 		& Quantum chromodynamics \\
ERE		& Effective range expansion &
QED		& Quantum electrodynamics \\ 
GEVP		& Generalized eigenvalue problem &
RFT		& Relativistic field theory \\
IAM 		& Inverse amplitude method &
SM 		& Standard model \\
irrep	& Irreducible representations &
$s$-OPE	& $s$-channel one-particle exchange \\
LEC		& Low-energy constant & 
SD		& Sinlget-doublet \\
LO		& Leading order &
SS		& Singlet-singlet \\ 
NGB 		& Nambu-Goldstone bosons & &

\end{tabular}

\begin{tabular}{llll}
\multicolumn{4}{l}{\normalsize\textbf{Part II - Cosmic string loops from lattice simulations}} \\
CMB		& Cosmic microwave background & 
NG		& Nambu-Goto \\
FLRW 	& Friedmann-Lemaître & & \\[-8pt]
 &-Robertson-Walker &
 RD		& Radiation domination \\
GW		& Gravitational wave &
TT		& Transverse-traceless \\
GWB		& Gravitational wave background &
UV		& Ultraviolet  \\
IR		& Infrared &
VOS 		& Velocity-dependent one-scale \\[0.5cm]
\end{tabular}

As of conventions, we work in natural units ($c=\hbar=1$) and use the mostly-minus signature for the metric. This means that the metric of Minkowski spacetime is $g_{\mu\nu}=\text{diag}(+1,-1,-1,-1)$. When indicating components of Minkowski vectors, we use Greek indices to refer to all its components, while Latin indices refer only to the spacial ones. We assume repeated indices in equations are summed over, unless otherwise stated. Also, spacial vectors are indicated using bold variables. 

%If no range is indicated in sums, it is assummed that it runs over all possible values of the index. For example, if $\bm{n}$ indicates a lattice coordinate, $\sum_{\bm{n}}$ runs over all possible values $\bm{n}$ can take in the lattice.

When considering matrix quantities describing scattering processes, in general, column indices refer to the initial state, while row indices correspond to the final state. Similarly, arguments related to the final state are indicated first than those related to the initial state, and time flows from right to left in Feynman diagrams. 

Lastly, we use NLO to refer only to the correction to the LO, rather than the full LO+NLO result. An analogous criteria is also used to denote higher orders in perturbation theory.

%%%%%%%%%%%%%%%%%%%%%%%%%%%%%%%%%%%%%%%%%%%%%%%%%%%%%%%%%%%%
% Preface
%%%%%%%%%%%%%%%%%%%%%%%%%%%%%%%%%%%%%%%%%%%%%%%%%%%%%%%%%%%%

\chapter*{Preface}
\addcontentsline{toc}{chapter}{Preface}
\markboth{Preface}{}

Field theory is the basic framework to describe high-energy physics. The Standard Model of particle physics is the paradigmatic example of a quantum field theory, which is the synthesis of quantum mechanics and special relativity. It describes three of the fundamental forces of nature---the electromagnetic, the weak and the strong forces---and has been successful at describing fundamental phenomena at an astonishing level of precision. Certain phenomena of the early universe can also be described by field theories, such as inflation, phase transitions and cosmic defects. In many cases, due to large occupation numbers, these phenomena can be investigated using classical-field-theory techniques.  %Another example of field theory is the standard cosmological mode. This is a classical field theory fundamented on the cosmological principle and the basis of general relativity, which has also succeeded at predicting many observed phenomena, such as the formation of the cosmic microwave background or the relative abundance of chemical elements.

In many regimes, quantum and classical field theories can be described by means of analytical techniques. Within the Standard Model, this is for example the case of quantum electrodynamics, which describes the electromagnetic interactions. The small size of the fine structure couling, $e^2/4\pi\approx 1/137$, allows for a perturbative expansion in powers of this quantity. Many processes occurring during the history of our universe can also be investigated analytically, by treating fluctuations of the metric or the matter fields as small perturbations.

However, the use of analytical techniques to study field theories is not always feasible. In the case of quantum chromodynamics (QCD), the theory that describes the strong interaction, the phenomenon of asymptotic freedom prevents perturbative techniques from being applicable at low energies, where the strong coupling grows large. This precludes analytical predictions, for instance, of the hadron spectrum. This is particularly true in relation to the existence and properties of resonances, which are hadronic states that only exist virtually during scattering processes. A longstanding question is whether QCD can predict the presence of such particles, which requires the use of non-perturbative techniques.

Analytical methods also fail to describe non-perturbative and non-linear phenomena that may take place in the early universe. This is, for instance, the case of particle production during and after inflation and also the case of the observable signatures of early-universe processes, such as the emission of gravitational waves (GWs). Making reliable predictions of these phenomena is of vital importance in view of current and projected GWs experiments. The detection of a background of GWs emitted by processes taking place during the early universe will open the door to the study of high-energy physics at scales well above those reachable by ground-based accelerators. However, a correct interpretation of a possible signal depends on reliable theoretical predictions, which in turn requires of alternative techniques to capture the non-linear dynamics.

The most successful approach to investigate beyond-analytical regimes in quantum and classical field theories is the use of lattice techniques. These are based on the formulation of the field theory on a discretized and finite volume, which allows to obtain predictions from numerical simulations. In the case of QCD, Monte Carlo methods allow to evaluate the path integral and extract information about physical observables, such as hadron masses or multiparticle scattering amplitudes. Regarding early-universe phenomena, the evolution of matter fields and the emission of GWs can be determined from the numerical solution of the equations of motion. %by studying numerically the corresponding equations of motion. %In many cases, these techniques can be complemented, for instance, by the use of effective models to describe the non-perturbative phenomena.

\begin{center}
%\pgfornament[width=0.4\textwidth]{89}
\large\adforn{21}\quad\adforn{11}\quad\adforn{49}
\end{center}

This doctoral thesis is devoted to the study of hadron scattering in QCD and the emission of particles and GWs from cosmic string loops that may arise in the early universe, using numerical lattice simulations. The dissertation consists of ten chapters, organized in two independent parts. 

\Cref{part:QCD} focuses on the study of two- and three-hadron scattering using lattice QCD techniques, complemented by analytical results from  chiral perturbation theory and the large $\Nc$ limit. The fundamentals of QCD are introduced in \cref{sec:QCD}, where lattice QCD, effective field theories and the large $\Nc$ limit are also discussed. The basics of multiparticle scattering are introduced in \cref{sec:hadrons}, with special emphasis on the quantization conditions that permit to study such processes in a finite volume.

The following five chapters present the results of my doctoral work on lattice QCD. They can be grouped in three separate lines of research. \Cref{sec:largeNpions,sec:largeNmesons} deal with the investigation of the $\Nc$ dependence of meson-meson scattering observables, and are mainly based on \rcite{Baeza-Ballesteros:2022azb} and ongoing work~\cite{Baeza-Ballesteros:largeNinprep}. \Cref{sec:pipipiKmatrix,sec:isospinKmatrix} focus on the study of three-pion interactions using chiral perturbation theory. More concretely, they summarize the results from \rrcite{Baeza-Ballesteros:2023ljl,Baeza-Ballesteros:2024mii}, where we determined the three-pion $K$-matrix up to next-to-leading order in chiral perturbation theory. Finally, \cref{sec:O3model} is devoted to the study of two- and three-particle interactions in the O(3) non-linear sigma model, commonly used as a toy model of QCD, with the aim of testing the relativistic-field-theory three-particle finite-volume formalism~\cite{Baeza-Ballesteros:O3inprep}.

\Cref{part:strings} of this thesis is dedicated to the study of the dynamics and GW emission from cosmic string loops using classical-field-theory lattice simulations. \Cref{sec:Cosmo} introduces the basics of classical field theory in an expanding universe, and explains how lattice simulations can be used to study the non-linear field dynamics, as well as the associated GW emission. \Cref{sec:global,sec:local} are focused on the application of these techniques to characterize the evolution and GW emission of decaying cosmic string loops. In particular, these two chapters present, respectively, the results from \rcite{Baeza-Ballesteros:2023say} for the case of global loops, and results from \rcite{Baeza-Ballesteros:stringsinprep} for local loops.

%%%%%%%%%%%%%%%%%%%%%%%%%%%%%%%%%%%%%%%%%%%%%%%%%%%%%%%%%%%%
% CONTENTS
%%%%%%%%%%%%%%%%%%%%%%%%%%%%%%%%%%%%%%%%%%%%%%%%%%%%%%%%%%%%
%\fancyfoot[C]{} %Do not have the number down anymore
\hypersetup{linkcolor=black} %Links in black for the contents only
\renewcommand{\baselinestretch}{0.76}\normalsize

\input{TesisSave.toc}
%\tableofcontents
\renewcommand{\baselinestretch}{1.0}\normalsize

\definecolor{linkcolour}{rgb}{0.85,0.15,0.15}
\hypersetup{linkcolor=linkcolour}\clearpage

\mbox{}\thispagestyle{empty}\clearpage

\mainmatter % To numerate with 1,2,3 ...
\renewcommand{\headrulewidth}{0.5pt} % To get the header line only from now on!

\lhead[{\bfseries \thepage}]{ \rightmark}
\rhead[ Chapter \thechapter. \leftmark]{\bfseries \thepage}

\titleformat{\part}
{\thispagestyle{empty}\vspace{10pt}\normalfont\huge}
{\tikzset{external/export next=false}
\begin{tikzpicture}[remember picture, overlay]
\settowidth{\imagewidth}{\includegraphics{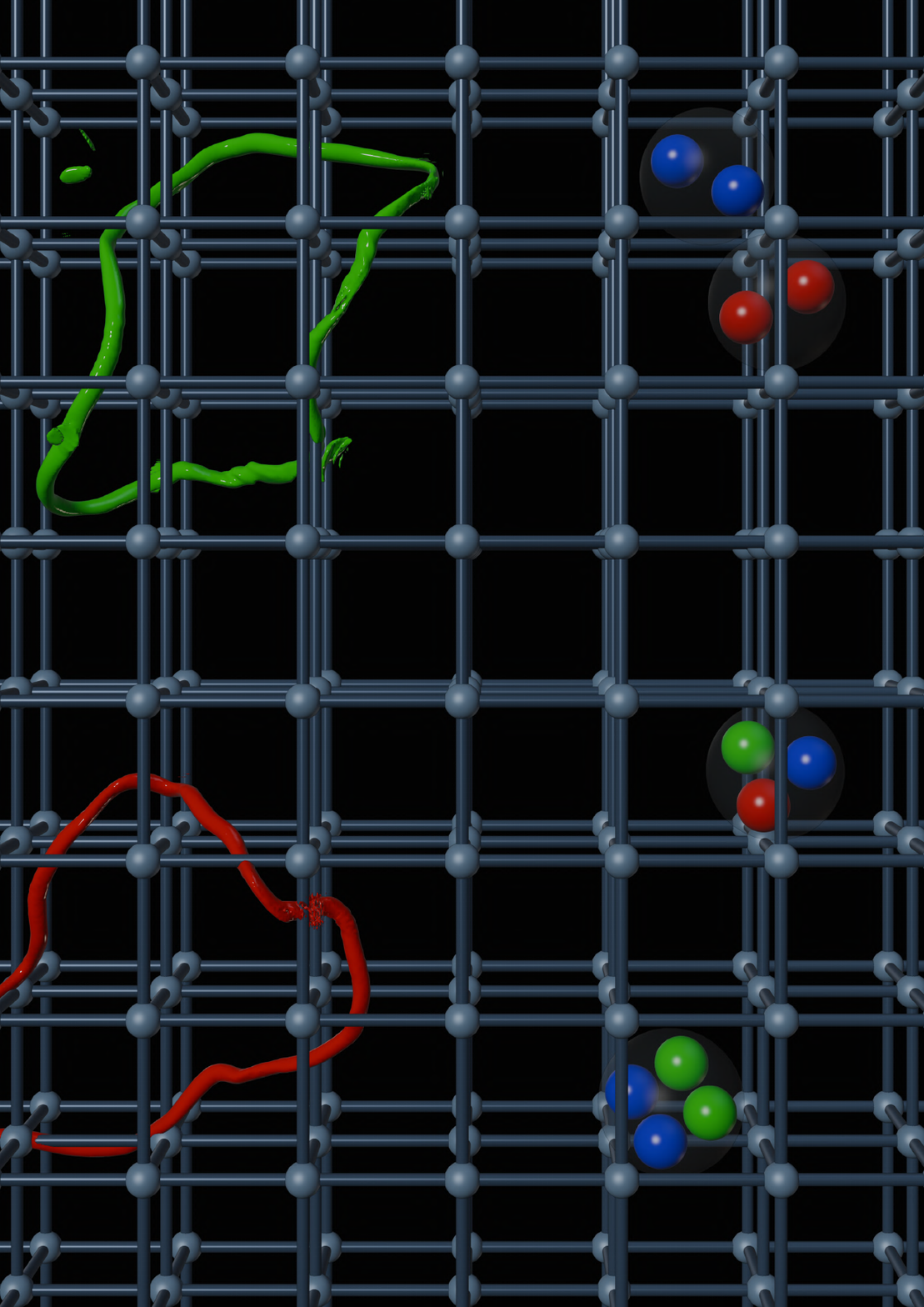}}
\node[anchor=west, inner sep=0pt] at (current page.west) {\includegraphics[width=0.4\paperwidth, height=1.03\paperheight, trim = {0.58\imagewidth{} 0 0.02\imagewidth{} 0}, clip=true]{Portada2.pdf}};
\node[align=left] at (10.5,5) {\partitlefont Part I \\[1.5ex] \partitlefont Hadron Interactions \\[0.5ex] \partitlefont from Lattice QCD};
\node[align=left, scale=0.6, font=\it] at (11.3,-8) {Tu voz pudo enternecerme,\\ tu presencia suspenderme,\\ y tu respeto turbarme.\\ ¿Quién eres? Que aunque yo aquí\\ tan poco del mundo sé, [...].};
\node[align=right, scale=0.6] at (13,-10.8) {\it La vida es sueño\\ Calderón de la Barca};
\end{tikzpicture}\filleft{\parbox[b][][b]{.38\textwidth}{$\,$}}}
{0pt}
{\Huge{\parbox[b][][t]{.83\textwidth}{\flushleft\hyphenpenalty=10000\partitlefont\phantom{QCD}\vspace{0.0cm}}}}[\vspace{0.5ex}\vspace{-0.7cm}]

\fancyfoot[C]{}
\part{Hadron interactions from lattice QCD}\label{part:QCD}
\chapter{Hadrons and quantum chromodynamics}
\label{sec:QCD}

The strong interaction is one of the four fundamental forces of nature. %, the others being the electromagnetic, weak and gravitational forces. The strong force
Its name comes from its strength: it is responsible for binding protons and neutrons in atomic nuclei, vastly overcoming the effect of electric repulsion. Quantum chromodynamics (QCD) is the quantum field theory that describes the strong force, and combined with the theory of electroweak interactions, conforms the Standard Model of particle physics.

The tale of the  strong interaction is now almost a century long.\footnote{See for example \rcite{Griffiths:111880} for a detailed history of strong interactions.} It started back in 1935 when Yukawa first proposed it as a new force responsible for binding protons and neutrons in atomic nuclei~\cite{Yukawa:1935xg}. He predicted that this force would be mediated by some massive particles, which he named mesons, of a mass around $100$ MeV. In 1947, the $\pi$ meson was discovered~\cite{Lattes:1947mw}, earning him a Nobel Prize in physics two years later. 
%The force, according to Yukawa, was mediated by some massive particles, which he named mesons, due to their predicted mass lying between that of electrons and protons. 
%The pi meson was eventually discovered in 1947, earning him the Nobel Prize in Physics two years later. 

Following this finding, a myriad of new subatomic particles were \mbox{discovered} in particle accelerators in the 1950s and 1960s, which were initially thought to be fundamental. It was soon realized by Gell-Mann~\cite{Gell-Mann:1961omu} and Ne'eman~\cite{NEEMAN1961222} independently that these particles could be organized in simple geometrical patterns according to their isospin, strange and charm quantum numbers, in what was called the \textit{Eightfold way}. This classification lead Gell-Mann and Zweig to propose the quark model~\cite{Gell-Mann:1964ewy,Zweig:1964jf} according to which all known subatomic particles were indeed composed of other more fundamental spin-$1/2$ ones, the quarks, $q$. They distinguished two types of composite particles, also known as \textit{hadrons}: \textit{mesons}, formed by a quark and an antiquark, $\overline{q}q$, and \textit{baryons}, composed of three quarks, $qqq$.

The quark model found rapid success, as it made possible to predict the existence of new hadrons that were observed in experiment shortly after~\cite{Gell-Mann:1953hzm,Barnes:1964pd}. The model also implied that, if quarks were to obey Pauli statistics, they should have an additional quantum number, which was named \textit{color}~\cite{Greenberg:1964pe,osti_6283678,GlashowSciAm1975}. This led to the proposal of QCD as the fundamental theory governing strong interactions~\cite{Fritzsch:1973pi}. As will be explained in this chapter, this is a non-Abelian SU(3) gauge theory with quarks and gluons as fundamental particles. These particles, however, can not be found as isolated states in nature. Due to the properties of asymptotic freedom and confinement, only colorless hadrons are observed. 

Fifty years after the proposal of QCD, it is still one of the most active fields of study within high energy physics. On the theory side, confinement and asymptotic freedom preclude the application of standard perturbative methods to study the hadron spectrum of QCD. Still, other alternative approaches exist, such as lattice QCD, the 't Hooft limit or the use of effective field theories, that make it possible to investigate the low-energy regime of QCD. On the experimental side, new composite particles are continuously being discovered and characterized. During the last decades some new states have been observed with properties that cannot be explained by a meson or baryon structure, but are expected to have a more exotic nature, such as tetraquarks~\cite{Belle:2003goh,Maiani:2004vq,LHCb:2020bls,LHCb:2020pxc,LHCb:2022sfr,LHCb:2022lzp,PDG:2020}, composed of two quarks and two antiquarks, $\overline{q}q\overline{q}q$. 

In this chapter the basics of QCD are reviewed, together with some of the most successful techniques that allow one to study its low-energy regime. In \cref{sec:QCD:introduction}, the theory of strong interactions is introduced and some of its main features are discussed. \Cref{sec:QCD:lattice} presents the lattice regularization of QCD, arguably the most successful first-principles approach to its low-energy regime. Finally, in \cref{sec:QCD:largeNc,sec:QCD:EFTs} two other techniques are considered: the limit of large number of colors and chiral perturbation theory, respectively.

%Since the proposal of QCD, many new hadrons have been discovered at characterized at experiments, and all of them could be organized following the principles of the quark model. Or at least that was the case until start of the early 21${}^\text{st}$ century, when some new particles that did not fit into this model began being found in particle accelerators. These are the so called exotic hadrons, which include tetraquarks, composed of two quarks and two antiquarks ($\overline{q}q\overline{q}q$), pentaquarks, formed by four quarks and one antiquark ($\overline{q}qqqq$) and hybrid states including a valence gluon ($g\overline{q}q$). {\jorge [Tengo que encontrar un lugar mejor para comentar esto y hablar sobre molecular states y los papers de Polosa.]}

%In this part of this thesis we ...
%Following the proposal of quantum chromodynamics 

% All hadrons (particles composed of quarks) appear a color singlets in nature. QCD has since become one of the building stones of the Standard Model. 

\section{Quantum chromodynamics}\label{sec:QCD:introduction}

Quantum chromodynamics is the quantum field theory that describes the strong interactions between quarks and gluons. It is a non-Abelian gauge theory based on a SU($\Nc$) Lie group, where $\Nc$ is the number of colors, which is the charge associated to the strong interaction. In nature, it can take three different values, $\Nc=3$: red, green and blue. 

The matter content of the theory are quarks and gluons. Quarks are represented by spin-1/2 fermion fields, $q(x)$, which transform under the fundamental irreducible representation (irrep) of SU($\Nc$) (while antiquarks, $\overline{q}(x)$, transform in the antifundamental one). They contain both spin and color indices, which are left implicit. In nature, quarks appear in $\Nf=6$ different flavors (up, down, strange, charm, bottom and top), which are distinguished by their couplings to the Higgs field. %having a different mass  and different interaction properties with the electroweak sector of the Standard Model. 

The second component of QCD are gluons, the mediators of the strong force. They are represented by a gauge field, $A_\mu(x)$, which transforms under the adjoint representation of the color group. This field can be expressed as a linear combination of the generators of the $\mathfrak{su}$($\Nc$) Lie algebra---the Gell-Mann matrices, $t^a$---with real coefficients, $A_\mu=A_\mu^at^a$, where the color indices are left implicit.

The dynamics of quarks and gluons are described by the QCD action,
\begin{equation}\label{eq:QCD:QCDactionMinkowski}
S_\text{QCD}[q,\overline{q},A]=\int\text{d}^4x\, \mathcal{L}_\text{QCD}[q,\overline{q}, A]\,,
\end{equation}
where the Lagrangian\footnote{We will abuse language and refer to the Lagrangian density simply as the Lagrangian.} is integrated over the whole Minkowski spacetime, $x=(t,\bm{x})$.%, and we use the mostly-plus metric, $g_{\mu\nu}=\text{diag}(-1,+1,+1,+1)$. 
The Langrangian density of QCD can be separated in two parts,
\begin{equation}\label{eq:QCD:QCDlagrangianMinkowski}
\mathcal{L}_\text{QCD}[q,\overline{q}, A]=\mathcal{L}_\text{F}[q,\overline{q}, A]+\mathcal{L}_\text{G}[A]\,.
\end{equation}
The first one contains all the fermionic terms, including their coupling to gluons,
\begin{equation}\label{eq:QCD:fermionlagrangianMinkowski}
\mathcal{L}_\text{F}[q,\overline{q},A]=\sum_{f=1}^{\Nf}\overline{q}_f\left(i\gamma_\mu D^\mu-m_f\right)q_f\,,
\end{equation}
where $\gamma_\mu$ are the Dirac gamma matrices, $\overline{q}=q^\dagger\gamma_0$, $m_f$ is the bare quark mass for flavor $f$, and 
\begin{equation}\label{eq:QCD:covariantderivativeQCD}
D_\mu=\partial_\mu-igA_\mu\,,
\end{equation}
is the covariant derivative, with $g$ the gauge coupling that characterizes the strength of the interactions between quarks and gluons. Note that we have written \cref{eq:QCD:fermionlagrangianMinkowski} as a sum over an arbitrary number of flavors, $\Nf$. While $\Nf=6$ in nature, it is convenient to keep the dependence on $\Nf$ explicit.

The second term of \cref{eq:QCD:QCDlagrangianMinkowski} contains the kinetic term for the gauge field and its self-interactions. It takes the simple form
\begin{equation}\label{eq:QCD:gaugelagrangianMinkowski}
\mathcal{L}_\text{G}[A]=-\frac{1}{2}\text{tr}\left[F_{\mu\nu}F^{\mu\nu}\right]\,,
\end{equation}
where the trace runs over the color indices and $F_{\mu\nu}=F_{\mu\nu}^a t^a$, with
\begin{equation}\label{eq:QCD:fieldstrengthtensorQCD}
F_{\mu\nu}^a=\partial_\mu A_\nu^a-\partial_\nu A_\mu^a +if^{abc}A_\mu^b A_\nu^c\,,
\end{equation}
is the field-strength tensor of the gauge field. Here we have introduced the structure constants of the $\mathfrak{su}(\Nc)$ algebra, $f^{abc}$ defined from
\begin{equation}\label{eq:QCD:structureconstants}
[t^a,t^b]=i f^{abc}t^c\,.
\end{equation}
As SU(3) is a non-Abelian group, $f^{abc}\neq0$ and so the QCD Lagrangian contains cubic and quartic self-interaction vertices for the gluon.

The action in \cref{eq:QCD:QCDactionMinkowski} is invariant under local gauge transformations,
\begin{equation}\label{eq:QCD:gaugetransformationsQCD}
\begin{array}{rl}
q(x)\rightarrow & \Omega(x) q(x)\,,\\
A_\mu(x)\rightarrow & \displaystyle\Omega(x) A_\mu(x) \Omega(x)^\dagger +\frac{i}{g}\partial_\mu\Omega(x) \Omega(x)^\dagger\,,
\end{array}
\end{equation}
where $\Omega(x)\in\text{SU}(3)$. In addition, the QCD action is invariant under $CP$ transformations, where $C$ refers to charge conjugation and $P$ to parity inversions. These two discrete symmetries act on the quark and gluon fields,
\begin{equation}\label{eq:CPtransformationsQCD}
\begin{array}{rlcrl}
q(t,\bm{x})\xrightarrow[]{\phantom{P}P\phantom{P}}& \gamma_0 q(t,-\bm{x})\,, &\quad\quad& q(t,\bm{x})\xrightarrow[]{\phantom{P}C\phantom{P}}& C^{-1} \overline{q}^\intercal(t,\bm{x})\,,\\
\overline{q}(t,\bm{x})\xrightarrow[]{\phantom{P}P\phantom{P}}&  \overline{q}(t,-\bm{x})\gamma_0\,, &\quad\quad& \overline{q}(t,\bm{x})\xrightarrow[]{\phantom{P}C\phantom{P}}& -q^\intercal(t,\bm{x}) C\,,\\
A_0(t,\bm{x})\xrightarrow[]{\phantom{P}P\phantom{P}}& A_0(t,-\bm{x})\,, &\quad\quad&  A_0(t,\bm{x})\xrightarrow[]{\phantom{P}C\phantom{P}}& - A_0^\intercal(t,\bm{x})\,,\\
A_i(t,\bm{x})\xrightarrow[]{\phantom{P}P\phantom{P}}& -A_i(t,-\bm{x})\,, &\quad\quad& A_i(t,\bm{x})\xrightarrow[]{\phantom{P}C\phantom{P}}& - A_i^\intercal(t,\bm{x})\,,\\
\end{array}
\end{equation}
where ${X}^\intercal$ denotes the transpose matrix of $X$, and $C$ is a matrix acting on spin indices defined implicitly via the relation $C\gamma_\mu C^{-1}=-\gamma_\mu^\intercal$.

It is possible to include one further term in the QCD Lagrangian in \cref{eq:QCD:QCDlagrangianMinkowski} which, while allowed by gauge symmetry, does not preserve $CP$ invariance. This is the so called $\theta$-\textit{term},
\begin{equation}\label{eq:thetalagrangianMinkowski}
\mathcal{L}_\theta[A]=-\theta\frac{g^2\Nf}{32\pi^2}\text{tr}\left[F_{\mu\nu}\tilde{F}^{\mu\nu}\right]\,,
\end{equation}
where $\tilde{F}^{\mu\nu}=\epsilon^{\mu\nu\rho\sigma}F_{\mu\nu}$ is the dual tensor of the field-strength tensor and $\theta$ is an unknown phase. This term can be written as a total derivative, and so does not contribute to perturbation theory at any order. Still, its integral over the whole spacetime takes an integer value, known as the \textit{topological charge}. Gauge configurations are classified in topological sectors, which contribute to QCD observables beyond perturbation theory. %Pilar sugiere "to the path integral", que no he mencionado aún.

In addition, as already mentioned, a non-zero $\theta$ parameter implies the violation of the $CP$ symmetry by the strong interactions. In particular, a neutron dipole moment would be induced. The experimental measurement of this observable make it possible to constrain $\theta<10^{-10}$~\cite{Baker:2006ts,Chupp:2017rkp}. The small size ofthe  measured $\theta$ angle is known as the \textit{strong $CP$ problem}. During the years, many different proposals have been made to solve it. One of the most intersecting ones is the inclusion of a new spontaneously broken U(1) symmetry---the \textit{Peccei-Quinn symmetry}~\cite{Peccei:1977hh,Peccei:1977ur}---with its corresponding Goldstone boson---the \textit{axion}---, as this new particle would also represents a dark matter candidate~\cite{Hui:2021tkt}. This hypothetical mechanism is discussed in the context of its cosmological implications in \cref{sec:global:QCDaxion}. %Recently, it has been argued that physical observables are really independent of $\theta$, and so QCD is $CP$ invariant~\cite{AI2021136616}. However, this goes in contradiction with recent lattice studies, see refs.~\cite{Shindler:2015aqa,Giusti:2018cmp,Dragos:2019oxn}.

\subsection{Asymptotic freedom and confinement}\label{sec:QCD:asymptoticfreedom}

At first glance, QCD seems very similar to the theory of electromagnetism,  quantum electrodynamics (QED). The only differences in the QCD action are the trace over color indices in \cref{eq:QCD:gaugelagrangianMinkowski} and the last term in \cref{eq:QCD:fieldstrengthtensorQCD}, which vanishes in QED. Still, the two theories are like night and day. 

A characteristic property of quantum field theories is the \textit{running} of the couplings, this is, the fact that the interaction couplings depend on the scale at which physics are probed. The coupling constant of QED, related to the electric charge of the electron $e$, is small at low energies or large scales, $e^2/4\pi\sim1/137$, and grows with the energy. This implies that QED can be studied with standard perturbative techniques. In contrast, the coupling constant of QCD decreases at small scales, i.e., $g(\Lambda)\rightarrow 0$ as $\Lambda\rightarrow\infty$, and becomes large at low energies. Thus, the low-energy properties of QCD cannot be understood in perturbation theory.

This characteristic property of QCD is known as \textit{asymptotic freedom}~\cite{Gross:1973id,Politzer:1973fx}, and gives rise to the rich phenomenology of the strong interaction. %Its importance was recognized in 2004, when the Nobel Prize in Physics was awarded to its discoverers, Gross, Wilczek and Politzer. 
One of its main implications is the fact that QCD can be studied perturbatively at high energies, where the coupling is small. In particular one can use perturbation theory to determine the running of the strong coupling at high energies. The beta function has been computed up to five loops in perturbation theory~\cite{vanRitbergen:1997va,Luthe:2017ttg}. At one-loop order it reads~\cite{Gross:1973id,Politzer:1973fx},
\begin{equation}\label{eq:QCD:oneloopbetafuntionQCD}
\beta(\alphas)=\mu\frac{\text{d}\alphas}{\text{d}\mu}=-\frac{\alphas^2}{2\pi}\beta_0+\mathcal{O}(\alpha_\text{s}^4)\,,
\end{equation}
with
\begin{equation}
\beta_0=\frac{11}{3}\Nc-\frac{2}{3}\Nf\,,
\end{equation}
where $\alphas=g^2/4\pi$. % and we leave explicit the dependence on number of colors, $\Nc$, and flavors, $\Nf$. 
The one-loop result for the beta function is universal, this is, does not depend on the regularization (this is also true for the two-loop-order result). 

From this result, one can observe that the theory is asymptotically free, with $\beta(\alpha_\text{s})<0$, as long as $\Nc\leq2\Nf/11$. This condition is well satisfied in nature, where $\Nc=3$ and $\Nf=6$.The result for the beta function allows one to determine a low-energy scale, $\Lambda_\text{QCD}$, at which the QCD coupling becomes infinite,

\begin{equation}\label{eq:QCD:QCDscaledefinition}
\LambdaQCD=\LambdaUV\,\text{exp}\left[-\frac{4\pi}{\alphas(\LambdaUV)}\right]\,,
\end{equation}
where $\LambdaUV$ is some reference high-energy scale. Using experimental results, one can estimate $\LambdaQCD\approx 300$ MeV. %This is the Landau pole of QCD. %, and roughly indicates the energies below which it is not possible to apply perturbative techniques.

This scale separates the high- and low-energy regimes of QCD. At high energies, $\alphas$ becomes small and QCD turns into a weakly interacting theory, in which quarks and gluons behave as quasi-free particles. This occurs for example in heavy-ion collisions and the early universe, where a quark-gluon plasma is formed~\cite{Letessier_Rafelski_2002}. In this regime, perturbation theory has been successfully used to make accurate predictions, which are in agreement to experimental results~\cite{PDG:2020}.

For energies below $\Lambda_\text{QCD}$, interactions are very strong, leading to the phenomenon of \textit{confinement}. Quarks and gluons do not exist in isolation, but instead form composite particles, the hadrons, which can only appear in nature as colorless states that transform under the singlet representation of the color group. In the quark model, hadrons can either be mesons, composed of a quark and an antiquark with the same color, or baryons formed of three quarks of different colors. Other possibly exotic hadrons, such as tetraquarks, pentaquarks or hybrid states have also been found in experiment~\cite{Belle:2003goh,LHCb:2020bls,LHCb:2020pxc,LHCb:2022sfr,LHCb:2022lzp,PDG:2020}. The mass of these composite states does not originate from the mass of its composing quarks (which only amounts for $\sim1\%$ of the mass in the case of the proton, for instance), but rather from the energy of the strong interaction itself. %Typically hadron masses are of the order of the QCD scale,
%\begin{equation}
%M_\text{had}\sim\Lambda_\text{QCD}\,.
%\end{equation}
The strong force is thus responsible for the majority of the (visible) mass of the universe.
 
In the low-energy regime, contrary to high energies, perturbation theory fails and one is forced to look for alternative approaches to QCD. During the last decades, many different techniques have been developed, which have shown different levels of predicting capability. In this chapter, we discuss three of them: the lattice regularization of QCD, the large $\Nc$ limit and effective field theories, which are the main tools guiding the first part of this doctoral dissertation. 

Before moving on, it is worth emphasizing that the proof of asymptotic freedom and confinement shown in this section is based on perturbative arguments. A fully non-perturbative demonstration has not been carried out, and a rigorous proof is one of the holy grails of mathematics~\cite{Jaffe:2000ne}. Still, a vast amount of experimental data has been gathered during the last decades, showing convincing evidence of these properties.

\newpage\subsection{Chiral symmetry and the hadron spectrum} \label{sec:QCD:chiralsymmetry}

To explain how quarks combine into hadrons and why these hadrons organize into the geometric patterns predicted by the quark model, we need to  introduce one last approximate symmetry of QCD, \textit{chiral symmetry}, and how it is spontaneously broken at low energies. %This will allow us to understand why hadrons organize according to irreps of SU($\Nf$) and why some of them, the pseudoscalar mesons, are much lighter than the others. 
%This symmetry is also the basis of one the most successful effective field theory of hadrons, chiral perturbation theory, which will be introduced in \cref{sec:QCD:chiralpertubationtheory}.
We consider first the massless limit of QCD, in which the fermion Lagrangian reads,
\begin{equation}\label{eq:QCD:masslessfermionlagrangian}
\mathcal{L}_\text{F}=\sum_{f=1}^{\Nf}\overline{q}_f i\gamma_\mu\left(\partial_\mu-igA^\mu\right)q_f\,.
\end{equation}
In this limit, QCD is invariant under the following global vector and axial transformations,
\begin{equation}\label{eq:sunsunu1u1}
G=\text{SU}(\Nf)_\text{V}\times \text{SU}(\Nf)_\text{A} \times \text{U}(1)_\text{V} \times \text{U}(1)_\text{A}\,,
\end{equation}
which defines the chiral symmetry group. This acts on quark fields as
\begin{equation}\label{eq:QCD:chiraltransformations}
\begin{array}{rlcrl}
q \xrightarrow[]{\text{SU}(\Nf)_{\text{V}}} & \text{exp}(i\alpha^a_\text{V} t^a)\,q\,, & \quad\quad\quad  & q \xrightarrow[]{\text{U}(1)_{\text{V}}} & \text{exp}(i\alpha_\text{V})\,q\,,\\
q \xrightarrow[]{\text{SU}(\Nf)_{\text{A}}} & \text{exp}(i\gamma_5\alpha^a_\text{A} t^a)\,q\,, & \quad\quad\quad  & q \xrightarrow[]{\text{U}(1)_{\text{A}}} & \text{exp}(i\gamma_5\alpha_\text{V})\,q\,.
\end{array}
\end{equation}
where $\gamma_5=i\gamma_0\gamma_1\gamma_2\gamma_3$, $t_a$ here are the generators of the SU($\Nf$) group acting in flavor space and quark fields are to be understood as a vector in the space of all flavors. While these transformation properties are reminiscent of gauge transformations shown in \cref{eq:QCD:gaugetransformationsQCD}, now we are considering global transformations, and so $\alpha_{\text{A,V}}^a$ and $\alpha_{\text{A,V}}$ are constants independent of the spacetime coordinates. The invariance of massless QCD under chiral transformations implies that a quark mass term cannot be generated by quantum corrections, and so quark masses only renormalize multiplicatively.

%The two vector We can first discuss the meaning of the non-Abelian transformations, which together form the so called \textit{chiral symmetry}. The conservation of this symmetry implies that quarks remain massless after quantum corrections. In the non-massless case, it implies that quark masses only get multiplicative quantum correction. 

At low energies, however, chiral symmetry is not realized. An axial SU($\Nf$) transformation would map a hadron to another one of opposite chirality, and so the hadron spectrum will be composed of pairs of mass-degenerate particles, which are not observed in nature. The reality is that chiral symmetry is spontaneously broken below the QCD scale to its vector component,
\begin{equation}\label{eq:QCD:spontaneouschiralsymmetrybreaking}
G\longrightarrow H=\text{SU}(\Nf)_\text{V}\times \text{U}(1)_\text{V}\,,
\end{equation}
 with an order parameter given by the quark condensate,
\begin{equation}\label{eq:QCD:orderparameterquarkcondensate}
\Sigma=\langle 0|\overline{q} q|0 \rangle\sim\Lambda_\text{QCD}\neq 0\,.
\end{equation}

The U(1) part of the unbroken symmetry group corresponds to the baryon number,\footnote{Note that baryon number, $B$, is not conserved in the electroweak sector of the Standard Model. Still, $B-L$ remains a global symmetry, where $L$ is the lepton number.} 
 while SU($\Nf)_\text{V}$ corresponds to the so called \textit{isospin symmetry}. Low-lying hadrons organize into different irreps of the latter, with definite isospin quantum numbers, as discussed later. These refer to total isospin, its third component, as well as strangeness and charmness.
%This unbroken symmetry is known as the isospin symmetry. We will later discuss how hadrons organize in the different irreducible representations of the isospin symmetry group.

According to Goldstone theorem~\cite{Nambu:1960,Goldstone:1961eq,Goldstone:1962}, a spontaneously broken continuous symmetry implies the existence of massless particles, known as Nambu-Goldstone bosons (NGB), in a number equal to that of the generators of the broken group. In the case of chiral symmetry, these particles are the pseudoscalar mesons, generated by acting on the vacuum with the Noether charges of the  SU($\Nf)_\text{A}$ symmetry. At the classical level, there is also another boson---the $\eta'$---associated to the $U(1)_\text{A}$ broken symmetry. However, as we will discuss below, this symmetry is additionally broken by quantum effects, i.e., it is anomalous~\cite{PhysRev.177.2426,Bell:1969ts}. 

Before that, we consider massive quarks.~Adding a mass term to \cref{eq:QCD:masslessfermionlagrangian}  explicitly breaks chiral symmetry. However, not all quarks have the same mass. In the case of up and down quarks, their masses are very similar and much smaller than $\LambdaQCD$, this is, $m_\text{u}\approx m_\text{d} \ll \LambdaQCD$. If we focus on $\Nf=2$ QCD, chiral symmetry is approximately realized. Hadrons are still organized according to the isospin irreps and pseudoscalar mesons (the pions, $\pi$) are associated to the spontaneous breaking of SU(2)${}_\text{A}$. Now, however, they are massive pseudo-Nambu Goldstone bosons (pNGB), with a mass
\begin{equation}
M_{\pi}^2\sim m_\text{u,d}\Lambda_\text{QCD}\,.
\end{equation}
which is still much smaller than that of other hadrons, such as the proton, $M_\pi\approx 135\,\text{MeV}\ll M_p\approx 935\,\text{MeV}$. Thus, chiral symmetry is key to understand the lightness of the pions in nature.

If one considers the strange quark, similar arguments also apply, as $m_\text{s}<\Lambda_\text{QCD}$. However, in this case the effects breaking chiral symmetry are larger and, as a result, the pseudoscalar mesons containing a strange quark (the kaons, $K$) are more massive than the pions, $M_K=435$ MeV. In the case of heavier quarks, such as the charm or the bottom, these argument does not hold, as $m_\text{c},m_\text{b}>\LambdaQCD$. However, if one considers a hypothetical scenario in which there are $\Nf$ light quarks, the conclusions presented here would still hold.  This is the base of one of the main research works presented in this doctoral thesis---see \cref{sec:largeNpions,sec:largeNmesons}.

We can now come back to the $\text{U}(1)_\text{A}$ symmetry, which is broken by the chiral anomaly~\cite{tHooft:1976snw,CALLAN1976334,Jackiw:1976}. The corresponding Noether current, $\cJ_\text{A}^\mu$ is not conserved, but couples to the topological charge of QCD,
\begin{equation}
\partial_\mu \cJ_\text{A}^\mu = -\Nf\frac{g^2}{16\pi^2}\text{tr}\left[F_{\mu\nu}\tilde{F}^{\mu\nu}\right]\,,
\end{equation}
which can also be derived from the change in the measure of the path integral under a U$(1)_\text{A}$ transformation~\cite{Fujikawa:1979ay}.%is obtained from studying the transformation of the path integral measure under a U(1)${}_\text{A}$ transformation~\cite{Fujikawa:1979ay}. 

The anomalous breaking of the U(1)${}_\text{A}$ symmetry implies that the associated pNGB, the $\eta'$ particle, is much more massive than the others pseudoscalar mesons~\cite{tHooft:1976rip} with a mass $M_{\eta'}=910$ MeV. In the limit of large number of colors---see \cref{sec:QCD:largeNc}---Witten and Veneziano derived a relation between the mass of the $\eta'$ and that of the pions~\cite{Witten:1979vv,Veneziano:1979ec}, the so-called \textit{Witten-Veneziano formula}, %Its mass is related to that of the other pseudoscalar mesons by the Witten-Veneziano formula~\cite{Witten:1979vv,Veneziano:1979ec},
\begin{equation}\label{eq:QCD:WittenVenezianoformula}
M_{\eta'}^2=2M_K^2-M_\eta^2+\frac{2\Nf}{F_\pi^2}\chi_\text{top}\,,
\end{equation}
where $F_\pi$ is the pion decay constant\footnote{We use the convention in which $\Fpi\approx92.3$ MeV in the real world.}, $M_\eta$ is the mass of the $\eta$ meson and $\chi_\text{top}$ is the topological susceptibility of the pure gauge theory. This is defined as an integral of the two-point function of the topological charge operator over the whole spacetime, evaluated in the pure Yang-Mills theory ($\Nf=0$),
\begin{equation}
\chi_\text{top}=\int \d^4 x \langle q(x)q(0)\rangle_{\Nf=0} \,,
\end{equation}
where $q(x)$ is the topological charge operator,
\begin{equation}
q(x)=\frac{g}{32\pi^2 \Nc}\Tr\left[F_{\mu\nu}\tilde{F}^{\mu\nu}\right]\,.
\end{equation}

As already mentioned, hadrons are classified in multiplets that correspond to the different irreps of the approximate SU($\Nf$)${}_\text{V}$ isospin symmetry. From \cref{eq:QCD:chiraltransformations}, it is clear that quarks transform in the fundamental irrep of the isospin group, while antiquarks transform in the antifundamental irrep. Hadrons are thus classified in the different irreps arising from the combination of quarks and antiquarks forming color singlets. Consider for example the real-world case with $\Nf=3$. Mesons are composed by a quark and an antiquark and so they form an octet and a singlet,
\begin{equation}
3\,\otimes \,\overline{3}=8\,\oplus \,1\,,
\end{equation}
where each irrep is labelled by its dimension and the overline refers to the antifundamental representation. In the case of pseudoscalar mesons, the octet is the Eightfold way~\cite{Gell-Mann:1961omu,NEEMAN1961222} and the singlet corresponds to the $\eta'$. This classification is also valid for other mesons formed of a quark and an antiquark, such as vector mesons. In the case of baryons, composed by three quarks, we get a singlet, two octets and a decuplet,
\begin{equation}
3\,\otimes\, 3\,\otimes \,3=10\,\oplus \,8\, \oplus \,8\, \oplus \,1\,.
\end{equation}
The proton and the neutron lie in one of the octets, while other baryons, such as the $\Delta$ baryons and the $\Omega^\text{\,-\,-}$  lie in the decuplet. This classification is the theoretical basis underlying the quark model.

\newpage\section{QCD on the lattice}\label{sec:QCD:lattice}

In QCD, as in any other quantum field theory, predictions of physical observables are extracted from the computation of expectation values of gauge-invariant operators. These can be determined using the path integral formalism, based on Feyman's approach to quantum mechanics~\cite{Feynman:1950}. The expectation value of some operator $O$ is computed as an integral over all possible field configurations,
\begin{equation}\label{eq:QCD:pathintegralMinkowski}
\langle O[q,\overline{q}, A]\rangle = \frac{1}{\mathcal{Z}}\int \cD q\,\cD\overline{q}\,\cD A\,O[q,\overline{q},A]\,\text{e}^{iS_\text{QCD}[q,\overline{q},A]}\,,
\end{equation}
where $\mathcal{Z}=\langle \mathbbm{1} \rangle$ is the partition function and $S_\text{QCD}$ is the QCD action from \cref{eq:QCD:QCDactionMinkowski}. % and the fields appearing in the integrand are classical functions. 

%Note this formalism allows to compute the expectation value of an operator $O[q},\hat{\overline{q}}, \hat{A}]$ that acts on the Hilbert space of the theory by integrating over possible values of a function of the classical fields, $O[q,\overline{q},A]$. 

The path-integral formalism provides a method to determine any physical observable. However, its practical utility is limited by its degree of complexity. %The integral needs to be performed over all possible values of the fields in every spacetime position. 
In many cases, physical predictions are obtained using perturbation theory, which is based on an asymptotic expansion in the interaction couplings. However, as we have seen in \cref{sec:QCD:asymptoticfreedom}, this is not an option for QCD at low energies.

The formulation of QCD on a discrete spacetime, better known as \textit{lattice QCD}, is an alternative approach to the theory of strong interactions that makes it possible to compute expectation values of the form of \cref{eq:QCD:pathintegralMinkowski} by numerical methods. It was first proposed by Kenneth Wilson in the 1970s~\cite{Wilson:1974sk,Wilson:1975id}, and has become the main approach to study the low-energy regime of QCD. For a detailed introduction to the topic see, for example, \rcite{Gattringer:2010zz}.

Lattice QCD relies on the computation of the path integral in a finite volume and Euclidean time via the discretization of the spacetime, that turns the infinite-dimensional path integral into a finite-dimensional one. %, in which one ``only'' integrates over all possible values of the fields in a finite number of lattice sites, $\Lambda$. 
Typically, we work with lattices of the form
\begin{equation}\label{eq:QCD:latticesites}
\Lambda\in\left\{a(n_0,n_1,n_2,n_3)\,|\,n_\mu\in\mathbbm{Z},0\leq n_0<T/a,0\leq n_i<L/a\right\}\,,
\end{equation}
where $L$ and $T$ are the space and time extents of the lattice, and $a$ is the lattice spacing, which also acts as a ultraviolet regulator of the theory. After rotating to Euclidean time, $t_\text{E}=-it$, and working in the all-plus Euclidean metric, $g_{\mu\nu}=\text{diag}(+1,+1,+1,+1)$, \cref{eq:QCD:pathintegralMinkowski} takes the form
\begin{equation}\label{eq:QCD:pathintegrallatticeEuclidean}
\langle O[q,\overline{q}, A]\rangle = \int\left[\prod_{x\in\Lambda} \text{d}q(x)\,\text{d}\overline{q}(x)\,\text{d} A(x)\right]\,O_\text{E}[q,\overline{q},A]\,\frac{1}{\mathcal{Z}}\,\text{e}^{-S_\text{QCD}^{\text{E},a}[q,\overline{q},A]}\,,
\end{equation}
where $S_\text{QCD}^{\text{E},a}$ denotes the discretized Euclidean QCD action---see \cref{sec:QCD:discretizedaction}---and $O_\text{E}$ indicates the Euclidean-time operator. The use of a finite volume also has the implication that the values of momenta allowed in the lattice are quantized, with their particular form depending on the choice of boundary. In the most common case of periodic boundary conditions, the allowed three-momenta are
\begin{equation}\label{eq:QCD:finitevolumemomenta}
\bm{k}=\frac{2\pi}{L}\bm{n}\,,
\end{equation}
where $\bm{n}\in\mathbbm{Z}^3$.

A direct evaluation of \cref{eq:QCD:pathintegrallatticeEuclidean} would still require of the computation of a large number of nested integrals. The introduction of an Euclidean time makes the exponential factor real, which can then be interpreted as a probability distribution, with normalization $\mathcal{Z}$. This allows the evaluation of the path integral using Markov chain Monte Carlo techniques~\cite{MonteCarlo,MonteCarlo2},
%If no further modification was to be done, the computation of the path integral would require the evaluation of a prohibitively large number of nested integrals. Here is where the use of a Euclidean time, $t_\text{E}=it$, comes in. It allows to rewrite the exponential of the path integral as a real term, and so allows to interpret it as a probability distribution. The path integral can then be evaluated using importance sampling methods,
\begin{equation}
 \langle O[q,\overline{q}, A]\rangle = \frac{1}{N_\text{conf}}\sum_{i=1}^{N_\text{conf}} O[q_i,\overline{q}_i,A_i]+\cO\left(N_\text{conf}^{-1/2}\right)\,,
\end{equation}
where the sum is performed over a number, $N_\text{conf}$, of field configurations, $\{q_i,\overline{q}_i,A_i\}$, distributed according to the probability distribution
\begin{equation}
\mathcal{P}\{q,\overline{q},A\}\sim \frac{1}{\mathcal{Z}}\text{exp}\left(-S_\text{QCD}^{\text{E},a}[q,\overline{q},A]\right)\,.
\end{equation}
%This is, the expectation value of $O$ is computed as an average over its value on field configurations generated following the distribution $\mathcal{P}$. 
Field configurations following this distribution are generated using different techniques. At the present time,  hybrid Monte Carlo algorithm is the state of the art for QCD computations~\cite{DUANE1987216}. In practice, generating configurations and subsequently computing observables is computationally very expensive and requires of the use of high-performance computing. %Theoretical, algorithmic and software developments during the last decades have allowed for very precised lattice computations, such as the hadron-vacuum-polarization contribution to the muon anomalous magnetic moment~\cite{Borsanyi:2020mff}.

The use of sampling techniques implies that numerical results have a statistical uncertainty that can be, in principle, systematically improved by increasing statistics. On the other hand, the systematic error introduced by a finite lattice spacing or the finite volume can be reduced by extrapolations to the continuum or to infinite volume, respectively.  A systematic uncertainty is unavoidable in these extrapolations and should be carefully quantified. It is usually said that lattice QCD is systematically improvable. % For example, one can tame discretization effects by simulating at several lattice spacings and extrapolating to the continuum. Similarly, statistical uncertainties can be reduced by increasing the number of configurations. 

%There is still the question of whether observables computed in an finite Euclidean spacetime, even if precise, can be used to extract information about infinite-volume Minkowski observables. In short, the answer is yes. However, the relation is not straighforward. This will be discussed in more detail in \cref{sec:QCD:latticeobservables,sec:hadrons:}.

Finally, it is worth mentioning that computations on the lattice are not restricted to real-world QCD. The use of lattice techniques allows one to vary parameters of the theory. For example, one can consider unphysical quark masses---typical lattice simulations are performed at heavier-than-physical pion mass due to a cheaper computational cost---or different number of flavors. It also makes it possible to simulate other theories such as simple toy models or possible theories beyond the Standard Model. In the work presented in this dissertation, we exploit this possibility, using the lattice to study QCD as a function of the number of colors---see \cref{sec:largeNpions,sec:largeNmesons}---and to explore the (1+1)-dimensional O(3) non-linear sigma model as a toy model for QCD---see \cref{sec:O3model}.

\subsection{The QCD lattice action}
\label{sec:QCD:discretizedaction}

The choice of a discretized action and operators plays a major role in lattice computations. It determines the symmetries of the discretized theory, and also has an impact on the size  of discretization corrections. While these can be accounted for with a continuum extrapolation, one wants to ensure the systematic error induced by such extrapolation is as small as possible. %cutoff effects are as small as possible.

%Before explaining how physical observables can be computed in the lattice, we comment on how the discretized lattice action is defined. While discretization effects can be accounted for by a continuum extrapolation, one wants to ensure these corrections are as small as possible. The size of these corrections depend on the choice of the observable and of the discrete lattice action. In addition, the lattice action also determines the symmetries of the discrete theory.

The simplest choice of a discretized QCD action is that of Wilson~\cite{Wilson:1974sk}. To guarantee gauge invariance, the $\mathfrak{su}$(3)-valued $A_\mu$ fields are substituted by SU(3)-valued \textit{link variables}, $U_\mu=\exp(ia gA_\mu)$. These are the parallel transporters between two adjacent point in the lattice. The Wilson gauge action is
\begin{equation}\label{eq:QCD:gaugewilsonactionlattice}
S_\text{G}[U]=\frac{\beta}{\Nc}\sum_{x\in\Lambda}\sum_{\mu<\nu}\text{Tr}[1-\Re\,P_{\mu\nu}(x)]\,,
\end{equation} 
where $\beta=2\Nc/g^2$ and  $P_{\mu\nu}(x)$ is the smallest Wilson loop, the \textit{plaquette},
\begin{equation}
P_{\mu\nu}(x)=U_\mu(x)U_\nu(x+a\hat{\mu})U_{\mu}(x+a\hat{\nu})^\dagger U_\nu(x)^\dagger\,,
\end{equation}
where $\hat{\mu}$ is the unit vector in the $\mu$ direction, and similarly for $\hat{\nu}$. The action in \cref{eq:QCD:gaugewilsonactionlattice} is equivalent to the continuum one in \cref{eq:QCD:gaugelagrangianMinkowski}, up to errors of order $\cO(a^2)$.

The fermionic Wilson action is given by
\begin{equation}\label{eq:QCD:fermionwilsonactionlattice}
S_\text{F}[q,\overline{q},U]=a^4\sum_{x,y\in\Lambda}\overline{q}(x)D_\text{W}(x,y)q(y)\,,
\end{equation}
where $D_\text{W}(x,y)$ is the Wilson-Dirac operator,
\begin{multline}\label{eq:QCD:Diracoperatorlattice}
D_\text{W}(x,y)=\left(m_f+\frac{4r}{a}\right)\delta_{xy}\\
\displaystyle -\frac{1}{2a}\sum_\mu\left(\frac{r}{2}-\gamma_\mu\right)U_\mu(x)\delta_{x+a\hat{\mu},y}+\left(\frac{r}{2}+\gamma_\mu\right)U_\mu^\dagger(x+a\hat{\mu}) \delta_{x-a\hat{\mu},y}\,,
\end{multline}
where $r$ is a real constant. This action recovers the continuum one in the $a\rightarrow0$ limit, although this discretization is not directly obtained from substituting the continuum covariant derivatives by gauge-transported finite differences. Instead, one also includes a $-(ra/2)\overline{q}\nabla^2 q$ operator, known as the Wilson term. The impact of this term can be seen from the quark propagator in the gauge-free case ($g=0$),
\begin{equation}
\displaystyle D_{\text{W,free}}^{-1}(x,y)= S_\text{free}(x,y)=\frac{1}{L^3T}\sum_{k}S_\text{free}(k)\text{e}^{ik(x-y)}\,,
\end{equation}
with 
\begin{equation}
S_\text{free}(k)=\frac{m_f+\frac{r}{2}\hat{k}^2-i\gamma_\mu\bar{k}^\mu}{\left(m_f+\frac{r}{2}\hat{k}^2\right)^2+\bar{k}^2}\,,
\end{equation}
where $\hat{k}_\mu=2\sin(k_\mu/2)$ and $\bar{k}_\mu=\sin k_\mu$. Without the Wilson term, $r=0$, the propagator has poles not only at the center of the first Brillouin zone, $k=0$, but also at the edges, where any $k_\mu=\pi/a$. This implies that the continuum theory describes $16=2^d$ fermions instead of one~\cite{Susskind:1976jm}, where $d=4$ is the number of spacetime dimensions. The fermion-doubling problem is solved by the addition of the Wilson term. Typically,  $r=1$ is chosen in lattice simulations. 

However, the addition of the Wilson term breaks chiral symmetry. This implies that discretization errors in \cref{eq:QCD:fermionwilsonactionlattice} are $\mathcal{O}(a)$, and also that the quark masses renormalize additively. On general grounds, the Nielsen-Ninomiya no-go theorem~\cite{Nielsen:1980rz,Nielsen:1981xu} states that it is not possible to find a local discrete Dirac operator with no doublers that respects continuum chiral symmetry. 

Other fermion discretization have also been proposed to deal with the lack of chiral symmetry. These include staggered fermions~\cite{Kogut:1974ag}, which possess an additional taste symmetry, domain-wall fermions~\cite{Kaplan:1992bt}, defined on a five-dimensional lattice, and non-local overlap fermions, which satisfy an exact modified chiral symmetry~\cite{Neuberger:1997fp}.

\subsubsection*{Improved actions}

To perform a controlled continuum extrapolation, one needs discretization effects to be small. This can be reached by going to very fine and computationally expensive lattices, but also by redefining the lattice action and operators so that cutoff effects appear only at higher orders. These are the so-called \textit{improved actions and operators}, and play a major role in reducing discretization effects, especially for the fermion action.%, which naively presents $\cO(a)$ discretization effects. 

A systematic method to reduce discretization effects is the \textit{Symanzik improvement program}~\cite{Symanzik:1983dc,Symanzik:1983gh}. The key idea is to include a complete set of higher-dimensional terms in the definition of the lattice action and operators, which vanish in the continuum. Tuning the coefficients of these terms allows one to systematically subtract cutoff corrections of order $a$, $a^2$ and so on.

In the case of the fermion Wilson action in \cref{eq:QCD:fermionwilsonactionlattice}, $\mathcal{O}(a)$ improvement can be obtained by adding a single dimension-five operator, known as the clover term~\cite{Sheikholeslami:1985ij},
\begin{equation}\label{eq:QCD:cloveractionimprovement}
S_\text{F}^\text{imp}=S_\text{F}+c_\text{sw}a^5\sum_{x\in\Lambda}\sum_{\mu<\nu}\overline{q}(x)\frac{1}{2}\sigma_{\mu\nu}F^\text{clover}_{\mu\nu}(x)q(x)\,,
\end{equation}
where $\sigma_{\mu\nu}=[\gamma_\mu,\gamma_\nu]/2i$, $F^\text{clover}_{\mu\nu}$ is a clover-discretized version of the gauge field-strength tensor and $c_\text{sw}$ is the Sheikloleslami-Wohlert coefficient. Note that in order to guarantee $\mathcal{O}(a^2)$ corrections, the value of $c_\text{sw}$ needs to be finely tuned. It can also be computed using lattice perturbation theory~\cite{Luscher:1996sc,Aoki:2003sj}. For the standard Wilson action, its value is
\begin{equation}
c_\text{sw}=1+0.2659g^2+\cO(g^4)\,.
\end{equation}
However, this perturbative result only ensures that the discretization effect can be at most $\mathcal{O}(a g)$. Full $\cO(a^2)$ improvement requires a non-perturbative tuning of $c_\text{sw}$~\cite{Luscher:1996ug,Luscher:1996vw,Jansen:1998mx}. In addition, one has to complement these changes of the action with a suitable modification of the operators~\cite{Bhattacharya:2005rb}. %Similar modifications need to be done to the operators to ensure $\mathcal{O}(a)$ improvement.

In the case of the gauge action, \cref{eq:QCD:gaugewilsonactionlattice} only presents $\cO(a^2)$ corrections. Still, it is possible to reduce the observed size of these effects by the addition of dimension-six operators. Some versions of the improved action are the Lüscher-Weisz~\cite{Luscher:1984xn} or the Iwasaki~\cite{Iwasaki:1985we,Iwasaki:1983iya} gauge actions, which include terms corresponding to other types of Wilson loops bigger than the plaquette, such as rectangles, saddles or clovers.

In parallel to the Symanzik improvement, other methods also help at reducing discretization effects. This is the case of a using a \textit{twisted mass}~\cite{Frezzotti:2000nk}---see \rcite{Shindler:2007vp} for a review. In the case of an even number of degenerate quarks of mass $m$, the twisted mass is implemented by changing the bare quark mass,
\begin{equation}
m\mathbbm{1}\rightarrow m\mathbbm{1} + i\mu \gamma_5 T\,,
\end{equation}
where $T$ is a diagonal traceless matrix acting in flavor space with diagonal entries equal to $\pm1$, and $m$ and $\mu$ are the bare real and twisted masses, respectively. In the continuum this modification can be undone by a chiral rotation of the quark fields. However, in a discrete theory that does not preserve chiral symmetry, the modification leads to an alternative regularization.

A relevant case of a twisted action is that of maximal twist~\cite{Frezzotti:2003ni}. This is achieved when the renormalized real quark mass becomes zero, which can be checked numerically from the partially-conserved axial current. The twisted mass then plays the role of the quark mass. In this limit, $\mathcal{O}(a^2)$ improvement is achieved for some observables and the vector current is protected against renormalization, which allows one to numerically determine the pion decay constant  from the correlator of two bare pseudoscalar currents. %without the need of renormalization factors,
%\begin{equation}\label{eq:QCD:twistedmassFpi}
%\Fpi=\frac{\sqrt{2}\mu \langle 0|P|\pi\rangle_\text{bare}}{M_\pi}\,,
%\end{equation}
%where $P0\overline{q}\gamma_5 q$ is a pseudoscalar current and ``bare indicates that the matrix element needs no renormalization factors.

The main disadvantage of using a twisted mass is that it explicitly breaks parity and isospin symmetry. This means pseudoscalar mesons become non-degenerate even in the case of degenerate quark masses. Although cutoff effects are $\mathcal{O}(a^2)$, they have been shown to become very large in certain cases~\cite{Buchoff:2008hh,Draper:2021wga}. These effects can be reduced by further including a clover term in the action~\cite{Becirevic:2006ii}---see \cref{eq:QCD:cloveractionimprovement}.
 
\subsection{Computing correlation functions}
\label{sec:QCD:correlationfunctions}
 
Once a lattice action is chosen, a chain of correlated configurations can be generated and used to compute the desired expectation values. Typically, these are the correlation function, or correlators, of products of operators defined at different time slices, $O=O(t_1)O(t_2)...$. In this dissertation, we focus on two-point correlation functions, in which operators are defined at two times slices, called the \textit{source} and the \textit{sink}. 

From the correlators, physical observables can be determined with some statistical error. To obtain the best estimate of this error, one needs to take into account the autocorrelations between field configurations~\cite{Wolff:2003sm}. A simpler alternative, which we use for the work presented in this dissertation, is to average correlated configurations into blocks which are assumed to be uncorrelated, and then use bootstrap~\cite{Bootstrap:1979} or jackknife resampling~\cite{Jackknife:1974} for the analysis. Note that, in this case, one must explicitly check that the result is not affected by the block size.  

How the correlation function is computed on each configuration depends on the operators themselves. In the case of purely gluonic operators, as well as for operators in scalar theories, one just evaluates the operators at different times, and computes the corelation function from their product.  %Statistical errors can be reduced by repeating the computation exploiting translational invariance in position and time.

When considering fermions, fermionic variables are first integrated over to rewrite the correlation function in terms of the fermion propagator, $S(x,y)$~\cite{Wick:1950ee}. As an example, let's consider a single $\pi^+$ interpolating operator with three-momentum $\bm{p}$,
\begin{equation}
\pi^+(t,\bm{p})=\sum_{\bm{x}}\text{e}^{-i\bm{p}\bm{x}}\overline{d}(t,\bm{x})\gamma_5u(t,\bm{x})\,,
\end{equation}
where the sum is performed over all lattice sites at fixed time slice. % Typically, the time slices where the initial- and final-state operators are defined are called \textit{source} and \text{sink}, respectively. 
The single-pion correlator can then be computed as
\begin{equation}\label{eq:QCD:singlepioncorrelationfunction}
\langle\pi^+(t,\bm{p})\pi^+(0,\bm{p})^\dagger\rangle=\sum_{\bm{x},\bm{y}}\text{e}^{i\bm{p}(\bm{y}-\bm{x})}\langle\text{Tr}[S(y,x)S^\dagger(y,x)]\rangle\,,
\end{equation}
where $x=(0,\bm{x})$ and $y=(t,\bm{y})$ denote lattice sites, the trace runs over the color and spin indices, and we have used $\gamma_5$-hermiticity\footnote{$\gamma_5 S(x,y)\gamma_5=S^\dagger(y,x)$.}
 to rewrite the right-hand side. Evaluating this correlator in a given configuration exactly would require to known $S(x,y)$ for all $x$ and $y$, which is not possible, as the Dirac operator possess $\mathcal{O}(10^{10})$ rows and columns.
 
To circumvent this limitation, one computes the inverse of the Dirac operator acting on some chosen source vector, $\psi(x)$, by solving the system of linear equations
\begin{equation}
\sum_{y,z} D(x,y)S(y,z)\psi(z)=\psi(x)\,.
\end{equation}
This usually involves some preconditioner~\cite{Smith1997,Luscher:2003qa} that alleviates the computational cost and then the iterative application of some Krylov solver~\cite{Saad:1986,Hestenes1952MethodsOC,Frommer:2014} until the desired precision is achieved. 

The choice of source plays a central role in the computation of the correlation functions. In the example of the single pion, the simplest option is to use a point source. This means setting a single entry of $\psi$ to one, associated to a fixed lattice site, color and spin. The computation of the desired correlation functions is achieved by combining several sources with other spin and color. Momentum can be selected by doing a Fourier transform in the sink coordinate, which corresponds to performing only the sum over $\bm{y}$ in \cref{eq:QCD:singlepioncorrelationfunction}. The statistical error of this observable can later be reduced by repeating the computation for different point sources at several lattice sites, for example over a coarse sublattice~\cite{Detmold:2019fbk}. %Momentum conservation is ensured by projecting to definite momentum at sink. The result can be improved by repeating the computation over multiple lattice sites, for example over a coarse sublattice~\cite{Detmold:2019fbk}.

Another option, only valid for meson operators, is to use \textit{stochastic sources}~\cite{Dong:1993pk,Wilcox:1999ab,Foley:2005ac}. In this case, the components of $\psi$ are either set to zero or to some random noise, $\xi(x)$, which enforces the correct sum over the source position, spin and color up to stochastic noise,
\begin{equation}
\xi_{s_1c_1}(x_1)\xi_{s_2c_2}(x_2)^\dagger=\delta_{x_1,x_2}\delta_{s_1,s_2}\delta_{c_1,c_2}+\mathcal{O}(N^{-1/2})\,.
\end{equation}
where $N$ is the number of non-zero elements of $\psi$. One possibility is to set the random noise in color and space, for a fixed time and spin, $\xi^i_{c}(\bm{x})$. This are the so-called \textit{time- and spin-diluted stochastic sources}. The computation is repeated for several stochastic sources on each configuration to reduce the stochastic noise. For example, using time- and spin-diluted stochastic sources, the correlator function of a pion at rest takes the form
\begin{equation}\label{eq:QCD:singlepioncorrelationfunctionwithstochasticsources}
\langle\pi^+(t)\pi^{+}(0)^{\dagger}\rangle=\sum_{i=1}^N\sum_{c_1,c_2}\sum_{\bm{x}_1,\bm{x}_2,\bm{y}}\left\langle\text{Tr}\left\{S(y,x_1)\xi^i_{c_1}(\bm{x}_1) S\left[(y,x_2)\xi^i_{c_2}(\bm{x}_2)\right]^\dagger\right\}\right\rangle\,,
\end{equation}
where the spin indices are left implicit and both $x_1$ and $x_2$ have $t=0$ time coordinate. In this expression, $i$ labels different realizations of the stochastic noise, highlighting the need to average over several independent sources to obtain a sensible result.
 
The use of stochastic sources also makes it possible to project to definite momentum the operators at source. This can be achieved by  multiplying each noise element of the source vectors by a Fourier factor, $\xi^{\bm{p}}(x)=\text{e}^{i\bm{px}}\xi(x)$. After multiplying two noise vectors, 
\begin{equation}
\xi_{s_1c_1}^{\bm{p}_1}(x_1)^\dagger\xi_{s_2c_2}^{\bm{p}_2}(x_2)=\text{e}^{i\bm{x}_1(\bm{p}_2-\bm{p}_1)}\delta_{x_1,x_2}\delta_{s_1,s_2}\delta_{c_1,c_2}+\mathcal{O}(N^{-1/2})\,.
\end{equation}
which is the Fourier factor of a meson with three-momentum $\bm{p}=\bm{p}_2-\bm{p}_1$. 
Naively, this requires to invert the Dirac operator for each value of the quark momentum. However for computations that aim to study operators with several momenta, one can reduce the number of required inversions by wisely choosing the quark momenta so that all the desired meson momenta can be reconstructed.

A more sophisticated option is to use smeared sources~\cite{Gusken:1989ad,ALEXANDROU199160,UKQCD:1993gym}, or the eigenvalues of the Laplacian or Heaviside operator of three-dimensionally smeared fields~\cite{HadronSpectrum:2009krc,Morningstar:2011ka}. This latter possibility is commonly known as \textit{distillation}.

\subsection{Physical observables from lattice correlators}
\label{sec:QCD:computationofobservables}

From the Euclidean correlation functions computed in a finite volume, we can determine finite-volume energies and matrix elements. %Now that we have shown how correlation functions are computed in the lattice, it is the moment to discuss how physical observables that can be compared to experimental observations can be determined. In particular, how they can be extracted from computations of QCD in a finite Euclidean spacetime.
If we focus on two-point correlation functions, their spectral decomposition takes the form
%The starting point of this discussion is the spectral decomposition of the expectation values (we will discuss later how these expectation values are computed). We will focus on the case of two-point functions, although the same conclusions will apply to $n$-point correlators. We will also consider for now $T\rightarrow\infty$. Consider the correlation function between some operator $O$ at two different times. Due to time-translation invariance, it can be decomposed as 
\begin{equation}\label{eq:QCD:spectraldecompositioninfiniteT}
\langle O(t)O^\dagger(0)\rangle = \frac{1}{\cZ}\Tr\left[\text{e}^{-TH}O(t)O^\dagger(0)\right]= \sum_n |\langle 0| O(0)|n\rangle|^2 \text{e}^{-E_n|T|}\,,
\end{equation}
where $H$ is the Hamiltonian operator and $\cZ=\Tr[\text{e}^{-TH}]$ is the partition function. In the last step, we have considered, for simplicity, infinite time extent, $T\rightarrow\infty$ and used time-translation invariance to set the source at $t=0$. The sum is performed over all states $|n\rangle$ in the Fock space of the theory, with energy $E_n$. Most of these states, however, have different quantum numbers from the studied operator, and so do not overlap onto it, $O(0)|n\rangle=0$. Note that set of all states in the theory is discrete, due to the theory being defined in a finite volume. Typically, one furthermore assumes that the states with non-zero overlap are non-degenerate.

%The correlation functions we compute in the lattice depend on Euclidean time. If we were able to determine them exactly for any $t$, one should be able to perform an analytic continuation to real time. In lattice computations, however, we only posses a finite number of data points with some error. The analytic continuation then becomes an ill-posed problem. 
If the correlation function could be determined for any Euclidean time with infinite precision, it would be possible to determine the real-time correlation functions by analytic continuation. However, as we only possess a finite number of noisy data points, this becomes an ill-posed problem. Still, the energies and matrix elements appearing in the spectral decomposition are physical quantities, provided the operators are properly renormalized. %intrinsic of the field theory, and can be used to determine information about Minkowski observables.

The values of these energies and matrix elements can be determined from fitting the lattice results. The fit is performed at late enough times, $t/a\gg1$, so that only the lowest-lying states survive, but early enough so that statistical noise is small~\cite{PARISI1984203,Lepage:1989hd}. Typically, the fit function is a single or a sum of exponentials. The fit is repeated for several fit ranges and the results are obtained from the ranges that show a plateau in the fitted results. % is chosen from where the results show a plateau. %This ensures that the fir is insensitive to the fit range and so the late-time hypothesis is verified.

The time range is typically chosen by visual inspection. Recently, it has also become common to use weights based on Bayesian arguments~\cite{Jay:2020jkz,Neil:2022joj,Frison:2023lwb} to average the results from different fit ranges. The set of results to average can be those obtained for all possible fits, or only a subset that are found to lie close to the plateau. In the first case, one expects that the averaged result coincides with the plateau. %, and so lead to an automatic identification. %allows to identify said plateaux automatically.%, either for all possible ranges or for those for which the results lie on a plateaux. 

The use of Bayesian weights allows one to obtain the final mean result as a weighted average and the final variance as the combination of a statistical and a systematic contributions. For example, let $\{a_i\}$ be the results for some quantity for different fit ranges, with variances $\{\sigma_{a,i}^2\}$ and corresponding weights $\{w_i\}$. The weighted mean is
\begin{equation}\vspace{-0.15cm}
\langle a \rangle = \sum_i a_i w_i\,,
\end{equation}
while the total variance becomes
\begin{equation}
\sigma_a^2=\sigma_{a,\text{stat}}^2+\sigma_{a,\text{syst}}^2\,,
\end{equation}
where the statistical and systematic contributions to the variance are, respectively,\vspace{-0.1cm}
\begin{equation}
\begin{array}{rl}
\sigma_{a,\text{stat}}^2 & \displaystyle=\sum_i\sigma_{a,i}^2\,,\\[-2pt]
\sigma_{a,\text{syst}}^2 & \displaystyle=\sum_i a_i^2 w_i - \left(\sum_i a_i w_i\right)^2\,.
\end{array}
\end{equation}
The simplest weight choice is
\begin{equation}\label{eq:QCD:weightsplateaux}
w_i\propto \exp\left[-\frac{1}{2}\left(\chi_i^2-2N_i+2N_\text{par}\right)\right]\,,
\end{equation}
where $\chi^2_i$ is the chi-square of the fit of sample $i$, $N_i$ is the number of fitted points and $N_\text{par}$ the number of parameters in the fit function. Note that this method also makes it possible to compare between different fit functions.

One of the obstavles in the extraction of finite-volume energies and matrix elements is that, in principle, all states of the Fock space contribute to the spectral decomposition. However, the operators can be chosen to minimize the number of states contributing. For example, one can use some operator having the same transformation properties under $C$ and $P$ as the state of interest. Also, one typically projects to definite total momentum. 

One also wants to define operators having the correct spin and angular momentum quantum numbers. However, the lattice itself breaks rotation invariance. Instead, the finite-volume theory has a discrete cubic symmetry, and operators can be projected into irreps, $\Lambda$, of the cubic group or of the relevant little group in case of non-zero total momentum, $\cG$. This is done using the formula~\cite{Dudek:2012gj,Morningstar:2013bda},
\begin{equation}\label{eq:QCD:cubicgroupprojection}
O^{\Lambda \lambda}(t) \propto \sum_{R\in\cG}\Gamma_{\lambda\lambda}^{(\Lambda)}(R) U_RO(t)U_R^\dagger\,,
\end{equation}
where $R$ denotes the elements of $\cG$, $U_R$ is the operator that applies the transformation $R$ to $O$ and $\Gamma^{(\Lambda)}(R)$ is the representation of $R$ in the \mbox{irrep $\Lambda$}. The index $\lambda$ indicates the component in those irreps that have more than one dimension, and is not summed over in \cref{eq:QCD:cubicgroupprojection}. Note that each irrep $\Lambda$ contains contributions coming from different total angular momenta~\cite{Dudek:2012gj,Morningstar:2013bda,Luscher:1990ck}.

Another of the limitations in the extraction of finite-volume results comes from using a single operator. In this case, one can only reliably extract the energies of the lowest lying state and and the fit range is limited by excited-state contamination. This problem can be sorted out by computing the matrix of correlators between a set of operators, $\left\{ O_i\right\}_{i=1}^N$,
\begin{equation}
C_{ij}(t)=\langle O_i(t)O_j^\dagger(0)\rangle\,.
\end{equation}
 This can then be used to solve a generalized eigenvalue problem (GEVP)~\cite{Luscher:1990ck,Blossier:2009kd},
\begin{equation}\label{eq:QCD:gevp}
C(t) v_n=\lambda_n(t,t_0)C(t_0)v_n\,.
\end{equation}
The eigenvalues of the GEVP are in general of the form
\begin{equation}
\lambda_n\propto\text{e}^{-E_n(t-t_0)}\left\{1+\mathcal{O}\left[\text{e}^{-(E_{m}-E_n)t}\right]\right\}\,,
\end{equation}
where $E_n$ indicates the $n$-th lowest finite-volume energy. Corrections to this asymptotic behaviour, as indicated, depend on the energy separation between energy levels~\cite{Luscher:1990ck}. However, if one solves the GEVP with $t\leq 2t_0$, these correction are dominated by  the distance to the $N+1$ state~\cite{Blossier:2009kd}, 
\begin{equation}
\lambda_n\propto\text{e}^{-E_n(t-t_0)}+\mathcal{O}\left[\text{e}^{-(E_{N+1}-E_n)t}\right]\,.
\end{equation}
The eigenvectors, on the other hand, are related to the overlaps of the operators onto the different states, 
\begin{equation}
C(t_0)v_n=Z_n^i \text{e}^{-E_n t_0}\,,
\end{equation}
where $Z_n^i=\langle0| O_i(0)|n\rangle$.

The finite-volume energies and matrix elements extracted from correlation functions can be used to constrain physical observables. In some cases, they are themselves very close to quantities of interest that can be measured experimentally. For example, if $O(t)$ is a single-pion operator with zero total momentum, the lowest energy is the finite-volume pion mass. From an all-order perturbation theory study~\cite{Luscher:1985dn}, it is known to be equal to its infinite-volume counterparts up to exponentially-suppressed volume corrections. In the case of the pion mass, these volume effects are known from chiral perturbation theory, which is introduced in \cref{sec:QCD:chiralperturbationtheory}, and can be subtracted. At leading order~\cite{Gasser:1987ah,Colangelo:2005gd},
\begin{equation}\label{eq:QCD:finitevolumepionmass}
\Mpi(L)=\Mpi\left[1+\frac{1}{2\Nf}\frac{\Mpi^2}{(4\pi\Fpi)^2}g_1(\Mpi L)\right]+...\,,
\end{equation}
with
\begin{equation}
g(x)=\sum_{\bm{n}\in\mathbbm{Z}^3}\frac{4}{|\bm{n}|x}K_1(|\bm{n}|x)\,,
\end{equation}
where $K_1(x) $ is the modified Bessel function of the second kind, which decays exponentially for $x\gg1$. Similar results also exist for the pion decay constant~\cite{Colangelo:2004xr,Colangelo:2005gd}. In the practice, several lattice computation are performed at different volumes and these perturbative results are used as an ansatz for infinite-volume extrapolations.

In other cases, however, finite-volume quantities are not directly related to physical observables. This is the case, for example, of multiparticle processes. Scattering amplitudes and decay widths are intrinsically real-time quantities defined in infinite volume, and so do not have an analogue in the Euclidean finite volume. Still, it is possible to constrain them using the finite-volume energy spectrum. The formalisms that make it possible to do so in the case of two- and three-particle systems is discussed in detail in \cref{sec:hadrons:particleinteractionslattice} and plays a central role in the work presented in this dissertation.

\subsection{Thermal effects}\label{sec:QCD:thermaleffectslattice}

In realistic simulations the time extent of the lattice is not infinite. It is not always possible to use lattices with $\Mpi T\gg 1$, and so one needs to take finite-$T$ effects into account. These are the so-called \textit{thermal effects}. For finite $T$, we can write the spectral decomposition as
\begin{multline}\label{eq:QCD:themalspectraldecomposition}
C(t)=\langle O(t)O(0)^\dagger\rangle =\\[5pt] \frac{1}{\cZ}\Tr\left[\text{e}^{-TH}O(t)O^\dagger(0)\right]=\frac{1}{\cZ}\sum_{n,m} |Z_{nm}|^2 \text{e}^{-E_n T}\text{e}^{(E_n-E_m)t}\,,
\end{multline}
where $\cZ=\Tr[\exp(-TH)]$ is the partition function, $H$ is the Hamiltonian operator and $Z_{nm}=\langle n| O(0)| m \rangle$. Compared to \cref{eq:QCD:spectraldecompositioninfiniteT}, the sum now runs over all possible pairs of states $(n,m)$, which leads to many new terms in the sum. If we assume Hermitian operators that have no vacuum quantum numbers, these terms can be classified in two groups.

First, there are terms in which one of the states is the vacuum, this is, $|n\rangle=|0\rangle$ or $|m\rangle=|0\rangle$. These correspond, respectively, to forward and backward propagation of the physical states of interest. The combined contribution to the correlation function takes the form
\begin{equation}\label{eq:QCD:backwardspropagatingthermaleffects}
C_{m=0}(t) \supset \frac{1}{\cZ}\sum_{n} |Z_{n0}|^2 \text{e}^{-E_n T/2}\cosh\left[E_n \left(t-\frac{T}{2}\right)\right] \,.
\end{equation}
These contributions are analogous to those in \cref{eq:QCD:spectraldecompositioninfiniteT}, but appear with a different functional dependence on the energies. Still, if these were the only terms present one could use the techniques presented in the previous section, including the GEVP, to extract the finite-volume energies, with the only difference being the use of a distinct fit function, i.e., a hyperbolic cosine instead of an exponential.

The second set of terms in \cref{eq:QCD:themalspectraldecomposition} are those for which both $n,m\neq 0$. These terms depend on the energy difference of two states, none of which is the state of interest. For example, if we are interested in two-particle states, the dominant contamination is related to single-particle states. If we assume one of these terms dominates, depending on some energy difference $\Delta E$,  and neglect the backpropagating contamination discussed above, the correlation function takes the form
\begin{equation}\label{eq:QCD:examplethermaleffect}
C(t)=\sum_n |Z_{n0}|^2 \text{e}^{-E_n t}+ B \text{e}^{-\Delta E t} \,,
\end{equation}
where $B$ is some unspecified amplitude. This contamination can be removed by performing a \text{weight-shifting}~\cite{Dudek:2012gj},
\begin{equation}\label{eq:QCD:shiftreweight}
C(t)\rightarrow C(t) - \text{e}^{\Delta E} C(t+1)\,.
\end{equation}
Multiplying the correlator by the exponential factor shifts the time dependence of the last term in \cref{eq:QCD:examplethermaleffect}, so that one can completely cancel it by computing the variation of the correlator. We note that this method can be extended to the case in which backpropagating thermal effects are also considered, although this sometimes needs of some approximations. This is discussed in \cref{sec:largeNmesons:finitevolumespectrum}.
%Note that further changes the function to which one needs to fit to extract the desired finite-volume energies. What is more, in the case of complex correlation matrices, this procedure may invalidate the premises to apply a GEVP.
%The idea behind this transformation is that the multiplication bu $\text{\Delta E

%GEVP DONE
%Thermal effects
%exponential FV DONE
%correlators, stochastic sources

\newpage\section{The 't Hooft limit of QCD}\label{sec:QCD:largeNc}

%Different approaches have been used during the last decades to investigate the low-energy regime of QCD. In this section, we focus on one of them, the large $N_\text{c}$ limit of QCD, which will be one of the main topics of this thesis. This limit has proven very useful in understanding the properties of QCD. It is worth mentioning that large $N$ expansion have been used to study other field theories such as the O($N$) non-linear sigma model.

The \textit{'t Hooft} or \textit{large $\Nc$ limit} of QCD  is an alternative approach to the low-energy regime of strong interactions~\cite{tHooft:1973alw}. It is the limit in which the number of colors is taken to infinity, $\Nc\rightarrow\infty$, while the number of quark flavors is kept constant, $\Nf=\text{const}$. The limit constitutes a simplification of the theory of strong interactions that keeps most of its non-perturbative features and allows one to make predictions as a power series in $1/\Nc$. For a review on the 't Hooft limit---see \rcite{Manohar:1998xv}.

To define a sensible $1/\Nc$ expansion, the coupling constant of QCD is rescaled, $g\rightarrow g/\sqrt{\Nc}$. The $\beta$-function becomes
\begin{equation}\label{eq:QCD:largeNbetafunction}
\beta(\alphas)=\mu\frac{\text{d}\alpha_\text{s}}{\text{d}\mu}=-\frac{\alpha_\text{s}^2}{2\pi}\left(\frac{11}{3}-\frac{2}{3}\frac{\Nf}{\Nc}\right)+\mathcal{O}(\alpha_\text{s}^3)\,,
\end{equation}
so $\beta(\alphas)<0$ at large $\Nc$, and so the large $\Nc$ theory is asymptotically free. This also means that we expect the theory to be confining at low energies. Indeed, QCD becomes a theory of non-interacting narrow resonances and glueballs at large $\Nc$~\cite{tHooft:1973alw,Witten:1979kh,Coleman:1980nk}.

The large $\Nc$ limit makes it possible to characterize the $\Nc$ and $\Nf$ dependence of physical observables, and can also be used to make qualitative and some quantitative predictions of hadron physics. However, quantitative predictions are not possible beyond leading order in $\Nc$, since subleading $\Nc$ corrections are hard to estimate analytically. The lattice regularization provides with a tool to study them from first principles by simulating at different $\Nc$---see \rcite{Hernandez:2020tbc} for a recent review. Studying these effects, in the context of two-mesons systems, has been one of the main focuses of my doctoral work---see \cref{sec:largeNpions,sec:largeNmesons}.

%Here, we will focus on the case in which all quarks have degenerate masses, although different masses are straightforward to include, as they do not affect the large $\Nc$ counting.

%Note that we consider all quark flavors to have degenerate masses. The non-degenerate case is straighforward to study, as quark masses do not affect arguments presented here. 

%

%In the 't Hooft limit, the coupling constant of QCD is rescaled as $g\rightarrow g/\sqrt{\Nc}$, which allows for a sensible $1/\Nc$ expansion. This limit keeps most of the qualitative features of QCD. If we consider the $\beta$-function,

\begin{figure}[t!]
\centering
    \includegraphics[width=0.7\textwidth]{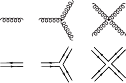} 
	\caption{Schematic representation of the gluon propagator and self-interacting vertices in the standard (top row) and double-line (bottom row) notation.}\label{fig:QCD:doublelinenotation}
\end{figure}

The $\Nc$ and $\Nf$ dependence of physical observables can be derived from the study of Feynman diagrams, rewritten using 't Hooft double-line notation that makes the color propagation explicit. The quark propagator is kept as a single directed line, while gluon propagators are substituted by two lines with opposite orientation.\footnote{This substitution is based on the U($\Nc$) gluon propagator. However, the difference between U($\Nc$) and SU($\Nc$) is in general $\mathcal{O}(\Nc^{-2})$.}
Interactions vertices are transformed in accordance---see \cref{fig:QCD:doublelinenotation}.
The $\Nc$ and $\Nf$ dependence of vacuum diagrams can then be determined. Each color loop amounts to a factor of $\Nc$, each coupling constant introduces a factor of $1/\sqrt{\Nc}$ and finally each internal quark loop adds a factor of $\Nf$. The scaling of a disconnected diagram is the product of the scaling of each separate connected piece.

%The $\Nc$ and $\Nf$ dependence of vacuum diagrams can then be trivially determined. Each color loop amounts to a factor of $\Nc$, each coupling constant introduces a factor of $1/\sqrt{\Nc}$ and finally each internal quark loop adds a factor of $\Nf$. The scaling of a disconnected diagram is the product of the scaling of each separate connected piece. %If we consider non-vacuum diagrams coming from operator correlation functions, this counting needs to be complemented by possible normalization factors included in the definition of such operators. Before turning into this point, however, it is worth mentioning about the geometric interpretation of this counting. 

The $\Nc$ counting can be given a geometric interpretation. Diagrams drawn using the double-line notation can be regarded as surfaces. Color loops represent polygons which are glued together by the gluon propagators, and quark loops, both internal and external, can be regarded as the boundaries of this surface. The $\Nc$ dependence of a given vacuum diagram is then $\Nc^\chi$, where $\chi$ is the Euler character of the surface. This number can be computed from the number of vertices, $V$, edges, $E$, and faces, $F$, of the surface, $\chi=V-E+F$. The leading order contributions originate from planar diagrams. Purely gluonic ones scale as $\mathcal{O}(\Nc^2)$, and those containing quarks are $\mathcal{O}(\Nc)$. It is commonly said that the 't Hooft limit of QCD is a limit of planar diagrams.

Alternative large $\Nc$ limits of QCD also exist. These include the Veneziano limit~\cite{Veneziano:1979ec}, in which $\Nc\rightarrow\infty$ while keeping the ratio $\Nf/\Nc$ constant, and a variation of the 't Hooft limit in which quarks are kept in the antifuntamental irrep of SU($\Nc$), that is isomorphic to the fundamental irrep for $\Nc=3$~\cite{Corrigan:1979xf,Armoni:2003gp,Armoni:2003fb,Armoni:2005wt}. The latter is relevant due to its relation to string theories~\cite{Angelantonj:1999qg,Armoni:1999gc}. Moreover, the large $\Nc$ limit has also been used to study other theories, such as (1+1)-dimensional QCD~\cite{tHooft:1974pnl} or QCD in the limit of heavy quarks~\cite{Witten:1979kh}, in which it makes quantitative computations. possible 

\subsection{Ordinary hadrons at large $\Nc$}

The nature and interactions of hadrons can be studied using the 't Hooft limit. With the large $\Nc$ counting introduced above, one can determine the $\Nc$ scaling of the chiral condensate,

\noindent\begin{equation}
\Sigma=-\langle\overline{q}(x)q(x)\rangle\propto \Nc\,.
\end{equation}
Therefore, spontaneous chiral symmetry breaking holds at large $\Nc$, and a low-lying spectrum of pseudoscalar mesons arises as the corresponding pNGB.

We can study the correlation function of mesons. Meson operators are fermion bilinears,
\begin{equation}
O_\text{M}(x)=\frac{1}{\sqrt{\Nc}}\overline{q}(x)\Gamma T q(x)\,,
\end{equation}
where $\Gamma$ is some combination of Dirac gamma matrices and $T$ is some matrix in flavor space. The $\sqrt{\Nc}$ normalization ensures that the meson correlator is of order unity, and so the mass of the mesons remains independent of $\Nc$, $M_\text{M}\propto\mathcal{O}(\Nc^0)$. One can in general study a connected $n$-meson correlation function,
\begin{equation}\label{eq:QCD:largeNmesoncorrelator}
\langle O_\text{M}(x_1)...O_\text{M}(x_n)\rangle_\text{conn}\sim \Nc^{1-n/2}\,.
\end{equation}
From this result, multiple conclusions can be extracted. First, the pion decay constant, defined as the two-point function of the axial current, scales as $F_\pi^2\propto\mathcal{O}(\Nc)$. Second, scattering processes are suppressed with $\Nc$. In particular, the two-meson scattering amplitude scales as $\mathcal{M}_{2}\propto\mathcal{O}(\Nc^{-1})$. Similarly, decay amplitudes that originate from three-point functions, are suppressed as $\Nc^{-1/2}$. One concludes that at large $\Nc$, QCD becomes a non-interacting theory of infinite stable resonances~\cite{tHooft:1973alw,Witten:1979kh,Coleman:1980nk}.\footnote{The fact that the number of resonance must be infinite can be seen from studying the logarithmic running of QCD correlation functions~\cite{Manohar:1998xv}.} Similar arguments also hold for glueballs.

It is interesting to comment on the singlet meson. At large $\Nc$ the axial anomaly is suppressed  as $\mathcal{O}(\Nf/\Nc$), and so the spontaneous chiral symmetry breaking pattern in \cref{eq:QCD:spontaneouschiralsymmetrybreaking} becomes
\begin{equation}\label{eq:QCD:largeNcsymmetrybreakingpattern}
G=\text{U}(\Nf)_\text{V}\times\text{U}(\Nf)_\text{A}\longrightarrow H=\text{U}(\Nf)_\text{V}\,.
\end{equation}
Thus, at large $\Nc$ the singlet mesons become mass-degenerate with the remaining mesons. In the case of pseudocalar mesons, this can be observed by looking at \cref{eq:QCD:WittenVenezianoformula}. After the $g\rightarrow g/\sqrt{\Nc}$ substitution and taking $\Nc\rightarrow\infty$, the $\eta'$ mesons becomes degenerate with the rest of pseudoscalars.

The 't Hooft limit can also be used to study baryons. 
%While the work presented in this part of the doctoral thesis will be focused on meson interactions, we can briefly comment about baryons in the large $\Nc$ limit. 
Contrary to mesons, the number of composing quarks of baryons grows with $\Nc$, as one needs to combine $\Nc$ quarks to create a color singlet. Thus, baryon masses grow as $M_\text{B}\propto\mathcal{O}(\Nc)$. Still, one can define large $\Nc$ counting rules for baryon correlation functions, with predictive power in the baryon sector~\cite{Witten:1979kh,Dashen:1993jt,Jenkins:1993zu,Carone:1993dz,Dashen:1994qi,Dai:1995zg}.

\newpage\subsection{Tetraquarks at large $\Nc$}\label{sec:QCD:largeNctetraquarks}

Finally, we can comment about exotic tetraquark states at large $\Nc$. The traditional picture, based on the views of Witten~\cite{Witten:1979kh} and Coleman~\cite{Coleman:1980nk}, is that these states do not exist in the 't Hooft limit. Their argument was based on the factorization property, according to which correlation functions of singlet operators factorize at large $\Nc$,
\begin{equation}
\langle O_1 O_2\rangle=\langle O_1\rangle \langle O_2\rangle +\mathcal{O}\left(\Nc^{-1}\right)\,.
\end{equation}
For example, in the case of tetraquarks, the correlation function of two local tetraquark operators, $\langle (\overline{q}q)^2(\overline{q}q)^2\rangle$, would decompose in the correlation functions of two non-interacting mesons.

This argument, however, was put in doubt by Weinberg~\cite{WeinbergTetra}. The connected and the disconnected parts of the four-quark operator correlation function may describe different phenomena, and one is not necessarily a subleading correction to the other. While the disconnected part represents two non-interacting mesons at large $\Nc$, tetraquarks could still arise from poles in the connected part, even if it is suppressed in $\Nc$. Therefore, it is not possible to rule out the existence of tetraquarks at large $\Nc$ and, if they were present, their decay width would scale as $\Gamma\sim1/\Nc$, just as ordinary resonances.

This last point was later revisited in \rcite{Knecht:2013yqa}, where it was shown that the scaling of $\Gamma$ would depend on the flavor composition of the tetraquark. In particular, it was argued that open-flavor tetraquarks (those which do not contain a quark-antiquark pair of the same flavor), should scale as $\Gamma\sim1/\Nc^2$. This result was also discussed in \rcite{Cohen:2014tg}, in which they further argued against the existence of these states in the large $\Nc$ limit. The authors of this last reference also studied the existence of tetraquarks in the large $\Nc$ limit with quarks in the antifundamental irrep~\cite{Cohen:2014via}. In this case, they managed to prove their existence with $\Gamma\sim 1/\Nc^2$. One of the objectives of the work presented in this dissertation is to study the possible existence of these resonances for varying $\Nc$ using lattice simulations---see \cref{sec:largeNmesons}.

\section{Effective field theories}\label{sec:QCD:EFTs}

Another widely used technique to describe QCD at low energies is that of effective field theories (EFTs). EFTs are quantum field theories that describe the dynamics of those states in a theory with masses below some cutoff scale, $\Lambda$,  at energies $E\ll \Lambda$. Even if the theory contains heavier states in the spectrum, their effect at these low energies appears in the form of higher-dimensional operators. EFTs are therefore non-renormalizable, but are predictive at fixed order in $E/\Lambda$.

The EFT Lagrangian is the most general Lagrangian consistent with the symmetries of the underlying theory, with which one is able to determine~\cite{Weinberg:1978kz} ``the most general matrix elements consistent with Lorentz invariance, quantum mechanics, unitarity, cluster decomposition and the assumed symmetries.'' The Lagrangian thus contains an infinite number of terms which appear multiplied by some unknown coefficients, known as \textit{Wilson coefficients} or \textit{low energy constants} (LECs). %Talk about LECs values and matching
 Despite having an infinite number of terms, EFTs still have predictive power. The operators in the Lagrangian can be organized depending on their energy dimension, so they contribute to processes at different orders in $E/\Lambda$. While this limits us to a finite precision, as more and more terms are needed to increase the accuracy of the result, EFTs are able to provide realistic predictions which can be compared against experiment. 
 
 %Another implication of the non-renormalizability of the theory is the need of an infinite number of counterterms to renormalize the theory. However, divergences appearing in loop diagrams can be absorbed by a redefinition of the fields and LECs from the effective Lagrangian~\cite{Weinberg_1995}. 

The use of EFTs is widely extended in theoretical physics. Fermi's theory has been used to describe weak decays~\cite{Fermi:1934hr}  and the Euler-Heisenberg's Lagrangian allows one to describe low-energy electron-photon scattering~\cite{heisenberg2006consequences}. EFTs are even used to treat the Standard Model as the low-energy limit of a more fundamental high-energy theory, the so-called Standard Model EFT~\cite{Brivio:2017vri}. In this thesis, we focus on the paradigmatic chiral perturbation theory~\cite{Weinberg:1978kz,PhysRevLett.17.616,Gasser:1983yg,Gasser:1984gg}, that describes the low-energy regime of QCD in terms of the lightest pseudoscalar mesons---see \rrcite{Pich:1995bw,Scherer:2002tk} for reviews. We also explain how this theory can be modified to study the large $\Nc$ limit of QCD. Further extensions exist that include baryons~\cite{JENKINS1991558,Bernard:1995dp}, heavy mesons~\cite{Burdman:1992gh} and even the effect of a lattice spacing and a twisted mass~\cite{Munster:2004wt,Sharpe:2004ny,Buchoff:2008hh} for application in lattice QCD.

\subsection{Chiral Perturbation Theory}\label{sec:QCD:chiralperturbationtheory}

Chiral perturbation theory (ChPT) is an EFT that describes the low-energy regime of QCD  in terms of the lightest non-singlet multiplet of pseudoscalar mesons. %As already commented in \cref{sec:QCD:chiralsymmetry}, these are the pseudo-Nambu-Goldstone bosons arising after spontaneous chiral symmetry breaking, see \cref{eq:QCD:spontaneouschiralsymmetrybreaking}. 
The ChPT Lagrangian is the most general $CP$-invariant Lagrangian consistent with spontaneous chiral symmetry breaking, see \cref{eq:QCD:spontaneouschiralsymmetrybreaking}. The basic building block of the ChPT Lagrangian is a representative element of the coset space, $G/H$, where the coordinates on this space represent the pNGB of the theory. A typical parametrization of this object is given in terms of the $\Nf\times \Nf$ unitary matrix,
\begin{equation}
U(x)=\text{exp}\left[i\frac{\phi(x)}{F}\right]\,,
\end{equation}
where $F$ is the bare pion decay constant and $\phi$ is a traceless matrix containing the pseudoscalar mesons. For example, for $\Nf=3$, it reads
\begin{equation}
\phi\equiv\begin{pmatrix}
\pi^0+\frac{1}{\sqrt{3}}\eta & -\sqrt{2}\pi^+ & -\sqrt{2} K^+ \\
\sqrt{2}\pi^- & -\pi^0 + \frac{1}{\sqrt{3}}\eta & -\sqrt{2}K^0 \\
\sqrt{2}K^- & \sqrt{2}\bar{K}^0 & -\frac{2}{\sqrt{3}}\eta
\end{pmatrix}\,.
\end{equation} 
Under a chiral transformation, $U(x)$ transforms non-linearly,
\begin{equation}
U(x)\rightarrow RU(x)L^\dagger\,,
\end{equation}
where $L,R\in\text{SU}(\Nf)$. This construction intrinsically reproduces the consequences of spontaneous chiral symmetry breaking. The ground state of the theory, $U_0=\mathbbm{1}$, is invariant under the unbroken group, that corresponds to vector transformations with $R=L$, but not under axial-vector ones, with $L=R^\dagger$. 

The most general chirally-invariant effective Lagrangian that can be build with the matrix $U(x)$ contains an infinite number of terms, which can be ordered in increasing powers of momentum, or, in other words, of the number of derivatives. \footnote{Note that the number of derivatives is always even due to parity conservation.} At lowest order, the Lagrangian is
\begin{equation}
\mathcal{L}_2=\frac{F^2}{4}\text{Tr}\left[\partial_\mu U^\dagger\partial^\mu U\right]\,,
\end{equation}
where the prefactor is chosen to ensure that kinetic terms for the mesons have the correct normalization. Expanding in terms of the pion field, one obtains
\begin{equation}
\mathcal{L} = \frac{1}{2}\partial_\mu\pi^0\partial^\mu\pi^0+\partial_\mu\pi^+\partial^\mu\pi^-+...
\end{equation}
Here, ... represents all the remaining terms from the expansion. These are an infinite number of interaction operators, involving all even powers of the pion field. This lowest-order Lagrangian thus makes it possible to describe interactions between any number of pNGB at tree level in terms of a single parameter, $F$. 

Up to this point, we have constructed an EFT for massless mesons. However, we would like to 
include the explicit breaking of chiral symmetry, and so the quark and meson masses. These are introduced in ChPT via a coupling to an external scalar classical field. Consider the mass term of the quarks in the QCD Lagrangian---see \cref{eq:QCD:fermionlagrangianMinkowski}---as a spurious field, $\chi$, that transforms under chiral transformations as
\begin{equation}
\chi\rightarrow R\chi L^\dagger\,,
\end{equation}
so that the mass term in the Lagrangian is chirally invariant. Then, we can build the EFT including this new spurious filed, which is taken to be a constant diagonal matrix containing the quark masses. For example, $\chi=\text{diag}(m_\text{u},m_\text{d},m_\text{s})$ for $\Nf=3$. 

A similar method makes if possible to include couplings to external pseudoscalar, vector and axial classical background fields, which can be used to study electromagnetic and weak interactions. In particular, an analysis of the axial current in the EFT allows one to identify $F$ as the pion decay constant, $\Fpi$, in de chiral limit. These external fields are also needed to define the Noether currents in the effective theory and the corresponding form factors. 

%One considers a spurious field, $\chi$, that transforms as $U(x)$,
%\begin{equation}
%\chi\rightarrow R\chi L^\dagger\,,
%\end{equation}
%and then build the most general effective Lagrangian containing $U(x)$ and $\chi$, which is taken to be a constant diagonal matrix containing the quark masses. For example, $\chi=\text{diag}(m_u,m_d,m_s)$ for $\Nf=3$. A similar method can be used to include coupling to an external pseudoscalar, vector and axial classical background fields, which can be used to study electromagnetic and weak interactions. In particular, a study of the axial current in the EFT allows to identify $F$ as the pion decay constant in de chiral limit. These external fields are also needed define the Noether currents in the effective theory, and to define the corresponding form factors. 

The addition of quark masses requires a consistent power counting. In ChPT, the quark mass is taken to be of the same order as the pion mass squared and the momentum squared,
\begin{equation}\label{eq:QCD:ChPTpowercounting}
\mathcal{O}(\delta)\sim\mathcal{O}(m_q)\sim\mathcal{O}(p^2)\sim\mathcal{O}(\Mpi^2)\,,
\end{equation}
where $\delta$ denotes the expansion parameter. 
Then, the most general Lagrangian at lowest order now contains a second term,
\begin{equation}
\mathcal{L}_2=\frac{F^2}{4}\text{Tr}\left[\partial_\mu U^\dagger\partial^\mu U\right]+\frac{BF^2}{2}\text{Tr}\left[\chi U^\dagger+\chi^\dagger U\right]\,,
\end{equation}
where $B$ is a LEC related to the chiral condensate. Expanding the Lagrangian in terms of the pion fields, one gets the new terms  
\begin{equation}
\mathcal{L}_2 \supset -\frac{1}{2}(m_\text{u}+m_\text{d})B\pi^0\pi^0+(m_\text{u}+m_\text{d})B\pi^+\pi^-+...
\end{equation}
from which the pion mass can be identified,
\begin{equation}
M_\pi^2=B(m_\text{u}+m_\text{d})\,,
\end{equation}
and justifies the power counting in \cref{eq:QCD:ChPTpowercounting}. We note that the actual range of  applicability of ChPT depends on how this small scale compares to the high energy scale at which we recover QCD. Typically, one takes $\Lambda$ to be the mass of the lowest-lying resonance, the $\rho(770)$, or uses a scale related to the pion decay constant, $4\pi\Fpi$. The latter is usually more convenient to compare against lattice computations, as then low energy observables are a function of
\begin{equation}\label{eq:QCD:chiralparameterdefinition}
\xi=\frac{\Mpi^2}{(4\pi\Fpi)^2}\,,
\end{equation}
commonly known as the \textit{chiral parameter}.

%At higher order, however, one needs to include additional terms in the Lagrangian to be consistent with the power counting scheme and properly renomalize loop divergencies~\cite{Gasser:1983yg}.

To perform computations beyond leading order (LO), higher order Lagrangian terms need to be included. At next-to-leading order (NLO), one-loop diagrams from $\mathcal{L}_2$ induce divergencies that can only be reabsorbed by operators of higher order~\cite{Weinberg_1995}. The fourth-order Lagrangian contains a larger number of terms. If one only considers purely mesonic operators, there are 9 different terms for generic $\Nf$,
\begin{equation}\label{eq:QCD:chptlagrangianL4}
\mathcal{L}_4=\sum_{i=0}^8 L_i(\mu) O_i[U(x),\chi,\partial_\mu]\,.
\end{equation}
Owing to the Cayley-Hamilton theorem~\cite{Hamilton1853,Cayley_2009,Frobenius1877}, the number of independent operators is reduced to 8 for $\Nf=3$ and to 4 for $\Nf=2$. Renormalized LECs, $L_i^\text{r}$, are defined after absorbing all one-loop divergencies. Note that LECs depend on the scale of the computation, $\mu$. Their value at two different scales $\mu_1$ and $\mu_2$ is related by
\begin{equation}\label{eq:QCD:LECsNfscaledependencegeneral}
L_i^\text{r}(\mu_2)=L_i^\text{r}(\mu_1)+\frac{\Gamma_i}{16\pi^2}\log\left(\frac{\mu_1}{\mu_2}\right)\,,
\end{equation}
where $\Gamma_i$ are rational constants~\cite{Gasser:1983yg}. In the $\Nf=2$ case, it is typical to define some scale-independent LECs, $\overline{\ell}_i$, via
\begin{equation}\label{eq:QCD:lecsNfscaledependence}
\ell_i^\text{r}(\mu) = \frac{\gamma_i}{32\pi^2}\left[\overline{\ell}_i +\log\frac{\Mpiphys^2}{\mu^2}\right]\,,
\end{equation}
where $\ell_i^r$ and $\gamma_i$ is the usual notation to denote the $\Nf=2$ LECs and their $\Gamma$-constants, respectively, and $\Mpiphys\approx 139.57$ MeV is the physical pion mass.\footnote{We will explicitly denote the physical pion mass and physical decay constant as $\Mpiphys$ and $\Fpiphys$, respectively, to distinguish them from their counterparts at non-physical quark masses, which is the typical scenario in lattice simulations.}

\subsection{ChPT at large $\Nc$}\label{sec:QCD:largeNChPT}

ChPT admits an extension to study meson interactions at large $\Nc$~\cite{Gasser:1984gg,DiVecchia:1980yfw,Rosenzweig:1979ay,Witten:1980sp,Kawarabayashi:1980dp,Leutwyler:1996sa,Herrera-Siklody:1996tqr,Kaiser:2000gs}. Recall from \cref{sec:QCD:largeNc} that in this limit, the spontaneous symmetry breaking pattern of chiral symmetry changes to that of \cref{eq:QCD:largeNcsymmetrybreakingpattern}. The singlet mesons becomes another pNGB and needs to be included into the ChPT meson matrix,
\begin{equation}
\tilde{U}(x)=U(x)\text{exp}\left(i\frac{\sqrt{2}\eta'(x)}{\sqrt{\Nf}F}\right)\,.
\end{equation}
In addition, $\Nc$ and the $\eta'$ mass are included in the power counting scheme,
\begin{equation}\label{eq:QCD:largeNChPTpowercounting}
\mathcal{O}(\delta)\sim\mathcal{O}(\Nc^{-1})\sim\mathcal{O}(\Metap^2)\,.
\end{equation}
This affects the order at which terms appear in the effective Lagrangian as LECs accompanying each operator scale as $\Nc^{1-r}$, with $r$ the number of flavor traces in the operator in question.

With this two modifications, one can construct the most general effective Lagrangian. The lowest order Lagrangian now has one additional term,
\begin{equation}
\mathcal{L}_2=\frac{F^2}{4}\text{Tr}\left[\partial_\mu \tilde{U}^\dagger\partial^\mu \tilde{U}\right]+\frac{BF^2}{2}\text{Tr}\left[\chi \tilde{U}^\dagger+\chi^\dagger \tilde{U}\right]+\frac{F^2}{4\Nf}M_0^2 X^2\,,
\end{equation}
where $X(x)=\log\det\tilde{U}+\theta$, with $\theta$ the QCD vacuum angle. From this Lagrangian, we observe $F_\pi^2\sim\mathcal{O}(\Nc)$ while $B,\chi_\text{top}\sim\mathcal{O}(\Nc^0)$. Moreover, this Lagrangian makes is possible to recover the Witten-Veneziano formula, \cref{eq:QCD:WittenVenezianoformula}, provided the new low-energy coupling obeys, $M_0^2=2\Nf\chi_\text{top}/\Nc$. %Note that the last term is not present in standard ChPT as $\det U=0$. 

In large $\Nc$ ChPT, terms analogous to those in \cref{eq:QCD:chptlagrangianL4} get split in two different orders of the chiral expansion according to the scaling with $\Nc$ of the respective LECs. For $\Nf\geq 4$ one has~\cite{Manohar:1998xv}
\begin{equation}\label{eq:QCD:largeNcscalingLECs}
\mathcal{O}(\Nc)\,: L_0, L_3, L_5, L_8\,, \quad\quad\quad \mathcal{O}(\Nc^0)\,: L_1, L_2, L_4, L_6, L_7\,. 
\end{equation}
Thus, at NLO only tree diagrams including the $\mathcal{O}(\Nc)$ LECs needs to be considered, while those with $\mathcal{O}(\Nc^0)$ LECs first enter at next-to-next-to-leading order (NNLO), together with one-loop diagrams from $\mathcal{L}_2$. New operators involving the $\eta'$ also appear at higher-order Lagrangians---see \rcite{Kaiser:2000gs}. 

Different phenomenological approaches have tried to determine the leading $\Nc$ dependence of the LECs. This is the case of the resonant chiral theory~\cite{Ecker:1988te}, in which ChPT is matched to a theory including heavy resonances and the LECs result from the exchange of these particles. The actual values of the LECs depend on the features of the assumed spectrum, as well as on the imposition of different positivity bounds.

Before concluding, two comments are in place. First, determining the $\Nc$ scaling of the LECs for $\Nf=2$ and $3$ is not so straightforward. Due to Cayley-Hamilton relations, some two-trace operators have enhanced $\Nc$ dependence. For example, at $\Nf=3$, both $L_1$ and $L_2$ scale as $\Nc$ and only the combination $L_1-2L_2$ is $\mathcal{O}(\Nc^0)$~\cite{Manohar:1998xv}. 

Second, although we use the same notation to denote the $SU(\Nf)$ and $U(\Nf)$ LECs, they do not take the same values. Nevertheless, it is possible to match both theories by integrating out the $\eta'$. Assuming $\Nf$ degenerate quark flavors, one finds~\cite{Peris:1994dh,Herrera-Siklody:1998dxd}

\noindent\begin{equation}\label{eq:QCD:matchingLECs}
\begin{array}{rl}
\displaystyle\left[L_6\right]_{\text{SU}(\Nf)}=&\displaystyle\left[L_6\right]_{\text{U}(\Nf)}+\frac{1}{16\Nf^2(4\pi)^2}(\lambda_0-1)\,,\\[10pt]
\displaystyle\left[L_7\right]_{\text{SU}(\Nf)}=&\displaystyle\left[L_7\right]_{\text{U}(\Nf)}-\frac{\Fpi^2}{16\Nf M_0^2}\,,\\[10pt]
\displaystyle\left[L_8\right]_{\text{SU}(\Nf)}=&\displaystyle\left[L_8\right]_{\text{U}(\Nf)}-\frac{\lambda_0}{8\Nf(4\pi)^2}\,,
\end{array}
\end{equation}
where $\lambda_0=\log(M_0^2/\mu^2)$. We note that the values of the SU($\Nf$) LECs depend implicitely on $\Nf$. Thus, quantities computed in traditional ChPT cannot be expanded in the standard way in the large $\Nc$ limit, only with positive powers of $\Nf$. In addition, in the SU($\Nf$) theory, $L_7$ presents an enhanced dependence on the number of colors, $\left[L_7\right]_{\text{SU}(\Nf)}\sim\mathcal{O}(\Nc^2)$~\cite{Peris:1994dh}.

\chapter[Hadron interactions]{Hadron interactions\phantom{y}}
\label{sec:hadrons}

QCD stands out from other quantum field theories due to its richness. At low energies, the phenomena of asymptotic freedom and confinement make a vast spectrum of hadrons emerge. The Particle Data Group~\cite{PDG:2020} summarizes the properties of all experimentally observed hadrons. These include mesons, baryons, and other observed states that are suspected to have more a more exotic structure. 

In spite of their large number, almost every hadron is unstable and cannot be directly observed in experiment. Even in the absence of electroweak interactions, most hadrons would rapidly decay into lighter ones, having a lifespan of less than $10^{-22}$ seconds in many cases. The properties of these unstable particles are instead inferred from the interactions of their byproducts.%, which are typically enhanced at energies close to the mass of one of these unstable particles. 

The short lifetime of many hadrons make the study of their properties from QCD incredibly challenging. The lattice, meanwhile, is in principle precluded from the study of real-time infinite-volume processes, as is the case of hadron interactions or decays.

In this chapter, we discuss how hadron interactions can be studied using ChPT and lattice QCD. In \cref{sec:hadrons:scatteringQCD} we introduce the main concepts used to describe two- and three-particle scattering. We then explain, in \cref{sec:hadrons:interactionsinChPT}, how ChPT can be used to study interactions of pseudoscalar mesons and how it can be modified to allow for the presence of resonances. Finally, \cref{sec:hadrons:particleinteractionslattice} explains the two-and three-particle finite-volume formalisms, that allow one to constrain infinite-volume scattering observables from the finite-volume multiparticle energy spectrum determined in lattice QCD computations. %The application of these formalisms will be a central part of this part of this doctoral thesis.

\newpage\section{Particle scattering in infinite volume}\label{sec:hadrons:scatteringQCD}

Experimentally, particle interactions can only be probed indirectly. The paradigmatic particle-physics experiment characterizes interactions between particles by studying the properties of the initial (ingoing) and final (outgoing) asymptotic states. Particles in both the initial and final states are assumed to be separated enough that they can be considered non-interacting. The process in which the particles of some initial state interact, leading to a different final one is known as a \textit{scattering process}. 

Theoretically, ingoing and outgoing states are defined asymptotically as a set of free non-interacting particles in the infinite past and future, respectively. We refer to these two states as $|i\rangle$ and $|f\rangle$, in this same order. The scattering matrix, or $S$-matrix, is defined as the matrix containing the transition probabilities between any two states of a given theory,%\footnote{We always use the first/second index of matrices to refer to the final/initial state.}
\begin{equation}
S_{fi}=\langle f|i\rangle\,.
\end{equation}
Therefore, it encodes all physical information that can be extracted from the quantum field theory. Due to conservation of probability, the $S$-matrix is unitary, $SS^\dagger=\mathbbm{1}$. 

Usually, one is interested in the non-trivial part of the scattering matrix, known as the scattering amplitude, $\cM$. If we consider a Lorentz invariant theory, as is the case of QCD, we define
\begin{equation}
S_{fi}=\delta_{fi}+i(2\pi)^4\delta^{(4)}(P_f-P_i)\cM_{fi}\,,
\end{equation}
where $P_{i}$ and $P_{f}$ refer to the total four-momentum of the initial and final states, respectively. %, and we use the standard normalization for a (3+1)-dimensional quantum field theory.

The analytic properties of the $S$-matrix play an essential role in the understanding of the underlying physics~\cite{eden2002analytic}. When considered as a function of the \textit{center-of-mass frame} (CMF) energy, the $S$-matrix contains branch cuts on the real axis, with branching points at the thresholds of multiparticle states. In addition, it also contains isolated poles. 

%As we have discussed, most hadrons are unstable, and so cannot be defined as part of asymptotic states. This means they do not form part of the Fock space of the theory. Instead, their existence and properties are inferred from the scattering of stable states. 

In the $S$-matrix, single-particle states appear as poles at some value of the energy in the CMF, $E^*$. Stable states correspond to poles on the real axis of the first Riemann sheet, and are either single hadrons, such as nucleons, or \textit{bound states} that appear close to some multiparticle threshold, such as the deuteron. These states can exist as asymptotically.

In addition, we have unstable states, which represent the majority of hadrons. These do not form part of the Fock space of QCD and cannot be defined asymptotically. Instead, their properties are inferred from the scattering of stable states. In the $S$-matrix, unstable states correspond to poles on the second Riemann sheet. If the pole lies in the lower complex plane, $E^*=M-i\Gamma/2$, the particle is known as a \textit{resonance}, with mass $M$ and decay width $\Gamma$. This is for example the case of the $\rho(770)$ resonance. %Physical resonances lie on the lower complex plane of the last Riemann sheet.   
If the pole instead lies on the real axis below some multiparticle threshold, $E^*=M$, the state is known as a \textit{virtual bound state}. %, with mass $M$ and binding energy $B=E_\text{thr}-M$, with $E_\text{thr}$ the energy of said multiparticle threshold. Bound states are either real or virtual, depending on whether they lie on the first or the second Riemann sheet associated to that multiparticle branch cut. Bound statesA well known example is the deuteron, a real bound-state of a proton and a neutron. 

\subsection{Two-particle scattering}\label{sec:hadrons:twoparticlescatteringinfinitevolume}

%The work presented on this part of the dissertation is centered on two- and three-particle interactions. It is thus convenient to study the properties of the scattering matrix in this case more in detail. In particular, we will consider the two cases separately, assuming no two-to-three interactions. We will also  focus on the case of particles of the same mass.

We now consider two-particle interactions, and focus on the case in which all particles have equal mass $m$. Let $\{k_1,k_2\}$ and $\{p_1,p_2\}$ be the ingoing and outgoing momenta of the particles in the initial and final states, respectively. The total momenta is then $P=k_1+k_2=p_1+p_2$. These are are all on shell, this is, they obey the relativistic dispersion relation, $p_a^2=k_a^2=m^2$. The two-particle scattering amplitude can then be written as a function of the Mandelstam variables,
\begin{equation}
s=(k_1+k_2)^2\,,\quad\quad t=(k_1-p_1)^2\,,\quad\quad u=(k_1-p_2)^2\,.
\end{equation}
These are related as
\begin{equation}
s+t+u=4m^2\,,
\end{equation}
and so the two-particle scattering amplitude, $\cM_2$, can be written as a function of only two independent variables.\footnote{This result stands from the difference between the number of free momenta components of the initial and final states and the number of generators of the Poincaré group, $12-10=2$.} These variables  are usually chosen to be $s$, related to the CMF energy, $s=E^{*2}$, and the scattering angle between the initial- and final-state particles three-momenta, $\theta$, defined from
\begin{equation}
t=-4q_2^{*2}\sin^2\frac{\theta}{2}\,,
\end{equation}
where $q_2^*=\sqrt{s/4-m^2}$ is the magnitude of the relative three-momentum in the CMF.

In rotational-invariant theories, as is the case of QCD, it is common to project $\cM_2$ to definite angular momentum, 
\begin{equation}
\cM_2(s,\theta)=\sum_{\ell',m'}\sum_{\ell,m} 4\pi Y^{*}_{\ell'm'}(\theta) Y_{\ell m}(\theta)\delta_{\ell'\ell}\delta_{m'm}\cM_{2,\ell}(s)\,.
\end{equation}
Here $\ell$ and $m$ are the total angular momentum and the corresponding azimuthal component of the initial state, and primed quantities refer to the final state. Note that for rotational invariant theories $\cM_2$ does not depend on the azimuthal angular momentum. 

For a given partial wave, the unitarity property of the $S$-matrix leads to constrains on the imaginary part of the scattering amplitude,
\begin{equation}\label{eq:hadrons:opticaltheorem}
\rho(s)|\cM_{2,\ell}(s)|^2=\Im\,\cM_{2,\ell}(s)\,,
\end{equation}
where we have introduced the two-particle phase-space factor,
\begin{equation}\label{eq:hadrons:phasespace}
\displaystyle\rho(s)=\frac{q_2^*}{16\pi s}\,.
\end{equation} 
\Cref{eq:hadrons:opticaltheorem} is known as the \textit{optical theorem}, and characterizes the analytic structure of $\cM_{2,\ell}$. It makes it possible to separate $\cM_{2,\ell}$ as 
\begin{equation}\label{eq:hadrons:Kmatrix}
\cM_{2,\ell}^{-1}=\cK_{2,\ell}^{-1}(s)-i\rho(s)\,,
\end{equation}
where $\cK_2$ is the two-particle $K$-matrix. %, that contains information about short-range interactions. 
Its components are meromorphic functions that take real values for physical kinematics (i.e., for $s>0$). \Cref{eq:hadrons:opticaltheorem,eq:hadrons:Kmatrix} also make clear the presence of a square-root branch cut in the two-particle scattering amplitude, which has a kinematic origin.

From $\cK_{2,\ell}$, one can define the scattering phase shift, $\delta_\ell$, that connects with quantum-mechanical scattering theory,
\begin{equation}
\cK_{2,\ell}^{-1}(s)=\rho(s)\cot\delta_\ell(s)\,.
\end{equation}
Close to threshold it is typical to expand the phase shift using an \textit{effective-range expansion} (ERE),
\begin{equation}\label{eq:hadrons:effectiverangeexpansion}
q_2^{*2\ell+1}\cot\delta_\ell(s)=\frac{1}{a_\ell}+r_\ell\frac{q_2^{*2}}{2}+...
\end{equation}
where $a_\ell$ is known as the \textit{scattering length} (although only $a_0$ has dimension of length) and $r_\ell$ is the \textit{effective range}. Note that we are using the convention for the scattering length that is positive for attractive interactions and negative for repulsive ones. The powers of $\qtwos$ on the left-hand-side of \cref{eq:hadrons:effectiverangeexpansion} are required to ensure $\cK_2$ is smooth at threshold, and so high partial waves are suppressed close to threshold. This is vital in making the study of multiparticle systems on the lattice viable---see \cref{sec:hadrons:QC2}.

If a bound state is present close to a $s$-wave two-particle threshold, the effective range and effective length  provide information about the nature of the state. Weinberg's criterium~\cite{Weinberg:1965zz,Matuschek:2020gqe} relates the probability of a bound state being a compact particle or a molecular state to its field renormalization factor, that can be computed as
\begin{equation}\label{eq:hadrons:Weinbergcriterium}
Z=1-\left[1-\frac{2r_0}{a_0}\right]^{-1/2}\,.
\end{equation}
A value of $Z$ close to unity indicates a state of compact nature, while a small value of $Z$ would point towards a hadronic molecule.

\newpage\subsection{Three-particle scattering}\label{sec:hadrons:infinitevolumethreeparticlescattering}

In the three-particle case, the description of a scattering process is more complicated. We let $\{k_1,k_2,k_3\}$ and $\{p_1,p_2,p_3\}$ be the momenta of the three particles in the initial and final state, respectively, and $P=k_1+k_2+k_3=p_1+p_2+p_3$ the total momenta. 

The three-particle scattering amplitude, $\cM_3$, can be written as a function of eight independent variables. However, in the line of the relativistic-field-theory (RFT) three-particle finite-volume formalism, to be introduced in \cref{sec:hadrons:threeparticlesfinitevolume}, we describe it redundantly in terms of 11 kinematic variables. These are the CMF energy, $E^*$, and five variables characterizing each of the initial and the final states. The dependence on $E^*$ is kept implicit. 

The three particles in the initial and final state are typically separated in an \textit{interacting pair} or \textit{dimer}, and a \textit{spectator}. Each state is described with the three-momentum of the spectator, $\bm{k}$ and $\bm{p}$ for the initial and final state, respectively, together with the direction of the relative momentum of the dimer particles in their CMF, denoted by $\hat{\bm{a}}_k$ and $\hat{\bm{a}}_p$ for the initial and final dimer, in this same order.\footnote{We use momentum subscripts to indicate that a quantity is defined in the CMF of the dimer associated to an spectator with that momentum.} 

Scattering quantities are typically projected to into partial waves of the dimers. For example, the three-particle scattering amplitude is projected as
\begin{equation}
\cM_3(\bm{p},\hat{\bm{a}}_p^*;\bm{k},\hat{\bm{a}}_k^*)=\sum_{\ell'm'}\sum_{\ell m}4\pi Y_{\ell'm'}^*(\hat{\bm{a}}_p^*)\cM_3(\bm{p},\bm{k})_{\ell'm';\ell m}Y_{\ell m}(\hat{\bm{a}}_k^*)\,,
\end{equation}
 We note that the partial-wave-projected three-particle scattering amplitude still depends on 11 variables, which are the three-momenta of the initial and final spectator, $\bm{k}$ and $\bm{p}$, the angular momentum indices of the initial and final dimer, $(\ell, m)$ and $(\ell',m'$), and the total energy, $E^*$.

It is common in the context of the formalism, for legibility, to leave the angular momentum indices implicit. Thus, $\cM_3(\bm{p},\bm{k})$ refers to the three-particle scattering amplitude projected to the partial waves of both the initial and final spectator. This same conventions are also kept for other quantities, for example, the scattering amplitude of the dimer or the corresponding phase space. We stress that, instead of indicating that these quantities depend on the Mandelstam variables of the corresponding dimer, we typically keep the three-momentum of the associated spectator as the argument. For instance, $\cM_2(\bm{k})$ has to be understood as the partial-wave projected two-particle scattering amplitude of the dimer, $\cM_2(\bm{k})\equiv\cM_{2,\ell}(s)$, with $s=(P-k_3)^2$ and $k=(\sqrt{\bm{k}^2+m^2},\bm{k}^2)$. This same notation holds for the phase space factor, defined in \cref{eq:hadrons:phasespace}, this is, we use $\rho(\bm{k})\equiv\rho(s)$.

As in the case of two particles, the analytic structure of $\cM_3$ is constrained by the unitarity of the $S$-matrix. One can rewrite $\cM_3$ as a function of both $\cM_2$, as pairwise scattering can happen within a three particle system, and a three-particle $K$-matrix that contains the information about short-range three-particle interactions. The definition of this quantity, however, is not unique and requires a prescription to separate successive two-particle scattering from short-range three-particle interactions. We here follow the approach of the RFT formalism~\cite{Hansen:2014eka,Hansen:2015zga}, in which a non-divergent or  \textit{divergence-free} (df) three-particle $K$-matrix is defined, called $\Kdf$. We also focus, for now, on the case of identical particles.

The first step to define $\Kdf$ is to remove possible divergences from $\cM_3$ that originate from successive two particle interactions~\cite{Rubin:1966zz,Brayshaw:1968yia,PhysRevA.16.2264,PhysRevA.16.2276}. The divergence-free three-particle scattering amplitude is introduced as,
\begin{equation}\label{eq:hadrons:divergencefreeamplitude}
\Mdf(\bm{p},\bm{k}) = \cM_3(\bm{p},\bm{k}) - \cS\left\{ \cD^\uu(\bm{p}, \bm{k})\right\}\,.
\end{equation}
Here $\cS$ indicates symmetrization over all nine possible assignments of initial and final momenta, and $\cD^\uu$ is the unsymmetrized subtraction term. This term satisfies an integral equation,
\begin{equation}\label{eq:hadrons:generalsubtraction}
    \cD^\uu(\bm p, \bm k) = 
        - \cM_2(\bm p) G^\infty(\bm p, \bm k) \cM_2(\bm k)
        - \int_r \cM_2(\bm p) G^\infty(\bm p, \bm r) \cD^\uu(\bm r, \bm{k})\,.
\end{equation}
where, recall, $\cM_2(\bm{q})$ denotes the partial-wave-projected scattering amplitude of a dimer associated to an spectator with three-momentum $\bm{q}$. Also we define $\int_r \equiv \int d^3 r/[2\omega_r (2\pi)^3]$, with $\omega_r =\sqrt{\bm r^2+m^2}$, and
\begin{equation}\label{eq:hadrons:Ginftydefinition}
    G^\infty(\bm p, \bm k)_{\ell' m';\ell m} =
        \left(\frac{ k_p^*}{q_{2,p}^*}\right)^{\!\ell'}
        \frac{4\pi Y_{\ell' m'}(\bm{\hat{k}}_p^*) H(x_p) H(x_k) Y^*_{\ell m}(\bm{\hat{p}}_k^*)}{b_{pk}^2-m^2 + i\epsilon}
        \left(\frac{ p_k^*}{q_{2,k}^*}\right)^{\!\ell}
        \,.
\end{equation}
Here, $b_{pk}=P-p-k$, with $k=(\omega_k,\bm{k})$ and $p=(\omega_p,\bm{p}$), $p_k^*$ refers to the magnitude of $\bm{p}$ on the CMF of the pair associated with spectator of momentum $k$, $q_{2,k}^*=|\hat{\bm{a}}_k^*|$, and similarly for $k_p^*$ and $q_{2,p}^*$, respectively. Finally $H(x)$ is any smooth cutoff function that vanishes for $x\leq0$ and is equal to unity for $x\geq 1$, with $x_q=(P-q)^2/4m^2$. The dependence on this arbitrary function sets a scheme, making  $\Kdf$ scheme-dependent. A typical choice for $0<x<1$ is~\cite{Hansen:2014eka}
\begin{equation}\label{eq:hadrons:standardcutoff}
H(x)=\text{exp}\left[-\frac{1}{x}\text{exp}\left(-\frac{1}{1-x}\right)\right]\,.
\end{equation}
%Note from \cref{eq:hadrons:Ginftydefinition} that $G^\infty$  by the same variables as $\cM_3$.

After eliminating the divergences in $\cM_3$, $\Kdf$ is obtained from another integral equation,
\begin{equation}\label{eq:hadrons:TandMdfintegralequation}
    \Mdf(\bm p, \bm  k) 
        = \cS\left\{\int_s \int_r 
            \cL^\uu(\bm p, \bm s) \cT(\bm s, \bm r) 
            \cR^\uu(\bm r, \bm k) \right\}\,,
\end{equation}
where
\begin{equation}
    \cL^\uu(\bm p, \bm k) 
        =  \left[ \frac13 + i \cM_2(\bm p) \rho(\bm p) \right] \bar\delta(\bm p - \bm k) 
        +i \cD^\uu(\bm p, \bm k) \rho(\bm k)\,,
    \label{eq:hadrons:Ldef}
\end{equation}
\begin{equation}\vspace{0.15cm}
    \cR^\uu(\bm p, \bm k)  
        = \bar\delta(\bm p - \bm k) \left[ \frac13 +i \rho(\bm p) \cM_2(\bm p)  \right] 
        +i \rho(\bm p) \cD^\uu(\bm p, \bm k)\,,
    \label{eq:hadrons:Rdef}
\end{equation}
with $\bar\delta(\bm p - \bm k) \equiv 2\omega_k (2\pi)^3 \delta^{(3)}(\bm p - \bm k)$, and
\begin{equation}\label{eq:hadrons:integralequationKdf}
    \cT(\bm p, \bm k) 
        = \Kdf(\bm p, \bm k)
        - \int_s \int_r \Kdf(\bm p, \bm s) \rho(\bm s) \cL^\uu(\bm s, \bm r) \cT(\bm r, \bm k)\,.
\end{equation}
\Cref{eq:hadrons:TandMdfintegralequation} separates the divergence-free amplitude into three factors. Both $\cL$ and $\cR$, often called \textit{decorators}, depend only on the two-particle scattering amplitude. Their role here is to remove from $\Mdf$ the effect of long-range pairwise interactions happening before or after short-range three-particle interactions take place. The remaining factor, $\cT$, contains the effect of these three-particle interactions. It is given by an integral equation, \cref{eq:hadrons:integralequationKdf}, which includes a factor of $\cL$, that represents possible long-range pairwise processes that can take place between successive short-range three-particle interactions. We finally note that, when solving this last equation, one must enforce $\Kdf$ to be symmetric under particle exchange. The study of the solutions of the integral equations relating $\Kdf$ to $\cM_3$ is an active field of study~\cite{Jackura:2020bsk,Dawid:2023jrj,Dawid:2023kxu}.

As in the case of the two-particle $K$-matrix, $\Kdf$ can be expanded about threshold. As it is a function of a larger number of kinematical variables, more than one term can appear at each order in the threshold expansion. In the case of three identical particles of mass $m$, these terms must be invariant under any permutations of the particle in the initial and final states, as well as under the exchange of the initial and final states themselves. Up to quadratic order it contains five different terms~\cite{Blanton:2019igq},
\begin{equation}\label{eq:hadrons:thesholdexpansionKdf}
\Mpi^2\Kdf=\cK_0+\cK_1\Delta+\cK_2\Delta^2+\KA\DA+\KB\DB+\cO(\Delta^3)\,,
\end{equation}
where
\begin{equation}\label{eq:hadrons:thesholdexpansionKdfingredients}
\Delta=\frac{P^2-9m^2}{9m^2}\,,\quad\DA=\sum_i(\Delta_i^2+\Delta_i^{\prime \,2})-\Delta^2\,,\quad\DB=\sum_{i,j} \tilde{t}_{ij\,2}-\Delta^2\,.
\end{equation}
Here, we have defined
\begin{equation}\label{eq:hadrons:tvariablesDeltavariables}
\tilde{t}_{ij}=\frac{(p_i-k_j)^2}{9m^2}\,,\quad\quad \Delta_i=\frac{(P-k_i)-4m^2}{9m^2}\,,\quad\quad \Delta_i^\prime=\frac{(P-p_i)-4m^2}{9m^2}\,.
\end{equation}
Note that these quantities are not all independent---see \rcite{Blanton:2019igq}. 

In general, one can determine the number of terms that appear at some order in the threshold expansion of $\Kdf$ using group-theory argument. For three mass-degenerate non-identical particles the allowed operators should transform under some irrep of the $G=(S_3\times S_3^\prime)\rtimes Z_2$ group, where $S_3^{(\prime)}$ is the permutation group with three elements, acting on the momenta of the initial (final) state, and $Z_2$ refers to exchange of the initial and final states. This is discussed in detail in app.~B of \rcite{Baeza-Ballesteros:2024mii}.

%{\jorge [Escribir sobre threshold expansion y comentar acerca de isospin generalization. Refs a Raúl y integral eqs. Añadir isospin channels en chap 1?]}
% comment on crossing symmetry and dispersion relations

\section{Meson interactions from ChPT}\label{sec:hadrons:interactionsinChPT}

%Obtaining analytical predictions about the scattering properties of hadrons directly from QCD is not possible. 
ChPT provides a tool to describe interactions between mesons using an effective Lagrangian, as explained in \cref{sec:QCD:chiralperturbationtheory}. If we consider isospin symmetry to be exact, pseudoscalar mesons organize in multiplets of the isospin group, SU($\Nf$)---see \cref{sec:QCD:chiralsymmetry}. What is more, interactions between multiple mesons can be classified in scattering channels, which correspond to different irreps of this same group. This means that both the initial and final scattering states lie on the same irrep, and so the scattering amplitude is (block-)diagonal in isospin.

Consider the simple $\Nf=2$ case. The pseudoscalar mesons are the pions, $\{\pi^+,\pi^0,\pi^-\}$, which form an isopin triplet with total isospin $I_\pi=1$. Scattering of two pions can happen in three scattering channels,
\begin{equation}
3\,\otimes\,3\,=\,5\,\oplus\, 3\,\oplus\, 1\,,
\end{equation}
corresponding to total isospin $I_{\pi\pi}=2$, $1$ and $0$, respectively. Isospin is conserved in scattering processes, and so the two-particle scattering amplitude is diagonal in isospin.

In the case of three pions, one first combines two of them into two-pion scattering channels and then adds the third particle~\cite{Hansen:2020zhy}, 
\begin{equation}
3\,\otimes\,3\,\otimes\,3\, = (5\,\oplus\, 3\,\oplus\, 1)\,\otimes\, 3 = 7\,\oplus\, (5\,\oplus \,5)\,\oplus\, (3\,\oplus\, 3\, \oplus \, 3) \,\oplus\, 1\,.
\end{equation} 
The combination of three isospin-one particles leads to seven different irreps, which can be organized into four scattering channels, corresponding to three-particle isospin $I_{\pi\pi\pi}=3,2,1$ and 0. 

The $I_{\pi\pi\pi}=2$ and 1 irreps have non-zero multiplicity. Typically, each copy is characterized by the isospin of the two-particle subsystem from which the three-pion irrep was created. These are also the possible isospin in which pairwise interactions can happen within each channel. In the case of the $\Ippp=2$ channel, pion-pion interactions can happen with $I_{\pi\pi}=2$ and $1$, while for $I_{\pi\pi\pi}=1$ all three two-particle channels are allowed. Note however, that the scattering amplitude is not diagonal in these subchannels, and so they mix.  In the $I_{\pi\pi\pi}=3$ and $\Ippp=0$ channels, on the other hand, pairwise interactions can only happen with $I_{\pi\pi}=2$ and 1, respectively.  

\subsection{Two-pion scattering in ChPT}\label{sec:hadrons:twopionsChPT}

ChPT can be used to obtain prediction for the pion-pion scattering amplitude. For the maximal isospin channel, the LO result is~\cite{Weinberg:1978kz}
\begin{equation}\label{eq:hadrons:scatteringamplitudetwopionsI2}
\cM_2^{\Ipp=2,\LO}=\frac{1}{\Fpi^2}(2\Mpi^2-s)\,.
\end{equation}
The scattering amplitude and the effective range obey
\begin{equation}
a_0^{\Ipp=2}=-\frac{\Mpi^2}{16\pi^2\Fpi^2}\,,\quad\quad\quad M_\pi^2a_0^{\Ipp=2}r_0^{\Ipp=2}=-3\,.
\end{equation}
These results, especially that for $a_0^{\Ipp=2}$, lead to very good agreement with experimental observations~\cite{FLAG:2021npn}. Higher-order results, which are known up to NNLO~\cite{Gasser:1983yg,Bijnens:1995yn}, only amount to some minor corrections. Note however, this is not true for every scattering amplitude. In some cases, NLO correction can be large, even bigger than the LO result. We will see an example of this when studying the three-pion $K$-matrix in \cref{sec:pipipiKmatrix,sec:isospinKmatrix}.

One of the limitations of ChPT, as we have already commented, is that it leads to scattering amplitudes that contain no poles. This limits its range of applicability to energies below that of the lowest lying resonance. One option to circumvent this hindrance is the introduction of additional fields in the chiral Lagrangian representing these resonances~\cite{Ecker:1988te,Ecker:1989yg}. This is the base of the resonant chiral theory, that allows one to determine the value of the LECs from integrating out these additional degree of freedom.

Another compelling alternative is the inverse amplitude method (IAM)~\cite{Truong:1988zp,Dobado:1989qm,Dobado:1992ha,Hannah:1995si,Dobado:1996ps}. The basis of this technique are the perturbative expansion of the partial-wave projected scattering amplitude,
\begin{equation}
\cM_{2,\ell}\approx\cM_{2,\ell}^{(0)}+\cM_{2,\ell}^{(1)}+\mathcal{O}(\delta^2)\,,
\end{equation} 
and the observation that, when working at some fixed order in the chiral expansion, unitarity conditions are only satisfied perturbatively,
\begin{equation}
\Im\cM_{2,\ell}^{(0)}=0\,,\quad\quad\Im\cM_{2,\ell}^{(1)}=\rho\cM_{2,\ell}^{(0)2}\,.
\end{equation}
This can be combined with dispersion relations to solve for the scattering amplitude. One obtains a result that satisfies perturbative unitarity exactly,
\begin{equation}
\cM_{2,\ell}\approx \frac{\cM_{2,\ell}^{(0)2}}{\cM_{2,\ell}^{(0)}-\cM_{2,\ell}^{(1)}}\,.
\end{equation}
At low energies, the standard ChPT expansion is recovered. However, it differs from the standard approach starting at $\mathcal{O}(\delta^2)$. This implies that the numerical values of the LECs in the unitarized theory will be similar, yet different, from those in standard ChPT~\cite{Truong:1988zp,Dobado:1992ha,Dobado:1996ps}. Note this procedure can be extended to include higher orders in the perturbative expansion~\cite{Dobado:1996ps}.

The amplitudes obtained with the IAM may present resonant poles, and so can be used to describe scattering processes in channels containing a resonant state, without the need of including them explicitly in the chiral Lagrangian.  For example, the IAM has been used to fit experimental data from meson scattering~\cite{Dobado:1992ha}, and also, in combination with large $\Nc$ arguments, to study the dependence of resonances with $\Nc$~\cite{Pelaez:2003ip,Pelaez:2005pi,Pelaez:2006nj}.

\subsection{Three-pion scattering at LO in ChPT}\label{sec:hadrons:ChPTthreepionsLO}
 
\begin{figure}[!b]
\tikzexternalize[--output-directory=./Chapters/Chapter_2/Figures/tikz/]
    \centering
    \begin{subfigure}{0.48\textwidth} 
        \centering
        \begin{tikzpicture}[xscale=\diagramxscale,yscale=\diagramyscale]
            \makeexternallegsshifted
            \coordinate (v1) at (+.5,-.5);
            \coordinate (v2) at (-.5,+.5);
            \draw[dprop] (k1) -- (v2) -- (p1);
            \draw[dprop] (k2) -- (v1) -- (v2) node [midway, above right] {$b$} -- (p2);
            \draw[dprop] (k3) -- (v1) -- (p3);
        \end{tikzpicture}
        \caption{OPE diagram}
        \label{fig:hadrons:pipipiLOOPE}
    \end{subfigure}
    \begin{subfigure}{0.48\textwidth}
        \centering
        \begin{tikzpicture}[xscale=\diagramxscale,yscale=\diagramyscale]
            \makeexternallegs
            \coordinate (v) at (0,0);
            \draw[dprop] (k1) -- (v) -- (p1);
            \draw[dprop] (k2) -- (v) -- (p2);
            \draw[dprop] (k3) -- (v) -- (p3);
        \end{tikzpicture}
        \caption{Contact diagram}
        \label{fig:hadrons:pipipiLOnOPE}
    \end{subfigure}
    \caption{
        Feynman diagrams contributing to $\cM_3$ at LO for maximal isospin.
        For diagram (a), there are  eight additional diagrams corresponding to the symmetrization of initial and final momenta. In all cases, time flows from right to left.}
    \label{fig:hadrons:pipipiLO}
\end{figure}
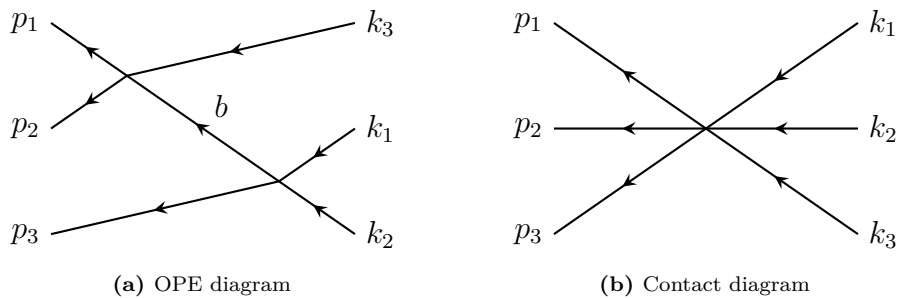

Three-pion scattering can too be described using ChPT, and scattering amplitudes are known up to NLO~\cite{Blanton:2019vdk,Bijnens:2021hpq,Bijnens:2022zsq}. These amplitudes can be separated in several parts, according to the topology of the Feynman diagrams that contribute to each of them. Here we focus on the maximal isospin case, $I_{\pi\pi\pi}=3$, at LO. The NLO case is studied in \cref{sec:pipipiKmatrix} and the remaining isospin channels are considered in \cref{sec:isospinKmatrix}. The isospin-three three-particle scattering amplitude, at LO, can be divided into two parts, corresponding to the two Feynman diagrams represented in \cref{fig:hadrons:pipipiLO}.
 
The first part is the so-called \textit{one-particle exchange} (OPE) amplitude, $\cM_3^{\LO,\OPE}$, which contains one-particle reducible diagrams that represent two consecutive pairwise interactions. The amplitude is given by the Feynman diagram in \cref{fig:hadrons:pipipiLOOPE}, after symmetrizing over the initial and final momenta. For a the choice of the interacting pair in \cref{fig:hadrons:pipipiLOOPE}, the unsymmetrized OPE amplitude is

\noindent\begin{equation}\label{eq:hadrons:OPEamplitude}
\cM_3^{\uu\LO,\OPE}(\bm{p}_3,\bm{k}_3)=-\cM^\LO_{2,\off}(\bm{p}_3)\frac{1}{b^2-\Mpi^2+i\epsilon}\cM^\LO_{2,\off}(\bm{k}_3)\,,
\end{equation}
where $\cM^\LO_{2,\off}$ is the LO two-particle amplitude in which one of the external particles is left off-shell, and $b=P-k_3-p_3$ is the momentum of the exchanged particle, which is the off-shell one. We recall that we indicate as the argument of the partial-wave-projected two-particle amplitude of dimer the three-momentum of the corresponding spectator and leave the angular momentum indices implicit. At LO, $\cM_{2,\off}$ is purely $s$-wave. For example, for the initial state it takes the form
\begin{equation}\label{eq:hadrons:M2offshellLO}
\cM_{2,\off}^{\LO}(\bm{k}_3)=\frac{1}{F_\pi^2}(t+u-2\Mpi^2)=\frac{1}{\Fpi^2}\left[(2\Mpi^2-s)+(b^2-\Mpi^2)\right]\,,
\end{equation}
where we have used the Mandelstam variables corresponding to the initial dimer,
\begin{equation}
s=(k_1+k_2)^2\,,\quad\quad t=(k_1-p_3)^2\,,\quad\quad u=(k_2-p_3)^2\,.
\end{equation}
%In the diagram in \cref{fig:hadrons:pipipiLOOPE}, the off-shell particle is the exchanged one, with momentum $b=P-k_3-p_3$. 
The off-shellness is introduced in the amplitude via the relation
\begin{equation}\label{eq:hadrons:offsellcondition}
s+t+u=3\Mpi^2-b^2\,.
\end{equation}
corresponding to the case in which the off-shell momentum has the same spacial components as its on-shell counterpart, but does not obey the on-shell dispersion relation, $b^2\neq\Mpi^2$. The total OPE amplitude is recovered from symmetrizing $\cM_3^{\uu\LO,\OPE}$ over all different assignments of momenta. %In other isospin channels, one needs to take into account the transformation properties under particle exchange when summing the different contributions, see \cref{sec:isospinKmatrix} and \rcite{Hansen:2020zhy}.

The second part of the $I_{\pi\pi\pi}=3$ amplitude is the non-OPE part, that contains all the remaining diagrams that are one-particle irreducible. At LO, this is just the diagram in \cref{fig:hadrons:pipipiLOnOPE},
\begin{equation}\label{eq:hadrons:amplitudeLOnonOPE}
\cM_3^{\LO,\nOPE}=\frac{1}{\Fpi^4}(-4P^2+18\Mpi^2)\,.
\end{equation}

The total amplitude is then the sum of the two parts
\begin{equation}
\cM^\LO_3=\cM_3^{\LO,\OPE}+\cM_3^{\LO,\nOPE}\,.
\end{equation}
Note, however, that this separation is arbitrary. The OPE contribution is defined from \cref{eq:hadrons:OPEamplitude}, but one is free to modify the off-shell part in \cref{eq:hadrons:M2offshellLO}, given there by the term proportional to $b^2-\Mpi^2$. The non-OPE part is the remainder of the full amplitude. The choice of a different parametrization of the pion fields may also lead to a different off-shell convention. %In this dissertation, we follow the amplitude definitions in \rcite{Bijnens:2021hpq}, and introduce the off-shellness via \cref{eq:hadrons:offsellcondition} Esto lo he quitado porque no estaba claro.

Using the results for $\cM^\LO_3$, it is possible to obtain predictions for the isospin-three three-pion $K$-matrix at LO in ChPT~\cite{Blanton:2019vdk,Baeza-Ballesteros:2023ljl}. While in general $\Kdf$ is obtained from $\cM_3$ via integral equations, these can be reduced to algebraic relations when working at fixed order in perturbation theory. 

Although in ChPT contributions are normally ordered as a power series in the pion mass and the momentum squared, here it is easier to study the powers of $1/\Fpi^2$. The scattering amplitudes are then $\cM_2^\LO=\cO(1/\Fpi^2)$ and $\cM_3^\LO=\cO(1/\Fpi^4)$. The subtraction needed to obtain the divergence-free amplitude can be obtained from \cref{eq:hadrons:generalsubtraction}. At LO, since $G^\infty=\cO(1)$,
\begin{equation}\label{eq:hadrons:subtractiontermLO}
\cD^{\uu\LO}(\bm{p},\bm{k})=-\cM_2^\LO(\bm{p})G^\infty(\bm{p},\bm{k})\cM_2^\LO(\bm{k})\,.
\end{equation}
so that the divergence-free amplitude becomes
\begin{equation}
\Mdf^\LO(\bm{p},\bm{k})=\cM_3^\LO(\bm{p},\bm{k})-\cS\left\{\cD^{\uu\LO}(\bm{p},\bm{k})\right\}\,,
\end{equation}
Note that in \cref{eq:hadrons:subtractiontermLO} two-pion amplitudes are on-shell. This term removes the divergencies arising when the exchanged particle in $\cM_3^{\LO,\OPE}$ goes on shell. The LO non-OPE part, by contrast, does not need any subtraction. 

In the $I_{\pi\pi\pi}=3$ channel pairwise interactions only happen with $I_{\pi\pi}=2$. At LO, the two-particle isospin-two amplitude is purely $s$-wave, and so we can substitute $G^\infty(\bm{p},\bm{k})=G_{ss}^\infty(\bm{p},\bm{k})$, where
\begin{equation}\label{eq:hadrons:Ginftyssdefinition}
G_{ss}^\infty(\bm p, \bm k)_{\ell' m'; \ell m} 
        \equiv \delta_{\ell' 0} \delta_{m' 0} \delta_{\ell 0} \delta_{m 0}\,
        \frac{H(x_p)H(x_k)}{b_{pk}^2 - \Mpi^2 + i\epsilon}\,.
\end{equation}
We can also set the cutoff functions to one, since all momenta in the subtraction in \cref{eq:hadrons:subtractiontermLO} are on-shell.

To compute $\Mdf^\LO$ it is convenient to separate $\cM_3^\LO$ into an OPE and a non-OPE part,
\begin{multline}
\Mdf^\LO(\bm{p},\bm{k})=\\
\cM_3^{\LO,\nOPE}(\bm{p},\bm{k})+\cS\left\{\cM_3^{\uu\LO,\OPE}(\bm{p},\bm{k})-\cD^{\uu\LO}(\bm{p},\bm{k})\right\}\,.
\end{multline}
The second term on the right-hand side is the divergence-free OPE amplitude. For the momenta asignment in \cref{fig:hadrons:pipipiLOOPE}, its unsymmetrized version is,
\begin{equation}
\Mdf^{\uu,\LO,\OPE}(\bm{p}_3,\bm{k}_3)=\frac{1}{\Fpi^4}\left[2p_1\cdot p_2+2k_1\cdot k_2-(b^2-\Mpi^2)\right]\,.
\end{equation}
After symmetrizing, one can combine with the non-OPE result in \cref{eq:hadrons:amplitudeLOnonOPE} to obtain,
\begin{equation}
\Mdf^\LO(\bm{p},\bm{k})=\frac{\Mpi^2}{\Fpi^4}\left(18+27\Delta\right)\,,
\end{equation}
where $\Delta$ is defined in \cref{eq:hadrons:thesholdexpansionKdfingredients}.

The final step to compute the three-particle $K$-matrix is to use \cref{eq:hadrons:integralequationKdf}, that relates the divergence-free amplitude to $\Kdf$. Both $\cL$ and $\cR$, given in \cref{eq:hadrons:Ldef,eq:hadrons:Rdef}, start at $\cO(1)$, 
\begin{equation}\label{sec:hadrons:decoratorsLO}
\cL^{\uu\LO}(\bm p, \bm k) 
= \cR^{\uu\LO}(\bm p, \bm k)
        = \frac13 \bar\delta(\bm p - \bm k)\,,
\end{equation}
and so $\cT^\LO=\cO(1/\Fpi^4)$, since it has the same order as $\Mdf$. From  \cref{eq:hadrons:integralequationKdf}, this implies $\Kdf^\LO=\cO(1/\Fpi^4)$, while the second term on the right-hand side is $\cO(1/\Fpi^8)$. Thus, $\cT^\LO=\Kdf^\LO$. 

We can finally put this into \cref{eq:hadrons:integralequationKdf} combined with the results in \cref{sec:hadrons:decoratorsLO} to finally get
\begin{equation}\label{eq:hadrons:KdfMdfrealtionLO}
\Mdf^\LO=\cS\left\{\frac{\Kdf^\LO}{9}\right\}=\Kdf^\LO\,,
\end{equation}
where the symmetrization yields a factor of nine since $\Kdf$ is symmetric, by definition, in the $\Ippp=3$ channel. Thus, we find
\begin{equation}\label{eq:hadrons:Kdfpipipiresult}
\Mpi^2\Kdf^\LO=\frac{\Mpi^4}{\Fpi^4}\left(18+27\Delta\right)\,.
\end{equation}
From this result, we can identify the coefficients of the threshold expansion in \cref{eq:hadrons:thesholdexpansionKdf}, finding $\Kiso=18\,(\Mpi/\Fpi)^4$ and $\Kisoone=27\,(\Mpi/\Fpi)^4$. These results at LO have been compared against lattice results, finding a large discrepancy. In \cref{sec:pipipiKmatrix} we study the size of NLO corrections, finding they are large and lead to much better agreement between ChPT and the lattice.

%{\jorge [Amplitude 2pi y scatt lenght/effr a LO. Amplitude 3pi y Kdf a LO. Comentar discrepancia lattice. IAM]}

\section{Particle scattering from the lattice}\label{sec:hadrons:particleinteractionslattice}

At first glance, the study of multiparticle interactions using lattice techniques may seem impossible. Scattering is a real-time process between asymptotic states in which particles are very far from each other, while lattice computations are performed in Euclidean time and a finite volume, where such asymptotic states cannot be defined. Using lattice simulations, we can only compute correlation functions, from which finite-volume energies can be extracted, as explained in \cref{sec:QCD:correlationfunctions}. %. If these have non-zero overlap on some multiparticle states, one is able to determine the finite-volume energies of these multiparticle states. 

However, from the finite-volume energies of multiparticle states one can extract information about the interactions between the asymptotic states of the theory. After all, both scattering amplitudes and the finite-volume energy spectrum are determined by the same Lagrangian. The relation between both is given by the so-called \textit{quantization conditions} (QCs), which have been developed for systems of both two-~\cite{Luscher:1986pf,Luscher:1990ux} and three-particles~\cite{Hansen:2014eka,Hansen:2015zga}. 

%As explained in \cref{sec:QCD:correlationfunctions}, this typically involves choosing a sufficient set of operators, $\{O_i\}$,  computing the corresponding correlation matrix, $C_{ji}(t)$, and using the GEVP to extract the finite-volume energies of stationary multiparticle states. To improve the extraction of energies, it is convenient to use opeators that 

%The finite-volume energy spectrum contains information about the interactions between the states of the theory---after all, both scattering amplitudes and the finite-volume energy spectrum are determined by the same Lagrangian. This is done using the so-called quantization conditions, which have been develop for systems of both two- and three-particles. 

%The two-particle formalism is already almost four decades old, and since has been extended to any possible two-particle process. In the case of three-particles, different approaches have been proposed in the last year

QCs are obtained from studying the relation between finite- and infinite-volume correlators. In finite volume, the correlator takes the form in \cref{eq:QCD:spectraldecompositioninfiniteT}. One can Fourier-transform the Euclidean time coordinate to Euclidean energy and then rotate it back to Minkowski energy, obtaining
\begin{equation}
C_L(E)=\sum_n \frac{2 iE_n|Z_n|^2}{E^2-E_n^2}\,,
\end{equation}
where we include a subscript ``$L$'' to indicate we are referring to the finite-volume correlator. 
The finite volume correlator thus presents simple poles at the finite-volume energies, $E_n$, with purely imaginary residues with a positive imaginary part. The QCs exploit this feature, by relating the finite-volume correlator to its infinite-volume counterpart, plus some corrections that depend on the scattering amplitudes and are singular at the finite-volume energies. 

Other alternative approaches to study multiparticle interactions on the lattice also exist. Two well-known examples are the HAL QCD method, which makes use of multiparticle potentials computed from spacial correlation functions to solve Schrödinger's equation~\cite{Ishii:2006ec,Aoki:2008hh,Aoki:2009ji,Aoki:2012tk}, and a recent technique that aims at extracting scattering amplitudes directly from the finite-volume correlator~\cite{Bruno:2020kyl}. %Tal vez, quitar el párrafo

\subsection{Two-particle quantization condition}\label{sec:hadrons:QC2}

The two-particle QC was first proposed for systems of two identical scalar particles in a seminal work by Lüscher~\cite{Luscher:1986pf,Luscher:1990ux}, and has since been extended to any possible two-particle process~\cite{Luscher:1991cf,Rummukainen:1995vs,Kim:2005gf,He:2005ey,Bernard:2008ax,Luu:2011ep,Briceno:2012yi,Briceno:2014oea,Gockeler:2012yj}. This includes higher partial waves, moving frames, coupled channels and arbitrary spin and masses.

\begin{figure}[t!]
\centering
\[\large
C_2(E)\hspace{0.1cm}=\hspace{0.1cm}C_{2,0}\hspace{0.1cm}+\hspace{0.1cm}
    \vcenter{\hbox{\includegraphics[height=1.2cm]{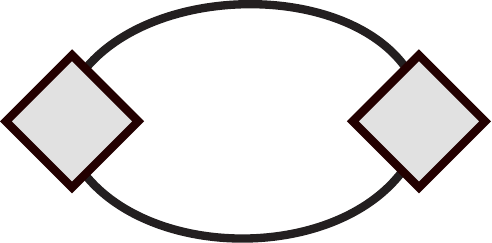}}}\hspace{0.1cm} +\hspace{0.1cm} \vcenter{\hbox{\includegraphics[height=1.2cm]{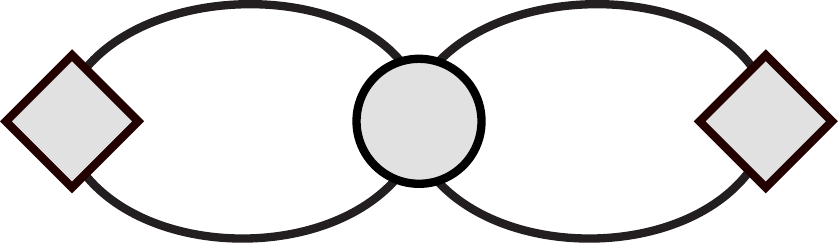}}} \hspace{0.1cm} +\hspace{0.1cm} \dots
\]
\caption{Schematic representation of the two-particle correlation function. Circles represent the Bethe-Slapeter kernels and squares are the operator insertions, while $C_{2,0}$ contains all diagrams with no two-particle $s$-channel loop.}\label{fig:hadrons:QC2}
\end{figure}

The starting point of the formalism is expressing the finite-volume correlation function as an all-orders expansion in perturbation theory, which is assumed to converge to the full theory. In the elastic energy region, the diagrams can be rewritten as a succession of smooth kernels---the \textit{Bethe-Salpeter kernels}---that contain all diagrams that are two-particle irreducible in the $s$-channel, joined by two-particle $s$-channel loops that are evaluated in finite volume---see \cref{fig:hadrons:QC2}. Using Poisson summation~\cite{Gelfand:105396}, it can be proved that the Bethe-Salpeter kernel is equal to its infinite-volume counterpart up to exponentially suppressed volume effects,\footnote{Technically, these corrections fall faster than any power law.} which are neglected. Two-particle $s$-channel loops are equal to their infinite-volume version plus an extra sum-minus-integral correction, which contains power-law finite volume effects arising when an intermediate two-particle state goes on shell.

After summing all the diagrams, the two-particle finite-volume correlator becomes
\begin{equation}\label{eq:hadrons:finitevolumecorrelatortwoparticles}
\displaystyle C_{2,L}(P)=C_{2,\infty}(P)-A_2(P)\frac{i}{F^{-1}(P,L)+\cK_2(E)}B_2(P)\,,
\end{equation}
where $A_2(E)$ and $B_2(E)$ are related to two-particle irreducible diagrams in the $s$-channel containing operator insertions, $C_\infty(P)$ is the infinite-volume correlation function, and $F$ is the aforementioned sum-minus-integral correction. The latter is a finite-volume geometric factor that only depends on the  total momentum and the size of the lattice. In the case of two identical scalar particles, it takes the form
\begin{multline}\label{eq:hadrons:QC2Fdefinition}
F(P,L)_{\ell'm';\ell m}\\
=\left[\frac{1}{L^3}\sum_{\bm{k}}-\text{PV}\int\frac{\text{d}^3k}{(2\pi)^3}\right]\frac{4\pi Y_{\ell' m'}(\hat{\bm{k}}^*)Y_{\ell m}^*(\hat{\bm{k}}^*)}{8\omega_k\omega_{Pk}(E-\omega_k-\omega_{Pk})}\left(\frac{k^*}{q_2^*}\right)^{\ell'+\ell}\,,
\end{multline}
where exponentially suppressed volume effects are neglected. Here, $Y_{\ell m}$ are the spherical harmonics, $q_2^*$ is the magnitude of the back-to-back momentum in the CMF, $\bm{k}^*$ is the $\bm{k}$ vector boosted to that frame, and we define the energies
\begin{equation}\label{eq:hadrons:omegadefinitionsQC2}
\omega_k=\sqrt{\bm{k}^2+m^2}\,,\quad\quad\quad\omega_{Pk}=\sqrt{(\bm{P}-\bm{k})^2+m^2}\,.
\end{equation}
Finally, the sum in \cref{eq:hadrons:QC2Fdefinition} is performed over all momenta allowed in the finite volume---see \cref{eq:QCD:finitevolumemomenta}. Note that the pole in the integral is regulated using the principal-value (PV) prescription---see \rcite{Davies:1996gee} for details on this prescription, and \rrcite{Luu:2011ep,Gockeler:2012yj,Morningstar:2017spu} for an efficient way to evaluate $F$.

The main implication of this result is that, since all $C_{2,\infty}$, $A_2$ and $B_2$ are smooth, the poles in the finite-volume correlator originate from zeros in the denominator of the last term. The energies at which these poles arise, which in turn are the finite-volume energies, are the solutions of the two-particle quantization condition,
\begin{equation}\label{eq:hadrons:quantizationconditiontwoparticles}
\det[F^{-1}(P,L)+\cK_2(E)]=0\,.
\end{equation}
Using this condition, finite-volume energies can be used to constrain infinite-volume scattering observables.

Some comments are in place. All the elements appearing in the quantization condition are matrices in angular momentum. This means $F$ and $\cK_2$ are, a priori, infinite dimensional. In the practice, they can be truncated at some maximum partial wave, $\ell_\text{max}$, since the contribution to the scattering amplitude near threshold is suppressed for large $\ell$---see \cref{eq:hadrons:effectiverangeexpansion}. Also, whereas $\cK_2$ is diagonal, $F$ mixes different partial waves, since rotational symmetry is broken in a cubic box. If one works instead in the basis of irreps of the cubic group or the corresponding little group, $F$ becomes block diagonal, and only a subset of partial waves are mixed---see~\rrcite{Luscher:1990ux,Morningstar:2013bda}.

Although not appearing explicitly in the QC, the $A_2$ and $B_2$ factors in \cref{eq:hadrons:finitevolumecorrelatortwoparticles} play an important yet subtle role. They project out those states on which the chosen operators have zero overlap, and so the determinant only runs over the relevant states. For example, if one is studying $\pi\pi$ scattering in the $\Ipp=2$ channel, $p$-channel elements of $F$ do not enter the quantization condition.

Finally, the applicability of \cref{eq:hadrons:finitevolumecorrelatortwoparticles,eq:hadrons:quantizationconditiontwoparticles} is restricted to the elastic region. For higher energies, intermediate states with three or more particles could go on-shell, leading to additional power-law finite-volume effects. For a system of two particles of mass $m$, the elastic region is, in general, $m<E^*<3m$, but the range is increased to $0<E^*<4m$ with $\mathbbm{Z}_2$ symmetry. Note that the formalism has been extended above the $3m$ barrier with the three-particle quantization condition---see \cref{sec:hadrons:threeparticlesfinitevolume}---and below the lower $m$ limit with the explicit consideration of one-particle exchange in the $t$- and $u$-channel~\cite{Raposo:2023oru,Meng:2023bmz,Bubna:2024izx}.

In some simple cases, it is possible to transform \cref{eq:hadrons:quantizationconditiontwoparticles} into an algebraic equation. For example, if we consider two identical particles of mass $m$ and only consider $s$-wave interactions, the two-particle QC takes the form
\begin{equation}\label{eq:hadrons:algebraicQC2}
k\cot\delta_0=\frac{2}{\gamma L\pi^{1/2}}\cZ_{00}^{\bm{P}}\left(\frac{q^* L}{2\pi}\right)\,,
\end{equation}
where $\gamma$ is the boost factor to the center-of-mass frame and $\cZ$ is the generalized Lüscher zeta function~\cite{Luscher:1986pf,Luscher:1990ux,Rummukainen:1995vs}.

In the limit of large $L$, \cref{eq:hadrons:quantizationconditiontwoparticles} can be expanded to obtain a relation between the energy shift, defined as the difference between the interacting and the free energy, $\Delta E=E-E_\text{free}$, and the scattering parameters. In the case of the ground state, $\Delta E_\text{thr}=E_\text{thr}-2m$ is related to the parameters in the ERE expansion, given in \cref{eq:hadrons:effectiverangeexpansion},

\noindent\begin{equation}\label{eq:hadrons:thresholdexpansionQC2}
\begin{array}{rl}
\displaystyle\Delta E_\text{thr} =  & \displaystyle-\frac{4\pi a_0}{mL^3}\left[1+\left(\frac{a_0}{\pi L}\right) \cI + \left(\frac{a_0}{\pi L}\right)^2 (\cI^2-\cJ)\right. \\[10pt]
& \displaystyle\left.-\left(\frac{a_0}{\pi L}\right)^3(-\cI^3+3\cI\cJ-\cK)+\frac{2\pi r_0 a_0^2}{L^3} + \frac{\pi a_0}{m^2L^3}\right]+\cO(L^{-7})\,, 
\end{array}
\end{equation}
where $\cI=-8.9136...$, $\cJ=16.5323...$ and $\cK=8.4019$ are numerical constants~\cite{Beane:2007qr}. The leading term in this relation was first obtained in \rcite{Huang:1957} for a model of hard spheres. The general quantum-field-theory result up to $\cO(L^{-5})$ was worked out in \rcite{Luscher:1986pf}, while the $\cO(L^{-6})$ correction, in which relativistic effects first appear, was computed in \rcite{Hansen:2015zta}. Note that this expansion is only expected to converge when $|a_0/L|\ll 1$. Similar relations also exists for excited states and for systems of more than two particles~\cite{Beane:2007es,Beane:2007qr,Hansen:2015zta,Romero-Lopez:2020rdq,Pang:2019dfe,Muller:2020vtt}. 

\subsection{Three-particle quantization condition}\label{sec:hadrons:threeparticlesfinitevolume}

%The two-particle formalism, while generalized to any possible two-particle process, is limited by the lowest three-particle threshold. Moreover, most resonances can decay into three hadrons, such as the $\omega(782)\rightarrow\pi\pi\pi$ meson or the roper resonance, $N(1440)\rightarrow N\pi, N\pi\pi$. 

The two-particle formalism is limited to energies below the lowest three-particle threshold. This limits its capability to study the hadron spectrum, as many resonances are known to mainly decay into three or more particles~\cite{PDG:2020}. During the last decade, three different versions of a three-particle QC have been developed to overcome this limitation: the RFT approach~\cite{Hansen:2014eka,Hansen:2015zga}, the finite-volume unitarity approach~\cite{Mai:2017vot,Mai:2017bge} and the non-relativistic effective-field-theory approach~\cite{Hammer:2017uqm,Hammer:2017kms}. 

 The first formalism to be developed was the RFT formalism, which has been generalized to systems including two-to-three processes~\cite{Briceno:2017tce}, two-particle resonances~\cite{Briceno:2018aml}, non-identical~\cite{Hansen:2020zhy} and non-degenerate~\cite{Blanton:2020gmf,Blanton:2021mih,Hansen:2024ffk} particles, three spin-1/2 particles~\cite{Draper:2023xvu}, and certain cases of coupled channels~\cite{Draper:2024qeh}. The other two formalisms were proposed at a later date and have seen less development. Note that all of them have been shown to be equivalent in the limits where they can be compared~\cite{Jackura:2019bmu,Blanton:2020jnm}.

In this dissertation, we focus on the RFT formalism, and discuss here the case of three-identical particles. The standard derivation of the QC is based on an all-order skeleton expansion similar to the one used for two-particles---see \rcite{Blanton:2020gha} for an alternative approach based on time-ordered perturbation theory. The three-particle finite-volume correlator is related to its infinite-volume counterpart as
\begin{equation}\label{eq:hadrons:finitevolumecorrelatorthreeparticles}
C_{3,L}(P)=C_{3,\infty}(P)+A_3(P)\frac{i}{F_3^{-1}(\cK_2,P,L)+\Kdf(E)}B_3(P)\,.
\end{equation}
Here $\Kdf$ is the three-particle divergence-free $K$-matrix, related to $\cM_3$ via the integral equations introduced in \cref{sec:hadrons:infinitevolumethreeparticlescattering}, $F_3$ is a geometric factor defined below that depends not only on the size of the box but also on two-particle interactions, $C_{3,\infty}$ is the infinite-volume correlator, and $A_3$ and $B_3$ are smooth factors related to three-particle operator insertions, that play an analogous role to $A_2$ and $B_2$ in the the two-particle case. 

\begin{figure}[!tp]
    \centering
    \begin{subfigure}{0.48\textwidth} 
    \centering
        \includegraphics[width=0.65\textwidth]{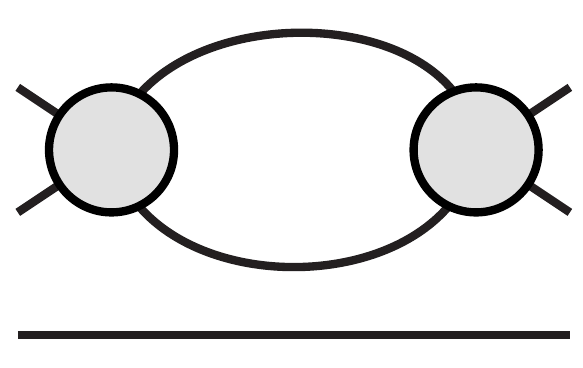}
        \caption{$\tilde{F}$ factor}
        \label{fig:hadrons:Ffactor}
    \end{subfigure}
    \begin{subfigure}{0.48\textwidth}
    \centering
       \includegraphics[width=0.65\textwidth]{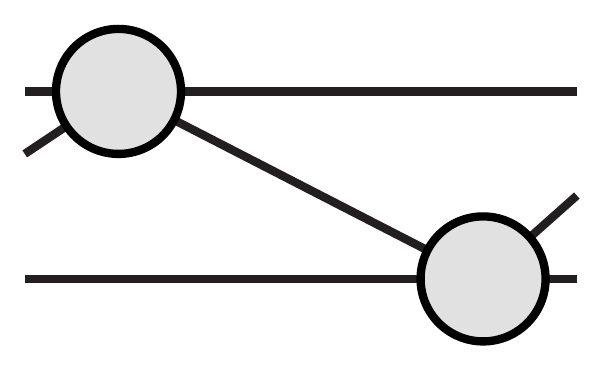}
        \caption{$G$ factor}
        \label{fig:hadrons:Gfactor}
    \end{subfigure}
    \caption{
        Diagrammatic representation of the factors appearing in the three-particle RFT formalism that lead to finite-volume power-law effects. Each line correspond to one particle and gray circles are Bethe-Salpeter kernels.}
    \label{fig:hadrons:FGfactor}
\end{figure}

%All the quantities appearing in \cref{eq:hadrons:finitevolumecorrelatorthreeparticles} belong to the space of three-particle states in the finite volume. 
Following the description made in \cref{sec:hadrons:infinitevolumethreeparticlescattering}, three-particle states are characterized with ($\bm{k} \ell m$) indices, where $\ell$ and $m$ refer to the angular momentum of the dimer, and $\bm{k}$ is the three-momentum of the spectator, which takes discrete values on the finite volume---see \cref{eq:QCD:finitevolumemomenta}. $F_3$ takes the form,
\begin{equation}\label{eq:hadrons:definitionF3threeparticles}
F_3=\frac{1}{3}\tilde{F}+\tilde{F}\frac{1}{\tilde{\cK}_2^{-1}-(\tilde{F}+G)}\tilde{F}\,.
\end{equation}
In this equation, $\tilde{F}$ and $\tilde{\cK}_2$ are modified versions of the geometric factor from the two-particle quantization condition, $F$, and two-particle $K$-matrix, $\cK_2$, respectively. The former originates from two-particle loops, diagramatically represented in \cref{fig:hadrons:Ffactor}, and contains power-law finite volume effects. It is diagonal on the spectator momenta but not on the angular momentum indices,\vspace{-0.2cm}
\begin{multline}\label{eq:hadrons:definitionFthreeparticles}
\tilde{F}(P,L)_{\bm{p}\ell'm';\bm{k}\ell m }=
\delta_{p k}H(x_k)\left[\frac{1}{L^3}\sum_{\bm{a}}-\text{PV}\int\frac{\text{d}^3a}{(2\pi)^3}\right]\\
\times\frac{4\pi Y_{\ell' m'}(\hat{\bm{a}}^*)Y_{\ell m}^*(\hat{\bm{a}}^*)H(x_a)H(x_{ka})}{8\omega_a \omega_{ka}(E-\omega_k-\omega_a-\omega_{ka})}\left(\frac{k^*}{q_k^*}\right)^{\ell'+\ell}\,,
\end{multline}
where $\omega_{ka}=\sqrt{m^2+(P-k-a)^2}$ and $x_{ka}=(P-k-a)^2/4m^2$, and other variables are introduced in \cref{sec:hadrons:infinitevolumethreeparticlescattering}. On the other hand, $\tilde{\cK}_2$  is diagonal in all the indices (here left implicit),
\begin{equation}\label{eq:hadrons:definitionK2threeparticles}
(\tilde{\cK}_2)^{-1}=\cK_2^{-1}+\rho(k)[1-H(k)]\,.
\end{equation}
Finally, $G$ is related to other type of power-law finite-volume effects coming from OPE-like diagrams, like that in \cref{fig:hadrons:Gfactor},
\begin{multline}\label{eq:hadrons:definitionGthreeparticles}
G(P,L)_{\bm{p}\ell'm';\bm{k}\ell m }=\\
\frac{1}{4\omega_k\omega_p L^3}\left(\frac{ k_p^*}{q_{2,p}^*}\right)^{\!\ell'}\frac{4\pi Y_{\ell' m'}(\hat{\bm{k}}^*_p)H(x_p)H(x_k)Y_{\ell m}^*(\hat{\bm{p}}^*_k)}{b^2-m^2}\left(\frac{ p_k^*}{q_{2,k}^*}\right)^{\!\ell}\,,
\end{multline}
In this equation, kinematical variables follow analogous definitions to  those presented in \cref{sec:hadrons:infinitevolumethreeparticlescattering}. The same holds true for the cutoff function, which in this case plays a twofold role. By removing high spectator momenta, situations in which the interacting pair energy would be on the left-hand cut, invalidating the formalism, are prevented. Also, it allows one to truncate the matrices at some $\bm{k}_\text{max}$, which is combined with truncation in angular momentum.

As in the case of the two-particle QC, three-particle energies lie at the zeros of the denominator in \cref{eq:hadrons:finitevolumecorrelatorthreeparticles}, leading to the three-particle QC,
\begin{equation}\label{eq:hadrons:quantizationconditionthreeparticles}
\det[F_3^{-1}(\cK_2;P,L)+\Kdf(P)]=0\,.
\end{equation}
This QC has a range of applicability that is limited, for general theories, to energies $2m<E^*<4m$. In the case of interactions with $\mathbb{Z}_2$ symmetry, as is the case of pions if isospin-breaking corrections are neglected, this range is extended to $m<E^*<5m$.
 
The QC in \cref{eq:hadrons:quantizationconditionthreeparticles} simplifies in the so-called isotropic approximation, in which only $s$-wave interactions are considered and $\Kdf$ is assummed to be only a function of the total energy, $E$. In this limit, \cref{eq:hadrons:quantizationconditionthreeparticles} becomes
\begin{equation}\label{eq:hadrons:isotropicKmatrix}
F_3^\text{iso}(E^*)=\langle 1|F_3|1\rangle = -\frac{1}{\Kdf^\text{iso}(E^*)}\,,
\end{equation}
where $|1\rangle$ is a vector with ones in all the entries. This limit was explored numerically in \rcite{Briceno:2018mlh}. A threshold expansion for the ground state energy shift has also been worked out in the context of the RFT formalism~\cite{Hansen:2016fzj}, and has been analytically tested in a simple scalar theory~\cite{Hansen:2015zta,Sharpe:2017jej}.
%This matrix equation ca also be simplified to an algebraic one in the so-called isotropic approximation. In this limit, only $s$-wave interactions are considered, and one assummes that 

Before concluding, we comment on the generalization of the RFT formalism to general three-pion isospin~\cite{Hansen:2020zhy}. While in the maximal isospin case the three pions can be treated as identical and \cref{eq:hadrons:quantizationconditionthreeparticles} can be applied, this is not true for the remaining channels. To properly include all channels, all elements in \cref{eq:hadrons:finitevolumecorrelatorthreeparticles} are promoted to $7\times7$ matrices, which are block diagonal in three-pion isospin channels, but not in the two-particle subchannels within each three-particle isospin. The non-zero entries of $\tilde{F}$, $\tilde{K}_2$ and $G$ are equal to those in \cref{eq:hadrons:definitionFthreeparticles,eq:hadrons:definitionK2threeparticles,eq:hadrons:definitionGthreeparticles}, respectively,  multiplied by some numerical factors related to Clebsh-Gordan coefficients. 
Moreover, $\Kdf$ obeys different symmetries under particle exchange. For example, while $\Kdf$ is symmetric under particle exchange in the $I_{\pi\pi\pi}=3$ channel, it becomes fully antisymmetric for $I_{\pi\pi\pi}=0$. These different transformation properties imply new forms of the threshold expansion, which are presented in \cref{sec:isospinKmatrix:thresholdexpansion}.

%{\jorge [Two and three-particle formalisms. Comentar isospin? Comentar cómo se procede en la lattice?]}

\clearpage
\makeatletter
\setlength{\@fptop}{0pt plus 1fil}
\setlength{\@fpbot}{0pt plus 1fil}
\makeatother
\chapter{Pion-pion scattering near threshold at large $N_\text{c}$}
\label{sec:largeNpions}

The large $\Nc$ limit of QCD, introduced in \cref{sec:QCD:largeNc}, has proven to have predictive power in the non-perturbative regime of the theory, and has been used by many phenomenological approaches to the low-energy regime of QCD---see \rcite{Manohar:1998xv} for a review. In some cases, however, large $\Nc$ predictions fail to reproduce experimental results. An example of this is the well-known $\Delta I=1/2$ puzzle~\cite{FUKUGITA1977237,Chivukula:1986du,Bardeen:1986vz,BARDEEN1986133,Bardeen:1986uz,Sharpe:1987cx,Pich:1995qp}. Large $\Nc$ predictions for the $K\rightarrow\pi\pi$ process predict a ratio $A_2/A_0=\sqrt{2}$ between the decay amplitude into two pions in the isospin-two and -zero channels. This result, however, completely disagrees with experimental measurements, $(A_2/A_0)_\text{exp}\approx22.4$.%  which is dominated by subleading $\Nc$ corrections~\cite{Donini:2016lwz,Donini:2020qfu}. 

The lattice regularization makes it possible to quantify such corrections by directly simulating at several values of $\Nc$. In this context, lattice techniques have been mainly used to study the hadron and glueball spectrum, as well as weak matrix elements, such as those involved in the $\Delta I=1/2$ rule~\cite{Bali:2013kia,Cordon:2014sda,DeGrand:2016pur,DeGrand:2017gbi,DeGrand:2020utq,DeGrand:2021zjw,Athenodorou:2021qvs,Donini:2016lwz,Hernandez:2019qed,Donini:2020qfu,Perez:2020vbn}---see~\rcite{Hernandez:2020tbc} for a recent review. Indeed, large subleading $\Nc$ corrections have been found in the case of non-leptonic kaon decays~\cite{Donini:2016lwz,Donini:2020qfu}. Lattice simulations also open the door to answer other timely questions, such as the possible existence of tetraquarks at large $\Nc$, discussed in \cref{sec:QCD:largeNctetraquarks}.

In this chapter, the results from \rrcite{Baeza-Ballesteros:2022azb,Baeza-Ballesteros:2021nxu} are summarized, in which we investigate the scattering of two pseudoscalar mesons as a function of $\Nc$. We work with $\Nf=4$ degenerate quark flavors, meaning all up, down, strange and charm quarks have the same mass. We consider $\Nc=3-6$ and $\Mpi\approx360-590$ MeV, and focus on two scattering channels. The same setup has previously been used in \rcite{Hernandez:2019qed} to study the $\Nc$ scaling of pion masses and decay constants, and in \rcite{Donini:2020qfu} to investigate the $\Delta I=1/2$ puzzle. 

Using lattice simulations, we measure two-particle finite-volume energies near threshold. In particular, we use Lücher's formalism to match our results to ChPT predictions including the $\eta'$, which we compute for the first time. From the matching of the lattice results and ChPT we study the $\Nc$ dependence of the relevant LECs from first principles.

This chapter is organized as follows. In \cref{sec:largeNpions:largeNChPT} we describe how the scattering of two pseudoscalar mesons in a theory with $\Nf=4$ can be classified in different isospin channels. We also use the large $\Nc$ limit and ChPT to obtain analytical predictions for the processes of interest. The lattice setup is presented in \cref{sec:largeNpions:laticesetup}, followed by lattice results in \cref{sec:largeNpions:energies}. In \cref{sec:largeNpion:resultsfitsChPT} we present the results of the fits to ChPT, and compare them to previous literature. The main conclusions of this work can be found in \cref{sec:largeNpions:conclusions}.

\section{Pion-pion scattering in $\Nf=4$ QCD}\label{sec:largeNpions:largeNChPT}

In a theory with $\Nf=4$ degenerate light quark flavors, spontaneous chiral symmetry breaking leaves an exact SU(4) isospin symmetry. The lightest pseudoscalar mesons appear as a 15-dimensional multiplet, plus a singlet representing the $\eta'$ particle,
\begin{equation}
4\,\otimes\,4\,=\,15\,\oplus\,1\,.
\end{equation}
The multiplet includes all pseudoscalar $\pi$, $K$, $D$, $D_s$ and $\eta$ mesons. Since all of them are degenerate in this scenario, we will refer to them generically as pions.

Two-pion scattering organizes in different scattering channels, corresponding to irreps of the isospin group. For $\Nf=4$, there is a total of seven scattering channels~\cite{Bijnens:2011fm},
\begin{equation}
15\,\otimes\,15\,=\,84\,\oplus\,45\,\oplus\,45\,\oplus\,20\,\oplus\,15\,\oplus\,15\,\oplus\,1\,.
\end{equation}
This same number of irreps holds for any $\Nf\geq 4$, while it reduces to six for $\Nf=3$ and three for $\Nf=2$---see \cref{sec:hadrons:twopionsChPT}. 

In this work, we focus on two of these scattering channels, which contain $s$-wave interactions and can be studied in the lattice without the need to compute quark propagators between two lattice sites with the same time coordinate. These are the 84- and the 20-dimensional irreps. The former is symmetric in quarks and antiquarks, and is equivalent to the $I_{\pi\pi}=2$ channel of two-flavor QCD. We refer to it as the $SS$ channel. The 20-dimensional irrep is antisymmetric in both quarks and antiquarks, and only exists for $\Nf\geq 4$. We call it the $AA$ channel. Representative states of each of these channels are, respectively, $|\pi^+\pi^+\rangle$ and $|D_s^+\pi^+\rangle-|D^+K^+\rangle$. Note that both channels contain states with four distinct quark flavors. Throughout this chapter, we denote a generic irrep using a $R\in\left\{SS,AA\right\}$ label.

The 45-dimensional irreps---which are degenerate, as one is the conjugate of the other---can also be studied in the lattice without the need of equal-time quark propagators. However, they only contain odd partial waves, and so their scattering amplitude vanishes at threshold. These channels are studied in \cref{sec:largeNmesons} as a function of the center-of-mass energy.

\subsection{Pion-pion scattering at large $\Nc$}

\begin{figure}[!bp]
    \centering
    \begin{subfigure}{0.48\textwidth} 
    \centering
        \includegraphics[width=0.9\textwidth]{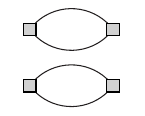}
        \caption{Disconnected ($D$) diagram, $\cO(\Nc^2)$}
        \label{fig:largeNpions:Dcontraction}
    \end{subfigure}
    \begin{subfigure}{0.48\textwidth}
    \centering
       \includegraphics[width=0.9\textwidth]{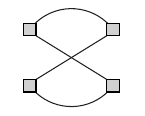}
        \caption{Connected ($C$) diagram, $\cO(\Nc)$}
        \label{fig:largeNpions:Ccontraction}
    \end{subfigure}
    \caption{
        Diagrammatic representation of those Wick contractions contributing to the $SS$ and $AA$ channels. Solid lines are quark propagators, while the squares represent pion insertions. %To study the large $\Nc$ dependence of each contraction one need to consider possible gluon lines and internal quark loops on these contractions. Diagrams with none contribute to the leading $\Nc$ term, which scale as indicated in teh subcaptions.
        }
    \label{fig:largeNpions:CDleadingcontractions}
\end{figure} 

The $\Nc$ and $\Nf$ scaling of the scattering amplitudes of the $SS$ and $AA$ channels can be determined from a perturbative analysis of the correlation functions. For both channels, these can be computed as a linear combination of two terms, corresponding to two topologies of the Wick contractions,
\begin{equation}\label{eq:largeNpions:correlatorCD}
C_{SS}=2(D-C)\,,\quad\quad\quad C_{AA}=2(D+C)\,,
\end{equation}
where $D$ and $C$ stand for the disconnected and connected contractions, respectively. They are diagrammatically represented in \cref{fig:largeNpions:CDleadingcontractions} and are explicitly expressed in \cref{eq:largeNpions:Wickcontractions} in terms of quark propagators.

One can determine the $\Nc$ and $\Nf$ scaling of these correlation functions. This requires to include possible gluon lines and internal quark loops to the diagrams in \cref{fig:largeNpions:CDleadingcontractions} and to use  the large $\Nc$ counting rules presented in \cref{sec:QCD:largeNc}. Diagrams with no additional quark or gluon lines contribute to the leading terms. The $D$ diagram, shown in \cref{fig:largeNpions:Dcontraction}, contains two color loops, and so is $\cO(1)$, while the $C$ diagram in \cref{fig:largeNpions:Ccontraction} is $\cO(\Nc^{-1})$, as it contains a single color loop. Note that both diagrams contain four insertions of pion operators, which we normalize with a $\Nc^{-1/2}$ factor. Including internal gluon lines in each disconnected piece of $D$ or in $C$ does not affect this scaling, and so such diagrams also contribute to the leading $\Nc$ term. 

From this result, one can determine the leading $\Nc$ dependence of other scattering observables. For example, the scattering length, defined in \cref{eq:hadrons:effectiverangeexpansion} can be expressed in terms of correlations functions of zero-momentum pions as
\begin{equation}\label{eq:largeNpions:scatteringlengthfromcorrs}
\Mpi a_0^R\propto \frac{C_R-C_\pi^2}{C_\pi^2}\,,
\end{equation}
where $C_\pi\sim\cO(1)$ is the single-pion propagator and, recall, $R\in\left\{SS,AA\right\}$ indicates some general two-pion channel. The single-pion propagator corresponds to one of the two disconnected pieces in \cref{fig:largeNpions:Dcontraction}, and its scaling was explicitly expressed in \cref{eq:QCD:largeNmesoncorrelator}. Note that it has the same topology as $C$, and so they have the same large $\Nc$ scaling.

\begin{figure}[!bp]
    \centering
    \begin{subfigure}{0.48\textwidth} 
    \centering
        \includegraphics[width=0.9\textwidth]{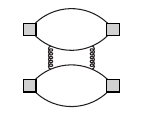}
        \caption{$\cO(\Nc^0)$}
        \label{fig:largeNpions:Dcontractionsub}
    \end{subfigure}
    \begin{subfigure}{0.48\textwidth}
    \centering
       \includegraphics[width=0.9\textwidth]{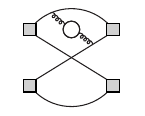}
        \caption{$\cO(\Nf\Nc^0)$}
        \label{fig:largeNpions:Ccontractionsub}
    \end{subfigure}
    \caption{
        Subleading $\Nc$ diagrams contributing to the Wick contractions in \cref{fig:largeNpions:CDleadingcontractions}. %, with their particular scaling given in the subcaptions. 
        Grey squares represent pion insertions, solid lines are quark propagators and curly lines represent gluons. }
    \label{fig:largeNpions:CDsubleadingcontractions}
\end{figure} 

The subtraction in the numerator of \cref{eq:largeNpions:scatteringlengthfromcorrs} implies that disconnected pieces in $D$ do not contribute to the scattering length. Non-factorizable contributions from $D$ correspond to subleading diagrams in which at least two gluons are exchanged between the two quark loops, and so are at most $\cO(\Nc^{-2})$---see, for example, \cref{fig:largeNpions:Dcontractionsub}. The leading contribution to the scattering length thus comes from $C$, which is $\cO(\Nc^{-1})$. Subleading corrections to this contraction come from diagrams exchanging a planar gluon that include an internal quark loop, and so scale as $\cO(\Nf/\Nc^2)$---see \cref{fig:largeNpions:Ccontractionsub}. 

All this discussion can be summarized as follow,
\begin{equation}
\begin{array}{rl}
C &= \displaystyle\frac{1}{\Nc}\left(a + b \frac{\Nf}{\Nc} \right) +\cO(\Nc^{-3}), \\
D - C_\pi^2 &=\displaystyle  \frac{c}{\Nc^2} +\cO(\Nc^{-3}), \\
C_\pi &= \displaystyle\frac{1}{\Nc}\left(d + e \frac{\Nf}{\Nc} \right) +\cO(\Nc^{-3}),
\end{array}
\end{equation}
where $a-e$ are numerical constants independent of $\Nc$ and $\Nf$, naturally expected to be $\cO(1)$. Combining these results with \cref{eq:largeNpions:correlatorCD,eq:largeNpions:scatteringlengthfromcorrs}, one derives
\begin{equation}\label{eq:largeNpions:NcNfscalingscatteringlength}
M_\pi a^{R}_0 = \pm \frac{1}{\Nc} \left( \tilde a + \tilde b \frac{\Nf}{\Nc} \mp \tilde c \frac{1}{\Nc}   \right) +\cO(\Nc^{-3}),
\end{equation}
where $\tilde{a}-\tilde{c}$ are linear combination of $a-e$ and the upper (lower) signs correspond to the $SS$ ($AA$) channels. Thus, large $\Nc$ arguments predict that one channel should be attractive (i.e., has positive scattering length), while the other is repulsive (negative scattering length), at least if subleading corrections are not abnormally large. Note that this same scaling is also predicted for other scattering quantities, such as the scattering amplitudes.

\subsection{Pion-pion scattering in SU($\Nf$) ChPT}

Scattering of two pions in a theory with $\Nf$ degenerate quarks can be described using standard ChPT, based on the SU($\Nf$) isospin group---see \cref{sec:QCD:chiralperturbationtheory}. Results for the scattering amplitudes are known up to NNLO~\cite{Bijnens:2011fm}. For the $SS$ and $AA$ channel, LO amplitudes are independent of $\Nf$,
\begin{equation}\label{eq:largeNpions:LOChPTscatteringamplitude}
\cM_2^{SS,\LO}=-\cM_2^{AA,\LO}=\frac{1}{\Fpi^2}(2\Mpi^2-s)\,.
\end{equation}
Using these results, the $s$-wave scattering lengths and effective ranges can be determined,
\begin{equation}\label{eq:largeNpions:LOChPTscatteringalength}
M_\pi a_0^{SS}=-\Mpi a_0^{AA}=-\frac{\Mpi^2}{16\pi^2\Fpi^2}\,,\quad\quad M_\pi^2a_0^Rr_0^R=-3\,.
\end{equation}
As anticipated, the results for the $SS$ channel are equal to that for the $I_{\pi\pi}=2$ channel presented in \cref{eq:hadrons:scatteringamplitudetwopionsI2} for $\Nf=2$. Note that LO ChPT predicts scattering lengths of opposite signs for the $SS$ and $AA$ channels, as expected from large $\Nc$ arguments. At LO, the $SS$ channel is predicted to be repulsive at threshold, while the $AA$ channel is expected to be attractive. Both scattering lengths also scale as $\cO(\Nc^{-1})$, since $\Fpi\sim\cO(\Nc^{1/2})$, in agreement with \cref{eq:largeNpions:NcNfscalingscatteringlength}.

At NLO, the amplitudes depend on linear combinations of the LECs from the fourth-order chiral Lagrangian in \cref{eq:QCD:chptlagrangianL4}. The NLO $s$-wave amplitudes---which do not include the LO results---read
\begin{multline}\label{eq:largeNpions:SSSUamplitudeChPT}
\frac{\Fpi^4}{\Mpi^4}\,\cM_{2,0}^{SS,\NLO} = 32L_{SS}+32q^2L_{SS}^\prime+\frac{128}{3}q^4L_{SS}^{\prime\prime} \\
\begin{array}{l}
\displaystyle +\frac{1}{4\pi^2}\left[-1-\frac{1}{\Nf^2}+\frac{1}{\Nf}+q^2\left(-3-\frac{\Nf}{18}\right)+q^4\left(-\frac{10}{3}-\frac{11\Nf}{27}\right)\right]\\[10pt]
\displaystyle +\frac{1}{4\pi^2}\left[-1-\frac{1}{\Nf^2}+\frac{1}{\Nf}+q^2\left(-3-\frac{\Nf}{6}\right)+q^4\left(-\frac{10}{3}-\frac{5N_\text{f}}{9}\right)\right]\ln{\frac{M_\pi^2}{\mu^2}} \\[5pt]
\displaystyle + \left(2+8q^2+8q^4\right)\bar{J}(s) + F_{SS}(q^2), 
\end{array}
\end{multline}\vspace{-1.1cm}

\begin{multline}\label{eq:largeNpions:AASUamplitudeChPT}
\frac{\Fpi^4}{\Mpi^4}\,\cM_{2,0}^{AA,\NLO} = -32L_{AA}+32q^2L_{AA}^{\prime}+\frac{128}{3}q^4L_{AA}^{\prime\prime} \\
\begin{array}{l}
\displaystyle +\frac{1}{4\pi^2}\left[-1-\frac{1}{\Nf^2}-\frac{1}{\Nf}+q^2\left(-3+\frac{\Nf}{18}\right)+q^4\left(-\frac{10}{3}+\frac{11\Nf}{27}\right)\right]\\[10pt]
\displaystyle +\frac{1}{4\pi^2}\left[-1-\frac{1}{\Nf^2}-\frac{1}{\Nf}+q^2\left(-3+\frac{\Nf}{6}\right)+q^4\left(-\frac{10}{3}+\frac{5\Nf}{9}\right)\right]\ln{\frac{M_\pi^2}{\mu^2}} \\[5pt]
\displaystyle + \left(2+8q^2+8q^4\right)\bar{J}(s) + F_{AA}(q^2). 
\end{array}
\end{multline}
Here $\bar{J}(x)$ is a loop integral---see~\rrcite{Passarino:1978jh,Scherer:2002tk}---and $q=q_2^*/\Mpi$ is the magnitude of the relative momentum normalized by the pion mass, 
\begin{equation}
q=\sqrt{\frac{s}{4\Mpi^2}-1}\,.
\end{equation}
We have also defined the following functions, which need to be evaluated numerically,
\begin{equation}
\begin{array}{rl}
F_{SS}(q^2)=&\displaystyle\int_{-1}^1\left[1+\frac{2}{\Nf^2}-\frac{2}{\Nf}+\frac{2\Nf}{3}+q^2\left(2+\frac{4N_\text{f}}{3}-2x\right) \right.\\[10pt]
&\displaystyle\left. + q^4 \left(2+\Nf-4x-\frac{4\Nf x}{3}+2x^2+\frac{\Nf x^2}{3}\right)\right] \bar{J}[t(x)]\,\d x,
\end{array}
\end{equation}
\begin{equation}
\begin{array}{rl}
F_{AA}(q^2)=&\displaystyle\int_{-1}^1\left[1+\frac{2}{\Nf^2}+\frac{2}{\Nf}-\frac{2\Nf}{3}+q^2\left(2-\frac{4\Nf}{3}-2x\right) \right.\\[10pt]
&\displaystyle\left. + q^4 \left(2-\Nf-4x+\frac{4\Nf x}{3}+2x^2-\frac{\Nf x^2}{3}\right)\right] \bar{J}[t(x)]\,\d x,
\end{array}
\end{equation}
with $t(x)=-2q_2^{*2}(1-x)$. Finally, we have defined the following linear combinations of LECs from the $\cL_4$ Lagrangian in \cref{eq:QCD:chptlagrangianL4},

\noindent
\begin{equation}\label{eq:largeNpions:SSLECs}
\begin{array}{rl}
L_{SS} &= L_0 + 2 L_1 + 2 L_2 + L_3 - 2 L_4 - L_5 + 2 L_6 + L_8\,,\\
L_{SS}^{\prime}&=4L_0+4L_1+6L_2+2L_3-2L_4-L_5\,,\\
L_{SS}^{\prime\prime}&=3L_0+2L_1+4L_2+L_3\,,
\end{array}
\end{equation} 
\begin{equation}\label{eq:largeNpions:AALECs}
\begin{array}{rl}
L_{AA} &= L_0 - 2 L_1 - 2 L_2 + L_3 +2 L_4 - L_5 - 2 L_6 + L_8\,,\\
L_{AA}^{\prime}&=-4L_0+4L_1+6L_2-2L_3-2L_4+L_5\,,\\
L_{AA}^{\prime\prime}&=-3L_0+2L_1+4L_2-L_3\,.
\end{array}
\end{equation} 
These results can be used to determine the scattering lengths, given in eqs.~(2.7) and~(2.8) of \rcite{Baeza-Ballesteros:2022azb}, and other scattering abservables.

The amplitudes in \cref{eq:largeNpions:LOChPTscatteringamplitude,eq:largeNpions:SSSUamplitudeChPT,eq:largeNpions:AASUamplitudeChPT} are expected to scale with $\Nc$ as given in \cref{eq:largeNpions:NcNfscalingscatteringlength}. As we have seen, this holds for LO terms, and is also straightforward to check for NLO terms without LECs. This means it must also hold for the LEC terms, as expected from \cref{eq:QCD:largeNcscalingLECs}. The combinations in \cref{eq:largeNpions:SSLECs,eq:largeNpions:AALECs} can thus be parametrized as a power series in $\Nc$, with common large $\Nc$ for both channels. For example, we write
\begin{equation}\label{eq:largeNpions:NcparametrizationLR}
L_R=\Nc L^{(0)}+L_R^{(1)}+\cO(\Nc^{-1})\,,
\end{equation}
and similarly for $L_R^\prime$ and $L_R^{\prime\prime}$. The scaling with $\Nf$ however, is not so clear. The chiral logarithms in \cref{eq:largeNpions:SSSUamplitudeChPT,eq:largeNpions:AASUamplitudeChPT} depend explicitly on inverse powers of $\Nf$, which are not expected from large $\Nc$ arguments. This does not mean the large $\Nc$ limit breaks, but is a consequence of considering the wrong effective theory. This $\Nf$ dependence is compensated by an analogous implicit dependence of the LECs. To consistently apply ChPT in the large $\Nc$ limit, one needs to include the singlet meson, the $\eta'$, which becomes degenerate with the pions. The symmetry group is therefore U$(\Nf)$ instead if SU($\Nf$), as we consider in the following.
% but instead that the LECs depends implicitly on $\Nf$ in a way that would cancel this $\Nf$ dependence al large $\Nc$. 

\subsection{Pion-pion scattering in U($\Nf$) ChPT}

At large $\Nc$, the $\eta'$ meson becomes degenerate with pions, and so needs to be included in the low-energy EFT. This is the basis of large $\Nc$ or U($\Nf$) ChPT, introduced in \cref{sec:QCD:largeNChPT}. The implicit $\Nf$ dependence of the LECs in the SU($\Nf$) theory is a result of integrating out this particle.

While several quantities have been computed in U($\Nf$) ChPT, this is not the case for meson scattering amplitudes, that we determined for the first time in \rcite{Baeza-Ballesteros:2022azb}. At LO, scattering amplitudes are equivalent to those in \cref{eq:largeNpions:LOChPTscatteringamplitude}. Contributions at NLO in the counting of large $\Nc$ ChPT, \cref{eq:QCD:ChPTpowercounting,eq:QCD:largeNChPTpowercounting}, are given by tree-level diagrams including $\cO(\Nc)$ LECs. However, NLO is in general insufficient, and one needs to go to NNLO.

\begin{figure}[!tp]
    \centering
    \begin{subfigure}{0.24\textwidth} 
    \centering
        \includegraphics[scale=0.24]{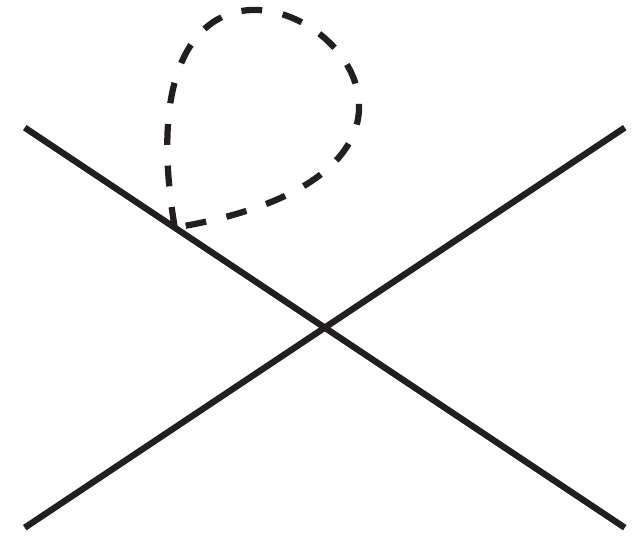} \caption{}\label{fig:largeNpions:loopmassrenorm}
    \end{subfigure}
    \begin{subfigure}{0.24\textwidth} 
    \centering
        \includegraphics[scale=0.24]{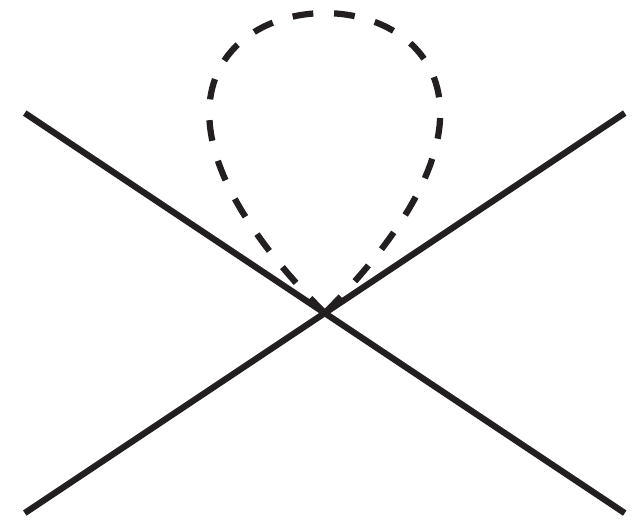} \caption{}\label{fig:largeNpions:tadpoleeta}
    \end{subfigure}
    \begin{subfigure}{0.24\textwidth} 
    \centering
        \includegraphics[scale=0.24]{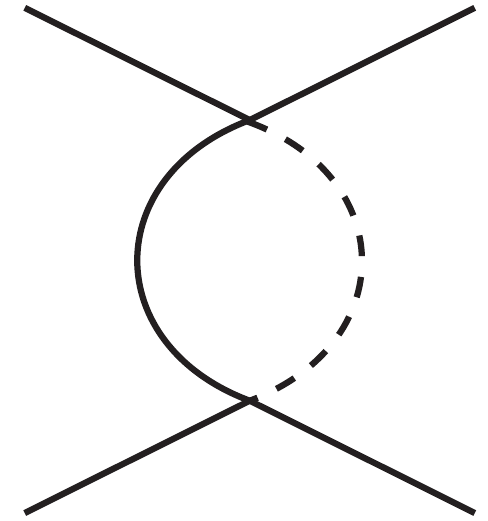} \caption{}\label{fig:largeNpions:loop1eta}
    \end{subfigure}
    \begin{subfigure}{0.24\textwidth} 
    \centering
        \includegraphics[scale=0.24]{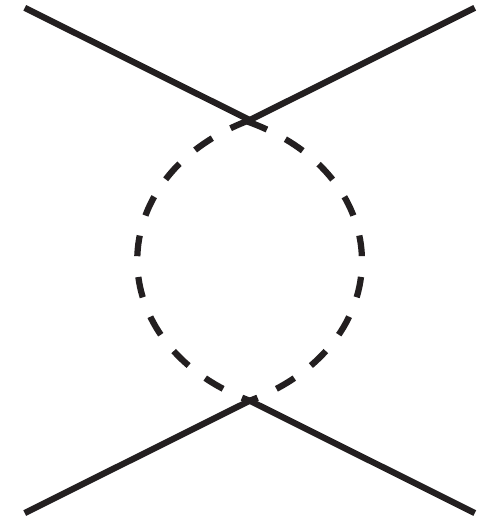} \caption{}\label{fig:largeNpions:loop2eta}
    \end{subfigure}
    \caption{
        Additional one-loop Feynman diagrams needed to study two-pion scattering in the $SS$ and $AA$ channels in large $\Nc$ ChPT. Solid lines depict pions, while dotted ones represent the $\eta'$.}
    \label{fig:largeNpions:etadiagrams}
\end{figure} 

At NNLO, loop diagrams start to contribute and effects from the $\eta'$ meson need to be included, represented in \cref{fig:largeNpions:etadiagrams}.  %This contributions originate from new diagrams that need to be added to the SU($\Nf$) result, represented in \cref{fig:largeNpions:etadiagrams}. 
In addition, one also needs to consider tree-level diagrams with two insertions of $\cO(\Nc)$ LECs. Overall, the full amplitudes up to NNLO can be computed as
\begin{equation}
\left[\cM_2^{R\text{,NNLO}}\right]_{\text{U}(N_\text{f})}  =  \left[\cM_2^{R\text{,NLO}}\right]_{\text{SU}(N_\text{f})} + \cM_K  + \Delta Z^2_{\eta'} \mathcal{M}_\text{LO}  + \mathcal{M}^\text{loop}_{\eta'}\,,
\end{equation}
where $\Delta Z^2_{\eta'}$ represents the additional mass renormalization due to $\eta'$ loops---diagram \ref{fig:largeNpions:loopmassrenorm}---and $\cM_{\eta'}^\text{loop}$ includes the contributions from diagrams~\ref{fig:largeNpions:tadpoleeta},~\ref{fig:largeNpions:loop1eta} and~\ref{fig:largeNpions:loop2eta}. Finally, $\cM_K$ accounts to tree level diagrams with products of two $\cO(\Nc)$ LECs.

The full scattering amplitudes up to NNLO in large $\Nc$ ChPT are 
 \begin{equation}\label{eq:largeNpions:SSUamplitudeChPT}
\begin{array}{rl}
 \left[\mathcal{M}_2^{SS,\text{NNLO}}\right]_{\text{U}(\Nf)}=&\displaystyle\left[\mathcal{M}_2^{SS,\text{NLO}}\right]_{\text{SU}(\Nf)} + \left(\frac{M_\pi^2}{F_\pi^2}\right)^3\left[K_{SS}+q^2K_{SS}'\right] \\[10pt]
 &-\displaystyle  \frac{2M_\pi^4}{F_\pi^4 \Nf}\left(1-\frac{2}{\Nf}\right)B_1(t)-\frac{2M_\pi^4}{F_\pi^4 \Nf^2}B_2(t)  + (t\leftrightarrow u)
\,,
\end{array}
\end{equation}
\begin{equation}\label{eq:largeNpions:AAUamplitudeChPT}
\begin{array}{rl}
 \left[\mathcal{M}_2^{AA,\text{NNLO}}\right]_{\text{U}(\Nf)}=&\displaystyle\left[\mathcal{M}_2^{AA,\text{NLO}}\right]_{\text{SU}(\Nf)} - \left(\frac{M_\pi^2}{F_\pi^2}\right)^3\left[K_{AA}+q^2K_{AA}'\right]  \\[10pt]
 &+\displaystyle \frac{2M_\pi^4}{F_\pi^4 \Nf}\left(1+\frac{2}{\Nf}\right)B_1(t)-\frac{2M_\pi^4}{F_\pi^4 \Nf^2}B_2(t)  + (t\leftrightarrow u)\,,
\end{array}
\end{equation}
where $B_1(x)$ and $B_2(x)$ are loop integrals corresponding to diagrams~\ref{fig:largeNpions:loop1eta} and~\ref{fig:largeNpions:loop2eta}, respectively, 
\begin{equation}
\def\arraystretch{1.5}
\begin{array}{rl}
B_1(z)=&\displaystyle\frac{1}{(4\pi)^2}\left\{\frac{1}{M_{\eta'}^2-M_\pi^2}\left(M_{\eta'}^2\log{\frac{M_{\eta'}^2}{\mu^2}}-M_\pi^2\log{\frac{M_\pi^2}{\mu^2}}\right)\right.\\[10pt]
&\displaystyle\left.+\int_0^1 \log\left[\frac{M_\pi^2x+M_{\eta'}^2(1-x)-x(1-x)z}{M_\pi^2x+M_{\eta'}^2(1-x)}\right]\right\}\text{d}x\,,
\end{array}
\end{equation}
\begin{equation}\label{eq:loopint2}
\displaystyle B_2(z)=\frac{1}{(4\pi)^2}+\frac{1}{(4\pi)^2}\log{\frac{M_{\eta'}^2}{\mu^2}}-\bar{J}\left(z\frac{M_\pi}{M_{\eta'}^2}\right)\,,
\end{equation}
with $\Metap$ given by the Witten-Veneziano formula in \cref{eq:QCD:WittenVenezianoformula}. Also, we have introduced $K_R$ and $K_R^\prime$, which  are products of $\mathcal{O}(N_\text{c})$ LECs. For example, we have
\begin{equation}\label{eq:largeNpions:LOKfactorUChPT}
 K_{SS}=  K_{AA}= \left[128(L_8-2L_5)^2\right]_{\mathcal{O}(N_\text{c}^2)} \,,
\end{equation} 
where $[...]_{\mathcal{O}(N_\text{c}^2)}$ refers to the leading $\cO(\Nc^2)$ contribution.
Moreover, note that although the linear combinations of LECs appearing in $\left[\mathcal{M}_2^{R,\text{NLO}}\right]_{\text{SU}(\Nf)}$ take the same form as in \cref{eq:largeNpions:SSLECs,eq:largeNpions:AALECs}, their numeric values may differ between the SU($\Nf$) and the U($\Nf$) theories.

Taking the large $\Nc$ limit, $\Metap\rightarrow \Mpi$, one can check the $1/\Nf$ factors appearing in the chiral logarithms cancel and the expected large $\Nc$ scaling from \cref{eq:largeNpions:NcNfscalingscatteringlength} is recovered. This same scaling must hold for the LEC terms, which can be parametrized as
\begin{equation}\label{eq:largeNpions:LECsparametrizationU4Nf}
\left[L_R \right]_{\text{U}(N_\text{f})} = N_\text{c} L^{(0)} + N_\text{f} L_\text{c}^{(1)} \mp L_\text{a}^{(1)} +\mathcal{O}(N_\text{c}^{-1})\,,
\end{equation}
where, again, the upper and lower signs correspond to the $SS$ and $AA$ channels, respectively. The correlated and anticorrelated terms, $L^{(1)}_\text{c}$ and $L^{(1)}_\text{a}$ are independent of $\Nf$. Comparing to \cref{eq:largeNpions:SSLECs,eq:largeNpions:AALECs}, we note
\begin{equation}\label{eq:largeNpions:LECsparametrizationU4Nfcombination}
\begin{array}{rl}
L_0 + L_3 - L_5 + L_8 &= N_\text{c} L^{(0)} + N_\text{f} L_\text{c}^{(1)}  +\mathcal{O}(N_\text{c}^{-1})\,, \\
2 L_1 + 2 L_2 - 2 L_4 + 2 L_6 &= L_\text{a}^{(1)} +\mathcal{O}(N_\text{c}^{-1})\,.
\end{array}
\end{equation}
Equivalent arguments hold for $L_R^\prime$ and $L_R^{\prime\prime}$, as well as for $K_R$ and $K_R^\prime$. Note that higher order $\Nc$ corrections only contribute at higher order in the power counting in \cref{eq:QCD:largeNChPTpowercounting}, and so truncating at this order is natural in U($\Nf$) ChPT.

We can also match U($\Nf$) and SU($\Nf$) ChPT in the $\Metap\gg\Mpi$ limit, obtaining a relation between the LECs in the two theories, as shown in \cref{eq:QCD:matchingLECs}. For example, for the $L_R^{(1)}$ terms in \cref{eq:largeNpions:NcparametrizationLR}, we get
\begin{equation}\label{eq:largeNpions:matchingSUUSS}
\begin{array}{rl}
\left[L^{(1)}_{SS}\right]_{\text{SU}(N_\text{f})}=&\displaystyle \left[L^{(1)}_{SS}\right]_{\text{U}(N_\text{f})}-\frac{1}{8N_{\text{f}}^2(4\pi)^2}(N_\text{f}\lambda_0-\lambda_0+1)\,,\\[10pt]
\left[L^{(1)}_{AA}\right]_{\text{SU}(N_\text{f})}=&\displaystyle \left[L^{(1)}_{AA}\right]_{\text{U}(N_\text{f})}+\frac{1}{8N_{\text{f}}^2(4\pi)^2}(1-N_\text{f}\lambda_0-\lambda_0)\,,
\end{array}
\end{equation}
with $\lambda_0=\log(M_0^2/\mu^2)$ and $M_0^2=\Metap^2-\Mpi^2$. This relation sheds light on the origin of the implicit $1/\Nf$ and $1/\Nf^2$ dependence of the SU($\Nf$) LECs.

\newpage\section{Lattice setup}\label{sec:largeNpions:laticesetup}

We study pion-pion scattering in the $SS$ and $AA$ channels using lattice simulations with $\Nc=3-6$. In all cases, we consider several values of the pion mass, $\Mpi=360-590$ MeV. Simulations are performed with the HiRep code~\cite{DelDebbio:2008zf,DelDebbio:2009fd} and we work with a lattice setup analogous to that of \rrcite{Hernandez:2019qed,Donini:2020qfu}. Configurations are generated using the Iwasaki gauge action with $\Nf=4$ flavors of dynamical clover-improved Wilson fermions. The value of $c_\text{sw}$ is determined from one-loop $\Nc=3$ results~\cite{Aoki:2003sj} boosted by the plaquette, which is kept constant for all $\Nc$ according to its expected leading $\Nc$ dependence~\cite{Hernandez:2019qed}. 

In this work, we compute the relevant correlation functions using two different regularizations in the valence sector: a unitary setup with the same action for sea and valence fermions, and a mixed-action setup~\cite{Bar:2002nr} with maximally-twisted clover-improved fermions for the valence quarks. In the latter case we tune the bare twisted mass, $a\mu_0$, to ensure the valence pion mass, $\Mpi^\text{v}$, matches its sea value, $\Mpi^\text{s}$, computed in the unitary setup. The use of maximally-twisted fermions allows us to achieve automatic $\cO(a)$ improvement, and also to determine $\Fpi$ without the need of renormalization constants~\cite{Shindler:2007vp},
\begin{equation}
\Fpi=\frac{\sqrt{2}\mu_0 \langle 0| \overline{q} \gamma_5 q|\pi\rangle}{\Mpi}\,.
\end{equation}
In addition, the combination of both regularizations is useful to study discretization effects, as physical observables must agree in the continuum limit.

A summary of the parameters used in the simulations is presented in \cref{tab:largeNpions:ensembles}, and in \cref{tab:largeNpions:massdecayxi} we summarize our results for single-pion quantities. Pion masses and decay constants are extracted from single-pion correlators, %, as describe in \cref{sec:QCD:computationofobservables}. % which we fit to a single cosh function, while the PCAC mass is determined from the axial-psudoscalar and the pseudoscalar-pseudoscalar correlation functions. 
%Both $\Mpi$ and $\Fpi$ have 
and are corrected to account for finite-volume effects estimated from ChPT---see \cref{eq:QCD:finitevolumepionmass}. We have also determined the physical lattice spacing of the simulations---see \rcite{Hernandez:2019qed} for details. The ``A'' ensembles have $a=0.075(2)$ fm and had previously been used in \rrcite{Hernandez:2019qed,Donini:2020qfu}. On the other hand, the ``B'' and ``C'' ensembles, generated for this work, have $a=0.065(2)$ fm and $a=0.059(2)$ fm, respectively. Assuming $\cO(a^2)$ scaling, the finest ensembles should present $\sim40\%$ smaller cutoff effects than the coarsest ones.

\begin{table}[p!]
\centering
\begin{tabular}{ccccccc}
\toprule
  Ensemble& $L^3 \times T$ &$\beta$ & $c_\text{sw}$ &$am^\text{s}$ & $am^\text{v}$ & $a \mu_0$  \\ \midrule 
3A10 & $20^3 \times 36$ &\multirow{ 5}{*}{1.778}&\multirow{ 5}{*}{1.69}& $-0.4040$ & $-0.4214$  & 0.01107   \\ %\cline{1-2} \cline{5-7} 
3A11 & $24^3 \times 48$ && & $-0.4040$ & $-0.4214$ & 0.01107   \\ %\cline{1-2} \cline{5-7} 
3A20 &$24^3 \times 48$ &&& $-0.4060$   & $-0.4196$ & 0.00781    \\ %\cline{1-2} \cline{5-7} 
3A30 &$24^3 \times 48$ &&& $-0.4070$   & $-0.4187$ & 0.00632   \\ %\cline{1-2}  \cline{5-7} 
3A40 &$32^3 \times 60$ &&& $-0.4080$   & $-0.4163$ & 0.00513  \\ \midrule 

3B10 & $24^3 \times 48$ &\multirow{ 2}{*}{1.820} &\multirow{ 2}{*}{1.66}& $-0.3915$  & $-0.4035$ & 0.00825 \\ %\cline{1-2}  \cline{5-7} 
3B20 & $32^3 \times 60$ & & & $-0.3946$ & $-0.4011$ & 0.00431     \\ \midrule   

3C10 & $24^3 \times 48$ &\multirow{ 2}{*}{1.850}&\multirow{ 2}{*}{1.64}& $-0.3817$ & $-0.3934$ & 0.00870  \\ %\cline{1-2}  \cline{5-7} 
3C20 & $32^3 \times 60$ & & & $-0.3847$ &$-0.3921$ & 0.00512      \\ \midrule   

4A10 &$20^3 \times 36$ &\multirow{ 4}{*}{3.570}&\multirow{ 4}{*}{1.69} &$-0.3735$&  $-0.4163$ & 0.00513 \\ %\cline{1-2}  \cline{5-7} 
4A20 &$24^3 \times 48$& && $-0.3752$ & $-0.3865$& 0.00844 \\ %\cline{1-2}  \cline{5-7} 
4A30 &$24^3 \times 48$& && $-0.3760$ & $-0.3865$& 0.00778 \\ %\cline{1-2} \cline{5-7} 
4A40 &$32^3 \times 60$& && $-0.3780$ & $-0.3851$& 0.00546\\ \midrule 
5A10 &$20^3 \times 36$& \multirow{ 4}{*}{5.969}&\multirow{ 4}{*}{1.69} &$-0.3458$ & $-0.3611$ &0.01225 \\ %\cline{1-2}  \cline{5-7} 
5A20 &$24^3 \times 48$& && $-0.3490$ & $-0.3611$& 0.00906   \\ %\cline{1-2} \cline{5-7} 
5A30 &$24^3 \times 48$& && $-0.3500$ & $-0.3607 $& 0.00824  \\ %\cline{1-2} \cline{5-7} 
5A40 &$32^3 \times 60$& && $-0.3530$ & $-0.3596 $& 0.00509   \\ \midrule 
6A10 &$20^3 \times 36$ &\multirow{ 4}{*}{8.974}&\multirow{ 4}{*}{1.69} &$ -0.3260$ & $-0.3415$ & 0.01298   \\ %\cline{1-2} \cline{5-7} 
6A20 &$24^3 \times 48$& && $-0.3300$ &$ -0.3414$ & 0.00956 \\ %\cline{1-2}  \cline{5-7} 
6A30 &$24^3 \times 48$& && $-0.3311$  & $-0.3414$ & 0.00803 \\ %\cline{1-2}  \cline{5-7} 
6A40 &$32^3 \times 60$& && $-0.3340$ & $-0.3409$ & 0.00542 \\ \bottomrule 
\end{tabular}
\caption{Summary of ensemble parameters used in this work. $L$ and $T$ indicate the number of points in the spatial and temporal extent of the lattices, respectively, $\beta$ is the gauge coupling, $c_\text{sw}$ is the Sheikloleslami-Wohlert coefficient, and $am^\text{s}$ is the bare mass of the Dirac operator in the sea sector. Finally, $am^\text{v}$ and $a \mu_0$ are bare mass and the bare twisted mass used in the mixed-action setup, in this same order.  }

\label{tab:largeNpions:ensembles}
\end{table}

\begin{table}[p!]

\centering
{
\begin{tabular}{ccccccc}
\hline
Ensemble& $aM_{\pi}^\text{s}$ & $aM_{\pi}^\text{v}$ & $aF_{\pi}$  & $\xi$ 	\\ \toprule
3A10 & 0.222(3) 		& 0.2211(23)	 & 0.0449(4) 	 & 0.154(6)	\\
3A11 & 0.2150(15) 	& 0.2185(10)	 & 0.0452(3) 	 & 0.148(3)	\\
3A20 & 0.1853(13) 	& 0.1830(8)	 & 0.0409(3) 	 & 0.1267(25)	\\
3A30 & 0.1611(16) 	& 0.1616(8)	 & 0.0378(3) 	& 0.1157(25)	\\
3A40 & 0.1419(10) 	& 0.1423(6)	 & 0.03577(14) 	& 0.1003(14)	\\  \midrule
3B10 & 0.1751(11) 	& 0.1764(9)	 & 0.03626(23) 	& 0.150(3)	\\
3B20 & 0.1189(8) 	& 0.1221(6)	 & 0.03121(13) 	& 0.0969(16)	\\  \midrule
3C10 & 0.1756(18) 	& 0.1759(18)	 & 0.0336(4) 	 	& 0.174(7)	\\
3C20 & 0.1308(13) 	& 0.1289(12)	 & 0.02866(24) 	& 0.128(4)	\\  \midrule
4A10 & 0.2044(13) 	& 0.2035(15)	 & 0.0521(4) 	   	& 0.0968(25)\\
4A20 & 0.1805(8) 	& 0.1799(6) 	 & 0.05103(16) 	& 0.0787(9)	\\
4A30 & 0.1707(7) 	& 0.1730(5)	 & 0.04952(19) 	& 0.0773(9)	\\
4A40 & 0.1399(8) 	& 0.1419(8) 	 & 0.0464(3) 	 	& 0.0593(12)	\\  \midrule
5A10 & 0.2125(11) 	& 0.2126(8)	 & 0.06154(22) 	& 0.0756(10)	\\
5A20 & 0.1803(6) 	& 0.1798(5)	 & 0.05846(24) 	& 0.0599(6)	\\
5A30 & 0.1707(6) 	& 0.1715(6)	& 0.0570(3) 	 	& 0.0573(9)	\\
5A40 & 0.1331(5) 	& 0.1330(4)	 & 0.05306(15) 	& 0.0398(3)	\\  \midrule
6A10 & 0.2147(7) 	& 0.2142(6)	 & 0.06874(21) 	& 0.0615(6)	\\
6A20 & 0.1798(6) 	& 0.1802(4)	 & 0.06582(21) 	& 0.0475(4)	\\
6A30 & 0.1685(7) 	& 0.1666(5)	 & 0.06324(23) 	& 0.0439(5)	\\
6A40 & 0.1353(3)		& 0.1347(5)	 & 0.05950(13)	& 0.0324(3)	\\  \bottomrule
\end{tabular}}
\caption{ Results for the pion mass for the unitary and mixed-action setups ($M_\pi^{\text{s}}$ and $M_\pi^\text{v}$, respectively), the pion decay constant, $\Fpi$, and the chiral parameter, $\xi$, as defined in \cref{eq:QCD:chiralparameterdefinition}. All quantities have their leading finite-volume effects corrected---see \cref{eq:QCD:finitevolumepionmass}.}
\label{tab:largeNpions:massdecayxi}
\end{table}

\newpage
\subsection{Extraction of finite-volume energies}

Two-pion ground-state energies are determined from two-point correlation functions,
\begin{equation}
C_R(t)=\langle O_R(t)O_R^\dagger(t)\rangle\,,
\end{equation}   
where we define two-pion operators for the channels of interest,
\begin{equation}
\begin{array}{rl}
O_{SS}(t)&=\pi^+(t)\pi^+(t)\,,\\
O_{AA}(t)&=\displaystyle\frac{1}{\sqrt{2}}\left[\pi^+(t)D_s^+(t)-K^+(t)D^+(t)\right]\,.
\end{array}
\end{equation}
Single-pion interpolators in these operators are projected to zero momentum. For example,
\begin{equation}
\pi^+(t)=\sum_i\bar{d}(\bm{x},t)\gamma_5 u(\bm{x},t)\,,
\end{equation}
and similarly for $K^+$, $D^+$ and $D_s^+$, with the corresponding quark flavors~\cite{PDG:2020}. As shown in \cref{eq:largeNpions:correlatorCD}, the correlation functions can be computed as a linear combination of the disconnected and the connected Wick contractions. On the lattice, these are evaluated, respectively, as
\begin{equation}\label{eq:largeNpions:Wickcontractions}
\begin{array}{rl}
D(t)&\displaystyle=\sum_{\bm{y}_1,\bm{y}_2} \sum_{\bm{x}_1,\bm{x}_2}\langle\Tr[S^\dagger(y_1,x_1)S(y_1,x_1)]\Tr[S^\dagger(y_2,x_2)S(y_2,x_2)]\rangle\,,\\[5pt]
C(t)&\displaystyle=\sum_{\bm{y}_1,\bm{y}_2} \sum_{\bm{x}_1,\bm{x}_2}\langle\Tr[S^\dagger(y_2,x_1)S(y_1,x_1)S^\dagger(y_1,x_2)S(y_2,x_2)]\rangle\,,
\end{array}\vspace{0.2cm}
\end{equation}
where  $x_i=(0,\bm{x}_i)$ and $y_i=(t,\bm{y}_i)$. We evaluate these contractions using time- and spin-diluted stochastic sources with $\mathbbm{Z}_2\times\mathbbm{Z}_2$ noise, and make use of time-translation invariance to average overmultiple equivalent locations of the source and the sink.

From the correlation functions, we are able to extract the ground-state energies. We take thermal effects into account, which become relevant around $t\sim T/2$---see \cref{sec:QCD:thermaleffectslattice}. Taking into account backwards propagation and that the leading thermal effects are time independent for zero total momentum, the two-pion correlation function asymptotically looks like
\begin{equation}
C_R(t)=A_R\cosh[E_R(t-T/2)]+B_R\,,
\end{equation}
where $A_R$ and $B_R$ are positive constants related to matrix elements of the two-pion interpolating operators, and $E_R$ is the corresponding two-pion energy. Instead of directly fitting the correlators, we consider the ratio~\cite{Umeda:2007hy,Feng:2009ij}
\begin{equation}\label{eq:largeNpions:ratiocreation}
R(t)=\frac{C_R(t+1)-C_R(t-1)}{C_\pi^2(t+1)-C_\pi^2(t-1)}\,,
\end{equation}
which eliminates the constant noise and has been empirically seen to reduce excited-states contamination~\cite{Fischer:2020jzp,Bulava:2022vpq}. At late times, the ratio behaves as
\begin{equation}\label{eq:largeNpions:ratiofunction}
R(t)=K_R\left[\cosh(\Delta E_R t')+\sinh(\Delta E_R t')\coth(2\Mpi t') \right]\,,
\end{equation}
where $\Delta E_R = E_R-2\Mpi$ is the energy shift, $t'=t-T/2$ and $K_R$ is a normalization constant that depends on the channel.

We extract the finite-volume energy shift from fits to \cref{eq:largeNpions:ratiofunction}, fixing $\Mpi$ to the value obtained from fitting the single-pion correlator and taking into account correlations between different times slices. Errors are determined using bootstrap, averaging the configurations on blocks several times larger than the autocorrelation time. We observe that our results are unaffected by the choice of the block size. Fits of the correlators are performed over the range $t\in[t_\text{min},T/2-1]$ and are repeated for several $t_\text{min}$. The energy shift is determined where the results show a plateaux, which we determine by visual inspection. Our results for the mixed-action setup are shown in \cref{tab:largeNpions:resultsenergya0}.

\begin{table}[p!]

\centering
{
\begin{tabular}{ccccc}
\toprule
Ensemble& $\Delta E_{SS}/M_\pi$ & $M_\pi a_0^{SS}$ & ${\Delta} E_{AA}/M_\pi$ & $M_\pi a_0^{AA}$ \\ \midrule   
3A10 & 0.100(4) 		& $-$0.494(17)	 &  $-$0.075(4) 	 & 0.73(4)     \\ 
3A11 & 0.0495(17) 	& $-$0.442(13)	 &  $-$0.0458(13) 	 & 0.72(3)     \\ 
3A20 & 0.082(4) 		& $-$0.416(14)	 &  $-$0.063(3) 	 & 0.57(4)     \\  
3A30 & 0.114(4) 		& $-$0.392(11)	 &  $-$0.082(7) 	 & 0.52(5)     \\   
3A40 & 0.0532(20) 	& $-$0.324(11)	 &  $-$0.050(3) 	 & 0.49(3)     \\  \midrule  
3B10 & 0.104(3) 		& $-$0.456(9)	 &  $-$0.093(3) 	 & 0.81(3)     \\   
3B20 & 0.087(3) 		& $-$0.324(10)	 &  $-$0.064(3) 	 & 0.390(24)   \\  \midrule  
3C10 & 0.120(6) 		& $-$0.503(19)	 &  $-$0.117(5) 	 & 1.02(4)     \\   
3C20 & 0.111(9) 		& $-$0.45(3) 	 &  $-$0.082(4) 	 & 0.64(4)     \\  \midrule  
4A10 & 0.0825(25) 	& $-$0.345(9) 	 &  $-$0.063(3) 	 & 0.445(25)   \\ 
4A20 & 0.0473(14) 	& $-$0.255(6) 	 &  $-$0.0410(15) 	 & 0.319(15)   \\  
4A30 & 0.0529(14) 	& $-$0.252(6)	 &  $-$0.0456(23) 	 & 0.317(20)   \\   
4A40 & 0.0269(13) 	& $-$0.179(9) 	 &  $-$0.0276(11) 	 & 0.236(11)   \\  \midrule 
5A10 & 0.0507(11) 	& $-$0.259(4)	 &  $-$0.0469(10) 	 & 0.356(10)   \\ 
5A20 & 0.0342(11) 	& $-$0.192(6)	 &  $-$0.0334(7) 	 & 0.250(6)    \\  
5A30 & 0.0381(12) 	& $-$0.186(6)	 &  $-$0.0323(15) 	 & 0.206(11)   \\   
5A40 & 0.0235(10) 	& $-$0.132(6)	 &  $-$0.0209(13) 	 & 0.141(9)    \\  \midrule 
6A10 & 0.0402(15) 	& $-$0.217(7)	 &  $-$0.0393(9) 	 & 0.295(8)    \\ 
6A20 & 0.0290(13) 	& $-$0.167(6)	 &  $-$0.0261(6) 	 & 0.189(5) 	   \\  
6A30 & 0.0301(16) 	& $-$0.139(7)	 &  $-$0.0308(16) 	 & 0.177(11)	   \\   
6A40 & 0.0171(5)		& $-$0.102(3)	 &  $-$0.0172(10)	 & 0.118(7)	   \\  \bottomrule  

\end{tabular}}
\caption{  Results for the two-pion energy shifts, $\mathrm{\Delta} E_R = E_R - 2M_\pi$ , for the $SS$ and $AA$ channels obtained in the mixed-action setup, together with the corresponding scattering lengths computed using \cref{eq:hadrons:thresholdexpansionQC2} to $\cO(L^{-5})$. }
\label{tab:largeNpions:resultsenergya0}
\end{table}

\section{Results for pion-pion scattering}\label{sec:largeNpions:energies}

Using the finite-volume energies, we determine the scattering lengths using the threshold expansion, given in \cref{eq:hadrons:thresholdexpansionQC2}. The results at $\cO(L^{-5})$ are presented in \cref{tab:largeNpions:resultsenergya0}, noting that the scattering lengths are roughly of the same magnitude for both channels with opposite signs. This agrees with expectations from the large $\Nc$ limit and ChPT, and indicates the $SS$ channel is repulsive, while the $AA$ channel is attractive. 

\begin{figure}[!p]
    \centering
    \begin{subfigure}{0.495\textwidth} 
    \centering
        \includegraphics[width=\textwidth]{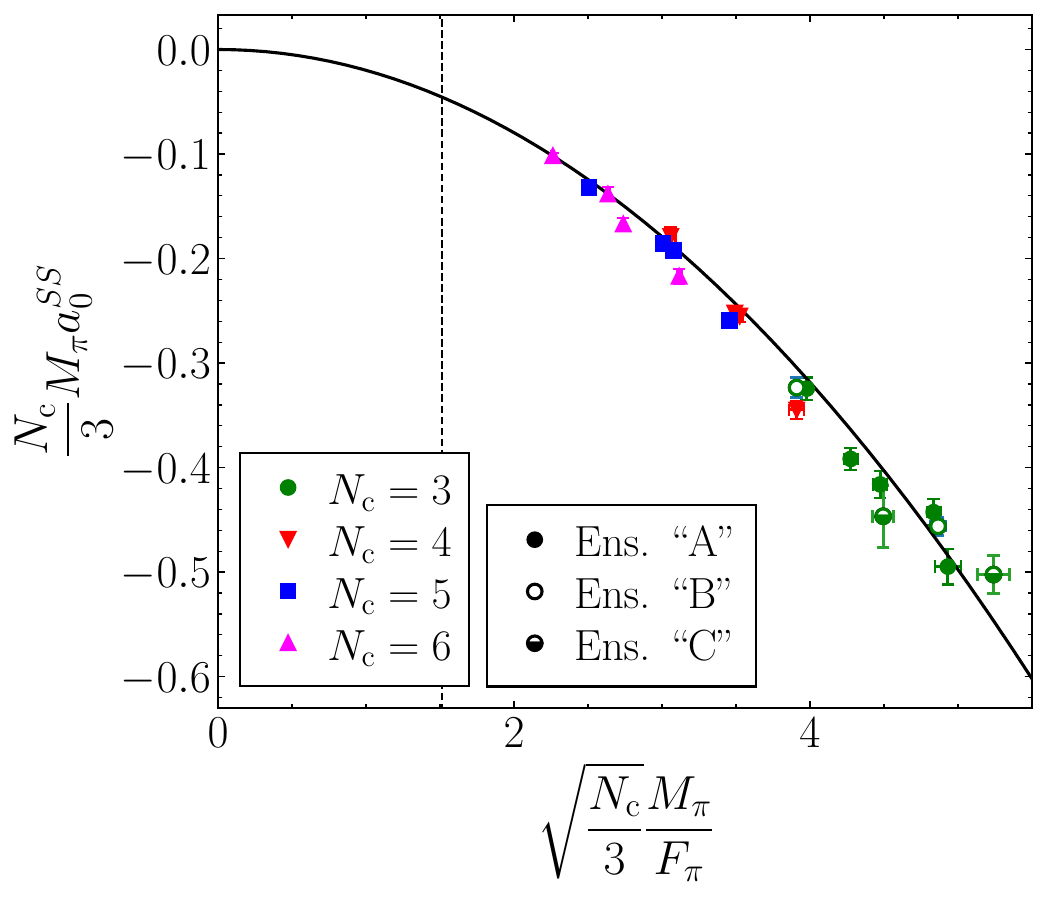}
        \caption{$SS$ channel}
        \label{fig:largeNpions:scatteringlengthSS}
    \end{subfigure}
    \begin{subfigure}{0.495\textwidth}
    \centering
       \includegraphics[width=0.99\textwidth]{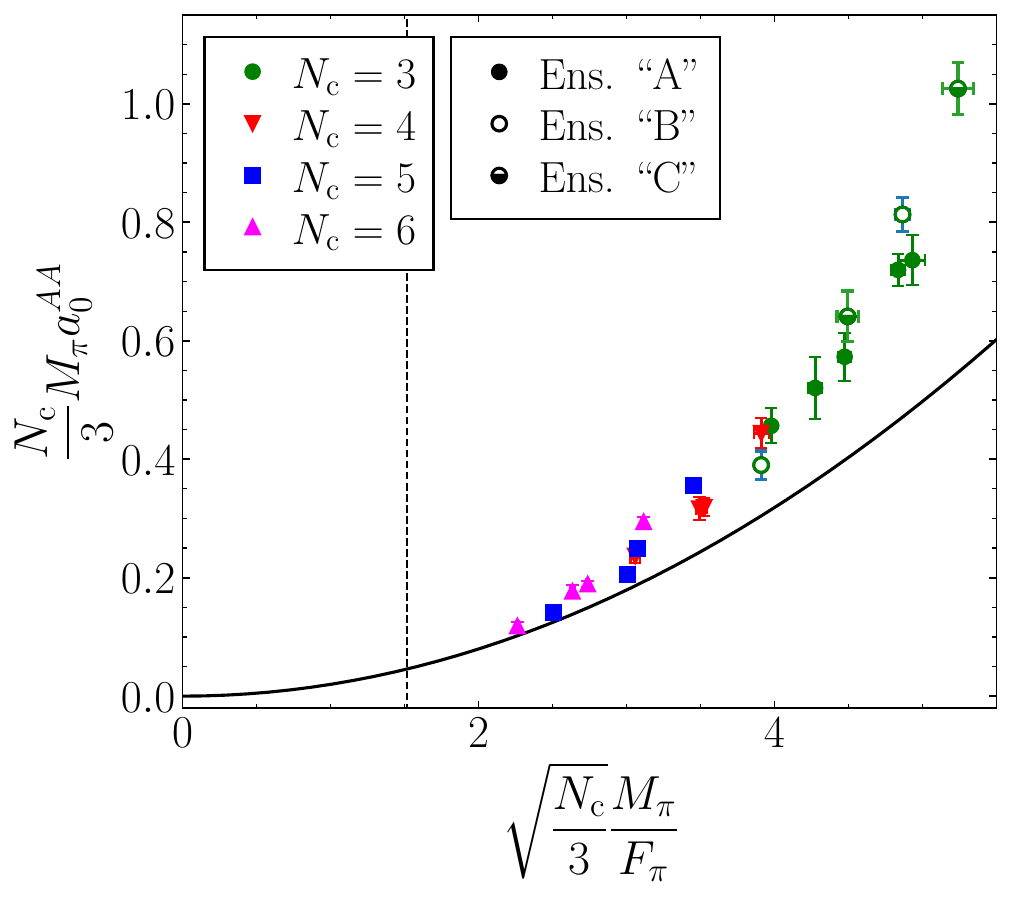}
        \caption{$AA$ channel}
        \label{fig:largeNpions:scatteringlengthAA}
    \end{subfigure}
    \caption{
        Results for the $s$-wave scattering length obtained using the threshold expansion to $\cO(L^{-5})$, together with LO ChPT predictions. Both axes are multiplied by a factor that eliminates leading $\Nc$ dependencies. The physical point is indicated with a vertical dashed line. }
    \label{fig:largeNpions:scatteringlength}\vspace{1.5cm}

    \centering
    \begin{subfigure}{0.495\textwidth} 
    \centering
        \includegraphics[width=\textwidth]{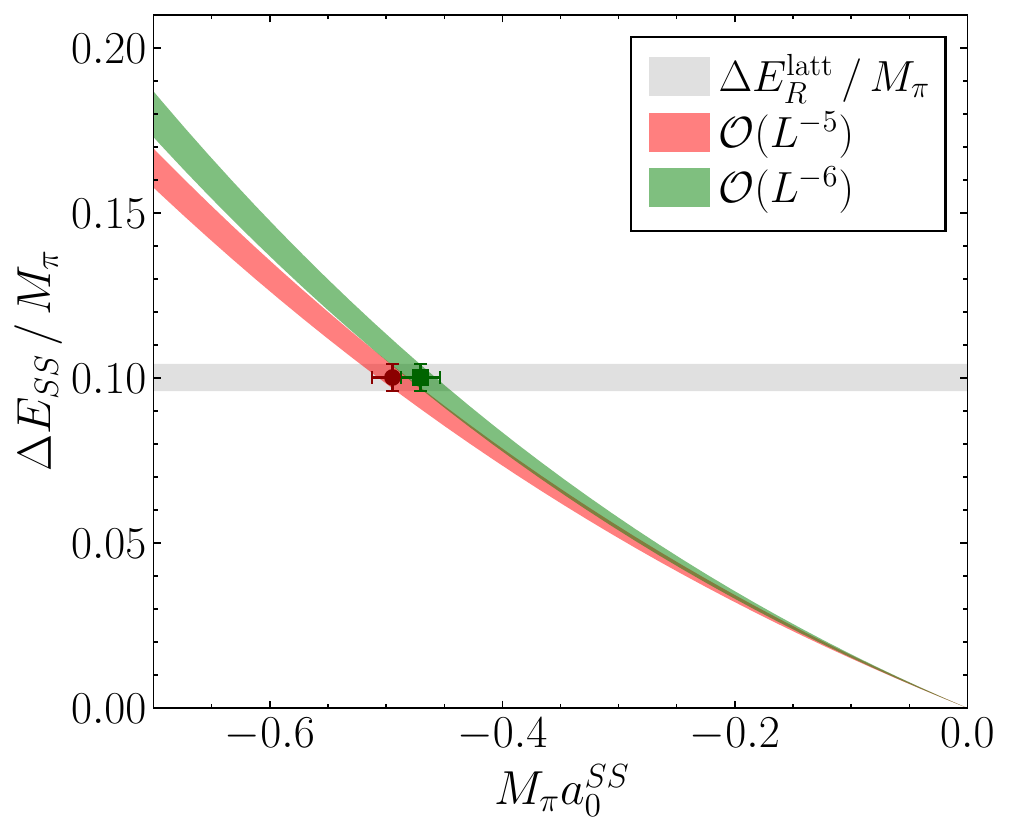}
        \caption{$SS$ channel, 3A10 ensemble}
        \label{fig:largeNpions:thresholdSS}
    \end{subfigure}
    \begin{subfigure}{0.495\textwidth}
    \centering
       \includegraphics[width=\textwidth]{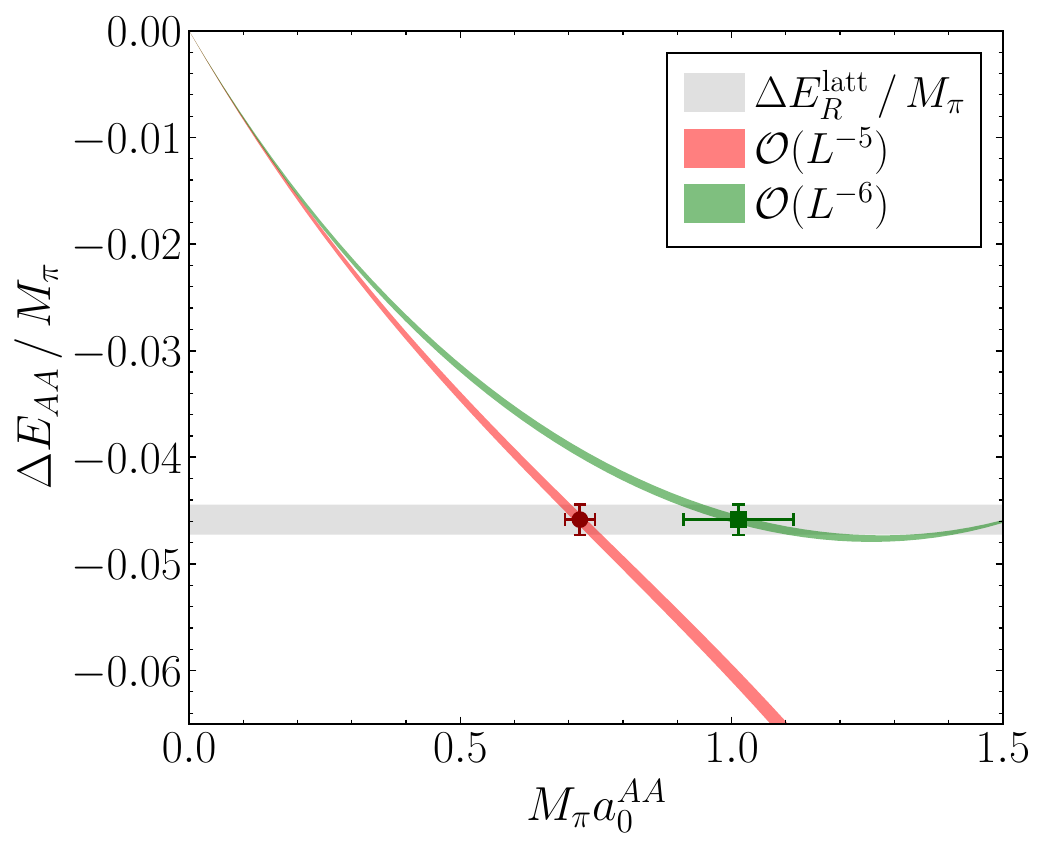}
        \caption{$AA$ channel, 3A11 ensemble}
        \label{fig:largeNpions:thresholdAA}
    \end{subfigure}
    \caption{
        Results for the scattering length obtained using the threshold expansion in \cref{eq:hadrons:thresholdexpansionQC2} up to $\cO(L^{-5})$ (red) and $\cO(L^{-6})$ (green). The gray band is the energy shift from the lattice, $\Delta E_R^\text{latt}$ and the points are the corresponding determinations of the scattering length. }
    \label{fig:largeNpions:threshold}
\end{figure}

In \cref{fig:largeNpions:scatteringlength} we compare the scattering length against the LO prediction from ChPT, given in \cref{eq:largeNpions:LOChPTscatteringalength}. We observe good agreement for the $SS$ channel, indicating that higher-order corrections are small. For the $AA$, on the other hand, such corrections seem sizable, although still smaller than the LO for the masses studied.

\subsection{Convergence of the threshold expansion}

The threshold expansion in \cref{eq:hadrons:thresholdexpansionQC2} is apriori only valid for $|a_0/L|\ll 1$, and so the truncation at $\cO(L^{-5})$ may introduce a non-negligible systematic error if $|a_0/L|\sim 1$. To estimate the size of such systematic errors, we compare results obtained at $\cO(L^{-5})$ and $\cO(L^{-6})$ using the mixed-action setup. For the $\cO(L^{-6})$ case, we use an estimate for the effective range based on LO predictions from ChPT, $M_\pi^2a_0^Rr_0^R=-3$.

For the $SS$ channel, both orders produce very similar results, with discrepancies smaller than one standard deviation. In \cref{fig:largeNpions:thresholdSS} this comparison is shown for the 3A10 ensemble. The gray horizontal band is the lattice result for $\Delta E_{SS}$, while the green and red bands represent the right-hand side of \cref{eq:hadrons:thresholdexpansionQC2} up to $\cO(L^{-5})$ and $\cO(L^{-6})$, respectively, with a width coming from the error of $\Mpi L$. %The points correspond to the results for $a_0$.

The case of the $AA$ channel is more complicated. For ensembles with small $\xi$ we observe reasonable agreement. However, this is not the case at larger $\xi$, and we even reach the case in which the expansion at $\cO(L^{-6})$ does not reproduce the finite-volume energy shift for any value of $a_0^{AA}$. In \cref{fig:largeNpions:thresholdAA} we show the same comparison as before for the 3A11 ensemble, with the same color code.  We observe how the convergence of the asymptotic expansion fails, with the $\cO(L^{-6})$ case barely yielding a result for the scattering length.

From this comparison, we conclude that some of our ensembles lie in a regime where the threshold expansion is not valid, and so may be affected by uncontrolled systematic effects. To circumvent this problematic, we make use of the full finite-volume formalism, given in \cref{eq:hadrons:algebraicQC2} for single-channel $s$-wave interactions,  in the remaining of this study.

\subsection{Discretization effects}

\begin{figure}[!h]
    \centering
    \begin{subfigure}{0.495\textwidth} 
    \centering
        \includegraphics[width=\textwidth]{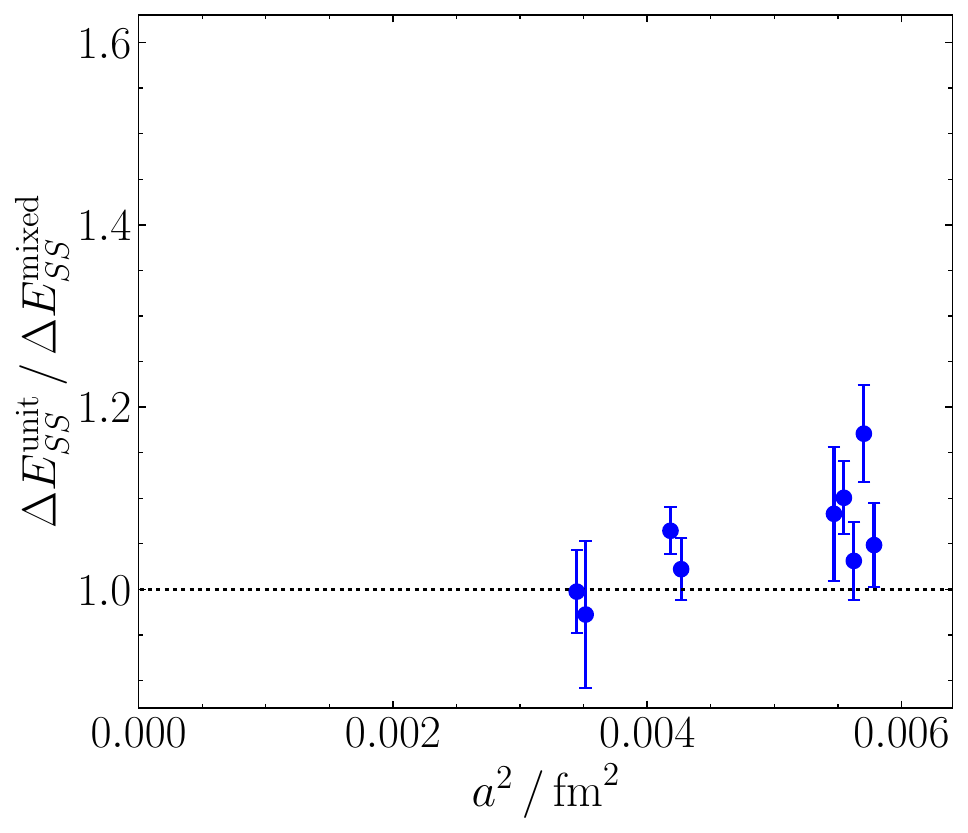}
        \caption{$SS$ channel}
        \label{fig:largeNpions:deltaEratioSS}
    \end{subfigure}
    \begin{subfigure}{0.495\textwidth}
    \centering
       \includegraphics[width=\textwidth]{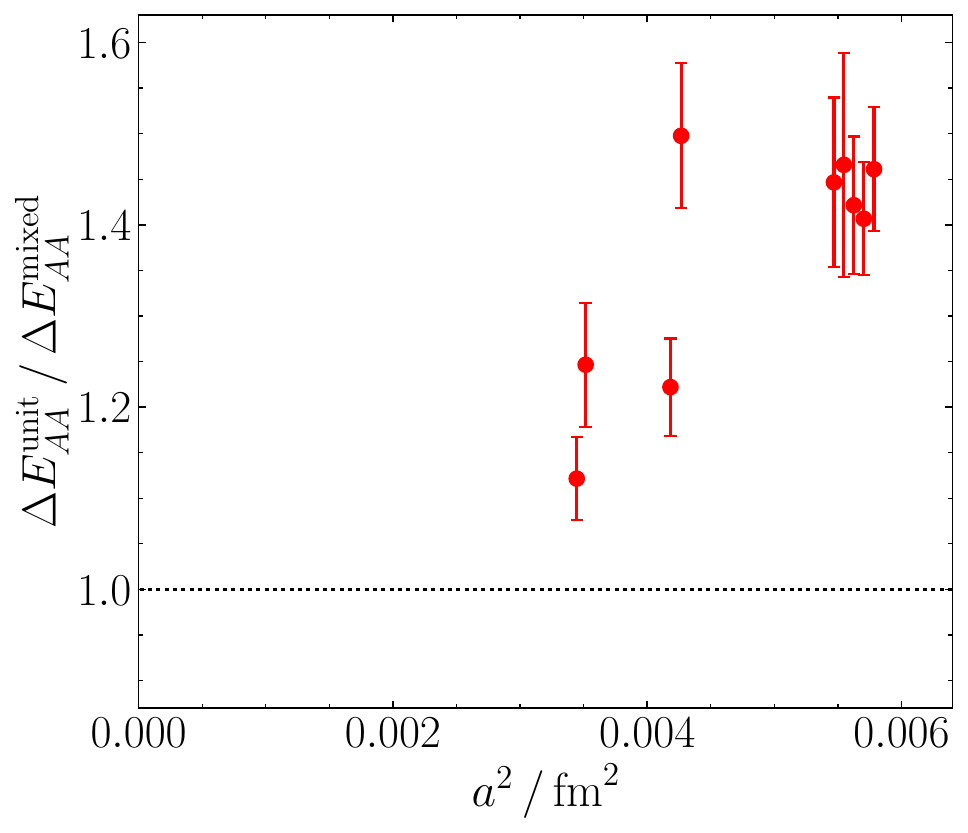}
        \caption{$AA$ channel}
        \label{fig:largeNpions:deltaEratioAA}
    \end{subfigure}
    \caption{
        Ratio  between the results for the energy shifts computed using a unitary and a mixed-action lattice setup, as a function of the lattice spacing. Results are only shown for $\Nc=3$ ensembles. }
    \label{fig:largeNpions:deltaEratio}
\end{figure}

Discretization effects are small for isospin-two pion-pion interactions~\cite{ETM:2015bzg}, while examples are known from the baryon sector in which they are significant~\cite{Green:2021qol}.  We studied their impact in our observables considering the $\Nc=3$ ensembles, which are available at different lattice spacings, and compare results obtained for the two fermion regularizations, which should be equal in the continuum. The ratio between the results for the energy shift is shown in \cref{fig:largeNpions:deltaEratio}, where the superscripts ``unit'' and ``mixed'' refer to the unitary and the mixed-action setups, respectively. In the $SS$ channel, we observe no significant cutoff effects, while large discrepancies are visible in the $AA$ channel, that seem to decrease for finer lattices. This is, to our knowledge, the first time large cutoff effects have been observed in the context of meson-meson scattering. 

To study the impact of discretization effects in the $AA$ channel in both regularizations, we perform a continuum extrapolation for  $q\cot\delta_0$, computed using \cref{eq:hadrons:algebraicQC2}. We recall that $\delta_0$ is the $s$-wave scattering phase shift  and we have defined $q=q_2^*/\Mpi$, with $q_2^*$ the relative momentum in the CMF. 

The continuum extrapolation must be performed along a line of constant physics. In our context, this is characterized by the value of $\xi$ and the relative CM momentum, $q$. Since our ensembles at different lattice spacing do not lie on the same line of constant physics, we need to interpolate/extrapolate to some reference value, ($q_\text{ref}^2, \xi_\text{ref}$), to perform a continuum extrapolation.

\begin{figure}[!p]
    \centering
    \begin{subfigure}{0.7\textwidth} 
    \centering
        \includegraphics[width=1\textwidth]{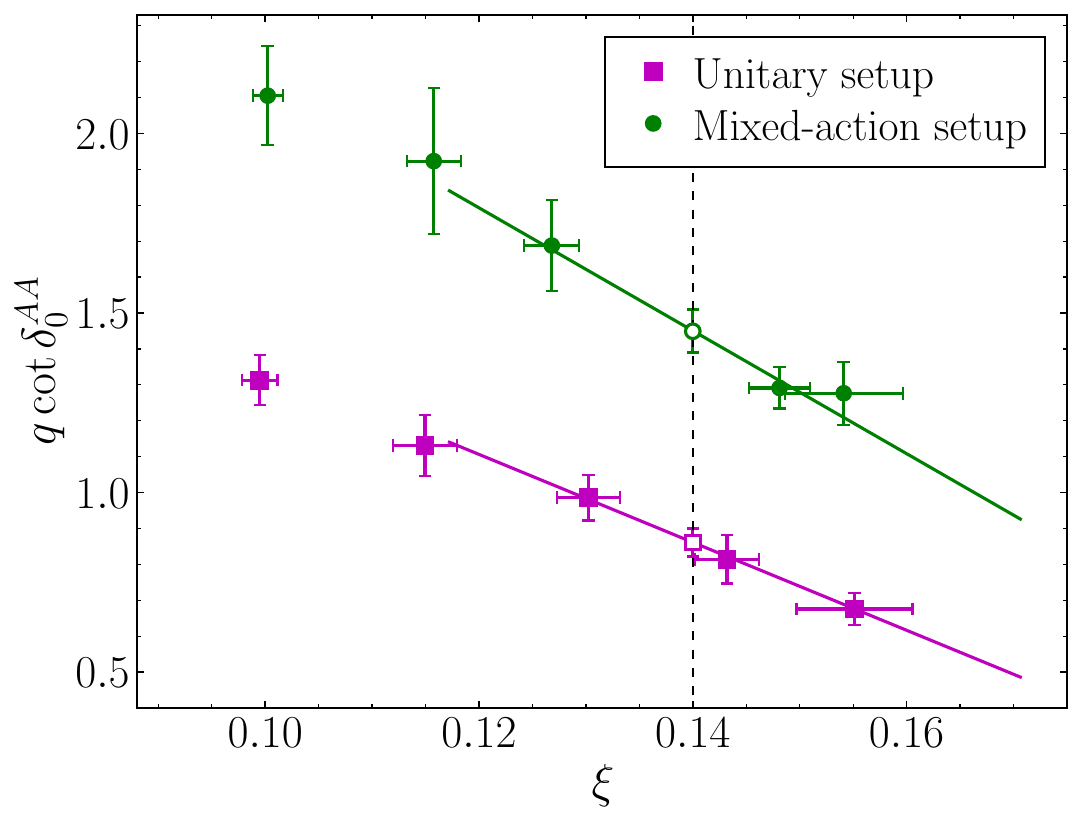}
    \end{subfigure}
    \caption{ Linear interpolation of $q\cot\delta_0$ to the reference value $\xi_\text{ref}=0.14$ (dashed line) for the ``3A'' ensembles at $q_\text{ref}^2=-0.08$. We show the results for the unitary (magenta squares) and the mixed-action (green dots) setups, and depict the result as an empty point. Only the three closest points are used for the interpolation. }
    \label{fig:largeNpions:xiinterpolation}\vspace{1cm}

    \centering
    \begin{subfigure}{0.7\textwidth} 
    \centering
        \includegraphics[width=1\textwidth]{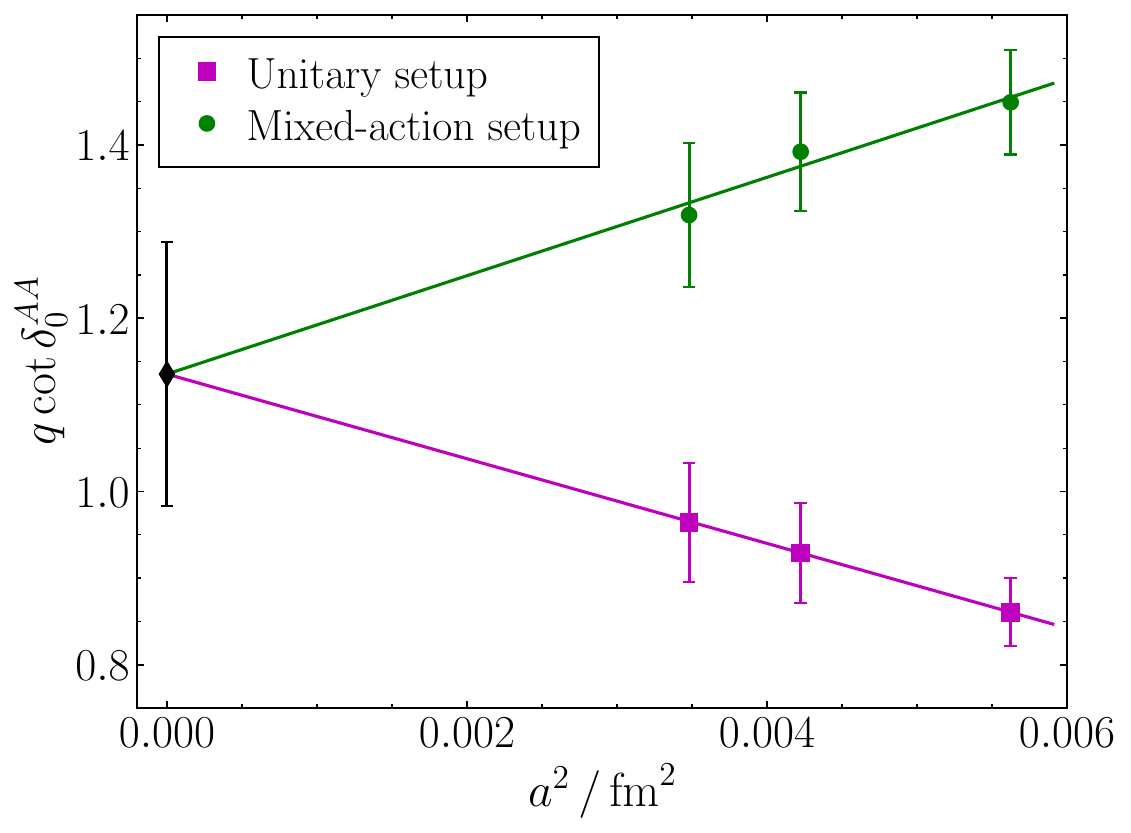}
    \end{subfigure}
    \caption{
        Constrained continuum extrapolation of $q\cot\delta_0$ computed from a unitary (magenta squares) and a mixed-action (green dots) setups at ($q_\text{ref}^2, \xi_\text{ref}=(-0.08,0.14)$. The continuum results is shown as a  black diamond.  }
    \label{fig:largeNpions:continuumextrapolation}
\end{figure}

We proceed as follows. First, for every $\Nc=3$ ensemble and both the unitary and the mixed-action result, we shift $q\cot\delta_0$ to $q_\text{ref}^2=-0.08$ using the ERE, \cref{eq:hadrons:effectiverangeexpansion}, to order $\cO(q^2)$. For the effective range, we use a prior based on LO ChPT---see \cref{eq:largeNpions:LOChPTscatteringalength}---with a conservative width, $\Mpi^2 r_0^{AA}a_0^{AA}\in[-5,-1]$. This prior introduces some systematic error in the shifted results, but this is usually much smaller than the original statistical error. Next, we interpolate linearly to $\xi_\text{ref}=0.14$ at fixed lattice spacing. We use three ensembles for the interpolation in the ``A'' case, as depicted in \cref{fig:largeNpions:xiinterpolation}, and both ensembles available for the ``B'' and ``C'' cases.

%The results are shown in \cref{fig:largeNpions:continuumextrapolation}. They are consistent with a universal continuum limit and the expected $\cO(a)$ improvement is observed in both regularizations. %Note, still, that cutoff effects seem to be of the same size in both case, although with opposite sign.

The results for both regularizations at fixed ($q_\text{ref}^2, \xi_\text{ref}$) are used to perform a continuum extrapolation, shown in \cref{fig:largeNpions:continuumextrapolation}. Our results are consistent with a universal continuum limit, and the expected $\cO(a)$ improvement is observed in both regularizations, since they seem to approach the continuum limit as $\cO(a^2)$. Similar conclusions also hold for other choices of $(q_\text{ref}^2,\xi_\text{ref})$.

We decide to include these discretization effects explicitly in the scattering amplitudes, which we modify as
\begin{equation}\label{eq:largeNpions:amplitudewithWcorrection}
\cM_2^{AA,\text{latt}}=\cM_2^{AA,\text{cont}}+32\pi^2a^2\xi W\,,
\end{equation}
where ``latt'' and ``cont'' superscripts refer to lattice and continuum quantities, respectively.
This modification is motivated by an extension of ChPT that includes the effect of the lattice spacing and a twisted mass---see app.~B from \rcite{Baeza-Ballesteros:2022azb} for details on this computation in our particular setup. In this effective theory, $W$ is a linear combination of new LECs and is expected to be independent of $\Nc$ at leading order, $W\sim\cO(\Nc^0)$. The results from this effective theory, however, do not explain why  discretization effects are larger in the $AA$ channel, compared to the $SS$ one. It is probably a consequence of the particular numerical values of the LECs in this effective theory. %Pilar pensó que esta últimas dos frases no se entendían bien

For the phase shift, \cref{eq:largeNpions:amplitudewithWcorrection} translates to
\begin{equation}
q\cot\delta_0^\text{AA,\text{latt}}=q\cot\delta_0^\text{AA,\text{cont}}\left(1-\frac{32\pi^2a^2\xi W}{\cM_{2,0}^{AA,\LO}}\right)\,.
\end{equation}
This expression is used to fit the results at finite lattice spacing for both regularizations imposing a common continuum limit. For the remaining of this work, we make use of the mixed-action setup. For that case, the continuum extrapolation allows us to determine
\begin{equation}\label{eq:largeNpions:Wpriorresult}
W_\text{mixed}=-42(29)\,\text{fm}^{-2}\,,
\end{equation}
which we use as an extra input when matching the $AA$ channel to ChPT.

\section{Matching to ChPT}\label{sec:largeNpion:resultsfitsChPT}

%We now match our lattice results for the finite-volume energies to predictions from ChPT. This allows us to determine the $\Nc$ dependence of the LECs, and also compare to results to previous literature.
Given a scattering amplitude, Lüscher's formalism allows one to obtain predictions of the corresponding finite-volume energies that can be matched to lattice results, constraining the parameters in the scattering amplitudes. In this section, we use the two-particle formalism to match our lattice results to SU(4) and U(4) ChPT perdictions, given in \cref{eq:largeNpions:AASUamplitudeChPT,eq:largeNpions:SSUamplitudeChPT,eq:largeNpions:AAUamplitudeChPT,eq:largeNpions:SSSUamplitudeChPT},  and constrain the $\Nc$ dependence of the LECs. In the $AA$ channel, we include cutoff effects as indicated in \cref{eq:largeNpions:amplitudewithWcorrection}.

\subsection{Fitting procedure}

To evaluate the ChPT predictions of the scattering amplitudes, we choose a value for the renormalization scale related to $4\pi\Fpi$ with no the leading $\Nc$ dependence,
\begin{equation}
\mu^2=\frac{3}{\Nc}(4\pi\Fpi)^2\,.
\end{equation}
Also, we use a value of $\Metap$ given by the Witten-Veneziano relation, \cref{eq:QCD:WittenVenezianoformula},
\begin{equation}\label{eq:largeNpions:renormalizationscale}
\frac{\Metap^2}{(4\pi\Fpi)^2}=\xi+\frac{a_0}{\Nc^2}\,,
\end{equation}
where $a_0\approx 6.5$ is determined using the topological susceptibility from \rcite{Ce:2016awn}. Finally, we set those LECs that appear in the amplitudes multiplied by some power of the momentum---$L_R^\prime$, $L_R^{\prime\prime}$ and $K_R^\prime$---to zero. We have checked that, as our results are close to threshold, we are mostly insensitive to the values of these LECs. Using the results for the amplitudes,we can obtain predictions for the finite-volume energies using Lüscher's formalism. For the evaluation of Lüscher's theta function we use the implementation from \rrcite{NPLQCD:2011htk,LuscherZetaGit}.

Due to non-negligible correlations between the finite-volume energies and $\xi$, an ordinary least-squares fit is not suitable for matching our results to ChPT. Instead, we use York regression~\cite{doi:10.1119/1.1632486} and define the $\chi^2$ as
\begin{equation}
\chi^2=\min_{\delta_i}[\bm{R}^\intercal V^{-1} \bm{R}]\,,
\end{equation}
where $\bm{R}$ is a vector containing our data with a specific form that depends on the channel of interest and $V$ is the corresponding covariance matrix. For example, for a fit to the $SS$ channel, $\bm{R}$ is composed by a succession of tuples $\bm{R}_i=(f(x_i+\delta_i)-y_i,\delta_i)$, where $\delta_i$ are real numbers which are minimized and $i$ labels each ensemble. Here, $x_i=\xi_i$, $y_i=E_{R,i}$ are our lattice results for the chiral parameter and finite-volume energies and $f$ are the ChPT predictions for the latter, obtained using the finite-volume formalisms, as described above. In this example, $V$ is block-diagonal, as ensembles are independent. %Pilar dice que es cumbersome

We also consider more complicated fits. For fits to the $AA$ channel, we constrain the $W$ factor in  \cref{eq:largeNpions:amplitudewithWcorrection}. We include in $\bm{R}$ the result for $W$ from \cref{eq:largeNpions:Wpriorresult} as a prior, and its correlations to the $\Nc=3$ ensembles in $V$, which is thus no longer block diagonal. We also perform simultaneous fits to both $SS$ and $AA$ channels, in which the tuples are enlarged
\begin{equation}
\bm{R}_i=(f^{SS}(x_i+\delta_i)-y^{SS}_i,f^{AA}(x_i+\delta_i)-y^{AA}_i,\delta_i)\,.
\end{equation}
Note we do not consider the error from $\Mpi L$ for the fits, as they are much smaller than the errors in $\xi$ and we have found them to have a negligible effect.

\subsection{Fits to ChPT}\label{sec:largeNpion:fitsChPT}

We first work at fixed $\Nc$ and fit each channel separately to SU(4) ChPT up to NLO and U(4) ChPT up to NNLO. We determine the values of $L_R$ from the fits, as well as $K_R$ and $W$ for fits to U(4) ChPT and the $AA$ channel, respectively. Results are shown in \cref{tab:largeNpions:fitresultsfixedNSS} for the $SS$ channel and \cref{tab:largeNpions:fitresultsfixedNAA} for the $AA$ channel. For the $SS$ channel, we observe that ChPT fails to describe the lattice results for the heaviest masses, and we opt not to include the $\Nc=3$ ensembles with $\xi\gtrsim 0.14$ in the fits. For fits to the $AA$ channel, on the other hand, we use the $W$ result in \cref{eq:largeNpions:Wpriorresult} as an additional prior.

\begin{table}[t!]
\centering
\begin{tabular}{cccccc}
\toprule
 \multirow{2}{*}{$N_\text{c}$} & \multicolumn{2}{c}{SU(4) ChPT} & \multicolumn{3}{c}{U(4) ChPT} \\
 & $L_{SS}\times10^{3}$ & $\chi^2\,/\,\text{dof}$ & $L_{SS}\times10^{3}$ & $K_{SS}\times10^{3}$ & $\chi^2\,/\,\text{dof}$ \\   \midrule
3 & $-1.85$(7) & 1.8/4 & $-2.5$(7) & $-0.5$(6) & 1.6/3\\
4 & $-1.79$(8) & 3.0/3 & $-0.9$(8) & 1.3(1.0) & 0.1/2\\
5 & $-1.83$(11) & 2.3/3 & $-2.2$(6) & $-0.3$(9) & 2.3/2\\
6 & $-2.10$(16) & 3.8/3 & $-1.7$(8) & 1.2(1.7) & 2.8/2\\ \bottomrule 
\end{tabular}
\caption{Results of fits of $\Delta E_{SS}$ to SU(4) and U(4) ChPT at fixed $N_\text{c}$. $N_\text{c}=3$ ensembles with $\xi \gtrsim 0.14$ are not fitted.}
\label{tab:largeNpions:fitresultsfixedNSS}
\centering\vspace{0.7cm}

\begin{tabular}{cccccccc}
\toprule
 \multirow{2}{*}{$N_\text{c}$}& \multicolumn{3}{c}{SU(4) ChPT} & \multicolumn{4}{c}{U(4) ChPT} \\ 
 & $L_{AA}\times10^{3}$ & $W$/fm$^{-2}$ & $\chi^2\,/\,\text{dof}$  & $L_{AA}\times10^{3}$ & $K_{AA}\times10^{3}$ & $W$/fm$^{-2}$ & $\chi^2\,/\,\text{dof}$ \\  \midrule
3 & $-2.4$(5) & $-72$(17) & 23.4/8 & 1.7(1.3) & 2.2(7) & $-39$(23) & 12.8/7 \\ 
4 & $-1.8$(1.2) & $-45$(30) & 1.2/3 & $-1.1$(2.7) & 0.5(2.3) & $-41$(32) & 1.2/2 \\
5 & $-4.2$(1.1) & $-75$(22) & 8.5/3 & 1.2(2.8) & 5.6(2.7) & $-41$(32) & 3.9/2 \\
6 & $-5.4$(1.5) & $-72$(24) & 5.4/3 & 0.1(3.2) & 6.6(3.3) & $-39$(33) & 1.6/2 \\ \bottomrule
\end{tabular}
\caption{Results of fits $\Delta E_{AA}$ to SU(4) and  U(4) ChPT at fixed $N_\text{c}$.}
\label{tab:largeNpions:fitresultsfixedNAA}
\end{table}

\begin{figure}[!h]
    \centering
    \begin{subfigure}{0.495\textwidth} 
    \centering
        \includegraphics[width=1\textwidth]{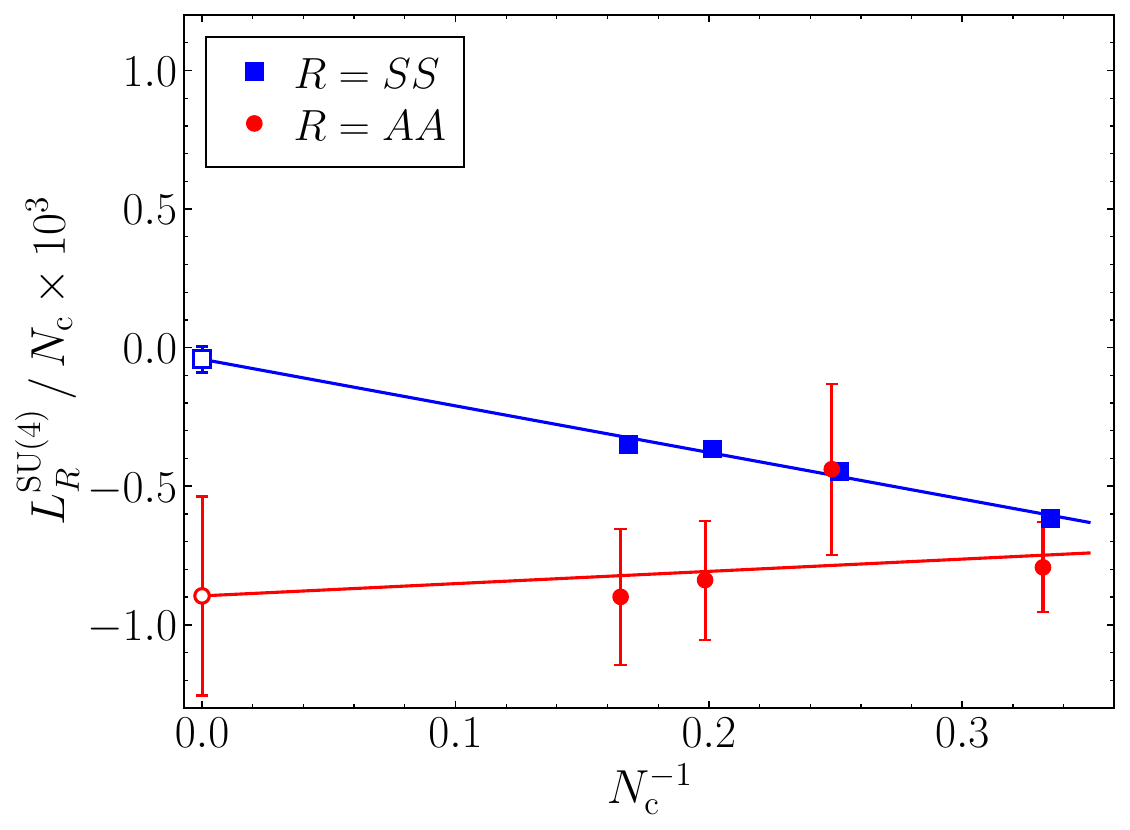}
        \caption{$L_R$ from SU(4) ChPT}
        \label{fig:largeNpions:LECfixedNfitsSU}
    \end{subfigure}
    \begin{subfigure}{0.495\textwidth}
    \centering
       \includegraphics[width=1\textwidth]{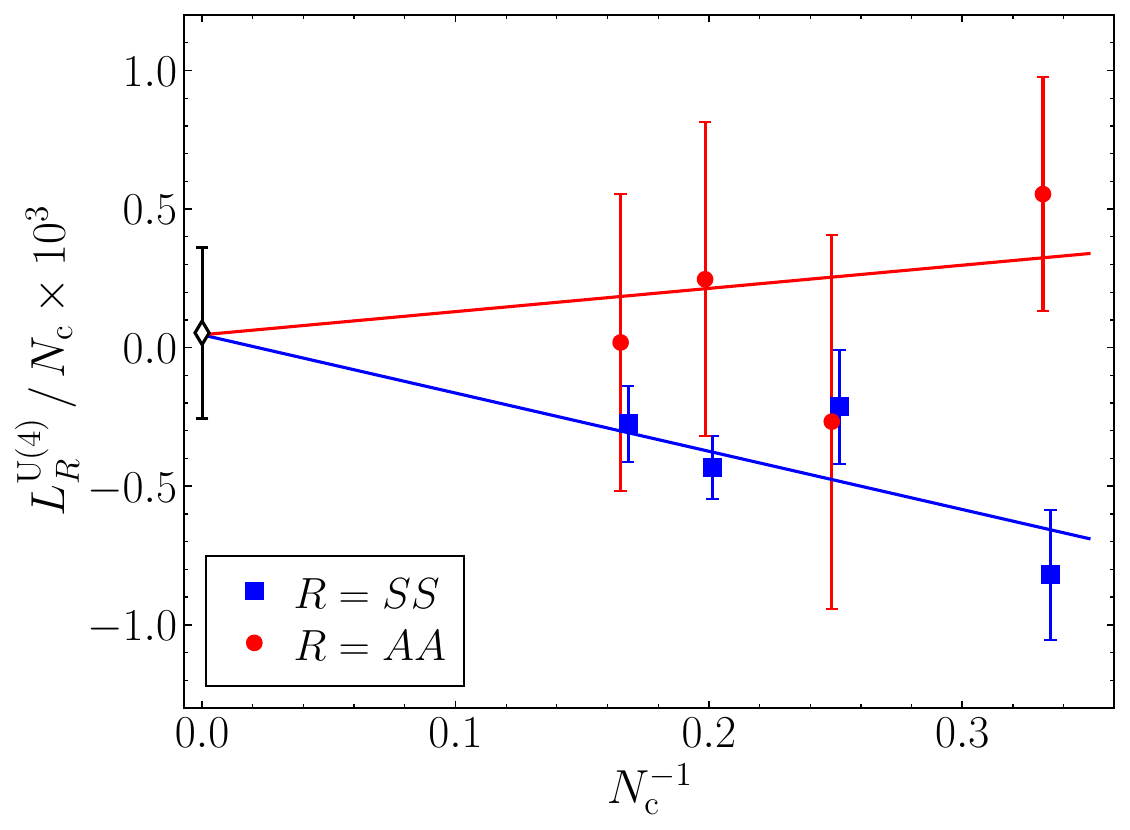}
        \caption{$L_R$ from U(4) ChPT}
        \label{fig:largeNpions:LECfixedNfitsU}
    \end{subfigure}\vspace{0.3cm}
    
    \begin{subfigure}{0.495\textwidth} 
    \centering
        \includegraphics[width=1\textwidth]{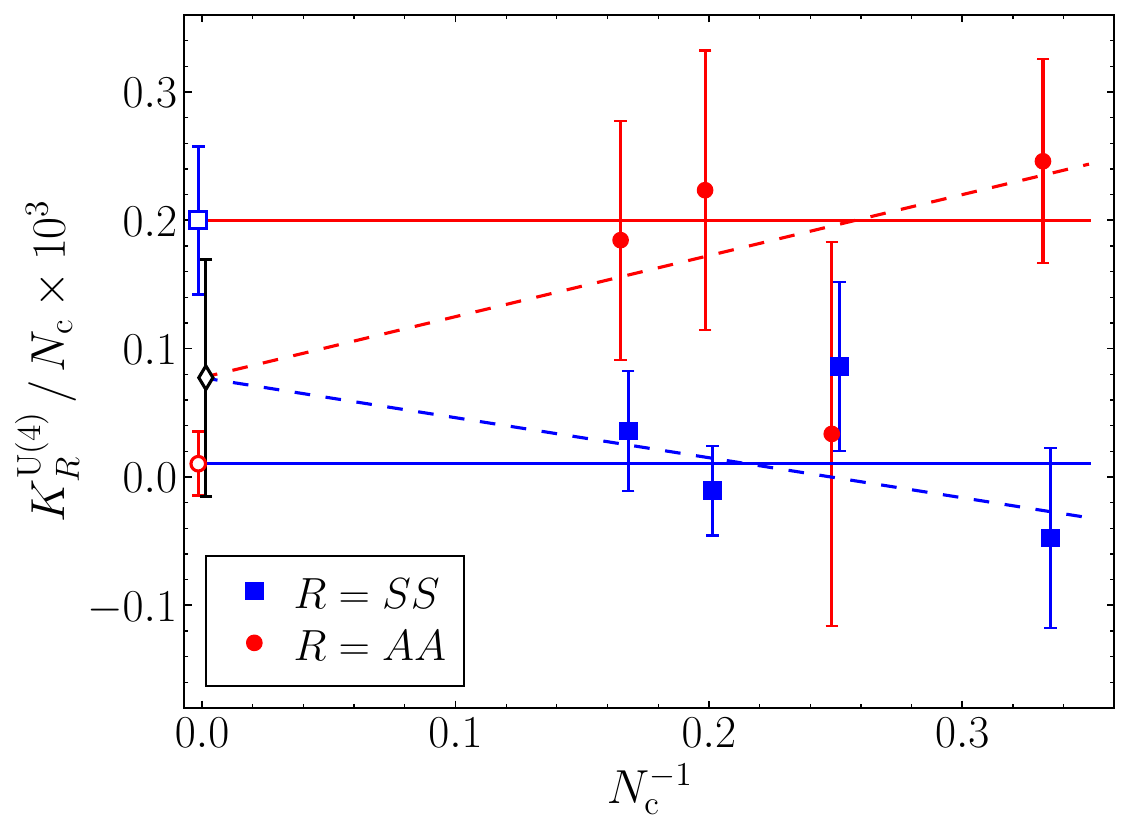}
        \caption{$K_R$ from U(4) ChPT}
        \label{fig:largeNpions:KLECfixedNfits}
    \end{subfigure}
    \caption{{\bf Top}: Results for the LECs from fits to ChPT at fixed $\Nc$ for the $SS$ (blue squares) and $AA$ (red dots) channels. The lines are fits to \cref{eq:largeNpions:NcparametrizationLR}, and the empty points are the large $\Nc$ prediction. In the U(4) theory, we have imposed a common large $\Nc$ result for both channels. {\bf Bottom}: Results for $K_R$ from fits to U(4) ChPT for the $SS$ (blue squares) and $AA$ (red dots) channels. The points are fitted to a different constant for each channel (solid lines) and to a constrained linear dependence (dashed lines).
         }
    \label{fig:largeNpions:LECfixedNfits}
\end{figure}

The results for $L_R$ are represented in \cref{fig:largeNpions:LECfixedNfitsSU,fig:largeNpions:LECfixedNfitsU}. We observe that both in the SU(4) and the U(4) theories they are well described by a leading and a subleading $\Nc$ dependence---see \cref{eq:largeNpions:NcparametrizationLR}. However, only the U(4) result reproduces the expected common large $\Nc$ behavior of both channels. Linear fits are performed separately for each channel in \cref{fig:largeNpions:LECfixedNfitsSU}, while in \cref{fig:largeNpions:LECfixedNfitsU} a common large $\Nc$ limit is imposed.

A similar study is presented in \cref{fig:largeNpions:KLECfixedNfits} for $K_R$ determined from fits to U(4) ChPT. This quantity should only include a leading $\Nc$ dependence at the order we work, which should be equal in both channels---see \cref{eq:largeNpions:LOKfactorUChPT}. We observe the fit results for each channel are consistent with a single $\cO(\Nc^2)$ term, but this is not equal for both channels. A universal large $\Nc$ limit is only  reproduced if subleading $\Nc$ corrections are included.

\begin{figure}[!t]
\vspace{0.5cm}

    \centering
        \includegraphics[width=1\textwidth]{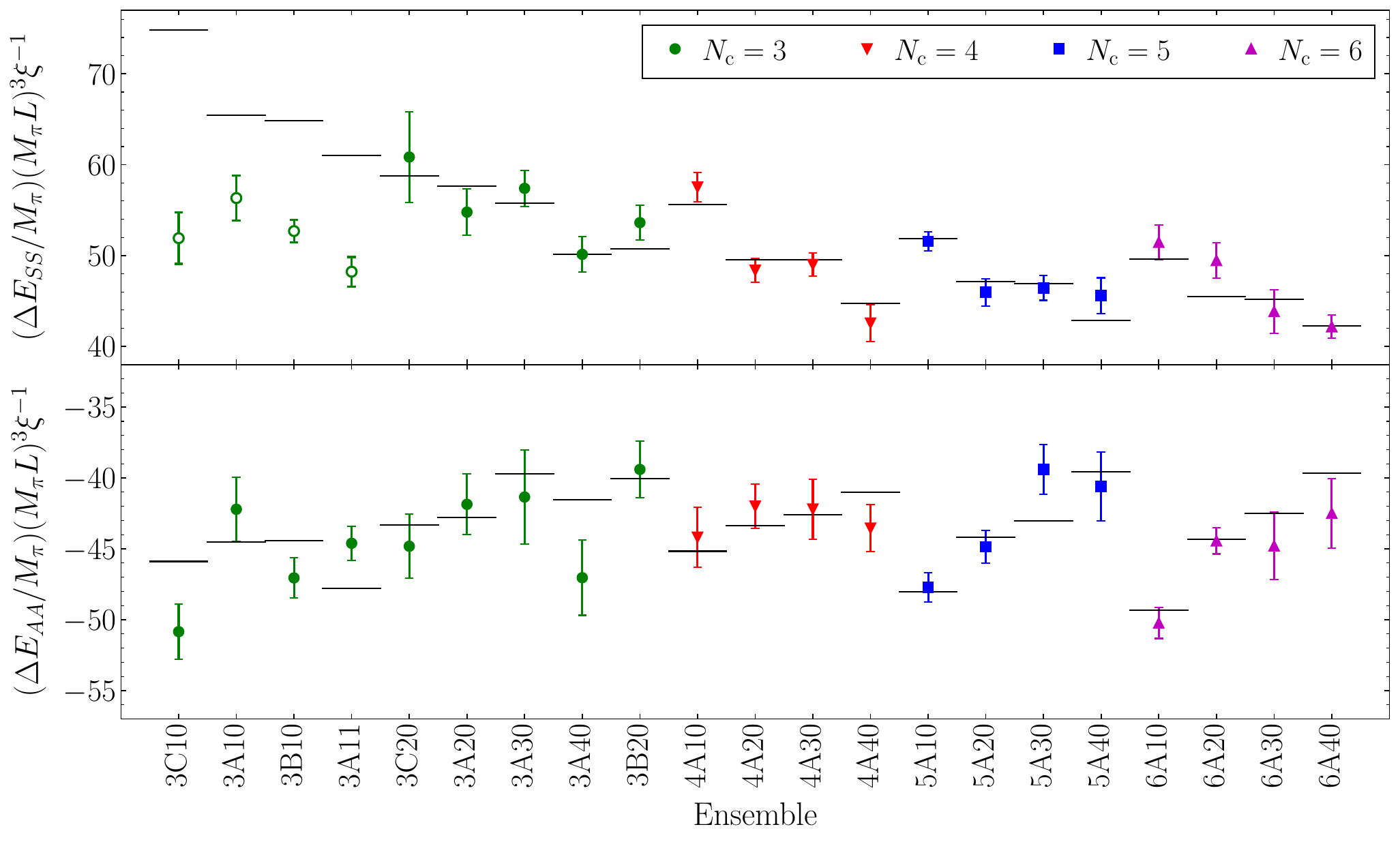}
    \caption{
        Results of a simultaneous chiral and $\Nc$ fit to U(4) ChPT predictions of both the $SS$ (top) and $AA$ (bottom) channels. Points represent our lattice results multiplied by a factor that eliminates leading chiral and $\Nc$ dependencies, and horizontal lines are the best fit predictions from U(4) ChPT, as summarized in \cref{tab:largeNpions:bothchannelglobalfits}. Empty points in the $SS$ channel are not fitted. Within each color, $\xi$ decreases from left to right.  }
    \label{fig:largeNpions:finalfit}
\end{figure}

We next perform fits including all $\Nc$ to both SU(4) and U(4) theories for both channels. We parametrize $L_R$ as  in \cref{eq:largeNpions:NcparametrizationLR} and include only a $\cO(\Nc^2)$ term for $K_R$, this is, $K_R=\Nc^2 K_R^{(0)}$. Results for separate fits to each channel are summarized in \cref{tab:largeNpions:singlechannelglobalfits}. Note that the results for $L^{(0)}$ are only compatible for the U(4) fit, which also leads to smaller values of the $\chi^2$. We finally perform a simultaneous fit of both channels to U(4) ChPT. We impose a common large $\Nc$ limit for $L_R$, but keep the $K_R$ terms different. The results are presented in \cref{tab:largeNpions:bothchannelglobalfits} and are illustrated in \cref{fig:largeNpions:finalfit}.

\begin{table}[h!]
\centering
\vspace{0.5cm}
\begin{tabular}{ccccccc}
\toprule
Channel & Fit & $L^{(0)}\times10^{3}$ & $L_R^{(1)}\times10^{3}$ & $K_R^{(0)}\times10^{5}$ & $W$/fm$^{-2}$ & $\chi^2\,/\,\text{dof}$  \\  \midrule
\multirow{2}{*}{$SS$} & SU(4) & $-0.04$(1.3) & $-1.70$(18) & --- & --- & 12.8/15  \\
 & U(4) & $-0.01$(7) & $-1.78$(20) & 1.2(2.5) & --- & 12.2/14 \\
\multirow{2}{*}{$AA$} & SU(4) & $-1.22$(19) & 0.8(4) & --- & $-94$(15) & 38.5/19 \\
 & U(4) & $-0.1$(4) & 1.8(4) & 21(5) & $-32$(23) & 22.5/18 \\
\bottomrule 
\end{tabular}
\caption{Fit results to SU(4) and U(4) ChPT for both the $SS$ and $AA$ channels.}
\label{tab:largeNpions:singlechannelglobalfits}\vspace{0.7cm}

\centering
\begin{tabular}{ccccccc}
\toprule
 $L^{(0)}\times10^{3}$ & $L^{(1)}_{SS}\times10^{3}$ & $L^{(1)}_{AA}\times10^{3}$ & $K_{SS}^{(0)}\times10^{5}$ & $K_{AA}^{(0)}\times10^{5}$ & $W$/fm$^{-2}$ & $\chi^2\,/\,\text{dof}$  \\  \midrule
$-0.02$(8) & $-1.79$(19) & 1.7(4) & 0.8(2.2) & 22(3) & $-31$(9) & 35.6/33  \\

\bottomrule 
\end{tabular}
\caption{Results from a simultaneous fit of both the $AA$ and $SS$ channels to U(4) ChPT, imposing a common large $\Nc$ limit for $L_R$. The goodness of the fit is illustrated in \cref{fig:largeNpions:LECfixedNfits}.}
\label{tab:largeNpions:bothchannelglobalfits}
\end{table}

\subsection{Discussion and comparison to previous literature}\label{sec:largeNpions:comparisonlitereatureandIAM}

From the results of the global fit of both channels to U(4) ChPT, given in \cref{tab:largeNpions:bothchannelglobalfits}, we observe that the leading $\Nc$ contribution to the LECs, $L^{(0)}$, is anomalously small. 
In addition, using \cref{eq:largeNpions:LECsparametrizationU4Nf,eq:largeNpions:LECsparametrizationU4Nfcombination}, we can  determine the $\Nf$ dependence of the subleading term,

\noindent\begin{equation}\label{eq:largeNpions:resultsdecomposedLECs}
\begin{array}{rl}
L_0+L_3-L_5+L_8=&-0.02(8)N_\text{c}-0.01(5)N_\text{f}+\mathcal{O}(N_\text{c}^{-1})\,,\\
L_1+L_2-L_4+L_6=&-0.88(10)+\mathcal{O}(N_\text{c}^{-1})\,.
\end{array}
\end{equation}
A striking result is that the leading $\Nc$ term is suppressed compared to subleading corrections, which dominates at low values of $\Nc$. It also seems that the subleading term proportional to to $\Nf$ is much smaller than that independent of the number of colors. However, note that these coefficients are scale-dependent, and this situation changes for other choices of the renormalization scale---see \cref{eq:QCD:LECsNfscaledependencegeneral}. %two subleading terms is much smaller than the other one, but we note that these are scale dependent, and the combination of LECs expected a priori to be dominant, seems to be suppressed in both the leading and subleading terms, compared to the a priori subleading one. %At this point it is worth commenting about the scale dependence of these results, since the subleading $\Nc$ parts of the LECs are scale dependent. We have studied the variation of the results in \cref{eq:largeNpions:resultsdecomposedLECs} under a large $\mu\rightarrow2\mu$ scale variation, and observed that the variation of the subleading corrections still remain much larger than the expected leading term.\jorge Discutir la scale dependence.}

Our results can be compared to predictions obtained in the resonant chiral theory (RChT)~\cite{Ecker:1988te,Ledwig:2014cla,Pich:2002xy}, in which the LECs at large $\Nc$ are saturated by the contributions obtained from integrating out low-lying resonances. In the single-resonance approximation, one obtains the same result for both the $SS$ and the $AA$ channel, $L_R^{\text{RChT}}=-0.07\cdot\Nc$~\cite{Ledwig:2014cla}. This agrees with our U(4) result at large $\Nc$, but fails at $\Nc=3$, where subleading effects dominate.

The $SS$ channel results  for $\Nc=3$ can moreover be compared to results available in the literature for the isospin-two channel with $\Nf<4$. To do so, we first determine $L_{SS}$ for the desired $\Nf$, using \cref{eq:largeNpions:resultsdecomposedLECs}. Then, we use \cref{eq:largeNpions:matchingSUUSS} to translate our results to the SU($\Nf$) theory and, finally, we change the renormalization scale to the one used in the literature.\footnote{Note that the running of the LECs is different between the SU($\Nf$)~\cite{Bijnens:2009qm} and U($\Nf$)~\cite{Kaiser:2000gs} theories.} 
Our ensembles have $\mu=1.40(12)$ GeV, determined averaging the quantity in \cref{eq:largeNpions:renormalizationscale} over all ensembles. We quote separately the error related to the change of scale.

For $\Nf=3$ we compare to results for the LECs coming from fits to experimental data. These are presented at the mass of the $\rho(770)$ resonance, $\mu=M_\rho=0.77$ GeV, 
\begin{equation}
\begin{array}{rll}
L_{\Ipp=2}^{\Nc=3,\Nf=3}&=-1.14(22)_\text{stat}(11)_\mu\times10^{-3}\quad & \text{this work}\,,\\
L_{\Ipp=2}^{N_\text{c}=3,\,N_\text{f}=3}&=-0.9(1.5)\times 10^{-3} \quad &\text{\rcite{Bijnens:2014lea} (table~1, col.~2)}\,,\\
L_{\Ipp=2}^{N_\text{c}=3,\,N_\text{f}=3}&=-0.7(3.2)\times 10^{-3} \quad &\text{\rcite{Gasser:1984gg}}\,,
\end{array}
\end{equation}
where the error in the last two is computed from adding in quadrature the error of the separate LECs contributing.

We also compare to lattice results for $\Nf=2$ at a scale $\mu=\sqrt{2}\Fpiphys\approx130.2$ MeV~\cite{FLAG:2021npn}. For $\Nf=2$, one typically quotes the quantity
\begin{equation}
\ell_{\Ipp=2}=512\pi^2L_{SS}^{\Nc=3,\Nf=2}\,.
\end{equation}
We obtain
\begin{equation}
\begin{array}{rll}
\ell_{\Ipp=2}&=4.3(1.2)_\text{stat}(0.5)_\mu,\quad\quad & \text{this work}\,,\\
\ell_{\Ipp=2}&=4.65(0.85)_\text{stat}(1.07)_\text{sys} \quad\quad &\text{\rcite{Feng:2009ij} }\,,\\
\ell_{\Ipp=2}&=3.79(0.61)_\text{stat}\left({}^{+1.34}_{-0.11} \right)_\text{sys} \quad\quad &\text{\rcite{ETM:2015bzg}}\,,
\end{array}
\end{equation}
where the results in \rrcite{Feng:2009ij,ETM:2015bzg} separately quote statistical and systematic errors.

The $AA$ channel cannot be compared to experimental results as it only exists for $\Nf\geq 4$. However, this channel shows attractive interactions and so one could wonder if it might present a resonant state that may be interpreted as a tetraquark. LHCb has recently reported on the finding of several tetraquark states. These include the $T_{cs0}^0(2900)$ in the mass spectrum of $D^-K^+$~\cite{LHCb:2020bls,LHCb:2020pxc}, and the $T_{c\bar{s}0}^{++}(2900)$ and $T_{c\bar{s}0}^{0}(2900)$ in the mass spectrum of $D_s^+\pi^+$ and $D_s^+\pi^-$~\cite{LHCb:2022sfr,LHCb:2022lzp}, respectively. In a theory with $\Nf=4$ degenerate quarks, all the aforementioned states would have the same flavor, spin and parity quantum numbers of the $AA$ channel.

\begin{figure}[!b]
    \centering
        \includegraphics[width=0.7\textwidth]{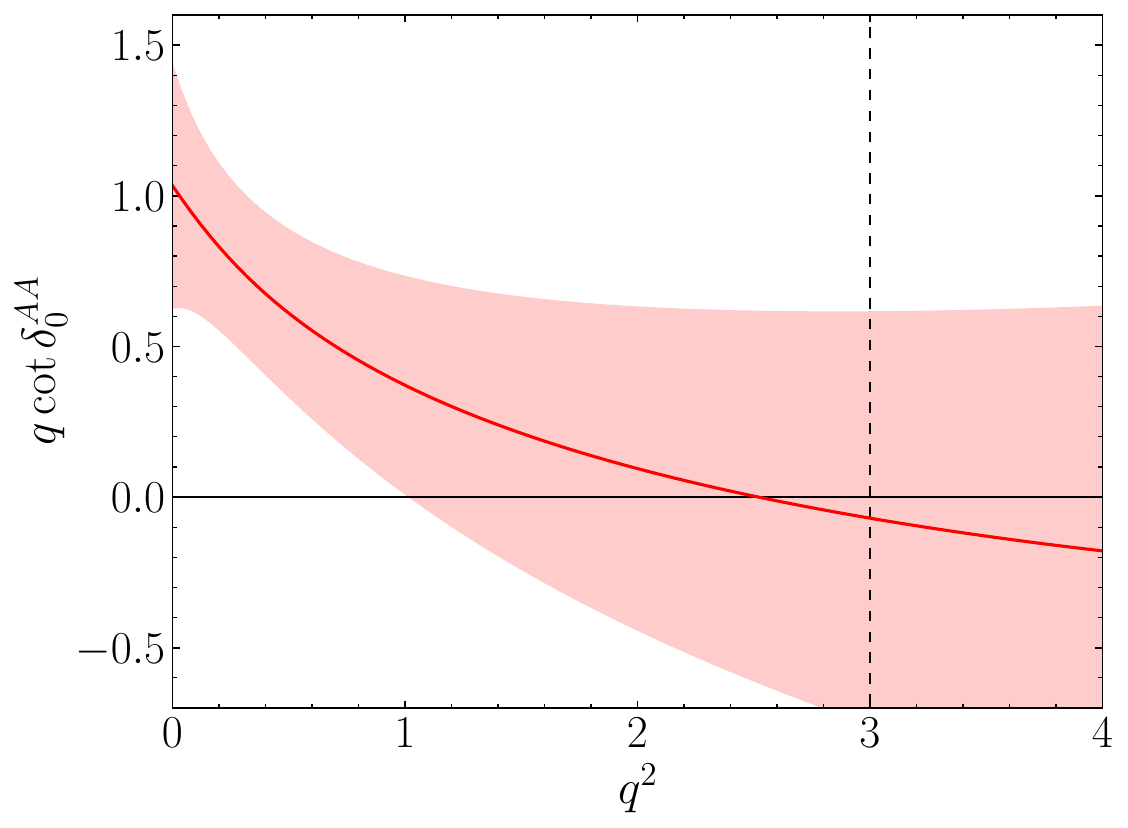}
    \caption{
        Unitarized result for $q\cot\delta_0$ for the $AA$ channel obtained applying the inverse amplitude method to the U($\Nf$) ChPT amplitude in \cref{eq:largeNpions:AAUamplitudeChPT}, with values of the parameters as defined in the text. We observe a change of sign, indicating the possible presence of a tetraquark resonance. }
    \label{fig:largeNpions:IAMexploration}
\end{figure}

To explore the possible existence of such a state at higher center-of-mass energy, we use the inverse amplitude method, described in \cref{sec:hadrons:twopionsChPT}. In \cref{fig:largeNpions:IAMexploration}, we represent the phase shift of the unitarized amplitude in the U(4) theory at fixed $\xi=0.1$ and $\Nc=3$, computed from \cref{eq:largeNpions:AAUamplitudeChPT}. We use our best-fit results from \cref{tab:largeNpions:bothchannelglobalfits} for $L_{AA}$ and $K_{AA}$, and let $L_{AA}^\prime$ and $L_{AA}^{\prime\prime}$ be zero, with errors of the size of $L_{AA}$. We also set $K_{AA}^\prime=0$. From this analysis, we conclude it is plausible that a tetraquark resonance is present in this channel, as indicated by $q\cot\delta_0$ crossing the horizontal axis from above. Note that while the LEC values differ between standard and unitarized ChPT, this difference is small and should not affect our qualitative conclusions.

\section{Conclusions}\label{sec:largeNpions:conclusions}

In this chapter, we have presented the results from \rrcite{Baeza-Ballesteros:2022azb,Baeza-Ballesteros:2021nxu}, in which pion-pion interactions near threshold are studied for varying number of colors, $\Nc$. We have used a theory with $\Nf=4$ degenerate quark flavors, and focused on two different scattering channels:  the $SS$ channel, which is analogous to the $\Ipp=2$ channel of two-flavor QCD, and the $AA$ channel, which is attractive and only exists for $\Nf\geq4$.

Running lattice simulations with $\Nc=3-6$ and $\Mpi=360-590$ MeV, we have determined the ground-state finite-volume energies for both channels of interest. The computations have been performed using two different regularizations of valence fermions: a unitary setup with Wilson fermions, and a mixed-action setup with a twisted mass in the valence quarks. By comparing both, which must coincide in the continuum, we have identified large discretization effects in the $AA$ sector, which are constrained by comparing results at various lattice spacing.

Using the two-particle finite-volume formalism, we  have matched our results to ChPT, finding that the correct large $\Nc$ behavior is only reproduced by U($\Nf$) ChPT predictions, which we have calculated up to NNLO. From this comparison, the leading and subleading $\Nc$ dependence of the relevant LECs have been constrained. The results indicate that the leading coefficient in the $\Nc$ is abnormally small compared to subleading corrections. %Our results for the $SS$ channel also are in good agreement with previous literature at $\Nc=3$ and $\Nf=2,3$.

One of the most interesting observations of this work is the attractive nature of the $AA$ channel. An exploratory study based on the inverse amplitude method suggests the possible existence of a tetraquark resonance at higher center-of-mass energies in this channel, which could be related to some experimental findings of tetraquark states in the LHCb experiment~\cite{LHCb:2020bls,LHCb:2020pxc,LHCb:2022sfr,LHCb:2022lzp}. In the next chapter, ongoing work in this direction is presented, in which we focus on the study of meson-meson interactions as a function of energy. One of our main goals is to explore the possible presence of a tetraquark state in the $AA$ channel, which may open the door to study the nature of tetraquark states as a function of the number of colors.

%{\jorge [Cosas que creo he olvidado: Comentario de K en funcion de L5, L8]}

%The amplitudes can be projected to $s$-wave and then used to obtain a prediction for $E_R$ as a function of $\xi$, $\Mpi L$ and the LECs. R
\chapter{Meson-meson scattering at large $N_\text{c}$}
\label{sec:largeNmesons}

The possible existence of tetraquarks in the large $\Nc$ limit has been a topic of controversy during the last decade---see \cref{sec:QCD:largeNctetraquarks}. Traditionally, such states were thought not to exist at large $\Nc$~\cite{Witten:1979kh,Coleman:1980nk}, but recent arguments pointed in the opposite direction~\cite{WeinbergTetra,Knecht:2013yqa,Cohen:2014tg}.  Lattice QCD provides us with a tool to shed light on this question from first principles, by simulating QCD with varying number of colors. In particular, the setup presented in \cref{sec:largeNpions}---with $\Nf=4$ degenerate light quark flavors---represents an interesting playground to explore the existence of tetraquarks at large $\Nc$. 

In a theory with $\Nf=4$ degenerate quark flavors, meson interactions happen in seven scattering channels---see \cref{sec:largeNpions:largeNChPT}.  Of particular interest for the search of tetraquarks is the $AA$ channel, which only exists for $\Nf\geq4$. It presents attractive interactions at low energies, and a study based on unitarized ChPT, presented in \cref{sec:largeNpions:comparisonlitereatureandIAM}, suggests that this channel may contain a resonance. Due to its flavor quantum number---some of its states have four open flavors---such a resonance would undoubtedly be a tetraquark.

The possible existence of a tetraquark in this channel is further supported by recent experimental findings. During the last year, LHCb has reported on the discovery of different tetraquark resonances. These include the $T_{cs0}^{0}(2900)$, observed in the mass spectrum of $D^-K^+$~\cite{LHCb:2020bls,LHCb:2020pxc}, and the $T_{c\bar{s}}^{++}(2900)$ and the $T_{c\overline{s}}^0(2900)$, found in the mass spectra of $D_s^+\pi^+$ and $D_s^+\pi^-$~\cite{LHCb:2022sfr,LHCb:2022lzp}, respectively. While in real-world QCD these states are expected to be part of two separate isospin triplets, they would all share the quantum numbers of the $AA$ channel in a theory with $\Nf=4$ degenerate flavors. It is worth mentioning that all these states have phenomenologically been described as vector-meson molecules~\cite{Molina:2022jcd}, since their masses lie close to the $D^*K^*$ and $D_s^* \rho$ thresholds.

Another interesting experimental finding is that of the $T_{c\overline{s}1}^0$ vector tetra-quark, observed in the mass spectrum of $D^-K^+$~\cite{LHCb:2020bls,LHCb:2020pxc}. In our model, this particle would be part of the 45 dimensional irreps, known as the $AS$ and $SA$ channels. These two irreps are related by charge conjugation and so share the same scattering amplitude. We thus only consider the $AS$ channel from here on. 

This chapter presents some results on the $\Nc$ dependence of meson-meson interactions for energies up to the four-pion threshold~\cite{Baeza-Ballesteros:largeNinprep,Baeza-Ballesteros:2024ogp}. In particular, we use the $\Nf=4$ setup presented in \cref{sec:largeNpions}, and focus on the $SS$, $AA$ and $AS$ channels. We analyze meson-meson interactions as a function of the energy below the four-pion threshold, and explore the possible presence of tetraquark states in the $AA$ and $AS$ channels. 

The chapter is organized as follows. In \cref{sec:largeNmesons:lattice}, the lattice setup used for this study is described, with particular emphasis on the set of operators used to compute the correlation functions. In \cref{sec:largeNmesons:finitevolumespectrum}, we explain how finite-volume energies are determined and present our results for the finite-volume spectra of all three channels for varying $\Nc$. In \cref{sec:largeNmesons:phaseshiftresults} the two-particle quantization condition is used to constrain the scattering phase shifts, and we study their dependence on the number of colors. We end with a brief conclusion in \cref{sec:largeNmesons:conclusions}.

\section{Lattice setup}\label{sec:largeNmesons:lattice}

To study meson-meson interactions as a function of $\Nc$, we use replicas of the 3A11, 4A10, 5A10 and 6A10 ensembles---see \cref{tab:largeNpions:ensembles} for a summary of the simulation parameters---and consider a unitary setup with clover-improved Wilson fermions. These ensembles have $\Nc=3-6$, as indicated by the first digit in their name, and share a very similar heavier-than-physical pion mass, $\Mpi\approx590$ MeV, and lattice spacing, $a=0.075$ fm. Lattice simulations are performed with the HiRep code~\cite{DelDebbio:2008zf,DelDebbio:2009fd}, as done for the work presented in \cref{sec:largeNpions}.

\subsection{Set of interpolating operators}

The study of meson-meson scattering on the lattice requires the determination of the two-particle finite-volume energy spectrum for the energy range of interest. As Lüscher's quantization condition is limited to energies at which only two-particle states can go on-shell, we are restricted to work below any three- or four particle elastic regime.\footnote{A formalism that allows one to study three-to-two processes exists---see \rcite{Briceno:2017tce}---but we restrict ourselves to the two particle sector.} 
 In our case, this roughly corresponds to energies below the four-pion inelastic threshold, $E_{4\pi}=4\Mpi$.\footnote{As we will see below, in our ensembles vector mesons are stable, and so the $\rho\pi\pi$ threshold lies slightly  below that of four pions.}% However, we expect interactions between pions and vertor mesons to be very small, and will neglect the  
 
To reliably determine all finite-volume energies,  as was explained in \cref{sec:QCD:computationofobservables}, one needs to compute a matrix of correlation functions,
\begin{equation}
C_{ij}(t)=\langle O_i(t)O_j^\dagger(0)\rangle\,,
\end{equation}
for a sufficiently large set of operators, $\{O_i\}$. This set must contain operators having significant overlap onto all the finite-volume states expected in the energy range of interest. For example, if one expects a resonant state to appear in the scattering process, the inclusion of an interpolating single-particle operator with the quantum numbers of such state may be necessary to properly determine the finite-volume spectrum~\cite{Dudek:2012xn}. 

For our work, we consider three types of interpolating operators, working at various values of the total three-momentum, $\bm{P}$. We use two types of two-meson operators, with the form of two pions, $\pi\pi$, or two vector mesons, $\rho\rho$, which are complemented by local tetraquark operators, $T$. The inclusion of $\rho\rho$ operators is required since in our ensemble vector mesons are stable bound states, with a mass $\Mrho \lesssim 2\Mpi$---see \cref{tab:largeNmesons:mesonmasses}---and so states of two vector mesons may lie below the four-pion threshold. It is also useful in the search for tetraquarks, since it has been argued~\cite{Molina:2022jcd} that the tetraquarks of interest could be molecular states of two vector mesons. %Note, we have not included axial or scalar mesons, which have a higher mass.

The starting point to construct two-particle interpolating operators are single-meson operators. These are constructed as quark-bilinears  projected to definite three-momentum $\bm{p}$,
\begin{equation}\label{sec:largeNmesons:singleparticleoperator}
O_M(\bm{p}, \Gamma; t)=\sum_{\bm{x}}\overline{q}_{f_1}(\bm{x}; t)\Gamma q_{f_2}(\bm{x}; t)\,\text{e}^{-i\bm{p}\bm{x}},
\end{equation} 
Here, $\Gamma$ is a matrix in Dirac space that characterizes the intrinsic transformation properties of the meson under rotations, parity and charge conjugation. Also, momentum projection is performed summing over the set of all lattice sites with fixed time coordinate $t$. Note the two fermions are allowed to have different flavors, $f_1$ and $f_2$, which we use to project to definite isospin.

Single-particle operators with zero total momentum are used to determine the mass of pseudoscalar ($\Gamma=\gamma_5$, to which we refer as $\pi$), scalar ($\Gamma=\mathbbm{1}$, $a_0$), vector ($\Gamma=\gamma_i$, $\rho$) and axial ($\Gamma=\gamma_5\gamma_i$, $a_1$) non-singlet mesons. The results are  summarized in \cref{tab:largeNmesons:mesonmasses} for all four ensembles. Note that these results only give an  accurate result for the pion mass. Other mesons have a non-zero overlap into two-particle states and so an exhaustive determination of their masses should come from a variational analysis that includes mixing with multiparticle states. Nevertheless, this approximate result shows that $\rho\rho$ operators must be included in the set of operators, as such states will appear close below  the four-pion threshold. Operators of two axial or scalar mesons, on the other hand, need not be included. We note that in the case of pions and vector mesons, we also determine the one-particle correlation functions at non-zero momentum, needed to compute ratios as presented in \cref{eq:largeNmesons:ratiodefinition}.

\begin{table}[t!]
\centering
\begin{tabular}{ccccc}
\toprule
  Ensemble& $\Gamma=\gamma_5\,(aM_\pi)$ & $\Gamma=\gamma_i\, (aM_\rho)$ & $\Gamma=\mathbbm{1}\,(aM_{a_0})$  & $\Gamma=\gamma_5\gamma_i\,(aM_{a_1})$  \\ \midrule 

3A11 & 0.2130(7) &  0.355(4) & 0.430(15) & 0.45(3) \\  
4A10 & 0.2026(6) &  0.396(8) & 0.409(24) & 0.55(3) \\  
5A10 & 0.2131(5) &  0.403(4) & 0.405(14) & 0.590(12) \\ 
6A10 & 0.2169(5) &  0.417(4) & 0.412(10) & 0.608(14) \\ \bottomrule 
\end{tabular}
\caption{Masses in lattice units of different types of non-singlet mesons, characterized by different choices of $\Gamma$ in \cref{sec:largeNmesons:singleparticleoperator}. The results are determined from the single particle correlator, neglecting interactions with multiparticle states.  }

\label{tab:largeNmesons:mesonmasses}
\end{table}

 %Using these, we are also able to obtain results for the energies of mesons with non-zero momentum, which we can use to test discretization effects on the continuum dispersion relation,
%\begin{equation}\label{eq:largeNmesons:dispersionrelation}
%E_{\bm{p}}^{\text{disp},2}=M^2+\bm{p}^2\,.
%\end{equation}
% This is shown in \cref{fig:largeNmesons:dispersionrelation} for the pion in the 3A11 ensemble. We note that for total energy of the order $aE_\pi\sim 1$, discretization effects start to be considerable.

Interpolating operators of two particles are build as the product of two single-meson operators,
\begin{equation}\label{eq:largeNmesons:pipirhorhoproduct}
\begin{array}{rl}
(\pi\pi)(\bm{p}_1,\bm{p}_2)&=O_M(\bm{p}_1,\gamma_5)O_M(\bm{p}_2,\gamma_5)\,,\\
(\rho_i\rho_j)(\bm{p}_1,\bm{p}_2)&=O_M(\bm{p}_1,\gamma_i)O_M(\bm{p}_2,\gamma_j)\,,
\end{array}
\end{equation} 
where $\bm{P}=\bm{p}_1+\bm{p}_2$. To study the scattering of two mesons, these operators are projected to definite flavor channel, as well as to irreps of the cubic group (for $\bm{P}=0$) or the relevant little group (for $\bm{P}\neq0$). For simplicity, we refer generically to the latter irreps as \textit{cubic-group irreps}. Note that flavor and cubic-group projections can be performed independently: the former only involves quark flavors, while the latter concerns momenta and spin.

The projection to flavor irreps determines the different Wick contractions contributing to the correlation function. For example, we consider pion-pion operators of the form

\noindent\begin{equation}\label{eq:largeNmesons:flavorstates}
\begin{array}{rl}
(\pi\pi)^{SS}(\bm{p}_1, \bm{p}_2)  =&\displaystyle   \pi^+(\bm{p}_1)D_s^+(\bm{p}_2)+K^+(\bm{p}_1)D^+(\bm{p}_2)+(\bm{p}_1,\leftrightarrow \bm{p}_2)\,,\\
 (\pi\pi)^{AA}(\bm{p}_1, \bm{p}_2)  =&\displaystyle \pi^+(\bm{p}_1)D_s^+(\bm{p}_2)-K^+(\bm{p}_1)D^+(\bm{p}_2) + (\bm{p}_1,\leftrightarrow \bm{p}_2, )\,,\\
(\pi\pi)^{AS}(\bm{p}_1, \bm{p}_2)  =&\displaystyle \pi^+(\bm{p}_1)D_s^+(\bm{p}_2)+K^+(\bm{p}_1)D^+(\bm{p}_2) - (\bm{p}_1\leftrightarrow \bm{p}_2)\,,\\
\end{array}
\end{equation}
and similarly for $\rho\rho$ operators, using the respective vector mesons ($\rho$, $K^*$, $D^*$ and $D_s^*$). These states can be used to determine the form of the correlation function, presented  in \cref{sec:largeNmesons:correlationfunctions} below. %, where we also show that the results for the $AS$ and $SA$ channels are expected to be the same. 
Note that the $SS$ and $AA$ states are even under particle exchange, while the $AS$ is odd. This implies the former two channels only contain even partial waves, while the latter contains odd partial waves.

Projection to definite cubic-group irreps is performed following \cref{eq:QCD:cubicgroupprojection}. This is simple for two-pion states, as one basically only needs to take into account possible momenta combinations, but  becomes cumbersome for two vector-mesons. The reason is that, as their name indicates, vector mesons possess an additional vector index. This means that for some given $\bm{p}_1$ and $\bm{p}_2$, nine different combinations of the vector indices $i$ and $j$ in $\rho\rho$ operators are possible---see the second line of \cref{eq:largeNmesons:pipirhorhoproduct}. Thus, for a fixed non-interacting energy---defined as the sum of the single-meson energy associated to each of these particles, $E^\free=E_{M,\bm{p}_1}+E_{M,\bm{p}_2}$---several different projections may exist to one same irrep. For example, if we consider the $\bm{P}=(0,1,1)$ frame with $|\bm{p}_1|^2=0$ and $|\bm{p}_2|^2=2$, three different $\rho\rho$ operators transform under the $A_1$ irrep,
\begin{equation}
\begin{array}{rl}
(\rho\rho)_1&=\rho_2([0,0,0])\rho_3([0,1,1])+\rho_3([0,0,0])\rho_2([0,1,1])\,,\\
(\rho\rho)_2&=\rho_2([0,0,0])\rho_2([0,1,1])+\rho_3([0,0,0])\rho_3([0,1,1])\,,\\
(\rho\rho)_3&=\rho_1([0,0,0])\rho_1([0,1,1])\,,
\end{array}
\end{equation}
where the numbers in square brackets are the three-momenta, which we express in this chapter in units of $2\pi/L$, with $L$ the lattice side. 

To systematically determine all $\rho\rho$ operators in each cubic-group irrep, we initially consider all possible operators with the same non-interacting energy, to which \cref{eq:QCD:cubicgroupprojection} is applied. The resulting projected operators are not all linearly independent, and so we need to determine a maximal set of linearly independent ones. We note that, while it is possible to construct two-vector-meson operators in almost all cubic-group irreps, we only focus on irreps containing pion-pion $s$-wave interactions for the $SS$ and $AA$, and $p$-wave interactions for the $AS$ channels, respectively. The number of operators considered in each irrep is shown in \cref{tab:largeNmesons:quantityofoperators}.

\begin{table}[t!]
\centering
\begin{tabular}{c>{\centering\arraybackslash}p{1.cm}>{\centering\arraybackslash}p{1.cm}>{\centering\arraybackslash}p{1.cm}>{\centering\arraybackslash}p{1.cm}>{\centering\arraybackslash}p{1.cm}>{\centering\arraybackslash}p{1.cm}}
\toprule
 \multirow{2}{*}{Irrep}     &    \multicolumn{3}{c}{$SS$ and $AA$ channels} &    \multicolumn{3}{c}{$AS$ channel} \\ 
  & $\pi\pi$ & $\rho\rho$ &  $T$ & $\pi\pi$ & $\rho\rho$ &  $T$ \\ \midrule 

$A_1^+(0)$ & 5 & 6  & 5 & -- & -- & -- \\ 
$T_1^-(0)$ & -- & --  & -- & 4 & 8 & 2 \\  
$A_1(1)$ & 5 &  9 & 7 & 5 & 9 & 2 \\  
$E(1)$ & -- &  -- & -- & 3 & 13 & 2 \\  
$A_1(2)$ & 8 & 18  & 9 & 6 & 12 & 2 \\ 
$B_1(2)$ & -- & --  & -- & 4 & 15 & 2 \\  
$B_2(2)$ & -- & --  & -- & 6 & 16 & 2 \\  
$A_1(3)$ & 5 & 7  & 7 & 5 & 7 & 2 \\ 
$E(1)$ & -- & --  & -- & 5 & 12 & 2 \\  
$A_1(4)$ & 5 & 8  & 7 & 2 & 3 & 2 \\ 
$E(1)$ & -- & --  & -- & 3 & 9 & 2 \\  \bottomrule
\end{tabular}
\caption{Number of operator of operators of each type used in each isospin channel and cubic-group irrep. The number in parenthesis indicates $|\bm{P}|^2$ in units of $2\pi/L$. We note that operators in $E$ and $T_1^+$ irreps have multiplicity two and three, respectively.   }

\label{tab:largeNmesons:quantityofoperators}
\end{table}

Two-particle operators are complemented by local tetraquark operators, which may be required to correctly determine the finite-volume energy spectrum if tetraquark states are present. These operators are constructed from the local product of two quark bilinears, projected to definite total momentum,

\noindent\begin{multline}
T_{\Gamma_1,\Gamma_2}(\bm{P},t)=\sum_{\bm{x}} T_{\Gamma_1\Gamma_2}(x)\,\text{e}^{-i\bm{P}\bm{x}}\\=\sum_{\bm{x}} \overline{q}_{f_1}(x)\Gamma_1 q_{f_2}(x)\overline{q}_{f_3}(x)\Gamma_2 q_{f_4}(x)\,\text{e}^{-i\bm{P}\bm{x}}\,,
\end{multline}
where $x=(t,\bm{x})$. We consider several combinations  $\{\Gamma_1,\Gamma_2\}$ for both scalar and vector channels, having the spin, $J$, parity, $P$ and charge conjugation, $C$, quantum numbers of interest. In particular, we use
\begin{equation}
\begin{array}{rl}
J^{PC}=0^{++}:&\{\gamma_5,\gamma_5\},\{\gamma_5\gamma_0,\gamma_5\gamma_0\},\{\mathbbm{1},\mathbbm{1}\},\{\gamma_i,\gamma_j\},\{\gamma_5\gamma_i,\gamma_5\gamma_i\}\,,\\
J^{PC}=1^{-+}:&\{i\gamma_5,\gamma_5\gamma_i\},\{\gamma_5\gamma_0,\gamma_5\gamma_i\}\,.
\end{array}
\end{equation}
Note the $i$ factor in the first combination of the $1^{-+}$ tetraquark. This is required to ensure the operator is correctly hermitian, and so the correlation matrix obeys $C^*(t)=C(T-t)$, with $T$ the time extent of the lattice, that we use to average our numerical results.

Tetraquark operators are also projected to flavor irreps, in analogy to  \cref{eq:largeNmesons:flavorstates}, and to definite cubic-group irreps using \cref{eq:QCD:cubicgroupprojection}. The total number of tetraquark operators considered for this work for each channel and irrep is presented in \cref{tab:largeNmesons:quantityofoperators}

\newpage\subsection{Computation of correlation functions}\label{sec:largeNmesons:correlationfunctions}

The correlation functions in the channels of interest can be computed as a linear combination of the connected and disconnected Wick contractions, diagrammatically shown in \cref{fig:largeNpions:CDleadingcontractions}, using different combinations of momenta and Dirac structures. We can write
\begin{equation}
\begin{array}{rl}
C^{SS}(t)&=D_1+D_2-C_1-C_2\,,\\
C^{AA}(t)&=D_1+D_2+C_1+C_2\,,\\
C^{AS}(t)&=D_1-D_2-C_1+C_2\,,
\end{array}
\end{equation}
where we have introduced
\begin{equation}
\begin{array}{rl}
D_1=&D(\bm{p}_1,\Gamma_1',\bm{p}_2,\Gamma_2';\bm{k}_1,\Gamma_1,\bm{k}_2,\Gamma_2)\,,\\
D_2=&D(\bm{p}_2,\Gamma_2',\bm{p}_1,\Gamma_1';\bm{k}_1,\Gamma_1,\bm{k}_2,\Gamma_2)\,,\\
C_1=&C(\bm{p}_1,\Gamma_1',\bm{p}_2,\Gamma_2';\bm{k}_1,\Gamma_1,\bm{k}_2,\Gamma_2)\,,\\
C_2=&C(\bm{p}_2,\Gamma_2',\bm{p}_1,\Gamma_1';\bm{k}_1,\Gamma_1,\bm{k}_2,\Gamma_2)\,.
\end{array}
\end{equation}
Here $(\bm{k}_1,\bm{k}_2)$ and $(\bm{p}_1,\bm{p}_2)$ are the momenta of the initial- and final-state particles, respectively, with $\bm{P}=\bm{k}_1+\bm{k}_2=\bm{p}_1+\bm{p}_2$, while $(\Gamma_1,\Gamma_2)$ and $(\Gamma_1',\Gamma_2')$ are the corresponding Dirac matrices. The connected and disconnected contractions take the general form, respectively,
\begin{multline}
D_1=\sum_{\bm{x}_1,\bm{x}_2}\sum_{\bm{y}_1,\bm{y}_2}\text{e}^{-i(\bm{p}_1\bm{y}_1+\bm{p}_2\bm{y}_2)}\text{e}^{i(\bm{k}_1\bm{x}_1+\bm{k}_2\bm{x}_2)}\Theta(\bm{x}_1,\bm{x}_2) \Theta(\bm{y}_1,\bm{y}_2)\\
\times\langle\Tr\left[\hat{\Gamma}_1'S(y_1,x_1)\tilde{\Gamma}_1S^\dagger(y_1,x_1)\right]\Tr\left[\hat{\Gamma}_2'S(y_2,x_2)\tilde{\Gamma}_2S^\dagger(y_2,x_2)\right]\rangle\,,
\end{multline}
\begin{multline}
C_1=\sum_{\bm{x}_1,\bm{x}_2}\sum_{\bm{y}_1,\bm{y}_2}\text{e}^{-i(\bm{p}_1\bm{y}_1+\bm{p}_2\bm{y}_2)}\text{e}^{i(\bm{k}_1\bm{x}_1+\bm{k}_2\bm{x}_2)}\Theta(\bm{x}_1,\bm{x}_2) \Theta(\bm{y}_1,\bm{y}_2)\\
\times\langle\Tr\left[\hat{\Gamma}_2'S(y_2,x_1)\tilde{\Gamma}_1S^\dagger(y_1,x_1)\hat{\Gamma}_1'S(y_1,x_2)\tilde{\Gamma}_2S^\dagger(y_2,x_2)\right]\rangle\,,
\end{multline}
where we define $\hat{\Gamma}=\gamma_5\Gamma$ and $\tilde{\Gamma}=\gamma_0\Gamma^\dagger\gamma_0\gamma_5$, and $\langle...\rangle$ indicates ensemble average. We have also introduced functions that generalize these contractions to both two-particle and tetraquark operators. For the initial state, we use

\noindent\begin{equation}
\Theta(\bm{x}_1,\bm{x}_2)=\left\{
\begin{array}{ll}
1 \quad\quad\quad& \text{Two-particle operator at source}\,,\\
\delta_{\bm{x}_1,\bm{x}_2}\quad & \text{Tetraquark  operator at source}\,,\\
\end{array}\right.
\end{equation}
and similarly for the final-state operator and $\Theta(\bm{y}_1,\bm{y}_2)$. 

To evaluate the correlation functions we use different techniques for each type of operator at source. At sink, on the other hand, we project to definite momentum summing over the whole lattice. Correlation functions with two mesons at source are computed using time- and spin-diluted $\mathbb{Z}_2\times\mathbb{Z}_2$ stochastic sources, and are averaged by repeating the computations on 36 different time slices for each configuration. On the other hand, correlation functions with a tetraquark operator at source are first computed using point sources localized in a sparse spacial lattice, 
\begin{multline}
\Lambda_\text{S}(t)=\left\{(t,s_1+dn_1,s_2+dn_2,s_3+dn_3)\right.\\\left.|\,n_i\in\mathbb{Z},0\leq n_i < L/d,0\leq s_i < d\right\}\,,
\end{multline}
where $s_i\in\mathbb{Z}$ are some randomly chosen offsets. The results are later projected to definite total momentum summing only over this sparse lattice,
\begin{equation}\label{sec:largeNmesons:sparseoperators}
T_\text{sp}(\bm{P})=\sum_{\bm{x}\in\Lambda_\text{S}(t)} T(\bm{x})\text{e}^{-i\bm{P}\bm{x}}\,.
\end{equation}  
We use a step $d=4$ for all our ensembles, and repeat the computation for two time slices separated by half of the lattice time extent, with different random offsets $s_i$.

After computing the matrix of correlation functions, we use the relations $C(t)=C^\dagger(t)$ and $C(T-t)=C^*(t)$ to average the numerical results. For all operators considered here, the matrix of correlators in the $SS$ and $AA$ channels is purely real. Thus, we set the imaginary part of our lattice results to zero, as it is just a numerical artifact of the use of stochastic sources for meson-meson operators. %The proof of this statement, however, is rather technical, and we defer it to \cref{app:largeNmesons:}. 
By contrast, the correlation function for the $AS$ channel is expected to be complex-valued. Still we empirically observe that, in general, the imaginary part of two-meson correlation functions to be much smaller than the real part. %However, $C^{AS}_{ij}=\pm C^{SA*}_{ij}$---see \cref{app:largeNmesons:}---with the sign depending on the particular $i$ and $j$ operators. This is also use to average the correlators.

\section{Finite-volume energy spectra}\label{sec:largeNmesons:finitevolumespectrum}

Using the matrix of correlators, $C(t)$, it is possible to determine the finite-volume energy spectrum beyond the ground state. This requires the application of variational methods, such as the GEVP---see \cref{eq:QCD:gevp}. The applicability of this technique, however, relies on the assumption that no thermal pollution is present. The only exception to this is the case of real correlation functions, for which backwards propagation can be allowed---see \cref{eq:QCD:backwardspropagatingthermaleffects}.

In our ensembles, general thermal pollution is present. As discussed in \cref{sec:QCD:thermaleffectslattice}, our correlators are expected to have the general form in \cref{eq:QCD:themalspectraldecomposition}. For example, the real part looks like

\noindent\begin{equation}\label{eq:largeNmesons:correlatorgenrealform}
\Re\,C(t)=\sum_n A_n\cosh(E_nt')+\sum_{m,n} B_{nm}\cosh(\Delta E_{mn}^\text{th} t')\,,
\end{equation}
where $\tilde{t}=t-T/2$, and $A_n$ and $B_{mn}$ are related to the correlator matrix elements, while the imaginary part is analogous with hyperbolic sine functions instead of hyperbolic cosines. In \cref{eq:largeNmesons:correlatorgenrealform}, the first sum contains the two-particle energies that we want to determine, while the second contains thermal effects that depend on an energy difference. For example, the dominant thermal pollution is related to the energy difference between two single-meson states,
\begin{equation}
\Delta E_{mn}^\text{th}\approx E_{M,\bm{q}_1}-E_{M,\bm{q}_2}\,,
\end{equation}
for momenta such that $\bm{P}=\bm{q}_1-\bm{q}_2$. The presence of this second term spoils the application of the GEVP, as it prevents the variational approach from converging close to the center of the lattice, where thermal effects may dominate the correlation function. 

To mitigate this effect, we use an extension of the shift-reweighting procedure presented in \cref{eq:QCD:shiftreweight} that takes into account that our correlation function behaves as a sum of hyperbolic functions rather than of exponentials. This is, we take into account backwards propagation in the correlator. We assume that a two-pion term dominates the second sum of \cref{eq:largeNmesons:correlatorgenrealform}, with energy $\Delta E^\text{th}=E_{\pi,\bm{p}_1}-E_{\pi,\bm{p}_2}$, and approximately eliminate it by redefining the correlator
\begin{equation}\label{eq:largeNmesons:shiftreweightcosh}
\tilde{C}(t)=\frac{1}{2}\left\{\frac{\cosh(\Delta E^\text{th} t')}{\cosh[\Delta E^\text{th} (t'_+)]}C(t+1)-\frac{\cosh(\Delta E^\text{th} t)}{\cosh[\Delta E^\text{th} (t'_-)]}C(t-1)\right\}\,,
\end{equation}
with  $t'_{\pm}=t'\pm 1$. The energies used to eliminate the thermal effects, $E_{\pi,\bm{p}}^\text{free}$, are single-pion energies computed using the continuum dispersion relation, 
 \begin{equation}\label{eq:largeNmesons:dispersionrelationsinglepion}
E_{\pi,\bm{p}}^\free=\sqrt{\bm{p}^2+M_\pi^2}\,.
\end{equation}
for different choices of the single-particle momenta, as shown in  \cref{tab:largeNmesons:thremalmomentumsubtractions}. In the case of the rest, $[0,1,1]$ and $[0,0,2]$ frames, this simply amounts to computing the derivative of the matrix of correlators. %\Cref{eq:largeNmesons:shiftreweightcosh} is a generalization of the method described \cref{eq:QCD:shiftreweight} that consider backwards propagation for for the thermal effects.

The procedure presented above removes dominant thermal effects in the real part of the matrix of correlators, but not  in the imaginary part. Thus, it completely eliminates the corresponding thermal contaminations for the $SS$ and $AA$ channels, but not for the $AS$ case, which contains a non-zero imaginary part. However, we empirically observe the imaginary part of the $AS$ channel correlators to be, in general, much smaller than its real counterpart, and so we expect this procedure to largely soften the effect of thermal pollution also in this channel. %We note alternative techniq have been tested, but lead to worst determination of the finite-volume energy.

\begin{table}[t!]
\centering
\begin{tabular}{ccc}
\toprule
  $\bm{P}$ & $|\bm{p}_1|^2$ & $|\bm{p}_2|^2$  \\ \midrule 

$[0,0,0]$ & 0 &  0 \\  
$[0,0,1]$ & 1 &  0 \\   
$[0,1,1]$ & 1 &  1 \\  
$[1,1,1]$ & 2 &  1 \\  
$[0,0,2]$ & 1 &  1 \\   \bottomrule 
\end{tabular}
\caption{Magnitude of the single particle momenta used to perform the shift-reweighting technique in \cref{eq:largeNmesons:shiftreweightcosh} that approximately eliminates leading thermal effects. Momenta are indicated in units of $2\pi/L$.  }

\label{tab:largeNmesons:thremalmomentumsubtractions}
\end{table}

After getting rid of the leading thermal effects, we use $\tilde{C}(t)$ to solve a GEVP. In particular, we use the \textit{single-pivot procedure}, in which the GEVP is only solved for the mean of the correlator at a single time $\td>t_0$,
\begin{equation}\label{eq:largeNmesons:singlepivotGEVP}
\tilde{C}(\td)v_n(t_0,\td)=\lambda(t_0,\td)\tilde{C}(t_0)v_n(t_0,\td)\,.
\end{equation}
Then, the resulting matrix of eigenvectors is used to rotate $\tilde{C}(t)$ for all times and samples. In other words, the matrix of eigenvectors determined from the GEVP in \cref{eq:largeNmesons:singlepivotGEVP} is used to uniquely define operators that have maximum overlap into a single finite-volume state. We note this procedure is numerically very stable. For our analysis, we use $(t_0,\td)=(3,5)$.

Another option is the so-called \textit{rolling-pivot approach}, in which the GEVP is solved for all $t$ in the mean of the correlator, and the eigenvectors at each $t$ are used to rotate $C(t)$ on each sample. We find that the application of this method leads to comparable results. 

Note that a priori the applicability of the GEVP to our matrices of correlators may seem incorrect, since we are defining different tetraquark operators at source and sink. At source, we use sparse operators, introduced in \cref{sec:largeNmesons:sparseoperators}, while we sum over all lattice sites at sink. This means that the matrix elements  $\langle 0|T_\text{sp}|n\rangle$ and $\langle 0|T|n\rangle$  are in general different. While both the sparse and the non-sparse operators have equal matrix elements for those state on which both overlap, the former operator also has non-zero overlaps with states having total momenta of the form $\bm{P}+2\pi \bm{n}/dL$, for $\bm{n}\in\mathbb{Z}^3$. However, the $A_n$ coefficient appearing in the spectral decomposition of the correlator---see \cref{eq:largeNmesons:correlatorgenrealform}---is the product of that matrix element and the one corresponding to the final-state operator, which does not overlap with those states. For example, if both the initial and the final state operators are a tetraquark,
\begin{equation}
A_n=\langle 0|T(0)|n\rangle \langle n|T_\text{sp}|0\rangle\,,
\end{equation} 
and similarly for any two-particle final operator. Thus, $A_n=0$ for those states on which the final-state operator does not overlap. By projecting over the whole lattice in the final states, we prevent possible excited-state contamination associated to the sparse operators.

\begin{figure}[!b]
    \centering
    \begin{tikzpicture}
  \node[] at (0,0) {\includegraphics[width=0.7\textwidth]{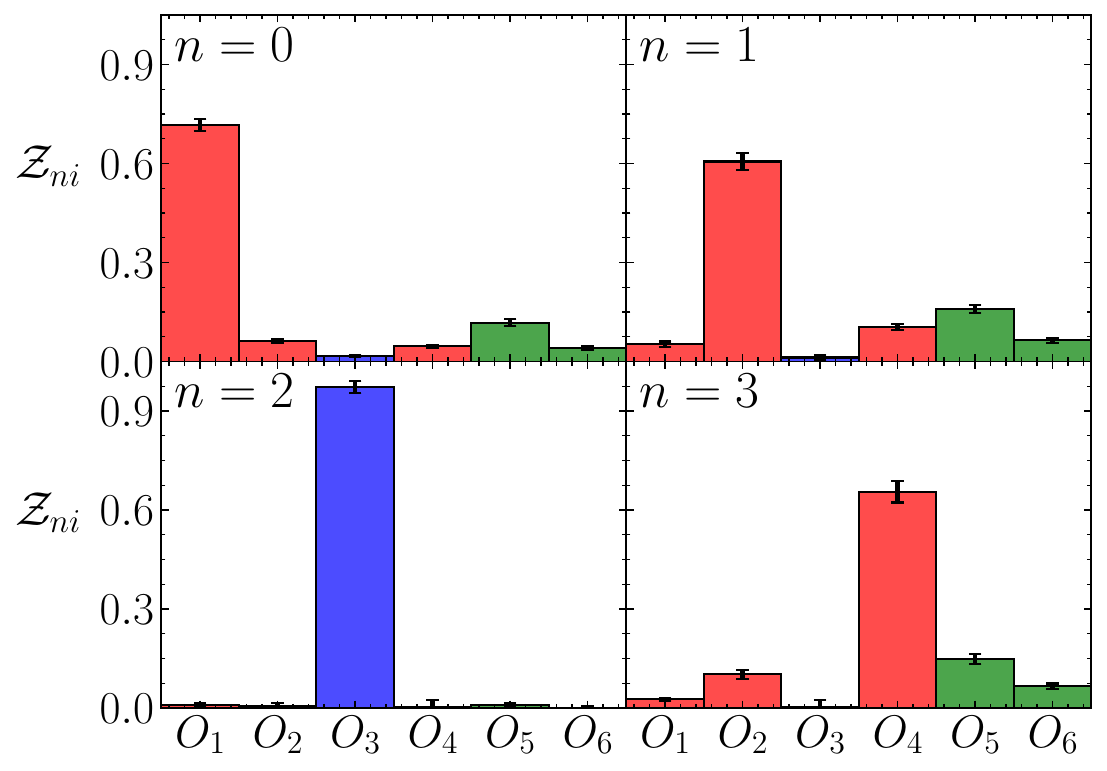}};
  
  \node[align=left]  at (6.2, -1.1) {
  $O_1=\pi(0)\pi(0)$\\[0.5ex]
  $O_2=\pi(1)\pi(1)$\\[0.5ex]
  $O_3=\rho(0)\rho(0)$\\[0.5ex]
  $O_4=\pi(2)\pi(2)$\\[0.5ex]
  $O_5=T_{\gamma_5\gamma_5}$\\[0.5ex]
  $O_6=\sum_i T_{\gamma_i\gamma_i}$};
 
\end{tikzpicture}
    \caption{
        Relative overlaps of the lowest-lying finite-volume states, $|n\rangle$, with the operators used to solve the GEVP---see \cref{eq:largeNmesons:overlaps}, for the rest-frame $A_1^+$ irrep for the $AA$ channel with $\Nc=3$. Different colors correspond to different types of operators: $\pi\pi$ (red), $\rho\rho$ (blue) and local tetraquarks (green). Numbers in parenthesis indicate the magnitude squared of the single-particle momentum, in units of $(2\pi/L)^2$. }
    \label{fig:largeNmesons:overlaps}
\end{figure}

The eigenvalues of the GEVP provide us with information about the overlap of the different operators with each finite-volume state. One can define the relative overlap of state $|n\rangle$ with the operator $O_i$ as
\begin{equation}\label{eq:largeNmesons:overlaps}
\cZ_{ni}=\frac{\langle 0| O_i|n\rangle}{\sum_j \langle 0| O_j|n\rangle}\,.
\end{equation}
An example of the relative overlaps is shown in \cref{fig:largeNmesons:overlaps} for a reduced set of states and operators in the rest-frame $A_1^+$ irrep for the $AA$ channel with $\Nc=3$. Each panel presents the relative overlaps of one state into the different operators.  We observe that, in general, finite-volume states have maximum overlap onto the operator with the closest associated free energy. For example, in the case of the $\pi(1)\pi(1)$ operator, where the number in parenthesis is the magnitude squared of the pion momentum, this would be two times the energy of a pion with $|\bm{p}|=1$. Also, we note that the $n=2$ state is predominantly a $\rho\rho$ state. Finally, no state has a dominant overlap into the tetraquark operator, which indicates that our states are all predominantly two-particle states. However, pion-pion operators seem to have a non-negligible overlap onto it, indicating that their inclusion into the operator base may help to better determine the pion-pion spectrum.

The eigenvalues of the GEVP are used to extract the lowest lying finite-volume energies, $E_n$. To improve this determination, we define ratios \mbox{generalizing} the method presented in \cref{eq:largeNpions:ratiocreation} to moving frames and excited states. Using the relative overlaps in \cref{eq:largeNmesons:overlaps}, we assign each eigenvalue to a different operator onto which the corresponding state has maximum overlap. %\footnote{If two states have maximum overlap onto the same operator, we assign it to the state with the larger relative overlap, and assign to the other state the operator onto which it has the second largest overlap, and so on}.
 For those eigenvalues associated to a two-meson operator of the form, $O_M(\bm{k}_1)O_M(\bm{k}_2)$, we define the following ratio,
\begin{equation}\label{eq:largeNmesons:ratiodefinition}
R(t)=\frac{\tilde{C}(t)}{\partial_0 [C_{M,\bm{k}_1}(t)C_{M,\bm{k}_2}(t)]}\,,
\end{equation}
where $C_{M,\bm{k}}$ refers to the single-meson correlation function for species $M\in\{\pi,\rho\}$ with momentum $\bm{k}$. This ratio is fitted assuming single- and two-particle correlation functions to be dominated by a single state. The single-particle correlator is approximated by
\begin{equation}\label{eq:largeNmesons:singlemesoncorrelatorform}
C_{M,\bm{k}}(t)=A\cosh\left(E^\text{latt}_{M,\bm{k}}t'\right)\,,
\end{equation}
where $E^\text{latt}_{M,\bm{k}}$ is the energy resulting from fitting the single-particle correlator $C_{M,\bm{k}}(t)$ and $A$ is some unknown amplitude. The two-particle correlation function is given in general in \cref{eq:largeNmesons:correlatorgenrealform}. If we consider a single state, and neglect thermal effects that are removed with \cref{eq:largeNmesons:shiftreweightcosh}, we can set $B_{nm}=A_{n\neq 1}=0$ and approximate,
\begin{equation}\label{eq:largeNmesons:twomesoncorrelatorform}
C(t)=A_1\cosh\left(E_\text{int}^\text{latt} t'\right)\,.
\end{equation}
This form of the two-particle correlator is introduced in \cref{eq:largeNmesons:shiftreweightcosh} to define $\tilde{C}(t)$, which is used in \cref{eq:largeNmesons:ratiodefinition} to define the fit function. The fit allows us to determine the finite-volume energy shift, defined as,
\begin{equation}
\Delta E_\text{int}=E_\text{int}^\latt-E_{M,\bm{k}_1}^\latt-E_{M,\bm{k}_2}^\latt\,.
\end{equation}
The total energy is finally reconstructed as
\begin{equation}
E_\inter=\Delta E_\inter + E_{M,\bm{k}_1}^\text{free}+ E_{M,\bm{k}_2}^\text{free}\,,
\end{equation}
where $E_{M,\bm{k}}^\text{free}$ is the single-meson energy computed using the continuum dispersion relation, \cref{eq:largeNmesons:dispersionrelationsinglepion}.
This definition of the finite-volume energy using the continuum dispersion relation, rather than the energies determined from the lattice,  has empirically been shown to reduce discretization effects on the determination of the finite-volume energies~\cite{Bulava:2022vpq,BaryonScatteringBaSc:2023zvt,BaryonScatteringBaSc:2023ori}.

\begin{figure}[!h]\vspace{1.cm}
    \centering
    \includegraphics[width=0.7\textwidth]{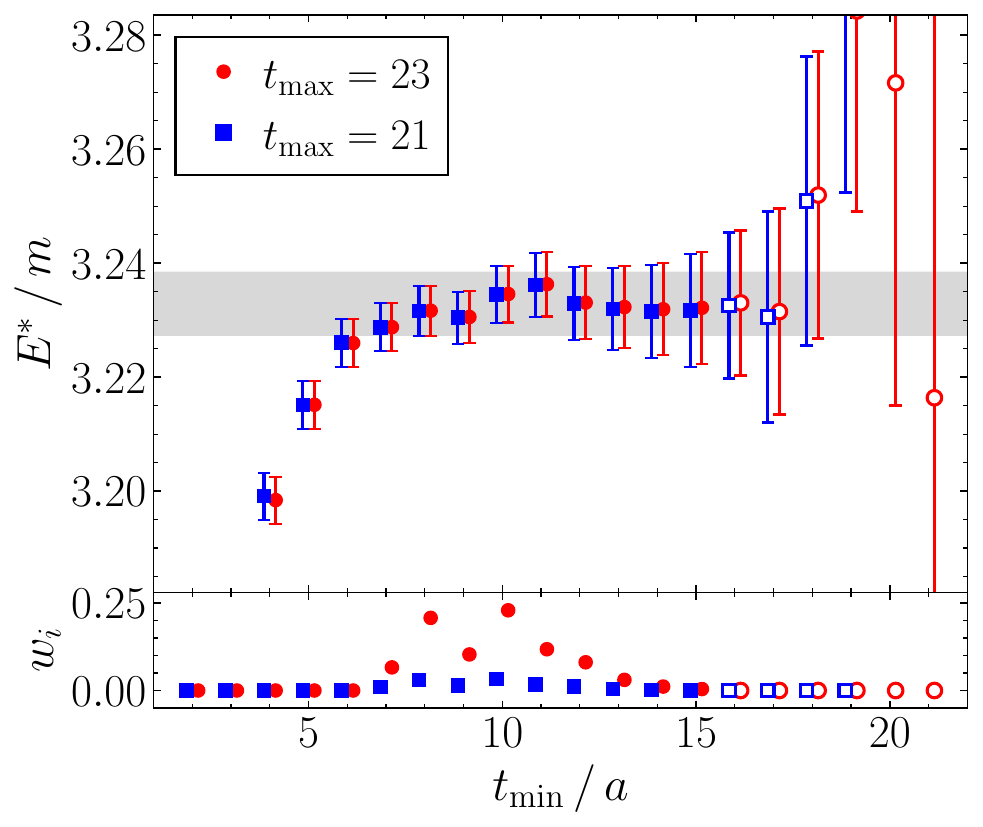}
    \caption{
        Best-fit results to \cref{eq:largeNmesons:ratiodefinition} for the ground-state energy in the $\bm{P}=[0,0,1]$ frame of the $AA$ channel with $\Nc=3$, for different values of $t_\text{min}$ and two choices of $t_\text{max}$. The final result (gray band) is obtained by averaging the results using the weights from \cref{eq:QCD:weightsplateaux}, in the bottom panel. Empty points are manually excluded from the average. }
    \label{fig:largeNmesons:plateaux}
\end{figure}

\begin{figure}[!p]
    \centering
    \includegraphics[height=0.89\textheight]{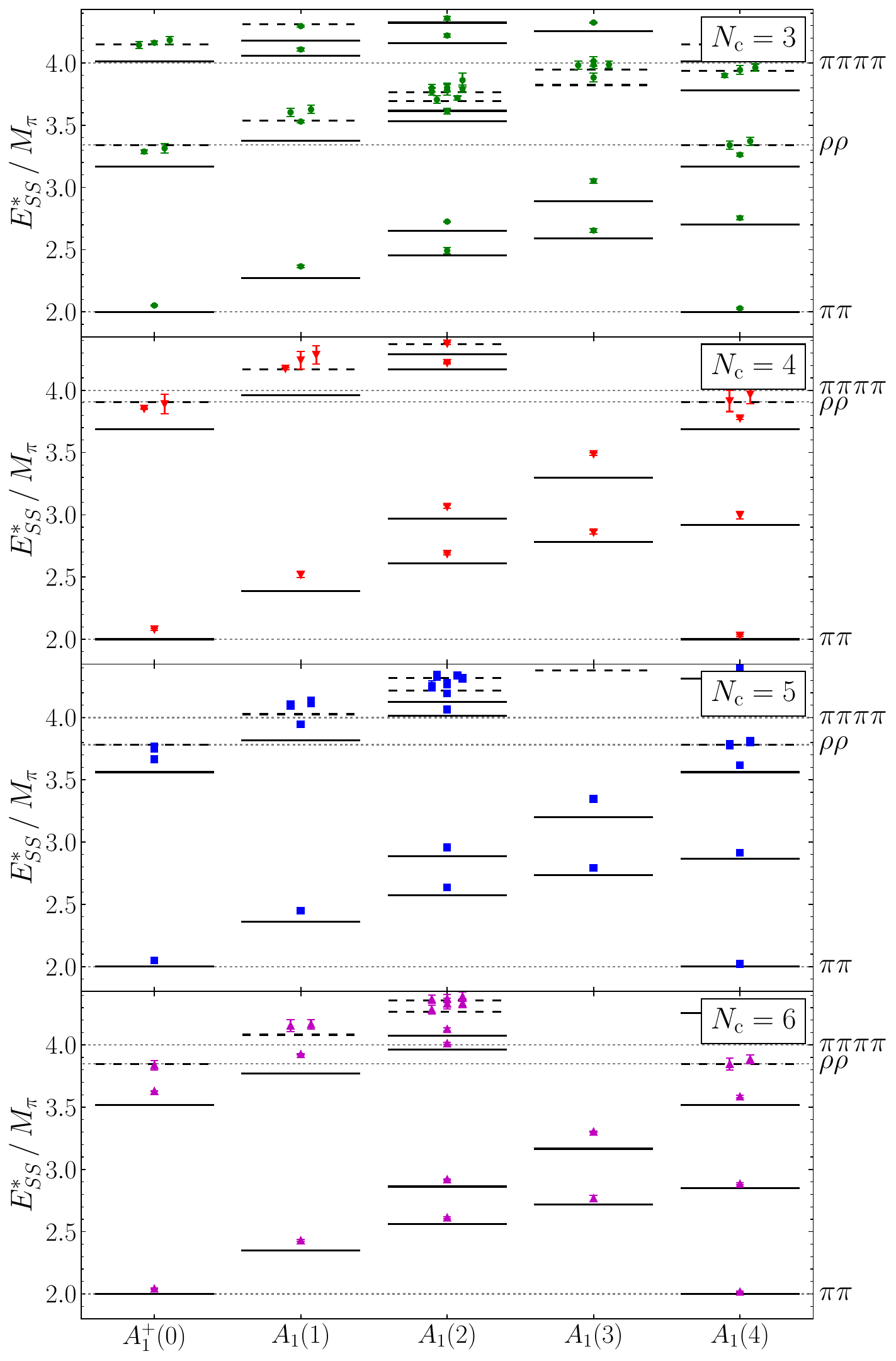}
    \caption{
        Finite-volume spectra for the $SS$ channel, extracted using the full set of operators. Each panel corresponds to a different $\Nc$ and each column represents a different irrep of the cubic group and momentum frame, with $|\bm{P}|^2$ indicated in parenthesis. Horizontal solid and dashed segments indicate non-interacting $\pi\pi$ and $\rho\rho$ energies, respectively, while gray dashes lines are relevant inelastic thresholds---see \cref{fig:largeNmesons:threshold} for a full list. }
    \label{fig:largeNmesons:energiesSS}
\end{figure}

\begin{figure}[!p]
    \centering
    \includegraphics[height=0.89\textheight]{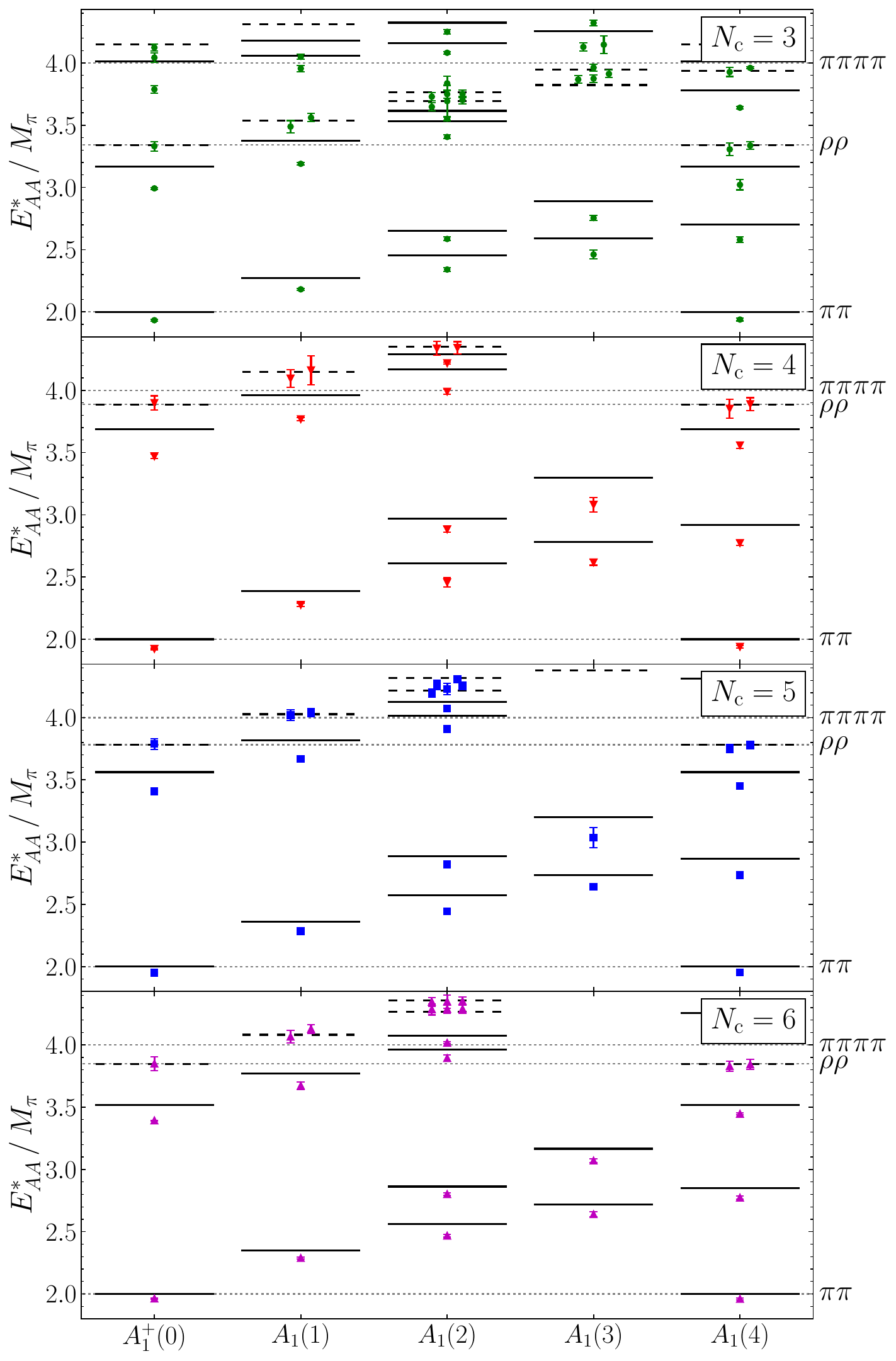}
    \caption{
        Same as \cref{fig:largeNmesons:energiesSS} for the $AA$ channel. }
    \label{fig:largeNmesons:energiesAA}
\end{figure}

\begin{figure}[!p]
    \centering
    \includegraphics[height=0.89\textheight]{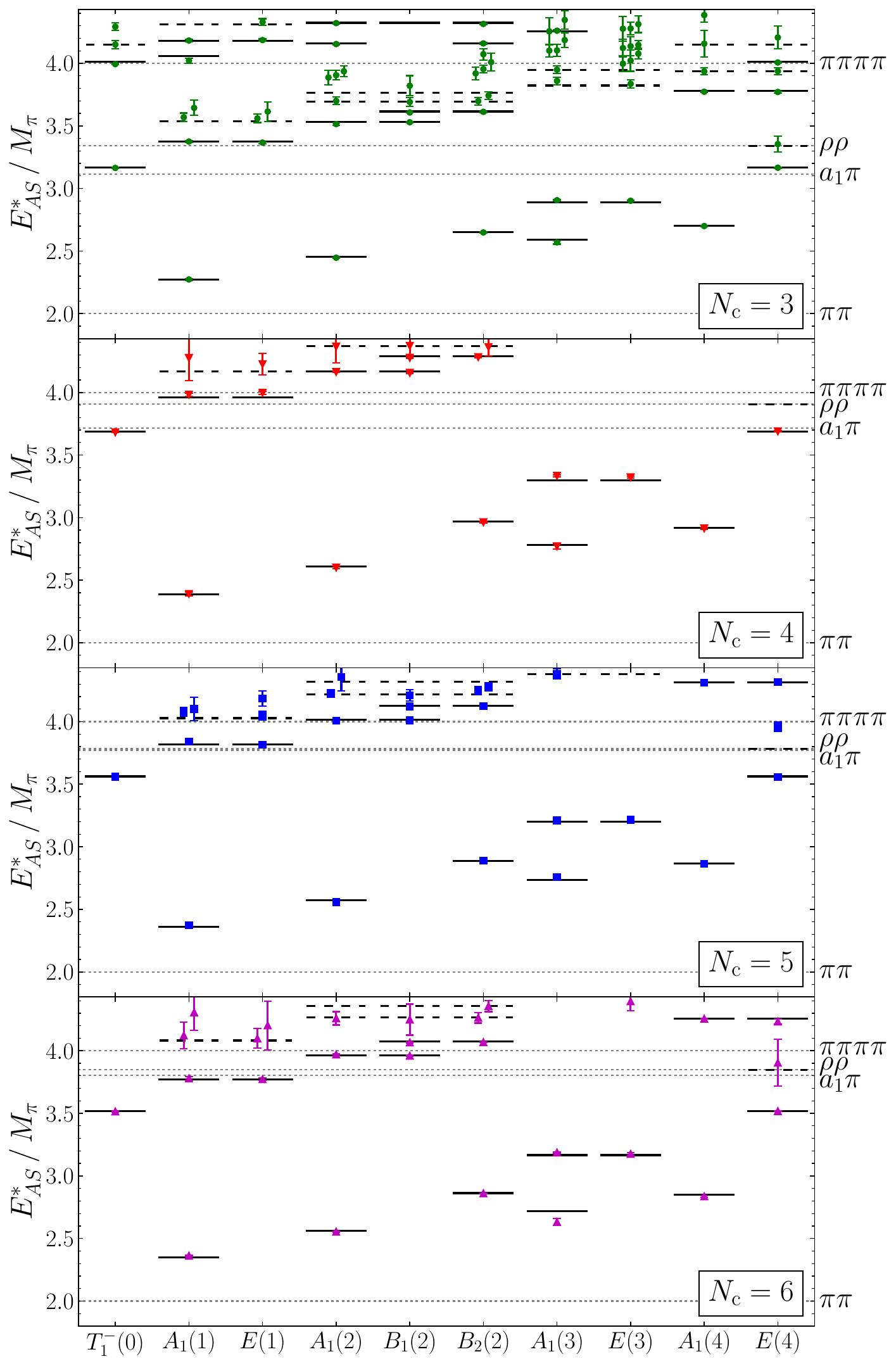}
    \caption{
        Same as \cref{fig:largeNmesons:energiesSS} for the $AS$ channel. }
    \label{fig:largeNmesons:energiesAS}
\end{figure}

\begin{figure}[!p]
    \centering
    \hspace{-0.cm}\includegraphics[width=0.88\textwidth]{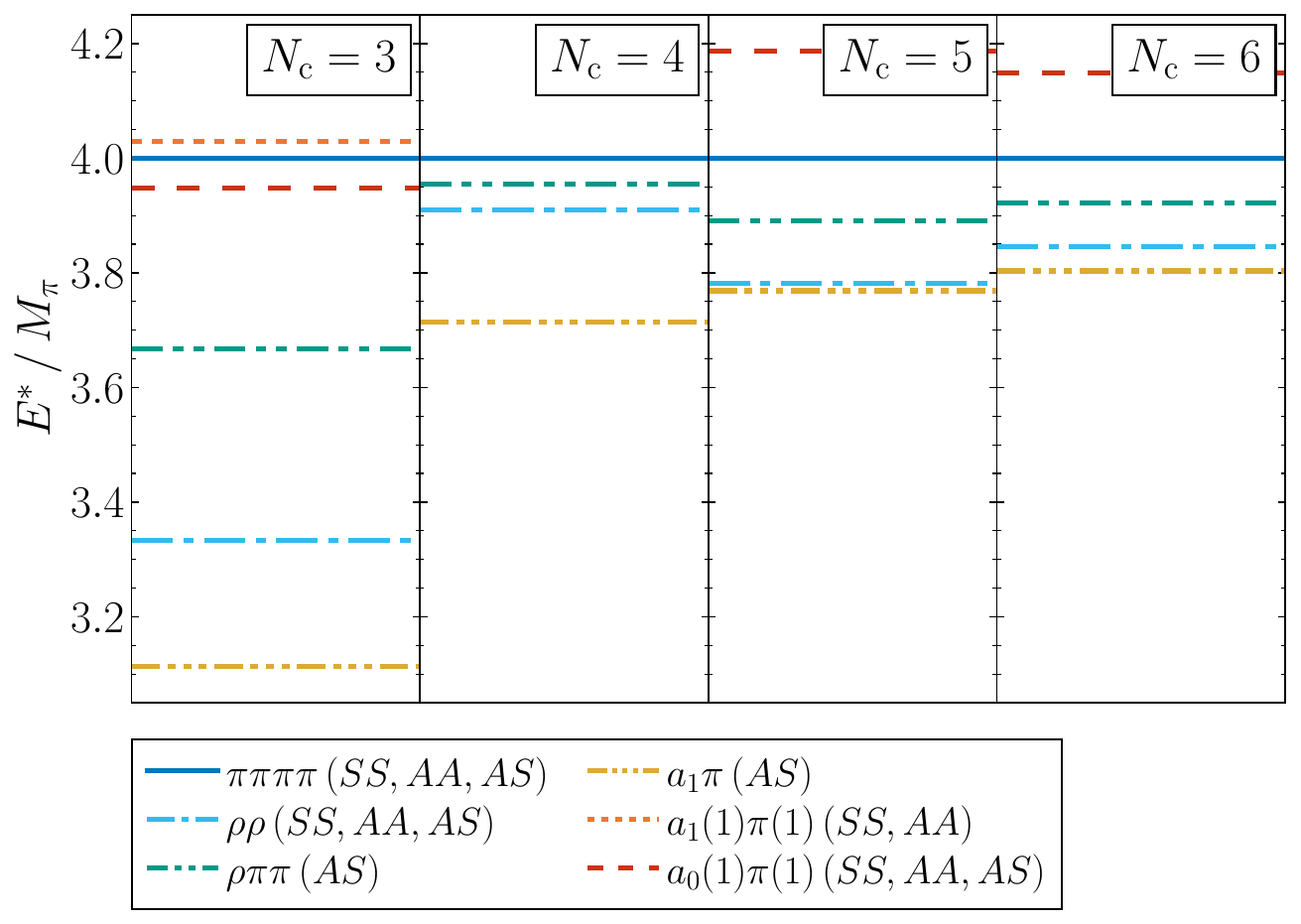}
    \caption{
       Relevant inelastic threshold in our ensembles, computed using the meson masses in \cref{tab:largeNmesons:mesonmasses}. We indicate in which channels these threshold are present in the figure legend.}
    \label{fig:largeNmesons:threshold}\vspace{1.cm}

    \centering
    \includegraphics[width=0.85\textwidth]{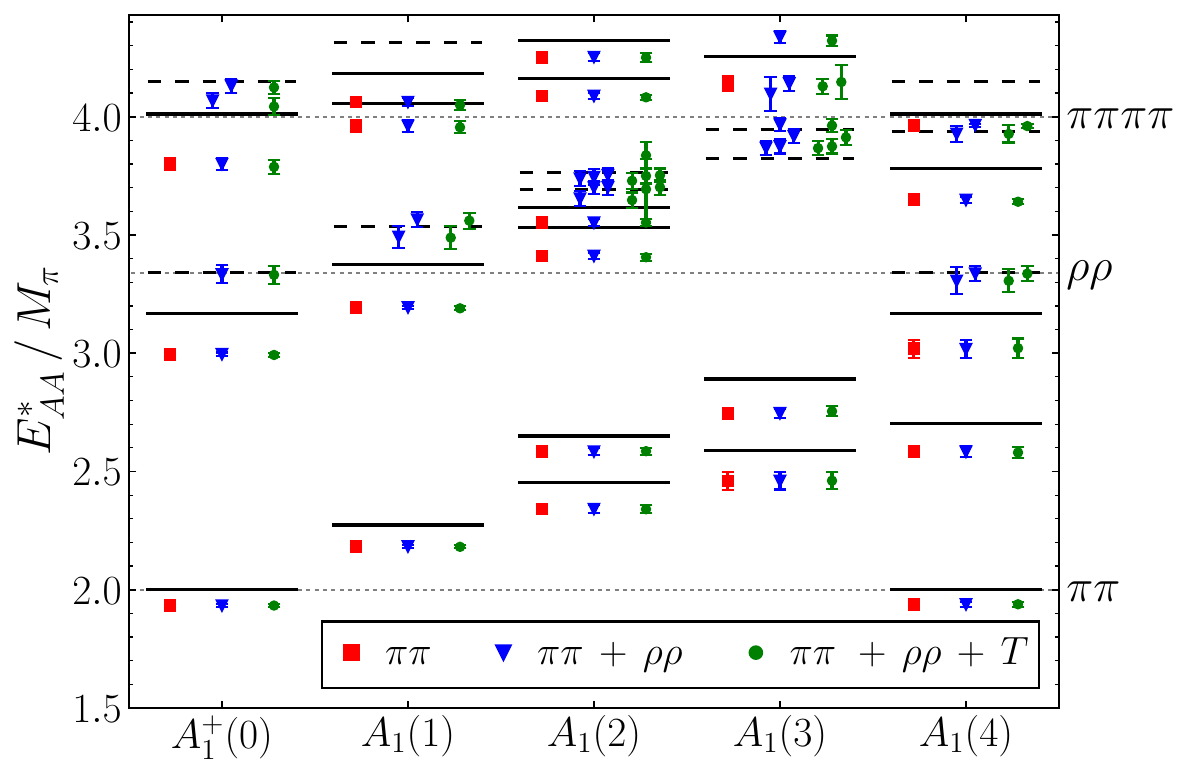}
    \caption{
        Results for the finite-volume spectrum of the $AA$ channel for $\Nc=3$ extracted using different choices of the operator set, as indicated in the figure legend.}
    \label{fig:largeNmesons:spectracomparison}
\end{figure}

This fit to \cref{eq:largeNmesons:ratiodefinition} is repeated for several fit ranges, $t\in[t_\text{min}, t_\text{max}]$, varying both the lower and the upper limit. The final results are obtained averaging the results in a range where they are observed to plateau, using the weights presented in \cref{eq:QCD:weightsplateaux}. An example of such determination is presented in \cref{fig:largeNmesons:plateaux}, and the complete results, obtained using the full basis of operators, are shown in \cref{fig:largeNmesons:energiesSS,fig:largeNmesons:energiesAA,fig:largeNmesons:energiesAS}, for the $SS$, $AA$ and $AS$ channels, in this same order. We also indicate in each case, the non-interacting $\pi\pi$ and $\rho\rho$ energies (solid and dashed black lines, respectively) and the some of the most relevant inelastic thresholds for each channel (gray dotted and dashed-dotted lines). A full summary of all the threshold relevant for each channel is presented in \cref{fig:largeNmesons:threshold}, computed using the meson masses in \cref{tab:largeNmesons:mesonmasses}. %Note that the $a_1\pi$ and the $\rho\pi\pi$ threshold is only indicated in the AS channel, since such states with zero relative momentum do not project to even parity.

%Those eigenvalues associated to states that are not assigned to a tetraquark operator can be fitted directly to \cref{eq:largeNmesons:shiftreweightcosh} using the expression for $C(t)$ in \cref{eq:largeNmesons:singletwomesoncorrelatorform}. In most cases, it is not possible to determine any energy from these results, or they this lead to energies well-above the four-pion inelastic threshold. Note however this is not the case in some of the frames of the AS channel, in which we are able to determine one additional state slightly below this threshold. These are presented in light gray in \cref{fig:largeNmesons:energiesASextra}. We believe these states are related to a two particle state formed by an axial and a pseudoscalar mesons. Such a state would project into vector irreps and has a non-interacting energy close to the determined ones. Due to the inconsistency of extracting these states, we opt not to include them and restrict our analysis in the $AS$ channel to energies below the this threshold.

%We have performed different check of consistency for these results for the $\Nc=3$ case. In \cref{fig:largeNpions:} we show the results for the finite-volume energy spectrum obatined for variations of the analysis procedure. This include the use of $(t_0,\td)=(5,7)$ to solve the GEVP, the uso of the rolling pivot approach, and the elimination of a number of elements of the operator set. 

Finally, we have studied the effect of varying the types of operators used to determine the matrix of correlation functions. The energy spectra of the $AA$ channel for $\Nc=3$ obtained for different choices of the operator set are presented in \cref{fig:largeNmesons:spectracomparison}. We observe how the inclusion of $\rho\rho$ operators leads to the determination of al large number of new energy-levels, related to states of two vector mesons. However, it has a minimal impact on states of two-pions. Similarly, further considering tetraquark operators also has an almost negligible effect on pion-pion states. Only for some excited states, the additional operators help in reducing the error. %A comparison of the finite-volume spectrum determined for three choices of the operator set for the  $AA$ channel with $\Nc=3$ is presented in \cref{fig:largeNmesons:spectracomparison}.%We observe that the inclusion of tetraquark operators increases the energy shift in both the $AA$ and the $AS$ channel, specially for those states close to the four-pion threshold. The inclusion of $\rho\rho$ operators, on the other hand, has a much more reduced impact.

%Comment on used deltaE for thermal effect
%Comment on single-pivot GEVP
%Comment on overlaps and ratio, and fitting function. Commenton reconstruction of delta E, that empirically is observed to reduce discretization effects
%Comment on using more or less ope

\newpage
\section{Extraction of infinite-volume scattering amplitudes}\label{sec:largeNmesons:phaseshiftresults}

Finite-volume energies can be used to constrain the infinite-volume scattering observables, using the two-particle quantization condition introduced in \cref{sec:hadrons:QC2}. In this chapter, we present results for pion-pion interactions. This means we restrict our analysis to those finite-volume states that have a maximum relative overlap into pion-pion operators. We also only focus on  the lowest partial waves in each channel. Furthermore, we consider all the states below the four-pion threshold, neglecting interactions with $\rho\rho$ and $\rho\pi\pi$ states, that we expect to be very small. %In any case, these should only be relevant for few states in the $\Nc=3$ ensemble, for which $\Mrho\sim 1.7\Mpi$.

In the case of the $SS$ and $AA$ channel, which contain only even partial waves, the quantization condition reduces to the algebraic relation given in \cref{eq:hadrons:algebraicQC2} if only $s$ wave is considered. In the case of the $AS$ channel, that contains odd partial waves, the QC can also be simplified if only $p$ wave is taken into account. Its form, however, depends on the particular cubic-group irrep and momentum frame---see \rrcite{Dudek:2012xn,Bulava:2016mks,Alexandrou:2017mpi}. For example, in the case of the $A_1$ and $E$ irreps of the $\bm{P}=[0,0,1]$ frame, it takes the form
\begin{equation}
\begin{array}{rl}
q_2^*\cot\delta_1^{A_1(1)}&=\displaystyle\frac{2}{\gamma L \pi^{1/2}}\left[\cZ_{00}^{\bm{P}}\left(\tilde{q}\right)+\frac{2}{\sqrt{5}q^2}\cZ_{20}^{\bm{P}}\left(\tilde{q}\right)\right]\,,\\
q_2^*\cot\delta_1^{E(1)}&=\displaystyle\frac{2}{\gamma L \pi^{1/2}}\left[\cZ_{00}^{\bm{P}}\left(\tilde{q}\right)-\frac{1}{\sqrt{5}q^2}\cZ_{20}^{\bm{P}}\left(\tilde{q}\right)\right]\,,\\
\end{array}
\end{equation}
where we have defined $\tilde{q}=q_2^*L/2\pi$, with $q_2^*$ is the magnitude of the relative momentum in the center-of-mass frame. 

\begin{figure}[!p]
    \centering
    \includegraphics[width=0.85\textwidth]{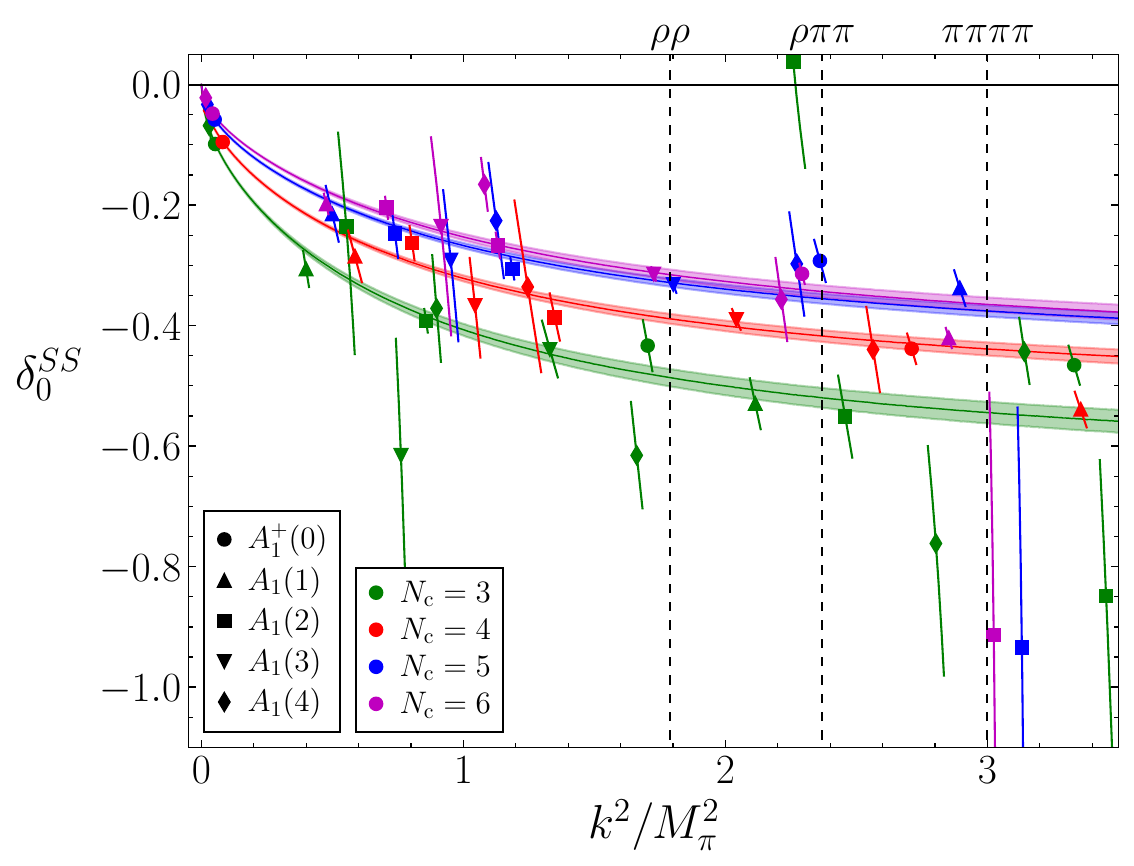}
    \caption{
        Results of the $s$-wave phase shift for the $SS$ channel, determined for different $\Nc$ and cubic group irreps, as indicated in the figure legends. Shaded regions indicate the best-fit results to a modified ERE---see \cref{eq:largeNmesons:mERE}.}
    \label{fig:largeNmesons:PSSS}\vspace{0.5cm}

    \centering
    \includegraphics[width=0.85\textwidth]{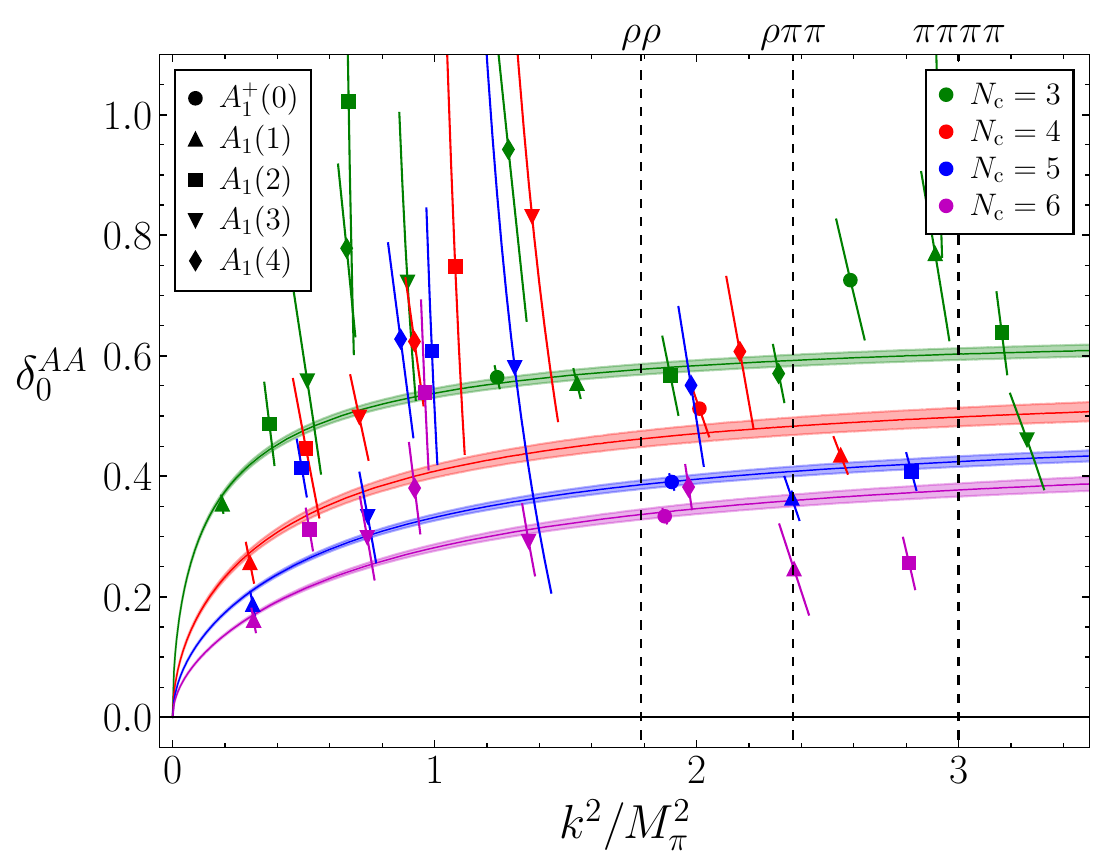}
    \caption{
       Same as \cref{fig:largeNmesons:PSSS} for the $AA$ channel.}
    \label{fig:largeNmesons:PSAA}
\end{figure}

\begin{figure}[!t]
    \centering
    \includegraphics[width=0.85\textwidth]{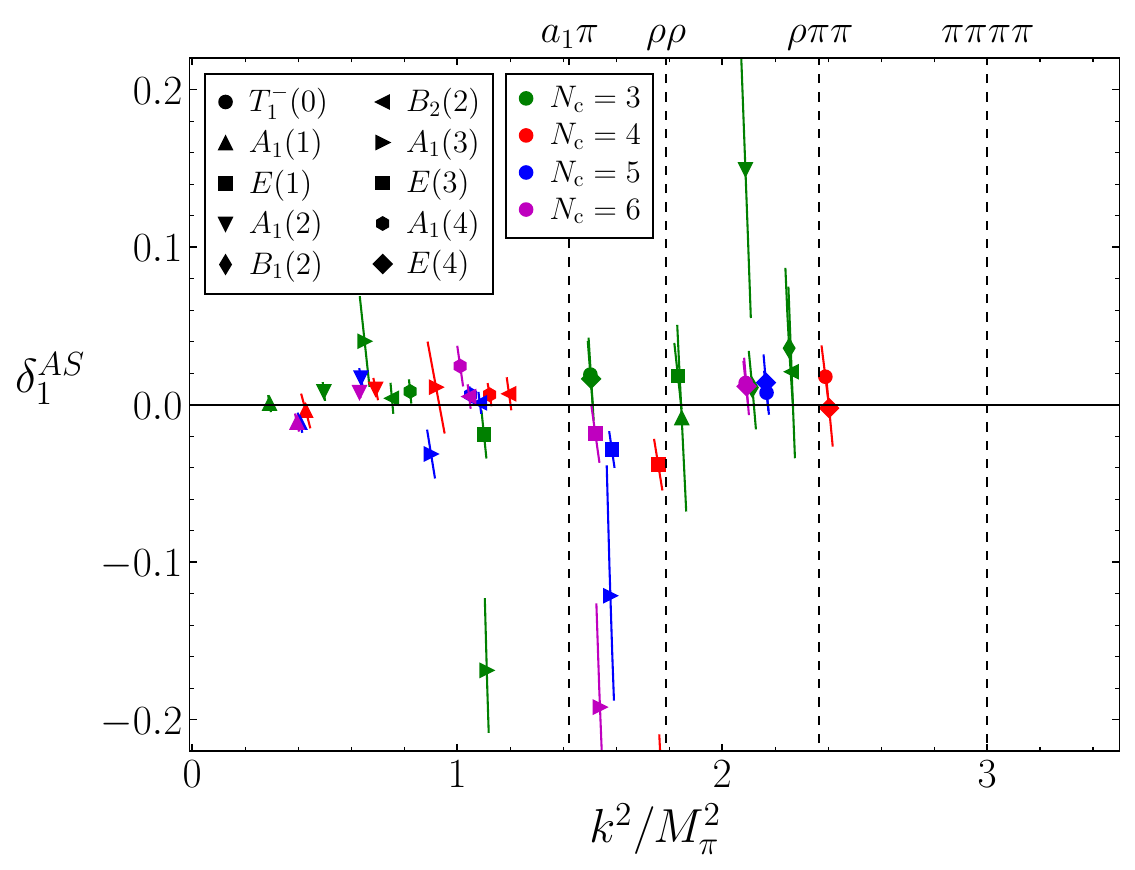}
    \caption{
       Results of the $p$-wave phase shift for the $AS$ channel determined for different $\Nc$ and  cubic group irreps, as indicated in the figure legends.}
    \label{fig:largeNmesons:PSAS}
\end{figure}

Using these simplified forms of the QC, we are able to extract a result for the scattering phase shift from each finite-volume energy level. The results are presented in \cref{fig:largeNmesons:PSSS,fig:largeNmesons:PSAA,fig:largeNmesons:PSAS} for the $SS$, $AA$ and $AS$ channels, respectively. We also indicate as vertical dashed lines the most relevant inelastic thresholds, computed using the lightest vector meson masses value among all ensembles, corresponding to the 3A11 ensemble. Note that in the $AS$ channel we additionally indicate the $a_1\pi$ threshold, where $a_1$ is the axial non-singlet meson, as this state may couple to two pions in odd partial waves. Other thresholds, are not indicated for legibility, but can be determined from the results in \cref{tab:largeNmesons:mesonmasses} and \cref{fig:largeNmesons:threshold}.%We note that two-particle operator of the $a_1\pi$ form has not been included in our set of operators, and so we cannot rely on 

For both the $SS$ (\cref{fig:largeNmesons:PSSS}) and $AA$ (\cref{fig:largeNmesons:PSAA}) channels, we observe that the phase shift grows in magnitude rapidly above threshold, but depends weakly o the energy above $k^2\gtrsim \Mpi^2$. The result for the $SS$ channel is negative, indicating repulsive interactions, while that of the $AA$ channel is positive, thus interactions are attractive. We note also that, for both channels, the phase shift, and so the strength of the interactions decreases with $\Nc$, as expected from large $\Nc$ arguments.  For the $AS$ channel (\cref{fig:largeNmesons:PSAS}), in contrast, the scattering phase shift remains small throughout the elastic region, indicating that interactions in this channel are very weak. This is consistent with ChPT expectations, which predict the scattering amplitude for this channel to only start at NLO---see \rcite{Bijnens:2011fm}.

We can use the QC to constrain the infinite-volume scattering amplitudes. Given some parametrization of the amplitude one can obtain a prediction for the finite volume energies. These predictions can be matched to the lattice results to determine the parameters of the amplitude.

We perform fits of our lattice results for the $SS$ and $AA$ channels to a modified ERE~\cite{Yndurain:2002ud,Pelaez:2019eqa},
\begin{equation}\label{eq:largeNmesons:mERE}
q_2^*\cot\delta_0=\frac{E^*M_\pi}{E^{*2}-2z^2}\left[B_0+B_1\frac{q_2^{*2}}{M_\pi^2}\right]\,,
\end{equation}
where $B_i$ and $z$ are parameters to be determined.
This is a modification of the standard effective-range expansion in \cref{eq:hadrons:effectiverangeexpansion}, that includes the $E^*$ kinematical factor in front and a denominator that reproduces the so-called Adler-zero~\cite{Adler,Adler2}, this is, the fact that ChPT predicts the scattering amplitude to have a zero below threshold.  We note that the parameters of the modified ERE can be related to those of the standard ERE expanding close to threshold,

\noindent\begin{equation}\label{eq:largeNmesons:mEREandERErelation}
M_\pi a_0=\frac{1}{B_0}\,,\quad\quad\quad M_\pi^2r_0a_0=2\frac{B_1}{B_0}-3\,,
\end{equation}
where we have used the LO ChPT prediction, $z=\Mpi$.

In \cref{tab:largeNmesons:fitresults}, we show the results of these fits at fixed $\Nc$, in which we set $z=M_\pi$. For $\Nc=3$ we have also tried other alternative fits with more free parameters, such as fitting  $z$ or one additional higher-order coefficient, but observed no significant improvement in the description of the lattice results. %We also compare to a fit to the standard ERE, and observe worse-quality results.

\begin{table}[b!]
\centering
\begin{tabular}{ccccc}
\toprule
 Channel     &    $\Nc$ &    $B_0$ & $B_1$ & $\chi^2\,/\,\text{dof}$\\  \midrule 

\multirow{4}{*}{$SS$} & 3 & $-2.38(4)$  & $-2.55(21)$ & $10.1/10=0.85$ \\
& 4 & $-3.09(8)$ & $-3.28(19)$ &  $6.1/8=0.76$ \\
& 5 & $-3.93(9)$ & $-3.81(23)$ & $7.9/8=0.98$ \\
& 6 & $-4.37(4)$ & $-3.84(22)$ & $13.5/8=1.68$ \\ \midrule
\multirow{4}{*}{$AA$} & 3 & $0.927(15)$  & $2.63(9)$ & $28.7/14=2.05$ \\
& 4 & $1.79(5)$ & 3.12(19) &  $24.1/10=2.41$ \\
& 5 & $2.53(6)$ & 3.63(17) & $30.2/10=3.02$ \\
& 6 & $3.37(7)$ & 3.94(20) & $29.1/10=2.91$ \\   \bottomrule
\end{tabular}
\caption{Results from the fits of the finite-volume energies to predictions from a modified ERE with $z=\Mpi$ fixed, as given in  \cref{eq:largeNmesons:mERE}. Fits are performed for each channel and value of $\Nc$ separately.  }

\label{tab:largeNmesons:fitresults}
\end{table}

\begin{figure}[!p]
    \centering
    \begin{subfigure}{0.495\textwidth} 
    \centering
        \includegraphics[width=1\textwidth]{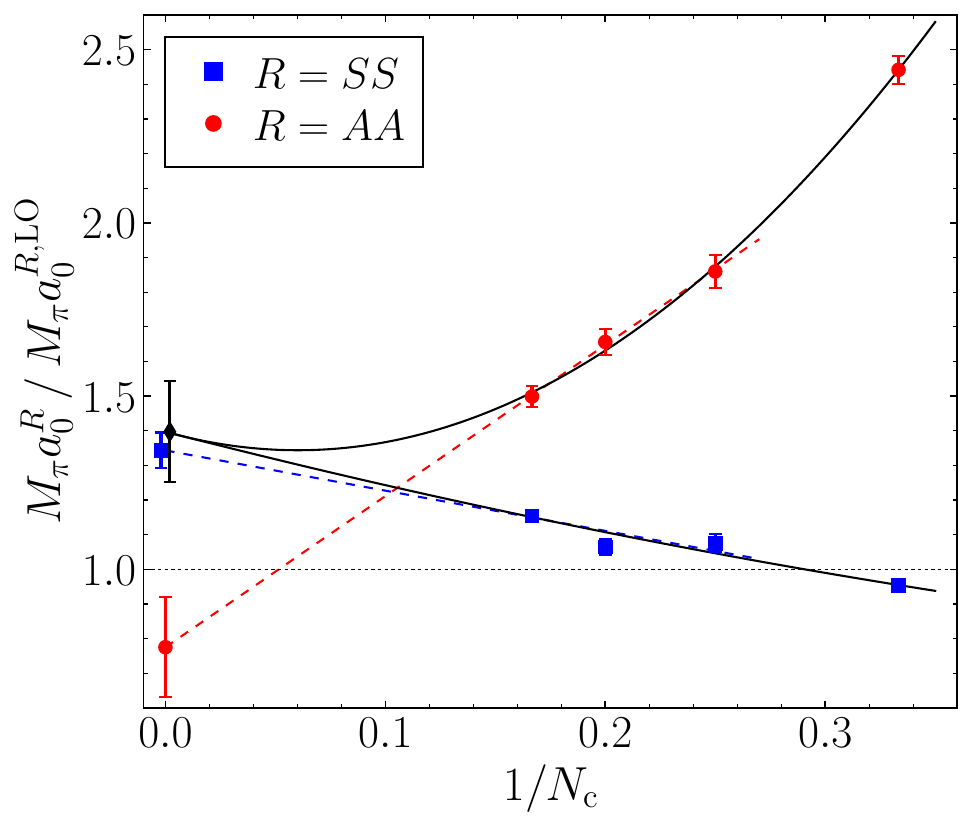}
    \end{subfigure}
     \begin{subfigure}{0.495\textwidth} 
    \centering
        \includegraphics[width=1\textwidth]{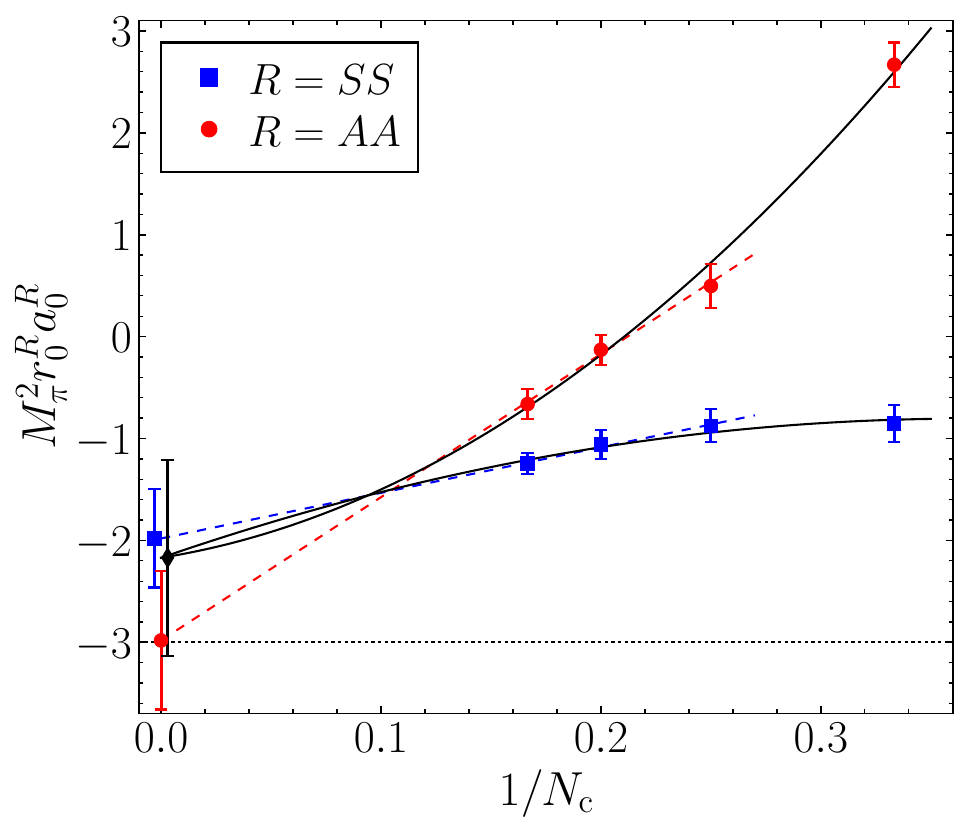}
    \end{subfigure}
    \caption{
         Results for the scattering length divided by LO ChPT predictions (left)---see \cref{eq:largeNpions:LOChPTscatteringalength}---and for the scattering range (right), for both the $SS$ and $AA$ channels. These are determined from the results of a fit to a modified effective range expansion, see \cref{eq:largeNmesons:mEREandERErelation}. Results for $\Nc=4-6$ are fitted to a linear relation (dashed lines) to extrapolate to the large $\Nc$ limit. We also indicate the result of a constrained large $\Nc$ extrapolation for all $\Nc=3-6$ results and including up to $\cO(\Nc^{-2})$ corrections (solid lines).}
    \label{fig:largeNmesons:fitresults}
\end{figure}

From these fits, we are able to extract information of the scattering length and effective range of the two channels, using the relations in \cref{eq:largeNmesons:mEREandERErelation}. The results are represented in \cref{fig:largeNmesons:fitresults} and summarized in \cref{tab:largeNmesons:fitresultsderived}. The results for the scattering length are divided by the LO ChPT prediction---see \cref{eq:largeNpions:LOChPTscatteringalength}---which removes leading $\Nc$ dependencies and we also expect to alleviate mass dependencies. These LO predictions are computed using the results for the pion decay constant computed for these ensembles in \rcite{Baeza-Ballesteros:2022azb}---see \cref{tab:largeNpions:massdecayxi}. Finally, we also indicate in \cref{fig:largeNmesons:fitresults} the LO ChPT prediction with horizontal dashed lines. %Finally, we note that the results are compatible with those obtained in 

\begin{table}[p!]
\centering
\begin{tabular}{cccccc}
\toprule
 Channel     &    $\Nc$ &    $M\pi a_0\,/\,M_\pi a_0^\LO$ & $\chi^2\,/\,\text{dof}$ & $M_\pi^2a_0r_0$ & $\chi^2\,/\,\text{dof}$\\  \midrule 

\multirow{5}{*}{$SS$} & 3 & $0.953(16)$ &  --- & $-0.85(18)$ & --- \\
& 4 & $1.075(26)$ & --- & $-0.88(16)$ & --- \\
& 5 & $1.065(24)$ & --- & $-1.06(14)$ & --- \\
& 6 & $1.155(10)$ & --- & $-1.24(10)$ & --- \\
& $\infty$ & $1.34(5)$ & $4.7/1 = 4.7 $ & $-2.0(5)$ & 0.05/1=0.05\\ \midrule
\multirow{5}{*}{$AA$} & 3 & 2.44(4) & --- & 2.67(22) & --- \\
& 4 & 1.86(5) & --- & 0.50(22) & --- \\
& 5 & 1.67(4) & --- & $-0.13(15)$ & --- \\
& 6 & 1.50(3) & --- & $-0.66(14)$ & --- \\
& $\infty$ & 0.78(14) & 0.08/1=0.08 & $-3.0(7)$ &  0.13/1 = 0.13 \\ \midrule  
$SS$+$AA$ & $\infty$ & 1.40(15) & $5.3//3=1.77$ & $-2.2(1.0)$ & 1.7/3=0.57 \\\bottomrule
\end{tabular}
\caption{Results for the scattering length, divided by LO ChPT predictions from \cref{eq:largeNpions:LOChPTscatteringalength}, and the effective range for both $SS$ and $AA$ channels, determined from the results in \cref{tab:largeNmesons:fitresults} using \cref{eq:largeNmesons:mEREandERErelation}. We also show the large $\Nc$ results obtained from a linear extrapolation of the $\Nc=4-6$ results for each channel separately, as well as from a constrained quadratic extrapolation of all $\Nc$ for both channels combined. }

\label{tab:largeNmesons:fitresultsderived}
\end{table}

 Using these results, we perform an extrapolation to large $\Nc$. We observe that in all cases the $\Nc=4-6$ data is well reproduced by a linear relation that includes leading and subleading $\Nc$ terms. However, it seems that $1/\Nc^2$ effects are significant for $\Nc=3$, which we thus do not include for the linear extrapolations. The results of the fit are represented as dashed lines in \cref{fig:largeNmesons:fitresults}, and the extrapolated predictions are presented in \cref{tab:largeNmesons:fitresultsderived} as a solid line. 
 
These large $\Nc$ results, however, are not consistent with large $\Nc$ expectations which predict the large $\Nc$ limit of the two quantities represented in \cref{fig:largeNmesons:fitresults} to be equal for both channels---see, for example, \cref{eq:largeNpions:NcNfscalingscatteringlength}. We have found, however, that our results, including those with $\Nc=3$, are consistent with a common large $\Nc$ limit if $1/\Nc^2$ effects are included. This fit, performed with a constrained large $\Nc$ limit, is also presented in \cref{fig:largeNmesons:fitresults}.

Both for $\Mpi a_0^R$ and $\Mpi^2 a_0^R r_0^R$, we observe that subleading $\Nc$ effects are larger for the $AA$ channel than for the $SS$ one. From the point of view of ChPT, these subleading effects are dominated by the chiral logarithms and other LEC-independent terms---see for example \cref{eq:largeNpions:SSSUamplitudeChPT,eq:largeNpions:AASUamplitudeChPT}. In the $AA$ channel, all such terms appear with the same sign, and so add up in the final contribution. In the $SS$ channel, on the other hand, different terms have different signs, and the total result is smaller. 

We find the large $\Nc$ result for the effective range to be consistent with the LO ChPT prediction, which is itself $\cO(\Nc^0)$. By contrast, there is a larger difference between large $\Nc$ results and LO ChPT for the scattering length. In the large $\Nc$ limit, subleading chiral corrections appear as a polynomial in the chiral parameter, $\xi=\Mpi^2/(4\pi\Fpi)^2$, with the coefficients being products of $\cO(\Nc)$ LECs. In our case, the chiral parameter is rather large, and so large chiral corrections could be expected.

\begin{figure}[!t]
    \centering
    \includegraphics[width=0.7\textwidth]{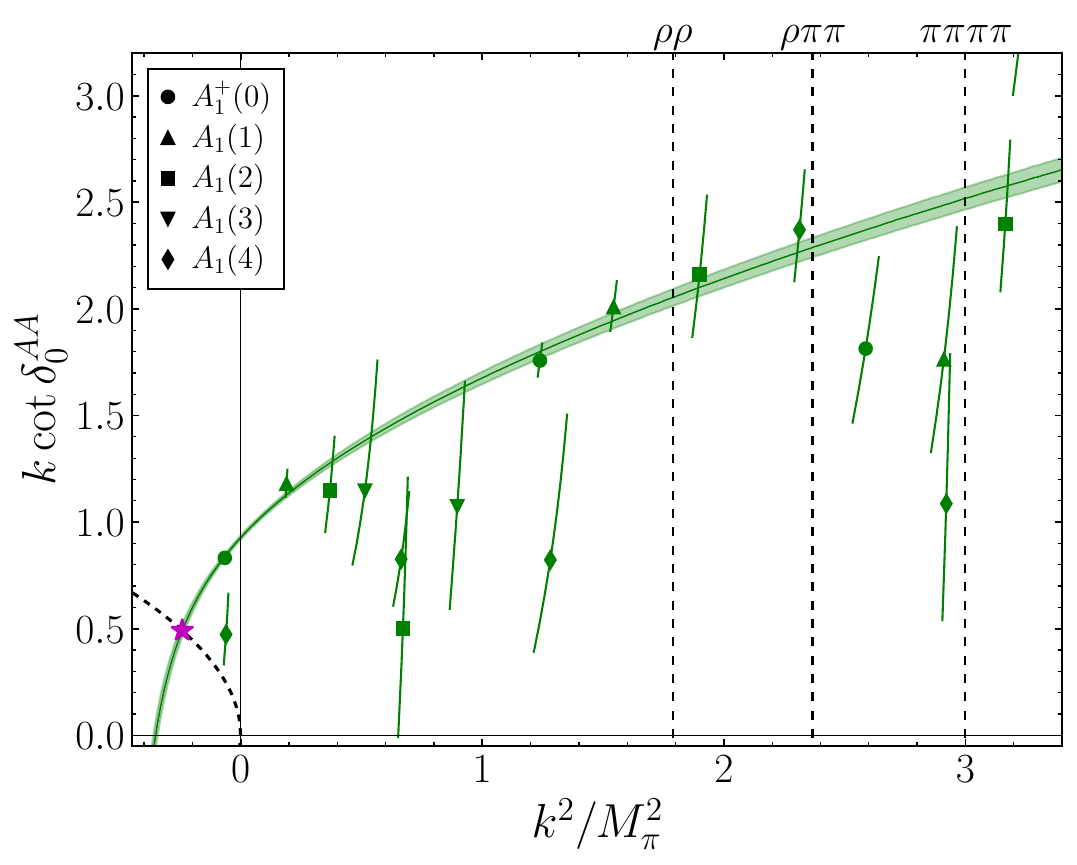}
    \caption{Scattering phase shift for the $AA$ channel with $\Nc=3$, together with the best-fit results to a modified ERE---see \cref{eq:largeNmesons:mERE}. We observe the presence of a virtual bound state, given by the solution to \cref{eq:largeNmesons:virtualstatesolution}, indicated by a star.}
    \label{fig:largeNmesons:kcotAA}
\end{figure}

Furthermore, while we observe no evidence of any pion-pion resonance in the $AA$ channel above threshold, the fir results for $\Nc=3$ are compatible with the presence of a virtual tetraquark bound state at $E_\text{bound}/\Mpi=1.741(13)$. This corresponds to a below-threshold pole in the scattering matrix on the real energy axis of the second Riemann sheet, given by the condition
\begin{equation}\label{eq:largeNmesons:virtualstatesolution}
q_2^*\cot\delta_0-\sqrt{-q_2^{*2}}=0\,,
\end{equation}
as illustrated with a star in \cref{fig:largeNmesons:kcotAA}. 
Using the results for the scattering length and effective range in \cref{tab:largeNmesons:fitresults}, and \cref{eq:hadrons:Weinbergcriterium}, we can determine the renormalization factor of the associated state, $Z=0.577(13)$. According to Weinberg's criterium in \cref{sec:hadrons:twoparticlescatteringinfinitevolume}, this quantity is related to the possibility of the state being compact or a hadronic molecule. In particular, $Z=1$ would indicate a purely compact state, while $Z=0$ corresponds to molecules. In our case, the observed virtual bound state seems to have both a compact and a molecular  component.
A strong confirmation of the existence of this bound state, however, requires to study its presence with respect to the use of different parametrizations of the scattering amplitude. We finally note that no such state is found for the remaining $\Nc$.

\newpage\section{Conclusions}\label{sec:largeNmesons:conclusions}

In this chapter, results on the study of meson-meson scattering as a function of the number of colors, $\Nc$, have been presented~\cite{Baeza-Ballesteros:largeNinprep,Baeza-Ballesteros:2024ogp}. We work in a theory with $\Nf=4$ degenerate quark flavor, and focus on three different scattering channels: the $SS$ channel, which is analogous to the isospin-two channel of two-flavor QCD, the $AA$ channel which only exists for $\Nf\geq 4$, and the $AS$ channel, that contain odd partial waves. The latter two are particularly interesting, since they have the quantum numbers which some recently found tetraquarks would be expected to have in a $\Nf=4$ world.

We have determined the finite-volume spectrum of all three channels for different momentum frames and cubic-group irreps, using two-particle operators with the form of two pions or two vector mesons, complemented by local tetraquark operators. For the $SS$ and $AA$ channels, we have used the quantization condition to constrain the pion-pion scattering amplitude, and characterized the $\Nc$ scaling of the scattering length and effective range, finding that they are compatible with the expected large $\Nc$ limit only if $\cO(\Nc^{-2})$ corrections are allowed. We have also observed evidence of a virtual bound state in the $AA$ channel for $\Nc=3$, at an energy $E_\text{bound}/\Mpi=1.741(13)$. For the $AS$ channel, on the other hand, we have found very weak interactions, which are consistent with ChPT, that predicts a zero scattering amplitude at LO.

The work presented in this chapter constitutes a further step in the study of subleading $\Nc$ effects in meson-meson scattering. Using the results presented in this chapter, we expect to constrain the $\Nc$ scaling of different combinations of LECs from ChPT, complementing the results presented in \cref{sec:largeNpions}. Also, further investigation is needed regarding the found virtual bound state in the $AA$ channel. In particular, its existence needs to be well-established by using several parametrizations of the scattering amplitudes, and also analyzing possible effects from higher partial waves and mixing with vector-meson states. It would also be compelling to analyze the dependence of this state on the quark mass, which could shed some light into the tetraquark resonances found experimentally. %Last but not least, our results for the finite-volume spectrum of two vector mesons opens the door to study their interactions, as well as couplings to states of two pions.

\chapter{The isospin-three three-pion $K$-matrix at NLO in ChPT}
\label{sec:pipipiKmatrix}

The study of multiparticle interactions on the lattice is an active field of research---see \rrcite{Briceno:2017max,Hansen:2019nir,Mai:2021lwb,Mai:2022eur} for recent reviews. Many two-particle systems have been analyzed in detail during the last decades, including a large number of two-meson systems, and also some meson-baryon and baryon-baryon examples. In the three particle case, the RFT finite-volume formalism, presented in \cref{sec:hadrons:threeparticlesfinitevolume}, allows to study many three-particle processes. Its practical use, however, has been mainly focused on three-meson systems at maximal isospin~\cite{Mai:2018djl,Horz:2019rrn,Blanton:2019vdk,Mai:2019fba,Culver:2019vvu,Fischer:2020jzp,Hansen:2020otl,Alexandru:2020xqf,Brett:2021wyd,Blanton:2021llb,Draper:2023boj}. %The most popular one is the system of three pions in the isospin-three channel, $\pi^+\pi^+\pi^+$, with states of three-kaons and kaons and pions have also been explored.

One of the most studied system is that of three pions at maximal isospin, $\pi^+\pi^+\pi^+$. A priori, one would expect ChPT to provide an accurate description of the system near threshold. At least, this is the case in the two-particle world, in which the two-pion scattering lengths are in very good agreement with LO ChPT~\cite{ETM:2015bzg,Mai:2019pqr,Fu:2017apw}. In the case of three pions, however, it was observed in \rrcite{Blanton:2019vdk,Blanton:2021llb} that lattice results for the three-pion divergence-free $K$-matrix, $\Kdf$, show a significant devitation from LO ChPT predictions, presented in \cref{sec:hadrons:ChPTthreepionsLO}. This discrepancy could be due to systematic effects neglected in the lattice determinations of $\Kdf$, or arise because of large higher-order ChPT corrections.

In this chapter, the results from \rcite{Baeza-Ballesteros:2023ljl} are presented, in which NLO ChPT contributions to the $I_{\pi\pi\pi}=3$ $K$-matrix were determined, %, which represents one of the works developed as part of my doctoral research. 
%This work focused on the second of the aforementioned process: the determination of the $3\pi^+$ $K$-matrix at NLO in ChPT. 
 starting from the three-pion NLO amplitudes from \rrcite{Bijnens:2021hpq,Bijnens:2022zsq}. % in a number of EFTs, including $\Nf=2$ ChPT, relevant to describe three-pion systems. 
As in the case of LO, the integral equations relating $\cM_3$ to $\Kdf$ reduce to an algebraic relation when working at NLO. This relation is presented in \cref{sec:pipipiKmatrix:algebraicrelation} and is later  used to determine the isospin-three $\Kdf$ up to quadratic order in the threshold expansion: the details of the computation are detailed in \cref{sec:pipipiKmatrix:computation} and the results are summarized in  \cref{sec:pipipiKmatrix:summary}. The conclusions of this work are presented in \cref{sec:pipipiKmatrix:conclusions}.

\newpage\section{Relating $\cM_3$ to $\Kdf$ at NLO in ChPT}\label{sec:pipipiKmatrix:algebraicrelation}

The three-particle divergence-free $K$-matrix is related to the three-particle scattering amplitude, $\cM_3$, by integral equations, introduced in \cref{sec:hadrons:infinitevolumethreeparticlescattering}. When working at fixed order in ChPT, these equations reduce to linear algebraic relations, as presented in  \cref{sec:hadrons:ChPTthreepionsLO} at LO. We now extend this relation to NLO. We note that by NLO we refer only to the next-to-leading order contribution, rather than to a complete LO+NLO result.

We recall that three-particle states can be kinematically described in two ways. In \rrcite{Bijnens:2021hpq,Bijnens:2022zsq}, scattering amplitudes are presented as a function of the momenta of all three initial- and final-state particles. We refer to these as $\{k_1,k_2,k_3\}$ and $\{p_1,p_2,p_3\}$, respectively, with $P=k_1+k_2+k_3=p_1+p_2+p_3$ the total momentum.\footnote{In reality, \rrcite{Bijnens:2021hpq,Bijnens:2022zsq} consider all six momenta incoming, but these two options can be trivially related by changing the sign of the final-state momenta.}  

On the other hand, the integral equations defining $\Kdf$ use the typical parametrization of the RFT formalism, which we now summarize. In this, a three-particle state is separated in a spectator and a two-particle dimer, and is described by the three-momentum of the spectator and the direction of the relative momentum of the dimer in its CMF. These quantities are called $\bm{k}$ and $\hat{\bm{a}_k}^*$ for the initial state, and $\bm{p}$ and $\hat{\bm{a}_p}^{'*}$ for the initial one, respectively. Typically, observables are projected to partial waves of the initial and final dimer. This means that, in general, scattering quantities are a function of the three-momenta of the initial and final spectators, $\bm{k}$ and $\bm{p}$, the angular-momentum indices of both states, $(\ell,m)$ and $(\ell',m')$, and the total energy in the center-of-mass frame of the three particles, $E^*$. In general, we leave the angular momentum indices, as well as the dependence on $E^*$, implicit. For example, the three-particle amplitude projected to the partial waves of both initial and final dimer is indicated as $\cM_3(\bm{p},\bm{k})$.

We stress that this same choice of arguments is maintained for two-particle quantities. For example, the two-particle scattering amplitude of the initial dimer, projected to the corresponding partial wave of this dimer, is indicated as $\cM_2(\bm{k})$. We note that this quantity depends on the Mandelstam variables of the initial dimer, $\cM_2(\bm{k})=\cM_{2,\ell}(s)$, with $s=(P-k)^2$ and $k$ the on-shell momentum of the initial spectator. 

Note the relation between both parametrizations of the kinematic variables of three-particle states is not fixed. Instead, all possible assignments of external momenta to the spectator and dimer usually need to be considered and combined in the determination of $\Kdf$. In the case of the $I_{\pi\pi\pi}=3$ channel, one needs to symmetrize aver all these possible combinations.
%, three particle states are describedIn the case of non-symmetrized quantities, $i=3$ refers to the quantity associated to the spectator. We can understand the power counting in the elements entering the integral equations as a series in $1/\Fpi^2$. 

%Before analyzing the 
We now derive the relation between $\cM_3$ and $\Kdf$ for three pions at maximal isospin at NLO in ChPT, the corresponding relation at LO was presented in \cref{sec:hadrons:ChPTthreepionsLO}. Recall this relation is obtained by studying the integral equations presented in \cref{sec:hadrons:infinitevolumethreeparticlescattering} as a power series in $1/\Fpi^2$. The scattering amplitudes at LO scale as $\cM_2^{\LO}=\cO(1/\Fpi^2)$ and $\cM_3^{\LO}=\cO(1/\Fpi^4)$, while at NLO they are $\cM_2^{\NLO}=\cO(1/\Fpi^4)$ and $\cM_3^{\NLO}=\cO(1/\Fpi^6)$. %$\Kdf$ will thus also be $\Kdf=\cO(1/\Fpi^6)$ and so we only need to keep terms up to this order in the integral equations. 

We first focus on the determination of the divergence-free amplitude at NLO, $\Mdf^\NLO$. Recall that at LO the unsymmetrized subtraction term in \cref{eq:hadrons:generalsubtraction} takes the form
\begin{equation}\label{eq:pipipiKmatrix:subtractiontermLO}
\cD^{\uu\LO}(\bm{p},\bm{k})=-\cM_2^\LO(\bm{p})G^\infty(\bm{p},\bm{k})\cM_2^\LO(\bm{k})\,,
\end{equation}
where $G^\infty$ is defined in \cref{eq:hadrons:Ginftydefinition}. At NLO, the subtraction term becomes,
\begin{equation}
\begin{array}{rl}\label{eq:pipipiKmatrix:subtractionNLOKdf}
    \cD^{\uu\NLO}(\bm{p}, \bm{k}) 
        = & \displaystyle- \cM_2^\LO(\bm{p}) G^\infty(\bm{p}, \bm{k}) \cM_2^\NLO(\bm{k}) \\
        & \displaystyle- \cM_2^\NLO(\bm{p}) G^\infty(\bm{p}, \bm{k}) \cM_2^\LO(\bm{k}) \\
        & \displaystyle + \int_r \cM_2^\LO(\bm p) G^\infty(\bm p, \bm r) \cM_2^\LO(\bm r) G^\infty(\bm r, \bm k) \cM_2^\LO(\bm k)\,,
\end{array}
\end{equation}
where we recall $\int_r=\int \text{d}^3r/[2\omega_r(2\pi)^3]$, with $\omega_r=\sqrt{\Mpi^2+\bm{r}^2}$.
 %Also here $\cM_2(\bm{k})$ and $\rho(\bm{k})$ are   the partial-wave-projected two-particle amplitude and the phase space factor, defined in \cref{eq:hadrons:phasespace}, of the initial dimer, which is associated to a  spectator of three-momentum $\bm{k}$. We stress that, while the different quantities associated to a two-particle, such as $\cM_2$ and the phase space factor depend on energyof these two particles, it is common in this context to indicate they depend on the three-momentum of the associated spectator. For example, $\cM_2(\bm{k})=\cM_2(s)$, where $s=(P-k)^2$ is the Mandelstam variable of the initial dimer. The possible dependence on the angular variables of the dimer, as well as the dependence on the total momentum, is left implicit.

In the case of the $I_{\pi\pi\pi}=3$ channel, we can substitute $G^\infty\rightarrow G^\infty_{ss}$, given in \cref{eq:hadrons:Ginftyssdefinition}, in the last line on \cref{eq:pipipiKmatrix:subtractionNLOKdf}, as $\cM_2^\LO$ is purely $s$-wave, see \cref{eq:hadrons:scatteringamplitudetwopionsI2}. On the other hand, $\cM_2^\NLO$ contains all even partial waves, and such substitution is not possible on the other terms. The divergence-free amplitude is defined after subtracting,
\begin{equation}\label{eq:pipipiKmatrix:MdfasfunctionofM}
\Mdf^\NLO(\bm{p},\bm{k})=\cM_3^\NLO(\bm{p},\bm{k})-\cS\left\{\cD^{\uu,\NLO}(\bm{p},\bm{k})\right\}\,,
\end{equation}
where $\cS$ denotes symmetrization over the initial and final momenta. 

The divergence-free amplitude is related to $\Kdf$ by an integral equation, that we reproduce here,
\begin{equation}\label{eq:pipipiKmatrix:TandMdfintegralequation}
    \Mdf(\bm p, \bm  k) 
        = \cS\left\{\int_s \int_r 
            \cL^\uu(\bm p, \bm s) \cT(\bm s, \bm r) 
            \cR^\uu(\bm r, \bm k) \right\}\,,
\end{equation}
where we recall that $\cS$ indicates that the quantity between brackets needs to be symmetrized over all permutations of momenta in the initial and final state, and $\cL$ and $\cR$ are decorators defined in \cref{eq:hadrons:Ldef,eq:hadrons:Rdef}. At LO and NLO in ChPT, they take the form
\begin{equation}\label{sec:pipipiKmatrix:decoratorsLONLO}
\begin{array}{rl}
\cL^{\uu\LO}(\bm p, \bm k)  = \cR^{\uu\LO}(\bm p, \bm k) & = \displaystyle\frac13 \bar\delta(\bm p - \bm k)\,,\\
\cL^{\uu,\NLO}(\bm{p},\bm{k})=\cR^{\uu,\NLO}(\bm{p},\bm{k}) & = \displaystyle i\cM_2^\LO(\bm{k})\rho(\bm{k})\overline{\delta}(\bm{p}-\bm{k})\,,
\end{array}
\end{equation}
where recall $\bar\delta(\bm p - \bm k) \equiv 2\omega_k (2\pi)^3 \delta^{(3)}(\bm p - \bm k)$. Combining the results in \cref{sec:pipipiKmatrix:decoratorsLONLO} into \cref{eq:pipipiKmatrix:TandMdfintegralequation} allows us to rewrite the latter as an algebraic relation,
\begin{multline}\label{eq:pipipiKmatrix:TandMdfintegralequationNLO}
\Mdf(\bm p, \bm  k) 
        = \cS\left\{
            \frac{1}{3} \cT^\LO(\bm p, \bm k) 
            \rho(\bm{k})\cM_2^\LO(\bm{k})\right. \\+ \left.\frac{1}{3} \cM_2^\LO(\bm{p}) \rho(\bm{p}) \cT^\LO(\bm p, \bm k) 
             +\frac{1}{9} \cT^\NLO(\bm p, \bm k)  \right\}\,.
\end{multline}
From here, it is clear that $\cT^\NLO=\cO(1/\Fpi^6)$.

The $\cT$ variable is related to $\Kdf$ via the integral equation
\begin{equation}\label{eq:pipipiKmatrix:integralequationKdf}
    \cT(\bm p, \bm k) 
        = \Kdf(\bm p, \bm k)
        - \int_s \int_r \Kdf(\bm p, \bm s) \rho(\bm s) \cL^\uu(\bm s, \bm r) \cT(\bm r, \bm k)\,.
\end{equation}
 Recalling $\cT^\LO=\Kdf^\LO=\cO(1/\Fpi^4)$, one finds that the last term on the right hand side of \cref{eq:pipipiKmatrix:integralequationKdf} is $\cO(1/\Fpi^8)$, and so can be neglected at the order we are working, meaning $\cT^\NLO=\Kdf^\NLO$, in analogy to LO. 

%We now focus on \cref{eq:hadrons:integralequationKdf}. The second term  from which we find that $\cT^\NLO=\cO(1/\Fpi^6)$, and so the second term in \cref{eq:hadrons:integralequationKdf} starts at $\cO(1/\Fpi^8)$, which we neglect. Thus, we have $\cT^\NLO=\Kdf^\NLO$, analogously to LO. 

%Finally, we focus on \cref{eq:hadrons:TandMdfintegralequation}. Since $\cT^\LO=\cO(1/\Fpi^4)$, the NLO contribution from \cref{eq:hadrons:TandMdfintegralequation} will be formed by three terms. One in which $\cT$ takes its NLO value, and the two decorators $\cR$ and $\cL$ are LO, see \cref{sec:hadrons:decoratorsLO}, and two other terms in which $\cT$ is LO and one of the decorators is NLO. Looking at \cref{eq:hadrons:Ldef,eq:hadrons:Rdef}, we observe that the NLO contribution to these decorators comes from the term proportional to $\cM_2$, which is purely imaginary, 
%\begin{equation}
%\cR^{\uu,\NLO}(\bm{p},\bm{k})=\cL^{\uu,\NLO}(\bm{p},\bm{k})=i\cM_2^\LO(\bm{k})\rho(\bm{k})\overline{\delta}(\bm{p}-\bm{k})\,.
%\end{equation}
%where $\bm{k}$ and $\bm{p}$ are the momenta of the initial and final spectators, respectively.
 %Also recall %label them with the momentum of the 
 %The dependence on the total momentum of the system is left implicit.

Finally, combining all these results into \cref{eq:pipipiKmatrix:TandMdfintegralequationNLO}, we find
\begin{multline}
    \Mdf^\NLO(\bm p, \bm k) 
         = \Kdf^\NLO(\bm p, \bm k)\\
        \displaystyle + \frac{i}{3} \cS\Big\{
            \Kdf^\LO(\bm p, \bm k) \rho(\bm k) \cM_2^\LO(\bm k) 
            +  \cM_2^\LO(\bm p) \rho(\bm p) \Kdf^\LO(\bm p, \bm k)
            \Big\}\,,
\end{multline}
where we have used that $\Kdf^\NLO$ is symmetric by definition, and so the symmetrization generates a factor of nine. If we further use the LO result for the three-particle $K$-matrix, given in \cref{eq:hadrons:KdfMdfrealtionLO}, $\Kdf^\LO=\Mdf^\LO$, this equation can be rewritten as
\begin{multline}\label{eq:pipipiKmatrix:Kdfrelationwithimeginaryterm}
    \Kdf^\NLO(\bm p, \bm k) 
         = \Mdf^\NLO(\bm p, \bm k)\\  
         \displaystyle - \frac{i}{3} \cS\Big\{
            \Mdf^\LO(\bm p, \bm k) \rho(\bm k) \cM_2^\LO(\bm k) 
            +  \cM_2^\LO(\bm p) \rho(\bm p) \Mdf^\LO(\bm p, \bm k) 
            \Big\}\,.
\end{multline}

The terms in brackets on the right hand side of \cref{eq:pipipiKmatrix:Kdfrelationwithimeginaryterm} are purely imaginary for physical kinematics, as $\cM_2^\LO$, $\cM_3^\LO$ and $\rho$ are all real. In contrast, $\Kdf$ is purely real. Therefore, one can simplify this relation
\begin{equation}\label{eq:pipipiKmatrix:KdfasarealpartofMdf}
\Kdf^\NLO=\text{Re}\Mdf^\NLO\,.
\end{equation}
The cancellation of the imaginary parts in \cref{eq:pipipiKmatrix:Kdfrelationwithimeginaryterm} can then be used as a cross-check of the formalism. Such analysis is presented in app.~E of \rcite{Baeza-Ballesteros:2023ljl}.

\newpage\section{Computation of $\Kdf$ at NLO}\label{sec:pipipiKmatrix:computation}

\begin{figure}[!b]
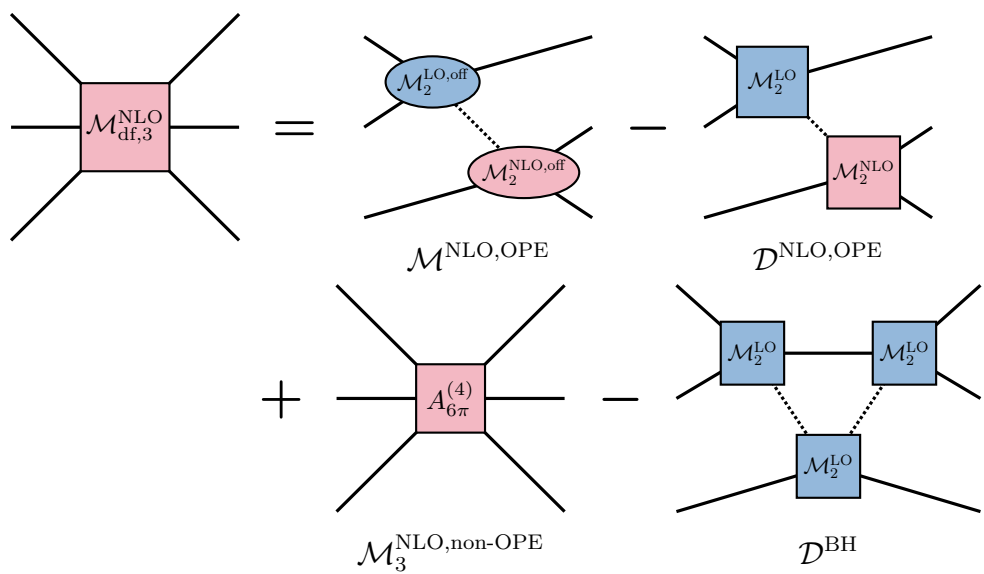


    \begin{multline*}
            \tikzineq[xscale=1.5,yscale=1.5]{
                \makeexternallegcoordinates
                \coordinate (v) at (0,0);
                \draw[sketch onshell prop] (k1) -- (v);
                \draw[sketch onshell prop] (k2) -- (v);
                \draw[sketch onshell prop] (k3) -- (v);
                \draw[sketch onshell prop] (v) -- (p1);
                \draw[sketch onshell prop] (v) -- (p2);
                \draw[sketch onshell prop] (v) -- (p3);
                \node[sketch onshell blob=NLOcolor, inner sep=-3pt]
                    (M) at (0,0) {\footnotesize $\Mdf^\NLO$};
            }
        \quad\scalebox{\sketchoperatorscale}{$\bm=$}\;\,
        \quad
        \underset{\rule{0pt}{1.5em}\displaystyle\cM^{\NLO,\OPE}}{%
            \tikzineq[xscale=1.5,yscale=1.2]{
                \makeexternallegcoordinates
                \coordinate (v1) at (+.4,-.5);
                \coordinate (v2) at (-.4,+.5);
                \draw[sketch onshell prop] (k1) -- (v2);
                \draw[sketch onshell prop] (k2) -- (v1) -- (k3);
                \draw[sketch onshell prop] (p3) -- (v1);
                \draw[sketch onshell prop] (p1) -- (v2) -- (p2);
                \draw[sketch offshell prop] (v1) -- (v2);
                \node[sketch offshell blob=NLOcolor, inner ysep=2pt, inner xsep=-1pt]
                    (NLO) at (v1) {\scalebox{.7}{$\cM_2^{\NLO,\off}$}};
                \node[sketch offshell blob=LOcolor, inner ysep=2pt, inner xsep=-1pt]
                    (LO) at (v2) {\scalebox{.7}{$\cM_2^{\LO,\off}$}};
            }
        }
        \quad\scalebox{\sketchoperatorscale}{$\bm-$}\quad
        \underset{\rule{0pt}{1.5em}\displaystyle\cD^{\NLO,\OPE}}{%
            \tikzineq[xscale=1.5,yscale=1.2]{
                \makeexternallegcoordinates
                \coordinate (v1) at (+.4,-.5);
                \coordinate (v2) at (-.4,+.5);
                \draw[sketch onshell prop] (k1) -- (v2);
                \draw[sketch onshell prop] (k2) -- (v1) -- (k3);
                \draw[sketch onshell prop] (p3) -- (v1);
                \draw[sketch onshell prop] (p1) -- (v2) -- (p2);
                \draw[sketch offshell prop] (v1) -- (v2);
                \node[sketch onshell blob=LOcolor, inner sep=0pt]
                    (LO) at (v2) {\scalebox{.7}{$\cM_2^{\LO}$}};
                \node[sketch onshell blob=NLOcolor, inner sep=-2pt]
                    (NLO) at (v1) {\scalebox{.7}{$\cM_2^{\NLO}$}};
            }
        }\\
        \quad\scalebox{\sketchoperatorscale}{$\bm+$}\quad
        \underset{\rule{0pt}{1.5em}\displaystyle\cM_3^{\NLO,\nOPE}}{%
            \tikzineq[xscale=1.5,yscale=1.5]{
                \makeexternallegcoordinates
                \coordinate (v) at (0,0);
                \draw[sketch onshell prop] (k1) -- (v);
                \draw[sketch onshell prop] (k2) -- (v);
                \draw[sketch onshell prop] (k3) -- (v);
                \draw[sketch onshell prop] (v) -- (p1);
                \draw[sketch onshell prop] (v) -- (p2);
                \draw[sketch onshell prop] (v) -- (p3);
                \node[sketch onshell blob=NLOcolor, inner sep=0pt]
                    (M) at (0,0) {\footnotesize $A^{(4)}_{6\pi}$};
            }
        }
        \quad\scalebox{\sketchoperatorscale}{$\bm-$}\quad
        \underset{\rule{0pt}{1.5em}\displaystyle\cD^\BH}{%
            \tikzineq[xscale=2,yscale=1.5]{
                \makeexternallegcoordinates
                \coordinate (v1) at (+.5,+.4);
                \coordinate (v2) at (-.5,+.4);
                \coordinate (v3) at (  0,-.6);
                \draw[sketch onshell prop] (k1) -- (v1);
                \draw[sketch onshell prop] (k2) -- (v1);
                \draw[sketch onshell prop] (k3) -- (v3);
                \draw[sketch onshell prop] (v2) -- (p1);
                \draw[sketch onshell prop] (v2) -- (p2);
                \draw[sketch onshell prop] (v3) -- (p3);
                \draw[sketch onshell prop] (v1) -- (v2);
                \draw[sketch offshell prop] (v1) -- (v3);
                \draw[sketch offshell prop] (v2) -- (v3);
                \foreach \i in {1,2,3}
                    \node[sketch onshell blob=LOcolor, inner sep=-1pt]
                        (M\i) at (v\i) {\scalebox{.7}{$\cM_2^\LO$}};
            }
        }
    \end{multline*}
    \caption{Schematic representation of \cref{eq:pipipiKmatrix:schematicequation}. Solid and dotted lines represent on- and off-shell pions, respectivey. Similarly, amplitudes in square boxes are fully on-shell, while those in ovals have one leg off-shell. Blue and pink colors correspond to LO and NLO, in this same order. Finally, $A^{(4)}_{6\pi}$ is the non-OPE part of the NLO amplitude, given in \rcite{Bijnens:2021hpq}.
        }
    \label{fig:pipipiKmatrix:diagramKdf}
\end{figure}

As we have shown, determining $\Kdf$ at NLO in ChPT just requires to determine the real part of the divergence-free NLO  amplitude. In the $\Ippp=3$ channel, the computation can be divided in three parts, each of which with different particularities. We write,
\begin{multline}\label{eq:pipipiKmatrix:schematicequation}
\Kdf^\NLO=\text{Re}\Mdf^\NLO =\\ \text{Re}\left\{\cM_3^{\NLO,\OPE}-\cD^{\NLO,\OPE}\right\}+\text{Re}\cM_3^{\NLO,\nOPE}-\text{Re}\cD^{\BH}\,,
\end{multline}
which is schematically represented in \cref{fig:pipipiKmatrix:diagramKdf}. 
The first term on the right-hand side is called the one-particle exchange (OPE) part of the amplitude, $\cM_3^{\NLO,\OPE}$, together with the corresponding subtraction, $\cD^{\NLO,\OPE}$, given by the first two lines of \cref{eq:pipipiKmatrix:subtractionNLOKdf}. The remaining of the amplitude is the non-OPE part, $\cM_3^{\NLO,\nOPE}$. It also requires a subtraction term, called the bull-head (BH) subtraction, $\cD^{\BH}$, corresponding to the last line in \cref{eq:pipipiKmatrix:subtractionNLOKdf}. Note that the real part of the non-OPE amplitude is smooth, and so is the real part of the BH subtraction. Thus both can be evaluated independently, which is not true for the imaginary parts. Finally, recall that this division into different pieces is unphysical, as it depends on the off-shell convention---see \cref{sec:hadrons:ChPTthreepionsLO} for a detailed discussion.

The computation of these three different terms will be presented in the remainder of this section. All of them can be determined almost analytically, the only numerical contribution comes from the BH subtraction, where the cutoff function, $H(x)$, plays a role.  We first present in \cref{sec:pipipiKmatrix:preliminaries} some preliminary results which are used throughout the computation. The details for the OPE, non-OPE and BH subtraction are then presented in \cref{sec:pipipiKmatrix:OPE,sec:pipipiKmatrix:nOPE,sec:pipipiKmatrix:BH}, in this same order. Finally, in \cref{sec:pipipiKmatrix:crosscheck} we comment on some numerical approaches used to cross-check the analytical results.

\subsection{Preliminaries to the computation}
\label{sec:pipipiKmatrix:preliminaries}

The main goal of this work is to determine the threshold expansion of $\Kdf$ up to quadratic order. The standard form of this expansion was presented in \cref{eq:hadrons:thesholdexpansionKdf},
\begin{equation}
\Mpi^2\Kdf=\Kiso+\Kisoone\Delta+\Kisotwo\Delta^2+\KA\DA+\KB\DB+\cO(\Delta^3)\,,
\label{eq:pipipiKmatrix:Kexpansion}
\end{equation}
where $\Delta$, $\DA$ and $\DB$ are defined in \cref{eq:hadrons:thesholdexpansionKdfingredients}. However, $\Kdf$ can also be expanded about threshold in other ways, which are useful throughout the computation presented in this chapter.

One alternative is to use kinematical variables that can expressed in terms of $\tilde{t}_{ij}$ variables, defined in \cref{eq:hadrons:tvariablesDeltavariables}. For example, we can write
\begin{equation}
\Mpi^2\Kdf = c_0+c_1 \cQ_0+ c_2 \cQ_1 + c_3 \cQ_2 + c_4 \cQ_3 + \cO(\Delta^3)\,,
\label{eq:pipipiKmatrix:alternativeexpansion}
\end{equation}
with kinematic operators,
\begin{equation}
\begin{array}{rll}
        \cQ_0&=\cS[\tij{11}]
            &= -2\Delta\,,\\
        \cQ_1&\equiv\cS[\tij{11}\tij{11}]
            &= \Delta^2+\DB\,,\\
        \cQ_2&\equiv\cS[\tij{11}\tij{12}+\tij{11}\tij{21}]
            &= \tfrac{1}{2}(2\Delta^2+\DA-2\DB)\,,\\
        \cQ_3&\equiv\cS[\tij{11}\tij{22}+\tij{21}\tij{12}]
            &=\tfrac{1}{2}(2\Delta^2-\DA+\DB)\,.
\end{array}
\end{equation}
where we have indicated the relations to those operators in \cref{eq:pipipiKmatrix:Kexpansion}. From here one can find a relation between the coefficients in both expressions. For example, $\Kisoone=-2c_1$.

Another option is to consider an unsymmetrized $K$-matrix, $\Kdf^\uu$, from which $\Kdf$ can be recovered after symmetrization, this is,
\begin{equation}
\Kdf=\cS\left\{\Kdf^\uu\right\}
\end{equation}
Under particle exchange $\Kdf^\uu$ has the same symmetries as a three-particle system with one particle different from the other two. The corresponding threshold expansion, which was first worked out in \rcite{Blanton:2021mih}, takes the following form
\begin{multline}
    \Mpi^2\Kdf^\uu 
        = c_0' + c_1' \Delta + c_2' \Delta_3^\text{S} + c_3' \tij{33} 
    \\
        + c_4' \Delta^2 + c_5' \Delta \Delta_3^\text{S} + c_6' \Delta \tij{33} 
        + c_7' \Delta_3 \Delta'_3 + c_8' (\Delta_3^\text{S})^2 + c_9' \Delta_3^\text{S} \tij{33}+ c_{10}' \tij{33}^{\,2}
    \\
        + c_{11}' \cQ_{--} + c_{12}' \cQ_{+-} + c_{13}' \cQ_{3-} + c_{14}' \cQ_{tu} + \cO(\Delta^3)\,,
        \label{eq:pipipiKmatrix:uuexpansion}
\end{multline}
where $\Delta_3$ and $\Delta_3'$ are defined in \cref{eq:hadrons:tvariablesDeltavariables}, and we have introduced $\Delta_3^\text{S}=\Delta_3+\Delta_3^\prime$. Here, $\cQ_X$ are operators that contribute to non-zero partial waves~\cite{Baeza-Ballesteros:2023ljl}. The only one of interest for this work is
\begin{equation}
\cQ_{tu} = \tij{13}\tij{23}+\tij{31}\tij{32}\,.
\end{equation}
 One can symmetrize the initial and final momenta to recover $\Kdf$ and relate the coefficients in \cref{eq:pipipiKmatrix:uuexpansion} to those in \cref{eq:pipipiKmatrix:Kexpansion}. For example, we find
\begin{equation}\label{eq:pipipiKmatrix:relationnormalunsymmetrizedexpansion}
        \KB = c_{10}' 
        + 9 c_{11}' + 3 c_{12}' - 6 c_{13}' - c_{14}'\,.
\end{equation}

\subsection{The subtracted OPE contribution}\label{sec:pipipiKmatrix:OPE}

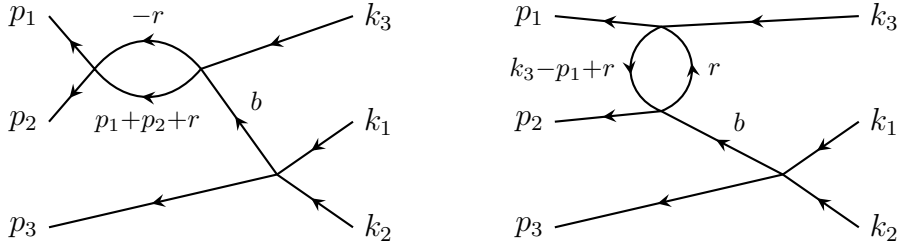
\begin{figure}[b!]
    \centering
    \begin{subfigure}{0.48\textwidth}
    \centering
        \begin{tikzpicture}[xscale=\diagramxscale,yscale=\diagramyscale]
            \makeexternallegsshifted
            \coordinate (v1) at (+.5,-.5);
            \coordinate (v2) at (  0,+.5);
            \coordinate (v3) at (-.7,+.5);
            \draw[dprop] (k1) -- (v2) to[bend right=60, looseness=1.5] (v3) -- (p1);
            \draw[dprop] (k2) -- (v1);
            \draw[dprop] (v1) -- (v2) node [midway, above right] {\footnotesize $b$};
            \draw[dprop] (v2) to[bend left=60, looseness=1.5] (v3) -- (p2);
            \draw[dprop] (k3) -- (v1) -- (p3);
            \draw (-.35,1) node {\footnotesize $-r$};
            \draw (-.35,0) node {\footnotesize $p_1{+}p_2{+}r$};
        \end{tikzpicture}
    \end{subfigure}
    \begin{subfigure}{0.48\textwidth}
    \centering
        \begin{tikzpicture}[xscale=\diagramxscale,yscale=\diagramyscale]
            \makeexternallegsshifted
            \coordinate (v1) at (+.5,-.5);
            \coordinate (v2) at (-.3,+.1);
            \coordinate (v3) at (-.3,+.9);
            \draw[dprop] (k1) -- (v3) -- (p1);
            \draw[dprop] (k2) -- (v1);
            \draw[dprop] (v1) -- (v2) node [midway, above right] {\footnotesize $b$};
            \draw[dprop] (v2) to[bend right=60] (v3) to[bend right=60] (v2) -- (p2);
            \draw[dprop] (k3) -- (v1) -- (p3);
            \draw (+.05,.5) node {\footnotesize $r$};
            \draw (-.95,.5) node {\footnotesize $k_3{-}p_1{+}r$};
        \end{tikzpicture}
    \end{subfigure}
    \caption{
        Examples of OPE NLO diagrams. In this case, $b=P-k_3-p_3$.}
    \label{fig:pipipiKmatrix:diagramsOPENLO}
\end{figure}

We first focus on the subtracted OPE contribution. This comes from diagrams such as those shown in \cref{fig:pipipiKmatrix:diagramsOPENLO}, together with the subtraction of divergencies that arise when the exchanged particle, of momentum $b=P-p_3-k_3$, goes on-shell. This contribution can be determined analytically only after this subtraction is included. 

To determine the OPE contribution to $\Kdf$ at NLO, we first determine the OPE contribution to the unsymmetrized $K$-matrix, $\Kdf^{\uu,\OPE}$. We consider the case in which the initial and final spectator have momentum $k_3$ and $p_3$, respectively, and later symmetrize the result. The unsymmetrized divergence-free amplitude takes the form
\begin{equation}
    \Mdf^{\uu\NLO,\OPE}(\bm p_3, \bm k_3)
        = \cM_3^{\uu\NLO,\OPE}(\bm p_3, \bm k_3)
        - \cD^{\uu\NLO,\OPE}(\bm p_3, \bm k_3)\,,\label{eq:pipipiKmatrix:MuuDuuNLOOPE}
\end{equation}
where the unsymmetrized amplitude and subtraction term are
\begin{equation}
    \cM_3^{\uu\NLO,\OPE}(\bm p_3, \bm k_3)
    =  -\cM_{2,\off}^\NLO(\bm p_3) \frac1{\bar{b}^2+i\epsilon} \cM_{2,\off}^\LO(\bm k_3) + \leftrightarrow\,,
    \label{eq:pipipiKmatrix:MuuNLOOPE}
\end{equation}
\begin{equation}
    \cD^{\uu\NLO,\OPE}(\bm p_3, \bm k_3)
        = -\cM_{2}^\NLO(\bm p_3) G^\infty(\bm p_3, \bm k_3) \cM_{2}^\LO(\bm k_3) 
        + \leftrightarrow \,.
    \label{eq:pipipiKmatrix:DuuNLOOPE}
\end{equation}
Here $\leftrightarrow$ denotes exchange between the LO and NLO amplitudes and we define $\bar b^2 \equiv b^2-\Mpi^2$. 

For the subsequent analysis, it is convenient to express the two-particle amplitudes, before projecting to partial waves, as a function of Mandelstam variables of the interacting pair. For example, we use $\cM_2(\bar{s},t,u)$ to refer to the scattering amplitude of the initial dimer, where we define
\begin{equation}
    \begin{gathered}
        \bar s = (k_1+k_2)^2-4\Mpi^2\,,
        \qquad t=(k_1-p_3)^2\,, 
        \qquad u=(k_2-p_3)^2\,,
    \end{gathered}
\end{equation}
all of which vanish at threshold. An analogous definitions hold for the final state, distinguished by primed variables. We note that, when off-shell, these variables obey the simple relation
\begin{equation}
\bar s +t +u = \bar s'+t'+u'=\bar b^2\,,
\end{equation}
with $\bar b^2=0$ on shell. We also note that we can set $\varepsilon=0$ in \cref{eq:pipipiKmatrix:MuuNLOOPE} and in the $G^\infty$ factor in \cref{eq:pipipiKmatrix:DuuNLOOPE}, defined in \cref{eq:hadrons:Ginftydefinition}, as the divergence that the $i\varepsilon$ term regulates is canceled by the subtraction. Also, as the LO amplitude is real and we are only interested in the real part of $\Mdf$, we can just work with the real part of the NLO result. Finally note that the cutoff factors in $G^\infty$ can be set to unity, as external particles are always on-shell.

The starting point to evaluate \cref{eq:pipipiKmatrix:MuuDuuNLOOPE,eq:pipipiKmatrix:MuuNLOOPE,eq:pipipiKmatrix:DuuNLOOPE} are the LO and NLO off-shell two-pion isospin-two amplitudes. These take the general form,
\begin{equation}
\Fpi^4 \cM_{2,\off}(\bar s,t,u) = A_{4\pi}(t,u,s)+A_{4\pi}(u,s,t)\,,
\end{equation}
where $A_{4\pi}$ is the four-pion amplitude~\cite{Bijnens:2021hpq}. At LO, it takes the form,
\begin{equation}\label{eq:pipipiKmatrix:twopionamplitudeLO}
\Fpi^2 A^{(2)}_{4\pi}(s,t,u)=s-\Mpi^2\,,
\end{equation}
while at NLO,
\begin{multline}\label{eq:pipipiKmatrix:twopionamplitudeNLO}
\Fpi^4 A_{4\pi}^{(4)}(s,t,u)  = 
(t-u)^2\left(-\frac{5}{36}\kappa -\frac{1}{6}L+\frac{1}{2}\ell_2^\text{r}\right) \\[5pt]
\begin{array}{l}
\displaystyle+ \Mpi^2 s\left(-\frac{2}{9}\kappa -\frac{2}{9}L-8\ell_1^\text{r}+2\ell_4^\text{r}\right) 
+ s^2 \left(-\frac{7}{12}\kappa -\frac{1}{2}L+2\ell_1^\text{r}+\frac{1}{2}\ell_2^\text{r}\right)\\[5pt]
\displaystyle+\Mpi^4 \left(\frac{13}{20}\kappa +\frac{7}{6}L+8\ell_1^\text{r}+2\ell_3^\text{r}-2\ell_4^\text{r}\right)
+\bar{J}(s) \left(\frac{1}{2}s^2 -\frac{1}{2}\Mpi^4\right)\\[5pt]
\displaystyle+\left[\frac{1}{6}\bar{J}(t)(2t^2-10\Mpi^2 t-4\Mpi^2s+st+14\Mpi^4)+(t\leftrightarrow u)\right]\,.
\end{array}
\end{multline}
Here $\kappa=1/16\pi^2$, $L=\kappa\log(\Mpi^2/\mu^2)$, with $\mu$ the renormalization scale, $\ell_i^\text{r}$ are the $\Nf=2$ LECs introduced in \cref{sec:QCD:chiralperturbationtheory}, and $\bar{J}$ are loop integrals~\cite{Passarino:1978jh,Scherer:2002tk}. Note that the LO amplitude only contains $s$-wave, while the NLO one contains all even partial waves. However, when working at quadratic order in the threshold expansion, only $\ell=0,2$ need to be considered.

The first step is to expand the off-shell amplitudes about threshold, including terms up to cubic order. The LO amplitude is easily found to be
\begin{equation}
\Fpi^2\cM_{2,\off}^\LO(\bar{s},t,u)=-2\Mpi^2-\bar{s}+\bar{b}^2\,.
\end{equation}
The NLO amplitude, on the other hand, requires expanding the $\bar J$ functions. Neglecting imaginary parts, one finds
\begin{align}\label{eq:pipipiKmatrix:Jexpansion}
    \frac{1}{\kappa}\text{Re}\bar J(4\Mpi^2+\bar s) &=
    2 - \frac{1}{2} \frac{\bar s}{\Mpi^2}+ \frac1{12} \frac{\bar s^2}{\Mpi^4} 
    - \frac{1}{60} \frac{\bar s^3}{\Mpi^6}+ \cO(\bar s^4)\,,
    \\
    \frac{1}{\kappa}\bar J(t) &=
    \frac{1}{6} \frac{t}{\Mpi^2} + \frac{1}{60} \frac{t^2}{\Mpi^4} + \frac{1}{420} \frac{t^3}{\Mpi^6} + \cO(t^4)\,.
\end{align}
The off-shell amplitude  then takes the schematic form
\begin{multline}\label{eq:pipipiKmatrix:expansiontwopionamplitude}
        \Fpi^4 \Re\cM_{2,\off}^\NLO(\bar s, t, u) 
        \displaystyle= e_0 \Mpi^4 + e_1 \Mpi^2 \bar s + e_2 \bar s^2 + e_3 \Mpi^2 \bar b^2 + e_4 \bar s \bar b^2 \\ \displaystyle+ e_5 (\bar b^2)^2 + e_{tu}  tu
        + e_6 \frac{\bar s^2 \bar b^2}{\Mpi^2} + e_7 \frac{\bar s (\bar b^2)^2}{\Mpi^2} 
        + e_8 \frac{(\bar b^2)^3}{\Mpi^6} + \tilde{e}_{tu} \frac{(\bar b^2- \bar s)}{\Mpi^6} tu+\dots\,,
\end{multline}
where $e_i$ are constants depending on $\kappa$, $L$ and the LECs, and ... indicates higher order terms. 

Next, we separate the NLO amplitude into different partial waves, needed to compute the subtraction term. %This is not needed for the LO amplitude, which is purely $s$-wave. 
Most of the terms in \cref{eq:pipipiKmatrix:expansiontwopionamplitude} only contribute to $s$-wave, the only exception being those containing $tu$ factors, which also contribute to $d$-wave. This factor can be decomposed as
\begin{equation}\label{eq:pipipiKmatrix:dwaveharmonicsdecomposition}
tu=\frac{1}{4}(\bar s-\bar b^2)^2 - 4(\bm{a}_k^*\cdot \bm{p}_k^*)^2\,,
\end{equation}
where $\bm{a}_k^*$ and $\bm{p}_k^*$ are the three momenta $\bm{k}_1$ and $\bm{p}_3$ boosted to the CMF of the initial-state dimer. Using the addition theorem of spherical harmonics~\cite{Arfken:379118}, one can decompose
\begin{equation}
    (\bm{ a}^{\,*}_{k}\cdot \bm p_k^*)^2 = q_{2,k}^{*2}  p_k^{*2} \left[ 
        \frac{8\pi}{15} \sum_{m} Y^*_{2 m}(\hat{\bm a}^{\,*}_{k}) Y_{2 m}(\hat{\bm p}_k^*) 
        + \frac{1}{3} 
        \right]\,,
\end{equation}
where we have introduced the magnitudes of $\bm{a}_k^*$ and $\bm{p}_k^*$, which can be expanded about threshold,
\begin{equation}\label{eq:pipipiKmatrix:expansionkinematicquantities}
\begin{array}{rl}
    |\bm a^{\,*}_k|^2 &\displaystyle = q_{2,k}^{*2} = \frac{1}{4} \bar s\,,\\[10pt]
    p_k^{*2} &\displaystyle = \frac{(s -  \bar b^2)^2}{4 s }- \Mpi^2=\frac{1}{4} \bar s - \frac{1}{2} \bar b^2 
    +  \frac{1}{16} (\bar b^2)^2 
    + \dots\,.
    \end{array}
\end{equation}
%Note that the expansion of the denominator in the second therm, $s=4\Mpi^2+\bar s$, rigorously limits the validity range of the expansion. 
From here, one can decompose the off-shell $tu$ term  into $s$- and $d$-wave parts,
\begin{equation}\label{eq:pipipiKmatrix:tudecomposition}
\begin{array}{rl}
\left[tu\right]_s &\displaystyle = \frac{1}{4}(\bar s-\bar{b}^2)^2-\frac{4}{3}q_{2,k}^{*2}  p_k^{*2}\,,\\[5pt]
\left[tu\right]_d &\displaystyle =  q_{2,k}^{\,*2}  p_k^{*2} \frac{8\pi}{15}  \sum_{m} Y^*_{2 m}(\hat{\bm a}^{\,*}_{k}) Y_{2 m}(\hat{\bm  p}_k^*) \,.
\end{array}
\end{equation}
Analogous results hold for the final-state $t'u'$ term.

The contribution from the $s$-wave part of the NLO amplitude to $\Kdf^{\uu,\NLO}$ is now easy to evaluate. We can set $G^\infty$ to $G^\infty_{ss}=1/\bar{b}^2$ and separate the relevant LO and NLO off-shell amplitudes as an on-shell part and a term proportional to $\bar{b}^2$,
\begin{equation}
\cM_{2,\off}=\cM_2+\bar b^2\delta\cM_2\,.
\end{equation}
Then, one finds
\begin{multline}
\Fpi^6\Kdf^{\uu\NLO,\OPE,s}=-\cM_2^\NLO\delta\cM_2^\LO \\- \delta\cM_2^\NLO\cM_2^\LO -\delta\cM_2^\NLO\delta\cM_2^\LO + \leftrightarrow\,.
\end{multline}
Substituting the results in \cref{eq:pipipiKmatrix:expansiontwopionamplitude}, one can rewrite this in terms of the operators in \cref{eq:pipipiKmatrix:uuexpansion}, obtaining the unsymmetrized $K$-matrix.

The contribution from the $d$-wave part of the NLO amplitude requires the use of $G_{20}^\infty$ and $G_{02}^\infty$, which include a barrier factor. Note these correspond to the $G^\infty_{\ell'\ell}$ factor when one of the two dimers is $d$ wave---see \cref{{eq:hadrons:Ginftydefinition}}. For example, if the $d$-wave interaction happens for the initial dimer ($\ell=2$), the barrier factor takes the form $G_{02}^\infty\propto (p_k^{*2}/q_{2,k}^{\,*2})^2$.
From \cref{eq:pipipiKmatrix:expansiontwopionamplitude,eq:pipipiKmatrix:tudecomposition}, and recalling that $p_k^{*2} = q_{2,k}^{\,*2}$ on shell, one note,
\begin{equation}\label{eq:pipipiKmatrix:dwavebarrierfactoreffect}
%\cM_{2,\ell=2}^\NLO(s)\left(\frac{p_k^{*2}}{q_{2,k}^{\,*2}}\right)^2=\cM_{2,\ell=2,\off}^\NLO(s)\,.
[tu]_{d,\on}\left(\frac{p_k^{*2}}{q_{2,k}^{\,*2}}\right)^2=[tu]_{d,\off}\,.
\end{equation}
This means the barrier factor in $G^\infty$ converts the on-shell $tu$ factor to its off-shell counterpart. 

For the term with $e_{tu}$ coefficient in \cref{eq:pipipiKmatrix:expansiontwopionamplitude}, this implies that the subtraction picks up the difference between the off- and on-shell values of the LO amplitude,
\begin{equation}
\Fpi^6\Kdf^{\uu\NLO,\OPE,d}\supset- \delta\cM_2^\LO e_{tu}\left([tu]_{d,\off}+[t'u']_{d,\off}\right)\,.
\end{equation}
To convert to the basis in \cref{eq:pipipiKmatrix:uuexpansion}, one substitutes
\begin{equation}\label{eq:pipipiKmatrix:dstusubstitution}
\left[tu\right]_d=tu-\left[tu\right]_s\,.
\end{equation}
The first term directly contributes to $\cQ_{tu}$ in \cref{eq:pipipiKmatrix:uuexpansion}, while the second, given in \cref{eq:pipipiKmatrix:tudecomposition}, only contributes to $s$-wave coefficients.

Slightly more complicated is the term with $\tilde{e}_{tu}$ in \cref{eq:pipipiKmatrix:expansiontwopionamplitude}, as one needs to consider $\bar{s}^2[tu]_d$ and $\bar{b}^2[tu]_d$ separately. The former follows the same lines as the $[tu]_d$ factor above, while the latter is off-shell by definition, and so has no associated subtraction term. The result is just given by substituting into \cref{eq:pipipiKmatrix:MuuNLOOPE} and keeping only the terms in the LO amplitude that are independent of momenta, $\cM^\LO=2/\Fpi^2+...$,
\begin{equation}
\Fpi^6\Kdf^{\uu\NLO,\OPE,d}\supset- 2 \tilde{e}_{tu}\left([tu]_{d,\off}+[t'u']_{d,\off}\right) \,.
\end{equation}
Again, by using \cref{eq:pipipiKmatrix:dstusubstitution} is is possible to determine the contribution to the different terms of the expansion in \cref{eq:pipipiKmatrix:uuexpansion}.

The subtracted-OPE contribution to $\Kdf$ at NLO is finally obtained after symmetrization of $\Kdf^{\uu\OPE,\NLO}$. Combining the $s-$ and $d-$wave contributions, we find,
\begin{align}
        \displaystyle\frac{\Fpi^6}{\Mpi^6}\,\Kiso^{\NLO,\OPE}
            &\displaystyle = 
            25 \kappa + 78 L - 72(8 \lrI + 6 \lrII + \lrIII - 2\lrIV) \,,\nonumber
        \\
        \displaystyle\frac{\Fpi^6}{\Mpi^6}\,\Kisoone^{\NLO,\OPE} 
            &\displaystyle = 
            \frac{6831}{20} \kappa + 372 L -18(74 \lrI + 67 \lrII - 14 \lrIV) \,,\nonumber
        \\
        \displaystyle\frac{\Fpi^6}{\Mpi^6}\,\Kisotwo^{\NLO,\OPE} 
            &\displaystyle =
            \frac{230481}{280} \kappa + 576 L -108(10 \lrI + 11 \lrII) \,,\label{eq:pipipiKmatrix:OPEresult}
        \\
        \displaystyle\frac{\Fpi^6}{\Mpi^6}\,\KA^{\NLO,\OPE} 
            &\displaystyle =
            - \frac{53199}{560} \kappa + 45 L + \frac{27}{2}(14 \lrI - 17 \lrII) \,,\nonumber
        \\
        \displaystyle\frac{\Fpi^6}{\Mpi^6}\,\KB^{\NLO,\OPE} 
            &\displaystyle =
            \frac{54171}{140} \kappa + 216 L - 324(2\lrI + \lrII) \,.\nonumber
\end{align}
%For a comparison, the LO given in \cref{eq:hadrons:}that can be compared to the LO results in \cref{eq:hadrons:}.

\subsection{The non-OPE contribution}\label{sec:pipipiKmatrix:nOPE}

The contribution from the non-OPE part of the NLO amplitude is more straightforward to compute, as it can be treated independently of the subtraction. For the three-pion $I_{\pi\pi\pi}=3$ channel, it takes the form
\begin{multline}\label{eq:fullamplitude}
    \cM_3^{\NLO,\nOPE}=\\ \begin{array}{l}
        \phantom{+} A^{(4)}_{6\pi}(k_1, -p_1, k_2, -p_2, k_3, -p_3)
        + A^{(4)}_{6\pi}(k_1, -p_2, k_2, -p_1, k_3, -p_3)
        \\
        + A^{(4)}_{6\pi}(k_1, -p_1, k_2, -p_3, k_3, -p_2)
        + A^{(4)}_{6\pi}(k_1, -p_2, k_2, -p_3, k_3, -p_1)
        \\
        + A^{(4)}_{6\pi}(k_1, -p_3, k_2, -p_1, k_3, -p_2)
        + A^{(4)}_{6\pi}(k_1, -p_3, k_2, -p_2, k_3, -p_1)\,,
        \end{array}
\end{multline}
where $A^{(4)}_{6\pi}(q_1,q_2,q_3,q_4,q_5,q_6)$ is the six-pion amplitude at NLO defined in eq.~(35) of \rcite{Bijnens:2021hpq} with all momenta ingoing. This amplitude can be decomposed into different terms,

\noindent\begin{equation}\label{eq:pipipiKmatrix:nonOPEamplitudedecomposition}
A^{(4)}_{6\pi}=A_\pi+A_L+A_l+A_J+A_C\,,
\end{equation}
each of which can be separately expanded around threshold. To identify the corresponding contributions to the coefficients in \cref{eq:pipipiKmatrix:Kexpansion}, we rewrite first the expanded amplitudes in terms of the operators in \cref{eq:pipipiKmatrix:alternativeexpansion}, and then relate them to the standard form of the expansion.

$A_{\pi}$, $A_L$ and $A_l$ contain terms proportional to $\kappa$, $L$ and $\ell_i^\text{r}$, and are by construction quadratic polynomials in products of the momenta. Thus, no expansion is needed, and the contributions to each coefficient in \cref{eq:pipipiKmatrix:Kexpansion} are straightforward to identify. The $A_J$ part contains $\bar{J}(q^2)$ functions, that need to be expanded up to quadratic order either about threshold, $q^2=4\Mpi^2$, or about $q^2=0$, using \cref{eq:pipipiKmatrix:Jexpansion}. 

\begin{figure}[!b]
    \centering
    \begin{subfigure}{0.48\textwidth}
    \centering
        \begin{tikzpicture}[xscale=\diagramxscale,yscale=\diagramyscale]
            \makeexternallegs
            \coordinate (v1) at (+.5,+.5);
            \coordinate (v2) at (-.5,+.5);
            \coordinate (v3) at (  0,-.7);
            \draw[dprop] (k1) -- (v1) -- (v2) node[midway, above] {\footnotesize $r$} -- (p1);
            \draw[dprop] (k2) -- (v1) -- (v3) -- (v2) -- (p2);
            \draw[dprop] (k3) -- (v3) -- (p3);
            \draw (+0.6,-.3) node {\footnotesize $k_1{+}k_2{-}r$};
            \draw (-0.6,-.3) node {\footnotesize $p_1{+}p_2{-}r$};
        \end{tikzpicture}
        \caption{BH diagram.}
        \label{fig:pipipiKmatrix:triangleloopregular}
    \end{subfigure}
    \begin{subfigure}{0.48\textwidth}
    \centering
        \begin{tikzpicture}[xscale=\diagramxscale,yscale=\diagramyscale]
            \makeexternallegs
            \coordinate (v1) at (+.5,+.5);
            \coordinate (v2) at (-.5,+.5);
            \coordinate (v3) at (  0,-.7);
            \draw[dprop=.8] (v1) .. controls ($ (k1)!.5!(v1) $) .. (p1);
            \draw[line width=3pt, white] (k1) .. controls ($ (p1)!.5!(v2) $) .. (v2);
            \draw[dprop=.2] (k1) .. controls ($ (p1)!.5!(v2) $) .. (v2);
            \draw[dprop] (v2) -- (v3) -- (v1);
            \draw[dprop] (k2) -- (v1) -- (v2) node[midway, above] {\footnotesize $r$}-- (p2)  ;
            \draw[dprop] (k3) -- (v3) -- (p3);
            \draw (+0.6,-.3) node {\footnotesize $p_1{-}k_2{+}r$};
            \draw (-0.6,-.3) node {\footnotesize $k_1{-}p_2{+}r$};
        \end{tikzpicture}
        \caption{``Crossed'' BH diagram.}
        \label{fig:pipipiKmatrix:triangleloopcrossed}
    \end{subfigure}
    \caption{
        Two configurations of the triangle-loop diagram that contribute to the isospin-three scattering amplitude. In total, 15 diagrams with the triangle topology contribute, of which nine correspond to the configuration~(a) and six to~(b). Note that only the former requires subtraction, given in \cref{eq:pipipiKmatrix:unsymmetrizedDBH1,eq:pipipiKmatrix:unsymmetrizedDBH2}.}
   \label{fig:pipipiKmatrix:truiangleloopdiagrams}
\end{figure}
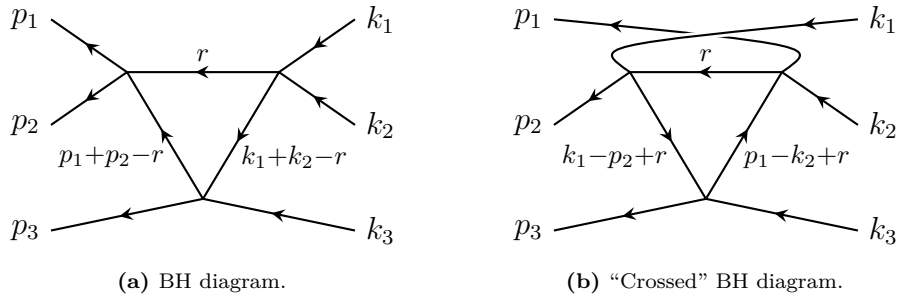

The last term, $A_C$, contains $C$ functions. These are related to triangle loop integrals arising in diagrams of the form of those shown in \cref{fig:pipipiKmatrix:truiangleloopdiagrams}, and depend on three pairs of momenta. As the real part of the amplitude is finite, it can be determined using a naive principal-value prescription~\cite{doi:10.1142/S0218202596000353}. %Comprobar si esta frase es cierta

The $C$ functions need to be expanded in two different cases, which we exemplify here for the simplest $C$ function, called $C_0$. We note the same techniques also hold for other $C$ functions, which contain powers of momentum in the denominator. In the Feynman-parameter representation, $C_0$ takes the form 
\begin{equation}
    C_0 = -\kappa\int_0^1 \text{d} x\,\text{d} y\,\text{d} z\,\frac{\delta(1-x-y-z)}{\Mpi^2-xy q_1^2 - yz q_2^2 - zx q_3^2}\,,
\end{equation}
where $q_1$, $q_2$ and $q_3$ are the total incoming momenta into each vertex of the triangle loop.

The first expansion occurs in the case all three $q_i^2$ small. This happens when the initial/final momenta enter the with the same topology as \cref{fig:pipipiKmatrix:triangleloopcrossed}. In that particular example $q_1=k_2-p_1$, $q_2=k_1-p_2$ and $q_3=k_3-p_3$.  The Feynman-parameter integral can be evaluated after expanding the denominator, obtaining,
\begin{multline}
    \frac{C_0}{\kappa}
    =-\frac{1}{2\Mpi^2}-\frac{1}{24\Mpi^4}\bigl(q_1^2+q_2^2+q_3^2\bigr)\\
    -\frac{1}{180\Mpi^6}\bigl(q_1^4+q_2^4+q_3^4+q_1^2 q_2^2+q_2^2 q_3^2+q_1^2 q_3^2\bigr)+\cdots
\end{multline}
Note that in this case the integral is finite, and we do not need any prescription to regulate divergencies.

The second expansion case originates from diagrams of the form of \cref{fig:pipipiKmatrix:triangleloopregular}, in which $q_1=k_1+k_2$, $q_2=-p_1-p_2$ and $q_3=k_3-p_3$. In this case, the threshold expansion corresponds to $q_3^2$ small and both $q_1^2$ and $q_2^2$ close to $4\Mpi^2$. This is equivalent to having small $\bar{s}_1=q_1^2-4\Mpi^2$ and $\bar{s}_2=q_2^2-4\Mpi^2$. In terms of these variables, the $C$ integral can be rewritten as
\begin{equation}
    C_0 
    = -\kappa\int_0^1 \text{d} x\,\text{d} y\,\text{d} z\,\frac{\delta(1-x-y-z)}{\Mpi^2(1-2y)^2-xy \bar s_1 - yz \bar s_2 - zx q_3^2}\,.
\end{equation}

As before, we first expand the denominator and then perform the Feynman integrals. These are trivial in the case of $x$ and $z$, but the result for the $y$ integral is divergent. We know that the real part is finite about threshold, and so can compute it using a principal-value (PV) prescription that discards the imaginary part. Integrals having singularities in the integration path are averaged over contours surrounding the singularity from above and below. For example,
\begin{multline}
    \PV\!\int_0^1\text{d} z\,\frac{1}{(1-2y)^n}
    = \frac{1}{2}\,\PV\!\int_{-1}^{1}\text{d} v\,\frac{1}{v^n}\\
    = \frac{1}{4}\int_\pi^0 \text{e}^{-in\theta}\,\text{d} \text{e}^{i\theta}
    + \frac{1}{4}\int_\pi^0 \text{e}^{in\theta}\,\text{d} \text{e}^{-i\theta}=
    \left\{\begin{array}{ll}
    0 & n\text{ odd\,,}\\
    -1/(n-1) & n\text{ even\,.}
    \end{array}\right.
\end{multline}
This allows us to obtain
\begin{multline}
     \frac{C_0}{\kappa}
    =\frac{1}{2\Mpi^2}+\frac{1}{\Mpi^4}\biggl(\frac{5}{72}q_3^2-\frac{1}{24}\bar s_1-\frac{1}{24}\bar s_2\biggr)\\
    +\frac{1}{\Mpi^6}\biggl[\frac{2}{225}q_3^4-\frac{1}{90}q_3^2(\bar s_1+\bar s_2)+\frac{1}{180}\bigl(\bar s_1^2+\bar s_2^2+\bar s_1\bar s_2\bigr)\biggr]+\cdots
\end{multline}
After expanding all $C$ functions about the relevant values, one can follow the same procedure as for the other terms in $A^{(4)}_{6\pi}$ to determine the contribution to the different coefficients in the threshold expansion.

Adding the contributions to $\Kdf^{\NLO,\nOPE}$ from all terms in \cref{eq:pipipiKmatrix:nonOPEamplitudedecomposition}, we obtain
    \begin{align}
        \frac{\Fpi^6}{\Mpi^6}\,\Kiso^{\NLO,\nOPE} 
            & = 
            14 \kappa + 33 L + 36 (8 \lrI +  \lrIII -2 \lrIV)\,,\nonumber
        \\
        \frac{\Fpi^6}{\Mpi^6}\,\Kisoone^{\NLO,\nOPE} 
            & = 
            -\frac{35}2 \kappa + 12 L + 36 (20 \lrI + \lrII - 4 \lrIV)\,,\nonumber
        \\
        \frac{\Fpi^6}{\Mpi^6}\,\Kisotwo^{\NLO,\nOPE} 
            & =
            -\frac{9747}{50} \kappa - 216 L + 324 (2 \lrI + \lrII)\,,\label{eq:pipipiKmatrix:nonOPE}
        \\
        \frac{\Fpi^6}{\Mpi^6}\,\KA^{\NLO,\nOPE} 
            & =
            \frac{576}5 \kappa - 54 L - 81 (2 \lrI - 3 \lrII)\,,\nonumber
        \\
        \frac{\Fpi^6}{\Mpi^6}\,\KB^{\NLO,\nOPE}
            & =
            -\frac{13797}{50} \kappa - 162 L + 243 (2 \lrI + \lrII)\,.\nonumber
    \end{align}
%The separate contributions of the different terms in \cref{eq:pipipiKmatrix:nonOPEamplitudedecomposition} can be found in \rcite{}.

\subsection{The bull's head subtraction contribution}\label{sec:pipipiKmatrix:BH}

The last term contributing to $\Kdf^\NLO$ is the BH part of the subtraction, which cancels divergencies arising from diagrams of the form of \cref{fig:pipipiKmatrix:triangleloopregular} when any of the loop particles goes on shell. Note these divergencies appear in the imaginary part, and so the real part of the subtraction is finite. However, while an expansion about threshold of this real part can be performed, its coefficients involve integrals including the cutoff function, $H(x)$, and so can only be computed numerically. %can only be expressed in term of numerical integrals, due to the presence of the cutoff function, $H(x)$.

Consider first the unsymmetrized subtraction, given by the last line in \cref{eq:pipipiKmatrix:subtractionNLOKdf}. For the momenta configuration in \cref{fig:pipipiKmatrix:triangleloopregular} it takes the form
\begin{equation}
    \displaystyle\cD^{\uu\BH}(\bm p_3, \bm k_3)  = -\frac{1}{\Fpi^6}(2p_1\cdot p_2)\, I(\bm p_3,\bm k_3) (2k_1\cdot k_2)\,,
   \label{eq:pipipiKmatrix:unsymmetrizedDBH1}
\end{equation}
with
\begin{equation}
   \displaystyle I(\bm p_3,\bm k_3)  = \int_r \frac{H(x_r) \left[(P-r)^2 - 2\Mpi^2\right]H(x_r)}{\left[(p_+-r)^2-\Mpi^2+i\epsilon\right]\left[(k_+-r)^2-\Mpi^2+i\epsilon\right]}\,,
   \label{eq:pipipiKmatrix:unsymmetrizedDBH2}
\end{equation}
where $x_r=(P-r)^2/(4\Mpi^2)$ and we have substituted $G^\infty\rightarrow G^\infty_{ss}$, given in \cref{eq:hadrons:Ginftyssdefinition}, since $\cM_2^\LO$ is purely $s$-wave. %Here we have used that only $s$-wave contributes to the LO two-particle amplitudes, and so no barrier factors are needed. Note that in this integral $r$ is on shell. 

%We would like to rewrite this expression to directly identify the contribution to the different terms in the threshold expansion in terms of, at most, numerical integrals. To do so, we start by 
We work in the CMF of the three-particle system, $P=(E^*,0)$, and set $\text{d}^3r=r^2\d r\,\d\cos\theta\,\d\phi$. To evaluate the angular integrals, we rewrite first
\begin{multline}
    (p_+-r)^2-\Mpi^2 = p_+^2-2p_+\cdot r \\ \hspace{-0.9cm}= p_+^2-2E_{p_+}\omega_r +2\bm{p}_+\cdot \bm r\equiv 4 \Mpi^2-4 \Mpi\omega_r+\Delta_{p_+}\,,
\end{multline}
where $p_+=(p_1+p_2)=(E_{p_+},\bm{p}_+)$ and we implicitly define $\Delta_{p_+}$. Similarly, we expand $(k_+-r)$ and define $\Delta_{k_+}$. These quantities are both $\cO(\sqrt{\Delta})$, and so we use them to expand the denominators about threshold,
\begin{equation}
    I(\bm p_3,\bm k_3) = \int_r H^2(x_r)\big[E^{*2}-2E^*\omega_r-\Mpi^2\big]\sum_{a,b=0}^\infty \frac{\Delta_{p_+}^a\Delta_{k_+}^b}{(4\Mpi^2-4\Mpi\omega_r)^{a+b+2}}\,,
\end{equation}
where the sum can be truncated at some convenient order. The angular integrals can be performed using integrals of the form 
\begin{equation}
    \int \text{d}\,\cos\theta\,\text{d}\phi\, (\bm{p}_+\cdot \bm r)(\bm{k}_+\cdot\bm r)
    = 4\pi\, \frac{r^2}{3}\, \bm{p}_+\cdot\bm{k}_+\,,
\end{equation}
and its generalizations~\cite{Baeza-Ballesteros:2023ljl}. 

At this point, one can expand the integrand about threshold, using $E=3\Mpi\sqrt{1+\Delta}$ and $x_r=(E^{*2}-2E^*\omega_r+\Mpi^2)/4\Mpi$, and identify the contribution to the different coefficients of the unsymmetrized threshold expansion, given in \cref{eq:pipipiKmatrix:uuexpansion}. This includes only terms with corresponding coefficients $c_0'-c_{10}'$. Symmetrizing over the initial and final momenta allows  us to obtain the results
    \begin{equation}
    \begin{array}{rl}
        \displaystyle \frac{\Fpi^6}{\Mpi^6}\,\Kiso^{\NLO,\BH} 
            & = 
            \displaystyle \frac{27}{2} H_{0,0} - \frac 94 H_{2,0}\,,
        \\[10pt]
        \displaystyle \frac{\Fpi^6}{\Mpi^6}\,\Kisoone^{\NLO,\BH}  
            &\displaystyle  = 
            \frac{117}{4} H_{0,0} - \frac{21}{8} H_{2,0} + \frac 34 H_{4,0} + \frac{189}{4} H_{0,1}\,,
        \\[10pt]
        \displaystyle \frac{\Fpi^6}{\Mpi^6}\,\Kisotwo^{\NLO,\BH}  
            &\displaystyle  =
            \frac{243}{160} H_{0,0} + \frac{2241}{320} H_{2,0} - \frac{423}{160} H_{4,0} \\[4pt]
            & \displaystyle \phantom{-}- \frac{369}{1280} H_{6,0} + \frac{5751}{64} H_{0,1} + \frac{567}{8} H_{0,2}\,,
        \\[10pt]
        \displaystyle \frac{\Fpi^6}{\Mpi^6}\,\KA^{\NLO,\BH}  
            &\displaystyle  =
            -\frac{891}{64} H_{0,0} + \frac{1161}{128} H_{2,0} - \frac{45}{64} H_{4,0} - \frac{9}{128} H_{6,0}\,,
        \\[10pt]
        \displaystyle \frac{\Fpi^6}{\Mpi^6}\,\KB^{\NLO,\BH}  
            & \displaystyle =
            -\frac{81}{320} H_{0,0} + \frac{297}{640} H_{2,0} + \frac{27}{160} H_{4,0} - \frac{27}{640} H_{6,0}\,.
    \end{array}\label{eq:pipipiKmatrix:DBHHfunctions}
    \end{equation}
Here, $H_{m,n}$ are given by integrals of the form
\begin{equation}
     H_{m,n} \equiv \frac{1}{\pi^2}\int_0^{1/\sqrt{3}}\text{d} z~\frac{\sqrt{1+z^2}}{z^{m}} \frac{\text{d}^n}{\text{d} x^n}\left[H^2(x)\right]\,.
    \label{eq:pipipiKmatrix:Hmnintegraldefinition}
\end{equation}
where $x=1-3z^3$, so the integration limits correspond $x=1$ and $x=0$. These terms originate after expanding $H(x)$, defined in \cref{eq:hadrons:standardcutoff}, about threshold and rewriting the integrals using a variable $z$ defined via $\omega_r=\Mpi(1+2z^2)$. These terms obey the recursion relation,

\noindent\begin{multline}
    H_{m,n+1}+H_{m-2,n+1} =\frac{1}{6} \left[(2-m)H_{m,n}-(m+1) H_{m+2,n}\right]\\
        -\frac{1}{6}\left[\left(f_{m-1}'(z)+f_{m+1}'(z)\right)\frac{\text{d}^n}{\text{d} x^n}H^2(x)\right]_0^{1/\sqrt3}\,,
    \label{eq:pipipiKmatrix:Hrelation}
\end{multline}
which has been used to simplify the results in \cref{eq:pipipiKmatrix:DBHHfunctions}. The functions $f_n$ are implicitly defined from 
\begin{equation}\label{eq:pipipiKmatrix:fdefinition}
    \frac{\text{d}}{\text{d} z}f_m(z) = \frac{1}{\pi^2} \frac{\sqrt{1+z^2}}{z^{m}}\,,
\end{equation}
independently of the integration constant in $f_m(z)$. Note they are singular at $z=0$ for $m>0$.

Note that $H(x)$ is non analytic at $x=0$ and $x=1$. Thus, the expansion is of asymptotic nature, not properly capturing the energy dependence in the upper limit of the integral. However, the expansion still captures very accurately the threshold behavior of $\cD^\BH$. We comment on this below, and also study the convergence numerically in \cref{sec:pipipiKmatrix:convergenceresults}, where we compare the threshold result up to quadratic order to the full $\Kdf$ evaluated numerically.

The evaluation of the $H_{m,n}$ factors in \cref{eq:pipipiKmatrix:DBHHfunctions} is troublesome. For $n=0$ and $m>0$, the integrand in \cref{eq:pipipiKmatrix:Hmnintegraldefinition} presents a singularity in the lower endpoint, so a naive application of the PV prescription is not valid. Instead, one can use the Hadamard finite-part prescription to regulate it~\cite{hadamard2003lectures}. 
To do so, we first use integration by parts
\begin{align}
    \label{eq:pipipiKmatrix:regularHn}
    H_{m,0} = f_m(z) H^2(x)\big|_{z=0}^{z=1/\sqrt3}-\int_{0}^{1/\sqrt3} \text{d} z\: f_m(z) (-6z)\, \frac{\text{d}}{\text{d} x}H^2(x)\,.
\end{align}
At $z=0$ the first term on the right-hand side of \cref{eq:pipipiKmatrix:regularHn} is singular for $m>0$, but not the second one since derivatives of $H$ vanish exponentially as $z\rightarrow 0$. The $z\rightarrow 1/\sqrt{3}$ limit vanishes identically, since $H$ and all its derivatives are zero in this case. The Hadamard finite part of $H_{m,0}$ is obtained by dropping the singular $z=0$ term and evaluating the integral, which yields a finite result.

The applicability of Hadamard finite-part prescription is not clear a priori. Most traditional proofs rely on the analyticity of the integrand, which is not valid here since $H(x)$ is not analytic. However, \rcite{costin2014foundational} presents a proofs that only requires smoothness of the integrands, which is satisfied in our case. It also requires $m>1$, but the only integral not satisfying this condition is $H_{0,0}$, which is convergent. The application of the prescription also relies on $\text{Re}\cD^{\uu\BH}$ being finite, but this is the case since all divergencies in $\cM_3^\uu$ appear in its imaginary part.

Note that the Hadamard finite-part prescription validates the Taylor expansion of $H(x)$ used to obtain \cref{eq:pipipiKmatrix:Hmnintegraldefinition}. The Taylor series converges for $0<x<1$, but the convergence is extremely poor in the vicinity of the essential singularities at the endpoints. However, taking the Hadamard finite partin \cref{eq:pipipiKmatrix:regularHn}, all remaining integrands contain derivatives of $H(x)$, which are exponentially suppressed near these endpoints. Thus, the result is insensitive to the regions were the convergence is poor.

There is one final simplification that can be done, which allows us to rewrite \cref{eq:pipipiKmatrix:Hmnintegraldefinition} in terms of a fully analytical parts plus some small cutoff-dependent correction that need to be evaluated numerically. The idea is to write $H(x)=1+\tilde{H}(x)$, so that
\begin{equation}
    H_{m,n} = \tilde{H}_{m,n} + \delta_{n,0}\,\int_0^{1/\sqrt{3}}\d z~\frac{1}{\pi^2}\frac{\sqrt{1+z^2}}{z^{m}}\,,
    \label{eq:pipipiKmatrix:Hmnanalytic}
\end{equation}
where $\tilde{H}_{m,n}$ is obtained by substituting $H(x)\rightarrow \tilde{H}(x)$. The second term on the right-hand side can be evaluated analytically using the Hadamard finite part prescription. Choosing $f_0(0)=0$ as the integration constant for $f_0$, the result of the integral is simply $f_m(1/\sqrt{3})$. \Cref{eq:pipipiKmatrix:Hrelation} can be rewritten for the $f_m\delta_{n,0}$ terms, 
\begin{equation}
    (m+1)f_{m+2}(z)+(m-2)f_m(z)=-(1+z^2)f_{m+1}'(z)\,,
    \label{eq:frelation}
\end{equation}
which can be used to determine all $f_m$ in terms of $\kappa$ and $f_0=f_0(1/\sqrt{3})=\frac{4}{3}\kappa(4+3\log3)$. 

Using this procedure, the results for the threshold coefficients of $\Kdf^{\NLO,\BH}$ are simplified,
    \begin{align}\label{eq:pipipiKmatrix:DBHsimplified}
        \displaystyle\frac{\Fpi^6}{\Mpi^6}\,\Kiso^{\NLO,\BH} 
            & \displaystyle= 
            96\kappa + 9f_0 + \Diso\,,\nonumber
        \\
        \displaystyle\frac{\Fpi^6}{\Mpi^6}\,\Kisoone^{\NLO,\BH}  
            & \displaystyle=
            296\kappa + 24 f_0 + \Disoone\,,\nonumber
        \\
        \displaystyle\frac{\Fpi^6}{\Mpi^6}\,\Kisotwo^{\NLO,\BH}  
            & \displaystyle=
            \frac{5661}{50}\kappa + \frac{621}{40}f_0 + \Disotwo\,,\nonumber
        \\
        \displaystyle\frac{\Fpi^6}{\Mpi^6}\,\KA^{\NLO,\BH}  
            & \displaystyle=
            -\frac{1764}5\kappa + \frac{135}{32}f_0 + \DisoA\,,\nonumber
        \\
        \displaystyle\frac{\Fpi^6}{\Mpi^6}\,\KB^{\NLO,\BH}  
            & \displaystyle =
            -\frac{612}{25}\kappa + \frac{189}{160}f_0 + \DisoB\,.
    \end{align}
Here $\cD_X$ are cutoff-depending numerical corrections coming from $\tilde{H}_{m,n}$, defined from the requirement that \cref{eq:pipipiKmatrix:DBHsimplified} equals \cref{eq:pipipiKmatrix:DBHHfunctions}. The values of the remainders, in the case of the standard cutoff in \cref{eq:hadrons:standardcutoff} are presented in \cref{eq:pipipiKmatrix:Dvalues}. A study of their dependence on the choice of the cutoff function is presented in app.~A of \rcite{Baeza-Ballesteros:2023ljl}. We remark that \cref{eq:pipipiKmatrix:DBHsimplified} cannot be directly obtained from \cref{eq:pipipiKmatrix:DBHHfunctions} by substituting $H_{m,n}\rightarrow f_m\delta_{n,0}$. Instead, one needs to perform this substitution before applying the simplifying relation given in \cref{eq:pipipiKmatrix:Hrelation}.

\newpage\subsection{Numerical evaluation}\label{sec:pipipiKmatrix:crosscheck}

The (almost) analytic results presented in the previous sections have been cross checked by direct numerical evaluation. The numeric results also allow us to test the convergence of the threshold expansion away from threshold, and to study the relative contributions from different partial waves, in the case of the subtracted OPE part. Such analysis are presented in \cref{sec:pipipiKmatrix:convergenceresults}.

\begin{table}[b!]
    \centering
    {\renewcommand{\arraystretch}{1.3}
    \begin{tabular}{c*{4}{@{$\quad\big($}r@{, }r@{, }r@{$\big)$}}}
        \toprule
        \multicolumn{1}{c@{$\quad\phantom{\big(}$}}{$a$}
                &    \multicolumn{3}{c}{$\bm{p}_1^{(a)}(p)$}
                &    \multicolumn{3}{c}{$\bm{p}_2^{(a)}(p)$}
                &    \multicolumn{3}{c}{$\bm{k}_1^{(a)}(p)$}
                &    \multicolumn{3}{c}{$\bm{k}_2^{(a)}(p)$}\\      
        \midrule
        1         &   $p$ &   0   &   0   
                  &   $-\tfrac12p$    &   $\tfrac{\sqrt3}{2}p$ &   0   
                  &   0   &   0   &   $-p$   
                  &   $\tfrac{\sqrt3}{2}p$ & 0 &   $\tfrac12p$
                \\[5pt]
        2         &   $p$ &   0   &   0   
                  &   $-\tfrac12p$    &   $\tfrac{\sqrt3}{2}p$ &   0   
                  &   $-p$ &   0   &   0   
                  &   $\tfrac12p$    &   $\tfrac{\sqrt3}{2}p$ &   0
                \\[5pt]
        3         &   $2p$ &   0   &   0   
                  &   $-p$    &   $\tfrac{\sqrt3}{2}p$ &   0   
                  &   0 &   0   &   $-2p$  
                  &  $\tfrac{\sqrt3}{2}p$ &  0 &  $p$ 
                \\
        \bottomrule
    \end{tabular}}
    \caption[]{
        Momenta families used in the numerical evaluation of $\Kdf$, all with zero total momentum. We only quote the spacial part of the four-momenta, with the remaining element constrained by the on-shell relation, $p_i^2=k_i^2=\Mpi^2$. Also, $\bm{k}_3$ and $\bm{p}_3$ can be inferred from $\bm P=\bm p_1+\bm p_2+\bm p_3 = \bm k_1+\bm k_2+\bm k_3 =  0$.}
    \label{tab:pipipiKmatrix:momentafamilies}
\end{table}

The numerical determination of the subtracted OPE is relatively simple. It requires to project the off- and on-shell two-pion amplitudes to partial waves, compute the subtraction with some finite $\varepsilon$ in the exchanged-particle propagator, including the correct barrier factors, and add all the partial waves with the relevant spherical harmonics. The computation of $\Mdf^\nOPE$ is also straightforward, as one can use known results for the $C$ functions in terms of dilogarithmic functions~\cite{Passarino:1978jh}.  

It is also possible to compute $\cD^\BH$ numerically, although this presents more complications. One option is to directly evaluate \cref{eq:pipipiKmatrix:unsymmetrizedDBH2}, but singularities in the integrand can lead to bad convergence. It is more convenient to first separate the integral into a convergent and a divergent part, and treat the latter one analytically, leaving it in a form that allows for easier numerical evaluation~\rcite{Baeza-Ballesteros:2023ljl}.

The numerical computation  of $\Kdf$ is performed for various chosen kinematical configurations. Each of these is a family of six on-shell four-momenta depending on a single parameter $p$ that vanishes at threshold. We refer to each of these families as $\cF_a$. The ones used in this work are summarized in \cref{tab:pipipiKmatrix:momentafamilies}. Numerically evaluating $\Kdf$ for different values of $p$ close to zero allows us determine the coefficients of a Taylor expansion of $\Kdf$ in $p^2$ as
\begin{equation}
\Kdf(\cF_a)=c_0^a+c_1^ap^2+c_2^ap^4+\cO(p^6)\,.
\end{equation}
If we consider the expansion of the kinematical operators appearing in the threshold expansion,
\begin{align}
\Delta(\cF_a)&=d_1^ap^2+d_2^ap^4+\cO(p^6)\,,\nonumber\\
\DA(\cF_a)&=d_\text{A}^ap^4+\cO(p^6)\,,\nonumber\\
\quad\DB(\cF_a)&=d_\text{B}^ap^4+\cO(p^6)\,,
\end{align}
one can note that a single family is sufficient to determine
\begin{equation}
\Kiso = c_0^a\,, \quad\quad\quad \Kisoone = c_1^a/d_1^a\,.
\end{equation}
For the quadratic-order coefficients, on the other hand, the results from three separate families need to be combined. One first constructs
\begin{equation}
Q = \begin{pmatrix} 
        (d_1^{a_1})^2 & d_\mathrm{A}^{a_1} & d_\mathrm{B}^{a_1} \\
        (d_1^{a_2})^2 & d_\mathrm{A}^{a_2} & d_\mathrm{B}^{a_2} \\
        (d_1^{a_3})^2 & d_\mathrm{A}^{a_3} & d_\mathrm{B}^{a_3} \\
    \end{pmatrix},\quad\quad V = \begin{pmatrix}
        c_2^{a_1}-c_1^{a_1} d_2^{a_1}/d_1^{a_1} \\
        c_2^{a_2}-c_1^{a_2} d_2^{a_2}/d_1^{a_2} \\ 
        c_2^{a_3}-c_1^{a_3} d_2^{a_3}/d_1^{a_3} \\
    \end{pmatrix},
\end{equation}
from which it is possible to obtain
\begin{equation}
\begin{pmatrix} 
        \Kisotwo & 
        \KA &
        \KB
    \end{pmatrix}^\intercal 
    = Q^{-1} V\,.
\end{equation}

\section{Results for $\Kdf$}\label{sec:pipipiKmatrix:summary}

Combining the analytical results determined in \cref{sec:pipipiKmatrix:computation}---see \cref{eq:pipipiKmatrix:OPEresult,eq:pipipiKmatrix:nonOPE,eq:pipipiKmatrix:DBHsimplified}---and including the LO results from \cref{eq:hadrons:Kdfpipipiresult}, the coefficients of the full LO+NLO isospin-three three-pion $K$-matrix are
    \begin{align}
        \Kiso 
            &= \MF^4 18
            &+& \MF^6 \biggl[- 3\kappa(35 + 12\log3) - \Diso + 111L + \elliso\biggr]\,, \nonumber
            \\
        \Kisoone 
            &= \MF^4 27   
            &+& \MF^6 \biggl[ - \frac{\kappa}{20}(1999 + 1920\log3) - \Disoone + 384L + \ellisoone \biggr]\,,\nonumber
            \\
        \Kisotwo 
            &= 
            &&  \MF^6 \biggl[\frac{207\kappa}{1400}(2923 - 420\log3) -\Disotwo + 360L +  \ellisotwo\biggr]\,,\nonumber
            \\
        \KA 
            &=
            && \MF^6 \biggl[\frac{9\kappa}{560}(21809 - 1050\log3) - \DisoA - 9L + \ellA\biggr]\,,\nonumber
            \\
        \KB 
            &= 
            && \MF^6 \biggl[\frac{27\kappa}{1400}(6698 - 245\log3) - \DisoB + 54L + \ellB\biggr]\,.\label{eq:pipipiKmatrix:results}%
    \end{align}
Here $\cD_X$ are the only cutoff-dependent parts, which need to be evaluated numerically, 
\begin{equation}
    \begin{gathered}
        \Diso       \approx -0.0563476589\,,\qquad
        \Disoone    \approx 0.129589681\,,\qquad
        \Disotwo    \approx 0.432202370\,,\\
        \DisoA         \approx 9.07273890\times10^{-4}\,,\qquad
        \DisoB         \approx 1.62394747\times10^{-4}\,.
    \end{gathered}
    \label{eq:pipipiKmatrix:Dvalues}
\end{equation}
Also, we have defined the following linear combinations of LECs,
\begin{equation}
    \begin{gathered}
        \elliso     = -288\lrI-432\lrII-36\lrIII+72\lrIV\,, \qquad
        \ellisoone  = -612\lrI-1170\lrII+108\lrIV\,, \\
        \ellisotwo  = -432\lrI-864\lrII\,, \qquad
        \ellA       = 27\lrI+\frac{27}{2}\lrII\,, \qquad
        \ellB       = -162\lrI-81\lrII\,.
    \end{gathered}
    \label{eq:pipipiKmatrix:LECscombinations}
\end{equation}
Note that, while both $\ell_i$ and $L$ depend on the renormalization scale, $\mu$, the complete results in  \cref{eq:pipipiKmatrix:results} are scale independent, and so is $\Kdf$.

\subsection{Comparison to lattice results}\label{sec:pipipiKmatrix:comparisonlattice}

The results in \cref{eq:pipipiKmatrix:results} can be compared to lattice determinations of $\Kdf$. Different works have studied the system of three pions at maximal isospin using the RFT formalism~\cite{Blanton:2019igq,Fischer:2020jzp,Hansen:2020otl,Blanton:2021llb}, finding similar qualitative disagreement with LO  ChPT predictions. This discrepancy can be partially explained by NLO ChPT corrections. In particular, we focus our comparison on the results of \rcite{Blanton:2021llb}, in which $\Kiso$, $\Kisoone$ and $\KB$ are determined for pion masses of $200$, $280$ and $340$ MeV. %More?

We use the following values for the scale-independent LECs, which were introduced in \cref{eq:QCD:lecsNfscaledependence},
\begin{equation}
    \bar\ell_1 = -0.4(6)\,,\qquad \bar\ell_2 = 4.3(1)\,,\qquad \bar\ell_3 = 3.07(64)\,,\qquad \bar\ell_4 = 4.02(45)\,.
    \label{eq:pipipiKmatrix:LECref}
\end{equation}
The values of $\bar\ell_1$ and $\bar\ell_2$ are determined by combining experiment, ChPT and dispersion relations~\cite{Colangelo:2001df}, while $\bar\ell_3$ and $\bar\ell_4$ come from averaging results from $\Nf=2+1$ lattice simulations~\cite{FLAG:2021npn}, based on \rrcite{MILC:2010hzw,Beane:2011zm,Borsanyi:2012zv,BMW:2013fzj,Boyle:2015exm}. We also take into account the correlations between $\bar\ell_1$ and $\bar\ell_2$ using the covariance matrix from \rcite{Colangelo:2001df},
\begin{equation}
    \mathrm{Cov}(\bar\ell_1,\bar\ell_2)=
    \begin{pmatrix}
    0.35 & -0.033\\
    -0.033 & 0.012
    \end{pmatrix}\,.
\end{equation}

For the comparison, we choose $\mu$ so the results in \cref{eq:pipipiKmatrix:results} depend only on $\Mpi/\Fpi$. This is achieved by taking $\mu=4\pi\Fpi$ in $L$ and approximating $\mu\approx 4\pi\Fpiphys$ in the definition of $\bar\ell_i$ given in \cref{eq:QCD:lecsNfscaledependence}. This different choice only affects the NNLO part, since $\Fpi$ does not depend on $\Mpi$ at LO, and allows us  to rewrite ChPT predictions only as a function of $\Mpi/\Fpi$. For example, $\Kiso$ can be rewritten as
\begin{multline}\label{eq:pipipiKmatrix:comparisonlattice}
    \Kiso 
        = \MF^4 18
        + \MF^6 \left[ - 3\kappa(35 + 12\log3) \right. \\ 
        \left.- \Diso + 111\kappa\log\left(\frac{\xi}{\xi_\text{phys}}\right) + \kappa\bar{\ell}_{(0)}\right]\,, 
\end{multline}
with $\bar{\ell}_{(0)}=-48\bar\ell_1-144\bar\ell_2+9\bar\ell_3+72\bar\ell_4$, $\xi\equiv\Mpi^2/(4\pi\Fpi)^2$ and $\xi_\text{phys}\equiv\Mpiphys^2/(4\pi\Fpiphys)^2$.

\begin{figure}[!p]
    \centering
    \begin{subfigure}{0.8\textwidth} 
    \centering
        \includegraphics[width=1\textwidth]{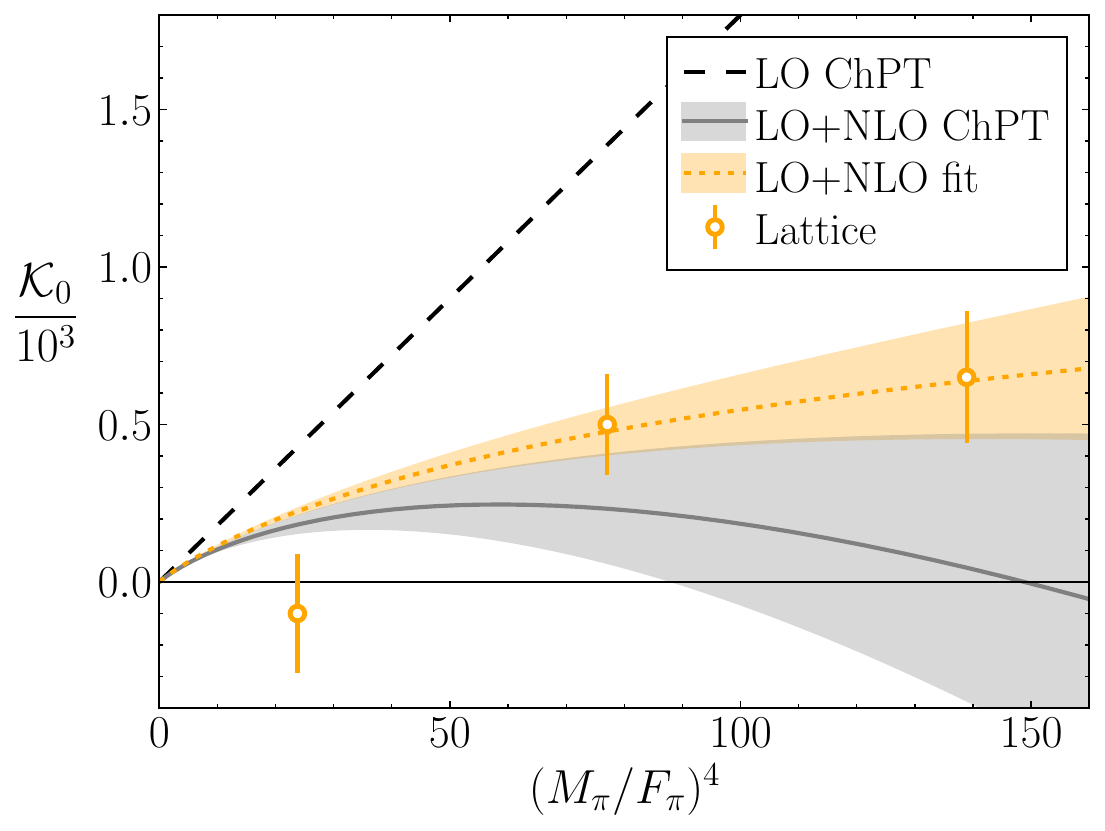}\\[0.5cm]
        
        \includegraphics[width=1\textwidth]{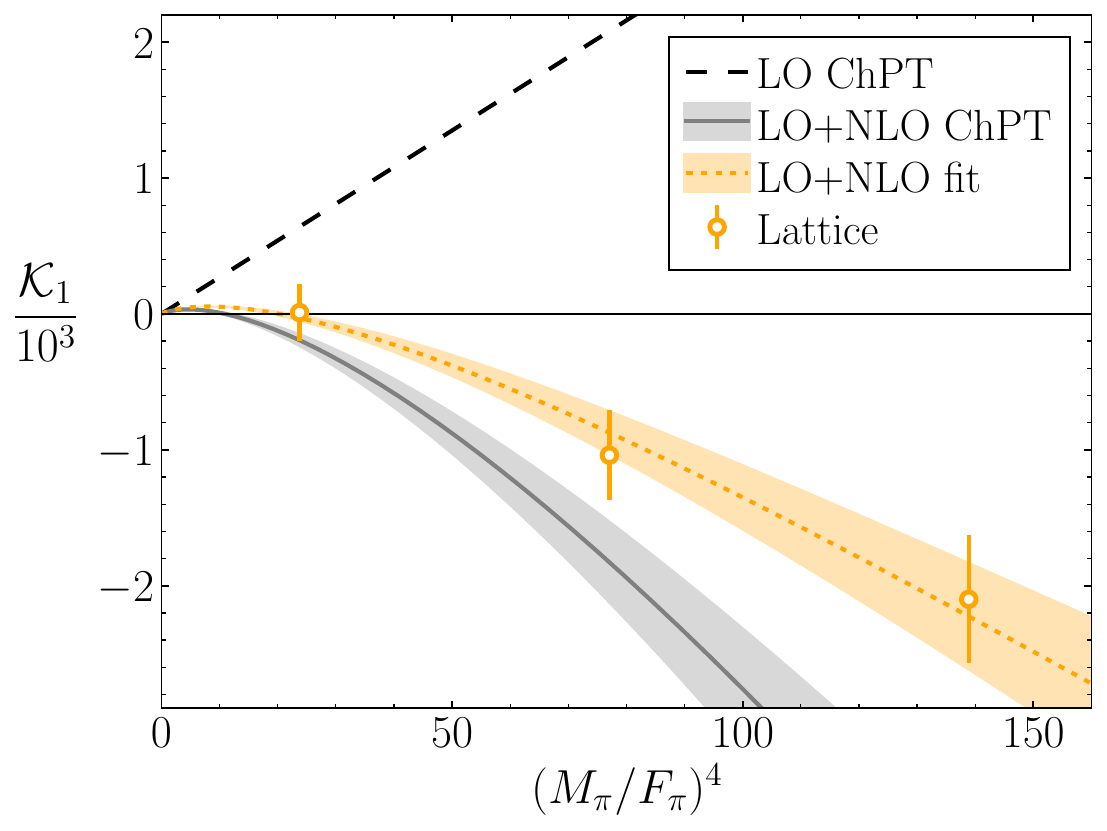}
    \end{subfigure}
    \caption{
         LO (dashed black line) and LO+NLO (grey line and band) ChPT predictions for $\Kiso$ (top) and $\Kisoone$ (bottom) as functions of $(\Mpi/\Fpi)^4$, using LECs from \rrcite{Colangelo:2001df,FLAG:2021npn} [see \cref{eq:pipipiKmatrix:LECref}].
        These predictions are compared to lattice results from \rcite{Blanton:2021llb} (orange points).
        We also present the best fit to the lattice data (dotted orange line and band). }
    \label{fig:pipipiKmatrix:K0K1chpt}
\end{figure} 

In \cref{fig:pipipiKmatrix:K0K1chpt}, we compare the LO+NLO ChPT predictions for $\Kiso$ and $\Kisoone$ to the lattice results. We also show the LO ChPT result for comparison. We observe how, in general, the discrepancy is reduced when NLO contributions are included. For $\Kiso$, it leads to smaller predictions that lie closer to the lattice results. For $\Kisoone$, on the other hand, the corrections are large, leading to a change of sign with respect to LO for all but small pion masses. The LO+NLO results for $\Kdf$ have the same sign and lie close to the lattice data. 

\begin{figure}[!h]
    \centering
    \begin{subfigure}{0.495\textwidth} 
    \centering
        \includegraphics[width=1\textwidth]{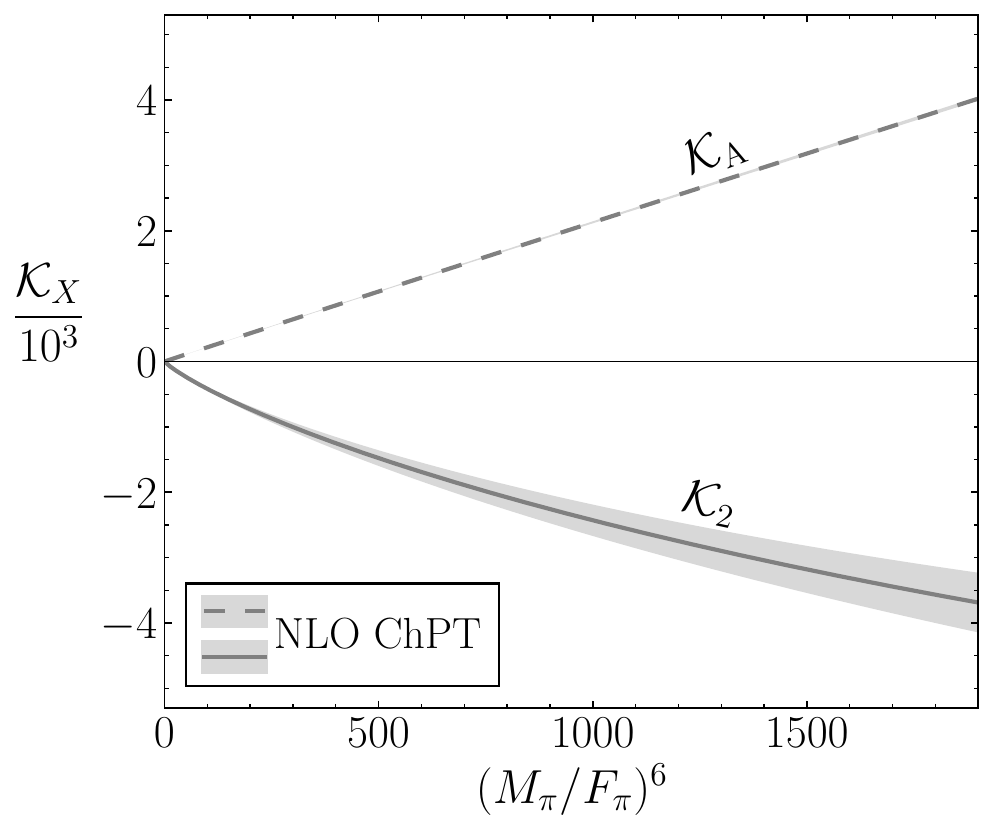}
    \end{subfigure}
     \begin{subfigure}{0.495\textwidth} 
    \centering
        \includegraphics[width=1\textwidth]{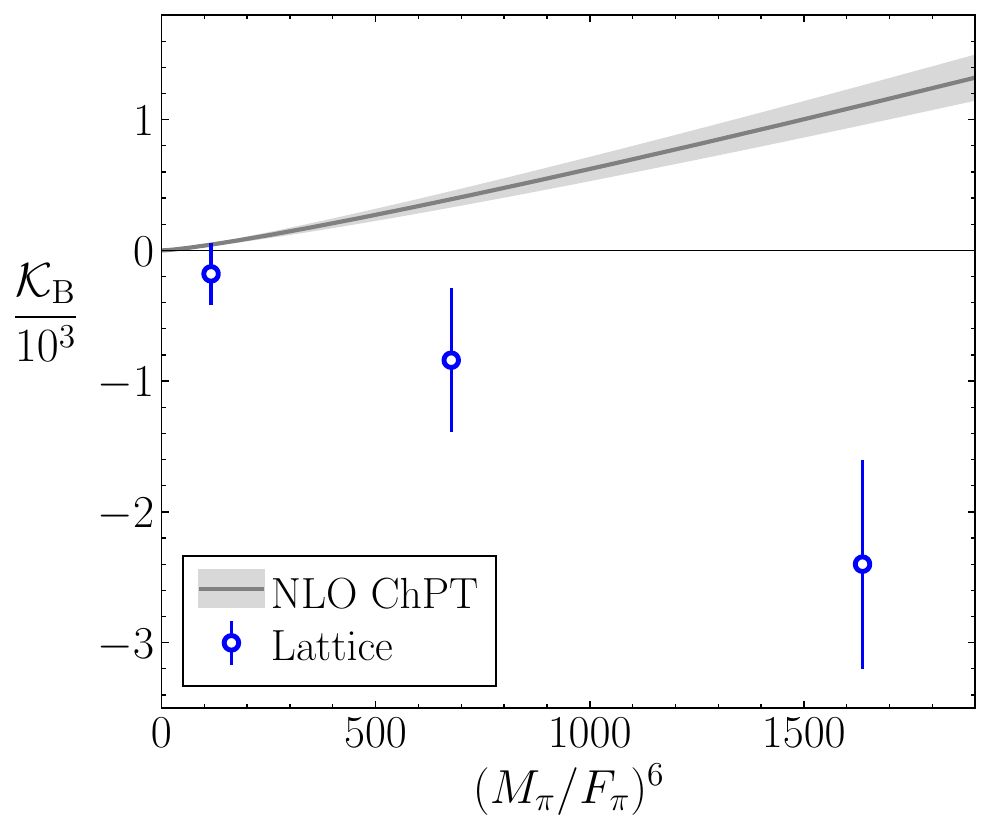}
    \end{subfigure}
    \caption{
         NLO ChPT predictions for $\Kisotwo$ and $\KA$ (left), and $\KB$ (right) as functions of $(\Mpi/\Fpi)^6$, using LECs from \rrcite{Colangelo:2001df,FLAG:2021npn} [see \cref{eq:pipipiKmatrix:LECref}].
        In the case of $\KB$, we compare to lattice results from \rcite{Blanton:2021llb} (blue points).}
    \label{fig:pipipiKmatrix:K2KAKBchpt}
\end{figure} 

To interpret these findings, two different perspectives can be taken. One could argue that, in view of the size of NLO corrections, the chiral expansion converges poorly in the mass region where lattice results lie and so ChPT cannot be trusted. A more optimistic interpretation is that, for some reason, NLO corrections are larger than the LO results. This could be for example due to the presence of new diagrams, such as triangle loops in \cref{fig:pipipiKmatrix:truiangleloopdiagrams}. %In both cases, anyways, the discrepancy between lattice and LO ChPT is resolved.

In view of the agreement between lattice results and the ChPT prediction, it is instructive to study how well the values of the LECs in \cref{eq:pipipiKmatrix:LECscombinations} can be constrained. From a single-parameter fit to the lattice data, shown in \cref{fig:pipipiKmatrix:K0K1chpt}, we obtain,
\begin{equation}
    \begin{alignedat}{2}
        \elliso     &= 1.55(11)\,,    & \qquad \chi^2/\text{dof} &= 2.93/2\,,\\
        \ellisoone  &= 4.09(25)\,,    & \qquad \chi^2/\text{dof} &= 0.36/2\,,\\
    \end{alignedat}
\end{equation}
where we show the $\chi^2$ of the fits and the corresponding number of degrees of freedom (dof). This results are to be compared to the phenomenological values,
\cref{eq:pipipiKmatrix:LECref},
\begin{equation}
    \begin{aligned}
        \elliso     &= 1.19(25)\,, \qquad
        \ellisoone   &= 2.71(46)\,.
    \end{aligned}
\end{equation}
The fits are shown in \cref{fig:pipipiKmatrix:K0K1chpt}. 

In \cref{fig:pipipiKmatrix:K2KAKBchpt} we show the NLO predictions for $\Kisotwo$, $\KA$ and $\KB$. Note that all these vanish at LO in ChPT. In the case of $\KB$, we compare to the lattice results from \rcite{Blanton:2021llb}. We observe a similar situation as between $\Kisoone$ and LO ChPT, with a large discrepancy between NLO predictions and lattice results of $\Kdf$, even showing opposite signs. It is possible that, similar to the $\Kisoone$ case, this discrepancy gets reduced by large NNLO corrections.

\subsection{Range of validity of the threshold expansion}\label{sec:pipipiKmatrix:convergenceresults}

In lattice studies, the three-particle $K$-matrix is typically parametrized using a threshold expansion, which is assumed to be valid for all energies below the first inelastic threshold. In the three-pion case, this corresponds to CMF energies below the mass of five pions, $E^*<5\Mpi$. We can use our results to test the actual convergence of the threshold expansion by comparing them to the full $\Kdf$ evaluated numerically using the techniques described in \cref{sec:pipipiKmatrix:crosscheck}. We here show the results for the first kinematical family in \cref{tab:pipipiKmatrix:momentafamilies}, with other families leading to analogous conclusions.

\begin{figure}[!p]
    \centering
    \begin{subfigure}{0.67\textwidth} 
    \centering
        \includegraphics[width=1\textwidth]{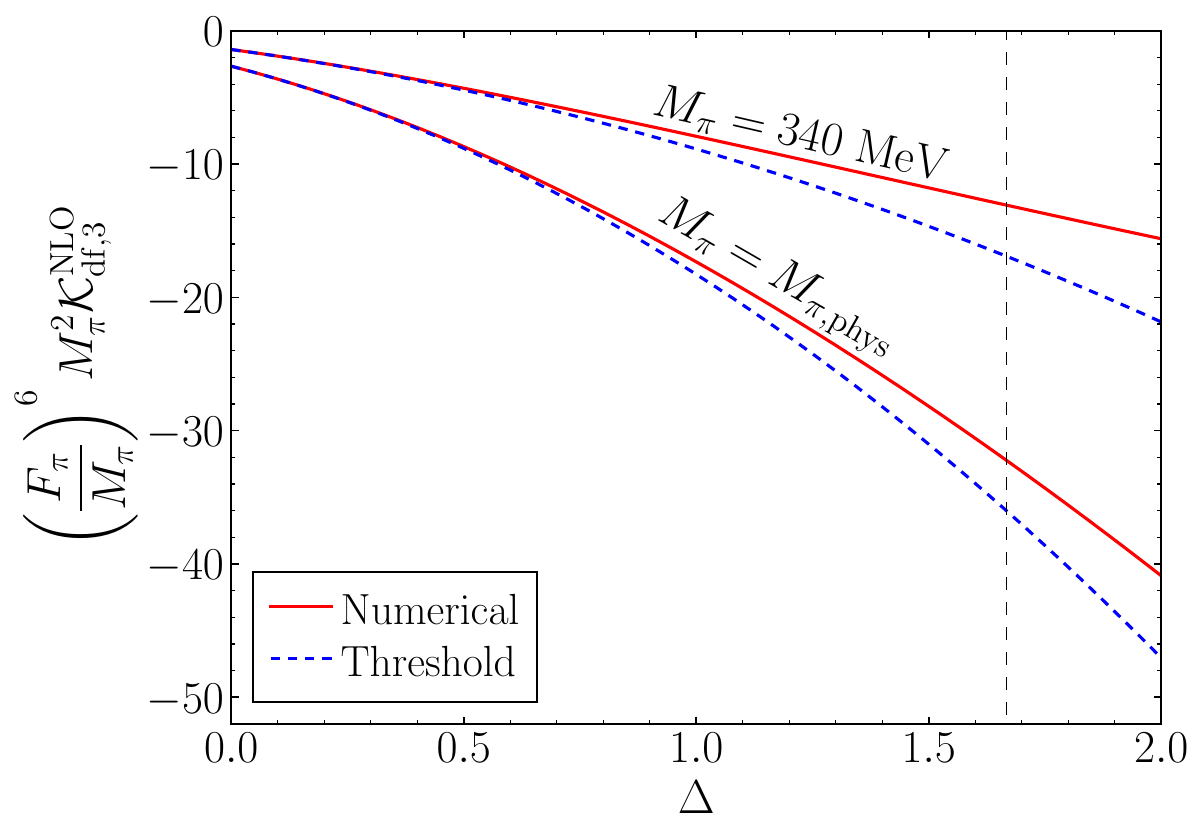}
    \end{subfigure}
    \caption{
         Comparison between numerical results and the threshold expansion for $\Kdf$, evaluated for the first momentum family in \cref{tab:pipipiKmatrix:momentafamilies}. The comparison is presented for two pion masses, with $\Mpi=340$ MeV corresponding to the heaviest pion mass used in \rcite{Blanton:2021llb}. The dashed vertical line indicates the inelastic threshold at $E^*=5\Mpi$.  }
    \label{fig:pipipiKmatrix:fullconvergence}\vspace{0.7cm}

    \centering
    \begin{subfigure}{0.47\textwidth} 
    \centering
        \includegraphics[width=1\textwidth]{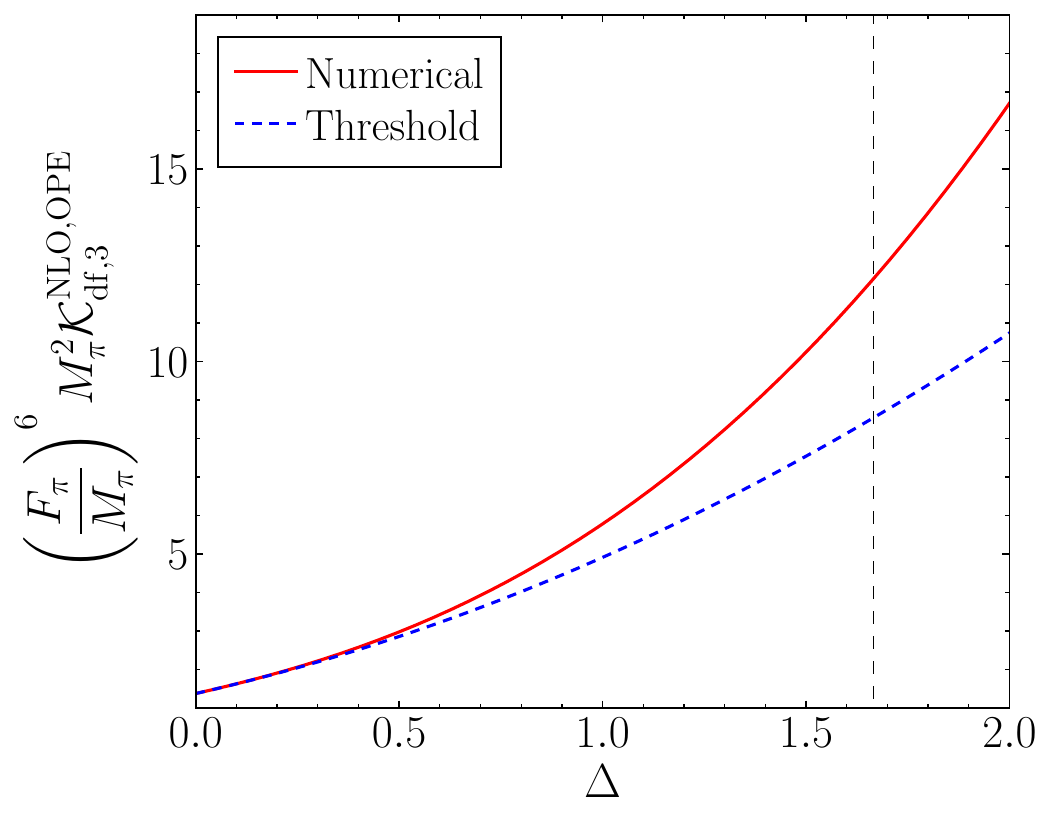}
    \end{subfigure}
     \begin{subfigure}{0.51\textwidth} 
    \centering
        \includegraphics[width=1\textwidth]{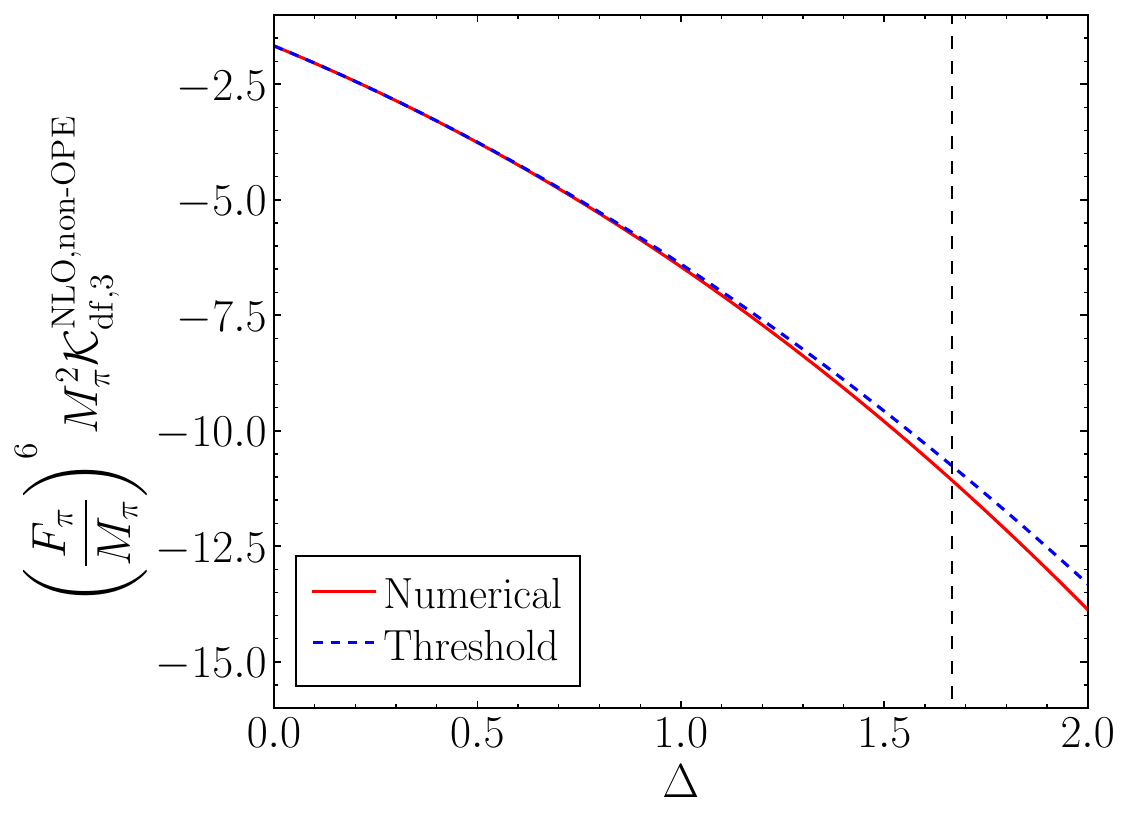}
    \end{subfigure}
    \begin{subfigure}{0.495\textwidth} 
    \centering
        \includegraphics[width=1\textwidth]{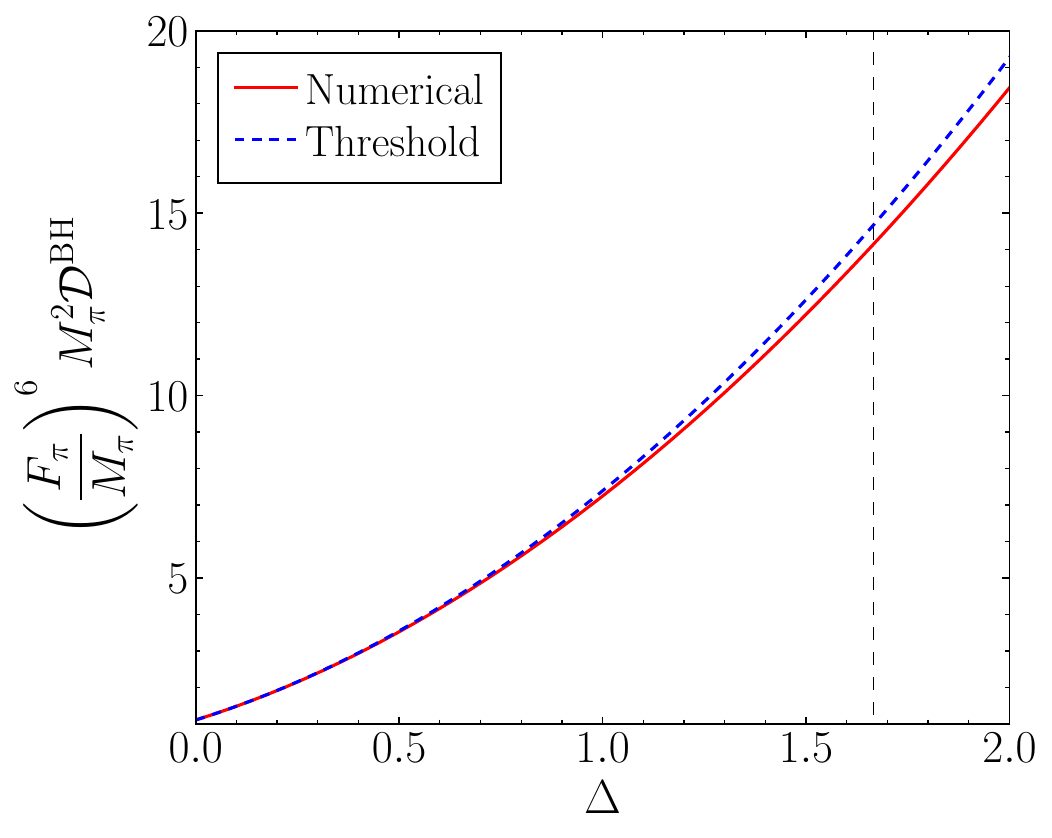}
    \end{subfigure}
    \caption{
         Comparison of the numerical and the threshold results for the different contributions to $\Kdf$---see \cref{eq:pipipiKmatrix:schematicequation}---evaluated for the first momentum family in \cref{tab:pipipiKmatrix:momentafamilies} for $\Mpi=340$ MeV.  The panels correspond to the OPE part (top left), non-OPE part (top right) and BH subtraction (bottom). The dashed vertical line indicates the inelastic threshold at $E^*=5\Mpi$.}
    \label{fig:pipipiKmatrix:partsconvergence}
\end{figure}

In \cref{fig:pipipiKmatrix:fullconvergence} we show this comparison for the full $\Kdf$ for physical pion mass and $\Mpi=340$ MeV, the heaviest value used in \rcite{Blanton:2021llb}. In both cases, we observe reasonable agreement between the full and the expanded results up to the inelastic threshold, where the discrepancy is $~10\%$ and $~20\%$ for physical mass and $\Mpi=340$ MeV, respectively. For the heavy pion mass, we present in \cref{fig:pipipiKmatrix:partsconvergence} a separate comparison for each of the terms in \cref{eq:pipipiKmatrix:schematicequation}. We observe good convergence for both the non-OPE and the BH subtraction part, at the $~5\%$ level in all the elastic regime, while the discrepancy is as big as $~30\%$ in the OPE part.

\begin{figure}[!tp]
    \centering
    \begin{subfigure}{0.7\textwidth} 
    \centering
        \includegraphics[width=1\textwidth]{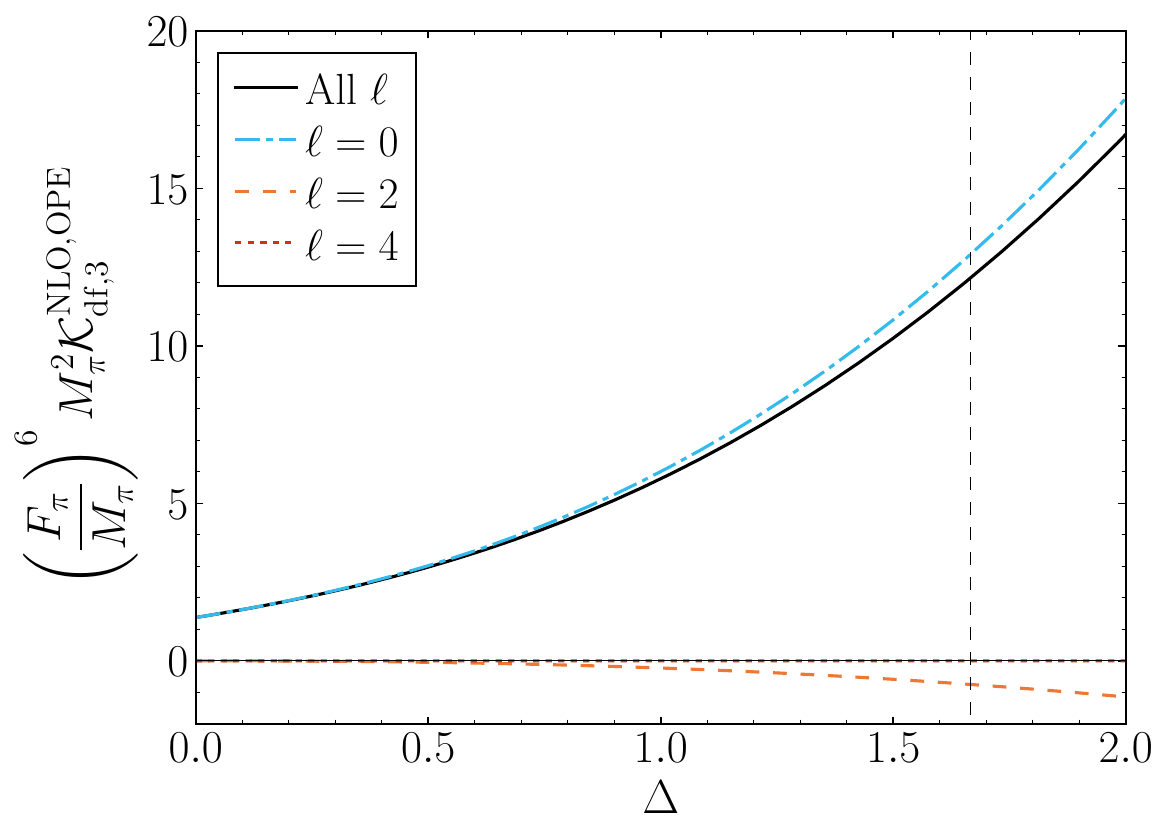}
    \end{subfigure}
    \caption{
         Comparison of contributions to $\Kdf^{\NLO,\OPE}$ from different partial waves of the NLO dimer amplitude, numerically evaluated for the first momentum family in \cref{tab:pipipiKmatrix:momentafamilies} with $\Mpi=340$ MeV. The black line is the full result including all partial waves, and the vertical dashed line is the inelastic threshold. Contributions from $\ell\geq 4$ are negligible.  }
    \label{fig:pipipiKmatrix:partialwaves}
\end{figure}

Finally, we also study the contribution of successive partial wave of the NLO amplitude to $\Kdf^\OPE$. While the threshold expansion only contains $s$ and $d$-waves, all even partial waves are present in the full result. However, as can be seen in \cref{fig:pipipiKmatrix:partialwaves} where the contribution of the first three nonzero partial waves is shown for $\Mpi=340$ MeV, there is rapid convergence with $\ell$. For $\ell\geq 4$, the contribution turns out to be negligible in the elastic region. A similar result holds for lighter pion masses.

\section{Conclusions}\label{sec:pipipiKmatrix:conclusions}

In this chapter the results from \rcite{Baeza-Ballesteros:2023ljl} have been presented, in which the divergence-free $K$-matrix of three pions at maximal isospin was computed at NLO in ChPT, motivated by the observed discrepancies between lattice results for $\Kiso$ and $\Kisoone$ and LO ChPT. Our main results are shown in \cref{eq:pipipiKmatrix:results,eq:pipipiKmatrix:Dvalues,eq:pipipiKmatrix:LECscombinations}. We have found that NLO corrections to these coefficients are indeed large and reduce the tension between lattice results and LO ChPT. This, however, may also be an indication that the convergence of the chiral expansion is rather poor for these quantities. %For the $\KB$ coefficient we observed a sign disagreement between lattice results and NLO ChPT, which 

In this work we developed a series of techniques to determine $\Kdf$ at NLO in ChPT analytically, up to some cutoff-dependent corrections that need to be evaluated numerically. We found that at NLO in ChPT the integral equations relating $\cM_3$ to $\Kdf$ simplify to a linear relation, $\Kdf^\NLO=\Re\cM_3^\NLO$, which can be separated into different non-singular contributions that we determined independently. This work opens the door to the computation of $\Kdf$ in ChPT for more complex systems, such as that of three pions at non-maximal isospin~\cite{Baeza-Ballesteros:2024mii}, which are presented in the next chapter. % In the next chapter I present the results from \rcite{Baeza-Ballesteros:2024mii}, in which we determined $\Kdf$ at NLO in ChPT for three pions at any isospin.

Finally, we have studied the convergence of the threshold expansion. We found good agreement between the expansion truncated at quadratic order and the full result, with corrections smaller than $20\%$ in the elastic region, $E*<5\Mpi$. This gives confidence in the usage of a threshold expansion to parametrize $\Kdf$ when extracting it from numerical lattice simulations. We also analyzed the contributions from different dimer partial waves to the OPE part, concluding that partial waves with $\ell>2$ are negligible.

\chapter[The three-pion $K$-matrix at NLO in ChPT]{The three-pion $K$-matrix at NLO in ChPT\phantom{y}}
\label{sec:isospinKmatrix}

Lattice studies of three particles have been mainly limited to pion and kaon states at maximal isospin~\cite{Mai:2018djl,Horz:2019rrn,Blanton:2019vdk,Mai:2019fba,Culver:2019vvu,Fischer:2020jzp,Hansen:2020otl,Alexandru:2020xqf,Brett:2021wyd,Blanton:2021llb,Draper:2023boj}. However, it is expected that more complicated systems will be investigated in the early future. Of particular interest are systems of three pions at non-maximal isospin, in which some resonances can be found that only decay into three particles, such as the $\omega(782)$ resonance~\cite{PDG:2020} in the isospin-zero channel.

%In the previous chapter, which covered the results of \rcite{Baeza-Ballesteros:2022azb}, the determination of the three-pion $K$-matrix at NLO in ChPT was presented for the isospin-three channel. We found large NLO corrections that reduced the observed discrepancy between lattice results and LO ChPT. Also, our results provided insight on the convergence of the chiral expansion for three-pion observables and served as a check for the RFT formalism.

In this chapter, the results from \rcite{Baeza-Ballesteros:2024mii} are presented, in which the three-pion $K$-matrix is determined up to NLO for all three-pion isospin channels. While the validity of ChPT at non-maximal isospin is limited by the presence of resonances,\footnote{This is especially true for heavier-than-physical pion masses, for which resonances lie closer to the two- or three-pion threshold, since the mass of resonances has been found to depend only weakly on $\Mpi$~\cite{Yu:2023xxf}.} these results can still prove useful to inspire parametrizations or put constrains on $\Kdf$ in the near-threshold energy region, and also contribute towards our understanding of the convergence of the chiral and threshold expansions.

The computation of $\Kdf$ for non-maximal isospin uses the same techniques developed for the isospin-three case presented in \cref{sec:pipipiKmatrix}, \mbox{starting} from the amplitudes in \rrcite{Bijnens:2021hpq,Bijnens:2022zsq}. However, it includes some new complications. These are related to new kinematic structures in the threshold expansion of $\Kdf$ in the different isospin sectors, $3\pi\rightarrow\pi \rightarrow 3\pi$ diagrams for isospin one, and $p$-waves in some of the channels in both LO and NLO amplitudes.

This chapter is organized as follows. In \cref{sec:isospinKmatrix:Kdfgeneralisospin} we discuss how the integral equations relating $\cM_3$ to $\Kdf$---see \cref{sec:hadrons:infinitevolumethreeparticlescattering}---are extended to include all isospin channels, and how they can be used to determine $\Kdf$ in ChPT. We also introduce in \cref{sec:isospinKmatrix:thresholdexpansionform} the threshold expansion of $\Kdf$ for the all isospin channels, which was first worked out in \rcite{Hansen:2020zhy}. The main steps of the computation of $\Kdf$ at LO and NLO are described in \cref{sec:isospinKmatrix:computationLONLO}. Finally, the results are presented in \cref{sec:isospinKmatrix:discussion}, where we also comment on the convergence of the threshold expansion

\section{$\Kdf$ for general isospin}\label{sec:isospinKmatrix:Kdfgeneralisospin}

\subsection{The flavor space}\label{sec:isospinKmatrix:flavorspace}

Three-pion interactions classify in four different isospin channels, $\Ippp=3,2,1$ and $0$---see \cref{sec:hadrons:interactionsinChPT}. Each of these channels has a different multiplicity, that corresponds to the different two-particle channels in which pairwise interactions can occur within the three-particle system. Both the $\Ippp=3$ and $\Ippp=0$ channels are one dimensional, as pairwise interactions only happen with two-particle isospin $\Ipp=2$ and $\Ipp=1$, respectively. The $\Ippp=2$ channel is two dimensional, containing both $\Ipp=2$ and $\Ipp=1$. Finally, the $\Ippp=1$ channel is three dimensional, and pairwise interactions occur in all three two-pion isospin channels.

The presence of different isospin channels implies that scattering quantities describing three-pion interactions are $7\times7$ matrices in flavor space---we neglect the third component of isospin, since interactions are diagonal in this quantum number. We indicate quantities that are matrices in flavor space using boldface font. For example, the general-isospin scattering amplitude is denoted as $\mM_3$, while $\mKdf$ is the three-pion $K$-matrix for all isospin. 

The particular form of observables depends on the choice of basis for the flavor basis. In the case of single-pion quantities, two different basis are useful. The \textit{charge basis}, $\{|\pi^+\rangle,|\pi^0\rangle,|\pi^-\rangle\}$, is composed of states of definite isospin, and so is suited for building multiparticle states. For the low-energy EFT computations, on the other hand, it is more common to use the \textit{flavor basis}, $\{|i\rangle\}$, in which the pion matrix is defined as $|\phi\rangle = \sum_i\sigma_i|i\rangle$, with $\sigma_i$ the Pauli matrices. These two basis are related as
\begin{equation}\label{eq:isospinKmatrix:singlepionbasisrelations}
    \ket{\pipm} = \mp\,\frac{\ket{1} \pm i\ket{2}}{\sqrt{2}}\,,\quad\quad\quad \ket{\pio} = \ket{3}\,,
\end{equation}
where we use the Condon--Shortley sign convention~\cite{Arfken:379118}.

Three-pion states are typically constructed combining single-pion states in the charge basis. Still, different basis of three-pion states can be chosen. One option is to work with states of definite three-particle isospin in which the flavor-space matrices become block-diagonal, with blocks of a size equal to the multiplicity of each isospin channel. This choice still does not fully fix the basis, as states can be rotated within each isospin block. A typical option, related to the finite-volume quantization condition~\cite{Hansen:2020zhy}, is to choose three-particle states with definite two-particle isospin between the first two particles of the state. This is the called the \textit{isospin basis},
\begin{equation}\label{eq:isospinKmatrix:isospinbasis}
    \ket{\pi\pi\pi}_{\mathrm I} = 
    \left(
    \begin{array}{l}
        \ket{\Pi\pi}_3   \vphantom{\Big(}\\
        \ket{\Pi\pi}_2   \vphantom{\Big(}\\
        \ket{\rho\pi}_2  \vphantom{\Big(}\\
        \ket{\Pi\pi}_1   \vphantom{\Big(}\\
        \ket{\rho\pi}_1  \vphantom{\Big(}\\
        \ket{\sigma\pi}_1\vphantom{\Big(}\\
        \ket{\rho\pi}_0\vphantom{\Big(}\\
    \end{array}
    \right)
    = 
    \left(
    \begin{array}{l}
       \displaystyle \tfrac1{\sqrt5}   \Big(\ket{\Pip\pim} + \sqrt3\ket{\Pio\pio} + \ket{\Pim\pip}\Big)          \\
       \displaystyle \tfrac1{\sqrt2}   \Big(\ket{\Pip\pim} - \ket{\Pim\pip}\Big)                                 \\
       \displaystyle \tfrac1{\sqrt6}   \Big(\ket{\rop\pim} + 2\ket{\roo\pio} + \ket{\rom\pip}\Big)               \\
       \displaystyle \tfrac1{\sqrt{10}} \Big(\sqrt3\ket{\Pip\pim} - 2\ket{\Pio\pio} + \sqrt3\ket{\Pim\pip}\Big)   \\
       \displaystyle \tfrac1{\sqrt2}   \Big(\ket{\rop\pim} - \ket{\rom\pip}\Big)                                 \\
       \displaystyle \vphantom{\Big(}       \ket{\sigma\pio}                                                     \\
       \displaystyle \tfrac1{\sqrt3}   \Big(\ket{\rop\pim} - \ket{\roo\pio} + \ket{\rom\pip}\Big)                \\
    \end{array}
    \right),
\end{equation}
where the subscript of the states  on the left-hand side indicates their three-particle isospin. On the right-hand side we have defined the two-particle states with definite isospin,
\begin{equation}
    \begin{gathered}
        \ket{\Pio}      =  \frac{\ket{\pip\pim} + \ket{\pim\pip} + 2\ket{\pio\pio}}{\sqrt6}\,,    \qquad
        \ket{\Pipm}     =  \frac{\ket{\pipm\pio} + \ket{\pio\pipm}}{\sqrt2}\,,                   \\ 
        \ket{\roo}      =  \frac{\ket{\pip\pim} - \ket{\pim\pip}}{\sqrt2}\,,                      \qquad
        \ket{\ropm}     =  \pm\frac{\ket{\pipm\pio} - \ket{\pio\pipm}}{\sqrt2}\,, \\
        \ket{\sigma}    =  \frac{\ket{\pip\pim} + \ket{\pim\pip} - \ket{\pio\pio}}{\sqrt3}\,,       
    \end{gathered}
\end{equation}
where the states in the top, central and bottom line have $\Ipp=2,1$ and 0, in this same order. Note that the name given to each state is reminiscent of the resonances present in each isospin channel. The $\rho(770)$ resonance appears in isospin-one pion-pion scattering, while the $\sigma(550)$ resonance is present in the isospin-zero channel.
%This is, two-pion states with $I_{\pi\pi}=0$ are called $|\sigma\rangle$ as this channel contains the $\sigma$ resonance, states with $I_{\pi\pi}=1$ are denoted as $\rho\rangle$, which is the resonance present in this channel, and states with $I_{\pi\pi}=2$, for which no resonance is present, are labelled $|\Pi\rangle$.

An alternative basis that keeps the block structure is the \textit{symmetric basis}. In this, the states within each isospin block are chosen to have definite transformation properties under the $S_3$ group, that describes permutations between the three pions. It is related to the isospin basis by a rotation of the $\Ippp=1$ block,
%This is the bast choice to work the threshold expansion, as the kinematical structures appearing in the expansion can be systematically studied based on the properties of this group (toether with time-reversal invariance). We called this the symmetric basis, related to the isospin basis as

\noindent\begin{equation}\label{eq:isospinKmatrix:symmetricbasis}
    \ket{\pi\pi\pi}_{\mathrm S} = 
    \left(
    \begin{array}{l}
        \ket{\chi_\text{s}}_3  \vphantom{\tfrac{\sqrt5}3}\\
        \ket{\chi_1}_2  \vphantom{\tfrac{\sqrt5}3}\\
        \ket{\chi_2}_2  \vphantom{\tfrac{\sqrt5}3}\\
        \ket{\chi_\text{s}}_1  \vphantom{\tfrac{\sqrt5}3}\\
        \ket{\chi_1}_1  \vphantom{\tfrac{\sqrt5}3}\\
        \ket{\chi_2}_1  \vphantom{\tfrac{\sqrt5}3}\\
        \ket{\chi_\text{a}}_0  \vphantom{\tfrac{\sqrt5}3}\\
    \end{array}
    \right)
    =
    \left(
    \begin{array}{l}
        \ket{\Pi\pi}_3  \vphantom{\tfrac{\sqrt5}3}\\
        \ket{\Pi\pi}_2  \vphantom{\tfrac{\sqrt5}3}\\
        \ket{\rho\pi}_2 \vphantom{\tfrac{\sqrt5}3}\\
        \hphantom{+}\tfrac23\ket{\Pi\pi}_1 + \tfrac{\sqrt5}3\ket{\sigma\pi}_1   \\
        -\tfrac{\sqrt5}3\ket{\Pi\pi}_1 + \tfrac23\ket{\sigma\pi}_1  \\
        \ket{\rho\pi}_1 \vphantom{\tfrac{\sqrt5}3}\\
        \ket{\rho\pi}_0  \vphantom{\tfrac{\sqrt5}3}\\
    \end{array}
    \right).
\end{equation}
Here $\ket{\chi_\text{s}}$ denotes a state that transforms under the one-dimensional trivial (symmetric) irrep of $S_3$, $\ket{\chi_\text{a}}$ refers to a state transforming under the alternating (antisymmetric) irrep, and $\{\ket{\chi_1},\ket{\chi_2}\}$ is a pair of states transforming under the two-dimensional standard irrep. Thus, states with $\Ippp=3$ transform under the trivial irrep, those with $\Ippp=2$ under the standard irrep, states with $\Ippp=1$ under a direct sum of the trivial and the standard irreps, and $\Ippp=0$ states under the alternating irrep.

To construct the elements of the quantization condition, \rcite{Hansen:2020zhy} makes use of  a different basis composed of states with zero electric charge. This is known as the \textit{charge basis},
\begin{equation}
    \ket{\pi\pi\pi}_{\mathrm C} = 
    \begin{pmatrix}
        \ket{\pim\pio\pip}  \\
        \ket{\pio\pim\pip}  \\
        \ket{\pim\pip\pio}  \\
        \ket{\pio\pio\pio}  \\
        \ket{\pip\pim\pio}  \\
        \ket{\pio\pip\pim}  \\
        \ket{\pip\pio\pim}  \\
    \end{pmatrix}.
    \label{eq:isospinKmatrix:chargebasisthreepions}
\end{equation}
It is related to the isospin and symmetric basis by rotation---see \rcite{Baeza-Ballesteros:2024mii}.%All these basis are related by some rotation. The particular matrices that relate the charge basis to the isospin and symmetry basis can be found in \rcite{}.

%\subsection{Relating $\mM_3$ to $\mKdf$}
\subsection{Building blocks of the integral equations}

When extended to general isospin, the building blocks of the integral equations that relate the scattering amplitude to the divergence-free $K$-matrix---see \cref{sec:hadrons:infinitevolumethreeparticlescattering}---become $7\times 7$ matrices in flavor space. This is the case of the two-particle amplitude that describes pairwise interactions between the first two particles of each state. In the charge basis, it is block diagonal,
\begin{equation}
    \label{eq:isospinKmatrix:flavorM2dimer}
    \m M_2 = 
        \text{diag}\left(
            \m M_2^+ \,,\, \m M_2^0 \,,\, \m M_2^-\right)\,,
\end{equation}
with blocks
\begin{align}
&\m M_2^0 =
        \begin{pmatrix}
            A_{4\pi}(s) + A_{4\pi}(t) &   -A_{4\pi}(s)   &   A_{4\pi}(s) + A_{4\pi}(u) \\[-5pt]
            -A_{4\pi}(s)   &   A_{4\pi}(s) + A_{4\pi}(t) + A_{4\pi}(u)  &    -A_{4\pi}(s)  \\[-5pt]
            A_{4\pi}(s) + A_{4\pi}(u) &   -A_{4\pi}(s)   &   A_{4\pi}(s) + A_{4\pi}(t) \\
        \end{pmatrix}\,,
    \nonumber\\    
    &\m M_2^+=\m M_2^- = 
        \begin{pmatrix}
            A_{4\pi}(t)    &   A_{4\pi}(u)    \\[-5pt]
            A_{4\pi}(u)    &   A_{4\pi}(t)    \\
        \end{pmatrix}\,,
\end{align}
where $A_{4\pi}$ is the four-pion amplitude in ChPT, given in \cref{eq:pipipiKmatrix:twopionamplitudeLO,eq:pipipiKmatrix:twopionamplitudeNLO} at LO and NLO, respectively. Here, we use $A_{4\pi}(s)=A_{4\pi}(s,t,u)$ as a shorthand notation, and similarly for $A_{4\pi}(t)$ and $A_{4\pi}(u)$, since $A_{4\pi}$ is symmetric in its last two arguments.

Similarly, the $G^\infty$ factor becomes a $7\times 7$ matrix,
\begin{equation}\label{eq:isospinKmatrix:Ginfty}
\mG^\infty_{\ell'\ell}=G^\infty_{\ell'\ell}\mT_G\,,
\end{equation}
where $\mT_G$ encodes the valid exchanges between the three-particle states, where the exchanged particle is the second one in the naming of the states. In the charge basis,
\begin{equation}
\setstacktabbedgap{6pt}
\setstackgap{L}{15pt}
    \m T_G = 
        \parenMatrixstack{
            \square      &\square      &\square      &\square      &\square      &\square      &\blacksquare \cr
            \square      &\square      &\square      &\square      &\blacksquare &\square      &\square      \cr
            \square      &\square      &\square      &\square      &\square      &\blacksquare &\square      \cr
            \square      &\square      &\square      &\blacksquare &\square      &\square      &\square      \cr
            \square      &\blacksquare &\square      &\square      &\square      &\square      &\square      \cr
            \square      &\square      &\blacksquare &\square      &\square      &\square      &\square      \cr
            \blacksquare &\square      &\square      &\square      &\square      &\square      &\square      
        },\qquad \square=0\,,\quad\blacksquare=1\,,
    \label{eq:Tgdef}
\end{equation}
where we use squares rather than numbers for legibility. This same flavor structure can be used, in combination with \cref{eq:isospinKmatrix:flavorM2dimer}, to build the OPE part of the three-pion scattering amplitude---see \cref{eq:isospinKmatrix:OPEcomputation}. %Note that $\mM_2$ can also be used to compute the OPE part of the amplitude, using a correlator $\mT_G/(\bar b^2+i\epsilon)$ for the exchanged particle.

In addition to working with matrices in flavor space, the symmetrization appearing in \cref{eq:hadrons:divergencefreeamplitude,eq:hadrons:TandMdfintegralequation} is modified to ensure the correct transformation properties under particle exchange, affecting momenta and flavor simultaneously. For example, the symmetrized divergence-free amplitude is defined from the unsymmetrized one as
\begin{equation}\label{eq:isospinKmatrix:symmetrization}
    \mMdf = \sum_{m=0}^2\sum_{n=0}^2 \big(\m R^m\big)^{\!\intercal} \mMdf^\uu\big(R^m\{p_i\},R^n\{k_i\}\big) \m R^n\,,
\end{equation}
where $R\{p_1,p_2,p_3\} = \{p_2,p_3,p_1\}$ is a cyclic permutation of the three momenta %(only cyclic permutations need to be considered fue to the symmetry of the dimer) 
and $\m R$ is the representation of the permutation in the space of three pions. In the charge basis it reads,
\begin{equation}
\setstacktabbedgap{6pt}
\setstackgap{L}{15pt}
    \m R = 
    \parenMatrixstack{
        \square      &\square      &\square      &\square      &\blacksquare &\square      &\square      \cr
        \square      &\square      &\square      &\square      &\square      &\square      &\blacksquare \cr
        \square      &\blacksquare &\square      &\square      &\square      &\square      &\square      \cr
        \square      &\square      &\square      &\blacksquare &\square      &\square      &\square      \cr
        \square      &\square      &\square      &\square      &\square      &\blacksquare &\square      \cr
        \blacksquare &\square      &\square      &\square      &\square      &\square      &\square      \cr
        \square      &\square      &\blacksquare &\square      &\square      &\square      &\square      
    }
    \,,\qquad \square=0\,,\quad\blacksquare=1\,.
\end{equation}
This same symmetrization procedure applies to any other unsymmetrized quantity as well.

\subsection{Relating $\mM_3$ to $\mKdf$}

\begin{figure}[t!]
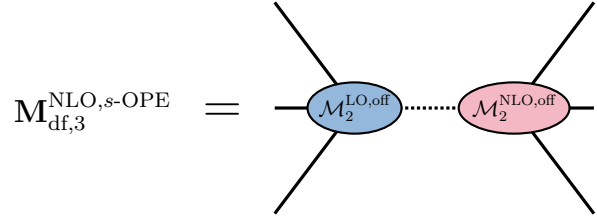

    \begin{equation*}
        {\displaystyle\mMdf^{\NLO,\sOPE}}\quad\scalebox{\sketchoperatorscale}{$=$}\quad
            \tikzineq[xscale=2.1,yscale=1.4]{
                \makeexternallegcoordinates
                \coordinate (v1) at (+.5,0);
                \coordinate (v2) at (-.5,0);
                \draw[sketch onshell prop] (k1) -- (v1);
                \draw[sketch onshell prop] (k2) -- (v1) -- (k3);
                \draw[sketch onshell prop] (p3) -- (v2);
                \draw[sketch onshell prop] (p1) -- (v2) -- (p2);
                \draw[sketch offshell prop] (v1) -- (v2);
                \node[sketch offshell blob=NLOcolor, inner ysep=2pt, inner xsep=-1pt]
                    (NLO) at (v1) {\scalebox{.7}{$\cM_2^{\NLO,\off}$}};
                \node[sketch offshell blob=LOcolor, inner ysep=2pt, inner xsep=-1pt]
                    (LO) at (v2) {\scalebox{.7}{$\cM_2^{\LO,\off}$}};
            }
    \end{equation*}
    \caption{
        Schematic representation of the $s$-channel OPE diagram, which needs to be added to \cref{fig:pipipiKmatrix:diagramKdf} to compute $\mKdf$. It only contributes to the  $\Ippp=1$ channel.
        External pions are on-shell, while the intermediate one is always off-shell, represented with a dotted line.
        Also, the amplitudes, both the LO and the NLO ones, have one leg off-shell.
        One also needs to consider an additional diagram with the $\LO$ and $\NLO$ amplitudes exchanged.
        }
    \label{fig:isospinKmatrix:diagramsOPE}
\end{figure}

As shown in \cref{sec:hadrons:ChPTthreepionsLO,sec:pipipiKmatrix:algebraicrelation} for the isospin-three channel, the relation between $\mM_3$ and $\mKdf$ becomes algebraic when working at LO and NLO in ChPT. At LO, the three-pion $K$-matrix is just the divergence-free amplitude, 
\begin{equation}
\mKdf^\LO = \mMdf^\LO\,,
\end{equation}
while at NLO,
\begin{equation}
\mKdf^\NLO = \Re \mMdf^\NLO\,.
\end{equation}
 As done for the $\Ippp=3$ case, it is convenient to subdivide the computation of the divergence-free amplitude in several parts. For the LO and NLO quantity, we write, respectively,
\begin{equation}
     \mMdf^\LO =  \left( \mMdf^{\LO,\OPE} -  \m D^{\LO,\OPE} \right) +   \mMdf^{\LO,\sOPE} + \mMdf^{\LO,\nOPE}\,, \label{eq:isospinKmatrix:LOseparation}
\end{equation}
\begin{multline}
     \text{Re} \mMdf^\NLO =  \left( \text{Re} \mMdf^{\NLO,\OPE} - \text{Re} \m D^{\NLO,\OPE} \right)\\[5pt]
      +  \text{Re} \mMdf^{\NLO\sOPE} + \text{Re} \mMdf^{\NLO,\nOPE} - \text{Re} {\m D}^{\NLO,\BH}\,. \label{eq:isospinKmatrix:NLOseparation}
\end{multline}
Note this decomposition is very similar to that used for the $\Ippp=3$ channels, diagramatically represented in \cref{fig:pipipiKmatrix:diagramKdf} at NLO. The main novelty is the so-called \textit{$s$-channel OPE} ($s$-OPE) contribution,  $\mMdf^{\sOPE}$, diagramatically represented in \cref{fig:isospinKmatrix:diagramsOPE}. This is only present in the $\Ippp=1$ channel, and is characterized by a single exchanged particle which carries all the momenta and isospin. The remaining terms in \cref{eq:isospinKmatrix:LOseparation,eq:isospinKmatrix:LOseparation} are the subtracted OPE, the non-OPE and the BH subtraction, in this same order, where the last one only appears at NLO. Each of the four contributions can be evaluated independently, as will be shown in \cref{sec:isospinKmatrix:computationLONLO}.

\newpage\subsection{Threshold expansion of $\mKdf$}\label{sec:isospinKmatrix:thresholdexpansion}\label{sec:isospinKmatrix:thresholdexpansionform}

Before moving to the computation of $\mKdf$, we discuss how it is expanded around threshold for the different isospin channels. These expansions were partially presented in \rcite{Hansen:2020zhy} for the first time. The result for three identical particles was presented in \cref{eq:hadrons:thesholdexpansionKdf}. This applies to the $\Ippp=3$ channel, since the corresponding pion states are symmetric under particle exchange. We reproduce it here for completion,
\begin{equation}
M_\pi^2\mKdf^{\Ippp=3}=\Kiso+\Kisoone\Delta+\Kisotwo\Delta^2+\KA\DA+\KB\DB+\cO(\Delta^3)\,,
\end{equation}
where $\Delta$, $\Delta_\text{A}$ and $\DB$ are defined in \cref{eq:hadrons:thesholdexpansionKdfingredients}. Note that, even if the $\Ippp=3$ flavor space is one-dimensional, we keep the boldface notation. % for the $K$-matrix. 

States with non-maximal isospin, on the other hand, present different transformation properties under particle exchange, which characterize the possible kinematic operators that can appear in the expansion. In app.~B of \rcite{Baeza-Ballesteros:2024mii}, we computed the number of independent operators appearing up to cubic order in the expansion using group theoretical arguments. In this section, we summarize the form of said operators up to quadratic order in the $\Ippp=2$ and $1$ channels, and up to cubic order for $\Ippp=0$.

%In this section we present the form of the threshold expansion for the remaining isospin channels, including terms up to quadratic order for $I=2,1$, and up to cubic order for $I=0$, since the constant and linear order vanish. We let the initial and final momenta be $\{k_i\}$ and $\{p_i\}$, respectively, where particles in the states above are assigned momenta in order. The total momenta is $P=k_1+k_2+k_3=p_1+p_2+p_3$. We will usually use a prime to refer to final-quantities.

%Before moving to the actual computation, we discuss about the form of the threshold expansion of $\mKdf$ for the different isospin channels.  States of three pions at non-maximal isospin have in general different transformation properties under particle exchange, that correspond to different irreps of the $S_3$ group. In App.~B of \rcite{} we presented an analysis based on group theoretical arguments that allowed us to determine the number of independent kinematical operators present at each order. Note that this expansion was first partially presented at \rcite{}, although we are changing the normalization here to make all quantities dimensionless.

The $\Ippp=2$ channel involves a two-dimensional flavor space, with states transforming under the standard irrep of $S_3$. Kinematic operators appearing in the threshold expansion are thus doublets transforming in this same irrep under particle exchange of either the initial or the final state. 

Following \rcite{Hansen:2020zhy}, we first introduce single-state doublets. At linear order in $\Delta$ we have, for the initial state,
\begin{equation}
    \boldsymbol{\xi}^\mu=(\xi_1^\mu,\xi_2^\mu)\,,
\end{equation}
where 
\begin{equation}
\xi_1=\frac{2k_3-k_1-k_2}{\sqrt6 \Mpi}\,,\quad\quad \xi_2=\frac{k_2-k_1}{\sqrt2 \Mpi}\,.
\end{equation}
At quadratic order, one can construct two independent tensor doublets, 
\begin{equation}
\boldsymbol{\xi}(S)^{\mu\nu}=\frac{\boldsymbol{\xi}^\mu P^\nu+\boldsymbol{\xi}^\nu P^\mu}{\Mpi}\,,
\end{equation}
\begin{equation}
\boldsymbol{\xi}(\bar{S})^{\mu\nu}=\left({\xi}_2^\mu{\xi}_2^\nu-{\xi}_1^\nu{\xi}_1^\mu,{\xi}_1^\mu{\xi}_2^\nu-{\xi}_2^\nu{\xi}_1^\mu\right)\,,
\end{equation}
and one scalar doublet,
\begin{equation}\label{eq:isospinKmatrix:xi2operator}
\boldsymbol{\xi}^{(2)}=\left( \xi_1^{(2)}, {\xi}_2^{(2)}\right)=-\frac2{9\Mpi}\boldsymbol{\xi}^\mu P_\mu\,,
\end{equation}
with
\begin{equation}
\xi_1^{(2)}=\frac{2\Delta_3-\Delta_1-\Delta_2}{\sqrt6}\,,\quad\quad \xi_2^{(2)}=\frac{\Delta_2-\Delta_1}{\sqrt2}\,.
\end{equation}

Using these quantities and their final-state counterparts, one can write down four operators with the correct transformation properties,
\begin{equation}
    \begin{array}{rl}
        \mXi{1} &\equiv \vxip[\mu]\otimes\vxi_\mu\,, \\
        \mXi{2} &\equiv \vxip[(2)]\otimes\vxi[(2)]\,, \\
        \mXi{3} &\equiv \tfrac{1}{\sqrt6}\big[\vxipSb[\mu\nu]\otimes\vxiS_{\mu\nu} + \vxipS[\mu\nu]\otimes\vxiSb_{\mu\nu}\big]\,,   \\
        \mXi{4} &\equiv \vxipSb[\mu\nu]\otimes\vxiSb_{\mu\nu}\,,
    \end{array}
\end{equation}
where $\otimes$ indicates tensor product. Combining them with $\Delta$, we find that the threshold expansion of $\mKdf$ in the isospin-two channel is,
\begin{equation}\label{eq:isospinKmatrix:thresholdexpansionI2}
    \Mpi^2 \mKdfI{2} = \Big(\KT + \KT[1]\Delta\Big)\:\mXi{1} + \sum_{n=2,3,4} \KT[n]\: \mXi{n} + \cO(\Delta^3)\,.
\end{equation}

In the $\Ippp=1$ channel, states transform under a direct sum of the standard and the trivial irrep of $S_3$. It is convenient to work in the symmetric basis, given in \cref{eq:isospinKmatrix:symmetricbasis}, putting the singlet as the first component.
\begin{equation}
    \mKdfI{1} = 
    \begin{pmatrix}
        \mKdfI{1,\SS}    &   \mKdfI{1,\SD}    \\
        \mKdfI{1,\DS}    &   \mKdfI{1,\DD}    \\
    \end{pmatrix}.
\end{equation}
The singlet-singlet (SS) term is equivalent to the $\Ippp=3$ case, \cref{eq:hadrons:thesholdexpansionKdf},
\begin{equation}
    \Mpi^2\mKdfI{1,\SS} = \KSS + \KSS[1]\Delta + \KSS[2]\Delta^2 + \KSSA\DA + \KSSB\DB\ + \cO(\Delta^3)\,,
\end{equation}
while the doublet-doublet (DD) is analogous to the $\Ippp=2$, \cref{eq:isospinKmatrix:thresholdexpansionI2},
\begin{equation}
    \Mpi^2\mKdfI{1,\DD} = \Big(\KDD + \KDD[1]\Delta\Big)\:\mXi{1} + \sum_{n=2,3,4} \KDD[n]\: \mXi{n} + \cO(\Delta^3)\,.
\end{equation}
The novelty are the operators entering the singlet-doublet (SD)---final-state singlet and an initial-state doublet---and doublet-singlet (DS) sectors. We consider the SD case, with analogous results holding for the DS. At linear order in $\Delta$, only one kinematic operator exists, $\bm{\xi}^{(2)}$, defined in \cref{eq:isospinKmatrix:xi2operator}. At quadratic order, we can construct four operators: $\Delta\,\bm{\xi}^{(2)}$ and  $\vxi[(4,n)] = \big(\xi^{(4,n)}_1,\;\xi^{(4,n)}_2\big)$ for $n=2,3,4$, where
\begin{equation}
    \begin{array}{ll}
       \displaystyle \xi_1^{(4,2)}={\frac{2\Delta_3^2 - \Delta_1^2 - \Delta_2^2}{\sqrt6}}\,,\quad\quad\quad &
           \displaystyle \xi_2^{(4,2)}={\frac{\Delta_2^2-\Delta_1^2}{\sqrt2}}\,,\quad \\[10pt]
       \displaystyle \xi_1^{(4,3)}={\sum_i\frac{(\tij{i1}+\tij{i2})\tij{i3} - 2\tij{i1}\tij{i2}}{\sqrt6}}\,,\quad\quad\quad &
           \displaystyle \xi_2^{(4,3)}={\sum_i\frac{(\tij{i2}-\tij{i1})\tij{i3}}{\sqrt2}}\,,\\[10pt]
        \displaystyle\xi_1^{(4,4)}={\sum_i\frac{2\tij{i3}^{\,2} - \tij{i1}^{\,2} - \tij{i2}^{\,2}}{\sqrt6}}\,,\quad\quad\quad &
           \displaystyle  \xi_2^{(4,4)}={\sum_i \frac{\tij{i2}^{\,2}-\tij{i1}^{\,2}}{\sqrt2}}\,,
    \end{array}
\end{equation}
where $\Delta_i$ and $\tilde{t}_{ij}$ are defined in \cref{eq:hadrons:tvariablesDeltavariables}.
The threshold expansion of $\mKdfI{1,\SD}$ is  then
\begin{equation}
    \frac{\Mpi^2}{\sqrt{30}}\mKdfI{1,\SD} = \Big(\KSD + \KSD[1]\Delta\Big)\vxi[(2)] + \sum_{n=2,3,4}\KSD[n]\vxi[(4,n)] + \cO(\Delta^3)\,,
\end{equation}
where we have pulled out a factor of $\sqrt{30}$ for convenience. $\mKdfI{1,\DS}$ is obtained by changing $k_i\leftrightarrow p_i$ and taking the transpose. Note, however, that due to time-reversal invariance of the $K$-matrix, the coefficients are equal in both SD and DS sectors, i.e., $\KSD[i]=\KDS[i]$.

Finally, the $\Ippp=0$ channel is one dimensional in flavor space. States in this channel are totally antisymmetric under particle exchange, and so the operators entering the threshold expansion must also be antisymmetric under permutation of initial- and final-state particles. There is no such operator at constant or linear order in $\Delta$, while we have a single operator at quadratic order and three independent ones at cubic order,
\begin{equation}
    \Mpi^2\mKdfI{0} = \Big(\KAS + \KAS[1]\Delta\Big)\DeltaAS{2} + \KAS[3]\DeltaAS{3}  + \KAS[4]\DeltaAS{4} + \cO(\Delta^4)\,,
    \label{eq:isospinKmatrix:thresholdexpansionI0}
\end{equation}
where the explicit form of the operators is

\noindent\begin{equation}
    \begin{array}{rl}
        \displaystyle\DeltaAS{2} &\equiv \displaystyle\sum_{\substack{i,j,k\\m,n,r}} \epsilon_{ijk}\epsilon_{mnr} \tij{im}\tij{jn}\,,\\[10pt]
       \displaystyle \DeltaAS{3} &\equiv \displaystyle\sum_{\substack{i,j,k\\m,n,r}} \epsilon_{ijk}\epsilon_{mnr} \tij{im}\tij{jn}\tij{kr}\,,\\[10pt]
        \displaystyle\DeltaAS{4} &\equiv \displaystyle\sum_{\substack{i,j,k\\m,n,r}} \epsilon_{ijk}\epsilon_{mnr} \tij{im}\tij{jn}\big(\tij{im} + \tij{jn}\big)\,.
    \end{array}
\end{equation}

\newpage\section{Computation of $\mKdf$} \label{sec:isospinKmatrix:computationLONLO}

The threshold-expansion coefficients of $\mKdf$ can be determined almost analytically, as was the case of the isospin-three channel presented in \cref{sec:pipipiKmatrix}. In this section, we explain the details of this computation at both LO and NLO in ChPT. We treat separately each of the terms in \cref{eq:isospinKmatrix:LOseparation,eq:isospinKmatrix:NLOseparation}. The calculation is very similar to that of the $\Ippp=3$ channel detailed in \cref{sec:pipipiKmatrix:computation}, with a few changes and additional complications. The complete results  at LO are presented in \cref{tab:isospinKmatrix:LOresults} including the separate contribution of each term in \cref{eq:isospinKmatrix:LOseparation}. The results at NLO are summarized in \cref{tab:isospinKmatrix:NLOresults} complemented by \cref{tab:isospinKmatrix:remainders}, while the separate contribution from each term in \cref{eq:isospinKmatrix:NLOseparation} can be found in \rcite{Baeza-Ballesteros:2024mii}. %We note that the results presented here have been cross-checked numerically, using an extension of the presented in \cref{}.

The computation of all the different contributions share some common lines, which in some cases differ from the methodology used in \cref{sec:pipipiKmatrix:computation}. In all cases, all contributions are expanded about threshold and symmetrized when required in the charge basis. Moreover, the coefficients of the threshold  expansion are only identified after symmetrization. This identification involves converting all products of momenta to $\tilde{t}_{ij}$ variables, and then identifying the coefficients of the threshold expansion. At LO this can be done by inspection, but requires to solve a system of linear equations at higher order. Note this differs from the approach taken, for example, for the $\Ippp=3$ OPE part in \cref{sec:pipipiKmatrix:OPE}, where the coefficients were worked out for the unsymmetrized $K$-matrix and then related to the symmetric result. However, the form of the expansion for the unsymmetrized $K$-matrix is not known at non-maximal isospin.

\subsection{The subtracted-OPE contribution} 

The subtracted-OPE contribution can be computed in a similar fashion to the $\Ippp=3$ channel, working first with unsymmetrized quantities. The unsymmetrized divergence-free OPE amplitude is,
\begin{equation}
    \mMdf^{\uu,\OPE} = -\m M_{2,\off} \frac{\m T_G}{\bar b^2+i\epsilon}\m M_{2,\off}
    + \sum_{\ell' \ell} \m M_{2,\ell'} \mG^\infty_{\ell' \ell} \m M_{2,\ell}\,,
    \label{eq:isospinKmatrix:OPEcomputation}
\end{equation}
where $\mM_2$ is given in \cref{eq:isospinKmatrix:flavorM2dimer} and $\mG^\infty_{\ell' \ell}$ in \cref{eq:isospinKmatrix:Ginfty}. Here recall $\bar{b}^2=b^2-\Mpi^2$, where $b$ is the momentum of the exchanged pion. Also $\m M_{2,\off}$ is the off-shell amplitude, with the external leg corresponding to the exchanged particle off-shell---see \cref{sec:hadrons:ChPTthreepionsLO}. The on-shell amplitude is recovered by setting $\bar b^2=0$.%, where $\bar b^2=b^2-\Mpi^2$ and $b$ is the momentum of the exchanged pion.

 %Instead of identifying the coefficients of the unsymmetrized threshold expansion, as what done for $Ippp$, we first symmetrize the full $\mMdf^{\uu,\OPE}$ using \cref{eq:isospinKmatrix:symmetrization} and later identify the coefficients of the symmetrized expansion.  %For example...

At LO, $\mMdf^{\uu,\OPE}$ is computed by substituting both $\mM_2$ factors by the LO ChPT prediction. This takes the general form
\begin{equation}
    \m M_2^\LO = \m k_0 + \m k_1\bar s + \m k_2 (t+u) + \m k_3 (t-u)\, ,
    \label{eq:Kmatrix:M2expansionLO}
\end{equation}
where $\m k_i$ are $7\times 7$ matrices of coefficients. % that, in the charge basis, come from substituting \cref{eq:hadrons:} into \cref{eq:isospinKmatrix:}. 
The key difference to the $\Ippp=3$ computation is that $p$-wave is present in this case. Thus, $s$- and $p$-waves need to be separated before performing the subtraction. At LO, this can be easily done taking the symmetric and antisymmetric parts, respectively,
\begin{equation}\label{eq:isospinKmatrix:amplitudeLOgenericexpansion}
    \begin{aligned}
        \m M_{2,s}(s,t,u) &= \m k_0 + \m k_2\bar s + \m k_2 (t+u)\,,\\[5pt]
        \m M_{2,p}(s,t,u) &= \m k_3 (t-u)\,.\\
    \end{aligned}
\end{equation}
The subtraction for the $s$-wave part is straightforward to compute. For the $p$-wave, the addition theorem of spherical harmonics allows us to write
\begin{equation}
    t-u = 4\bm p^*_k\cdot\bm a^*_k 
        = 4 p^*_k q^*_{2k} \bigg[
            \frac{4\pi}{3}\sum_m Y^*_{1m}(\hat{\bm a}^*_k) Y_{1m}(\hat{\bm p}^*_k)
        \bigg]\,,
        \label{eq:isospinKmatrix:pwaveharmonicsdecomposition}
\end{equation}
where recall $p^*_k$ is the magnitude of the momentum of the final spectator in the CMF of the initial dimer, $a^*_k$ is the relative CMF momentum of the particles in the dimer and $q_{2k}^*=|\bm{a}^*_k|$. An analogous relation also holds for the final state. As was the case of $d$-waves in \cref{eq:pipipiKmatrix:dwavebarrierfactoreffect}, the on- and off-shell $t-u$ terms are related by the barrier factors within $\mG^\infty$,
\begin{equation}
    \m [t-u]_{\off} = [t-u]_{\on} \biggl(\frac{p^*_k}{q^*_{2k}}\biggr)\,,
    \label{eq:isospinKmatrix:offonrelationpwave}
\end{equation}
and similarly for the final state. At LO, this implies the full on-shell $p$-wave amplitude is set off shell by the barrier factor. This is not the case for the NLO case, which we consider below, where $t-u$ appears multiplied by other factors which may vanish on shell, such as $\bar{b}^2$.

Taking \cref{eq:isospinKmatrix:offonrelationpwave} into account, it is convenient to rewrite the off-shell amplitude,
\begin{equation}
    \m M_{2,\off}^\LO = \m M^{\LO}_{2,s} + \bar b^2\:\delta\m M^{\LO}_{2,s}  + \m M^{\LO}_{2,p,\off}\,,
\end{equation}
where $s$-wave contributions are separated into an on-shell and an off-shell part, given by the first and second term, respectively, while the last term is purely $p$-wave. 
Multiplying the last term by the barrier factor within $\mG^\infty$, puts it on shell, and so \cref{eq:isospinKmatrix:OPEcomputation} becomes
\begin{equation}
    \mMdf^{\uu,\OPE,\LO} =  -\delta\m M^{\LO}_{2,s} \m T_G   \m M_{2, \off}^\LO  - \m M_{2, \off}^\LO  \m T_G  \,\delta\m M^{\LO}_{2,s}  + \bar b^2\delta\m M^{\LO}_{2,s} \m T_G  \,\delta\m M^{\LO}_{2,s}\,.
\end{equation}
The LO OPE contribution to the coefficients in the threshold expansion can be identified after substitution and symmetrization. The results are shown in \cref{tab:isospinKmatrix:LOresults} in \cref{sec:isospinKmatrix:discussion}.

We now turn to the NLO computation. The two-particle amplitude at NLO is obtained plugging \cref{eq:pipipiKmatrix:twopionamplitudeNLO} into \cref{eq:isospinKmatrix:flavorM2dimer}. To compute the threshold expansion up to quadratic order, we need to expand this amplitude up to cubic order,
\begin{multline}\label{eq:isospinKmatrix:NLOamplitudegeneralexpanded}
        \m M^\NLO_2 = \m a_1 +  \m b_1 \bar s + \m  b_2 (t+u) +  \m b_3 (t-u) \\
	\begin{array}{l}           
             +\,  \m c_1 \bar s^2 +  \m c_2 \bar s (t+u)   +  \m c_3 \bar s (t-u)  +  \m c_4  (t+u)^2 +  \m c_5 (t+u)(t-u)  +  \m c_6 tu \\
            +\, \m d_1 \bar s^3  + \m d_2 \bar s^2 (t+u)+ \m d_3 \bar s^2 (t-u) + \m d_4 \bar s (t+u)^2 + \m d_5 \bar s (t+u)(t-u) \\
            +\, \m d_6 \bar s  tu + \m d_7 (t+u)^3 + \m d_8 (t+u)^2(t-u) + \m d_9 (t+u)tu + \m d_{10} (t-u)tu\\
            +\,\cO(\bar s^4,t^4,u^4)\,,
    \end{array}
\end{multline}
where $\m a_i$, $\m b_i$, $\m c_i$ and $\m d_i$ are $7\times 7$ matrices of constant coefficients that depend on $\Mpi$, $\Fpi$ and the LECs, $\ell_i$. Note this requires expanding some $\bar{J}$ functions---see \cref{eq:pipipiKmatrix:Jexpansion}. 

The NLO contribution to $\mKdf$ is obtained using \cref{eq:isospinKmatrix:OPEcomputation} considering  the initial vertex at LO and the final one at NLO, and vice versa. The  computation requires to separate the amplitude in different partial waves and to perform the subtraction, using \cref{eq:isospinKmatrix:offonrelationpwave} and \cref{eq:pipipiKmatrix:dwavebarrierfactoreffect} to absorb the barrier factors in $p$- and $d$-wave on-shell amplitudes, respectively. This computation can be divided in multiple parts, grouped according to their peculiarities. In most cases, the computation is straightforward, but in other cases some subtleties arise.

For instance, the terms with $\m c_5$, $\m d_5$ and $\m d_8$ are pure $p$-wave, and even after accounting for barrier factors some off-shell parts remain. More interesting is the term with $\m d_{10}$ coefficient. As it is of cubic order, it only contributes in combination to the LO $\m k_0$ term, meaning only the off-shell part of the $\m d_{10}$ term survives the subtraction. This, however, contributes to both $p$ and $f$ waves ($\ell=3$). Combining \cref{eq:pipipiKmatrix:dwaveharmonicsdecomposition,eq:isospinKmatrix:pwaveharmonicsdecomposition}, and using the identity
\begin{equation}
    (\bm a^*_p \cdot \bm k_p^* )^3 = q_{2,p}^{*3} k^{*3} \left[ 
        \frac{3}{5}  \frac{4\pi}{3} Y^*_{1m}( \hat{\bm a}^*_p) Y_{1m}( \hat{\bm k}^*_p) 
        + \frac{2}{5} \frac{4\pi}{7} Y^*_{3m}( \hat{\bm a}^*_p) Y_{3m}( \hat{\bm k}^*_p)  
        \right],
	\label{eq:isospinKmatrix:fwaveharmonicsdecomposition}
\end{equation}
we can decompose $(t-u)tu$ into an $p$-wave and an $f$-wave part. The latter cancels completely, while the $p$-wave part looks as
\begin{equation}
    [ (t-u)tu ]_p =  \frac{1}{4}\left[ (\bar s - \bar b^2)^2 - \frac{48}{5} q_{2,p}^{*2} k^{*2} \right]  q_{2,p}^{*} k^{*}  \frac{16\pi}{3} Y^*_{1m}( \hat{\bm a}^*_p) Y_{1m}( \hat{\bm k}^*_p)\,.
\end{equation}
%The on-shell amplitude is obtained from setting $\bar b^2=0$ and $q_{2,p}^{*2} = k^{*2}$. 
The total contribution is finally obtained taking into account the $p$-wave barrier factor and subtracting. %From here, we symmetrize and determine the contributions to the coefficients in the threshold expansions.

Lastly, we can comment on the computation of the cubic order of $\mKdfI{0}$, as one may think that an expansion of $\mM^\NLO_2$ up to quartic order is required. However, the only two-particle isospin present in the $\Ippp=0$ channel is $\Ipp=1$. At LO only the $\m k_3$ term in \cref{eq:isospinKmatrix:amplitudeLOgenericexpansion} contributes, while the only relevant terms at NLO are those with $\m d_3$, $\m d_5$, $\m d_8$ and $\m d_{10}$ coefficients in \cref{eq:isospinKmatrix:NLOamplitudegeneralexpanded}.

\subsection{The $s$-channel OPE contribution}

The second term in \cref{eq:isospinKmatrix:LOseparation,eq:isospinKmatrix:NLOseparation} is the $s$-channel OPE. This only occurs in the $\Ippp=1$ channel and needs no subtraction since the exchanged particle is always off-shell for $E^*>3\Mpi$, with momentum $b=P$. In the charge basis, the exchanged particle is a neutral pion, $\pi^0$.

 The $s$-channel OPE part of the amplitude can be always factorized as
\begin{equation}
    \mMdf^{\sOPE} = -\bm v(p_1,p_2,p_3)\frac{1}{P^2 - \Mpi^2}\bm v^\dag(k_1,k_2,k_3)\,.
\end{equation}
Here $\bm{v}$ is a column vector containing the $\pi\pi\pi\rightarrow \pio$ amplitudes. In the charge basis, they can be computed from 
\begin{equation}
    \begin{aligned}
        \Fpi^2\cM_2[\pio(k_1)\pio(k_2)\pio(k_3)\to\pio(P)] &= A_{4\pi}(s)+A_{4\pi}(t)+A_{4\pi}(u)\,,\\[5pt]
        \Fpi^2\cM_2[\pip(k_1)\pio(k_2)\pim(k_3)\to\pio(P)] &= -A_{4\pi}(t)\,,       
    \end{aligned}
\end{equation}
plus permutations of the three-particle state. %Here $s_{ij}=(k_i+k_j)^2$.%, and $A$ is given in \cref{eq:hadrons:} at LO and in \cref{eq:pipipiKmatrix:} at NLO. 
As for the OPE, LO contribution to $\mKdfI{1}$ are obtained when both vertices are  LO, while the NLO contribution is obtained in the case one vertex is LO and the other is NLO, and vice versa.

In addition to the vertices, the computation of the coefficients of the threshold expansion requires expanding the propagator,
\begin{equation}
    \frac{1}{P^2 - \Mpi^2} = \frac{1}{ \Mpi^2 (8  - 9 \Delta)}  = \frac{1}{8\Mpi^2}\left[1 - \frac{9}{8}\Delta + \frac{81}{64}\Delta^2 + \cO(\Delta^3)\right]\,.
\end{equation}
Note that this formally sets the radius of convergence of the threshold expansion to $|\Delta|<9/8$. We analyze the implications of this limited convergence radius in \cref{sec:isospinKmatrix:discussion}, where we compare exact numerical  to threshold-expanded results of $\mKdf$. %We also note that a similar effect happens in the OPE and the BH terms, when expanding denominators, see for example \cref{eq:isospinKmatrix:expansionkinematicquantities}. 
Also notice that the $s$-channel OPE is the only part of the amplitude which, at LO, contributes to terms of the threshold expansion beyond linear order. The separate results at LO are presented in \cref{tab:isospinKmatrix:LOresults}  in \cref{sec:isospinKmatrix:discussion}. 

\newpage\subsection{The non-OPE contribution}

The non-OPE amplitude is the remainder of the total three-particle amplitude after subtracting the OPE and $s$-OPE parts. As its real part is finite, its contribution to $\mKdf$ can be computed independently of the associated subtraction. At LO, this contribution can be evaluated in the charge basis using,
\begin{equation}
    \begin{aligned}
        \cM_3^{\LO,\nOPE}[\pio\pio\pio\to\pio\pio\pio] &= 27\Mpi^2\,,\\[5pt]
        \cM_3^{\LO,\nOPE}[\pio\pio\pio\to\pip\pio\pim] &= 5\Mpi^2 - 3s'_{13} - t_{12}-t_{22}-t_{32}\,,\\[5pt]
        \cM_3^{\LO,\nOPE}[\pip\pio\pim\to\pip\pio\pim] &= -6\Mpi^2 + s_{13}+s'_{13} + t_{11}+2t_{22}+t_{33}\,,
    \end{aligned}
\end{equation}
where $s_{ij}=(k_i+k_j)^2$, $s_{ij}'=(p_i+p_j)^2$ and $t_{ij}=(p_i-k_j)^2$. After rotating to the symmetric basis, the coefficients can be directly identified. They are summarized in \cref{tab:isospinKmatrix:LOresults}  in \cref{sec:isospinKmatrix:discussion}.

At NLO, the amplitude can be easily computed in the charge basis using the results from \rcite{Bijnens:2021hpq}. The general-isospin non-OPE amplitude at NLO includes some new types of $C$ functions compared to the $\Ippp=3$ case, but all of them can be expanded about threshold using the same techniques presented in \cref{sec:pipipiKmatrix:nOPE}.

\subsection{The BH subtraction}

The BH subtraction only appears at NLO. It cancels divergencies happening in the non-OPE amplitude, associated to triangle diagrams as those in \cref{fig:pipipiKmatrix:truiangleloopdiagrams}. These divergencies happen in the imaginary part of the non-OPE amplitude, and so the real part of the subtraction is finite and can be computed independently of $\mM_3^{\nOPE}$. As in \cref{sec:pipipiKmatrix:BH}, the starting point is the unsymmetrized subtraction with the momentum assigned as in \cref{fig:pipipiKmatrix:triangleloopregular},
\begin{equation}
    \mD^{\uu,\BH}(\bm p_3,\bm k_3) = \int_r \m M^\LO_{2}(\bm p_3) \mG^\infty(\bm p_3,\bm r) \m M^\LO_{2}(\bm r) \mG^\infty(\bm r,\bm k_3) \m M^\LO_{2}(\bm k_3)\,,
    \label{eq:isospinKmatrix:Duudefinition}
\end{equation}
where recall $r=(\omega_r,\bm{r})$ is on-shell, with $\omega_r=\sqrt{\bm{r}^2+\Mpi^2}$.

The main difference with the $\Ippp=3$ computation is the presence of new kinematic structures and $p$-waves in the LO amplitudes. Before expanding, it is convenient to rewrite \cref{eq:isospinKmatrix:Duudefinition} in terms of integrals that can be addressed with techniques similar to those presented in \cref{sec:pipipiKmatrix:BH}. As a first step, we rewrite 
\begin{align}
    \mD^{\uu,\BH}(\bm p_3, \bm k_3) = \frac1{\Fpi^6}\int_r H^2(x_r)
     \m N(\bm p_3,\bm k_3,\bm r)\,  D^{-1}(\bm p_3,\bm k_3,\bm r) \,.
    \label{eq:isospinKmatrix:Duunumeratordenominator}
\end{align}
Here, $D$ is the same denominator as in \cref{eq:pipipiKmatrix:unsymmetrizedDBH2}, equal for all initial- and final-states of the charge basis,
\begin{equation}
D(\bm p_3,\bm k_3,\bm r) =  \big[(P-p_3-r)^2-\Mpi^2+i\epsilon\big]\big[(P-k_3-r)^2-\Mpi^2+i\epsilon\big]\,.
\end{equation}
The numerator, $\m N$, captures the isospin dependence. Its elements can be written as a product of three factors, $\n{ijk}\equiv n_i(p_3)n_j(r)n_k(k_3)$, with $i,j,k\in\{0,1,2\}$---see \rcite{Baeza-Ballesteros:2024mii} for the exact expression. Each $n_i(q)$ incorporates the effect of one factor of $\mM_2^\LO(\bm{q})$ in \cref{eq:isospinKmatrix:Duudefinition} and the associated barrier factors from $\mG^\infty$, and depends on the momenta of the corresponding spectator. In particular, $n_0(q)$ is related to $A_{4\pi}(s_q)$, $n_1(q)$ from $A_{4\pi}(t_q)-A_{4\pi}(u_q)$, and $n_2(q)$ from $A_{4\pi}(t_q)+A_{4\pi}(u_q)$, where $s_q$, $t_q$ and $u_q$ are the on-shell Mandelstam variables of a dimer with spectator of momentum $q$.

Both $n_0(q)$ and $n_2(q)$ are $s$-wave, and so do not contain barrier factors. They are,
\begin{equation}
\begin{array}{rl}
n_0(q)&= s_q-\Mpi^2\,,\\
n_2(q)&=t_q+u_q-\Mpi^2\,.
\end{array}
\end{equation}
By contrast, $n_1(q)$ is proportional to $t_q-u_q$. It contains $p$-waves and the corresponding barrier factors, and can be rewritten using \cref{eq:isospinKmatrix:offonrelationpwave}. This leaves a simple result when the spectator is a external particle that is on shell,
\begin{equation}
n_1(p_3)=-2(p_1-p_2)\cdot r\,,
\end{equation}
and similarly for $n_1(k_3)$. However, if the spectator has momentum $r$, a more lengthy development is required, which leaves
\begin{equation}\label{eq:isospinKmatrix:expansionnfactorinnervertex}
n_1(r)=4\bigg[\frac{p_{3}\cdot(P-r)\;k_{3}\cdot(P-r)}{(P-r)^2} - p_3\cdot k_3\bigg]\,.
\end{equation}

Taking this decomposition into account, \cref{eq:isospinKmatrix:Duunumeratordenominator} can be rewritten in terms of 27 integrals of the form
\begin{equation}
    I_{ijk}(\bm p,\bm k) \equiv \frac{1}{\Fpi^6}\int_r H^2(x_r)  n_{ijk}(\bm p,\bm k,\bm r)\,D^{-1}(\bm p,\bm k,\bm r) \,,\qquad i,j,k\in\{0,1,2\}\,,
\end{equation}
corresponding to the 27 different possible numerators. To each of these integrals we can apply similar techniques to those in \cref{sec:pipipiKmatrix:BH}, and expand them in terms of variables
\begin{equation}\label{eq:isospinKmatrix:Hmnpdefinition}
    H_{m,n,p} \equiv \frac{1}{\pi^2}\int_0^{1/\sqrt3} \d z\,\frac{\sqrt{1+z^2}}{z^m x^p} \frac{\d^n}{\d x^n}\big[H^2(x)\big]\,,
\end{equation}
where recall $x=1-3z^2$ and $z$ is defined from $\omega_r=\Mpi(1+2z^2)$. The main difference to the $H_{m,n}$ functions in \cref{eq:pipipiKmatrix:Hmnintegraldefinition} is the new $p$ index relative to the power of $x$ in the denominator. This originates due to the $1/(P-r)^2$ factor in $n_1(r)$---see \cref{eq:isospinKmatrix:expansionnfactorinnervertex}. These functions obey a recursive relation,
\begin{equation}
   H_{m,n,p} = 3H_{m-2,n,p} + H_{m,n,p-1} = \frac{1}{3}\big(H_{m+2,n,p}-H_{m+2,n,p-1}\big)\,,
\end{equation}
in addition to \cref{eq:pipipiKmatrix:Hrelation} for $p=0$. Note that the $p=0$ results are the $H_{m,n}$ functions used for the $\Ippp=3$ channel in \cref{sec:pipipiKmatrix:BH}.

%After symmetrization, one can determine the result for the threshold expansion in terms of where $p=0$ corresponds to the case treated in \cref{sec:pipipiKmatrix:}. The results for the threshold expansion after symmetrization are shown in Table.~8 of %\rcite{}, which have been simplified using the relation
%\begin{equation}
%    H_{m,n,p} = 3H_{m-2,n,p} + H_{m,n,p-1} = \tfrac13\big(H_{m+2,n,p}-H_{m+2,n,p-1}\big)\,,
%\end{equation}
%and \cref{eq:pipipiKmatrix:} for $p=0$. 

To evaluate the integrals, we regularize them using the Hadamard finite-part prescription. For $n=0$ we obtain
\begin{equation}\label{eq:isospinKmatrix:Hmnpevaluation}
    H_{m,0,p} =
        \displaystyle\rule{0pt}{2em}\int_0^{1/\sqrt3}\d z\; 6zf_{m,p}(z) \frac{\d}{\d x}\big[H^2(x)\big]\,,
\end{equation}
where we define
\begin{equation}
    \frac{\d}{\d z}f_{m,p}(x) = \frac{1}{\pi^2}\frac{\sqrt{1 + z^2}}{z^m x^p}\,.
\end{equation}
These results can be directly evaluated numerically.
For $n\neq 0$, on the other hand, there is no need to use the Hadamard finite-part prescription. Derivatives of $H(x)$ are exponentially suppressed at the endpoints of the integral, which cancels the divergences there.

As in \cref{sec:pipipiKmatrix:BH}, it is possible to obtain an analytic result by approximating $H(x)=1$, in which case all $H_{m,n,0}$ are reduced to $f_{m,0}(1/\sqrt{3})\delta_{n,0}$. In reality, this approximation is more complicated for $H_{m,n,p}(x)$, as it does not converge for $p>0$ due to a pole in the upper limit of  \cref{eq:isospinKmatrix:Hmnpdefinition}, at which $x=0$. This divergence can also be regularized using the the Hadamard finite-part prescription, which is equivalent to dropping terms with $p>0$. The complete result for $H_{m,n,p}$ can then be separated into this analytical approximation and a remainder, $\tilde{H}_{m,n,p}$,
\begin{equation}
    H_{m,n,p} = \tilde H_{m,n,p} + f_{m,0}(1/\sqrt3)\delta_{n,0}\delta_{p,0}\,.
\end{equation}
Here $\tilde H_{m,n,p}$ is the only cutoff-dependent part, that vanishes for $H(x)=1$, and needs to be evaluated numerically. 

The complete contributions to the coefficients of the threshold expansion of $\mKdf$ can then be written as
\begin{equation}\label{eq:isospinKmatrix:cutoffterms}
    \cK_X = \cK_X^{[\tilde H_{m,n,p}=0]} - \cD_X\,,
\end{equation}
where $\cK_X^{[\tilde H_{m,n,p}=0]}$ can be expressed analytically, while $\cD_X$ needs to be evaluated numerically. Their values are listed in \cref{tab:isospinKmatrix:remainders}. Typically, these numeric corrections are a small fraction of the full result. Note however, this is not always the case, as happens for the SS sector of the $\Ippp=1$ channel. %A detailed study of the relative size of these numeric corrections, compared to the full BH subtraction, is presented in app.~A of \rcite{Baeza-Ballesteros:2024mii}.

\newpage\section{Results for $\mKdf$}\label{sec:isospinKmatrix:discussion}

\begin{table}[b!]
    \centering
    
    {\renewcommand{\arraystretch}{1.2}
        
    \begin{tabular}{clrrrr}
        \toprule
		& \multirow{2}{*}{$\cK_X$}     &    \multicolumn{4}{c}{$(\Fpi/\Mpi)^4\times\cK_X$}
        \\
            &   
            &   \multicolumn{1}{c}{Total}
            &   \multicolumn{1}{c}{OPE}
            &   \multicolumn{1}{c}{$s$-OPE}
            &   \multicolumn{1}{c}{non-OPE}
        \\
        \midrule
        \multirow{2}{*}{$\Ippp=3$}
        &   $\Kiso$  &   18                  &   36              &   0                   &   $-18$ \\[4pt]
        &   $\Kisoone$    &   27                  &   63              &   0                   &   $-36$ \\
        \midrule
        \multirow{1}{*}{$\Ippp=2$}
        &   $\KT   $      &   $\frac{9}{2}$        &   $\frac{21}{2}$   &   0                   &   $-6$  \\
        \midrule
        \multirow{7}{*}{$\Ippp=1$}
        &   $\KSS  $      &   $-\frac{111}{8}$     &   $-54 $            &   $-\frac{135}{8} $    &   57  \\[4pt]
        &   $\KSS[1] $    &   $-\frac{1137}{64}$   &   $-27$             &   $-\frac{945}{64} $   &   24  \\[4pt]
        &   $\KSS[2] $    &   $-\frac{135}{512}$   &   0               &   $-\frac{135}{512}$   &   0   \\[4pt]
        
        &   $\KSD  $      &   $-\frac{3}{8} $      &   $-9 $             &   $-\frac{27}{8}$      &   12  \\[4pt]
        &   $\KSD[1]$     &   $\frac{27}{64} $     &   0               &   $\frac{27}{64}$      &   0   \\[4pt]
        
        &   $\KDD $       &  $ \frac{1}{2} $       &   $\frac{21}{2}$   &   0                   &   $-10$ \\[4pt]
        &   $\KDD[2]$     &   $-\frac{81}{4} $      &   0               &   $-\frac{81}{4} $      &   0   \\
        \midrule
        \multirow{1}{*}{$\Ippp=0$}
        & \multicolumn{5}{c}{There are no $\Ippp=0$ contributions at LO}\\
        \bottomrule
    \end{tabular}}

\caption{
        LO ChPT results for $\mKdf$, together with the separate contributions from each term in \cref{eq:isospinKmatrix:LOseparation}. There is no BH subtraction or cutoff dependence at this order.}
    \label{tab:isospinKmatrix:LOresults}
\end{table}

\begin{figure}[!hp]
    \centering
    \begin{subfigure}{\textwidth} 
    \centering
        \includegraphics[width=1\textwidth]{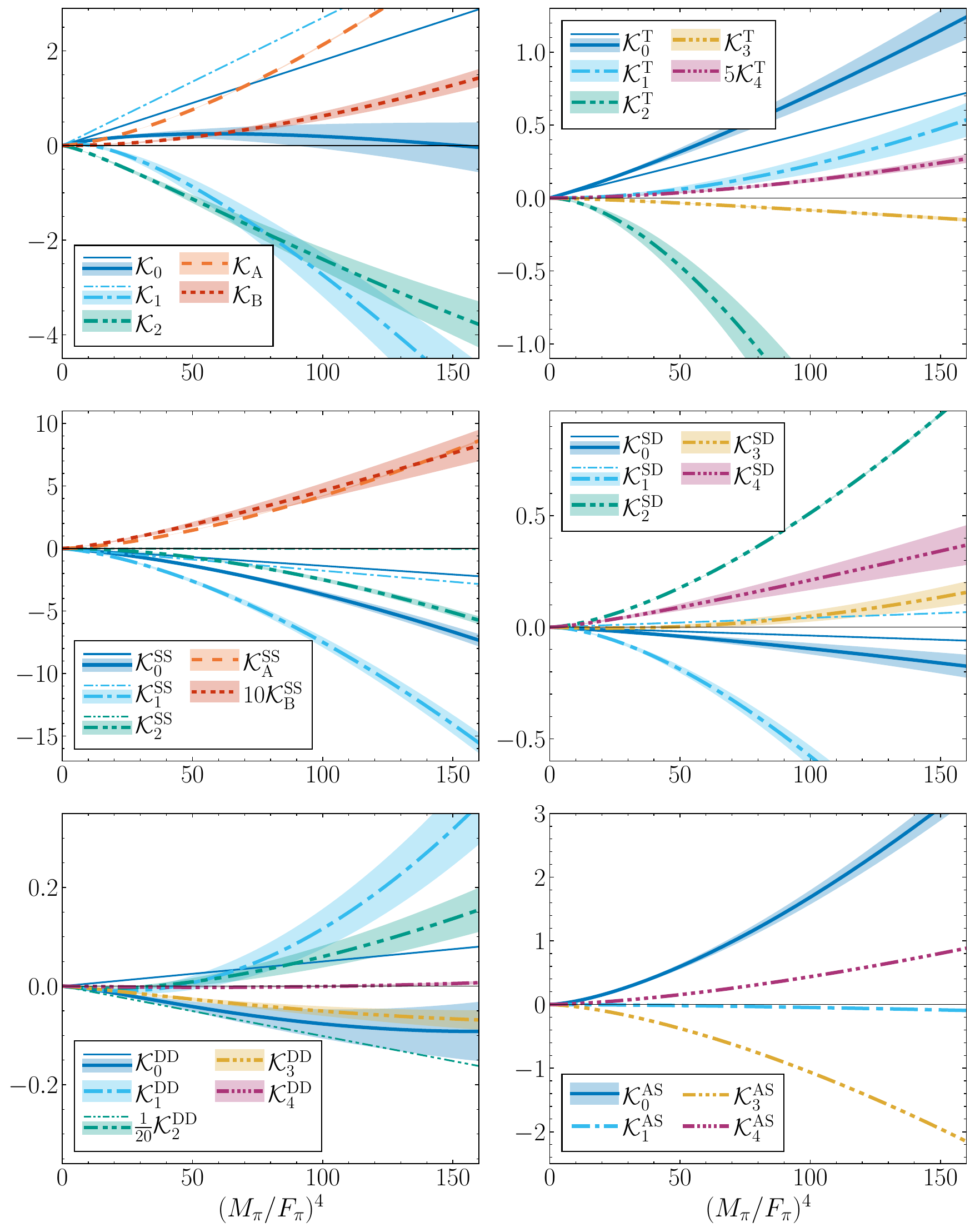}
    \end{subfigure}
    \caption{
         LO (thin lines) and LO+NLO (thick lines and bands) ChPT predictions for the threshold-expansion coefficients of $\Kdf$ as a function of $(\Mpi/\Fpi)^4$, divided by a factor of $10^3$. Bands represent the uncertainties arising from the LECs---see  \cref{eq:pipipiKmatrix:comparisonlattice}. Each panel corresponds to a different isospin and irrep of the permutation group, for the $\Ippp=1$ channel. Some coefficients are rescaled for legibility. We note $\cK^\text{AS}_{1}$, $\cK^\text{AS}_{3}$ and $\cK^\text{AS}_{4}$ have no error, as they are independent of the LECs. }
    \label{fig:isospinKmatrix:results}
\end{figure}

The complete results for $\mKdf$ are summarized in \cref{tab:isospinKmatrix:LOresults} at LO, which also include the separate contribution from each term in \cref{eq:isospinKmatrix:LOseparation}, and in \cref{tab:isospinKmatrix:NLOresults} at NLO, complemented by the cutoff-dependent remainders in \cref{tab:isospinKmatrix:remainders}. These remainders are computed for the standard choice of the cutoff function, \cref{eq:hadrons:standardcutoff}. The full LO+NLO results are also shown  in \cref{fig:isospinKmatrix:results} as a function of $(\Mpi/\Fpi)^2$, where we use the values of the LECs as given in \cref{eq:pipipiKmatrix:LECref} and a renormalization scale $\mu=4\pi\Fpi$, as done for example \cref{eq:pipipiKmatrix:comparisonlattice}. Note that the represented error bands (shaded regions) arise from the LECs, while we make no attempt at estimating higher-order corrections from ChPT. Some coefficients, moreover, do not have errors as they are independent of the LECs. Note the results for the $\Ippp=3$ channel coincide with those presented in \cref{sec:pipipiKmatrix:summary}, and we reproduce them here for completion.

We can compare LO to LO+NLO results for those coefficients for which the LO is non-zero. As was the case for the $\Ippp=3$ channel, we in general observe bad convergence of the threshold expansion, with large corrections coming from NLO. This effect is thus generic for three-pion processes. Also note these corrections are larger at heavier-than-physical pion masses, while they are somehow smaller close to the physical point.

\begin{table}[tp]
    \centering
   
    \hspace{-.3cm}
    \setlength{\tabcolsep}{1pt}
    \begin{tabular}{rl@{\hspace{0.4cm}}rrcrcrcrcrcrcr}
    \toprule
    & $\cK_X$ & \multicolumn{14}{c}{$(\Fpi/\Mpi)^6\times\cK_X^{[\tilde H_{m,n,p}=0]}$} \\
    \midrule
                &$\Kiso$            &    $-$ $\kappa\big($ & 105                       &+& $36 \logthree\big)$                    &+& 111L                    &$-$& 288 $\lr1$                 &$-$& 432$\lr2$                 &$-$& 36$\lr3$              &+& 72$\lr4$              \\
        &$\Kisoone$         &    $-$ $\kappa\big($ & $\tfrac{1999}{20} $         &+& $96 \logthree\big)$                    &+& 384L                    &$-$& 612$\lr1$                 &$-$& 1170$\lr2$                &&                      &+& 108$\lr4$             \\
        &$\Kisotwo$         &      $\kappa\big($ & $\tfrac{605061}{1400}$      &$-$& $\tfrac{621}{10}\logthree\big)$       &+& 360L                    &$-$& 432$\lr1$                 &$-$& 864$\lr2$                 &&                      &&                      \\
        &$\KA$              &      $\kappa\big($ & $\tfrac{196281}{560}$       &$-$& $\tfrac{135}{8}\logthree\big)$        &$-$& 9L                      &+& 27$\lr1$                  &+& \lrfrac[27]{2}2         &&                      &&                      \\
        &$\KB$              &      $\kappa\big($ & $\tfrac{90423}{700}$        &$-$& $\tfrac{189}{40}\logthree\big)$       &+& 54L                     &$-$& 162$\lr1$                 &$-$& 81$\lr2$                  &&                      &&                      \\
        \midrule
        &$\KT$              &    $-$ $\kappa\big($ & $\tfrac{59113}{3240} $      &+& $\tfrac{1009}{144}\logthree\big)$     &&                          &$-$& 90$\lr1$                  &$-$& 9$\lr2$                   &&                      &+& 18$\lr4$              \\
        &$\KT[1]$           &      $\kappa\big($ & $\tfrac{9486697}{453600}$   &$-$& $\tfrac{989}{480}\logthree\big)$      &+& \Lfrac[53]{2}           &$-$& \lrfrac[195]{2}1        &$-$& \lrfrac[123]{4}2        &&                      &&                      \\
        &$\KT[2]$           &    $-$ $\kappa\big($ & $\tfrac{1248031}{7200} $    &+& $\tfrac{5641}{320}\logthree\big)$     &$-$& \Lfrac[171]{2}          &+& \lrfrac[837]{2}1        &+& \lrfrac[189]{4}2        &&                      &&                      \\
        &$\KT[3]$           &      $\kappa\big($ & $\tfrac{23833}{33600} $     &$-$& $\tfrac{317}{960}\logthree\big)$      &+& \Lfrac[27]{4}           &$-$& \lrfrac[45]{4}1         &$-$& \lrfrac[117]{8}2        &&                      &&                      \\
        &$\KT[4]$           &      $\kappa\big($ & $\tfrac{332981}{75600} $    &$-$& $\tfrac{59}{960}\logthree\big)$       &+& \Lfrac[5]{3}            &$-$& 5$\lr1$                   &$-$& \lrfrac[5]{2}2          &&                      &&                      \\
        \midrule
        &$\KSS$             &    $-$ $\kappa\big($ & $\tfrac{1955}{8}      $     &+& $\tfrac{369}{4}\logthree\big)$        &$-$& \Lfrac[1237]{8}         &+& 342$\lr1$                 &+& 438$\lr2$                 &$-$& \lrfrac[57]{2}3     &$-$& \lrfrac[111]{2}4    \\
        &$\KSS[1]$          &    $-$ $\kappa\big($ & $\tfrac{191089}{320} $      &+& $\tfrac{993}{8}\logthree\big)$        &$-$& \Lfrac[24\,439]{64}     &+& \lrfrac[2637]{4}1       &+& \lrfrac[4125]{4}2       &+& \lrfrac[45]{16}3    &$-$& \lrfrac[1137]{16}4  \\
        &$\KSS[2]$          &    $-$ $\kappa\big($ & $\tfrac{34274101}{89600}$   &+& $\tfrac{33957}{320}\logthree\big)$    &$-$& \Lfrac[119\,505]{512}   &+& \lrfrac[8811]{32}1      &+& \lrfrac[18\,027]{32}2   &$-$& \lrfrac[405]{128}3  &$-$& \lrfrac[135]{128}4  \\
        &$\KSSA$            &      $\kappa\big($ & $\tfrac{1102239}{2240}$     &$-$& $\tfrac{19575}{128}\logthree\big)$    &+& \Lfrac[273]{8}          &$-$& \lrfrac[297]{4}1        &$-$& \lrfrac[261]{4}2        &&                      &&                      \\
        &$\KSSB$            &    $-$ $\kappa\big($ & $\tfrac{521271}{5600}$      &+& $\tfrac{13419}{640}\logthree\big)$    &$-$& 36L                     &+& 108$\lr1$                 &+& 54$\lr2$                  &&                      &&                      \\
        \midrule
        &$\KSD$             &      $\kappa\big($ & $\tfrac{10853}{160}  $      &$-$& $\tfrac{255}{8}\logthree\big)$        &+& \Lfrac[23]{16}          &+& 36$\lr1$                  &$-$& \lrfrac[39]{2}2         &$-$& \lrfrac[9]{4}3      &$-$& \lrfrac[3]{2}4      \\
        &$\KSD[1]$          &      $\kappa\big($ & $\tfrac{643087}{8960}$      &$-$& $\tfrac{3543}{64}\logthree\big)$      &$-$& \Lfrac[1647]{128}       &+& \lrfrac[585]{8}1        &$-$& \lrfrac[9]{8}2          &+& \lrfrac[81]{32}3    &+& \lrfrac[27]{16}4    \\
        &$\KSD[2] $         &      $\kappa\big($ & $\tfrac{166953}{2240}$      &+& $\tfrac{513}{128}\logthree\big)$      &+& \Lfrac[21]{4}           &&                          &$-$& \lrfrac[63]{4}2         &&                      &&                      \\
        &$\KSD[3]$          &      $\kappa\big($ & $\tfrac{27783}{320}$        &$-$& $\tfrac{3699}{128}\logthree\big)$     &+& \Lfrac[75]{4}           &+& \lrfrac[81]{4}1         &$-$& \lrfrac[531]{8}2        &&                      &&                      \\
        &$\KSD[4]$          &      $\kappa\big($ & $\tfrac{109539}{5600}$      &$-$& $\tfrac{11097}{320}\logthree\big)$    &$-$& \Lfrac[39]{2}           &+& \lrfrac[297]{4}1        &+& \lrfrac[171]{8}2        &&                      &&      \\ \bottomrule
    \end{tabular}
     \caption{
        Full $\NLO$ results for $\Kdf$ up to quadratic order in the threshold expansion (cubic for $\Ippp=0$). The cutoff-dependent remainders defined in \cref{eq:isospinKmatrix:cutoffterms}, $\cD_X$, are not included. They are listed in  \cref{tab:isospinKmatrix:remainders}.}
    \label{tab:isospinKmatrix:NLOresults}
\end{table}

\begin{table}[!p]
    \centering
   
    \hspace{-.3cm}
    \setlength{\tabcolsep}{1pt}
    \begin{tabular}{rl@{\hspace{0.4cm}}rrcrcrcrcrcrcr}
    \toprule
    & $\cK_X$ & \multicolumn{14}{c}{$(\Fpi/\Mpi)^6\times\cK_X^{[\tilde H_{m,n,p}=0]}$} \\
    \midrule
        &$\KDD$             &      $\kappa\big($ & $\tfrac{49121}{3240}$       &+& $\tfrac{1259}{144}\logthree\big)$     &+& 28L                     &$-$& 54$\lr1$                  &$-$& 63$\lr2$                  &&                      &+& 2$\lr4$               \\
        &$\KDD[1]$          &      $\kappa\big($ & $\tfrac{11178103}{453600}  $&+& $\tfrac{4279}{480}\logthree\big)$     &+& \Lfrac[265]{6}          &$-$& \lrfrac[149]{2}1        &$-$& \lrfrac[381]{4}2        &&                      &&                      \\
        &$\KDD[2]$          &      $\kappa\big($ & $\tfrac{27345737}{50400}  $ &$-$& $\tfrac{11869}{320}\logthree\big)$    &$-$& \Lfrac[123]{2}          &+& \lrfrac[1251]{2}1       &+& \lrfrac[459]{4}2        &&                      &$-$& 81$\lr4$              \\
        &$\KDD[3]$          &      $\kappa\big($ & $\tfrac{150229}{11200}   $  &+& $\tfrac{449}{320}\logthree\big)$      &+& \Lfrac[45]{4}           &$-$& \lrfrac[217]{12}1       &$-$& \lrfrac[593]{24}2       &&                      &&                      \\
        &$\KDD[4]$          &      $\kappa\big($ & $\tfrac{212299}{75600}  $   &+& $\tfrac{83}{320}\logthree\big)$       &+& \Lfrac[25]{9}           &$-$& \lrfrac[7]{3}1          &$-$& \lrfrac[43]{6}2         &&                      &&                      \\
        \midrule
        &$\KAS$             &      $\kappa\big($ & $\tfrac{2721}{20}      $    &$-$& $\tfrac{81}{2}\logthree\big)$         &&                          &$-$& 162$\lr1$                 &+& 81$\lr2$                  &&                      &&                      \\
        &$\KAS[1]$          &      $\kappa\big($ & 132                       & &             $\big)$                    &&                          &&                          &&                          &&                      &&                      \\
        &$\KAS[3]$          &    $-$ $\kappa\big($ & $\tfrac{164673}{1120}$      &$-$& $\tfrac{2187}{32}\logthree\big)$      &&                          &&                          &&                          &&                      &&                      \\
        &$\KAS[4]$          &      $\kappa\big($ & $\tfrac{28863}{448}$        &$-$& $\tfrac{3645}{128}\logthree\big)$     &&                          &&                          &&                          &&                      &&                      \\
        \bottomrule
    \end{tabular}
     \caption*{\textbf{Table 6.2 (cont.). }
        Full $\NLO$ results for $\Kdf$ up to quadratic order in the threshold expansion (cubic for $\Ippp=0$). The cutoff-dependent remainders defined in \cref{eq:isospinKmatrix:cutoffterms}, $\cD_X$, are not included. They are listed in  \cref{tab:isospinKmatrix:remainders}.}

    \vspace{1cm}

    \centering
    
    {\renewcommand{\arraystretch}{1.2}
    \begin{tabular}{c@{\quad}l@{\quad\quad\quad}c@{\quad}l}
	\toprule
	$\cD_0$ &  $-0.056\,347\,6589$ & $\cD^\text{SS}_0$ &       $\phantom{-}1.213\,748\,64$ \\[2pt]
	$\cD_1$ &  $\phantom{-}0.129\,589\,681$ & $\cD^\text{SS}_1$ &       $\phantom{-}4.737\,727\,30$ \\[2pt]
	$\cD_2$ &  $\phantom{-}0.432\,202\,370$ & $\cD^\text{SS}_2$ &       $\phantom{-}2.098\,947\,60$ \\[2pt]
	$\cD_\text{A}$ &  $\phantom{-}0.000\,907\,273\,890$ & $\cD^\text{SS}_\text{A}$ &       $-2.393\,448\,70$ \\[2pt]
	$\cD_\text{B}$ &  $\phantom{-}0.000\,162\,394\,747$ & $\cD^\text{SS}_\text{B}$ &       $-1.089\,982\,49$ \\ 
	\midrule
	$\cD^\text{T}_0$ &  $-0.007\,042\,111\,64$ & $\cD^\text{SD}_0$ &       $\phantom{-}0.060\,539\,453\,1$ \\[2pt]
	$\cD^\text{T}_1$ &  $-0.095\,869\,747\,4$ & $\cD^\text{SD}_1$ &       $\phantom{-}0.558\,130\,406$ \\[2pt]
	$\cD^\text{T}_2$ &  $-0.264\,963\,303$ & $\cD^\text{SD}_2$ &       $-0.105\,910\,881$ \\[2pt]
	$\cD^\text{T}_3$ &  $\phantom{-}0.021\,650\,723\,1$ & $\cD^\text{SD}_4$ &       $-0.135\,426\,533$ \\[2pt]
	$\cD^\text{T}_4$ &  $-0.001\,531\,207\,94$ & $\cD^\text{SD}_4$ &       $-0.349\,051\,891$ \\ 
	\midrule
	$\cD^\text{AS}_0$ &  $-0.301\,063\,917$ & $\cD^\text{DD}_0$ &       $\phantom{-}0.048\,482\,775\,8$ \\[2pt]
	$\cD^\text{AS}_1$ &  $\phantom{-}0.881\,880\,013$ & $\cD^\text{DD}_1$ &       $-0.316\,388\,524$ \\[2pt]
	$\cD^\text{AS}_3$ &  $\phantom{-}0.607\,228\,425$ & $\cD^\text{DD}_2$ &       $-1.906\,375\,12$ \\[2pt]
	$\cD^\text{AS}_4$ &  $-0.227\,122\,084$ & $\cD^\text{DD}_4$ &       $\phantom{-}0.034\,410\,564\,7$ \\[2pt]
					 &  					 & $\cD^\text{DD}_4$ &       $-0.017\,668\,886\,1$ \\ \bottomrule
        
    \end{tabular}}
    \caption{
        Summary of the numerical cutoff-dependent remainders from the BH subtraction. They are computed using the standard cutoff in \cref{eq:hadrons:standardcutoff}.}
    \label{tab:isospinKmatrix:remainders}
\end{table}

%Similarly to the $I=3$, which is represented in the top left panel of \cref{fig:isospinKmatrix:results}, we bad convergence is observed of the threshold expansion, with large corrections coming from NLO. 

\begin{figure}[!hp]
    \centering
    \begin{subfigure}{\textwidth} 
    \centering
        \includegraphics[width=1\textwidth]{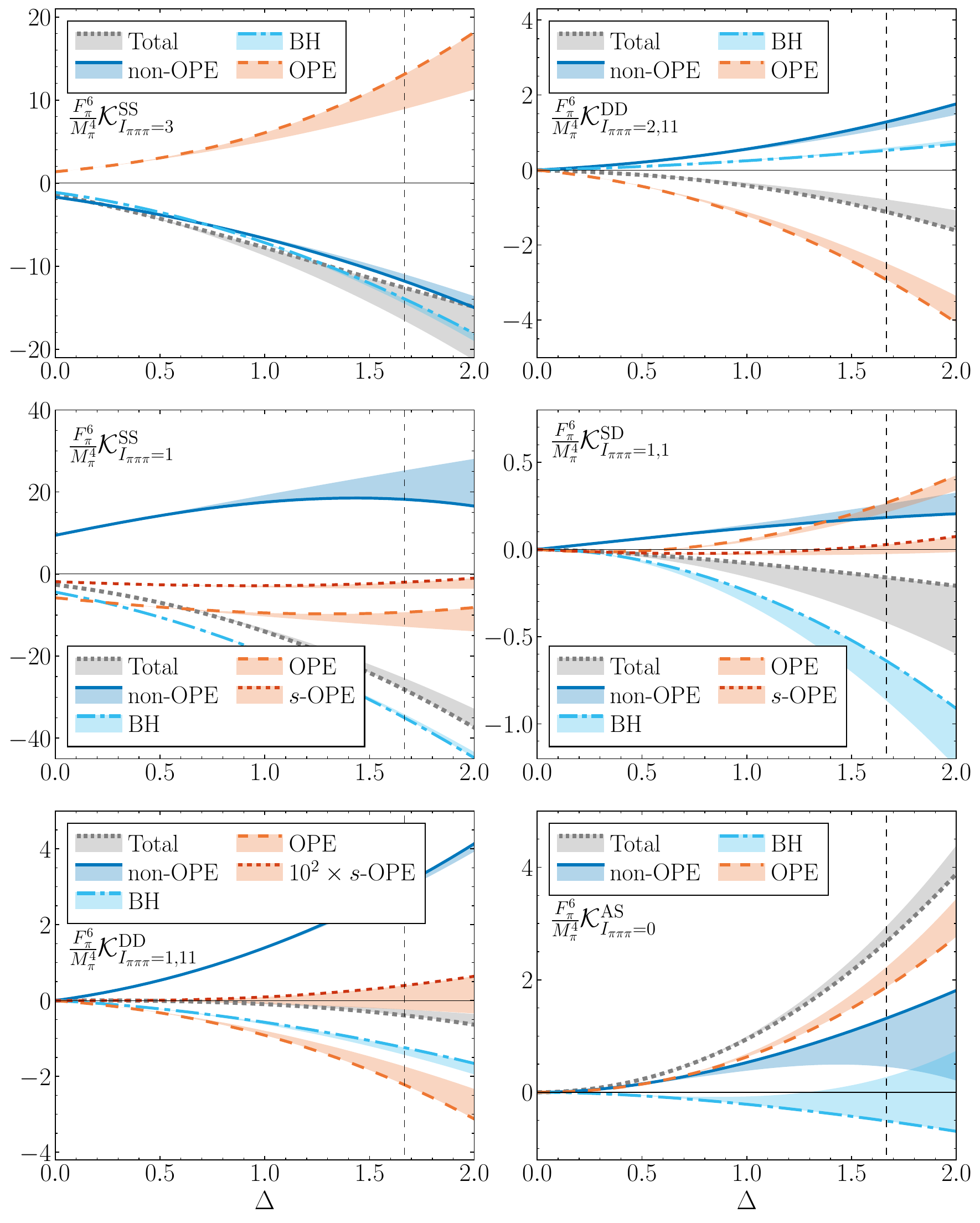}
    \end{subfigure}
    \caption{
         Convergence of the threshold expansion at $\Mpi=340$ MeV, for various components of $\mKdf^\NLO$ (LO is omitted) in the symmetric basis using the kinematic \mbox{configuration} in \cref{eq:isospinKmatrix:kinematicconfiguration}. Results are represented for the total $K$-matrix as well as for the separate contributions according to \cref{eq:isospinKmatrix:NLOseparation}, where ``BH'' refers to $\mK^\BH=-\mD^\BH$. Lines represent the threshold expansion, and the the width of the bands is the difference to the exact results, which thus correspond to the other end of the band. All values are divided by a \mbox{factor  $10^3$}. Vertical lines represent the five-pion threshold. }
    \label{fig:isospinKmatrix:convergence}
\end{figure}

We also study the convergence of the threshold expansion, comparing our threshold-expanded results to exact numeric determinations. This comparison is shown in \cref{fig:isospinKmatrix:convergence} for six of the 15 non-zero coefficients of $\mKdf$ in the symmetric basis. We take $\Mpi=340$ MeV and use the momenta configuration,
\begin{equation}\label{eq:isospinKmatrix:kinematicconfiguration}
\begin{array}{ll}
\bm{k}_1=p\left(1,0,0\right)\,,\quad\quad&\bm{p}_1=p\left(\frac{1}{2},\frac{3}{4},\frac{\sqrt{3}}{4}\right)\,,\\[7pt]
\bm{k}_2=p\left(-\frac{3}{4},1,0\right)\,,\quad\quad&\bm{p}_2=p\left(-\frac{\sqrt{3}}{2}-\frac{3}{8},\frac{\sqrt{3}}{4}-\frac{9}{16},-\frac{3\sqrt{3}}{16}+\frac{1}{4}\right)\,,\\[7pt]
\bm{k}_3=p\left(-\frac{1}{4},-1,0\right)\,,\quad\quad&\bm{p}_2=p\left(\frac{\sqrt{3}}{2}-\frac{1}{8},-\frac{\sqrt{3}}{4}-\frac{3}{16},-\frac{\sqrt{3}}{16}-\frac{1}{4}\right)\,,\\
\end{array}
\end{equation}
with $p$ a real parameter. This configuration is asymmetric enough that all relevant elements in $\mKdf$ are non-zero. 

We observe good convergence for all components up to $\Delta\lesssim 1$, and also that in many cases the convergence of the total result is better than that of the separate terms in \cref{eq:isospinKmatrix:NLOseparation}. Note that the formal limitation of the series to $|\Delta|<8/9$, originating from the $s$-OPE contribution, does not seem to break the expansion. The $s$-OPE part is generally small and converges poorly (rather than diverging) above that limit.
We found the non-presented elements of $\mKdf$ to show similar convergence, which is also true for other kinematical configurations and values of the pion mass. 

\begin{figure}[!hp]
\phantom{aa}
    \centering
    \begin{subfigure}{\textwidth} 
    \centering
        \includegraphics[width=1\textwidth]{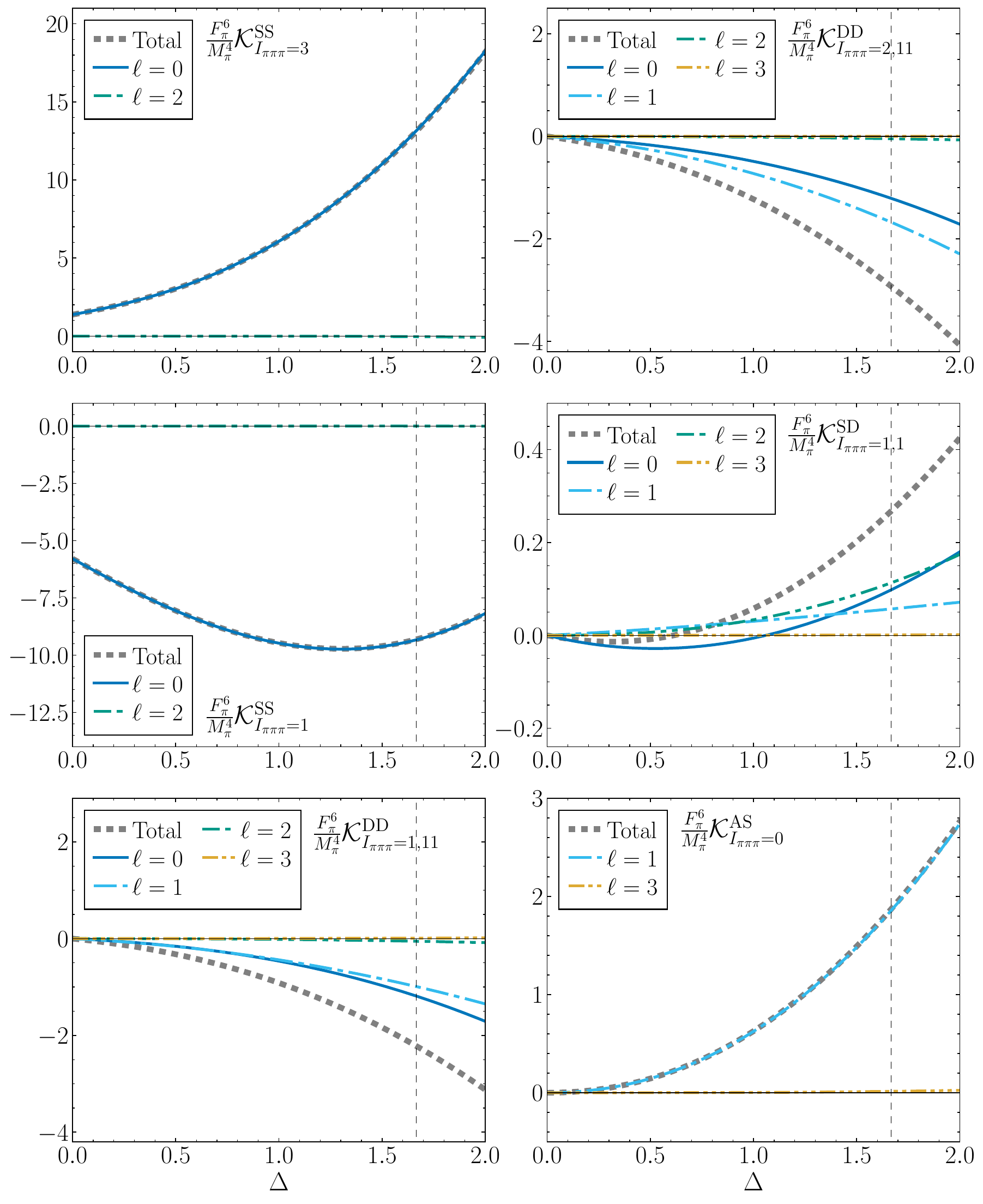}
    \end{subfigure}
    \caption{
         Comparison of contributions to $\mKdf^{\NLO,\OPE}$ from different dimer partial waves in $\mM_2^\NLO$, evaluated numerically at $\Mpi=340$ MeV using the kinematic configuration in \cref{eq:isospinKmatrix:kinematicconfiguration}. All values are divided by a \mbox{factor $10^3$}. Partial waves that are identically zero, as well as the negligibly small $\ell>3$, are omitted. }
         \phantom{aa}
    \label{fig:isospinKmatrix:partialwaves}
\end{figure}

Finally, we analyze the convergence of the NLO OPE part as higher partial waves of the NLO vertex are included. In \cref{fig:isospinKmatrix:partialwaves}, we represent the contributions of different partial waves of the NLO two-particle amplitude to the OPE part of $\mKdf$ for the same six coefficients, kinematic configuration and pion mass used in \cref{fig:isospinKmatrix:convergence}. We observe our result is dominated by the lowest partial waves, with negligible contributions above $\ell= 2$, the highest one captured by the threshold expansion.

\section{Conclusions}

In this chapter, the work from \rcite{Baeza-Ballesteros:2024mii} has been presented, in which the three-pion $K$-matrix was determined for general isospin up to NLO in ChPT. This generalized the work from \rcite{Baeza-Ballesteros:2023ljl} presented in \cref{sec:pipipiKmatrix}. Our results show that NLO corrections are in general large compared to LO results, especially at heavier-than-physical pion mass. The bad convergence of the chiral expansion seems thus to be a generic feature of all three-pion isospin channels. On a more positive side, we found that the threshold expansion truncated at quadratic order provides a good description of the full $K$-matrix in the elastic region.

We expect that the results from this work will prove valuable in the future. In particular, it will be of major interest to compare them against future lattice results for three pions at non-maximal isospin. This work also opens the door to an extension to more complicated systems. An example of this could be systems of three kaons or pions and kaons, for which lattice results are already available~\cite{Blanton:2021llb,Draper:2023boj}. However, the NLO ChPT amplitudes would need to be worked out first.
\chapter{Two- and three-particle scattering in the O(3) model}
\label{sec:O3model}

The study of three-particle interactions on the lattice has undergone tremendous progress during the past decade. On the theoretical side, different formalisms have been developed. The relativistic-field-theory (RFT) approach, introduced in \cref{sec:hadrons:threeparticlesfinitevolume}, has been extended to a large number of three-particle systems, such as non-identical mesons, three nucleons or processes containing resonances. Its practical application top QCD, however, has been limited to states of three mesons at maximal isospin~\cite{Blanton:2019vdk,Fischer:2020jzp,Hansen:2020otl,Blanton:2021llb,Draper:2023boj}, which present very weak interactions. By contrast, the formalism has never been applied to a system for which analytical solutions are known%, that allows to compare lattice results against analytical predictions and learn more about the intricacies of the formalism
---see however \rcite{Garofalo:2022pux} for an application to a scalar theory with a three-particle resonance. 

%three-particle formalism, introduced in \cref{sec:hadrons:threeparticlesfinitevolume}, has been extended to a large number of systems, including resonant processes, non-identical and non-degenerate mesons os three nucleons. However, its usage has been mainly limited to simple systems of pions and kaons at maximal isospin

A commonly used toy model for QCD is the (1+1)-dimensional O(3) non-linear sigma model. It is an integrable model---i.e., it allows for analytic predictions of the scattering matrix---which presents very strong interactions. The model shares many of the non-perturbative features of QCD, such as asymptotic freedom and a low-energy spectrum of isospin-one particles. In addition, the lower dimensionality allows for cheap numerical computations. This characteristics make the O(3) model ideal to investigate the RFT formalism, as lattice results could be compared against analytical predictions. Moreover, the strength of the interactions in this model makes it possible to explore a regime of the formalism which has not been studied before.

In this chapter, we present results on the study of two- and three-particle interactions in the (1+1)-dimensional O(3) non-linear sigma model~\cite{Baeza-Ballesteros:O3inprep,Baeza-Ballesteros:2022bsn}. In the two-particle sector, we focus on the isospin-two and -one channels, which we compare against analytical predictions. In the case of three particles we take a first step to test the RFT formalism. We determine the finite-volume energy spectra in the isospin-three, -two and -zero channels, and compare the results against predictions made assuming a zero three-particle $K$-matrix, $\Kdf=0$, using the RFT, that we have adapted to 1+1 dimensions.

% we compare against analytical predictionsWe focus on the isospin-two and isospin-one channel for two particles, and the isospin-three, -two and -zero channels for three-particles. Our aim is to test the formalism by comparing lattice results against analytical predictions available for the O(3) model~\cite{Zamolodchikov:1977nu,Zamolodchikov:1978xm}. Moreover,  we intend to learn about the intricacies of the formalism that may arise due to the very strong interactions in the  model.

The chapter is organized as follows. In \cref{sec:O3model:theory}, we introduce the O(3) model and review the analytical results for the scattering matrix. The (1+1)-dimensional version of the two- and three-particle finite-volume formalisms is presented in \cref{sec:O3model:finitevolumeformalisms}, followed by a description of the lattice techniques used to determine the finite-volume spectrum in \cref{sec:O3model:lattice}. \Cref{sec:O3model:resultsenergies} then presents our results for the two- and three-particle finite-volume energies, which we compare in \cref{sec:O3model:results} to analytical predictions, setting $\Kdf=0$ in the three-particle sector.  We finalize with a short conclusion in \cref{sec:O3model:conclusion}.

\section{The (1+1)-dimensional O(3) non-linear sigma model}\label{sec:O3model:theory}

The (1+1)-dimensional O(3) non-linear sigma model---\textit{O(3) model} from here on---has commonly served as a sandbox of QCD due to important qualitative similarities of the two theories. For example, it has been used to study the two-particle scattering formalism~\cite{Luscher:1990ck} and to test a novel technique to extract spectral functions form Euclidean correlators~\cite{Bulava:2021fre}. 

The Minkowski action for the O(3) model is
\begin{equation}
S[\boldsymbol{\sigma}]=\frac{\beta}{2} \int\d^2x\,\partial_\mu\boldsymbol{\sigma}(x)\cdot\partial^\mu\boldsymbol{\sigma}(x)\,,
\end{equation}
where $\beta$ is a dimensionless coupling constant, $x=(t,\bm{x})$ is a coordinate in the the (1+1)-dimensional Minkowski spacetime,\footnote{We use boldface to indicate the spacial component of Minkowski vectors, even if there is only one spacial dimension, to distinguish from the full Minkoski vector.} and $\boldsymbol{\sigma}$ is a three-component field of unit length, $\boldsymbol{\sigma}(x)\cdot\boldsymbol{\sigma}(x)=1$, this is, $\boldsymbol{\sigma}(x)\in S^2$. %Note that we use boldface for the spacial vectors, even if they are one-dimensional, to differentiate from the full Minkowski coordinate.

The O(3) model, like QCD, is  asymptotically free and has a mass gap, $m$~\cite{Karowski:1977th,Belavin:1975fg}. %This means that the only excitations in the model are massive with mass $m$~\cite{Karowski:1977th,Belavin:1975fg}.  
In addition, it presents a global O(3) symmetry. Combined, these features imply a low energy spectrum consisting of three particles of mass $m$ transforming in the fundamental representation of the $\mathfrak{o}(3)$ algebra. These properties have been proven in the large $N$ limit of the O($N$) model~\cite{Bardeen:1976zh,Brezin:1976qa}, and are assumed to hold for all $N$.

While these features are shared by all O($N$) models, the $N=3$ case shows further similarities with two-flavor QCD. As commented, the O(3) model presents a low-lying spectrum of particles transforming under the fundamental irrep of $\mathfrak{o}(3)$, while two-flavor QCD has a SU(2) global isospin symmetry and pions transform under the adjoint irrep of the associated Lie algebra. However, these two groups are isomorphic, $\text{O}(3)\cong\text{SU}(2)$, and also are the fundamental irrep of the $\mathfrak{o}(3)$ algebra and adjoint irrep of $\mathfrak{su}(2)$. Thus, like QCD, the O(3) model presents a low-energy spectrum of isospin-one particles.
%Similar theories, such as other O($N$) models, share the qualitative features of the O(3) model and are also integrable, see \rcite{}, but do not present this particular parallelism with QCD. 

The presence of an isospin-one multiplet in the theory implies \mbox{multiparticle} states are organized in the same scattering channels as pions in QCD---see \cref{sec:hadrons:interactionsinChPT}. In particular, two-particle interactions can occur with two-particle isospin $\Iss=2$, 1 and 0, while three-particle scattering happens with three-particle isospin $\Isss=3$, $2$, $1$ and $0$. Recall that the $\Isss=2$ and $\Isss=1$ channels have multiplicity two and three, respectively, which we characterize by the possible two-particle channels in which pairwise interactions can occur. These are the two-particle $\Iss=2$ and $\Iss=1$ channels for $\Isss=2$, and all three two-particle channels for $\Isss=1$ (this is, $\Iss=2,1$ and $0$). %In the work presented here, we focus on the isospin-two and -one channels for two-particles and the isospin-three, -two and -zero channels for three particles.

%Posiblemente quiera reordenar las secciones 1 y 2 y mover este texto 

To study multiparticle interactions, it is key to understand the spacial symmetries of the (1+1)-dimensional theory. Due to the lower dimensionality, the Poincaré group is reduced to contain only three generators: two related to time and spatial translations, and another one related to relativistic boosts. There is however no generator related to spatial rotations, as there is only one single spacial dimension. Instead, the SO(3) rotation group in 3+1 dimensions gets reduced to a discrete $\mathbb{Z}_2$ parity group. 

This has important implications when studying scattering. In the (3+1)-dimensional world, interactions preserving rotation invariance are decomposed into different partial waves. In 1+1 dimensions, instead, scattering observables are separated into parity-odd and parity-even sectors. In the case of two-particles, this division is redundant, as each definite isospin contains only one sector. For example, two-particle isospin-two states which are even under particle exchange only contain partity-even interactions, while $\Iss=1$ states are odd under particle exchange, and thus contain parity-odd interactions.

In the three-particle case, the situation is a bit more complicated as all channels can be projected to both parity sectors. This plays a major role in the application of the three-particle finite-volume formalism and the determination of lattice interpolating operators, since states with zero total momentum need to be projected to definite parity---see \cref{sec:O3model:QC3,sec:O3model:operators}. 

\subsection{The integrable $S$-matrix}\label{sec:O3model:integrability}

Theories that allow for an analytical determination of the scattering $S$-matrix, usually based on the existence of an infinite number of symmetries, are said to be \textit{integrable}. Most known examples of integrable models live in 1+1 dimensions, such as the O(3) model. In contrast, it is believed that no interacting integrable theories exist in the (3+1)-dimensional world, since such a number of conservation laws would imply the absence of interactions.% due to the Coleman-Mandula theorem~\cite{ColemanMandula:1967}.

The existence of an infinite number of conservation laws imposes very restrictive conditions into possible interactions in 1+1 dimensions. The main consequence is the property of \textit{factorization}~\cite{Karowski:1977th}, which implies that the $S$-matrix of three or more particles factorizes into products of successive two-particle $S$-matrices. The infinite number of symmetries also implies no particle production, this is, that the number of particles is preserved~\cite{Polyakov:1977vm}, and that the initial and final sets of momenta are equal, with momenta only being reshuffled. 

In the particular case of the O(3) model, the two-particle $S$-matrix is determined by combining factorization with unitarity and crossing symmetry. The result, which we present below, was first determined in \rrcite{Zamolodchikov:1977nu,Zamolodchikov:1978xm}, where the factorization property was assumed. The existence of an infinite number of non-local symmetries in the O(3) model was proven later in \rrcite{Luscher:1977uq,Buchholz:1978fv,Buchholz:1985xs}.%, which were shown to lead to the same result for the two-particle $S$-matrix and the absence of particle production.

Here, we briefly review the determination of the two-particle $S$-matrix from \rrcite{Zamolodchikov:1977nu,Zamolodchikov:1978xm}. We let $\{k_1,k_2\}$ and $\{p_1,p_2\}$ be the momenta of the two particles in the initial and final states. Similarly, we denote the flavors of the two incoming particles as $(a,b)$ and those of the two outgoing ones as $(c,d)$. Recall that states in the flavor basis, $\{|\sigma^1\rangle, |\sigma^2\rangle, |\sigma^3\rangle\}$ are related to those with definite isopin, $\{|\sigma^+\rangle, |\sigma^0\rangle, |\sigma^-\rangle\}$, by the relations analogous to those in \cref{eq:isospinKmatrix:singlepionbasisrelations}. 

The factorizable $S$-matrix in the flavor basis takes the general form
\begin{multline}\label{eq:O3model:Smatrixgeneral}
S_{cd;ab}(p_1,p_2;k_1,k_2)=\delta^2(p_1-k_1)\delta^2(p_2-k_2)\\
\times\left[\delta_{cd}\delta_{ab}\sigma_1(s)+\delta_{ca}\delta_{db}\sigma_2(s)+\delta_{da}\delta_{cb}\sigma_3(s)\right]\,,
\end{multline}
where $s=(k_1+k_2)^2$ is the usual Mandelstam variable. %Note that we impose that the initial and final sets of momenta to be equal, due to the existence of an infinite set of conservation laws.
The $\sigma_i$ factors are meromorphic functions that can be analytically determined from the properties of unitarity, crossing symmetry and factorization. It is common to work in terms of the relative rapidity of the two particles, $\theta=\theta_1-\theta_2$, where $\theta_i$ is the rapidity of particle $i$,
\begin{equation}
s=2m^2(1+\cosh\theta)\,.
\end{equation}
This transforms physical values of $s$, $s>4m^2$, into real values of $\theta$, while values with $0<s<4m^2$ are mapped to the imaginary $\theta$ axis, with $0<\Im(\theta)<\pi$. Finally, values with $s<0$ are turned to complex $\theta$ values with $\Im(\theta)=\pi$ fixed and varying real part.

\begin{figure}[!t]
    \centering
    \begin{subfigure}{0.495\textwidth} 
    \centering
        \includegraphics[width=\textwidth]{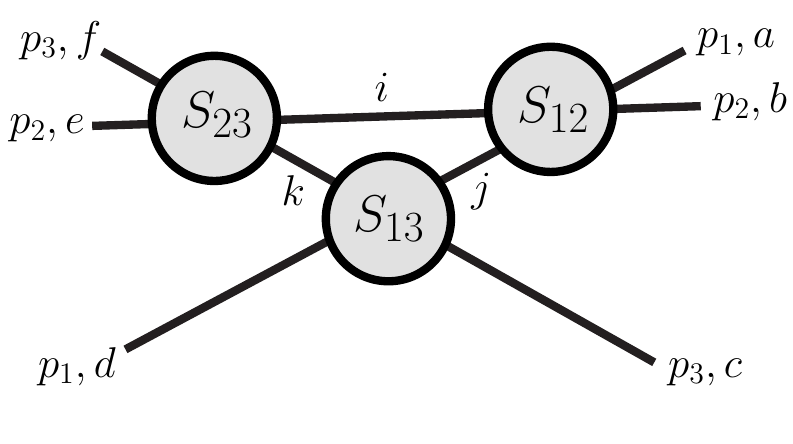}
    \end{subfigure}
    \begin{subfigure}{0.495\textwidth}
    \centering
       \includegraphics[width=1\textwidth]{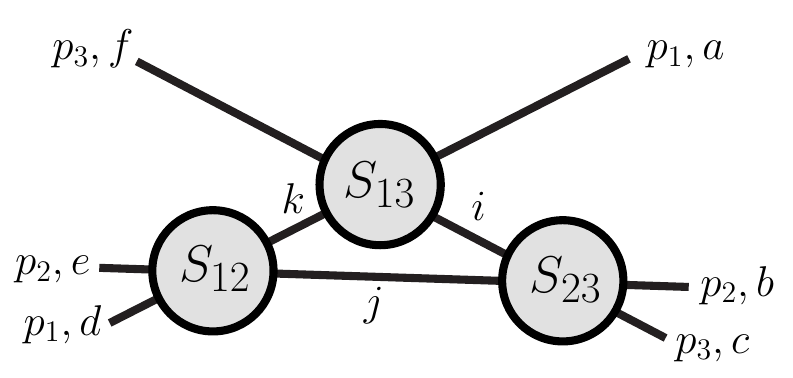}
    \end{subfigure}
    \caption{
        Diagramatic representation of the two equivalent ways of computing the three-particle $S$-matrix in a factorizable theory. This equivalence leads to the Yang-Baxter equations, \cref{eq:O3model:factorization}. Note we have imposed that the final set of momenta is identical to the initial one, which is not true for the flavor indices. We use $S_{ab}=S(\theta_{ab})$, where $\theta_{ab}$ is the relative rapidity of particles $a$ and $b$, and leave the flavor indices implicit. }
    \label{fig:O3model:Smatrix}
\end{figure}

Unitarity and crossing symmetry translate into the following conditions on the $S$-matrix, respectively,
\begin{equation}\label{eq:O3model:unitarity}
S(\theta)S^\dagger(-\theta)=\mathbbm{1}\,,
\end{equation}
\begin{equation}\label{eq:O3model:crossingsymmetry}
S(\theta)=S^\dagger(i\pi-\theta)\,,
\end{equation}
where the Hermitian conjugate of the $S$-matrix acts on initial- and final-state indices separately,
\begin{equation}
S_{cd;ab}^\dagger(p_1,p_2;k_1,k_2)=S_{dc;ba}(p_2,p_1;k_2,k_1)\,.
\end{equation}
These conditions are complemented by the so-called \textit{Yang-Baxter equation}~\cite{McGuire:1964zt,Yang:1967bm}, which incorporates the property of factorization. A multiparticle $S$-matrix is computed as a product of successive two-particle $S$-matrices and the final result must be independent of the order in which two-particle interactions occur. In the case of three-particles, this can happen in two ways, represented in \cref{fig:O3model:Smatrix}. If we label the three particles as ``1'', ``2'' and ``3'', and the initial and final flavor indices as $\{a,b,c\}$ and $\{d,e,f\}$, respectively, the Yang-Baxter equation looks like
\begin{equation}\label{eq:O3model:factorization}
\begin{array}{rl}
S_{def;abc}^\text{3-part}(\theta_{12},\theta_{23},\theta_{31})&=S_{ef,jk}(\theta_{23})S_{dk,ci}(\theta_{13})S_{ij;ab}(\theta_{12})\\
& =S_{kf;aj}(\theta_{31})S_{de;ki}(\theta_{12})S_{ij;bc}(\theta_{23})\,,
\end{array}
\end{equation}
where sum over repeated $i,j,k$ indices is implicit and $\theta_{nm}$ denotes the relative rapidity of particle $n$ and $m$.

\Cref{eq:O3model:unitarity,eq:O3model:crossingsymmetry,eq:O3model:factorization} can be translated into some conditions for the $\sigma_i(\theta)$ functions. The simplest solution to these equations is believed to correspond to the O(3) model,

\noindent\begin{align}
\sigma_2(\theta)&=\displaystyle\frac{\theta(i\pi-\theta)}{(2\pi i-\theta)(i\pi+\theta)}\,,\nonumber\\[5pt]
\sigma_1(\theta)&=\displaystyle-\frac{2\pi i}{\theta}\sigma_2(\theta)\,,\\[5pt]
\sigma_3(\theta)&=\displaystyle-\frac{2\pi i}{i\pi-\theta}\sigma_2(\theta)\,.\nonumber
\end{align}
%which are expected to hold for the O(3) model, based on comparison with the O($N$) model at large $N$~\cite{}.
In general, \cref{eq:O3model:unitarity,eq:O3model:crossingsymmetry,eq:O3model:factorization} lead to a more general results, as $\sigma_2$ above may be multiplied by an arbitrary number of the so-called Castillejo-Dalitz-Dyson poles~\cite{Castillejo:1955ed}. The result with one such pole is expected to correspond to the Gross-Neveu model~\cite{GrossNeveu,Zamolodchikov:1978xm}.

\subsection{Two- and three-particle $S$-matrix}

Using the analytic results for the $S$-matrix, one can obtain the scattering amplitude for the two- and three-particle channels of interest. The projection can be performed using one particular state of each channel, as the scattering amplitude is independent of the third component of isospin. For example, we can consider states with maximal third isospin component. In the isospin basis, they are,
\begin{align}
|\Iss=2\rangle&=|\sigma^+\sigma^+\rangle\,,\nonumber\\[5pt]
|\Iss=1\rangle&=\displaystyle\frac{1}{\sqrt{2}}\left(|\sigma^+\sigma^0\rangle-|\sigma^0\sigma^+\rangle\right)\,,\\[5pt]
|\Iss=0\rangle&=\displaystyle\frac{1}{2}\left(2|\sigma^0\sigma^0\rangle-|\sigma^+\sigma^-\rangle-|\sigma^-\sigma^+\rangle\right)\,,\nonumber
\end{align}
where we leave the momentum dependence implicit.

\begin{figure}[!b]
    \centering
    \begin{subfigure}{0.7\textwidth} 
    \centering
        \includegraphics[width=1\textwidth]{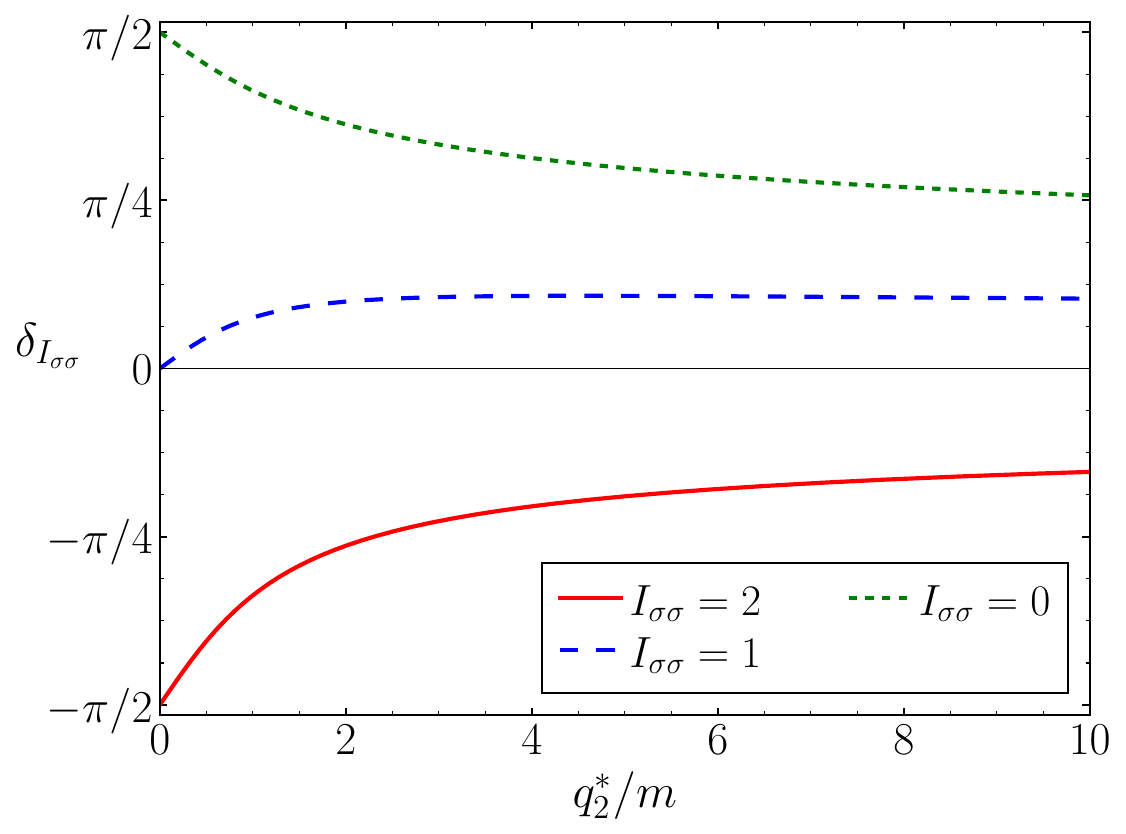}
    \end{subfigure}
    \caption{
         Analytic results for the two-particle scattering phase shift in all isospin channels. We focus on the $\Iss=2$ and $\Iss=1$ channels in this work.  }
    \label{fig:O3model:scatteringphaseshift}
\end{figure}

The two-particle $S$-matrix for each isospin channel then takes the form
\begin{align}\label{eq:O3model:TwoparticleSmatrixresult}
S_{\Iss=2}(\theta)&=\displaystyle\frac{\theta+2\pi i}{\theta - 2\pi i}\,,\nonumber\\[5pt]
S_{\Iss=1}(\theta)&=\displaystyle\frac{(\theta+2\pi i)(\theta-i\pi)}{(\theta - 2\pi i)(\theta+i\pi)}\,,\\[5pt]
S_{\Iss=0}(\theta)&=\displaystyle\frac{\theta-i\pi}{\theta+i\pi}\,.\nonumber
\end{align}
From here, the scattering phase shift can be determined, 
\begin{equation}\label{eq:O3model:Twoparticlephaseshiftresult}
S_{\Iss}(\theta)=\exp\left[2i\delta_{\Iss}(\theta)\right]\,.
\end{equation}
For all two-particle channels, the exact results for $\delta_{\Iss}$ are presented in \cref{fig:O3model:scatteringphaseshift}, as a function of the magnitude of the relative two-particle momentum, $q_2^*$. Note that for all the channels, the phase shift decays logarithmically at large energies, as expected in an asymptotically-free theory.%~\cite{eden2002analytic}.

These results also make it possible to determine the two-particle $K$-matrix, similarly to the (3+1)-dimensional case,%, although no angular momentum appears in this case (and we leave parity implicit),
\begin{equation}\label{eq:O3model:Kmatrixdefinition}
\cK_{2,\Iss}=\rho(s)\cot\delta_{\Iss}(s)\,,
\end{equation}
where $\rho(s)$ is the phase space factor. In (1+1)-dimensions, it takes the form
\begin{equation}\label{eq:O3model:phasespace1plus1}
\rho(s)=\frac{1}{8E^*q_2^*}\,,
\end{equation}
where $E^*=\sqrt{s}$ is the total energy in the CMF frame.% and $q_2$ is the magnitude of the relative momentum.

The Yang-Baxter equation, \cref{eq:O3model:factorization}, also enables us to determine the three-particle $S$-matrix, which can then be projected to any channel of interest. For example, for the $\Isss=3$ channel, one can use the state,
\begin{equation}
|\Isss=3\rangle=|\sigma^+\sigma^+\sigma^+\rangle\,.
\end{equation}
%where we label the three-particle state by its three-particle isospin and the two-particle isospin of the dimer from which the state is constructed.
The $S$-matrix for this channel then is,
\begin{equation}
S_{\Isss=3}(\theta_{12},\theta_{23},\theta_{31})=S_{\Iss=2}(\theta_{12})S_{\Iss=2}(\theta_{23})S_{\Iss=2}(\theta_{31})\,,
\end{equation}
as one could have naively expected since pairwise interactions in this channel can only happen with $\Iss=2$. 
This result can be used to determine the three-particle scattering amplitude, $\cM_3$, and the divergence-free $K$-matrix, $\Kdf$, using a (1+1)-dimensional version of the integral equations presented in \cref{sec:hadrons:infinitevolumethreeparticlescattering}. % in 3+1 dimensions. This determination, however, is left for future work.

\newpage\section{QCs in 1+1 dimensions}\label{sec:O3model:finitevolumeformalisms}

Quantization conditions in 1+1 dimensions are derived in a similar way as their (3+1)-dimensional counterparts---see \cref{sec:hadrons:particleinteractionslattice}. The main novelty is related to the absence of angular momentum. % Instead, one needs to take into account the existence of even and odd parity sectors. 
Spherical harmonics in \cref{eq:hadrons:QC2Fdefinition,eq:hadrons:definitionFthreeparticles,eq:hadrons:definitionGthreeparticles} are substituted by the corresponding functions in one dimension: the identity function for even parity ($p=0$) and the sign function for odd parity ($p=1$),%\footnote{The harmonic functions in $1+1$ dimensions are simply the constant and the linear function.}
\begin{equation}
Y_{p}(\bm{q})=\left\{\begin{array}{ll}
1\,,\quad\quad\quad&\text{even parity}\,,\\
\sign(\bm{q})=\bm{q}/|\bm{q}|\,,\quad\quad\quad&\text{odd parity}\,,\\
\end{array}\right.
\end{equation}
where we indistinctly use $\bm{q}$ to refer to the single-component spacial momentum vector and the component  itself.
In addition, infinite-volume integrals and finite-volume sums in \cref{eq:hadrons:QC2Fdefinition,eq:hadrons:definitionFthreeparticles} now run only over one spacial dimension. In the case of the finite volume and working with periodic boundary conditions, momenta sums run over the finite-volume set,
\begin{equation}\label{eq:O3model:finitevolumemomentumset}
\bm{q}=\frac{2\pi}{L}\bm{n}\,,
\end{equation}
where $\bm{n}\in\mathbb{Z}$ and $L$ is the spacial size of the lattice.

\subsection{Two-particle formalism}\label{sec:O3model:QC2}

The two-particle formalism, introduced in \cref{sec:hadrons:QC2}, is vastly simplified in the (1+1)-dimensional world~\cite{Briceno:2021aiw}. The $F$ geometric factor takes the form,
\begin{equation}\label{eq:O3model:twodimFdefinition}
\tilde{F}(P,L)_{p'p}=\left[\frac{1}{L}\sum_{\bm{k}}-\text{PV}\int\frac{\d k}{2\pi}\right]\frac{ Y_{p'}(\hat{\bm{k}}^*)Y_{p}^*(\hat{\bm{k}}^*)}{8\omega_k\omega_{Pk}(E-\omega_k-\omega_{Pk})}\left(\frac{k^*}{q_2^*}\right)^{p'+p}\,,
\end{equation}
where, recall, $P=(E,\bm{P})$ is the total momentum, $q_2^*$ is the magnitude of the relative momentum in the CMF and $k^*$ is the magnitude of $\bm{k}$ boosted to the CMF. Also, $\omega_k$ and $\omega_{Pk}$ are defined in \cref{eq:hadrons:omegadefinitionsQC2}, and $p$ and $p'$ denote the parity of the initial and final state, respectively.

A priori, this is a $2\times2$ matrix in parity space. However, note that $F_{01}=F_{10}=0$ as they have an odd integrand. Moreover, $F_{00}=F_{11}$ up to exponentially suppressed volume effects, which we neglect. The remaining $F_{00}$ coefficient can be analytically evaluated~\cite{Briceno:2021aiw},%Posiblemente explicar más
\begin{equation}\label{eq:O3model:twodimF00definition}
F(P,L)_{00}=\frac{\rho(s)}{2}\left\{\cot\left[\frac{L\gamma(q_2^*+\omega_q^\star\beta)}{2}\right]+\cot\left[\frac{L\gamma(q_2^*-\omega_q^\star\beta)}{2}\right]\right\}\,.
\end{equation}
Here, $\omega_q^*=E^*/2=\sqrt{q_2^{*2}+m^2}$ is the single-particle energy in the CMF, with $E^*=\sqrt{s}$ the total energy in the CMF, and $\beta=\bm{P}/E$ and $\gamma=E/E^*$ are the boost factors to that frame.

The result in  \cref{eq:O3model:twodimF00definition} can be combined with \cref{eq:O3model:Kmatrixdefinition}, so the two-particle QC, given in  \cref{eq:hadrons:quantizationconditiontwoparticles}, becomes an algebraic relation~\cite{Briceno:2021aiw},
\begin{equation}
\cot\delta(s)+\frac{1}{2}\left\{\cot\left[\frac{L\gamma(q_2^*+\omega_q^\star\beta)}{2}\right]+\cot\left[\frac{L\gamma(q_2^*-\omega_q^\star\beta)}{2}\right]\right\}=0\,.
\end{equation}
This QC was first worked out  for $\bm{P}=0$ in \rcite{Luscher:1990ck}, in which case it takes the simple form
\begin{equation}
2\delta(s)=-q_2^*L\,\,\mod 2\pi\,.
\end{equation}
%which can also be derived from a quantum-mechanical point of view.

\subsection{RFT Three-particle formalism}\label{sec:O3model:QC3}

To study three-particle interactions in the O(3) model, we adapt the RFT formalism to the (1+1)-dimensional world. Contrary to the two-particle case, the RFT formalism does not simplify to an algebraic relation in 1+1 dimensions, but is still a matrix equation of the form of \cref{eq:hadrons:quantizationconditionthreeparticles}. Recall we describe three-particle states as composed of a spectator  and a two-particle dimer. The different quantities appearing in the (1+1)-dimensional QC are matrices in the momenta of the initial and final spectator, as well as the parity---instead of its angular momentum---and the two-particle isospin of the dimer. As we have commented above, different two-particle channels project to a single parity sector, and so the last two indices are redundant.

The main two building blocks of the three-particle QC that need to be modified in 1+1 dimensions are $\tilde{F}$ and $G$, introduced in \cref{eq:hadrons:definitionFthreeparticles,eq:hadrons:definitionGthreeparticles}, respectively. $\tilde{\cK}_2$ takes the same form as in \cref{eq:hadrons:definitionK2threeparticles}, with the phase-space factor being that in \cref{eq:O3model:phasespace1plus1}, while $\Kdf$ is still a dense matrix in all indices.

We consider first the case of three identical particles; the extension to general isospin will be discussed below.
%We comment now on the changes in the case of identical particles, and detail below how they are extended to include different isospin channels. 
The (1+1)-dimensional $\tilde{F}$ factor is defined in a very similar way to \cref{eq:O3model:twodimFdefinition}, with the main difference being the appearance of some factors of the cutoff, $H(x)$, introduced in \cref{eq:hadrons:standardcutoff}
\begin{multline}\label{eq:O3model:twodimFdefinitionthreeparticles}
F(P,L)_{\bm{k}'p',\bm{k}p}
=\delta_{\bm{k}'\bm{k}}H(x_k)\left[\frac{1}{L}\sum_{\bm{a}}-\text{PV}\int\frac{\d a}{2\pi}\right]\\[5pt]
\times\frac{ Y_{p'}(\hat{\bm{a}}^*)Y_{p}^*(\hat{\bm{a}}^*)H(x_a)H(x_{ka})}{8\omega_k\omega_{Pk}(E-\omega_k-\omega_{Pk})}\left(\frac{a^*}{q_2^*}\right)^{p'+p}\,,
\end{multline}
where recall $\bm{a}^*$ indicates the momentum of one of the dimer particles in their CMF, $a^*$ is the magnitude of this vector, and we define $x_a=(P-a)^2/4m^2$, $x_k=(P-k)^2/4m^2$ and $x_{ka}=(P-k-a)^2/4m^2$.
 The two factors of $H$ with $\bm{a}$-dependent arguments, however, take only values different from one for non-singular values of the integrand, and so setting them to one everywhere only introduces exponentially suppressed volume effects, that we neglect. Thus, we can use \cref{eq:O3model:twodimF00definition} to write, %Posiblemente explicar más
\begin{equation}
\tilde{F}(P,L)_{\bm{k}'p',\bm{k}p}=\delta_{\bm{k}'\bm{k}}\delta_{p'p}H(x_k)F(P-k,L)_{00}\,.
\end{equation}

The $G$ factor is also modified, although its final form remains essentially the same as in 3+1 dimensions,
\begin{multline}\label{eq:O3model:definitionGthreeparticlesoneplusone}
G(P,L)_{\bm{k}'p';\bm{k}p }=\\
\frac{1}{4\omega_k\omega_p L}\left(\frac{ k_{k'}^*}{q_{2,k'}^*}\right)^{\,p'}\frac{ Y_{p'}(\hat{\bm{k}}^*_{k'})H(x_{k'})H(x_k)Y_{p}(\hat{\bm{k}}^{\prime*}_k)}{b^2-m^2}\left(\frac{ k_k^{\prime*}}{q_{2,k}^*}\right)^{\,p}\,,
\end{multline}
where recall $k_{k'}^*$ refers to the magnitude of the spectator momentum, $\bm{k}$, in the rest frame of the dimer associated to a spectator of momentum $\bm{k}'$, and $q_{2,k}^*$ is the magnitude of the momentum of the interacting pair associated to a spectator of momentum $\bm{k}$ in its CMF, and similarly for $k_k^{\prime*}$ and $q_{2,k'}^*$. Also $x_{k'}$ is defined similarly to $x_k$.

These quantities can be used to construct $F_3$ as in  3+1 dimensions---see \cref{eq:hadrons:definitionF3threeparticles}. The three-particle QC, \cref{eq:hadrons:quantizationconditionthreeparticles}, makes it possible to either constrain the values of $\Kdf$ from the finite volume energies, or to predict the finite-volume spectrum from an analytic result of $\Kdf$. %Note that now the determinant in the QC runs over  

A particularly useful exercise is the determination of the three-particle finite-volume spectrum %using the results for the two-particle $S$-matrix in \cref{eq:O3model:TwoparticleSmatrixresult}, 
assuming $\Kdf=0$. In this case, the QC simplifies,
\begin{equation}\label{eq:O3model:QC3zeroKmatrix}
\det\left[\tilde{\cK}_2^{-1}-\tilde{F}-G\right]=0\,.
\end{equation}
A comparison of such predictions, obtained using the scattering matrices in \cref{eq:O3model:TwoparticleSmatrixresult}, to lattice results for the finite-volume energies would provide insight on whether $\Kdf=0$ in the O(3) model. A priori, one may believe this to be the case, as factorization implies three-particle scattering is the result of successive two-particle interactions. However, the integral equations relating the scattering amplitude to $\Kdf$---see \cref{sec:hadrons:infinitevolumethreeparticlescattering}---may not preserve the factorization property. This highlights the unphysical nature of $\Kdf$, as it may take non-zero values even in the absence of short-range three-particle interactions. This comparison is presented in \cref{sec:O3model:resultsthreeparticles} below.

The extension of the RFT formalism to general isospin follows the same lines in 3+1 and 1+1 dimensions~\cite{Hansen:2020zhy}. All variables of the formalism become matrices in flavor space, and we use boldface to refer to them. In the in the symmetric basis, defined in  \cref{eq:isospinKmatrix:symmetricbasis}, they are block diagonal in the different isospin channels, each block having a size equal to the possible two-particle channels in which pairwise interactions can occur. %Within each block, we  consider a basis of three-particle states with definite two-particle isospin between the first two. This is the so-called \textit{isospin basis},  in the context of QCD.

Both $\tilde{\mK}_2$ and $\tilde{\mF}$ are diagonal within each three-particle isospin block, with the latter taking the value corresponding to the two-particle isospin of the corresponding state. For example, in the $\Isss=2$ channel,
\begin{equation}
\tilde{\mK}_2^{\Isss=2}=\begin{pmatrix}
\tilde{\cK}_2^{\Iss=2} & 0 \\
0 & \tilde{\cK}_2^{\Iss=1}
\end{pmatrix}\,.
\end{equation}
The $\mG$ term, on the other hand, is a dense matrix within the block, with all its elements equal to \cref{eq:O3model:definitionGthreeparticlesoneplusone} multiplied by some factor related to Clebsch-Gordan coefficients. For the three channels on which we focus,
\begin{equation}
\mG^{\Isss=3}=G=-\mG^{\Isss=0}\,,
\quad\quad\mG^{\Isss=2}=\displaystyle\frac{G}{2}\begin{pmatrix}
-1 & -\sqrt{3}\\
-\sqrt{3} & 1
\end{pmatrix} \,,
\end{equation}
Finally, $\mKdf$ is also dense within each block, and can be expanded around threshold for each three-particle isospin channel as presented in  \cref{sec:isospinKmatrix:thresholdexpansion}. 

Before moving on, a final comment is in order. When obtaining predictions for the finite-volume energies in the rest frame, $\bm{P}=0$, one needs to separate between parity-even and parity-odd states. The building blocks in the QC can be block-diagonalized in these two sectors, which can therefore be studied separately. 

Consider the basis of matrices in which the building blocks of the QC live, which is just the external product of two analogous basis for the initial and final state, defined for a fixed total energy. For the former, for example, such a base is composed of states $|\bm{k},p\rangle$ with spectator momentum $\bm{k}$ in the finite-volume set, \cref{eq:O3model:finitevolumemomentumset}, and dimer parity $p$. Remember the parity of the dimer and its isospin are completely redundant, and so we do not indicate the latter one. %Remember this partity refers to the dimer, not to the whole three-particle state.

In the rest frame, the spectator momentum runs from $-\bmk_\text{max}$ to $\bmk_\text{max}$, where $\bm{k}_{\text{max}}$ is the highest momentum such that $H(x_{k_\text{max}})>0$. States with $\bm{k}\neq 0$ can be grouped into pairs $(|-\bm{k},p\rangle,|\bm{k},p\rangle)^\intercal$, which can then be rotated to definite three-particle parity-even and -odd sectors,
\begin{equation}
\begin{pmatrix}
|\bm{k},\text{even}\rangle \\
|\bm{k},\text{odd}\rangle
\end{pmatrix}
=\frac{1}{\sqrt{2}}
\begin{pmatrix}
1 & (-1)^p \\
-1 & (-1)^p
\end{pmatrix}
\begin{pmatrix}
|-\bm{k},p\rangle \\
|\bm{k},p\rangle
\end{pmatrix}\,.
\end{equation}
States with $\bm{k}=0$ are intrinsically even or odd, depending on the parity of the dimer. For example, for the $\Isss=3$ channel the $\bm{k}=0$ state is even, as dimer interactions happen with $\Iss=2$.%, which is itself even.

%A final comment is in place about the $\bm{P}=0$ frame, as in this the building blocks of the three-particle formalism can be block-diagonalized into 
%Comentar sobre projection for P=0

\newpage\section{Lattice setup}\label{sec:O3model:lattice}

Two- and three-particle finite-volume energies are determined using numerical lattice simulations. We employ the standard discretized Euclidean action,
\begin{equation}\label{eq:O3model:starndarddiscretizedaction}
S_\text{E}[\boldsymbol{\sigma}]=-\frac{\beta}{2}\sum_{x\in\Lambda}\sum_\mu\boldsymbol{\sigma}(x)\cdot\boldsymbol{\sigma}(x+a\hat{\bm{\mu}})\,,
\end{equation}
where $x=(t,\bm{x})$ denotes a site of the (1+1)-dimensional lattice, $\Lambda$, with time and spacial extent $T$ and $L$, respectively, and $a$ is the lattice spacing. Also $\hat{\bm{\mu}}$ is a unit vector in the direction $\mu$.

\subsection{Single- and three-cluster algorithms}

To generate configurations and evaluate single- and multi-particle correlation functions we use a cluster algorithm. This is a collective update algorithm applicable to spin systems that overcomes critical slowing down---this is, the exponential increase of autocorrelation times, especially in topological observables, as one approaches the continuum---and, when used to measure correlation functions, improves the signal to noise ratio~\cite{SwendsenWang,Wolff1,Wolff2,Wolff3,Wolff4,Wolff5}. It was first proposed in \rcite{Wolff1} for a single cluster, %extending the work in \rcite{SwendsenWang} done for the Potts model
and then generalized to two clusters in \rcite{Luscher:1990ck} to study two-particle correlation functions. In this work, we further generalize it to three clusters to be able to investigate three-particle interactions.% single- and double-cluster versions of the algorithm e

Beginning with the single-cluster algorithm, consider a field configuration composed by a set of spins, $\boldsymbol{\sigma}(x)\in S^2$, in every lattice site, $x\in\Lambda$. To update the configuration, we choose a random unit vector $\bm{r}\in S^2$ and a ``seed'' lattice site. From this site, a cluster, $C_r$, is grown: for each site $x\in C_r$, we consider all of its non-cluster nearest neighbors, $y\notin C_r$, which are added to the cluster with probability,
\begin{equation}
p_\text{add}=1-\exp\left\{\min\left[-2\beta\sigma_r(x)\sigma_r(y),0\right]\right\}\,,
\end{equation} 
where $\sigma_r(x)=\bm{r}\cdot\boldsymbol{\sigma}(x)$. This process is repeated for each added site and non-rejected neighbors until all neighbors have been considered. Then, the configuration is updated by modifying all spins in $C_r$,
\begin{equation}\label{eq:O3model:clusterflip}
\boldsymbol{\sigma}(x)\rightarrow\boldsymbol{\sigma}(x)-2\sigma_r(x) \bm{r}\,.
\end{equation}
%One can prove that this algorithm is ergodic, reversible and obeys detail balance, and so defines as valid Markov chain of configurations.

The cluster can also be used to obtain improved estimators of single-particle correlation functions with reduced noise. The idea is to define fixed-time interpolators with definite momentum by projecting only within the cluster,
\begin{equation}
\boldsymbol{\sigma}^{C_r}(t,\bm{p})=\sum_{\bm{x}\in C_r(t)}\boldsymbol{\sigma}(x)\text{e}^{i\bm{p}\bm{x}}\,,
\end{equation}
where $C_r(t)$ denotes all sites in $C_r$ with time coordinate $t$. This field is then used to compute the desired result,
\begin{equation}\label{eq:O3model:singleclustertwopointfunction}
C_{\text{2pt}}(t,\bm{p})=\langle\boldsymbol{\sigma}^{C_r}(t,p)\cdot\boldsymbol{\sigma}^{C_r\,*}(0,p)\rangle= 3\langle\sigma^{C_r}_r(t,\bm{p})\sigma^{C_r\,*}_r(0, p)\rangle\,,
\end{equation}
where we have used that the theory is invariant under global rotations of the spins. Here $\langle\cdot\rangle$ indicates that we average over multiple configurations, and also use time-translation invariance to average over multiple equivalent time separations. The single-particle correlation function with $\bm{p}=0$ can then be used, for example, to extract the single particle mass, as explained in \cref{sec:QCD:computationofobservables}.

It is worth discussing why the use of this algorithm improves the signal-to-noise ratio. %The basic idea is that, when a cluster is created, points within the cluster are correlated in the direction given by $\bm{r}$, while they are decorrelated from points outside the cluster. 
To make it more clear, we consider an alternative update algorithm in which, instead of using one single cluster, we cover the full lattice with $\Ncluster$ non-overlapping clusters, $C_{r,i}$, such that
\begin{equation}
\Lambda=\bigcup_{i=1}^\Ncluster C_{r,i}\,,\quad\quad\quad\quad \bigcap_{i=1}^\Ncluster C_{r,i}=\varnothing\,.
\end{equation}
These clusters are generated in succession using the same vector $\bm{r}$ and ``seed'' sites chosen fromthe sites not belonging to any of the already generated clusters.
The configuration is updated by flipping each of the cluster as in \cref{eq:O3model:clusterflip} with probability $1/2$. Thus, this process may lead to $2^\Ncluster$ possible updated configurations, all with the same propability. 

Naively, the single-particle correlator is computed using single-particle operators projected over the whole time slice,
\begin{equation}
C_{\text{2pt}}(t,\bm{p})=3\sum_{\bm{y}\in\Lambda(t)}\sum_{\bm{x}\in\Lambda(0)}\text{e}^{-i\bm{p}(\bm{x}-\bm{y})}\langle\sigma_r(y)\sigma_r(x)\rangle\,,
\end{equation} 
which is averaged over multiple realizations of the field. In particular, consider the average of  over all $2^\Ncluster$ possible configurations resulting from the updating step described above. Within each cluster, $\sigma_r$ has always the same sign, while spins in different clusters appear with equal and opposite signs an equal number of times in this set of configurations. Therefore, contributions where $x$ and $y$ belong to different clusters average to zero, and only lead to higher variances. In \cref{eq:O3model:singleclustertwopointfunction}, instead, we only sum over a single cluster, avoiding variance increases. The single-cluster algorithm is a simplification of this update process, in which we only generate and update one cluster, which is used to measure observables.

We note, however, that even for single-cluster observables there is still a grow in the uncertainty at long ranges. This responds to the finite-size of the clusters. Clusters grow up to a size of order the correlation length, $m^{-1}$, and non-zero measurements for longer time separations are scarce. %we note that measurements in the cluster algorithm are usually performed as configurations are generated, and many consecutive measurements need to be averaged to obtain a sensible result.

As one can observe from \cref{eq:O3model:singleclustertwopointfunction}, the single-cluster algorithm is suitable to study single-particle correlation functions, but not multiparticle states. In \rcite{Luscher:1990ck}, the algorithm was generalized to two clusters and in this work we  extend it to three. The basic idea of the three-cluster algorithm is to choose three random orthonormal vectors, uniformly sampled from the unit sphere, $\bm{r}, \bm{u}, \bm{v}\in S^2$, so that $\bm{r}^2=\bm{u}^2=\bm{v}^2=1$ and $\bm{r}\cdot\bm{u}=\bm{u}\cdot\bm{v}=\bm{v}\cdot\bm{r}=0$. %This selection is typically performed by choosing first some $\bm{r}$ uniformly drawn from the unit sphere, then sampling $\bm{u}$ from the set of states orthogonal to $\bm{r}$ and finally randomly choosing $\bm{v}$ between the two options orthogonal to both $\bm{r}$ and $\bm{u}$. 
%We note that one cannot choose the last vector as $\bm{v}=\bm{r}\times\bm{u}$, as this would break detailed balance. 
After drawing the three vectors, one also randomly selects three (not-necessarily different) ``seed'' sites, which are used to grow three independent clusters, $C_r$, $C_u$ and $C_v$, following the same procedure as in the single cluster case. The three cluster are allowed to overlap, and we also note that their growth does not interfere with each other, due to the orthogonality of the vectors. The clusters are finally updated following \cref{eq:O3model:clusterflip} and used to measure cluster-improved versions of the correlation functions of interest.

Any of the three clusters alone can be used to measure the single-particle correlation functions as in \cref{eq:O3model:singleclustertwopointfunction}, while correlation functions of two and three particles require the use of two and three clusters, respectively. The relevant two- and three-particle correlation functions needed to study the channels of interest, called $A_i$ ($i=1,2$) and $B_j$ ($j=1...6$), are presented in \cref{fig:O3model:contractions}. Two examples of how they are evaluated using the clusters are
\begin{equation}
\begin{array}{rl}
A_3&\propto\langle\sigma_r^{C_r}(\tau, \bm{p}_2)\sigma_u^{C_u}(\tau,\bm{p}_1)\sigma_u^{C_u\,*}(0,\bm{k}_2)\sigma_r^{C_r\,*}(0,\bm{k}_1)\rangle\,.\\[0pt]
B_2&\propto\langle\sigma_r^{C_r}(\tau,\bm{p}_3)\sigma_u^{C_u}(\tau,\bm{p}_2)\sigma_v^{C_v}(\tau,\bm{p}_1)\sigma_r^{C_r\,*}(0,\bm{k}_3)\sigma_v^{C_v\,*}(0,\bm{k}_2)\sigma_u^{C_u\,*}(0,\bm{k}_1)\rangle\,,
\end{array}
\end{equation}
where $\bm{k}_i$ refers to the momenta of initial-state particles while $\bm{p}_i$ refers to final state ones. The normalization of these correlation functions is not relevant for this work, as it does not affect the determination of the finite-volume energies. 
%The complete list of correlation functions that we compute and how they are used to project to definite isospin channels is presented in \cref{fig:O3model:} and discussed in the next subsection. 
Note that none of the correlation functions computed in this work contains equal-time correlators. This means that their evaluation requires all two or three clusters to  overlap in time. In the practice, the measurements need to be averaged over a large number of configurations. Correlation functions with equal-time contractions could be measured without this limitation, but they do not benefit from the noise reduction.

Finally, we highlight that the study of interactions of four or more particles in the O(3) model would not profit of the use of cluster-improved estimators. For this model, only three orthogonal directions can be defined and so one is limited to three independent clusters. One, however, could generalize the cluster algorithm for other O($N$) models with $N>3$.

\newpage\subsection{Two- and three-particle correlation functions}\label{sec:O3model:operators}

\begin{figure}[t!]
   \centering
\begin{subfigure}[c]{0.395\linewidth}
\centering%
             \begin{minipage}{0.38\textwidth}
\includegraphics[width=\textwidth]{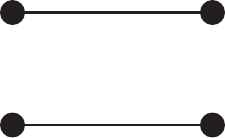} 

       \centering \scalebox{1.}{$A_2$}
\end{minipage} \hspace{0.2 cm}
\begin{minipage}{0.38\textwidth}
\includegraphics[width=\textwidth]{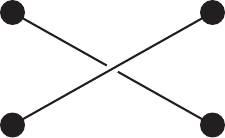} 

        \centering\scalebox{1.}{$ A_3$}
\end{minipage}

\caption{Two-particle contractions}
\label{fig:contraction2part}
\end{subfigure}
\begin{subfigure}[c]{0.595\linewidth}
\centering%
\begin{minipage}{0.25\textwidth}
\includegraphics[width=\textwidth]{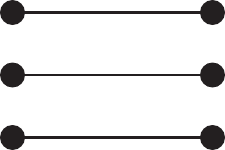}

       \centering ${B_1}$
\end{minipage} \hspace{0.1 cm}
\begin{minipage}{0.25\textwidth}
\includegraphics[width=\textwidth]{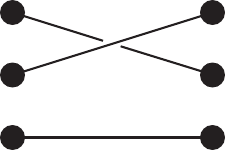} 

       \centering ${B_2}$
\end{minipage} \hspace{0.1 cm}
\begin{minipage}{0.25\textwidth}
\includegraphics[width=\textwidth]{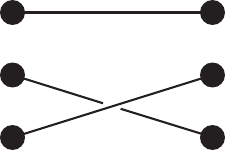} 

       \centering ${B_3}$
\end{minipage} \vspace{0.3cm}

\begin{minipage}{0.25\textwidth}
\includegraphics[width=\textwidth]{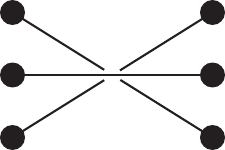} 

       \centering ${B_4}$
\end{minipage} \hspace{0.1 cm}
\begin{minipage}{0.25\textwidth}
\includegraphics[width=\textwidth]{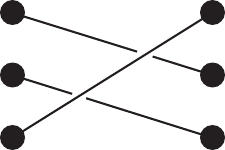} 

       \centering ${B_5}$
\end{minipage} \hspace{0.1 cm}
\begin{minipage}{0.25\textwidth}
\includegraphics[width=\textwidth]{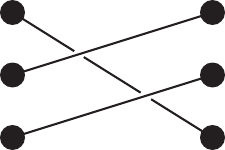} 

       \centering ${B_6}$
\end{minipage} 
\caption{Three-particle contractions}
\label{fig:O3model:contraction3part}
\end{subfigure}
 \caption{Diagramatic representation of the Wick contractions needed for the computation of two- and three-particle correlation functions in this work. Initial and final states are represented at the right and left of the diagrams, respectively, and the corresponding momenta are $\{k_i\}$ and $\{p_i\}$, assigned to the vertices in order from top to bottom.}
   \label{fig:O3model:contractions}
\end{figure}

In this work, we focus on the isospin-two and -one channels for two particles and the isospin-three, -two and -zero channels for three particles. The corresponding correlations functions can be evaluated as a linear combination of different Wick contractions that contain no product of fields evaluated at equal times, represented diagrammatically in \cref{fig:O3model:contractions}. In the case of two particles,
\begin{equation}
\begin{array}{rl}
C_{\Iss=2}&=A_2-A_3\,,\\
C_{\Iss=1}&=A_2+A_3\,.
\end{array}
\end{equation}

Three-particle correlation functions, on the other hand, are matrices in flavor space, with a multiplicity equal to that of the isospin channel. Isospin-three and -zero channels are one-dimensional, with correlation function
\begin{equation}
\begin{array}{rl}
C_{\Isss=3}&=B_1+B_2+B_3+B_4+B_5+B_6\,,\\
C_{\Isss=0}&=B_1-B_2-B_3-B_4+B_5+B_6\,.
\end{array}
\end{equation}
The isospin-two channel, having multiplicity two, is more complicated. We work in the isospin basis, introduced in  \cref{eq:isospinKmatrix:isospinbasis}, and let the two first particles of the states---the two upper dots in each state in \cref{fig:O3model:contraction3part}---represent the dimer. The $\Isss=2$ correlation function then takes the form
\begin{equation}\label{eq:O3model:I2correlationmatrix}
C_{\Isss=2}=\begin{pmatrix}
C_{2,2} & C_{2,1} \\
C_{1,2} & C_{1,1}
\end{pmatrix}\,,
\end{equation}
where $C_{i,j}$ is the correlation function for a $\Isss=2$ state where the initial and final dimer have two-particle isospin $j$ and $i$, respectively,
\begin{equation}
\begin{array}{rl}
C_{2,2}=& \displaystyle B_1+B_2-\frac{1}{2}B_3-\frac{1}{2}B_4-\frac{1}{2}B_5-\frac{1}{2}B_6\,,\\[2pt]
C_{2,1}=& \displaystyle \frac{\sqrt{3}}{2}\left(B_3-B_4+B_5-B_6\right)\,,\\[5pt]
C_{1,2}=& \displaystyle \frac{\sqrt{3}}{2}\left(B_3-B_4-B_5+B_6\right)\,,\\[2pt]
C_{1,1}=& \displaystyle B_1-B_2+\frac{1}{2}B_3+\frac{1}{2}B_4-\frac{1}{2}B_5-\frac{1}{2}B_6\,.
\end{array}
\end{equation}
Note that, if one chooses the two last particles of the state to form the dimer, for example, the correlation function would be different, but will be related to the one in \cref{eq:O3model:I2correlationmatrix} by a rotation of the basis. %The same happens if one considers the same assignments of momenta to each operator in different orders.

To extract the finite-volume spectrum, we determine the correlation functions over a set of two- and three-particle operators characterized by different choices of momenta, $\sigma(\bm{q}_1)\sigma(\bm{q}_2)$ and $\sigma(\bm{q}_1)\sigma(\bm{q}_2)\sigma(\bm{q}_3)$, respectively. %The non interacting energies associated to these operators are
%\begin{equation}
%\begin{array}{rl}
%E^\free_{2part}&=\sqrt{\bm{q}_1^2+m^2}+\sqrt{\bm{q}_2^2+m^2}\,,\\
%E^\free_{3part}&=\sqrt{\bm{q}_1^2+m^2}+\sqrt{\bm{q}_2^2+m^2}+\sqrt{\bm{q}_3^2+m^2}\,.
%\end{array}
%\end{equation}
We consider total-momentum frames $\bm{P}=0$, 1, 2 and 3, in units of $2\pi/L$, and use all possible combinations of momenta so that the corresponding free energies in the CMF lie below $4m$ for two particles and $5m$ for three. Although no particle production occurs in the O(3) model, as discussed in \cref{sec:O3model:integrability}, we choose these values as our upper cutoffs. Therefore, we are  only able to reliably extract the finite-volume energy below these limits. 

  Note that not every operator with arbitrary momenta can be projected to every isospin channel. In the case of two particles, operators with $\bm{q}_1=\bm{q}_2$ do not project to the $\Iss=1$ channel, which is antisymmetric under particle exchange. A similar thing happens in the three-particle isospin-zero channel, which is fully antisymmetric under particle exchange. In this case, only operators with all three momenta different have non-zero projection.

The $\Isss=2$ case is more complicated, as it contains states in which the dimer has isospin two or one. Those states with all three momenta equal, $\bm{q}_1=\bm{q}_2=\bm{q}_3$, do not project to any of the two subchannels. States with two identical momenta only project to a single state. If $\bm{q}_1=\bm{q}_2\neq\bm{q}_3$, this is the state with an isospin-two dimer. Otherwise, both $\Iss=2$ and $\Iss=1$ states can be constructed, but turn out to be proportional once one takes into account the transformation properties of the states.

Finally, we can also comment on three-particle states with zero total momentum, which can be projected to even or odd parity, or both. In the isospin-three channel, all operators project to even parity, but only those having all three momenta different from zero project to the odd sector. In the isospin-zero channel, the contrary situation happens. Lastly, in the isospin-two channel all operators can be projected to both sectors.

\newpage\subsection{Lattice simulations}

\begin{figure}[!tp]
    \centering
    \begin{subfigure}{0.7\textwidth} 
    \centering
        \includegraphics[width=1\textwidth]{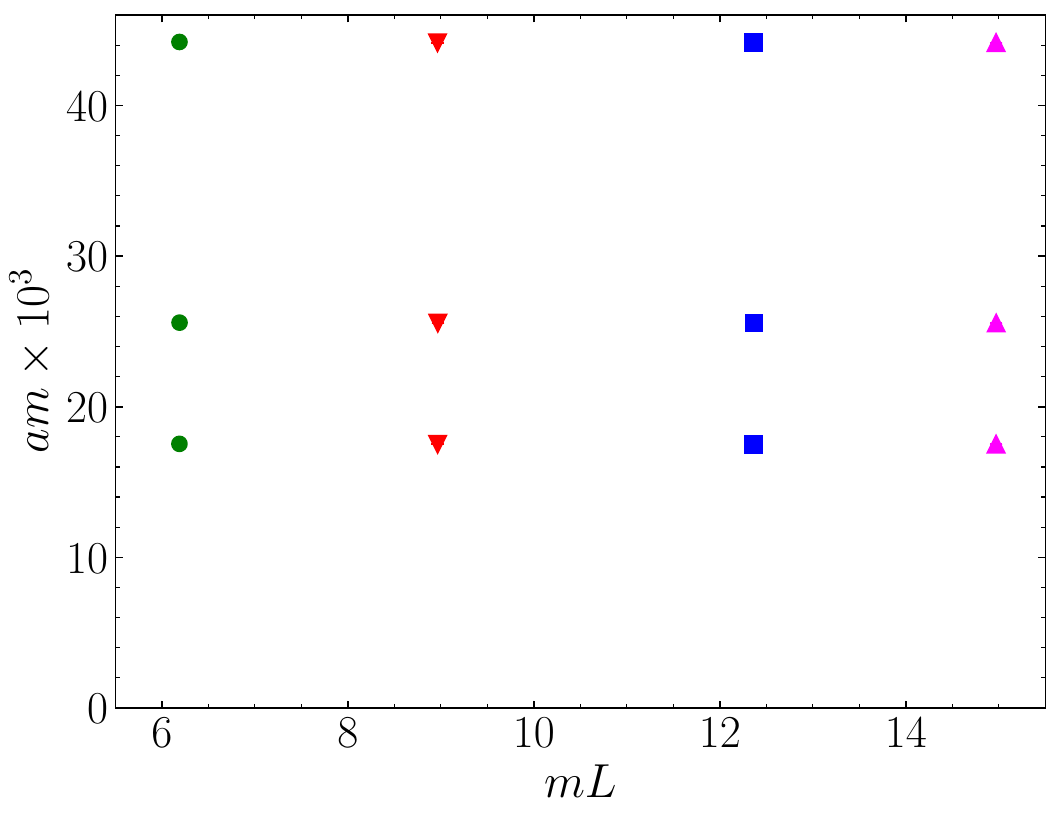}
    \end{subfigure}
    \caption{
         Summary of the simulations used in this work. All have $mT\approx 20.4$ to avoid thermal effects.   }
    \label{fig:O3model:ensemblesummary}
\end{figure}

Numerical simulations are performed using the \texttt{o3\_cluster} code, an early version of which was shared with us by J. Bulava~\cite{Bulava:2022}. We generate a total of 12 ensembles, with four values of $mL$, each at three lattice spacings, corresponding to different choices of $\beta$. Both  $L$ and $T$ are finely tuned to be able to extrapolate the finite-volume energies directly to the continuum. We set $mL>6$ in all ensembles, so that exponentially suppressed finite-volume effects are negligible, and also $mT\sim 20.4$ to be safe of thermal effects. The parameters used for the simulations, together with the result for the single-particle masses, are presented in \cref{tab:O3model:parameters} and summarized in \cref{fig:O3model:ensemblesummary}. %We also present the results for the finite-volume mass, and its value after finite-volume exponentially suppressed volume effect are corrected, $m_\infty$. %We have checked that exponential corrections are completely negligible at our level of precision except for the ``A'' ensembles, for which they amount to a single unit in the last shown digit.

\begin{table}[b!]

\centering
{
\begin{tabular}{ccccc}
\hline
Ensemble& $L\times T$ & $\beta$ & $am$  & $mL$\\ \toprule %& $mL$ & $mT$ 	\\ \toprule
A1 & $140\times 463$ 	& 1.63	 & 0.0442101(28)	 & 6.1894(4) \\
A2 & $242\times 801$ 	& 1.72   & 0.0255795(13)	 & 6.1902(3) \\
A3 & $353\times 1169$ 	& 1.78	 & 0.0175327(9)	 & 6.1891(3) \\\midrule%	 
B1 & $203\times 464$ 	& 1.63	 & 0.0441608(28)	 & 8.9646(5) \\
B2 & $351\times 802$ 	& 1.72   & 0.0255473(15)	 & 8.9671(5)	 \\
B3 & $512\times 1170$ 	& 1.78	 & 0.0175109(9) 	 & 8.9656(5)	 \\\midrule
C1 & $280\times 464$ 	& 1.63	 & 0.0441573(25)	 & 12.3640(7) \\
C2 & $484\times 802$ 	& 1.72   & 0.0255503(13)	 & 12.3664(6) \\
C3 & $706\times 1170$ 	& 1.78	 & 0.0175095(10)	 & 12.3617(7) \\\midrule
D1 & $339\times 464$ 	& 1.63	 & 0.044159(3)	 & 14.9699(10)	 \\
D2 & $586\times 802$ 	& 1.72   & 0.0255486(16)	 & 14.9715(9)	 \\
D3 & $855\times 1170$ 	& 1.78	 & 0.0175092(9) 	 & 14.9704(8)	 \\\bottomrule
\end{tabular}}
\caption{ Summary of the parameters used for the lattice simulations, together with the results for the mass.}
\label{tab:O3model:parameters}
\end{table}

We generate $N_\text{rep}=256$, $512$ and $1024$ replicas, for the coarsest, intermediate and finest ensembles, respectively. In all cases, the lattices are randomly initialized and we perform $12.8\times10^6$ thermalizing three-cluster updates. Following these, we measure the relevant correlation functions over  $N_\text{meas}=4\times10^6$, $2\times10^6$ and $10^6$ successive updates for the coarsest, intermediate and finest ensembles, respectively. Results are averaged over each replica, and we use jackknife for the analysis.

%We have determined two- and three-particle finite-volume energies for $\bm{P}=0,1,2$ and 3. We have consider all possible combinations of momenta which correspond to non-interacting energies below the $4m$ for two particles and $5m$ for three. Although no particle production occurs in the O(3) model, as discussed in \cref{}, we have chosen these as our upper cutoff. We can thus reliably extract the finite-volume energy below these energies, but possible results above these are not reliable. 

\newpage\section{Results for finite-volume energies}\label{sec:O3model:resultsenergies}

\begin{figure}[!p]
    \centering
    \begin{subfigure}{0.495\textwidth} 
    \centering
        \includegraphics[width=\textwidth]{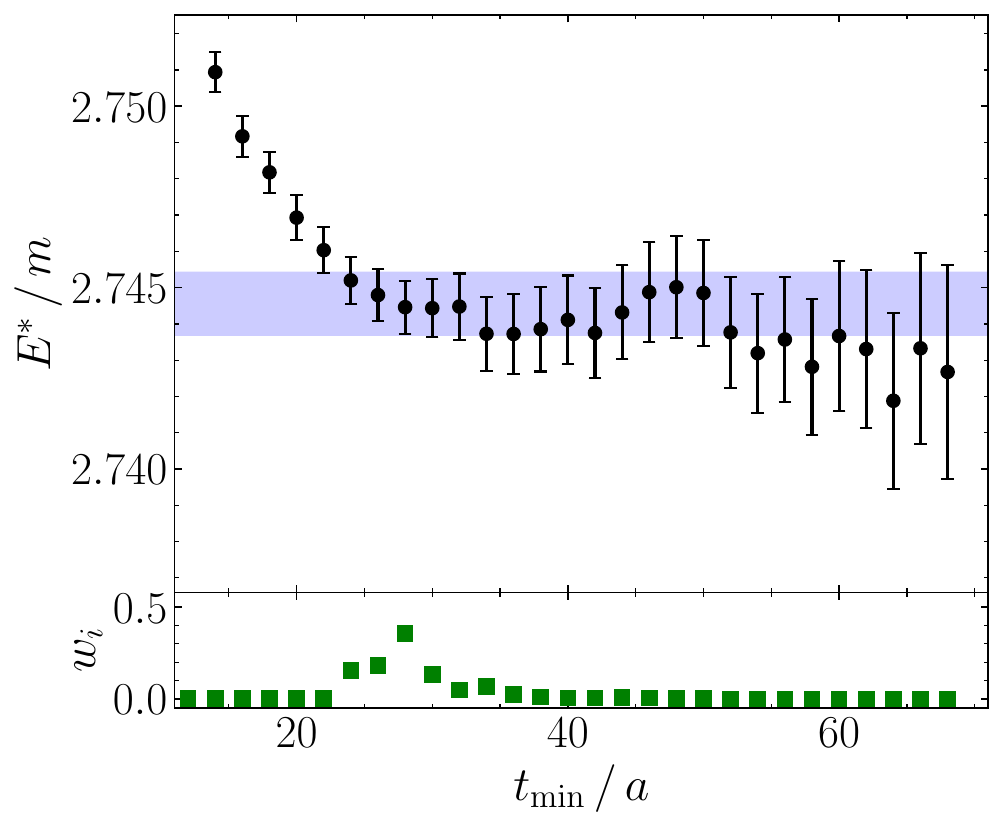}
        \caption{First excited state in the $\Iss=2$ channel}
        \label{fig:O3model:plateaux2}
    \end{subfigure}
    \begin{subfigure}{0.495\textwidth}
    \centering
       \includegraphics[width=\textwidth]{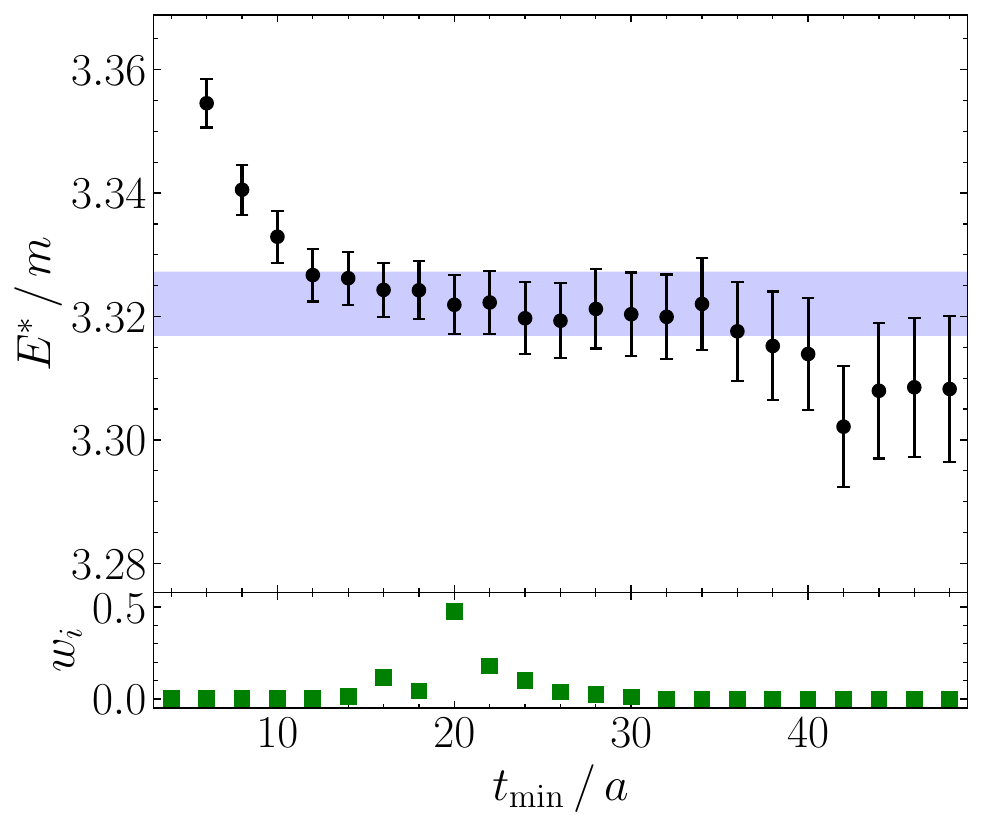}
        \caption{Ground state in the $\Isss=3$ channel}
        \label{fig:O3model:plateaux3}
    \end{subfigure}
    \caption{
        Best-fit results to a single exponential of one of the generalized eigenvalues, $\lambda_n$, of the correlation matrix of the two-particle isospin-two channel (left) and three-particle isospin-three channel (right).  Both cases correspond to the rest-frame and the B1 ensemble. Fits are performed with varying $\tmin$ and fixed $\tmax$ and the final result is obtained by averaging the results using weights based on the Akaike Information Criterion---see \cref{eq:QCD:weightsplateaux}---presented in the lower pannels. }
    \label{fig:O3model:plateaux}\vspace{1.5cm}

    \centering
    \begin{subfigure}{\textwidth} 
    \centering
        \includegraphics[width=1\textwidth]{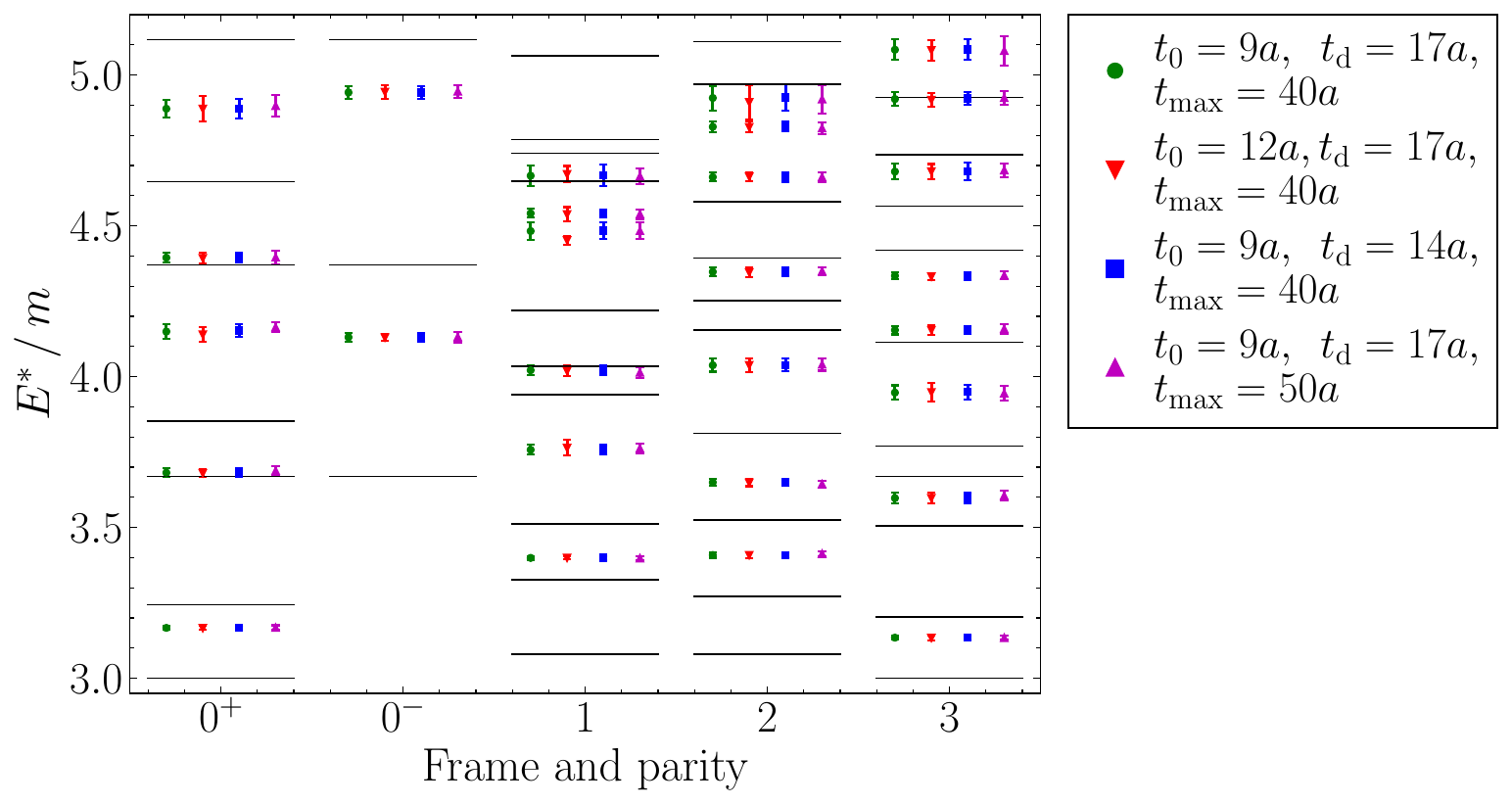}
    \end{subfigure}
    \caption{
         Comparison of the finite-volume energies for the $\Isss=3$ channel in the C1 ensemble for different choices of the parameters used for the analysis. We observe variations differences between different choices of the parameters.   }
    \label{fig:O3model:comparisonGEVP}
\end{figure}

Two- and three-particle matrices of correlators are used to solve a GEVP---see \cref{eq:QCD:gevp}---using the single-pivot approach described in \cref{sec:largeNmesons:finitevolumespectrum}. The resulting eigenvalues are fitted to a single exponential,
\begin{equation}
\lambda_n(t)=A\text{e}^{-E_n t}\,,
\end{equation}
to extract the finite-volume energies. This fit is repeated for different fit ranges, $t\in[t_\text{min}, t_\text{max}]$, with varying $\tmin$ and fixed $\tmax$, and the results are averaged using the weights given in \cref{eq:QCD:weightsplateaux}. In most cases, the weighted average lies where the results show a plateau, although in some other cases we need to restrict the range over which $\tmin$ is allowed to vary. Two examples of the fit results are shown in \cref{fig:O3model:plateaux}. We observe that our results are stables under changes of $\tmax$, $t_0$ and $\td$, introduced in \cref{eq:largeNmesons:singlepivotGEVP}. This is shown in \cref{fig:O3model:comparisonGEVP} for the $\Isss=3$ energies in the C1 ensemble, observing in general very minor variations. %Comentar acerca del step?

\begin{figure}[!tp]
    \centering
    \begin{subfigure}{\textwidth} 
    \centering
        \includegraphics[width=1\textwidth]{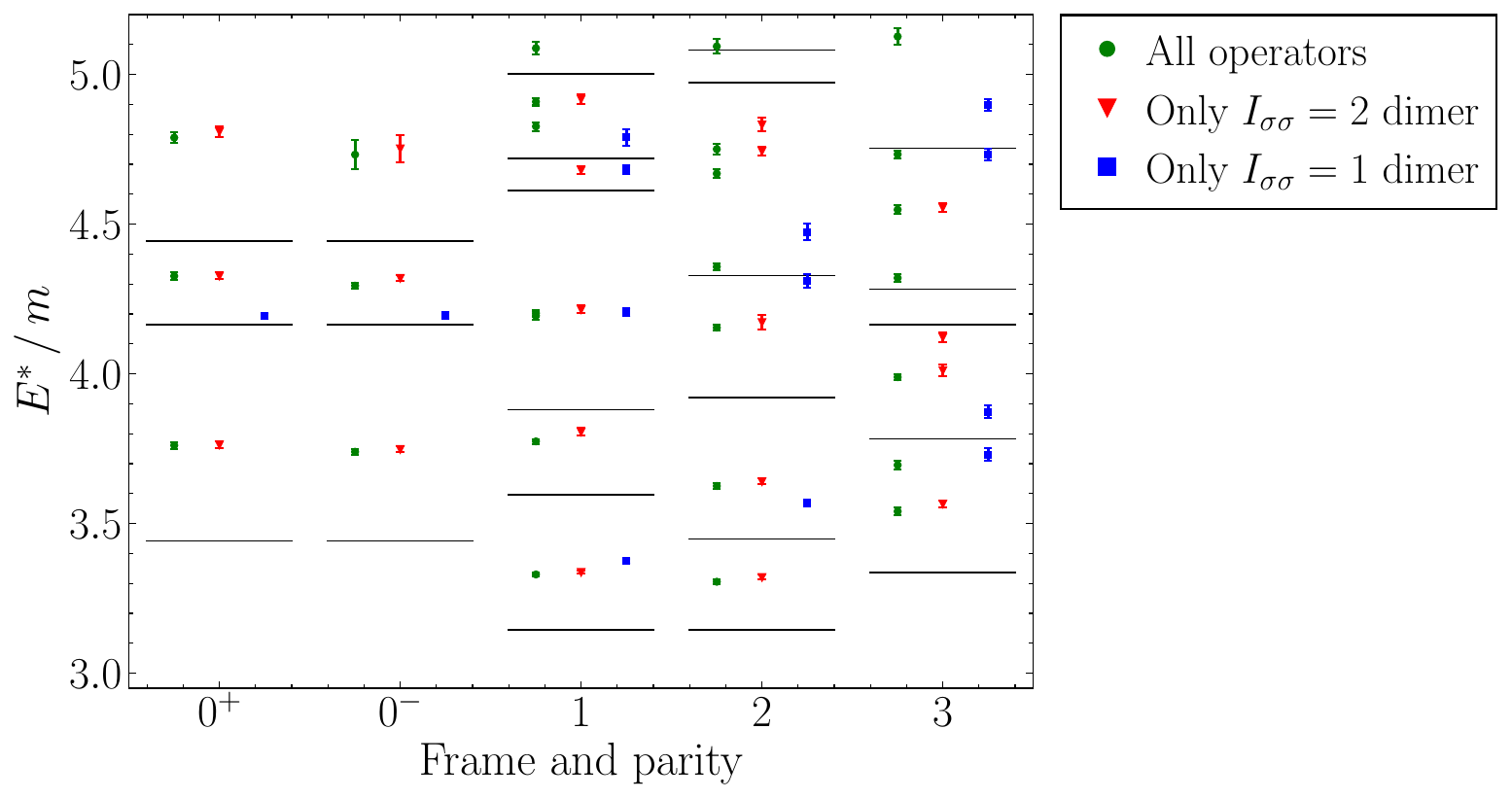}
    \end{subfigure}
    \caption{
         Comparison of the finite-volume energies for the $\Isss=2$ channel in the B1 ensemble determined using different sets of operators to compute the matrix of correlators.   }
    \label{fig:O3model:comparisonOperators}
\end{figure}

We also study the stability of our results under a change on the set of operators used to compute the matrix of correlation functions. We observe a negligible dependence of the states of interest on the inclusion of operators with large associated free energy. 

More interesting is to investigate, in the case of the $\Isss=2$ channel, the effect of operators in which the dimer is projected to $\Iss=2$ or $\Iss=1$. In \cref{fig:O3model:comparisonOperators}, we compare the finite-volume energies obtained using the full correlation matrices (green dots) to those that are obtained if we only use operators in which the dimer is in the isospin-two (red triangles) or isospin-one (blue squares) channel. We observe major differences between the determined spectra. This, however, was to be expected, as selecting the first two particles as the dimer is an arbitrary choice, and a difference selection would lead to operators that are linear combinations of the ones we use. Also note that the combined number of states obtained using only operators with $\Iss=2$ or $\Iss=1$ is larger than the set of states obtained with the full set of operators. This is due to the fact that, for some momenta combinations, operators defined from both values of $\Iss$ are proportional.

\begin{figure}[!b]
    \centering
    \begin{subfigure}{0.495\textwidth} 
    \centering
        \includegraphics[width=\textwidth]{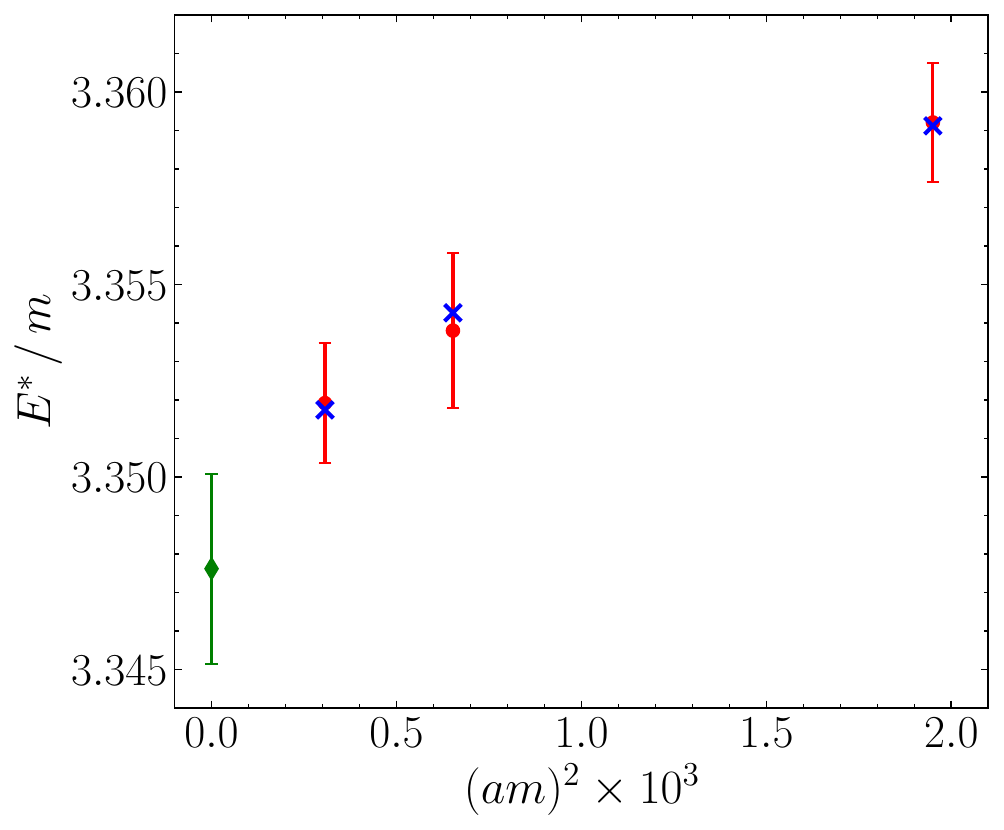}
        \begin{minipage}{0.8\textwidth}
        \centering
        \caption{First excited state for $\bm{P}=2\pi/L$ in the $\Iss=1$ channel.}
        \end{minipage}
        \label{fig:O3model:extrapolations1}
    \end{subfigure}
    \begin{subfigure}{0.495\textwidth}
    \centering
       \includegraphics[width=\textwidth]{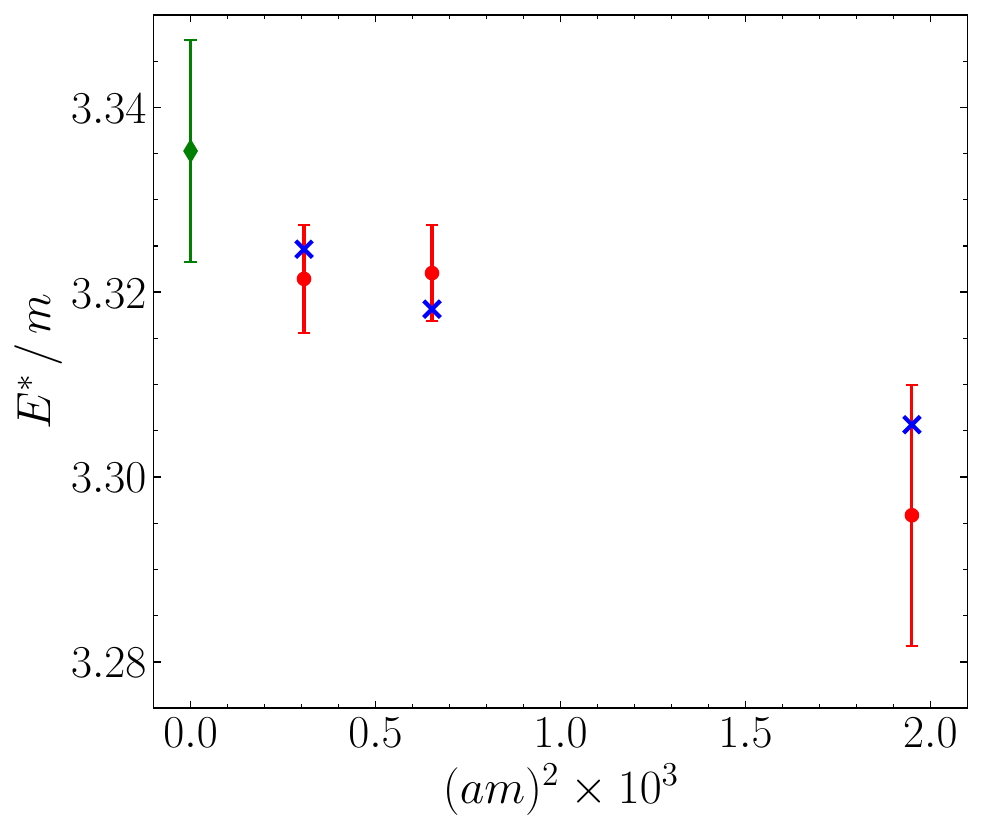}
        \begin{minipage}{0.8\textwidth}
        \centering
        \caption{Ground state for $\bm{P}=0$ with positive parity in the $\Isss=3$ channel.}
        \end{minipage}
        \label{fig:O3model:extrapolations2}
    \end{subfigure}
    \caption{
        Example of continuum extrapolations of  finite-volume energies in the $B$ ensembles. Blue crosses represent the best-fit predictions from \cref{eq:O3model:contibnuumextrapolationexpression} including only $k\leq2$ terms. }
    \label{fig:O3model:extrapolations}
\end{figure}

Using our finely-tuned ensembles, we extrapolate our results for the finite-volume energies directly to the continuum. The O(3) model is well-known to present large logarithmic discretization effects~\cite{Balog:2009yj,Balog:2009np}. These have been studied in detail and on-shell quantities, $Q$, such as finite-volume energies, are known to present the following asymptotic behavior,
\begin{equation}\label{eq:O3model:contibnuumextrapolationexpression}
Q(am)=Q(0)+C \beta^3(am)^2\left[1+\sum_{k=1}^\infty c_k\beta^k\right]+\cO(a^4)\,,
\end{equation}
where $C$ and $c_{\geq3}$ depend on the particular observable, while $c_1$ and $c_2$ are universal for on-shell quantities, and can be computed from two- and three-loop integrals in lattice perturbation theory, respectively. For the standard action in \cref{eq:O3model:starndarddiscretizedaction}, they are~\cite{Balog:2009np}
\begin{equation}
c_1=-1.13861509\,,\quad\quad\quad c_2=-0.4881\,,
\end{equation}
where the first in determined analytically, while the second is estimated numerically in \rcite{Balog:2009np}. 
To extrapolate our results to the continuum, we consider only the $k=1$ and $k=2$ terms, and fit $Q(0)$ and $C$ for each energy level, using the three available lattice spacings. Two examples of such extrapolations are shown in \cref{fig:O3model:extrapolations}. 

Our results of the continuum-extrapolated two-particle finite-volume energies are shown in \cref{sec:O3model:results} in \cref{fig:O3model:energiestwoparticlesI2} and \cref{fig:O3model:energiestwoparticlesI1} for the isospin-two and -one channels, respectively. The results for three particles are shown in \cref{fig:O3model:energiesthreeparticlesI3}, \cref{fig:O3model:energiesthreeparticlesI2} and \cref{fig:O3model:energiesthreeparticlesI0} for $\Isss=3$, 2 and 0, in this same order. These figures also show the free energies and predictions obtained using the quantization conditions, as discussed in the next section.

Before moving on, though, it is worth commenting about the qualitative strength of the interactions in this model. Looking at the two-particle results in the $\Iss=2$ channel, we observe that in many cases the energy shift with respect to the corresponding free energies is positive  and of comparable size to the separation between consecutive free energies. This implies interactions are repulsive and very strong. The isospin-one-channel, on the other hand, is less strongly interacting and attractive. 

These conclusions allow us to qualitatively understand the observed spectra in the three-particle case, which is dominated by pairwise interactions, with short-range three-particle interactions having small effect. In the $\Isss=3$ channel, that contains only $\Iss=2$ pairwise scattering, interactions are repulsive and so strong that finite-volume states have energy shifts larger than the separation between free states. The situation is similar in the $\Isss=2$ channel, although the interaction strength seem to be somewhat smaller. Note however the presence of a larger number of states below $E^*=5m$, due to the dimensionality of the channel. Finally, the $\Isss=0$ shows attractive and not-so-strong interactions, which can be understood since pairwise interactions only happen with $\Iss=1$.

\section{Comparison to analytical predictions}\label{sec:O3model:results}

\subsection{Two-particle energies}

\begin{figure}[!p]
    \centering
    \begin{subfigure}{\textwidth} 
    \centering
        \includegraphics[width=\textwidth]{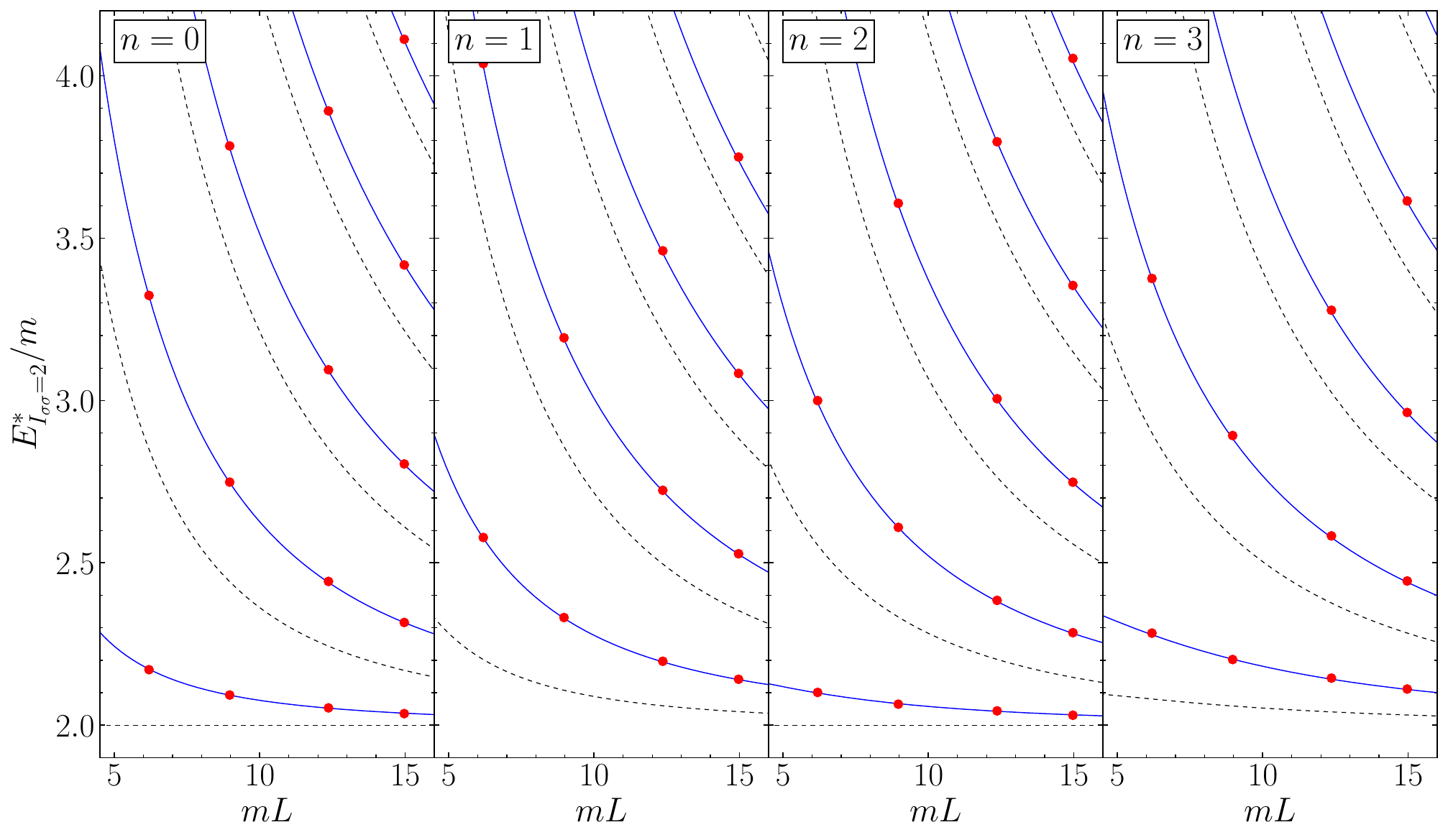}
        \caption{$\Iss=2$ channel}
        \label{fig:O3model:energiestwoparticlesI2}
    \end{subfigure}\vspace{0.4cm}
    \begin{subfigure}{\textwidth}
    \centering
       \includegraphics[width=\textwidth]{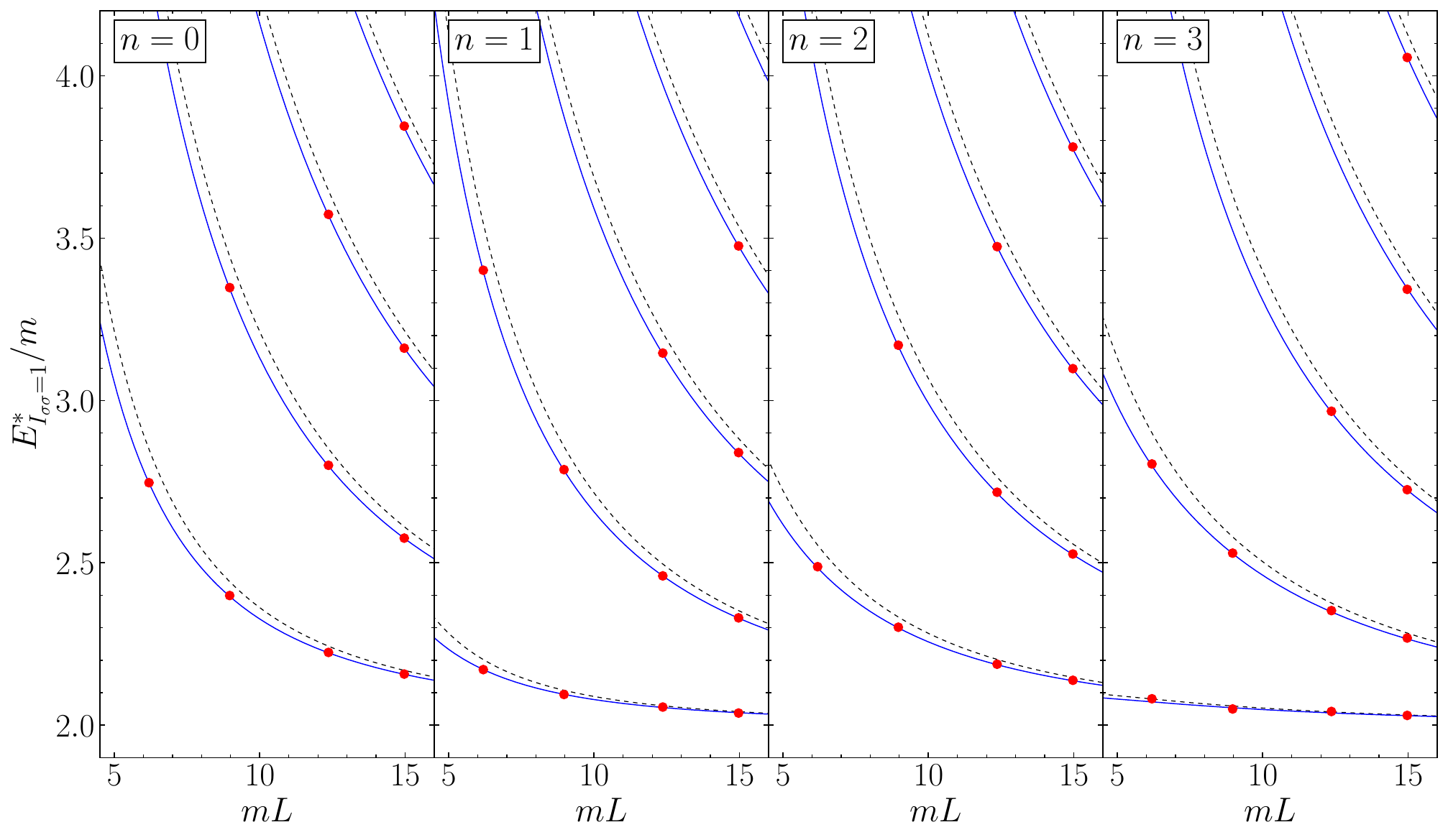}
        \caption{$\Iss=1$ channel}
        \label{fig:O3model:energiestwoparticlesI1}
    \end{subfigure}
    \caption{
        Results for the continuum-extrapolated finite-volume energies of two particles. Lattice results (red dots) are represented together with analytical predictions (solid blue lies) and non-interacting energies (dashed black lines). Each panel corresponds to a different momentum frame, with $\bm{P}=2\pi n/L$. }
    \label{fig:O3model:energiestwoparticles}
\end{figure}

Analytical results for two-particle finite-volume energies can be computed in the O(3) model using the two-particle QC introduced in \cref{sec:O3model:QC2}, together with the analytical predictions of the scattering phase shift in \cref{eq:O3model:TwoparticleSmatrixresult,eq:O3model:Twoparticlephaseshiftresult}. These are shown as solid blue lines in \cref{fig:O3model:energiestwoparticlesI2,fig:O3model:energiestwoparticlesI1}, together with the continuum-extrapolated lattice results (red dots). We also indicate, as dashed lines, the non-interacting energies. Finally, note that no inelastic threshold is indicated, since particle production is forbidden in factorizable theories.

%Qualitatively, one observes that the $\Iss=2$ channel presents repulsive interactions, while interactions in in the $\Iss=1$ channel are attractive. It is of particular interest the strength of the $\Iss=2$ interactions, which can be observed in the size of the separation between the non-interacting and the interacting finite-volume energies. 

We observe very good agreement between lattice results and numerical predictions. We quantify this by computing the $\chi^2$ between both,
\begin{equation}
\begin{array}{rl}
\chi^2\,/\,\dof=54.6\,/\,47=1.16 & \quad\quad\quad\quad (\Iss=2\text{ channel})\,,\\
\chi^2\,/\,\dof= 45.9\,/\,42=1.09& \quad\quad\quad\quad (\Iss=1\text{ channel})\,.\\
\end{array}
\end{equation}
We stress these are not the results from a fit to the lattice, but instead a direct comparison between them and analytical expectations. We also note that states with $E^*>4m$ are not considered to compute the $\chi^2$ as their energies may not be well determined due to the scarcity of operators associated to free energies above this limit. These results root our confidence on the simlation and analysis procedure used, as described in \cref{sec:O3model:lattice,sec:O3model:resultsenergies}.

\subsection{Three-particle energies }\label{sec:O3model:resultsthreeparticles}

Analytical results for three particles could, a priori, be compared against analytical predictions. In the case of the RFT formalism, this would require first to analytically determine the three-particle $K$-matrix, $\Kdf$, solving a (1+1)-dimensional version of the integral equations introduced in \cref{sec:hadrons:infinitevolumethreeparticlescattering}. 

In this work we take a first step in this direction, by comparing our lattice results against predictions obtained under the assumption that $\Kdf=0$. A priori, this could seem like a reasonable assumption, since three-particle interactions in the O(3) model are the result of successive two-particle scattering. However, as we have already discussed, the factorization property is not preserved by the integral equations, and indeed, the definition of $\Kdf$ depends on the prescription used to remove the divergencies---in the case of the RFT formalism, the choice of the cutoff function. A comparison of the lattice results to $\Kdf=0$ predictions will thus provide insight on to which extent $\Kdf\neq0$, and with what statistical significance this can be claimed. Exploring the quantization condition in the simple $\Kdf=0$ case also allows us to learn about the intricacies of the RFT formalism in a theory with very strong interactions. Recall the applications of the RFT formalism to QCD has been restricted to three-meson systems at maximal isospin, in which interactions are weak.

Obtaining analytical predictions of the finite-volume energies involves solving the QC. Under the assumption that $\Kdf=0$, this reduces to finding all solutions to \cref{eq:O3model:QC3zeroKmatrix} in the range of energies of interest, in this case $E^*\leq 5m$. A reliable option to use the eigenvalues, $\lambda_n$, of $ \tilde{\cK}_2^{-1}-\tilde{F}-G$. The finite-volume energies correspond to the location of the simple roots of any of the eigenvalues that cross zero from above~\cite{Blanton:2019igq}. We recall that the number of eigenvalues is equal to the size of the matrices in the QC, which is given by the number of possible spectator momenta in the finite-volume set so that $(P-k)^2>0$. Thus, the size of the matrices depends on the energy.

We face several complications while finding the solutions to the QC. One of them is related to the lower dimensionality of the system. At large $mL$, free energies are expected to approach threshold as $E^\text{free}\sim (mL)^{-2}$, while the finite-volume energy shifts scale as $\Delta E\sim (mL)^{-1}$ in one spacial dimension---it scales as $\Delta E\sim (mL)^{-3}$ in 3+1 dimensions, see for example \cref{eq:hadrons:thresholdexpansionQC2}. This means that as one increases $mL$, interacting energies cross the free energies corresponding to higher states. At these crossing, some elements in $\tilde{F}$ diverge, leading to numerical complications. Another difficulty arises when a solution lies just above from the opening of a new shell, this is, close to an energy where the size of the matrices in the QC is increased. In these cases, we find that the condition number of our matrices becomes large, leading again to numerical issues. 

\begin{figure}[!p]
    \centering
    \begin{subfigure}{0.7\textwidth} 
    \centering
        \includegraphics[width=\textwidth]{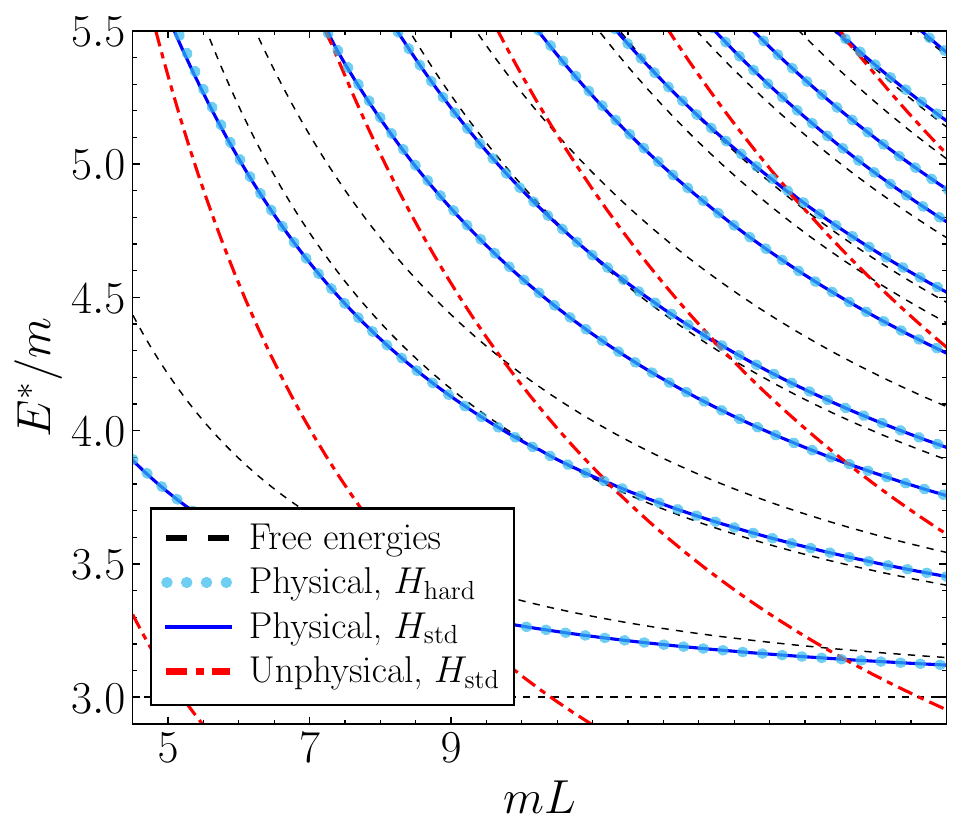}
    \end{subfigure}
    \caption{
       Results for the finite-volume energies assuming $\Kdf=0$ for the $\Isss=3$ channel in the rest frame with positive parity, obtained using two different cutoff functions. We observe the presence of unphysical solutions (red dashed lines) with the standard cutoff choice, \cref{eq:hadrons:standardcutoff}, which decay faster than expected and do not converge to threshold. Physical results (blue solid lines and dots), on the other hand, show negligible differences between the two cutoff choices. }
       \label{fig:O3model:unphysicalsolutionsmL}\vspace{0.6cm}

    \centering
    \begin{subfigure}{0.7\textwidth}
    \centering
       \includegraphics[width=\textwidth]{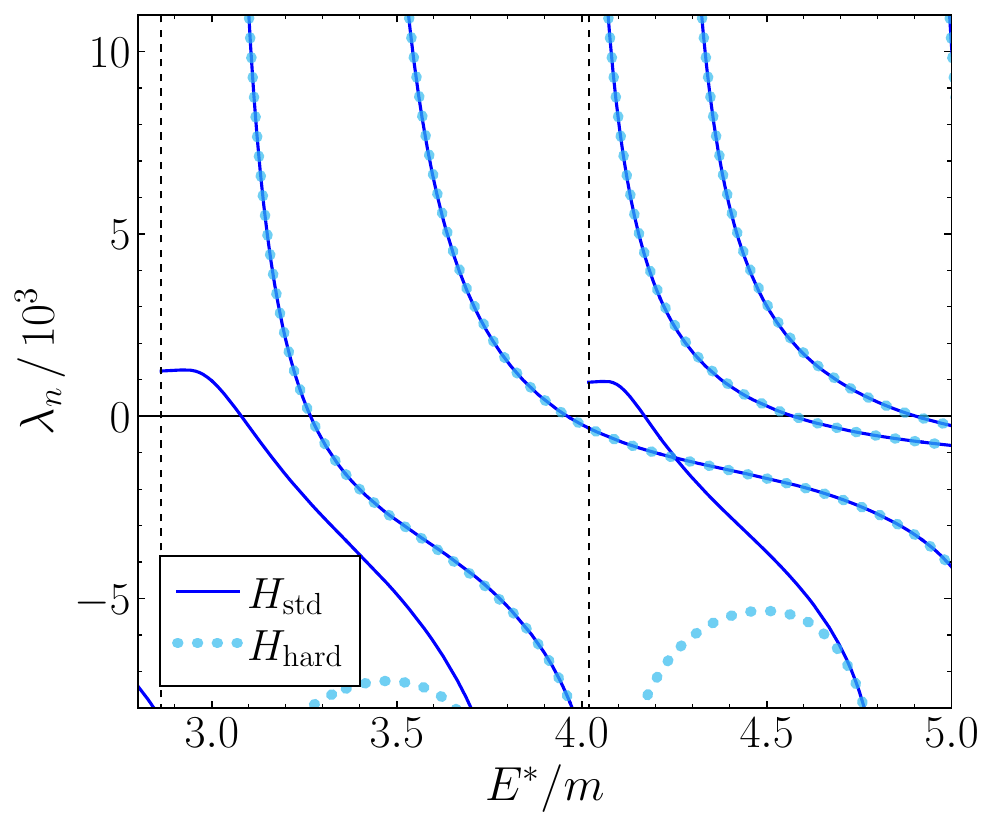}
    \end{subfigure}
    \caption{
         Eigenvalues of $\tilde{K}_2^{-1}-\tilde{F}-G$ for the $\Isss=3$ channel rest frame with positive parity,  for $mL=10$, presented for the standard (blue solid lines) and the hard cutoff (blue dots). Vertical dashed lines indicate the energies at which the size of the matrix is increased. }
    \label{fig:O3model:eigenvalues}
\end{figure}

However, probably the most concerning problem we come across is the appearance of unphysical solutions. These show the correct physical properties expected from a solution to the QC, i.e., they are simple zeros of the eigenvalues that cross zero from above~\cite{Blanton:2019igq}. %\footnote{The eigenvalues of $F_3$ must cross zero from below at physical solutions}. 
However, they decay faster than free energies with $mL$ and do not  converge to any value at large $mL$. The full set of solution is presented for varying $mL$ in \cref{fig:O3model:unphysicalsolutionsmL}. Red lines correspond to the unphysical ones, while the physical ones are indicated as solid blue lines. We also represent the free energies in dashed black.

More interestingly, we find that the unphysical solutions disappear when we change the cutoff function. In \cref{fig:O3model:unphysicalsolutionsmL}, we also present as blue dots the finite-volume energies obtained using a hard cutoff,
\begin{equation}\label{eq:O3model:hardcutoff}
H_\text{hard}(x_k)=\Theta[(P-k)^2]\,,
\end{equation}
with  $\Theta$ the Heaviside step function, instead of the standard choice in \cref{eq:hadrons:standardcutoff}, which we call here $H_\text{std}$ for clarity. Using the hard cutoff, we observe no presence of the unphysical solutions, while the difference in the determination of physical energies is negligible. In \cref{fig:O3model:eigenvalues}, we present the values of $\lambda_n$ as a function of the energy  for both cutoff choices for the positive-parity rest frame in the isospin-three channel using $mL=10$. Additional solutions appear when the standard cutoff is used, at energies slightly above the opening of new matrix shells, denoted by the vertical dashed lines. We note that this effect is observed to happen in both the isospin-three and -two channels.

While the hard cutoff function is not smooth, we find it to work well numerically, and opted to use it to determine the predictions for the finite-volume energies. We note that it has already been used to explore analytical continuation of the solutions to the integral equations~\cite{Dawid:2023jrj}.

\begin{figure}[!p]
    \centering
    \begin{subfigure}{\textwidth} 
    \centering
        \includegraphics[width=\textwidth]{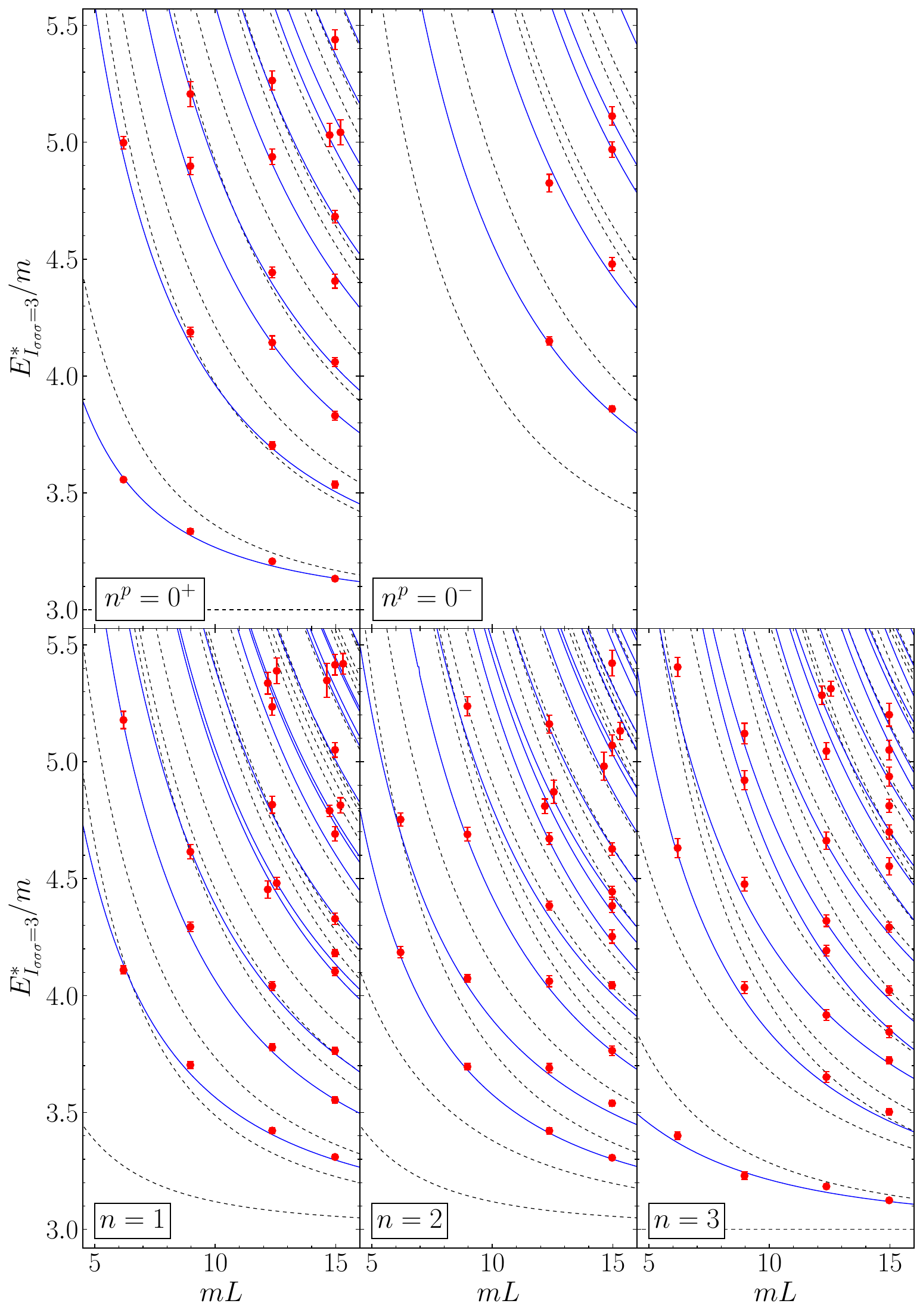}
    \end{subfigure}
    \caption{
        Results for the continuum-extrapolated finite-volume energies in the $\Isss=3$ channel. Lattice results (red dots) are represented together with analytical predictions made assuming $\Kdf=0$ and using the hard cutoff in \cref{eq:O3model:hardcutoff} (solid blue lines), and non-interacting energies (dashed black lines). Each panel corresponds to a different momentum frame, with $\bm{P}=2\pi n/L$. We also present the two parity sectors, $p$, for the $n=0$ case. }
    \label{fig:O3model:energiesthreeparticlesI3}
\end{figure}

\begin{figure}[!p]
    \centering
    \begin{subfigure}{\textwidth} 
    \centering
        \includegraphics[width=\textwidth]{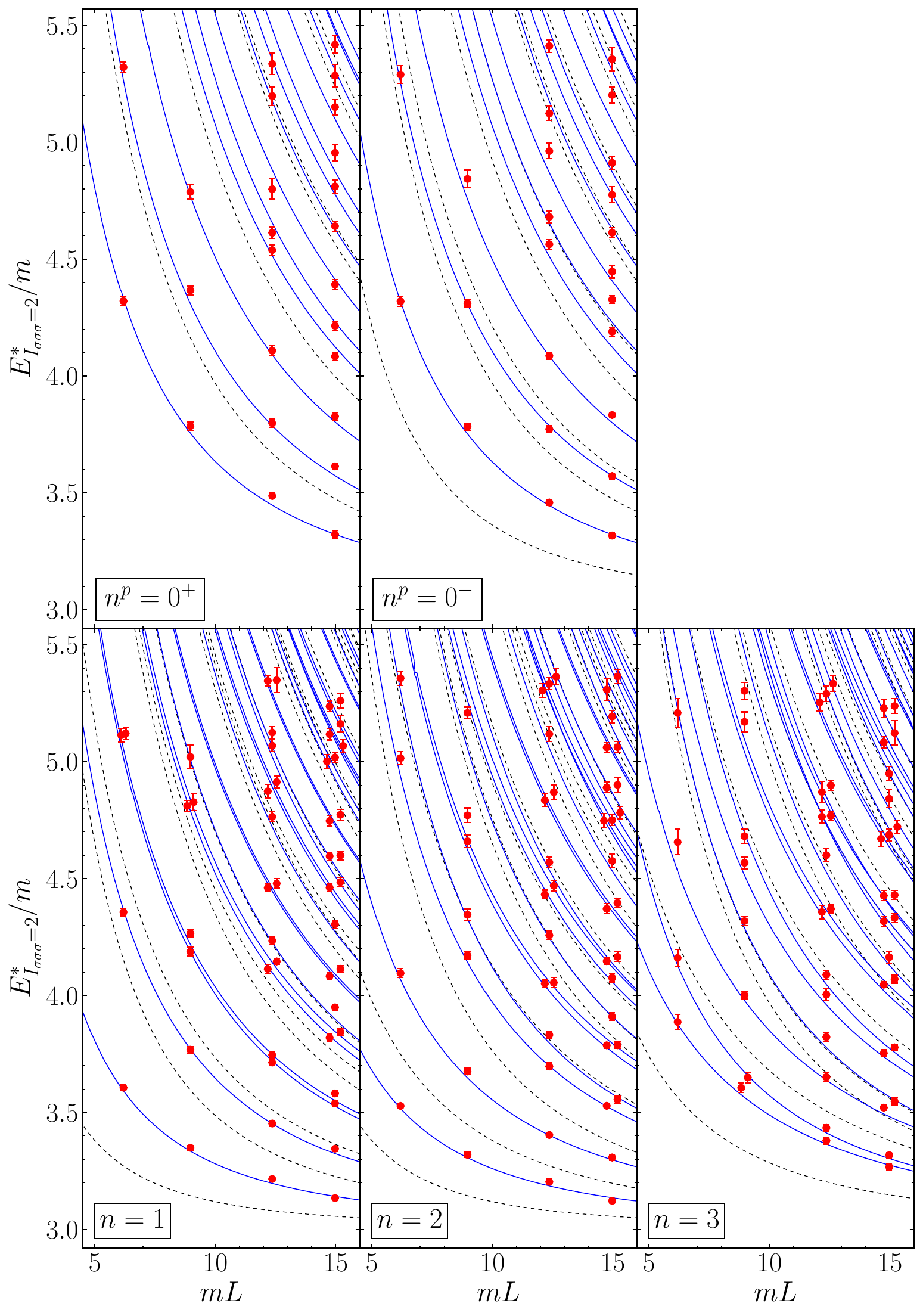}
    \end{subfigure}
    \caption{
        Same as \cref{fig:O3model:energiesthreeparticlesI3} for the $\Isss=2$ channel.}
    \label{fig:O3model:energiesthreeparticlesI2}
\end{figure}

\begin{figure}[!p]
    \centering
    \begin{subfigure}{\textwidth} 
    \centering
        \includegraphics[width=\textwidth]{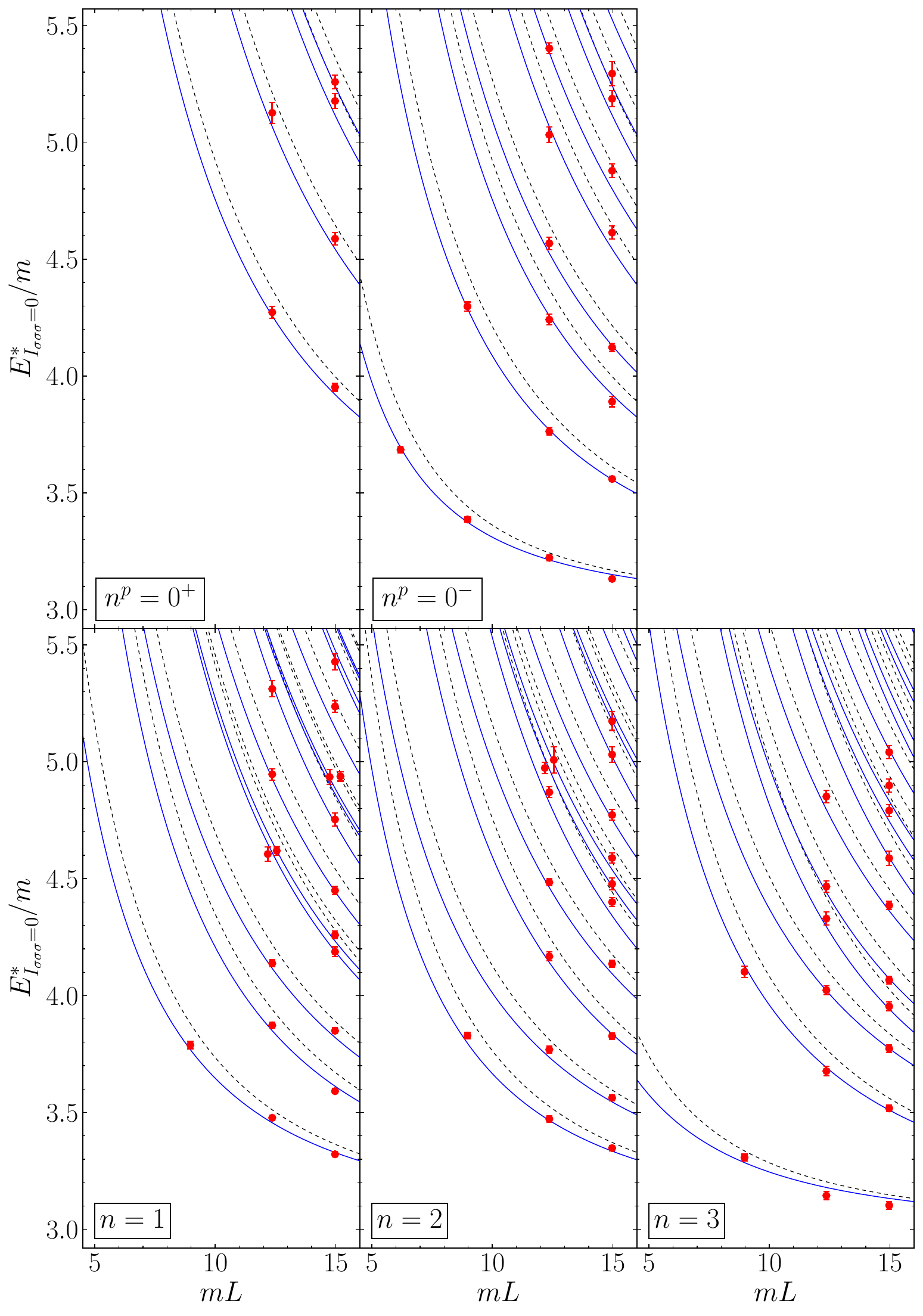}
    \end{subfigure}
    \caption{
        Same as \cref{fig:O3model:energiesthreeparticlesI3} for the $\Isss=0$ channel.}
    \label{fig:O3model:energiesthreeparticlesI0}
\end{figure}

Predictions of the finite-volume energies computed using the two-particle amplitudes from \cref{eq:O3model:TwoparticleSmatrixresult} and assumming $\Kdf=0$ are presented as blue lines in \cref{fig:O3model:energiesthreeparticlesI3,fig:O3model:energiesthreeparticlesI2,fig:O3model:energiesthreeparticlesI0} for the isospin-three, -two and -zero channels, respectively. We also present the lattice results (red dots) and the non-interacting energies (black dashed lines). We observe how the finite-volume spectra are qualitatively well reproduced by $\Kdf=0$ predictions, whith only minor deviations.

To quantify the size of these differences, we compute the $\chi^2$ between the predictions and the lattice results with $E^*<5m$, as the set of operators used to compute correlation matrices only allow us to reliably extract the finite-volume spectrum up to this energy,
\begin{equation}
\begin{array}{rl}
\chi^2\,/\,\dof=215.0 \,/\, 83=2.59 & \quad\quad\quad\quad (\Isss=3\text{ channel})\,,\\
\chi^2\,/\,\dof= 602.7 \,/\, 152=3.96& \quad\quad\quad\quad (\Isss=2\text{ channel})\,,\\
\chi^2\,/\,\dof= 122.9 \,/\, 63=1.95& \quad\quad\quad\quad (\Isss=0\text{ channel})\,.
\end{array}
\end{equation}
These results, while not showing a strong evidence of a non-zero $\Kdf$, point in that direction. We believe that the agreement with lattice results could be improved by allowing for a non-zero $K$-matrix, and especially by using an analytic determination of $\Kdf$. The results from the comparison also evidence the difficulty of accurately determining $\Kdf$ from lattice simulations. Since three-particle interactions are dominated by pairwise scattering, a precise determination of $\Kdf$ requires of a vast level of precision in the finite-volume energies, well in the subpercent level.

\begin{figure}[!tp]
    \centering
    \begin{subfigure}{0.7\textwidth} 
    \centering
        \includegraphics[width=1\textwidth]{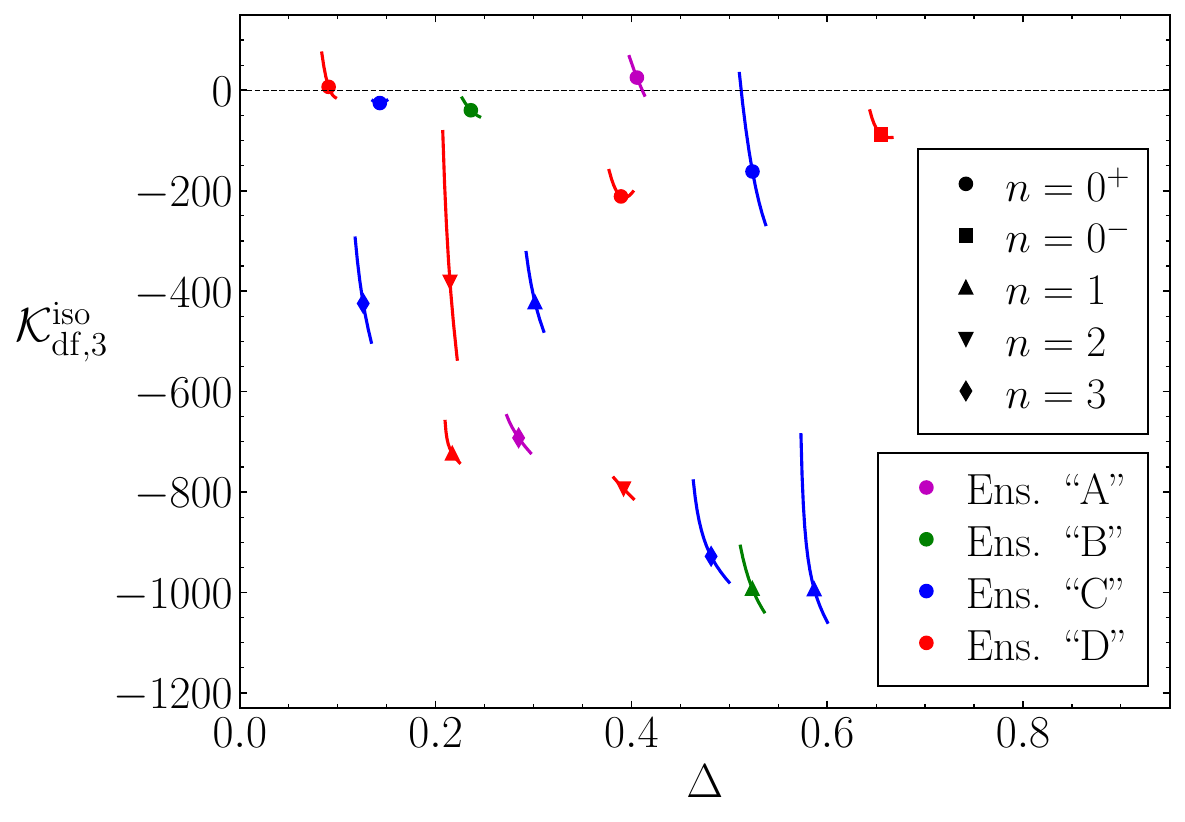}
    \end{subfigure}
    \caption{
         Results for the isotropic $K$-matrix in the $\Isss=3$ channel, computed using the standard cutoff for those states with $E^*<4m$.   }
    \label{fig:O3model:isotropicKmatrix}
\end{figure}

Finally, we compute the isotropic $K$-matrix, $\Kdf^\text{iso}$, in the $\Isss=3$ channel, following \cref{eq:hadrons:isotropicKmatrix}. %A final study we perform, this time in the $\Isss=3$ channel, is the computation of the isotropic $K$-matrix, defined in \cref{eq:hadrons:isotropicKmatrix}. 
The results, computed only for states with $E^*<4m$ and using the standard cutoff, are presented in \cref{fig:O3model:isotropicKmatrix}, as a function of $\Delta=(P^2-9m^2)/9m^2$. We observe how the points show a significant dispersion, with many of them taking values different from zero. This suggests that indeed $\Kdf\neq 0$ in the O(3) model (otherwise all points would lie close to the $\Kdf^\text{iso}=0$ line) and also that the isotropic approximation does not properly reproduce the scattering $K$-matrix (since no trend is observed).

%To obtain analytical predictions under the assumption that $\Kdf=0$, we need to find all the solutions to \cref{eq:O3model:QC3zeroKmatrix} that lie in the range of energies of interest. To make the numerical analysis more stable, it is convenient to multiply the all the matrices in \cref{eq:O3model:QC3zeroKmatrix} by a diagonal matrix $\cQ$ on the right and the left, which has entries 
%\begin{equation}
%\cQ_{\bm{k}'p',\bm{k}p}=\delta_{\bm{k}'\bm{k}}\delta_{p'p} q_{2,k_i}^{*,p}\,.
%\end{equation}
%This eliminated the denominators in the barrier factors in $G$, preventing numerical instabilities when said denominators vanish. Note that this is a complex-valued matrix as $q_{2,k}$ becomes complex when the corresponding dimer has a CMF energy below $2m$.

% Use of one cutoff or the other
% Figure of eigenvalues
% Fit to Kdf

\section{Conclusions}\label{sec:O3model:conclusion}

The O(3) model is commonly used as a toy model of QCD, as both theories share many features, such as asymptotic freedom and a low-energy spectrum of isospin-one particles. In addition, the O(3) model is integrable, allowing for analytical predictions of the scattering matrix. In this chapter, results on the study of two- and three-particle scattering in the O(3) model have been presented, with which we ultimately aim to test the RFT three-particle formalism by comparing lattice results to exact analytical predictions.

We have determined a large number of two- and three-particle finite-volume energies using a three-cluster generalization of the cluster update algorithm, working at four values of the physical volume at three lattice spacings each. Finite-volume energies have been directly extrapolated to the continuum by comparing the results at all three values of the lattice spacing. We have found very good agreement between lattice results for two-particle energies and analytical predictions. In the case of three-particles, we have compared our results to analytical determinations made under the assumption that $\Kdf=0$, obtained using the RFT formalism, which we have extended to 1+1 dimensions. We have found that these predictions reproduce well the structure of the lattice results, but the discrepancy between analytical and lattice results suggests a non-zero $K$-matrix. We have also observed the presence of unphysical solutions to the QC, which appear only for some choices of the cutoff function.

The work presented in this chapter represents a first step towards a test of the three-particle QC, based on comparing lattice results to exact analytical predictions available in the O(3) model. Following this work, we plan to use our results for the finite-volume energies to constrain the values of $\Kdf$. Eventually, we aim at comparing these to an analytical determination obtained from the factorizable $S$-matrix.

\cleardoublepage
\makeatletter
\setlength{\@fptop}{0pt}
\setlength{\@fpbot}{0pt plus 1fil}
\makeatother

\titleformat{\part}
{\thispagestyle{empty}\vspace{30pt}\normalfont\huge}
{
\tikzset{external/export next=false}
\begin{tikzpicture}[remember picture, overlay]
\node[anchor=west, inner sep=0pt] at (current page.west) {\includegraphics[width=\paperwidth, height=1.\paperheight]{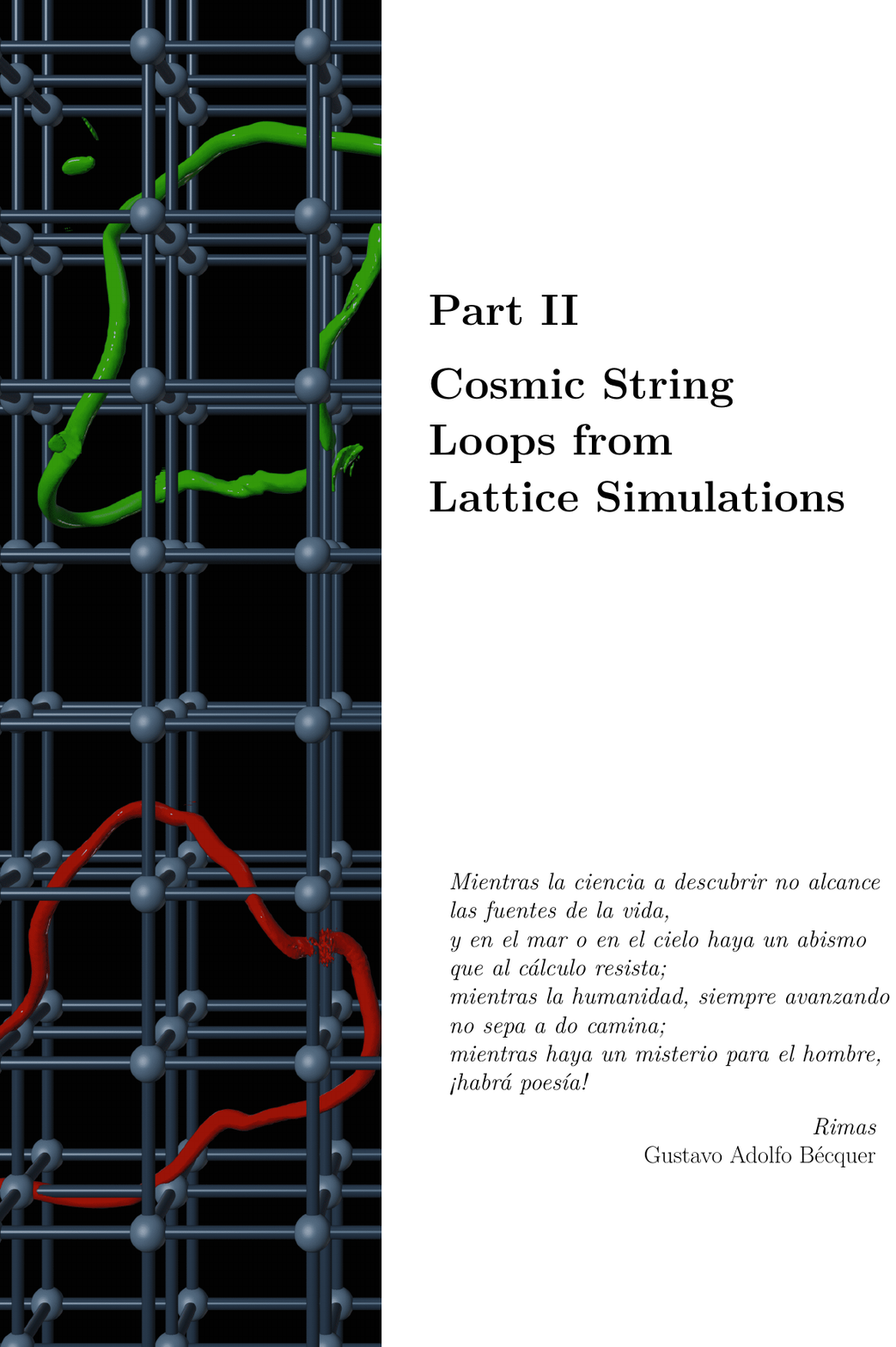}};
\end{tikzpicture}\filleft{\parbox[b][][b]{.38\textwidth}{$\,$}}
}
{0pt}
{\Huge{\parbox[b][][t]{.83\textwidth}{\flushleft\hyphenpenalty=10000\partitlefont $\,$\vspace{0.0cm}}}}[\vspace{0.5ex}\vspace{-0.7cm}]

\part{Cosmic strings loops from lattice simulations}\label{part:strings}
\chapter{Lattice techniques for cosmology}
\label{sec:Cosmo}

%The $\Lambda$ cold dark matter ($\Lambda$CDM) model is the standard model of cosmology.\footnote{See for example \rcite{Weinberg:2008zzc,Baumann_2022} for introductions to the model.} It was formulated in the late 1990s, and has been widely successful in describing most experimental observations, such as the relative abundance of elements in the universe~\cite{PDG:2020} or the cosmic microwave background (CMB)~\cite{Planck:2018nkj}.% and the formation of structure.

%Following the $\Lambda$CDM model, 
The history of the universe can be divided in two epochs, the early and the late universe, separated approximately when the cosmic microwave background (CMB) was released. Our knowledge of the early universe, especially of the very first second, is slim. The most accepted picture is that energies were reached well beyond the electroweak scale and, therefore, phenomena happening possibly involved physics beyond the Standard Model. This implies that the study of this very first moments of our universe is an open window to yet undiscovered particle physics. %On the other hand, the late universe is colder and much less interesting from the particle physics perspective.

%The early universe is characterized by high temperatures and energies. In particular, during the very first seconds of the universe, energies were reached well above the electroweak scale, and so phenomena occurring during this early stages possibly involved physics beyond the Standard Model. This implies that the study of this very first moments of our universe is an open window to yet undiscovered particle physics. On the other hand, the late universe is colder and much less interesting from the particle physics perspective.

%and a colder late universe that is mainly characterized by the formation of structure. During the very first seconds of the universe, energies were reached well above the electroweak scale, and phenomena happening during this time involved physics beyond the Standard Model. Therefore, the study of the very early universe is an open window to yet undetected particle physics.

%The experimental study of the early universe, however, is cumbersome. 
Electromagnetic detectors, though, can only observe the universe after the release of the CMB. %, and so the study of the early universe requires of novel techniques. 
Certain processes that may occur in the first second of the universe are characterized by non-linear dynamics, and many of them are expected to emit gravitational waves that could be observed today as a stochastic background. Due to the weakness of the gravitational interaction, that background would retain information about the particle-physics process producing the gravitational radiation.% interactions between GWs and matter, such a  background would retain its original features, related to the particle physics process that produced said gravitational radiation. 

A proper interpretation of a potential measurement of primordial GWs, however, requires of accurate predictions. The non-linear nature of the phenomena that produce the GWs usually implies that analytical methods have limited applicability. Instead, to fully capture the non-linear dynamics, one is forced to rely on numerical computations.

In this chapter, we review the basics of field theory in an expanding background, and how it can be simulated on the lattice. \Cref{sec:Cosmo:LCDM} explains how non-linear phenomena in the early universe can be investigated using classical-field-theory techniques. Then, \cref{sec:Cosmo:lattice} discusses how such processes can be simulated on the lattice, and how we do it in the \CosmoLattice package. Finally, \cref{sec:Cosmo:GW} reviews how GWs are simulated on the lattice. %The procedure to simulate the emission of GWs detailed in this section was implemented in \CosmoLattice as part of my doctoral work~\cite{GWmodule:2022,GWmodule:2023}.

%In \cref{sec:Cosmo:LCDM} we briefly present the basis of the $\Lambda$CDM model, describe the inflationary paradigm, and explain how phenomena occurring during this early epoch can be studied using classical field theory. I then explain in \cref{sec:Cosmo:lattice} how such processes can be simulated on the lattice, and in particular how we do it in the \CosmoLattice code, which I have used for the study of early-universe phenomena. Finally, in \cref{sec:Cosmo:GW}, I review the basics of GW production and how it can be numerically studied on the lattice. The procedure to simulate the emission of GWs detailed in this section was implemented in \CosmoLattice as part of my doctoral work~\cite{GWmodule:2022,GWmodule:2023}.

\newpage\section{The expanding universe}
\label{sec:Cosmo:LCDM}

The foundations of the standard cosmological model are the theory of general relativity and the \textit{cosmological principle}. This principle states that, statistically, the universe is isotropic and homogeneous, this is, it is independent of direction and position. The cosmological principle constrains the metric of the universe, $g_{\mu\nu}$, to be the so-called \textit{Friedmann-Lemaître-Robertson-Walker} (FLRW) \textit{metric}~\cite{Robertson:1935zz}. Using spherical spacial coordinates, it takes the form
\begin{equation}
\d s^2=g_{\mu\nu}\d x^\mu \d x^\nu =
\d t^2 - a^2(t)\left[\frac{\d r^2}{1+kr^2}+r^2(\d \theta + \sin^2\theta\d \phi^2)\right]\,,
\label{eq:Cosmo:metric}
\end{equation}
where $a(t)$ is the scale factor, that determines the time evolution of physical spacial distances, and $k$ is the spatial curvature, that characterizes the geometry of the universe. It can be either positive, zero or negative, for a closed, flat and open universe, respectively. In the case of a flat universe, as is supported by observations, the metric can be rewritten in Cartesian coordinates  as
\begin{equation}\label{eq:Cosmo:metricflat}
\d s^2 =\d t^2 - a^2(t)(\d x^2+\d y^2 + \d z^2)\,.
\end{equation} 
In some cases, it is convenient to conformally redefine the time coordinate to general $\mathit{\alpha}$-\textit{time}, $\eta$, defined from, $ \d t =a^{\alpha}(\eta)\d \eta$, where $\alpha$ is some real parameter. In particular, $\alpha=0$ corresponds to the standard \textit{cosmic time}, while $\alpha=1$ is the so-called \textit{conformal time}, which we denote as $\tau$. Throughout the remaining of this dissertation, we use the symbol $f^\prime=\d f/\d \eta$ to indicate derivative with respect to general $\alpha$-time.

%The FRLW metric accepts some transformations. First, it is invariant under a rescaling $a\rightarrow ca$, $r\rightarrow r/c$ and $k\rightarrow c^2 k$, for any constant $c$. This allows to set the scale factor $a(t_0)=1$ at any time, $t_0$, typically today. Second, one can perform a conformal redefinition of the time variable to  general $\alpha$-time, $ \d t =a^{\alpha}(\eta)\d \eta$. In particular $\alpha=0$ corresponds to the standard cosmic time, while $\alpha=1$ is the so-called \textit{conformal time}, which we denote as $\tau$. We will use the symbol $f^\prime=\d f/\d \eta$ to indicate derivative with respect to general $\alpha$ time.

From Einstein field equations, one can obtain the evolution equations of the scale factor. These are the so-called \textit{Friedmann equations}~\cite{Fridman2008},
\begin{equation}\label{eq:Cosmo:firstFriedman}
\cH^2=\left(\frac{a'}{a}\right)^2=a^{2\alpha}\frac{\rho}{3\mpl^2}\,,
\end{equation}
\begin{equation}\label{eq:Cosmo:secondFriedman}
\frac{a^{\prime\prime}}{a}=\frac{a^{2\alpha}}{6\mpl^2}[(2\alpha-1)\rho-3p]\,,
\end{equation}
where $\rho$ and $p$ are the background energy density and pressure generated by the contents of the universe, $\mpl=1/\sqrt{8\pi G}\approx 2.44\times10^{18}$ GeV is the reduced Planck mass, with $G$ the Newtonian constant of gravitation, and $\cH=a^\prime/a$ is the Hubble rate in $\alpha$-time.%, which indicates the rate of expansion of the universe.

Based on the cosmological principle, the energy density and pressure of matter can be described as a perfect fluid that obeys a barotropic equation of state, $\rho=wp$, with $w$ constant.
%The energy density and pressure of matter can be described in several ways. The simplest option is to consider the universe to be filled by a perfect fluid, with energy density $\rho$ and pressure $p$, that obey some constant equation of state, $w=\rho/p$. %with stress-energy tensor
%\begin{equation}\label{eq:Cosmo:perfectfluid}
%T_{\mu\nu}=(\rho+p)u_\mu u_\nu+pg_{\mu\nu}\,,
%\end{equation}
%where here $\rho$ and $p$ refer to the energy density and pressure of the fluid, and $u_\mu$ is the velocity of an inertial observer with respect to the background fluid.
%Typically, one assumes the background fluid obeys a constant equation of state, $w=\rho/p$. 
In the case of radiation, non-relativistic matter and vacuum energy, $w=1/3, 0$ and $-1$, in this same order. The Friedmann equations then lead to
\begin{equation}\label{eq:Cosmo:Friedmansolutionsinglefluid}
a(\eta)=a_0\left[1+\frac{\cH_0}{p}\left(\eta-\eta_0\right)\right]^r\,,
\end{equation}
where
\begin{equation}
r=\frac{2}{3(1+w)-2\alpha}\,,
\end{equation}
and $\cH_0$ and $a_0$ are the Hubble rate and scale factor at some reference time $\eta_0$. At late enough time, $\eta\gg \cH_0^{-1}$, the scale factor scales roughly as $a(\eta)\propto\eta^r$. For example, for a radiation-dominated universe, it grows linearly with conformal time, $a(\tau)\propto \tau$. %This result can be complemented by the continuity equation of the fluid, which determines the evolution of $\rho$ and $p$.

%Another option is to consider the universe filled by some classical fields, with energy density and pressure determined from volume averages. This scenario is sometimes called \textit{self-consistent expansion}, and is presented in detail in \cref{sec:Cosmo:fields} in the case of scalar and Abelian gauge fields.

%\subsection{The inflationary early universe}

%The FRLW metric constitutes the foundation of the $\Lambda$CMD model, and successfully explains many experimental observations, such as the age of the universe---round 13.9 billion years,---the relative abundance of chemical elements and the mean temperature of the CMB. What is more, if one includes possible inhomogeneities of the metric and the background fluid, the $\Lambda$CDM model also explains the formation of structure~\cite{Bernardeau:2001qr} and the inhomogeneities observed in the CMB~\cite{Kovac:2002fg}.

The standard cosmological model is successful at explaining many observed phenomena, but it has some shortcomings. One of the most notorious examples is the so-called \textit{horizon problem}. According to the standard cosmology, the universe has a finite age, and so it should be divided in causally disconnected patches. Observationally, however, such patches have the same properties, in apparent contradiction with causality. Another unexplained aspect is the origin of the fluctuations that lead to the formation of structure. 

All these problems are solved by the inclusion of an initial phase of accelerated expansion, $a''>0$, called \textit{inflation}~\cite{Guth:1980zm,Kazanas:1980tx,Starobinsky:1980te,Sato:1980yn,Linde:1981mu}. This phase should last for a long enough time. This is typically measured by the number of times the scale factor increases by a factor of $\text{e}$, also known as \textit{e-foldings}. In order to explain observations, inflation must last for around 60 e-foldings.

\subsection{Field-theory description}\label{sec:Cosmo:fields}

Inflation and other subsequent processes can be generally described in the language of field theory. In many cases, it is possible to work in the \textit{semi-classical approximation}, in which quantum corrections can be neglected due to large occupation numbers, and one can make use of classical-field-theory techniques. 

To better understand this approximation we can first consider the well-known example of black-body radiation. At low temperatures, the occupation number of the different frequencies is small, and the radiation follows the quantum Planck's law. However, as temperature grows, the occupation number of low-frequency modes becomes large, and the low-frequency tail of Planck's law can be approximated by the classical Boltzmann distribution. Note, however, that this does not apply to very high-frequency modes, which still behave in a quantum regime.

An analogous situation happens in the case of certain early-universe phenomena, such as possible phase transitions, the formation of cosmic defects or the emission of GWs. The modes relevant for the process have large occupation, and thus can be described in the semiclassical approximation. This limit, in contrast, cannot be applied to modes with a low occupation number. %However, these are in general exponentially suppressed and have a negligible impact on the dynamics.% of the processes under study. %Quantum effects can be neglected based on the \textit{semi-classical approximation}: due to the high temperatures, the occupation number is large and the quantization of the fields becomes a negligible correction.

The starting point of any classical-field-theory model is the action of the matter fields. For the purpose of this dissertation, we consider a model consisting of a real scalar, $\phi$, a complex scalar, $\varphi$, which can also be expressed in terms of two real scalars, $\varphi=(\phi_1+i\phi_2)/\sqrt{2}$, and a U(1) Abelian gauge field, $A_\mu$, that couples to the complex field. The techniques that we present here, nonetheless, can be generalized to include several copies of each field, scalar multiplets and non-Abelian gauge fields---see \rcite{Figueroa:2020rrl} for a review.

For the model we consider, the action takes the form
\begin{equation}\label{eq:Cosmo:fieldlagrangian}
S=\int\d^4 x\,\sqrt{-g} \left[\frac{1}{2}\partial_\mu\phi\partial^\mu\phi+(D_\mu\varphi)(D^\mu\varphi)^*-\frac{1}{4}F_{\mu\nu}F^{\mu\nu}-V(\phi,\varphi)\right]\,,
\end{equation}
where $g=\text{det}(g_{\mu\nu})$ is the determinant of the FLRW metric, \cref{eq:Cosmo:metricflat}, and Greek indices are contracted using this metric. In this equation, $V(\phi,\varphi)$ is a general potential density depending on the scalar fields, and we have defined the covariant derivative and the field-strength tensor, respectively,
\begin{equation}
D_\mu=\partial_\mu-ieQ A_\mu\,,\quad\quad\quad F_{\mu\nu}=\partial_\mu A_\nu-\partial_\nu A_\mu\,,
\end{equation}
where $e$ is the gauge coupling and $Q$ is the charge of $\varphi$ under the U(1) gauge symmetry. In analogy to  QCD---see \cref{eq:QCD:gaugetransformationsQCD}---this action is invariant under local gauge transformations,
\begin{equation}\label{eq:Cosmo:gaugetransformations}
\begin{array}{rl}
\phi(x) & \longrightarrow \phi(x)\,,\\
\varphi(x) & \longrightarrow \text{exp}\left[-i e Q \alpha(x)\right]\,\varphi(x)\,,\\
A_\mu(x) & \longrightarrow A_\mu(x)-\partial_\mu\alpha(x)\,,
\end{array}
\end{equation}
where $\alpha(x)$ is some arbitrary function. To study the dynamics of the fields, we work in the temporal gauge, in which $A_0=0$. Note that while $A_i$ is still gauge-dependent, one can define gauge-independent observables: the  \textit{electric} and the \textit{magnetic fields}, respectively,
\begin{equation}\label{eq:Cosmo:electricmagneticfield}
E_i=F_{0i}\,,\quad\quad\quad B_i=\frac{1}{2}\epsilon_{ijk}F^{jk}\,,
\end{equation}
with $\epsilon_{ijk}$ the Levi-Civitta symbol.

The evolution of the fields is governed by the classical equations of motion. These are obtained from minimizing the action, \cref{eq:Cosmo:fieldlagrangian},  with respect to the fields. They read
\begin{align}
\phi^{\prime\prime}+(3-\alpha)\frac{a^\prime}{a}\phi^\prime-a^{-2(1-\alpha)}\partial_i\partial_i\phi & =-a^{2\alpha}\frac{\d V}{\d\phi}\,,\label{eq:Cosmo:fieldseomsinglet}\\
\varphi^{\prime\prime}+(3-\alpha)\frac{a^\prime}{a}\varphi^\prime-a^{-2(1-\alpha)}D_i D_i\varphi & =-a^{2\alpha}\frac{\d V}{\d\varphi}\,,\label{eq:Cosmo:fieldseomcomplex}\\
F_{0i}^\prime+(1-\alpha)\frac{a^\prime}{a}F_{0i}-a^{-2(1-\alpha)}\partial_j F_{ji} & = a^{2\alpha}J_i\,,\label{eq:Cosmo:fieldseomgauge}\\
\partial_i F_{0i} & =a^2J_0\label{eq:Cosmo:fieldseomgauss}\,,
\end{align}
where we define the U(1) current density,
\begin{equation}\label{eq:Cosmo:complexscalarcurrent}
J_\mu= 2e Q\Im[\varphi^*D_\mu \varphi]\,.
\end{equation}
It is worth mentioning that \cref{eq:Cosmo:fieldseomgauss} is the Gauss' law, which is not a dynamical equation of the fields, but rather a constraint equation. 

\Cref{eq:Cosmo:fieldseomsinglet,eq:Cosmo:fieldseomcomplex,eq:Cosmo:fieldseomgauge} allow one to study the dynamics of the fields in an expanding background. The evolution of the scale factor can be fixed, given by some background fluid---see \cref{eq:Cosmo:Friedmansolutionsinglefluid}---if the matter fields are treated as spectators, or be induced by the fields themselves. This latter case is known as \textit{self-consistent expansion}, in which the second Friedmann equation, \cref{eq:Cosmo:secondFriedman}, is regarded as the equation of motion of the scale factor, while the first Friedmann equation, \cref{eq:Cosmo:firstFriedman}, can be treated as a constrain equation, extending energy conservation to expanding backgrounds. 

In the case of self-consistent expansion, the source of the Friedmann equations are the background energy density and pressure generated by the matter fields, which are determined from volume averages. The total energy density and pressure can be decomposed into the energy-density components of the different fields, %The stress-energy tensor is given by
%\begin{multline}\label{eq:Cosmo:stressenergytensorfields}
%T_{\mu \nu} = -\frac{2}{\sqrt{g}}\frac{\delta(\sqrt{g} \mathcal{L})}{\delta g^{\mu \nu}}
%= g_{\mu \nu} \mathcal{L} - 2 \frac{\delta \mathcal{L}}{\delta g^{\mu \nu}}\\
%\begin{array}{rl}
%= & - g_{\mu \nu} \left\{g^\ab\left[ (D_{\alpha} \varphi_i ) (D_{\beta} \varphi_i )^* + \frac{1}{2} (\partial_{\alpha} \phi_i ) (\partial_{\beta} \phi_i )\right] 
%+ \frac{1}{4}g^{\alpha\delta}g^{\beta\lambda}  F_{\alpha \beta} F_{\delta\lambda} + V \right\} \\
%+ & \left\{ 2 {\rm Re}\left[(D_{\mu} \varphi_i) (D_{\nu} \varphi_i )^*\right] + (\partial_{\mu} \phi_i) (\partial_{\nu} \phi_i ) \right\}  + g^{\alpha\beta}  F_{\mu\alpha} F_{\nu\beta} \,. 
%\end{array}
%\end{multline}
%From here, the local energy density and pressure are
\begin{equation}\label{eq:Cosmo:energydensitypressurefields}
\begin{array}{rl}
\rho=&\displaystyle\rhok^\phi+\rhok^\varphi+\rhog^\phi+\rhog^\varphi+V+\rhok^\text{U(1)}+\rhog^\text{U(1)}\,,\\[5pt]
p=&\displaystyle\rhok^\phi+\rhok^\varphi-\frac{1}{3}\left(\rhog^\phi+\rhog^\varphi\right)-V+\frac{1}{3}\left(\rhok^\text{U(1)}+\rhog^\text{U(1)}\right)\,,
\end{array}
\end{equation}
where $V$ is the potential energy density,  and we define the kinetic, $\rhok$, and gradient, $\rhog$, energy densities of the scalar and gauge fields,
\begin{equation}
\begin{array}{rlcrl}
\displaystyle\rhok^\phi&\displaystyle=\frac{1}{2a^{2\alpha}}\phi^{\prime\,2}\,, & \quad\quad\quad & \displaystyle\rhog^\phi&\displaystyle=\frac{1}{2a^2}\sum_{i} (\partial_i\phi)^2\,,\\
\displaystyle\rhok^\varphi&\displaystyle=\frac{1}{a^{2\alpha}} |\varphi^\prime|^2\,, & \quad\quad\quad & \displaystyle\rhog^\varphi&\displaystyle=\frac{1}{a^2}\sum_{i} |D_i\varphi|^2\,,\\
\displaystyle\rhok^\text{U(1)}&\displaystyle=\frac{1}{2a^{2+2\alpha}}\sum_i F_{0i}^2\,, & \quad\quad\quad & \displaystyle\rhog^\text{U(1)}&\displaystyle=\frac{1}{2a^4}\sum_{i,j<i} F_{ij}^2\,.\\
\end{array}
\end{equation}
The kinetic and gradient energy density of the Abelian field, in the last line, are also called electric, $\rho_E$, and magnetic, $\rho_B$, energy densities, respectively, since they depend on the electric and magnetic field---see \cref{eq:Cosmo:electricmagneticfield}. %To evaluate the Friedman equations, one uses volume averages of the quantities in \cref{eq:Cosmo:energydensitypressurefields}, which defines the background energy density and pressure.

The last ingredient needed to study the classical field dynamics are some \textit{initial conditions} from which the fields are evolved. These should be chosen to represent the physics of interest. For scalar fields and their derivatives, the initial conditions typically consist of a homogeneous mode, on top of which some fluctuations following a given spectrum are added. For example, studies of particle production at the end of inflation set some homogeneous value for the scalar field that drives the inflation, while other daughter fields are set to zero. On top of these homogeneous values, initial fluctuations are added resembling those of quantum origin.  Gauge fields are generally set to zero, while the corresponding derivatives are chosen so that the Gauss constrain, \cref{eq:Cosmo:fieldseomgauge}, is initially verified~\cite{Figueroa:2015rqa}. This means the initial magnetic energy is set to zero, while there is some amount of electric energy.

\section{Classical field theory on the lattice}\label{sec:Cosmo:lattice}

Analytical methods can be used to study early-universe processes when field fluctuations are small, by treating these fluctuations as small perturbations. However, this approach is not always possible. In some regimes, early-universe phenomena present large non-linearities and the only option to capture them is the use of classical-field-theory lattice simulations, performed using a discretized and dimensionless version of the equations of motion. 
%Given a system consisting of some field content and their initial conditions, three ingredients are needed to perform numerical simulations: a lattice, a discretized and dimensionless version of model, and an evolution algorithm. We now comment on these points. %For a review on numerical algorithms used to study early universe cosmology, see \rcite{Figueroa:2020rrl}. 
In this section we describe how lattice techniques can be used to study the early universe. These are the basis of the work presented in \cref{sec:global,sec:local}, which makes use of the \CosmoLattice code~\cite{Figueroa:2020rrl,Figueroa:2021yhd}.

\subsection{Lattice definition and Fourier transform}

A lattice is a discretized and finite representation of space. We consider a cubic three-dimensional lattice,
\begin{equation}
\Lambda = \{\bm{n}=(n_1,n_2,n_3)\,|\,n_i\in\mathbbm{Z}\,,\,0\leq {n}_i < N\}\,,
\end{equation}
where $N$ is the number of sites per dimension.\footnote{The same techniques presented here can be extended to any number of dimensions.} We define the lattice spacing, $\delta x$, as the comoving distance between consecutive sites. In the context of an FLRW universe, lattice sites have fixed comoving coordinates, $\delta x \bm{n}$, and physical distances are obtained multiplying the comoving ones by the scale factor. The comoving side of the lattice is thus $L=\delta x N$. Finally, continuum functions, $f(\bm{x})$ are defined on the lattice as functions $f(\bm{n})$ so that $f(\bm{n})=f(\bm{x})$ when $\bm{x}=\delta x \bm{n}$.\footnote{We distinguish continuum functions, $f(\bm{x})$, from functions defined on the lattice, $f(\bm{n})$, by their argument.} Note we work with periodic boundary conditions in all directions, meaning $f(\bm{n}+L\hat{\bm{i}})=f(\bm{n})$ with $\hat{\bm{i}}$ a unit vector in the $i$ direction.

The finite volume and lattice spacing  limit the range of scales that can be studied on the lattice, as they set the longest and shortest distances that can be probed. To better understand this limitation, it is convenient to first introduce the \textit{discrete Fourier transform} and the \textit{reciprocal lattice}. Given some real function $f(\bm{n})$, its discrete Fourier transform is defined as
\begin{equation}\label{eq:Cosmo:DFT}
f(\tn)=\sum_{\bm{n}\in\Lambda}\text{e}^{-\frac{2\pi i}{N}\tn \bm{n}}f(\bm{n})\,,
\end{equation}
which we distinguish from the original field by the tilde in its argument. Similarly, we also define the inverse Fourier transform as
\begin{equation}\label{eq:Cosmo:inverseDFT}
f(\bm{n})=\frac{1}{N^3}\sum_{\tn\in\LambdaR}\text{e}^{\frac{2\pi i}{N}\tn \bm{n}} f(\tn)\,,
\end{equation}
where $\LambdaR$ refers to the reciprocal lattice, defined in \cref{eq:Cosmo:reciprocallatticedefinition} below. The Fourier transform of a real field is in general complex valued. However, it is also Hermitian, $f(-\tn)=f^*(\tn)$, and so the number of independent components of $f(\tn)$ equals that of $f(\bmn)$. %In the case of a complex field this holds true for its complex and imaginary parts, separately.

The Fourier transform of a field takes values on the so-called \textit{reciprocal lattice}, that contains all Fourier modes of the system, this is, all wavelengths that can be captured by our lattice. It is defined as 
\begin{equation}\label{eq:Cosmo:reciprocallatticedefinition}
\LambdaR = \left\{\tn=(\tilde{n}_1,\tilde{n}_2,\tilde{n}_3)\,|\,\tilde{n}_i\in\mathbbm{Z}\,,\,-\frac{N}{2} < \tilde{n}_i \leq \frac{N}{2}\right\}\,,
\end{equation}
so that each site corresponds to a comoving momentum,
\begin{equation}
\bm{k}=\frac{2\pi}{L}\tn\,.
\end{equation}
The reciprocal lattice makes it clear the existence of a minimum momentum that can be probed. This is called the \textit{infrared} (IR) \textit{momentum}, $\kIR=2\pi/L$, which  represents the lattice spacing of the reciprocal lattice.  Similarly, there is a maximum ultraviolet (UV) momentum that can be probed in each direction, $\kUV=\pi/\delta x$,  and also a maximum momentum that can be captured on the full lattice, $k_\text{max}=\sqrt{3}\pi/\delta x$. Note that all modes with $|\bm{k}|\leq \kUV$ lie within a sphere of radius $\kUV$ that fits entirely within the reciprocal lattice, and so are lattice-isotropic, i.e., isotropic up to discretization effects. This is not the case for $|\bm{k}|> \kUV$, for which the modes are not isotropically distributed.

Using the Fourier transforms of the fields, one can define  their  \textit{power spectrum}. In the continuum, the power spectrum of a real field, $f(x)$, denoted as $\Delta_f$, is defined via
\begin{equation}\label{eq:Cosmo:ensembleaverage}
\langle f^2(\bmx) \rangle = \int  \Delta_f(k)\,\d \log k\,,
\end{equation}
where $f(\bmk)$ is the continuum Fourier transform of $f(\bmx)$, $k=|\bmk|$ and $\langle f^2(\bmx) \rangle$ denotes the ensemble average of $f^2(\bmx)$, this is, the average over multiple realizations of the field. The power spectrum is related to the two-point function of the field,
\begin{equation}
\langle f(\bmk) f^*(\bmk^\prime)\rangle = (2\pi)^3\frac{2\pi^2}{k^3}\Delta_f(k) \delta^3(\bm{k}-\bm{k}^\prime)\,,
\end{equation}
with $f^*$ the complex conjugate of $f$.

On the lattice, the ensemble average is substituted by a volume average,
\begin{equation}\label{eq:Cosmo:volumeaverage}
\langle f^2(\bmn)\rangle_V=\frac{1}{N^3}\sum _{\bm{n}\in\Lambda} f^2(\bm{n})=\frac{1}{N^6}\sum_{\tn\in\LambdaR}|f(\tn)|^2\,,
\end{equation}
where in the second step we have used the discrete Fourier transform of the field---see \cref{eq:Cosmo:DFT}. The sum over Fourier modes can be decomposed in a sum over all modes within spherical shells, $R_l$ and the sum over all shells. We define spherical shells as
\begin{equation}
R_l=\left\{\tn\in\Lambda_\text{R} \,|\, l-\Delta\tilde{n}_l^-\leq |\tilde{n}|< l+\Delta\tilde{n}_l^+\right\}\,,
\end{equation} 
where $l$ is the mode that labels the shell, and $\Delta\tilde{n}_l^-$ and $\Delta\tilde{n}_l^+$ are the width of the shell from $l$ downwards and upwards, respectively. While shells need not be labeled by their mean or have all equal width, it is common to choose  $\Delta\tilde{n}_l^-=\Delta\tilde{n}_l^+=1/2$, called the \textit{canonical binning}. \Cref{eq:Cosmo:volumeaverage} can then be written as
\begin{equation}
\langle f^2(\bmn)\rangle_V=\frac{1}{N^6}\sum_l\#_l\langle |f(\tilde{n})|^2\rangle_{R_l}\,,
\end{equation}
where $\langle ...\rangle_{R_l}$ is the average value over the spherical shell $R_l$, and $\#_l$ denotes the exact number of modes within $R_l$. From here, the lattice power spectrum is introduced,
\begin{equation}
\langle f^2(\bmn)\rangle_V= \sum_l \Delta \log k(l)\, \Delta_f(l)\,.
\end{equation}
A common option is to define
\begin{equation}\label{eq:Cosmo:latticepowerspectrum}
\Delta_f(l) = \frac{k(l)\delta x}{2\pi N^5}\#_l \langle |f(\tilde{n})|^2\rangle_{R_l}\,,
\end{equation}
where $k(l)=l\kIR$. Alternative definitions of the lattice power spectrum also exist---see \rcite{PSmodule:2023}. It is worth mentioning that the number of modes within each shell is often approximated as $\#_l\approx 4\pi l^2$. However, this is not correct for the lowest and highest modes on the lattice, for which the discretization and the finiteness of the reciprocal lattice becomes relevant, respectively, and the multiplicity significantly differs from this approximation.

Finally, we note that the substitution of the ensemble average in \cref{eq:Cosmo:ensembleaverage} by a volume average in \cref{eq:Cosmo:volumeaverage} is only justified if the number of modes within each bin is large. This is true in most cases, but for very IR or UV modes it can lead to significant systematic effects, in the same sense as the cosmic variance~\cite{Somerville:2003bq}. Thus, one wants all the relevant scales of the problem to be well encompassed within the lattice.

\subsection{A dimensionless discretized model}\label{sec:Cosmo:latticemodel}

To numerically study the equations of motion, they must be rewritten in terms of dimensionless variables, and discretized. The first requirement is simple to achieve. In $\mathcal{C}$osmo$\mathcal{L}$attice~\cite{Figueroa:2020rrl,Figueroa:2021yhd}, the dimensionless model is defined using two physical scales: one related to a typical field amplitude, $\fstar$, and another related to a typical frequency of the system, $\omegastar$. We then rescale the fields, coordinates and momenta as
\begin{equation}
\begin{array}{ccccc}
\tphi = \phi / \fstar\,, & \quad\quad & \tvarphi = \varphi/\fstar\,, & \quad\quad & \tA_\mu=A_\mu/\omegastar\,,\\
\tx_i = \omegastar x_i\,, & \quad\quad & \teta = \omegastar \eta\,, & \quad\quad & \tilde{k}_i = k_i / \omegastar\,.
\end{array}
\end{equation}
%Note in particular the different definition of the dimensionless gauge field compared to scalar fields. This is required to keep gauge invariance of the theory, and leads to the appearance of $\fstar/\omegastar$ factors in some of the numerical equations of motion. 
Ideally, the choice of $\fstar$ and $\omegastar$ ensures that typical field values are $\cO(1)$, minimizing the effects of numerical errors.

The second step is to discretize the equations of motion. The case of scalar fields is simple, as one can naively substitute continuum derivatives by \textit{finite differences}. Basic options for the finite differences are, for example,
\begin{equation}
\begin{array}{rl}
[\Delta^{\pm}_i f](\bm{n})&=\displaystyle\frac{\pm f(\bm{n}\pm \hat{\bm{i}})\mp f(\bm{n})}{\delta x}\,, \\[5pt]
[\Delta_i^0 f ](\bm{n}) &= \displaystyle\frac{ f(\bm{n}+ \hat{\bm{i}}) - f(\bm{n}- \hat{\bm{i}})}{2\delta x}\,,
\end{array}\label{eq:Cosmo:finitedifferences}
\end{equation}
which correspond to \textit{forward}/\textit{backward} and \textit{neutral derivatives}, respectively, evaluated on the lattice sites. These can directly be substituted into the equations of motion, although this procedure requires some caution to make sure all terms are defined at the same lattice site and time. Another possibility, in some cases, is to perform the substitution in the action, from which the discrete equations of motion are derived.

Associated to a lattice derivative, one can define a \textit{lattice momentum}, $\bm{k}_\text{L}$, using the Fourier transform
\begin{equation}
[\Delta_i f](\tn)=-i{k_\text{L},i}(\tn)f(\tn)\,.
\end{equation}
For the lattice derivatives in \cref{eq:Cosmo:finitedifferences}, defined on the lattice sites, one gets,
\begin{equation}\label{eq:Cosmo:latticeMomentum}
\begin{array}{rl}
k_{\text{L},i}^{\pm} & = \displaystyle 2\,\text{exp}\left(\mp i\frac{\pi\tilde{n}_i}{N}\right)\frac{\sin(\pi\tilde{n}_i/N)}{\delta x}\,,\\[10pt] k_{\text{L},i}^0 & \displaystyle= \frac{\sin(2\pi\tilde{n}_i/N)}{\delta x}\,,
\end{array}
\end{equation}
which are complex and real, respectively.

The discretization of gauge fields is a bit more cumbersome, as it needs to preserve gauge invariance. The main difference is that covariant derivatives are substituted as a whole by parallel-transported finite differences. For example, we can define the forward parallel-transported lattice derivative as
\begin{equation}\label{eq:Cosmo:latticemomenta}
D_i^+ \varphi \longrightarrow \frac{U_i(\bm{n}+\hat{i})\varphi(\bm{n}+\hat{i}) - \varphi(\bm{n})}{\delta x}\,,
\end{equation}
where
\begin{equation}
U_i(\bm{n})=\text{exp}[-i e Q \delta x A_i(\bm{n})]\,,
\end{equation}
is the gauge parallel transporter along the lattice link, also known as link variable, defined in analogy to the QCD case---see \cref{sec:QCD:discretizedaction}. This ensures our action is invariant under discrete gauge transformations, obtained from discretizing the derivative in the gauge transformation of $A_\mu$ in \cref{eq:Cosmo:gaugetransformations}.

The field-strength tensor of the gauge field admits several discretizations.  The simplest option is the \textit{non-compact formulation}, in which one works with the $A_\mu$ field as the degrees of freedom, and discretizes $F_{\mu\nu}$ by substituting the continuum derivatives with finite differences. 

Another option is the so-called \textit{compact formulation}, analogous to that used in QCD---see \cref{eq:QCD:gaugewilsonactionlattice}. In this case the link variables become the dynamical degrees of freedom and the field-stress tensor term of the action, $F_{\mu\nu}F^{\mu\nu}/4$ is substituted by
\begin{equation}
\frac{1}{2\delta x^4 e^2 Q^2}\sum_{\mu\nu}\left[1-\Re(\cP_{\mu\nu})\right]\,,
\end{equation}
where $\cP_{\mu\nu}$ is the \textit{plaquette},
\begin{equation}
\cP_{\mu\nu}(\bm{n})=U_\mu(\bm{n})U_\nu(\bm{n}+\hat{\mu})U^*_\mu(\bm{n}+\hat{\nu})U^*_\nu(\bm{n})\,.
\end{equation}
When using the compact formulation, one needs to accordingly redefine the relevant energy components. In particular, the components of the stress-energy tensor are recovered from

\noindent\begin{equation}\label{eq:Cosmo:hybridFmunu}
F_{\mu\nu}=\frac{i}{2eQ\delta x^2}(\cP_{\mu\nu}-\cP_{\mu\nu}^*)\,,
\end{equation}
while the  magnetic energy is
\begin{equation}
E_B=\frac{1}{a^4\delta x^4 e^2Q^2}\sum_{i,j<i}[1-\Re(\cP_{ij})]\,.
\end{equation}
%Note this approach can only be applied to the action, from which the discrete equations of motion are derived. 
We note that, in the case of non-Abelian fields, only the non-compact formulation makes it possible to keep gauge invariance~\cite{Figueroa:2020rrl}.

Another option for Abelian fields is a \textit{hybrid formulation}. In this case, the action is written in terms of the links, as in the compact case, but the $A_\mu$ field is kept as  the  dynamical degree of freedom. 

\subsection{Solving the field dynamics}

To numerically evolve the equations of motion of the fields, they need to be expressed in a way that is suited for a numerical evolution algorithm. The equations are typically a system of coupled non-linear second-order differential equations. To solve them numerically, it is convenient to rewrite the system in a Hamiltonian scheme. This  requires to define some \textit{conjugate momenta} to express each equation as a system of two first-order differential equations. While this is typically done by choosing the time derivative as the conjugate momentum, in the context of an expanding background it is convenient to include some power of the scale factor. This makes it possible  to eliminate the friction terms---those proportional to derivatives of the scale factor in \cref{eq:Cosmo:fieldseomsinglet,eq:Cosmo:fieldseomcomplex,eq:Cosmo:fieldseomgauge}---from the equation of motion.

We can consider, for example, a real scalar field in the continuum. Defining the conjugate momenta as $\pi_\phi=a^{3-\alpha}\phi^\prime$, the equation of motion, \cref{eq:Cosmo:fieldseomsinglet}, becomes a system of two first-order differential equations,
\begin{equation}
\begin{array}{rl}
\pi_\phi^\prime &\displaystyle= -a^{3+\alpha}\frac{\d V}{\d\phi}+a^{1+\alpha}\partial_i\partial_i\phi\,,\\
\phi^\prime &\displaystyle= a^{-(3-\alpha)}\pi_\phi\,.
\end{array}
\end{equation}
A similar result holds for complex scalars.
In the case of gauge fields, on the other hand, the conjugate momentum is defined as $\pi_{A,i}=a^{1-\alpha}A^\prime_i$, and the equation of motion in the temporal gauge becomes
\begin{equation}
\begin{array}{rl}
\pi_{A,i}^\prime &= a^{1+\alpha}J_i+a^{\alpha-1}\partial_j F_{ji}\,,\\[-3pt]
A_i^\prime &= a^{-(1-\alpha)}\pi_{A,i}\,.
\end{array}
\end{equation}
In the discretized model, an analogous procedure is used.
%These same transformations are performed in the discretized model. %As before, we note that these redefinitions can be performed in the action or directly in the equations of motion, and either in the continuum or the discrete theory.

After the dimensionless, discretized equations of motion are rewritten in the Hamiltonian scheme, they can be evolved using some numerical evolution scheme. The simplest ones suited for the systems presented in this thesis are either the \textit{leapfrog} or the \textit{Verlet algorithms}~\cite{Verlet1967}, which fall in the class of symplectic algorithms. This class of algorithms is ideal to study problems with conservative forces, as they preserve the volume of the phase space to $\cO(\delta \eta^2)$ accuracy, with $\delta\eta$ the time step used in the numerical evolution. Mathematically, this is related to Liouville's theorem~\cite{Liouville1838,Gibbs_2010}. In the case of non-conservative forces, one can use some non-symplectic integrator, such as the Runge-Kutta algorithms~\cite{Runge1895,Kutta1901}. Note that in all cases the time step must obey  the Courant stability condition~\cite{Courant:1928lij}, this is, $\delta\eta<\delta x/\sqrt{3}$, related to the hyperbolic form of the equations of motion of the fields.

\section{Emission of gravitational waves}\label{sec:Cosmo:GW}

Gravitational waves (GWs) are the transverse-traceless (TT) part tensor perturbations of the metric.\footnote{See \rcite{Maggiore:2007ulw} for an extensive introduction to the topic of GWs.}  They are represented by a symmetric tensor field, $h_{ij}=h_{ji}$, that depends on the coordinates of spacetime, $h_{ij}=h_{ij}(\eta,\bm{x})$, and is transverse and traceless, this is,
\begin{equation}
\partial_i h_{ij}=0\,,\quad\quad\quad h_{ii}=0\,.
\end{equation}
This two properties imply GWs represent only two independent degrees of freedom. On a FLRW background, GWs are defined as
\begin{equation}
\d s^2 = a^{2\alpha}(\eta)\d\eta^2-a^2(\eta)\left[\delta_{ij} + h_{ij}(\eta,\bm{x})\right]\d x^i\d x^j\,,
\end{equation}

The emission of GWs by early universe processes can usually be studied in the \textit{linearized gravity regime}, $|h_{ij}|\ll 1$. In the work presented in this dissertation, the backreaction of the GWs into the matter fields is neglected. Also, we make \textit{passive use of gravity}, this is, we compute the emission of GWs from the matter fields, but do not subtract them from the energy budget of the source fields. %as these only appear at higher order in the perturbations, and so the emission of GWs does not diminish the energy budget of the fields. In general, this is an assumption that needs to be checked a posteriori.

Using Einstein equations, one can obtain the equations of motion for the $h_{ij}$ fields. This requires to separate the equation into a homogeneous background part, corresponding to the Friedmann equations, and an equation of motion for the perturbations. Moreover, one needs to evaluate the stress-energy tensor of the matter fields using the perturbed metric. The resulting equation of motion for the GWs is~\cite{Caprini:2018mtu}

\noindent\begin{equation}\label{eq:Cosmo:eomcontinuumGW}
h_{ij}^{\prime\prime}+(3-\alpha)\frac{a^\prime}{a} h_{ij}^\prime-a^{-2(1-\alpha)}\partial_k\partial_k h_{ij}=-a^{2\alpha}\frac{2}{\mpl^2}\Pi_{ij}^\TT\,,
\end{equation}
where the source of GWs, $\Pi_{ij}^\TT$, is the spatial-spatial part of the TT anisotropy tensor, $\Pi_{\mu\nu}$, which thus obeys
\begin{equation}
\partial_i \Pi_{ij}^\TT=0\,,\quad\quad\quad \Pi_{ii}^\TT=0\,.
\end{equation} 
In the context of \cref{eq:Cosmo:fieldlagrangian}, $\Pi_{ij}^\TT$ takes the form~\cite{figueroa2010phenomenology}
\begin{equation}\label{eq:Cosmo:TTanisotropictensor}
\Pi_{ij}^\TT=\left\{\frac{1}{a^2}\partial_i\phi\partial_j\phi+\frac{2}{a^2}\Re\left[D_i\varphi(D_j\varphi)^*\right]-\frac{1}{a^{2\alpha}}E_iE_j-\frac{1}{a^2}B_i B_j\right\}^\TT\,,
\end{equation}
where $\left\{...\right\}^\TT$ indicates the TT part of the expression between the brackets, and $E_i$ and $B_i$ are, respectively, the components of the electric and magnetic fields, defined in \cref{eq:Cosmo:electricmagneticfield}. 

The TT part of the anisotropic tensor can be defined from an effective tensor, $\Pi_{ij}^\eff$, containing only those terms of the full $\Pi_{ij}$ that have a non-zero TT projection,
\begin{equation}\label{eq:Cosmo:effectiveanisotropictensor}
\Pi_{ij}^\eff=\frac{1}{a^2}\partial_i\phi\partial_j\phi+\frac{2}{a^2}\Re\left[D_i\varphi(D_j\varphi)^*\right]-\frac{1}{a^{2\alpha}}E_iE_j-\frac{1}{a^2}B_i B_j\,.
\end{equation}
From here, the source of GWs can be recovered by projecting to the TT component. This projection is a very non-local operation in coordinate space, but becomes linear in Fourier space~\cite{Misner:1973prb},
\begin{equation}\label{eq:Cosmo:TTprojectionPItensor}
\Pi_{ij}^\TT(\bmk)=\Lambda_{ij,lm}(\hbmk)\Pi_{lm}^\eff(\bmk)\,,
\end{equation}
where the TT projector is
\begin{equation}
\Lambda_{ij,lm}(\hbmk)=P_{il}(\hbmk)P_{jm}(\hbmk)-\frac{1}{2}P_{ij}(\hbmk)P_{lm}(\hbmk)\,,
\end{equation}
with 
\begin{equation}
P_{ij}(\hbmk)=\delta_{ij}-\hat{k}_i\hat{k}_j\,.
\end{equation}
and $\hat{k}_i=k_i/|\bm{k}|$.

Note that $\Pi_{ij}^{\TT}$ depends on spatial derivatives of the fields, and so the production of GWs is tied to the presence of inhomogeneities in the system. In the early universe, non-linear phenomena that develop large non-linearities may lead to the production of a vast amount of GWs with a particular spectral distribution that depends on the producing process. After the source ends, these GWs propagate freely until today, redshifting their spectral features, and could be measured as a stochastic GW  background~\cite{Caprini:2018mtu}. Thus, a detection of a stochastic background of GWs would provide precious information of the processes emitting the GWs.

GWs detectors can probe the power spectrum of a GW background, the particular shape of which depends on the phenomena that emitted the gravitational radiation. The GW energy density can be defined from an average over a volume, $V$, that encompasses all the relevant scales of the problem,
\begin{equation}
\rho_\GW(\eta)=\frac{\mpl^2}{4a(\eta)^{2\alpha}}\langle h_{ij}^\prime(\eta,\bm{x})h_{ij}^\prime(\eta,\bm{x})\rangle=\int \frac{\d\rho_\GW}{\d \log k}\d \log k\,.
\end{equation}
Here, we have defined the \textit{GW energy-density power spectrum},
\begin{equation}
\frac{\d\rho_\GW}{\d\log k}=\frac{\mpl^2 k^3}{8\pi^2a(\eta)^{2\alpha} V}\int\frac{\d\Omega_k}{4\pi} h_{ij}^\prime(\eta,\bm{k})h_{ij}^{\prime\,*}(\eta,\bm{k})\,,
\end{equation}
where $\d \Omega_k$ refers to the solid-angle measure in Fourier space. Typically, one defines the \textit{fractional GW energy-density power spectrum}, 
\begin{equation}
\Omega_\GW=\frac{1}{\rho_\text{c}}\frac{\d\rho_\GW}{\d\log k}\,,
\end{equation}
where $\rho_\text{c}=3\cH^2\mpl^2$ is the critical energy density of the universe. In the case of self-consistent expansion, this is indeed the total energy density of the matter fields, $\rho_\text{c}^\text{self-cons}=\rho$, given in \cref{eq:Cosmo:energydensitypressurefields}. When working in Minkowski spacetime, on the contrary, no critical energy density can be defined, and one usually uses the total energy density of the matter fields.% Similar definitions using the total energy density of the matter fields are also used, specially when working in Minkowski spacetime where no critical energy density can be defined.

\subsection{Simulating the emission of GWs}\label{sec:Cosmo:GWsimulation}

The TT nature of GWs makes their study on the lattice complicated. The source of \cref{eq:Cosmo:eomcontinuumGW} is the TT part of the anisotropic tensor, and its computation would require to go back and forth to Fourier space in every time step of the simulation. It is true that fast Fourier transform algorithms exist that allows the computation of Fourier transform in $\cO(N^3\log N)$ time,\footnote{In \CosmoLattice we use the FFTW~\cite{FFTW} and PFFT~\cite{PFFT} libraries for the computation of fast Fourier transforms.} 
 where $N$ is the number of points per dimension of the lattice. However, this remains much more expensive that standard evolution steps, which scale as $\cO(N^3)$.
 
In \rcite{Garcia-Bellido:2007fiu}, a workaround was proposed to overcome this limitation. The TT projection, \cref{eq:Cosmo:TTprojectionPItensor}, and the equation of motion for GWs, \cref{eq:Cosmo:eomcontinuumGW}, are both linear operations on the fields. Thus, we can work with six unphysical non-TT  fields, $u_{ij}=u_{ji}$, related to the $h_{ij}$ fields as
\begin{equation}
h_{ij}(\bmk, \eta)=\Lambda_{ij,kl}(\hbmk)u_{kl}(\bmk, \eta)\,,
\end{equation}
which are sourced by the effective anisotropic tensor in \cref{eq:Cosmo:effectiveanisotropictensor},
\begin{equation}\label{eq:Cosmo:eomcontinuumuGW}
u_{ij}^{\prime\prime}+(3-\alpha)\frac{a^\prime}{a} u_{ij}^\prime-a^{-2(1-\alpha)}\partial_k\partial_k u_{ij}=-a^{2\alpha}\frac{2}{\mpl^2}\Pi_{ij}^\eff\,.
\end{equation}
Thus, one can evolve these $u_{ij}$ fields without the need to TT-project the source. Only when the GW energy density needs to be determined, the fields are Fourier transformed and projected to the TT component,
\begin{equation}
\Omega_\GW(\eta)=\frac{\mpl^2 k^3}{8\pi^2\rho_\text{c}a(\eta)^{2\alpha} V}\int\frac{\d\Omega_k}{4\pi} u^\prime_{ij}(\eta,\bm{k})\Lambda_{ij,kl}u_{kl}^{\prime\,*}(\eta,\bm{k})\,.
\end{equation}
%The GW fields in coordinate space can also be recovered by Fourier-transforming back after the TT projection. 
Note that the use of this technique has a drawback, as one is forced to work with six unphysical degrees of freedom $u_{ij}$, instead of the two physical ones. However, the gain in computing time is notorious.

The lattice implementation of \cref{eq:Cosmo:eomcontinuumuGW} follows a similar approach to that presented in \cref{sec:Cosmo:lattice} for scalar fields. The $u_{ij}$ fields are dimensionless and we define program variables as
\begin{equation}
\tilde{u}_{ij}=\left(\frac{\mpl}{\fstar}\right)^2u_{ij}\,,
\end{equation}
to eliminate the factor of $\mpl^2$ in the right-hand side of the equation of motion, \cref{eq:Cosmo:eomcontinuumuGW}.
The continuum equation of motion for these fields then reads
\begin{equation}\label{eq:Cosmo:eomcontinuumprogramuGW}
\tilde{u}_{ij}^{\prime\prime}+(3-\alpha)\frac{a^\prime}{a} \tilde{u}_{ij}^\prime-a^{-2(1-\alpha)}\tilde{\partial}_k\tilde{\partial}_k \tilde{u}_{ij}=-2a^{2\alpha}\tilde{\Pi}_{ij}^\eff\,.
\end{equation}
To solve it numerically, we rewrite it in a Hamiltonian scheme, using conjugate momentum variables $(\pi_{\tilde{u}})_{ij}=a^{3-\alpha}\tilde{u}_{ij}^\prime$,
\begin{equation}
\begin{array}{rl}
\pi_{\tilde{u},ij}^\prime &= 2a^{1+\alpha}\tilde{\Pi}_{ij}^\eff+a^{1+\alpha}\tilde{\partial}_k\tilde{\partial}_k\tilde{u}_{ij}\,,\\
\tilde{u}_{ij}^\prime &= a^{-(3-\alpha)}\pi_{\tilde{u},ij}\,.
\end{array}
\end{equation}
This is then discretized by substituting the continuum derivatives by finite differences. %We note that the $\Pi_{ij}^\eff$ tensor is naively defined to live in the middle of plaquettes, but this misalignment only leads to $\cO(\delta x^2)$ errors.

As we have commented above,  $u_{ij}$ fields are projected in Fourier space to the TT components to measure physical observables. However, the definition of the TT projector on the lattice carries some subtleties~\cite{Figueroa:2011ye}. On the lattice, this projection leads to fields that are transverse with respect to some choice of the discrete derivative, and so the particular form of the projector depends on the choice of derivative. As the lattice momentum associated to discrete derivatives can in general be complex---see \cref{eq:Cosmo:latticemomenta}---one defines a complex version of the projector
\begin{equation}\label{eq:Cosmo:projectorGWLambdalattice}
\Lambda_{ij,kl}(\hbmn)=P_{ik}(\hbmn)P_{jl}^*(\hbmn)-\frac{1}{2}P_{ij}(\hbmn)P_{kl}^*(\hbmn)\,,
\end{equation}
with 
\begin{equation}\label{eq:Cosmo:projectorGWPlattice}
P_{ij}(\hbmn)=\delta_{ij}-\hat{k}_{\text{L},i}^*\hat{k}_{\text{L},j}\,,
\end{equation}
where $\hat{k}_{\text{L},i}=k_{\text{L},i}/|\bm{k}_\text{L}|$. This projector obeys, among others, the following set of properties,
\begin{equation}\label{eq:Cosmo:projectorGWpropertieslattice}
\begin{array}{ll}
    \text{1}) ~k_{\text{L},i}^\pm P^{\pm}_{ij} = 0\:, \quad \quad\quad  &\text{2)}~ (k_{\text{L},i}^\pm)^{*}P^{\pm}_{ij} \neq 0\:, \vspace{0.2cm}\\
    \text{3)}~k_{\text{L},j}^\pm P^{\pm}_{ij} \neq 0 \:,\quad \quad \quad &\text{4)}~(k_{\text{L},j}^\pm)^*P^{\pm}_{ij} = 0 \:,\vspace{0.2cm} \\
    \text{5)}~ {P^{\pm}_{ij}}^* = P^{\pm}_{ji}\:, \quad \quad \quad
    &\text{6)}~P^{\pm}_{ij}(-{\tilde{\bm n}}) = P^{\pm}_{ji}({\tilde{\bm n}})\:, \vspace{0.2cm} \\
     \text{7)}~P^{\pm}_{ij}P^{\pm}_{jk} = P^{\pm}_{ik}\:, \quad \quad \quad
     &\text{8)}~ P^{\pm}_{ij}P^{\pm}_{kj} \neq P^{\pm}_{ik} \:.
\end{array}
\end{equation}
For example, these properties indicate that the projector is hermitian (\mbox{property} 5) and idempotent (property 7). For a real lattice momentum, as is the case for neutral finite differences---see \cref{eq:Cosmo:latticeMomentum}---the projector becomes real and these properties simplify~\cite{Figueroa:2011ye,GWmodule:2022}. %We note that it is also possible to define the GW fields to live in the center of the plaquettes, but these requires to use a more general version of the TT projector, see \rcite{Figueroa:2011ye}.

Using the projected fields, one can measure the power spectrum of the fractional GW energy density. On the lattice, this is defined similarly to other field power spectra---see \cref{eq:Cosmo:latticepowerspectrum}. For example,
\begin{equation}
\Omega_{\rm GW} (\tbmn ,\teta) = \dfrac{1}{\tilde{\rho}_\text{c}}\dfrac{ \tilde{k}(l)}{(8 \pi a^{2\alpha})}\left(\dfrac{\delta \tilde{x}}{N^5}\right)\left(\dfrac{f_*}{\mpl}\right)^2 \#_{l} a^{-2(3-\alpha)} \langle \tilde{T}(\tbmn,\teta)\rangle_{R(l)}\,,  
\end{equation}
with
\begin{equation}
\begin{array}{rl}
T(\bmn,\eta)&=u^\prime_{ij}(\bmn,\eta)\Lambda_{ij,kl}(\bmn)u_{kl}^{\prime\,*}(\bmn,\eta)\\[5pt]
&=\Tr[{\m P}{\m u}^\prime {\m P}{\m u}^{\prime\,*}]-\frac{1}{2}\Tr[{\m P}{\m u}^\prime]\Tr[{\m P} {\m u}^{\prime\,*}]\,,
\end{array}
\end{equation}
where ${\m P}$ and ${\m u}$ are matrices with entries $P_{ij}(\bm{n})$ and $u_{ij}(\bmn,\eta)$, respectively, and we have used \cref{eq:Cosmo:projectorGWLambdalattice,eq:Cosmo:projectorGWPlattice,eq:Cosmo:projectorGWpropertieslattice}.

The implementation of the techniques presented in this section in $\mathcal{C}$osmo$\mathcal{L}$attice, for both scalar and Abelian gauge theories, was done as part of the doctoral research presented in this dissertation~\cite{GWmodule:2022,GWmodule:2023}.

\makeatletter
\setlength{\@fptop}{0pt plus 1fil}
\setlength{\@fpbot}{0pt plus 1fil}
\makeatother

\chapter{GW emission from cosmic string loops: global case}
\label{sec:global}

Phase transitions taking place during the early universe may lead to the formation of \textit{cosmic defects}~\cite{Kibble:1976sj,Kibble:1980mv,Vilenkin:1984ib,Vilenkin:2000jqa}. These are metastable structures in which one or more fields are trapped far from the true vacuum of a theory due to topological constrains. Defects forming in the early universe may contain a large amount of energy and  leave observable cosmological imprints after decaying.

The characteristics of the defects are related to the topological structure of the vacuum manifold, $\cM$, of the underlying field theory, which corresponds to the possible value the field can take in the equivalent vacua of the theory. This manifold can be identified with the coset space of the associated symmetry breaking pattern, $G\rightarrow H$, this is, $\cM=G/H$.
 
One of the better-known examples of defects are \textit{cosmic strings}. These are one-dimensional topological defects that arise in theories in which the vacuum manifold has the topology of a circle, $\cM\cong S^1$,  characterized by having a non-trivial first homotopy group, $\pi_1(S^1)=\mathbbm{Z}$.  %Cosmic strings are one-dimensional topological defects that arise in theories containing a vacuum manifold that is not simply connected. %They are predicted by a variaety of field-theory and superstring early Universe

The formation of networks of cosmic strings in the early universe is predicted by a variety of field-theory and superstring early-universe scenarios~\cite{Hindmarsh:1994re,Copeland:2009ga,Copeland:2011dx,Vachaspati:2015cma}. These networks are composed of long (infinite) strings that stretch along the observable universe, and \textit{loops}. The former are expected to decay mainly via the production of loops due to self-intersection, while loops emit particles and gravitational waves (GWs), leading to the formation of a stochastic GW background (GWB)~\cite{Vilenkin:1981bx,Hogan:1984is,Vachaspati:1984gt}. This could potentially be detected by current and future GW experiments, including LIGO and VIRGO~\cite{Abbott:2017mem,LIGOScientific:2021nrg}, pulsar-timing-array collaborations, that recently announced evidence of a GWB signal~\cite{NANOGrav:2023gor,Antoniadis:2023ott,Reardon:2023gzh,Xu:2023wog}, and next-generation detectors, such as LISA~\cite{Audley:2017drz}.

A correct interpretation of a positive signal of a GWB requires of accurate predictions of the hypothetical sources of GWs. In the case of strings, these predictions are typically based on the Nambu-Goto (NG) approximation~\cite{nambu1971lectures,Goto:1971ce}, in which strings are considered to be infinitely thin, and the decay into particles is neglected. However, the latter has been argued to be a vital decay channel for some types of string loops~\cite{Hindmarsh:2017qff,Hindmarsh:2021mnl}, and an in-depth study requires of field-theory simulations.

In this chapter, the results from \rcite{Baeza-Ballesteros:2023say} are presented. We use lattice simulations to study the decay of cosmic string loops in theories containing a complex singlet scalar field with a Mexican-hat-like potential, the so-called \textit{global strings}. We consider both particle and GW emission simultaneously, for loops with an initial length up to 1700 times their core width. %We found, as expected, that GW roduction is very suppressed compared to the emission of particles, with a ratio between the emission powers of $\Pphi /\PGW\approx\cO(10)\left(v^2/\mpl\right)^2\ll 1$, where $v$ is the vacuum expectation value of the broken symmetry. These results were found to be independent of the shape, length and angular momentum of the loops, with no indication that this would change for longer loops.
In \cref{sec:global:cosmicstrings} we introduce global strings and  present some analytical results. We also discuss the NG approximation. We then detail in \Cref{sec:global:simulations} how lattice simulations of cosmic strings are performed. \Cref{sec:global:observables} introduces the different observables used to study the string dynamics, and \cref{sec:global:initialconditionnetworks,sec:global:initialconditionsartificial} explain the procedure we use to generate the string loops. Our results on particle and GW emission are summarized in \cref{sec:global:results}, and we finalize in \cref{sec:global:conclusions} with some brief conclusions.

\section{Global cosmic strings in the early universe}\label{sec:global:cosmicstrings}

The simplest model leading to the formation of cosmic strings consists of a single complex scalar field, $\varphi=(\phi_1+i\phi_2)/\sqrt{2}$, where $\phi_{1}$ and $\phi_2$ are real scalar fields, with action
\begin{equation}\label{eq:global:complexscalaraction}
S=\int\d^4 x\sqrt{-g}\big[(\partial_\mu\varphi)^*\partial^\mu\varphi-V(\varphi)\big]\,,
\end{equation}
where  $V(\varphi)$ is the potential
\begin{equation}\label{eq:global:complexscalarpotential}
V(\varphi)=\lambda\left(|\varphi|^2-\frac{v^2}{2}\right)^2\,.
\end{equation}
Here, $\lambda$ is a dimensionless coupling constant and $v$ is the vacuum expectation value of the complex field. This model is invariant under a U(1) global symmetry,
\begin{equation}
\varphi\rightarrow e^{i\alpha}\varphi\,,
\end{equation}
with $\alpha$ a constant value. %In conformal time, which we use for the work presented in this chapter, the equation of motion for the field reads,
%\begin{equation}\label{eq:global:equationofmotion}
%\varphi^{\prime\prime}+2\frac{a^\prime}{a}\varphi^\prime-\nabla^2\varphi=-a^2\lambda(|\varphi|^2-v^2)\varphi\,.
%\end{equation}

The model in \cref{eq:global:complexscalaraction} presents a symmetric and a broken phase. The former is characterized by a vacuum $\langle \varphi\rangle =0$ that realizes the U(1) symmetry, while in the broken phase the vacuum becomes $\langle \phi_1^2+\phi_2^2\rangle=v^2$,  spontaneously breaking the global symmetry. The vacuum manifold after symmetry breaking has the topology of a circle, $\cM\cong S^1$, and so may lead to the formation of cosmic strings after the phase transition.

The formation of the strings may happen  via the so-called \textit{Kibble mechanism}~\cite{Kibble:1976sj}. In a thermal environment, $\varphi$ does not realize the potential in \cref{eq:global:complexscalarpotential}, but instead some finite-temperature effective potential that includes corrections depending on the temperature, $T$,
\begin{equation}
V_\text{eff}(\varphi,T)=V(\varphi)-\frac{\lambda}{3}T^2\,.
\end{equation}
At high temperatures, the minima of this effective potential corresponds to the symmetric phase, but as the temperature falls below some critical value, $T_\text{c}=\sqrt{3} v$, the field falls into the true vacuum and the global U(1) symmetry is spontaneously broken. %This represents a second order phase transition.

If this phase transition happens before the end of inflation, $\varphi(x)$ takes a homogeneous value within a significant fraction of the observable universe today. However, if the transition happens after the end of inflation, causally disconnected patches of the universe may fall into different realizations of the true vacuum. Once they get in causal contact at a later time, cosmic strings must form to ensure continuity of the complex field.

%The formation of the string is represented pictorically in \cref{fig:global:cosmicstring}. We consider the case in which different points of the observables universe, which are initially causally disconnected, fall into different points of the true vacuum, so that there is a closed path that goes around the false vacuum. In order for the field to be continuum inside this path, some points will be force to lie in the the false vacuum, thus containing a large amount of energy.

Cosmic strings are thus topologically protected from decaying. The corresponding topological invariant is the \textit{winding number}, $k\in\mathbbm{Z}$, measured over any closed path. The winding number over some path quantifies the number of times the phase of $\varphi$ completes a full $2\pi$ rotation along that path. Its conservation implies that long strings can only disappear if they collide with another string. Loops, even if formed by string segments with non-zero winding number, globally have $k=0$, and so can collapse.  Note that while the winding number is conserved, the length of the strings can vary. Infinite strings can emit part of their length as small loops, and curved regions in any type of strings can be smoothen up via the emission of particles and GWs.

It is possible to study the behavior of the theory close to the true vacuum, by expanding $\varphi(x)$ in terms of a radial, $\chi(x)$, and an angular, $\theta(x)$, degrees of freedom,
\begin{equation}\label{eq:global:fieldexcitationsdecomposition}
\varphi(x)=\frac{v+\chi(x)}{\sqrt{2}}\text{e}^{i\theta(x)}\,.
\end{equation}
In these new variables, the action takes the form
\begin{multline}
S=\int\d^4 x\sqrt{-g}\left[\frac{1}{2}\partial_\mu \chi\partial^\mu \chi+\frac{1}{2}\left(1+\frac{\chi}{v}\right)\partial_\mu\theta\partial^\mu\theta\right.\\\left.-\lambda v^2\chi^2-\lambda v^3 \chi^3 - \frac{1}{4}\lambda \chi^4\right]\,.
\end{multline}
Thus, after spontaneous symmetry breaking, the theory becomes that of a massive scalar field with mass $m_\chi=\sqrt{2\lambda} v$, and a massless field with $m_\theta=0$.

\subsection{The QCD axion}\label{sec:global:QCDaxion}

One of the best-known scenarios that may lead to the formation of cosmic strings are \textit{axion models}~\cite{Weinberg:1977ma,Wilczek:1977pj}. These aim at solving the strong $CP$ problem, introduced in \cref{sec:QCD:introduction}, but also represent a viable explanation of dark matter~\cite{Hui:2021tkt}. 

Axion models extend the SM by adding a complex scalar field with an anomalous $U(1)$ symmetry, known as the Peccei-Quinn symmetry~\cite{Peccei:1977hh,Peccei:1977ur}. The potential in \cref{eq:global:complexscalarpotential} is the simplest example having such symmetry. In this case, the QCD axion, $a$, corresponds to the $\theta$ fluctuation in \cref{eq:global:fieldexcitationsdecomposition}. After spontaneous symmetry breaking of the anomaluos symmetry, the QCD axion develops a coupling to the gluon topological term,
\begin{equation}\label{eq:global:axionthetaterm}
\cL_\text{axion}[a,A]\supset \frac{g^2 a \Nf}{32\pi^2 f_a}\text{tr}\left[F_{\mu\nu}\tilde{F}^{\mu\nu}\right]\,,
\end{equation}
where $f_a$ is the axion decay constant, and $\Nf$, $g$, $F_{\mu\nu}$ and $\tilde{F}_{\mu\nu}$ are the number of quark flavors, the QCD gauge coupling, the gluon field-strength tensor and its dual counterpart, respectively. This term is analogous to the theta-term from QCD---see \cref{eq:thetalagrangianMinkowski}---and so both can be combined via a redefinition of the axion field $a\rightarrow \tilde{a}=a-\theta f_a$. In addition to \cref{eq:global:axionthetaterm}, the axion also develops a mass $m_a=\chi_\text{top}/ f_a^{2}$, with $\chi_\text{top}$ the topological susceptibility of QCD, as well as a potential $V[a,\theta]$ that reaches a minimum when $\tilde{a}=0$. This justifies the small upper bound of the observed CP violation in QCD, thus solving the strong $CP$ problem.

Furthermore, QCD axions are very weakly interacting~\cite{Kim:1979if,Shifman:1979if,Zhitnitsky:1980tq,Dine:1981rt}, and so they also represent a viable dark-matter candidate~\cite{Preskill:1982cy,Abbott:1982af,Dine:1982ah}. Closely-related dak-matter models are the so-called \textit{axion-like theories}~\cite{Svrcek:2006yi,Arvanitaki:2009fg}. These also include an anomalous spontaneously broken U(1) symmetry, but do not aim at solving the strong $CP$ problem, and so $f_a$ and $m_a$ can take a wider range of values. %These include an anomalous spontaneously broken U(1) symmetry, in analogy to axion models. 

Both the QCD axion and axion-like models may lead to the formation of cosmic strings if the anomalous U(1) symmetry is spontaneously broken after the end of inflation. The study of the cosmological implications of cosmic strings makes it possible to impose bounds on the parameters of the theories, such as the mass of the axion~\cite{Gorghetto:2020qws,Buschmann:2021sdq}.

\subsection{Analytical solutions for cosmic strings}

Investigating the dynamics of cosmic strings requires in general of field-theory numerical computations. However, one can gain insight on the properties of the strings by analytically studying some simple configurations. Here, we consider the global version of the so-called \textit{Nielsen-Olsen vortex}~\cite{NIELSEN197345}, which corresponds to a static infinite straight string. For simplicity, we restrict ourselves to Minkowski spacetime, with scale factor $a=1$. %Note however that the solution for expanding are analogous, with the coordinates representing the physical one, $r\rightarrow ar$. %While infinitely straight strings are not expected in nature, 
 %This result should represent an accurate of string segments with large curvature radius.

\begin{figure}[!h]
    \centering
    \begin{minipage}{0.495\textwidth} 
    \centering
        \includegraphics[width=\textwidth]{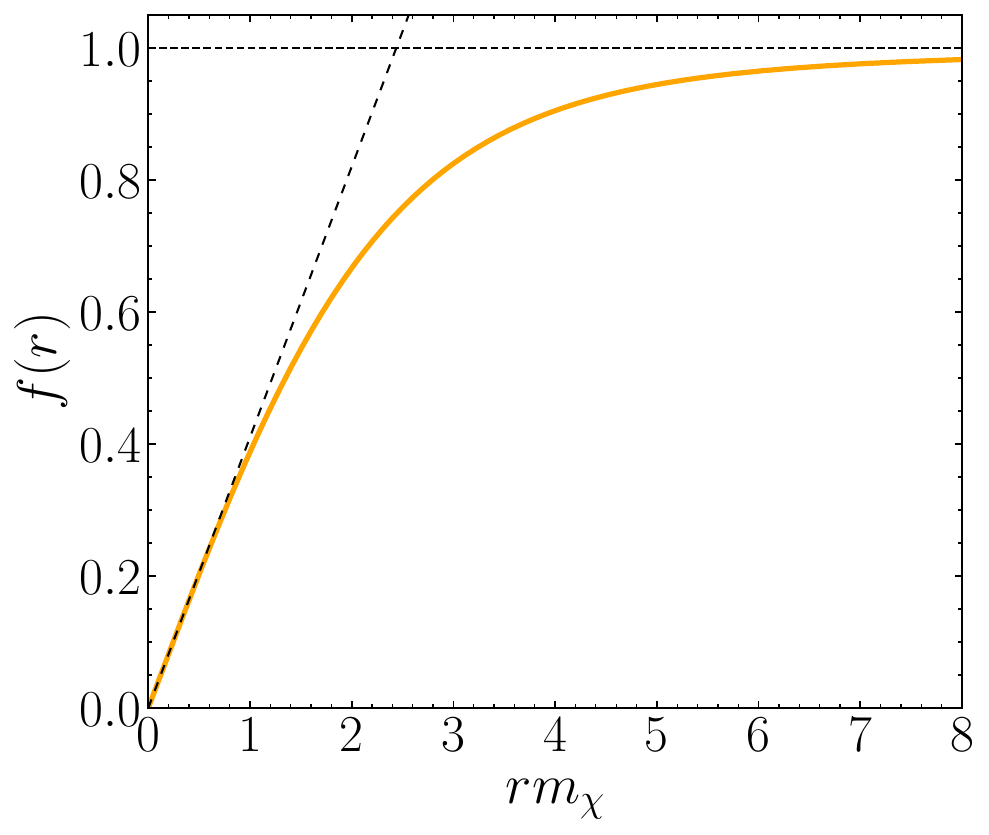}
    \end{minipage}
    \begin{minipage}{0.495\textwidth} 
    \centering
        \includegraphics[width=\textwidth]{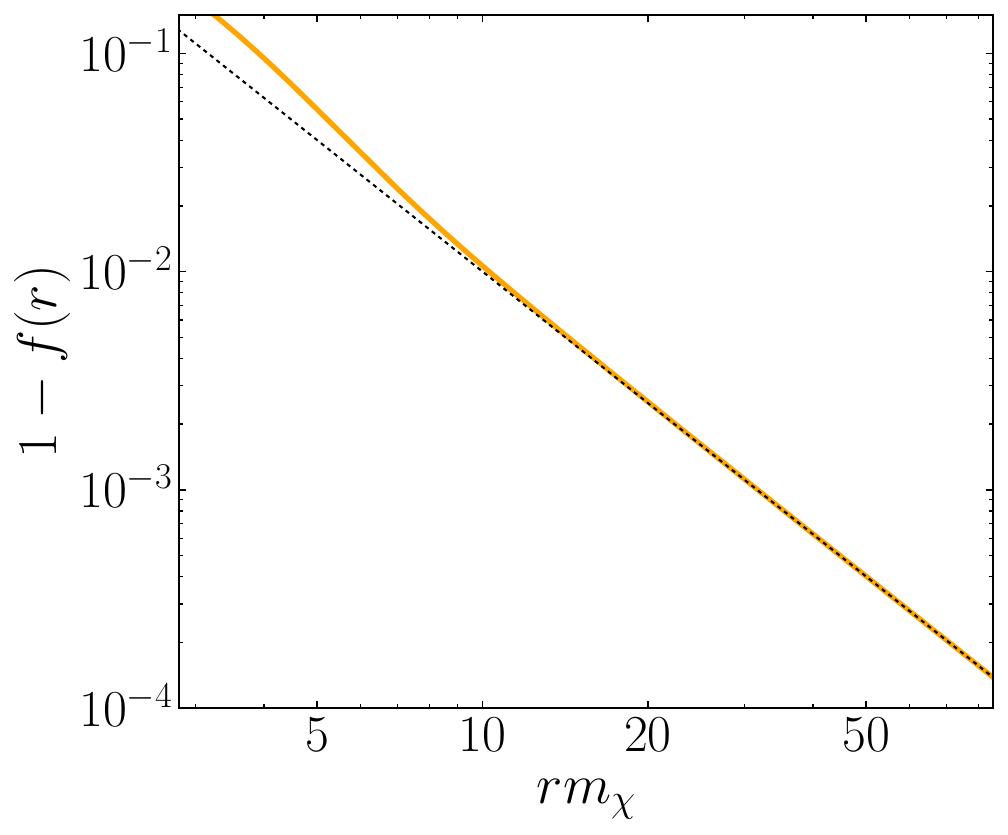}
    \end{minipage}
    
    \caption{
        Radial profile of the global NO vortex with $k=1$. We represent the short-range (dashed line, left panel) and the long-range (dotted line, left panel, in logarithmic scale) approximations. }
    \label{fig:global:NOprofile}
\end{figure}

Using cylindrical coordinates, $(r,\theta, z)$, and assuming the string lies on the $z$-axis, we can write the complex field as
\begin{equation}\label{eq:global:NOvortex}
\varphi_{\NO}=f(r)\frac{v}{\sqrt{2}}\text{e}^{ik\theta}\,,
\end{equation}
where $k\in\mathbbm{Z}$ is the winding number of the string and $f(r)$ is its radial profile. Substituting this into the equation of motion for the complex scalar field, \cref{eq:Cosmo:fieldseomcomplex}, one obtains
\begin{equation}\label{eq:global:NOvortexequation}
\frac{\d^2 f}{\d \tilde{r}^2}+\frac{1}{\tilde{r}}\frac{\d f}{\d \tilde{r}}+k^2\frac{f}{\tilde{r}^2}- f(f^2-1)=0\,,
\end{equation}
where $\tilde{r}=\sqrt{\lambda}v r$.
This equation has been solved numerically using relaxation methods and imposing the boundary conditions $f(0)=0$ and $f(\infty)=1$. The results for $k=1$are shown in \cref{fig:global:NOprofile}. From here, one notes that strings have a typical core radius of the order $\rc\sim m_\chi^{-1}$. This result can be generalized to a homogeneous expanding background changing $r\rightarrow ar$. This means that it is the physical width of the string which remains constant as the universe expands, rather than its comoving counterpart, $w_\text{c}= \rc/a$, which decreases.

It is interesting to study the short- and long-range behaviour of $f(r)$. At short distances, $r \ll \mh^{-1}$, the last term in \cref{eq:global:NOvortexequation} can be neglected and $f(r)$ is found to grow linearly, $f(r)\propto r$. At long separation, $r \gg \mh^{-1}$, on the other hand, the field approaches the vacuum quadratically, $f(r)=1-\cO(r^{-2})$. 

This behavior of the field has some implications for the dynamics of global strings. The total energy density of the string is
\begin{equation}
\rho(r)=\frac{1}{2}\left(\frac{\d f}{\d r}\right)^2+\frac{\lambda}{4}(f^2-1)^2+\left(\frac{k f}{r}\right)^2\,.
\end{equation}
The string tension, $\mu$, representing the energy of the string per unit length, is obtained from integrating this energy density over all $r$ and $\theta$. The first two terms correspond to the gradient and potential energy of the massive field, and the result from integrating them is a finite contribution, $\mu_0$, that dominates up to distances of a few core radii. The last term is related to the massless field, and leads to a logarithmically divergent contribution to the tension that dominates at long distances,
\begin{equation}
\mu_\theta(R) \approx 2\pi\int_{\rc}^R\left(\frac{1}{r}\frac{\d \varphi_\NO}{\d \theta}\right)^2 r\d r=2\pi v^2\log\left(\frac{R}{\rc}\right)\,,
\end{equation}
where $R$ is some cutoff scale, which in an early universe scenario would be related to the curvature of real strings or to the Hubble radius, $H^{-1}$. The total string tension  is then
\begin{equation}\label{eq:global:logdivergenttension}
\mu=\mu_0+\mu_\theta(R)\approx\mu_0+2\pi v^2\log\left(\frac{R}{\rc}\right)\,.
\end{equation}
The logarithmic divergent tension implies that interactions between global strings are long-ranged, with a force that scales roughly as $\sim r^{-2}$. %This force is, in the broad sense, mediated by the massless particles. 

Before concluding, we note that the NO solution can be used to create other string configurations. For example, one can consider a boosted NO vortex or use the \textit{product ansatz}~\cite{Vilenkin:2000jqa} to create a multistring solution, valid as long as the strings are sufficiently separated. However, the study of more complicated configurations requires of numerical lattice simulations.

\newpage\subsection{The Nambu-Goto approximation}\label{sec:global:NG}

Numerical simulations of cosmic-string networks are restricted to a limited dynamical range. Cosmological string scenarios may reach a separation of scales of the order $\log(\mh/H)\sim{70}$~\cite{Klaer:2017qhr}, while current lattice simulations are limited to $\log(\mh/H)\lesssim 9$~\cite{Gorghetto:2018myk,Hindmarsh:2019csc,Hindmarsh:2021vih}, from which the results need to be extrapolated. 
An alternative approach to study cosmic strings is the use of an effective description, in which the strings are represented by infinitely-thin objects. 

To understand this effective description, we can start by neglecting the massless modes in the theory. The dynamics of strings is then described using the Nambu-Goto action~\cite{nambu1971lectures,Goto:1971ce},
\begin{equation}
S_\NG=-\mu\int \d^2\sigma\sqrt{-\gamma}\,,
\end{equation}
where  $\sigma=(\sigma^0,\sigma^1)$ are the coordinates of the string worldsheet and ${\gamma=\text{det}(\gamma_{ab})}$, with
\begin{equation}
\gamma_{ab} = g_{\mu\nu}\frac{\d x^\mu}{\d \sigma^a}\frac{\d x^\nu}{\d \sigma^b}\,,
\end{equation}
the metric on the worlsheet, being $x$ the spacetime coordinates of the string core. The NG effective action can be recovered from field theory by integrating out the massive degrees of freedom, and remains valid as long as the curvature radius of the string is much bigger than its core width.

From this action, an equation of motion for the strings can be derived. In Minkowski spacetime one can write it in the form of a wave equation~\cite{Vilenkin:2000jqa},
\begin{equation}
\ddot{x}-x^{\prime\prime}=0\,,
\end{equation}
where $\dot{x}=\partial x/\partial \sigma^0$ and $x^\prime=\partial x/\partial \sigma^1$. This equation is complemented by the gauge conditions 
\begin{equation}
\dot{x}\cdot x^\prime=0\,,\quad\quad\quad \dot{x}^2+x^{\prime\,2}=0\,,
\end{equation}
and by setting $\sigma_0$ to be equal to the Minkowski time coordinate. A general solution can be expressed as a combination of right- and left-moving waves. This implies, for example, that the dynamics of loop have period $T=L_\text{str}/2$, where $L_\text{str}$ is the proper length of the string,
\begin{equation}
L_\text{str}=\int \d \sigma^1\,.
\end{equation}
The traveling modes on the string may present discontinuities in $x^\prime$ and even infinite values. These features are typically known as \textit{kinks} and \textit{cusps}, respectively.

In the NG approximation, string loops can only decay via the emission of GWs. In particular they emit gravitational radiation at harmonic frequencies, $\omega_n=2\pi n/T_\text{NG}=4\pi n/L$ ($n=1,2...$), with individual power, 
\begin{equation}\label{eq:global:GWraditionharmonicNG}
P_n=\frac{8\pi\Gamma \mu^2}{\mpl^2} \frac{n^{-q}}{\zeta(q)}\,,
\end{equation}
where $\Gamma$ is a dimensionless number, and $q$ is a parameter that takes different values depending on the feature that dominates the GW emission~\cite{Vachaspati:1984gt,Vilenkin:2000jqa}: $q=4/3$ for cusps, $q=5/3$ for kinks and $q=2$ for kink-kink collisions. The value of $\Gamma$ can be analytically computed for some specific string configuration~\cite{Burden:1985md,Garfinkle:1987yw}, or determined from numerical simulations in more general cases. In particular, \rcite{Blanco-Pillado:2017oxo} gets, from averaging over many simulations, that the distributions of values of $\Gamma$ is peaked at $\Gamma\sim 50$. The total emission power of the loop is~\cite{Weinberg:1972kfs}
\begin{equation}\label{eq:global:NGpredictionGWs}
\PGW=\sum_{n=1}^\infty P_n=\frac{8\pi\Gamma\mu^2}{\mpl^2}\,.
\end{equation}

Note that \cref{eq:global:GWraditionharmonicNG} is based on an asymptotic expansion of the tensor wave form, and so is a priori only valid for $n\gg1$. At low $n$ the structure of the entire loop becomes important~\cite{Burden:1985md}. However, the total power emitted is independent of this assumption---see \rcite{Vilenkin:2000jqa}. It is thus common to consider \cref{eq:global:GWraditionharmonicNG} valid for all harmonics. %From numerical simulations, $\Gamma\sim 50$~\cite{Blanco-Pillado:2017oxo}.

The NG approximation has been used as the basis of numerical studies of cosmic strings~\cite{Allen:1990tv,Vanchurin:2005pa,Vanchurin:2005yb,Ringeval:2005kr,Olum:2006ix,Blanco-Pillado:2011egf}. It is also the basis of other empirical approximations to string networks. This is for example the case of the \textit{velocity-dependent one-scale} (VOS) \textit{model}~\cite{Martins:1996jp,Martins:2000cs,Martins:2018dqg}, which describes the statistical distribution of infinite strings and their velocities as a function two quantities: the mean string separation, $\xi$, and the average mean-square velocity, $v_\text{rms}^2$. 

The VOS model predicts that evolving string networks reach a stable \textit{scaling regime} in which $\xi$ grows linearly with conformal time and $v_\text{rms}^2$ remains constant. In the case of NG strings, this constant value is $v^2_{\text{rms,}\NG}=1/2$, but the model has also been used to describe field-theory strings~\cite{Hindmarsh:2019csc,Correia:2019bdl,Hindmarsh:2021vih}, which show smaller results for $v^2_\text{rms}$. Note this model only describes the statistical properties of networks of infinite strings. The loss of energy from long strings via the emission of loops is encoded in some of the empirical parameters of the VOS model.% finally that this model does not include the dynamics from loops, but instead the loss of energy of long strings coming from the emission of loops is encoded in some of its empirical parameters.  

Beyond the NG model, it is also possible to construct an effective theory for the strings that includes massless modes, which is better suited to describe global strings. % can also be taken into account in an effective theory a priori more suitable for global strings.
This is the so-called \textit{Kalb-Ramond model}~\cite{KalbRamond1974}, in which massless modes are represented by an antisymmetric tensor field coupled to the string. This model makes it possible to obtain the logarithmic-divergent tension in \cref{eq:global:logdivergenttension}. If naively we substituted this tension in \cref{eq:global:NGpredictionGWs}, the emission of GW would be enhanced at cosmological scales. However, the validity of \cref{eq:global:NGpredictionGWs} is not clear in this case. In addition, the Kalb-Ramond model also predicts that the total emission power of loops into massless radiation takes the form,
\begin{equation}
P_\theta = \Gamma_\theta v^2\,,
\end{equation}
with $\Gamma_\theta\sim 100$~\cite{VilenkinVachaspati1987,Garfinkle:1987yw}. 

\section{Lattice simulations of global strings}\label{sec:global:simulations}

The zero-width approximation, presented in the previous section, disregards the internal structure of strings. The emission of massive particles is neglected, and even if massless modes are included, the effective description fails in the high curvature regions, which are those expected to dominate the radiation of particles and GWs. To fully capture the field-theory nature of cosmic strings, we need lattice simulations. While their applicability to study full networks is limited by the dynamical range, one can focus on individual strings or features. These could be, for example, oscillations of infinite strings~\cite{Drew:2019mzc,Drew:2022iqz}, kink-kink collisions~\cite{Drew:2023ptp} or the decay of isolated loops~\cite{Matsunami:2019fss,Saurabh:2020pqe}. In this chapter, we focus on this last case.

The use of lattice simulations to study the decay of global string loops follows the same basics as any other field-theory early-universe simulation. We describe the particular string observables that we measure in \cref{sec:global:observables}, and present the procedure used to generate two types of loops that we study, which we call \textit{network} and \textit{artificial loops}, in \cref{sec:global:initialconditionnetworks,sec:global:initialconditionsartificial}, respectively. We consider a Minkowski background %\footnote{We do consider evolution in a radiation-dominated background for the generation of some of the initial conditions, see \cref{eq:global:initialconditionnetworks}, but our final results are all obtained in Minkowski spacetime.} 
and work in dimensionless program variables, as introduced in see \cref{sec:Cosmo:latticemodel}, defined with
\begin{equation}
\fstar=v\,,\quad\quad\quad \omegastar=\sqrt{\lambda}v\,.
\end{equation}
We recall that dimensionless variables are indicated by adding a tilde to its dimensionful version. For example, if $L$ denotes the string length, $\tilde{L}=\sqrt{\lambda} v L$.

\subsection{Global-string observables}\label{sec:global:observables}

To monitor the length and the dynamics of the strings, we determine the location of the string cores from those plaquettes having a non-zero winding number. A plaquette with lattice coordinate $\bm{n}$ in the $ij$-plane, has winding number~\cite{Vachaspati:1984dz,Rajantie:1998vv},

\noindent\begin{equation}\label{eq:global:windingplaquette}
W_{ij}(\bm{n})=\frac{1}{2\pi}\left[Y_i(\bm{n})+Y_j(\bm{n}+\hat{\bm{i}})-Y_i(\bm{n}+\hat{\bm{j}})-Y_j(\bm{n})\right]\,,
\end{equation}
where 
\begin{equation}
Y_{i}(\bm{n})=\left[\theta(\bm{n}) - \theta(\bm{n}+\hat{\bm{i}})\right]_\pi\,,
\end{equation}
is the  variation of the phase along each link of the plaquette. In this equation, $[\alpha]_\pi$ sets the angle $\alpha$ in the range $-\pi<\alpha\leq \pi$. The determination of the coordinates of the string core can be refined by using the values of the field at the plaquette vertices to infer its exact location on the plaquette~\cite{Fleury:2015aca,Drew:2019mzc}, but we do not use these ttheechniques in the work presented here.

Using the total number of pierced plaquettes, $N_\text{w}$, we can estimate the total comoving length of the string in the lattice frame,
\begin{equation}\label{eq:global:lengthwindingdefinition}
L_\text{w}=\frac{2}{3}\delta x N_\text{w}\,,
\end{equation}
where the $2/3$ factor takes into account the Manhattan effect~\cite{Fleury:2015aca}, i.e., the fact that after entering some lattice cell, strings may leave it through any of the other five faces. %Note this definition of winding does not capture strings with winding number higher than one, nor cases in which two strings pierce the same plaquette. However, we have not observed any indication of these configurations happening in our simulations.

In addition, the coordinates of the pierced plaquettes allow us to have real-time insight on the structure and shape of the string configuration. For example, we can use these results to determine if an isolated loop has been found. This, however, requires to reconstruct the strings from the measurement of the coordinates of the pierced plaquettes. %The procedure we use for this reconstruction is diagramatically represented in \cref{fig:global:diagramjoining}. 

Knowing the sign of the winding on each plaquette, one can determine an unordered list of string segments, $s_i=\{\bm{n}_i^\text{ini},\bm{n}_i^\text{fin}\}$. Each one starts at the center of some cell, with coordinate $\bm{n}_i^\text{ini}$, and ends at the center of one of the adjacent cells, of coordinate $\bm{n}_i^\text{fin}$. Note that both loops and infinite strings are closed in a periodic lattice, meaning that for every cell $\bm{n}$ where some segment $i$ ends, another segment $j$ starts, $\bm{n}_i^\text{fin}=\bm{n}_j^\text{ini}$. Reconstructing the string implies identifying, for each segment $s_i$, the segment $s_j$ that follows.

\begin{figure}[!h]
    \centering
    \begin{minipage}{\textwidth} 
    \centering
        \includegraphics[width=\textwidth]{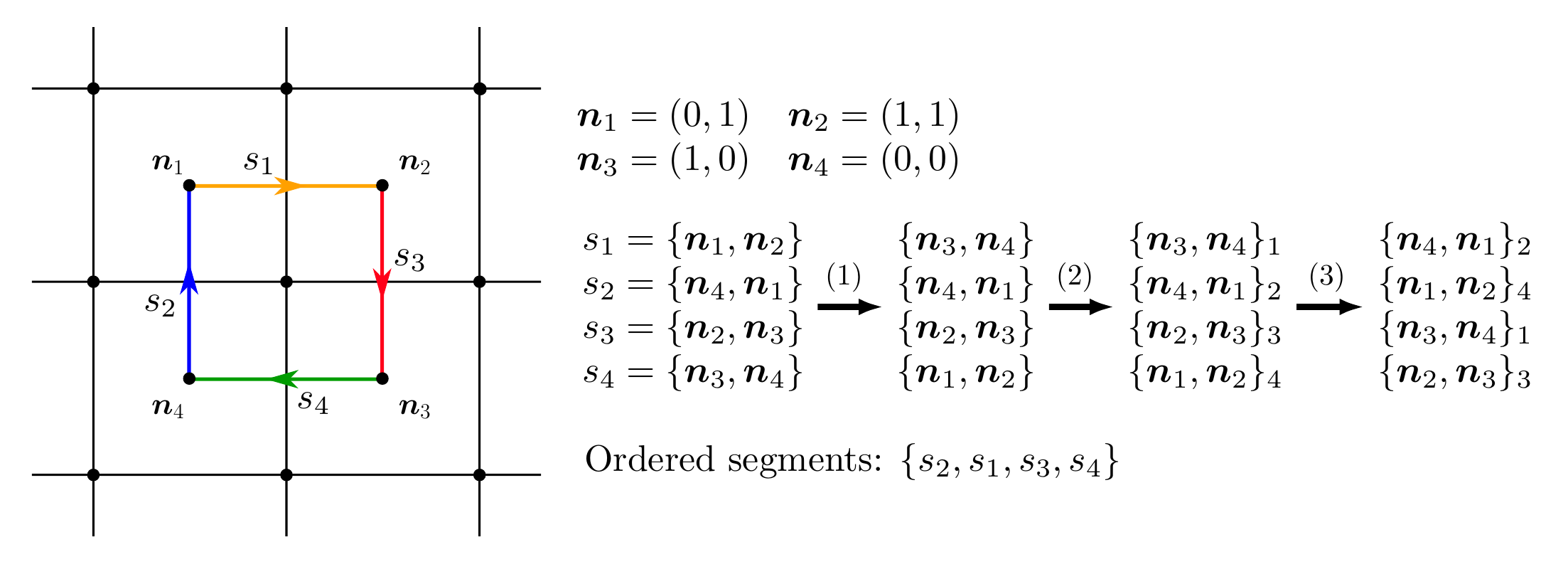}
    \end{minipage}
    
    \caption{
        Representation of the algorithm used to join the segments of strings determined from the pierced plaquettes, in the case of a four-segment loop. }
    \label{fig:global:diagramjoining}
\end{figure}

Naively, one would order the segments iterating over the full list and, for each segment $s_i$, finding the segment $s_j$ that follows it. This algorithm, however, would have a time complexity $\cO(N_\text{w}^2)$, representing a huge bottleneck for the simulations. To overcome this limitation, we propose an algorithm that scales as $\cO(N_\text{w}\log N_\text{w})$. This is explained now, complemented diagrammatically with \cref{fig:global:diagramjoining}, with a simple example of a string consisting of four segments.

The first step is to order the segments according to their final coordinate, $\bm{n}^\text{fin}$. This can be done, for example, comparing first the $x$ component, followed by the $y$ component and finally the $z$ one. %In the example shown the ordering would be $\{\bm{n}_4,\bm{n}_1,\bm{n}_3,\bm{n}_2\}$.  
The second step consists on indexing each segment with an integer number, $k$, that indicates their position in the current ordered list. % string. The final coordinate of each segment is no longer needed, and so we do not indicate it. 

In a third step, we order again the indexed segments according to their initial coordinate, $\bm{n}^\text{ini}$, following the same criteria as in the first step. In this newly-ordered list, the index of each segment indicates the position, within the current list, of the  next segment. In our example, for instance, the segment with initial coordinate $\bm{n}_4$ has index $k=2$, indicating that the following string in the second in the newly-ordered list, in this case that with initial coordinate $\bm{n}_1$. From this, one can move to the following segment in the string, until the initial segment is reached again, closing the string. Then, if there are segments that have still not been assigned to any string, one considers another of these segments and reconstructs the corresponding string and so on. Thus, using this ordered indexed list, reconstructing the strings becomes trivial.

%Our algorithm takes advantage of the fact strings are closed and so the full collection of $\bm{n}_i$ and $\bm{m}_i$ are equal. First, we order the pairs according to the final coordinate.\footnote{This ordering can be done, for example, comparing first the $x$ component, then the $y$ component and finally the $z$ one.}
%Then, we substitute $\bm{m}_i\rightarrow k$, where $k$ is the index of the pair in the ordered array. The collection $\{(\bm{n}_i, k)\}$ is then ordered according to $x_i$. The resulting array has the property that the given any $(\bm{n}_i,k)$ pair, the subsequent coordinate of the string is at position $k$ on this set. One can then trivially reconstruct all strings, which only requires to redraw the starting point when one string is closed.

%\begin{algorithm}
%\caption{Euclid’s algorithm}\label{alg:global:orderstrings}
%\begin{algorithmic}[1]
%\Procedure{Euclid}{$a,b$}\Comment{The g.c.d. of a and b}
%\State $r\gets a\bmod b$
%\While{$r\not=0$}\Comment{We have the answer if r is 0}
%\State $a\gets b$
%\State $b\gets r$
%\State $r\gets a\bmod b$
%\EndWhile\label{euclidendwhile}
%\State \textbf{return} $b$\Comment{The gcd is b}
%\EndProcedure
%\end{algorithmic}
%\end{algorithm}

We also measure the energy components of the strings. To avoid including free radiation and prevent issues with the logarithmically-divergent tension, we make use of a weight function, $W(\varphi)$, that is peaked at the string core. We use a weight related to the potential,
\begin{equation}\label{eq:global:weightfunction}
W(\varphi)=\frac{4 V(\phi)}{\lambda v^4} \Theta\left(\frac{v^2}{2}-|\varphi|^2\right)\,,
\end{equation}
normalized such that $W(0)=1$. Here $\Theta$ is the Heaviside function that ensures we are not including field values with $|\varphi|^2>v^2/2$. Using this weight, we can measure the total kinetic, gradient and potential energy component of the string. In an expanding background, with $a$ the scale factor and working in conformal time, they take the form
\begin{equation}\label{eq:global:stringenergycomponentsweighteddefinition} 
\begin{array}{rlrl}
E_\text{K,str}&=\displaystyle a\int W(\varphi)|\varphi'|^2\d^3 x\,,\quad\quad\quad &
E_\text{G,str}&=\displaystyle a\sum_i\int W(\varphi)|\partial_i\varphi|^2\d^3 x\,,\\[10pt]
E_\text{V,str}&=\displaystyle a^3\int W(\varphi)V(\varphi)\d^3 x\,, & &
\end{array}
\end{equation}
with the Minkowski results corresponding to the $a=1$ case.
The sum of all three components is the total energy of the string,
\begin{equation}\label{eq:global:stringenergyweighteddefinition} 
E_\text{str}=E_\text{K,str}+E_\text{G,str}+E_\text{V,str}\,.
\end{equation}
The string angular momentum is also determined~\cite{Saurabh:2020pqe},
\begin{equation}\label{eq:global:stringJweighteddefinition} 
J_{\text{str},i}=-\frac{a^2}{2}\varepsilon_{ijk}\int W(\varphi)\left[\Delta x_j\left(\dot{\varphi}\partial_k\varphi^*-\dot{\varphi}^*\partial_k \varphi\right)\right]\d^3x\,,
\end{equation}
where $\Delta x_i=x_i-x^\text{geom}_i$ are the relative coordinates with respect to the geometric center of the string, $\bm{x}^\text{geom}$, that we determine using the pierced plaquettes,
\begin{equation}
\bm{x}^\text{geom}=\frac{\delta x}{N_\text{w}} \sum_i^{N_\text{w}} \bm{n}_\text{pierced}\,,
\end{equation}
where $\bm{n}_\text{pierced}$ are the coordinates of the pierced plaquettes. Note that we can only reliably estimate the angular momentum of strings with a total size smaller than the lattice volume, i.e., those that do not warp around the periodic boundary, as the definition of $\Delta x_i$ is otherwise ill-defined. Program variables corresponding to the string energy  and angular momentum are defined, respectively, as
\begin{equation}
\tilde{E}_\str=(\sqrt{\lambda}/v)E_\str\,,\quad\quad\quad\tilde{J}_i=\lambda J_i\,.
\end{equation}

Using the weighted energy components, it is possible to obtain estimates of the proper comoving length of the strings and its mean squared velocity~\cite{Hindmarsh:2019csc,Hindmarsh:2021vih},
\begin{equation}
\begin{array}{rl}
L_\text{str}&=\displaystyle\frac{1}{a}\frac{E_\str+f_V E_\text{L,str}}{\mu(1+f_V)}\,,\\[10pt]
v_\text{rms}^2&=\displaystyle\frac{E_\str+E_\text{L,str}}{E_\str + f_VE_\text{L,str}}\,,
\end{array}
\end{equation}
where $E_\text{L,str}=E_\text{K,str}-E_\text{G,str}-E_\text{V,str}$ is the weighted Lagrangian energy of the string and $\mu$ and $f_V$ are the weighted tension and the fraction of weighted potential energy of the NO vortex,
\begin{equation}
\begin{array}{rl}
\mu&=a^2\displaystyle\int W(\varphi)\left[\frac{1}{2}\left(\frac{\d f}{\d r}\right)^2+\frac{\lambda}{4}(f^2-1)+\left(\frac{k f}{r}\right)^2\right]\d^2 x\,,\\[10pt]
f_B&=\displaystyle\frac{\lambda a^2}{4\mu}\int W(\varphi)(f^2-1)^2\d^2 x\,,
\end{array}
\end{equation}
with $f$ the profile function introduced in \cref{eq:global:NOvortex}.
We compute both these quantities for our choice of the weight function, \cref{eq:global:weightfunction}, finding $\mu=1.7824v^2$ and $f_V=0.3689$ for $k=1$.  Other independent estimators of the mean square velocity can also be defined---see \rcite{Hindmarsh:2021vih}. We note $L_\text{str}$ is approximately related to $L_\text{w}$ in \cref{eq:global:lengthwindingdefinition} as
\begin{equation}
L_\text{w}=L_\str \langle \gamma^{-1}\rangle\,,
\end{equation}
where $\gamma$ is the usual boost factor. The average of $\gamma^{-1}$, however, is in general not equal to $(1-v_\text{rms}^2)^{1/2}$.

Finally, we also study the emission of massive and massless particles from the string loops. Separating the radial and the angular modes as $\varphi(x)=\rho(x)/\sqrt{2}\times\exp[i\theta(x)]$, the energy density of $\varphi$ can be divided between the two,
\begin{equation}\label{eq:global:energyperturbativemodes}
\begin{array}{rl}
\rho_\rho&=\displaystyle\frac{1}{2}\dot{\rho}^2+\frac{1}{2}\partial_i \rho\partial_i \rho+\frac{\lambda}{4}(\rho^2-v^2)^2\,,\\[5pt]
\rho_\theta&=\displaystyle\frac{\rho^2}{2}\left[\dot{\theta}^2+\partial_i \theta\partial_i \theta\right]\,.
\end{array}
\end{equation}
To compute the gradients and time derivatives appearing in these expressions, it is convenient to work with the full complex scalar field~\cite{Drew:2019mzc}. For example, recalling we define $\varphi=(\phi_1+i\phi_2)/\sqrt{2}$, we can compute
\begin{equation}\label{eq:global:timederivativeangularmode}
\dot{\theta}=\frac{\phi_1\dot{\phi_2}-\phi_2\dot{\phi_1}}{\sqrt{2}v|\varphi|}\,,
\end{equation} 
and similarly for $\dot{\chi}$, $\partial_i\theta$ and $\partial_i\chi$. 

We also study the power spectrum of both massive and massless modes and of the associated energy components. For example, in the case of massless modes, the energy per Fourier mode is defined as~\cite{Saurabh:2020pqe} 
\begin{equation}\label{eq:global:energymasslessmodespowerspectrum}
\rho_\theta(k)=\frac{v^2}{2}\left[ |\dot{\theta}({\bm k})|^2 + k^2|\theta({\bm k})|^2\right]\,.
\end{equation}

%\subsection{Initial conditions}\label{sec:global:initialconditions}

%We consider two types of string loops generated using two different techniques, which we now describe. We refer to these two types of loops as \textit{network} and \textit{artificial loops}. Relevant stages of the initialization procedure for both types of loops are presented in in \cref{fig:global:stringsnapshots}.  %They are generated using different initial conditions, which we now describe, that lead to different features in the strings.

\subsection{Generation of network loops}\label{sec:global:initialconditionnetworks}

\begin{figure}[!b]
    \centering
    \begin{subfigure}{0.45\textwidth} 
    \centering
        \includegraphics[width=1\textwidth]{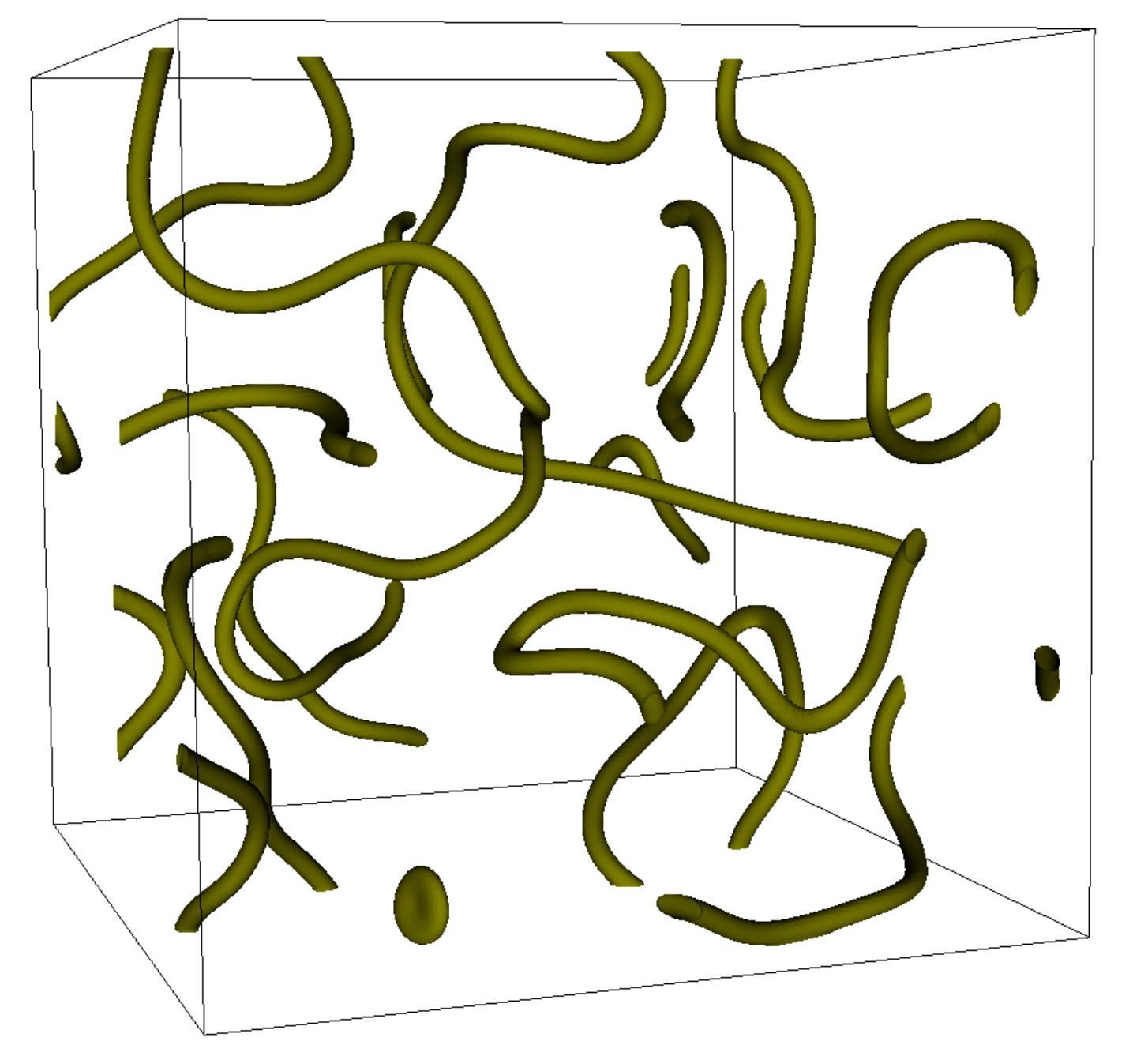}
        \caption{End of diffusion.}
        \label{fig:global:stringsnapshotsN1}
    \end{subfigure}\hspace{0.5cm}
    \begin{subfigure}{0.45\textwidth}
    \centering
       \includegraphics[width=1\textwidth]{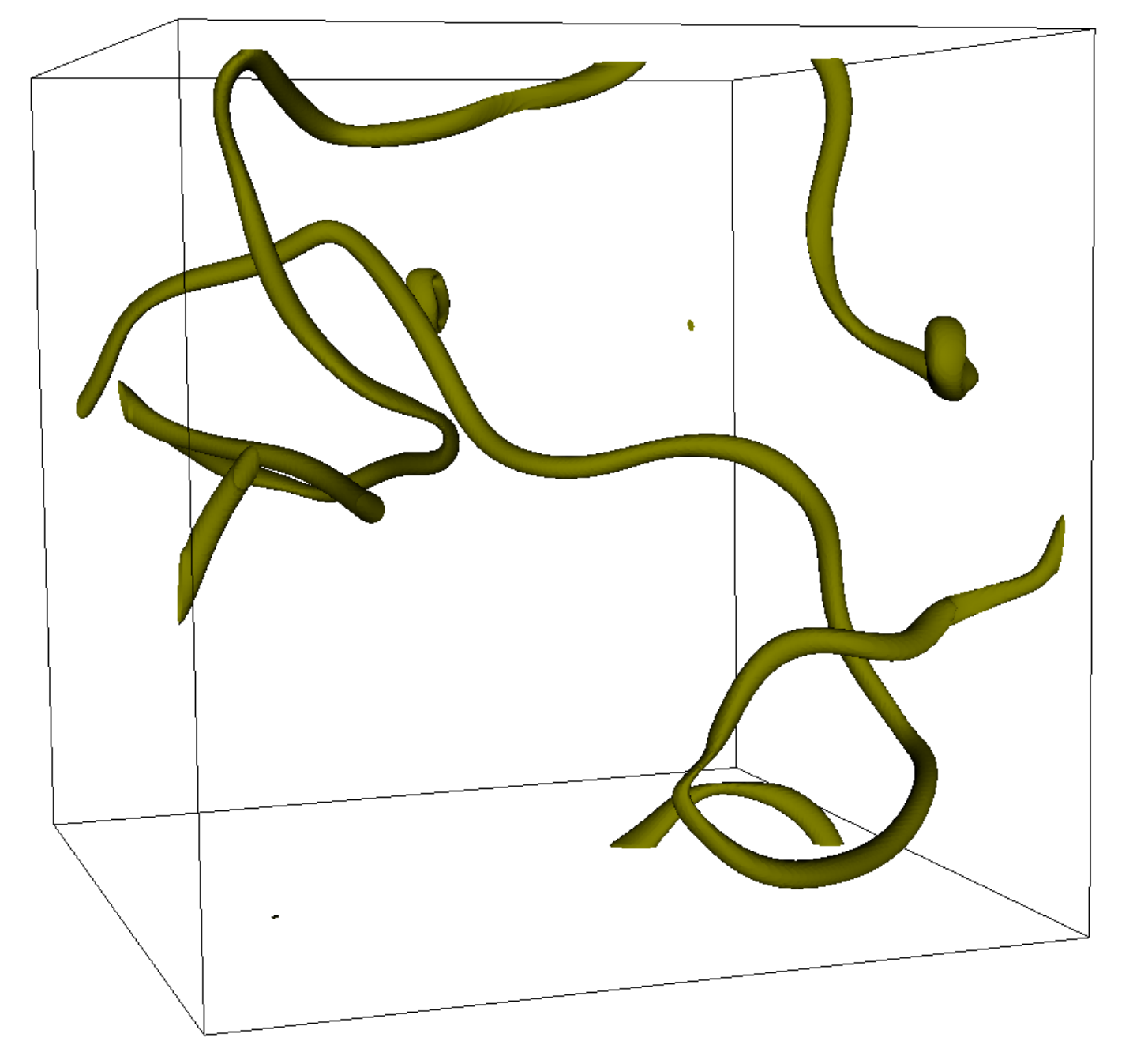}
        \caption{End of extra-fattening.}
        \label{fig:global:stringsnapshotsN2}
    \end{subfigure}\\[0.3cm]
    
    \begin{subfigure}{0.45\textwidth} 
    \centering
        \includegraphics[width=1\textwidth]{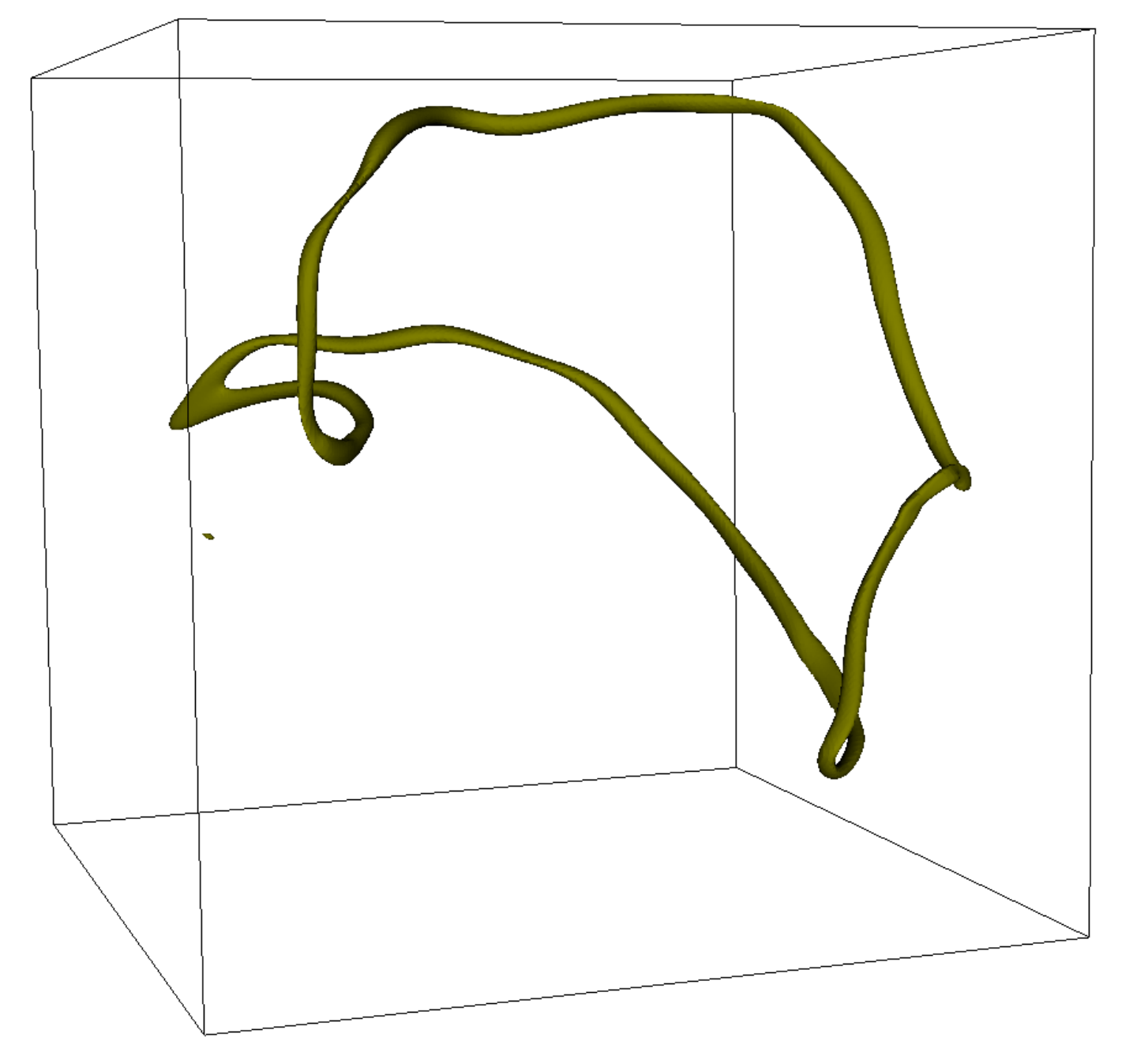}
        \caption{Moment when an isolated loop is found.}
        \label{fig:global:stringsnapshotsN3}
    \end{subfigure}
    \caption{Snapshots of $|\varphi^2|=0.1v^2$ surfaces of a network loop generated with $\tilde{\ell}_\str=5$ in a box with $\tilde{L}=64$, $\delta\tilde{x}=0.25$. 
         }
    \label{fig:global:stringsnapshotsN}
\end{figure}

We consider two types of string loops for this work. The first type are called \textit{network loops}. They are generated from the decay of string networks that are close to the scaling regime, and are expected to have shapes and features that resemble physical configurations expected in the early universe.  

The string networks that lead to the loops are generated following the procedure from \rrcite{Hindmarsh:2019csc,Hindmarsh:2021vih}. Simulations are initialized with a random Gaussian field in Fourier space with variance given by a power spectrum,
\begin{equation}\label{eq:global:PSnetworkloops}
\Delta_{\phi_i}(k)=\frac{k^3 v^2 \ell_{\str}^3}{\sqrt{2\pi}}\exp\left(-\frac{1}{2}k^2\ell_{\str}^2\right)\,,
\end{equation}
that is normalized such that $\langle\phi_1^2\rangle +\langle \phi_2^2 \rangle=v^2$---see \cref{eq:Cosmo:ensembleaverage}. Here, $\ell_\str$ is a correlation length that controls the density of the network. The time derivative of the scalar field is initially set to zero. 

The configuration resulting from the previous step is initially too energetic. To remove the excess energy, we evolve the complex field following a diffusion process,
\begin{equation}
\sqrt{\lambda} v \varphi^\prime_i-\partial_j\partial_j\varphi_i=-2\lambda\left(|\varphi|^2-\frac{v^2}{2}\right)\varphi\,.
\end{equation} 
We make this period last for $20$ units of program time, which we found to be enough to leave a smooth configuration. An example of the resulting network is presented in \cref{fig:global:stringsnapshotsN1}. 

After the diffusion phase, the string network is evolved in a radiation-dominated (RD) background. The equation of motion of the field, working in conformal time $\tau$, is

\noindent\begin{equation}
\varphi^{\prime\prime}+2\frac{a^\prime}{a}\varphi^\prime-\partial_i\partial_i\varphi=-2a^2\lambda \left(|\varphi|^2-\frac{v^2}{2}\right)\varphi\,,
\end{equation}
where $\varphi'=\d\varphi/\d\tau$. In addition, the scale factor is fixed to evolve as $a(\tau)=\tau/\tau_0$, where we use $\tau_0=70/\sqrt{\lambda}v$ in our simulations. This evolution of the scale factor is the late-time approximation of \cref{eq:Cosmo:Friedmansolutionsinglefluid} in a RD universe. The background expansion is maintained for a half-box-light-crossing time, $\Delta\tau_\HL=L/2$, with $L$ the side of the lattice, which is the time it takes a beam of light to cross half of the lattice side. 

Evolving the string network in an expanding universe, however, comes with a drawback. The string width  is constant in physical units, and so the comoving width of the strings reduces with time, $w_\text{c}=r_\text{c} a^{-1}$. This means that as the universe expands, less and less lattice sites are contained inside of the string core and we loose resolution of the string structure. 

\begin{figure}[!t]
    \centering
    \begin{minipage}{0.7\textwidth} 
    \centering
        \includegraphics[width=\textwidth]{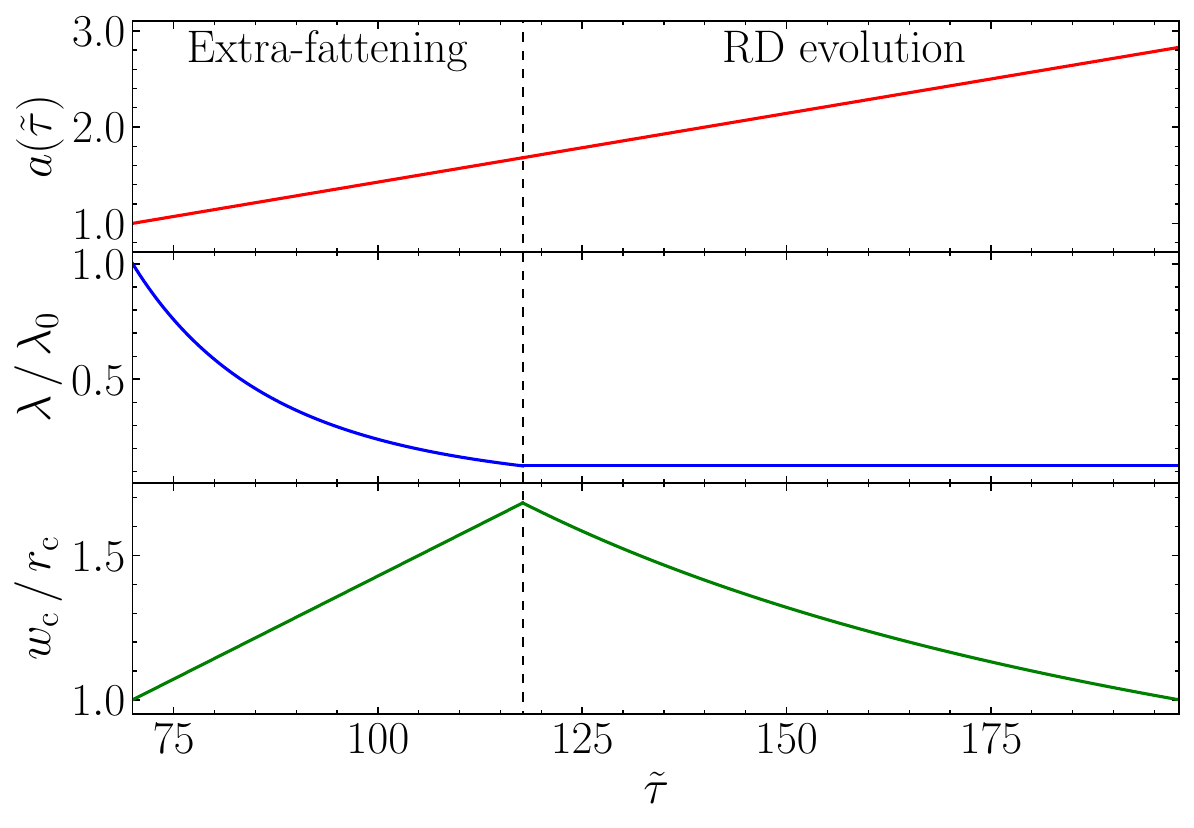}
    \end{minipage}
    
    \caption{
        Evolution of the scale factor (top), the $\lambda$ parameter (center) and the core width of the string (bottom) during the extra-fattening phase and the subsequent evolution in a RD background, separated by a vertical dashed line. Both $\lambda$ and $w_\text{c}$ are normalized by their initial values at the end of diffusion. }
    \label{fig:global:diagramaticevolution}
\end{figure}

To prevent this loss of resolution, we employ the resolution-preserving approach from \rcite{Press:1989yh}, known as \textit{extra-fattening}. The idea is to perform an initial phase of evolution during which the coupling constant $\lambda$ is promoted to a time-dependent parameter, $\lambda = a^{-4}\lambda_0$, with $\lambda_0$ its value at the end of diffusion, which we use to define program variables, $\omegastar=\sqrt{\lambda_0}v$. This change implies that the comoving string width grows in time, $w_\text{c}\propto a$. The extra-fattening phase is taken to last for $\Delta\tau_\text{ef}=\sqrt{\tau_0(\Delta \tau_\HL+\tau_0)}$, after which standard physical evolution in a RD background follows. This guarantees that at time $\tau_0+\Delta\tau_\HL$ the comoving string width is equal to its value at the end of diffusion, $w_\text{c}(\tau_0+\Delta\tau_\HL)=w_\text{c}(\tau_0)$. An example of the evolution of the scale factor, $\lambda$ parameter and string width is represented in \cref{fig:global:diagramaticevolution}, for a simulation with $\tilde{L}=256$ and $\delta\tilde{x}=0.25$. We show an example of a network at the end of the extra-fattening phase in  \cref{fig:global:stringsnapshotsN2}. It is worth mentioning that this approach differs from the so-called \textit{fattening} in which $\lambda\rightarrow a^{-2}\lambda$ is kept during the whole simulation so that the comoving string radius is kept constant. %A different alternative to deal with the loss of resolution would is the use of adaptative mesh refinement~\cite{Drew:2019mzc,Buschmann:2021sdq}, which we do not consider in this work.

\begin{figure}[!t]
    \centering
    \begin{minipage}{0.495\textwidth} 
    \centering
        \includegraphics[width=0.95\textwidth]{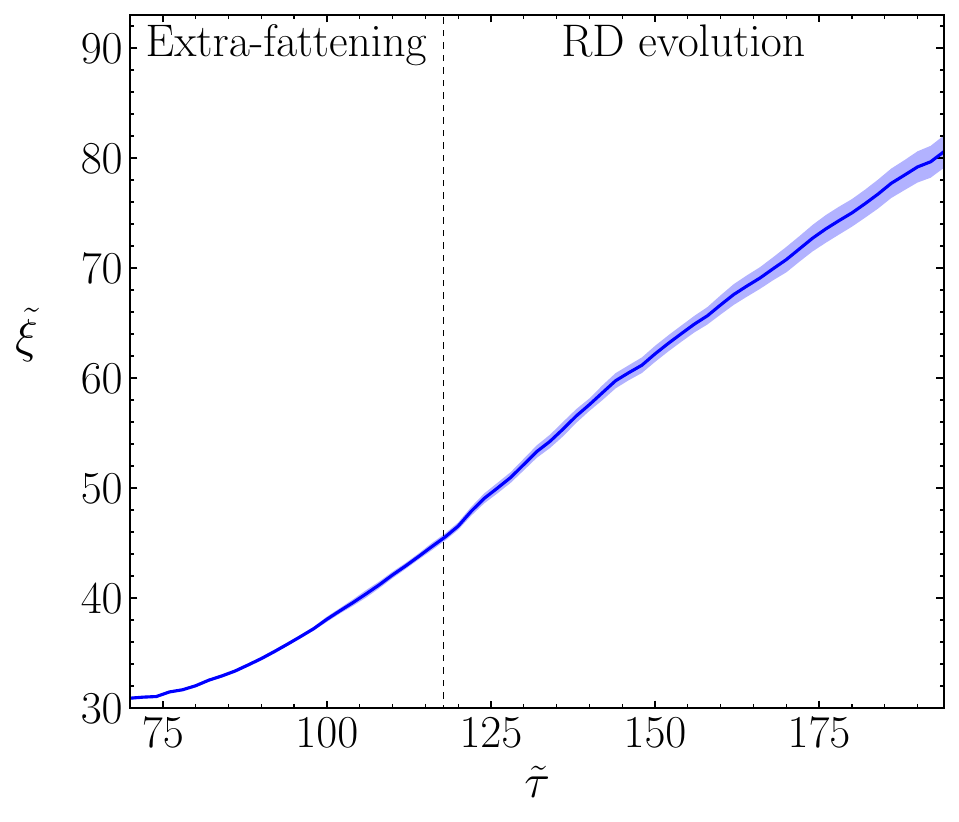}
    \end{minipage}
    \begin{minipage}{0.495\textwidth} 
    \centering
        \includegraphics[width=\textwidth]{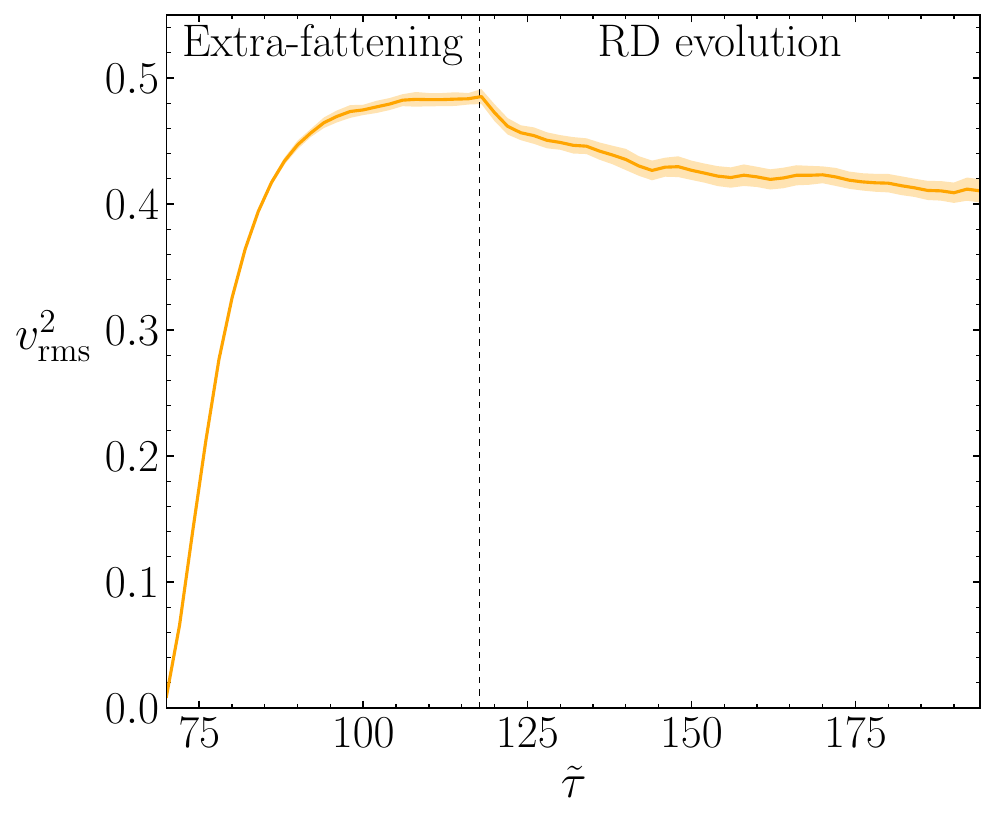}
    \end{minipage}
    
    \caption{
        Time-evolution of the comoving mean string separation (left) and the mean-squared velocity (right) of a network of global strings averaged over 10 independent realizations. The networks are generated with $\tilde{\ell}_\str=15$ and simulations are performed with $\tilde{L}=256$ and $\delta\tilde{x}=0.25$. The vertical dotted line indicates the end of extra-fattening and the shaded region corresponds to one standard deviation.}
    \label{fig:global:networkscaling}
\end{figure}

After evolving the network in an expanding background for a time $\Delta \tau_\HL$, that includes the extra-fattening phase during the initial $\Delta\tau_\text{ef}$, networks are found to be close to the scaling regime. This can be seen in \cref{fig:global:networkscaling}, where we represent $\xi=(L^3/L_\text{str})^{1/2}$ and $v^2_\text{rms}$ averaged over 10 independent network realizations. One observe that $\xi$ approaches a linear scaling, $\xi\propto \tau$, while $v_\text{rms}^2$ approaches a constant around $v_\text{rms}^2\sim0.4$, which is slightly smaller than the NG expectation. Note that, due to the limited size of our simulations, we are not able to completely reach the expected scaling regime---see however \rrcite{Hindmarsh:2019csc,Hindmarsh:2021vih} for bigger simulations. %, and so remain agnostic about possible logarithmic corrections to scaling~\cite{Gorghetto:2018myk,Hindmarsh:2019csc,Gorghetto:2020qws,Hindmarsh:2021vih,Buschmann:2021sdq}.

At this point, however, the network has not had enough time to decay into a single isolated loop. We thus let it evolve longer in a Minkowski background ($a=1, \dot{a}=0$) for a maximum time of $\Delta \tau_\HL/2$, which we find sufficient for the network to lead to a single loop. We find that $\sim 35\%$ of our simulations have decayed into a single loop after this time, with the remaining ones forming multiple loops or infinite strings, and hence not being suitable for our study. An example of an isolated network loop is shown in \cref{fig:global:stringsnapshotsN3}.

Once we find an isolated loop, the emission of GWs is turned on and we study the evolution of the loop, still in Minkowski background, until it disappears, this is, until the moment when no pierced plaquettes are measured during some consecutive time units. It turns out that only $\sim 40\%$ of the isolated loops could be used for this study. We discarded those that self-intersected forming infinite strings or that were much longer that the box size. Overall, only $\sim10\%$ of the simulations of network loops were used for this study.

\subsection{Generation of artificial loops}\label{sec:global:initialconditionsartificial}

Artificial loops are generated from the intersection of two pairs of parallel infinite boosted strings, following the procedure in \rcite{Saurabh:2020pqe}. This gives us control over the initial conditions at the cost of a more artificial squared shape of the loops. We consider one pair of infinite strings parallel to the $z$-axis and the other aligned with the $x$-axis. We will refer with a subscript ``1'' to those quantities related to the $z$-axis pair and with a ``2'' subscript to quantities related to the $x$-axis one.

\begin{figure}[!b]
    \centering
        \includegraphics[width=0.55\textwidth]{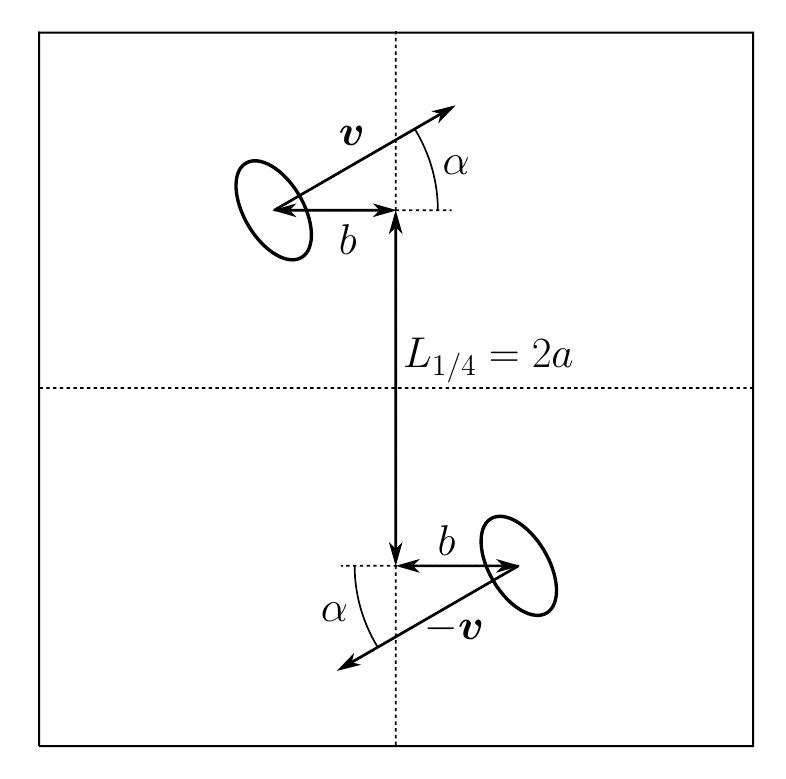}
    \caption{Schematic representation of the relevant variables used for the creation of a pair of parallel boosted strings. We have exagerated $b$ compared to the actual simulations, which use $\rc\lesssim b\ll a =\Lonefourth/2$.
         }
    \label{fig:global:initialpair}
\end{figure} 

We explain how the pair parallel to the $z$ axis is generated---an analogous procedure applies for the other pair. The relevant quantities used to generate each pair are depicted in \cref{fig:global:initialpair}. We start from the NO vortex solution, $\varphi^{(k)}_\NO$, given in \cref{eq:global:NOvortex}, where $k=\pm 1$ denotes the winding number of the string. The static NO solution can be boosted with velocity $\bm{v}_1=v_1(\sin\alpha_1,\cos\alpha_1)$ in the $(x,y)$-plane, $\varphi_{{\bm v}_1}^{(\pm)}(x,y;t)=\varphi_\text{NO}^{(\pm)}(x',y')$, where 
\begin{equation}\label{eq:global:relativisticboost}
\begin{array}{rl}
   x'&=-\gamma_1 v_1 s_1 t + [1+(\gamma_1-1)s_1^2] x+(\gamma_1-1)s_1c_1y\,,\\
   y'&= -\gamma_1 v_1 c_1 t + (\gamma_1-1)s_1c_1x+[1+(\gamma_1-1)c_1^2]y\,, 
\end{array}
\end{equation}
with $c_1=\cos\alpha_1$, $s_1=\sin\alpha_1$ and $\gamma_1=(1-v_1^2)^{1/2}$. Here $(x',y')$ are the coordinates in the rest frame, where the string is static, and $(x,y,t)$ refer to the lattice frame. %The time derivative of the boosted string can be computed by subsequent application of the chain rule. %These results allow to compute the derivative of the field. 
%The time derivative of the boosted field con be computed applying the chain rule,
%\begin{equation}
%\varphi_1'(x,y;t)=-\gamma_1 v_1 s_1\frac{\partial\varphi_\text{NO}^{(\pm)}(x',y')}{\partial x'}-\gamma_1 v_1 c_1\frac{\partial\varphi_\text{NO}^{(\pm)}(x',y')}{\partial y'}\,,
%\end{equation}
%where the derivatives of the NO solution with respect the Cartesian coordinates in their rest frame is computed by a subqequent application of the chain rule.

A pair of strings parallel to the $z$-axis is constructed using the product ansatz on two strings with opposite winding, $k=\pm 1$, and velocities, $\pm\bm{v}_1$, 
\begin{multline}\label{eq:global:productansatzsinglepair}
    \varphi_1(x,y;t)=\displaystyle\frac{1}{v}\varphi_{{\bm v}_1}^{(+)}\left[x-\left(\frac{L}{2}+a_1\right),y-\left(\frac{L}{2}+b_1\right);t\right] \\
     \displaystyle \times \varphi_{-{\bm v}_1}^{(-)}\left[x-\left(\frac{L}{2}-a_1\right),y-\left(\frac{L}{2}-b_1\right);t\right]\,,
\end{multline}
where $a_1$ and $b_1$ refer, respectively, to the displacement of the string in the $x$ and $y$ directions with respect to the center of the box---see \cref{fig:global:initialpair}. Note that, for this work, we consider symmetric configurations in which both strings have the same velocity magnitude and displacements. 

\begin{figure}[!t]
    \centering
    \begin{subfigure}{0.495\textwidth} 
    \centering
        \includegraphics[width=1\textwidth]{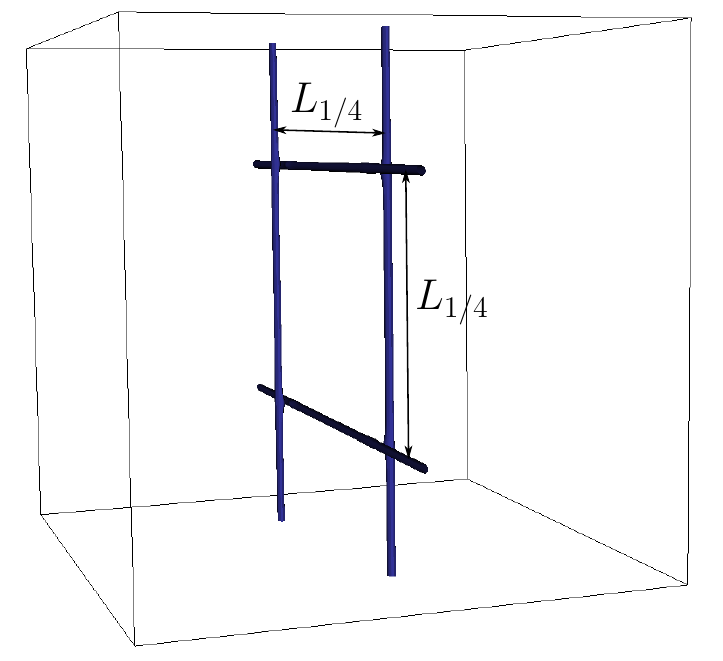}
        \caption{Beginning of the simulation.}
        \label{fig:global:stringsnapshotsA1}
    \end{subfigure}\\[0.3cm]
    \centering
    \begin{subfigure}{0.45\textwidth}
    \centering
       \includegraphics[width=1\textwidth]{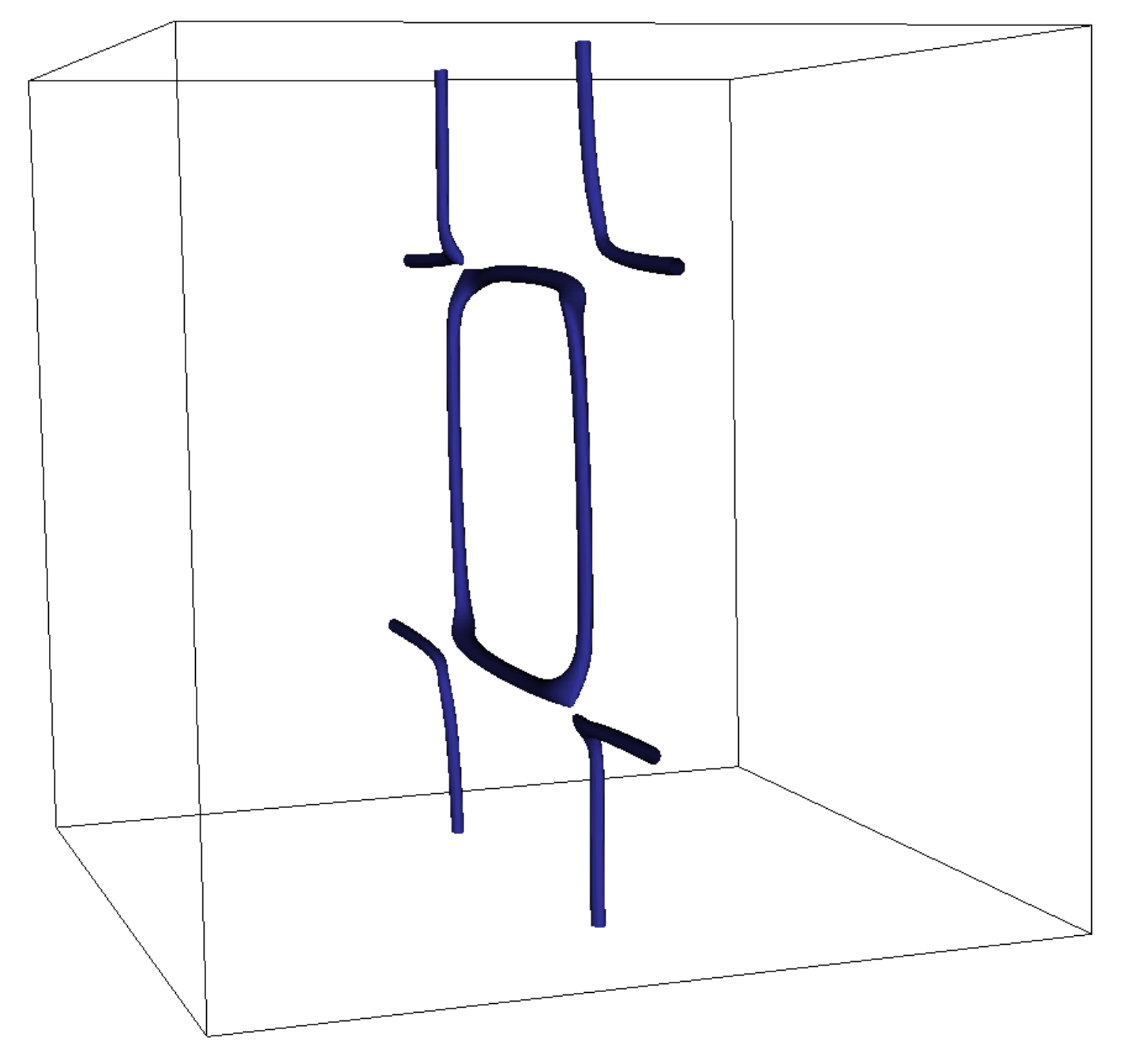}
        \caption{Instant before isolating the inner loop.}
        \label{fig:global:stringsnapshotsA2}
    \end{subfigure}\hspace{0.5cm}
    \begin{subfigure}{0.45\textwidth} 
    \centering
        \includegraphics[width=1\textwidth]{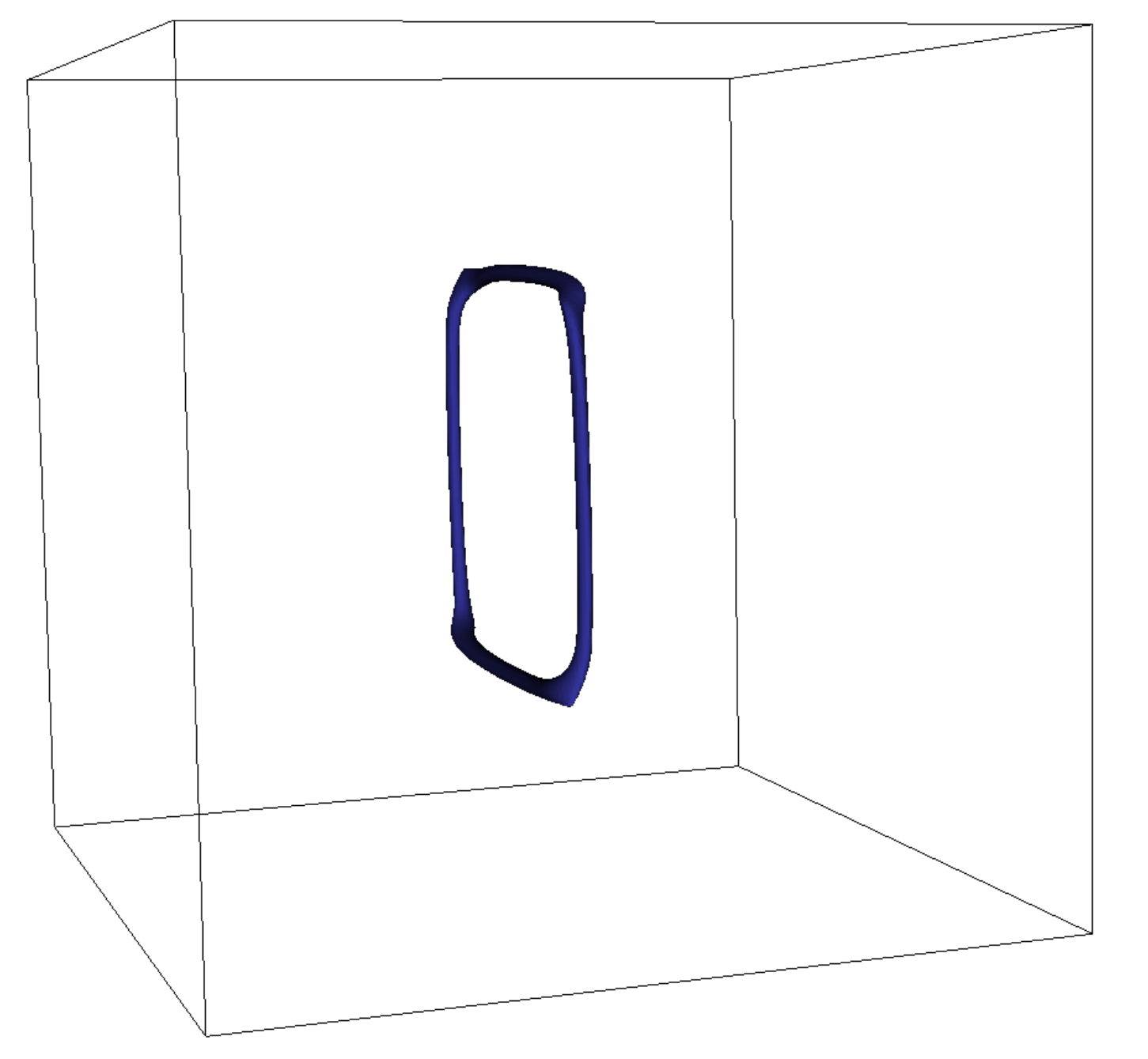}
        \caption{Instatn after isolating the inner loop.}
        \label{fig:global:stringsnapshotsA3}
    \end{subfigure}
    \caption{Snapshots of $|\varphi^2|=0.06v^2$ surfaces of an artificial loop generated with $v_1=0.6$, $v_2=0.8$ and $\sin\alpha=0.5$ in a box with $\tilde{L}=64$, $\delta\tilde{x}=0.25$. 
         }
    \label{fig:global:stringsnapshotsA}
\end{figure}

The resulting field and its derivative, evaluated at $t=0$, are modified following the procedure presented in \rcite{Saurabh:2020pqe} to incorporate them in a periodic lattice. The complete initial configuration for our simulations is obtained using the product ansatz again in two perpendicular string pairs, parallel each one to the $z$ and the $x$ axis,
\begin{equation}\label{eq:global:initialconfiguration}
\varphi(x,y,z) = \varphi_1(x,y;t=0) \times \varphi_2(z,y;t=0)\,.    
\end{equation}
An example of the resulting configuration is shown in \cref{fig:global:stringsnapshotsA1}. The time derivative of the fields can be computed as
\begin{equation}
\dot{\varphi}=\dot{\varphi}_1\varphi_2+\varphi_1\dot{\varphi}_2\,,
\end{equation}
where $\dot{\varphi}_i$ can be determined from successive application of the chain rule.
%The time derivative of the field can be obtained by successive application of the chain rule.

We consider, in general, boost velocities for the two pairs with different magnitude, $v_1\neq v_2$ and $v_2\geq v_1$, but the same angle, $\alpha_1=\alpha_2=\alpha$. We also take the strings to initially lie almost in the $y=L/2$ plane. We set $b_1=0$ and $b_2=2/\sqrt{\lambda}v$, so that the strings are  separated  enough that the product ansatz applies, but also the intersection between the strings happens soon in the simulation. In addition, we consider $a_1=a_2$ to be a significant fraction of the box size, typically $L/4$ or $L/6$.  The approximate separation of the two strings within each pair is denoted $\Lonefourth=2a$---see \cref{fig:global:stringsnapshotsA1}.

The initial field configuration in \cref{eq:global:initialconfiguration} is evolved in a Minkowski background, and the four strings soon intersect forming two square-shaped loops, an \textit{inner loop} with initial length approximately $L_\text{w}=4\Lonefourth$, and an \textit{outer loop}. Initially the two loops are almost planar and the inner one has its center close to that of the lattice. Shortly after forming, the two-loops start to shrink due to particle emission and separate from each other. %Note that i

\begin{figure}[!t]
    \centering
        \includegraphics[width=0.6\textwidth]{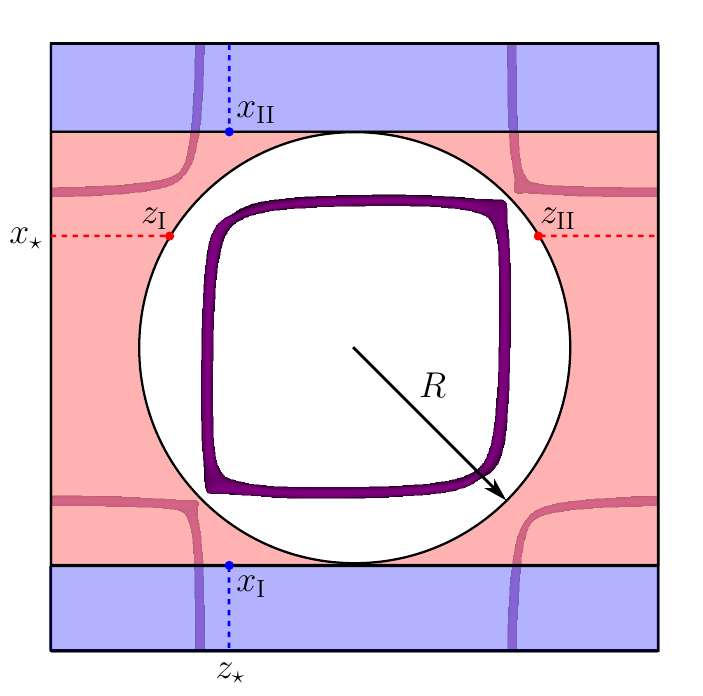}
    \caption{Schematic representation of the relevant variables used to isolate the inner loop in artificial-loop simulations. The red and blue regions outside the cylinder are substituted by a smooth configuration in two separate steps.
         }
    \label{fig:global:isolation}
\end{figure}

We have developed a method to isolate the inner loop. %, which we consider to be less affected by the initial conditions. 
After the loops form, we let them evolve until their separation is a fraction (we choose a 15\%) of the radius of the inner loop, defined as the maximum distance to the center of the box from any point of the string core. We then consider a cylinder of radius $R$ and axis parallel to the $y$-axis, so that it encompasses the inner loop and its surface lies midway between the two loops. The field and its derivative outside the cylinder are then substituted by a smooth field configuration containing no loops. The cylinder and other quantities relevant for the isolation procedure, which we introduce below, are pictorically represented in \cref{fig:global:isolation}

This substitution is performed in two steps, repeated for each fixed value of the $y$ coordinate, which we call $y_\star$. First, we consider the points $(x, y_\star, z)$ outside the cylinder with $|L/2-x|<R$, represented by the red region in \cref{fig:global:isolation}. Each value of $x=x_\star$ defines a line that intersects the cylinder twice, at $z_\RNum{1}$ and $z_\RNum{2}>z_\RNum{1}$. The field and its derivative outside the cylinder are substituted by a smooth field configuration: phases are linearly interpolated using the values at $z_\RNum{1}$ and $z_\RNum{2}$, while their magnitudes are substituted by some periodic ansatz based on the long-range behavior of the NO solution,
\begin{multline}
\displaystyle |\varphi(x_\star,y_\star,z)|=\\[5pt]\left\{\begin{array}{l}
\displaystyle\frac{\xi_\RNum{1}}{(\Delta-z+z_\RNum{1})^2}+\frac{\xi_\RNum{2}}{(L + z + \Delta - z_\RNum{2})^2}+\frac{v}{\sqrt{2}}\,,\\[10pt]
\displaystyle\frac{\xi_\RNum{1}}{(L-z+\Delta+z_\RNum{1})^2}+\frac{\xi_\RNum{2}}{(\Delta+z-z_\RNum{2})^2}+\frac{v}{\sqrt{2}}\,, 
\end{array}
\right.\quad
\begin{array}{l}
 \displaystyle 0\leq z < z_\RNum{1}\,,\phantom{\frac{A_\RNum{1}}{\Delta^2}}\\[10pt]
 \displaystyle z_\RNum{2}<z<L\,,\phantom{\frac{A_\RNum{1}}{\Delta^2}}
\end{array}
\end{multline}
\begin{equation}
\displaystyle |\dot{\varphi}(x_\star,y_\star,z)|=\left\{\begin{array}{l}
\displaystyle\frac{\pi_\RNum{1}}{(\Delta-z+z_\RNum{1})^3}+\frac{\pi_\RNum{2}}{(L + z + \Delta - z_\RNum{2})^3}\,, \\[10pt]
\displaystyle\frac{\pi_\RNum{1}}{(L-z+\Delta+z_\RNum{1})^3}+\frac{\pi_\RNum{2}}{(\Delta+z-z_\RNum{2})^3}\,, 
\end{array}
\right.\quad
\begin{array}{l}
 \displaystyle 0\leq z < z_\RNum{1}\,,\phantom{\frac{A_\RNum{1}}{\Delta^2}}\\[10pt]
 \displaystyle z_\RNum{2}<z<L\,,\phantom{\frac{A_\RNum{1}}{\Delta^2}}
\end{array}
\end{equation}
where $\xi_{\RNum{1},\RNum{2}}$ and $\pi_{\RNum{1}, \RNum{2}}$ are constants that ensure continuity at the surface of the cylinder, and $\Delta$ is a length parameter we set to $\Delta=2/\sqrt{\lambda}v$. 

In a second step, we consider points with $|L/2-x|\geq R$ ( represented by the blue region in \cref{fig:global:isolation}) and proceed analogously working at fixed $z=z_\star$. Both steps are repeated for all values of $y_\star$. An example of a loop configuration before and after isolating the inner loop is shown in  \cref{fig:global:stringsnapshotsA2,fig:global:stringsnapshotsA3}, respectively. After the inner loop is isolated, we turn on the emission of GWs and study the loop until it decays.

We finally remark that we have studied the effect of varying $\Delta$, as well as the isolation time and the size of the cylinder. Our results are  insensitive, within errors, to such changes, as long as, at the isolation time, both loops are enough separated from each other and the cylinder surface is kept at a distance from both loops. %Results for the evolution of the length of the inner loop are shown in \cref{fig:global:} for different choices of these parameters. We observe these results are insensitive to these changes, as long as at the isolation time both loops are enough separated from each other and the cylinder surface is kept separated from the two loops.

\section{Results}\label{sec:global:results}

We now present our results on loop decay into particles and GWs. We characterize the particle emission power in \cref{sec:global:resultsparticles} and study in \cref{sec:global:resultsspectra} the spectral distribution of the massive and massless modes produced by the string. Then, we analyze the GW emission power in \cref{sec:global:resultsGWs}. We stress that the energy radiated in form of GWs is not subtracted from the strings, since we neglect backreaction of the GWs into the matter fields. We justify this assumption self-consistently later. This implies that loops in our simulations decay only due to the emission of particles. %We also study in \cref{sec:global:resultsspectra} the spectral distribution of the massive and massive modes radiated by the string.

\subsection{Loop decay into particles}\label{sec:global:resultsparticles}

We characterize the lifetime of the loops, $\Delta t_\dec$, as a function of their initial size, $L_0$,  energy, $E_\text{str,0}$, and angular momentum, $J_0=|\bm{J}_0|$. We analyzed 23 different network loops with length-to-width ratios up to $ L_0/\rc\lesssim 1700$, and 49 artificial loops with $L_0/\rc\lesssim 800$. The parameters used in the simulations of network loops are summarized in \cref{tab:global:simulationparams}. For artificial loops, for each set of boost parameters in \cref{tab:global:parametersartificial}, we run simulations for all the lattices in \cref{tab:global:simulationsartificial}

\begin{table}[p!]
\centering
\begin{tabular}{>{\centering\arraybackslash}p{2.7cm}>{\centering\arraybackslash}p{2.75cm}>{\centering\arraybackslash}p{0.7cm}>{\centering\arraybackslash}p{1cm}}
\toprule
 $\tilde{L}$ & $\delta\tilde{x}$ & $\tilde{\ell}_\text{str}$ & $\tilde{L}_0$   \\ \midrule 
 256 & 0.25 & 25 & 868.3 \\
 256 & 0.25 & 20 & 924.3 \\
 256 & 0.25 & 20 & 889.7 \\
 $320/\sqrt{2}\approx226.3$ & $0.25/\sqrt{2}\approx0.177$ & 22 & 1162.5 \\
 $256/\sqrt{2}\approx181.0$ & $0.25/\sqrt{2}\approx0.177$ & 15 & 1071.5 \\
 $256/\sqrt{2}\approx181.0$ & $0.25/\sqrt{2}\approx0.177$ & 15 & 684.0 \\
 $256/\sqrt{2}\approx181.0$ & $0.25/\sqrt{2}\approx0.177$ & 15 & 371.2 \\
 144 & 0.25 & 12 & 557.0 \\
 144 & 0.25 & 12 & 296.3 \\
 144 & 0.1875 & 12 & 605.0 \\
 144 & 0.1875 & 12 & 553.5 \\
 144 & 0.125 & 12 & 633.6 \\
 $192/\sqrt{2}\approx135.8$ & $0.25/\sqrt{2}\approx0.177$ & 15 & 728.6 \\
 $192/\sqrt{2}\approx135.8$ & $0.25/\sqrt{2}\approx0.177$ & 15 & 699.8 \\
 $192/\sqrt{2}\approx135.8$ & $0.25/\sqrt{2}\approx0.177$ & 15 & 582.2 \\
 $192/\sqrt{2}\approx135.8$ & $0.25/\sqrt{2}\approx0.177$ & 15 & 462.4 \\
 $192/\sqrt{2}\approx135.8$ & $0.25/\sqrt{2}\approx0.177$ & 15 & 405.2 \\
 128 & 0.25 & 10 & 433.2 \\
 $128/\sqrt{2}\approx90.5$ & $0.25/\sqrt{2}\approx0.177$ & 8 & 260.7 \\
 $128/\sqrt{2}\approx90.5$ & $0.25/\sqrt{2}\approx0.177$ & 8 & 234.1 \\
 $128/\sqrt{2}\approx90.5$ & $0.25/\sqrt{2}\approx0.177$ & 8 & 215.2 \\
 $128/\sqrt{2}\approx90.5$ & $0.25/\sqrt{2}\approx0.177$ & 8 & 88.2 \\
 $128/\sqrt{2}\approx90.5$ & $0.25/\sqrt{2}\approx0.177$ & 8 & 60.3 \\ \bottomrule
\end{tabular}
\caption{Summary of the simulation parameters used to study the decay of network loops, together with the initial length of the isolated loops estimated from the number of pierced plaquettes---see \cref{eq:global:lengthwindingdefinition}.}
\label{tab:global:simulationparams}
\end{table}

\begin{table}[t!]

\centering
\begin{tabular}[t]{>{\centering\arraybackslash}p{0.7cm}>{\centering\arraybackslash}p{0.7cm}>{\centering\arraybackslash}p{1cm}}
\toprule
  $v_1$ & $v_2$ & $\sin\alpha$   \\\midrule
  0.9 & 0.9 & 0.4 \\
  0.9 & 0.6 & 0.4 \\
  0.9 & 0.3 & 0.4 \\
  0.9 & 0.0 & 0.4 \\
  0.6 & 0.6 & 0.4 \\
  0.6 & 0.3 & 0.4 \\
  0.3 & 0.3 & 0.4 \\ \bottomrule
\end{tabular}
\caption{Sets of boost parameters (rows) used to generate artificial loops.}
\label{tab:global:parametersartificial}\vspace{0.2cm}

\begin{tabular}[t]{>{\centering\arraybackslash}p{2.7cm}>{\centering\arraybackslash}p{2.75cm}>{\centering\arraybackslash}p{2.7cm}}
\toprule
 $\tilde{L}$ & $\tilde{L}_{1/4}$ & $\delta\tilde{x}$  \\ \midrule 
 $50/\sqrt{2}\approx35.4$  & $25/\sqrt{2}\approx17.7$ & $0.25/\sqrt{2}\approx0.177$ \\
 $100/\sqrt{2}\approx70.7$ & $50/\sqrt{2}\approx35.4$ & $0.25/\sqrt{2}\approx0.177$\\
 $150/\sqrt{2}\approx106.1$ & $75/\sqrt{2}\approx53.0$ & $0.25/\sqrt{2}\approx0.177$\\
 $200/\sqrt{2}\approx141.4$ & $100/\sqrt{2}\approx70.7$ & $0.25/\sqrt{2}\approx0.177$\\
 $250/\sqrt{2}\approx176.8$ & $125/\sqrt{2}\approx88.4$ & $0.25/\sqrt{2}\approx0.177$\\
 $300/\sqrt{2}\approx212.1$ & $150/\sqrt{2}\approx106.1$ & $0.25/\sqrt{2}\approx0.177$\\
 $400/\sqrt{2}\approx282.8$ & $200/\sqrt{2}\approx141.4$ & $0.25/\sqrt{2}\approx0.177$\\\bottomrule 
\end{tabular}
\caption{Sets of lattice parameters used for artificial-loop simulations. For each family of boost parameters in \cref{tab:global:parametersartificial}, a simulation is performed using each set in this table.}
\label{tab:global:simulationsartificial}

\end{table}

\begin{figure}[!p]
    \centering
    \begin{subfigure}{0.495\textwidth} 
    \centering
        \includegraphics[width=\textwidth]{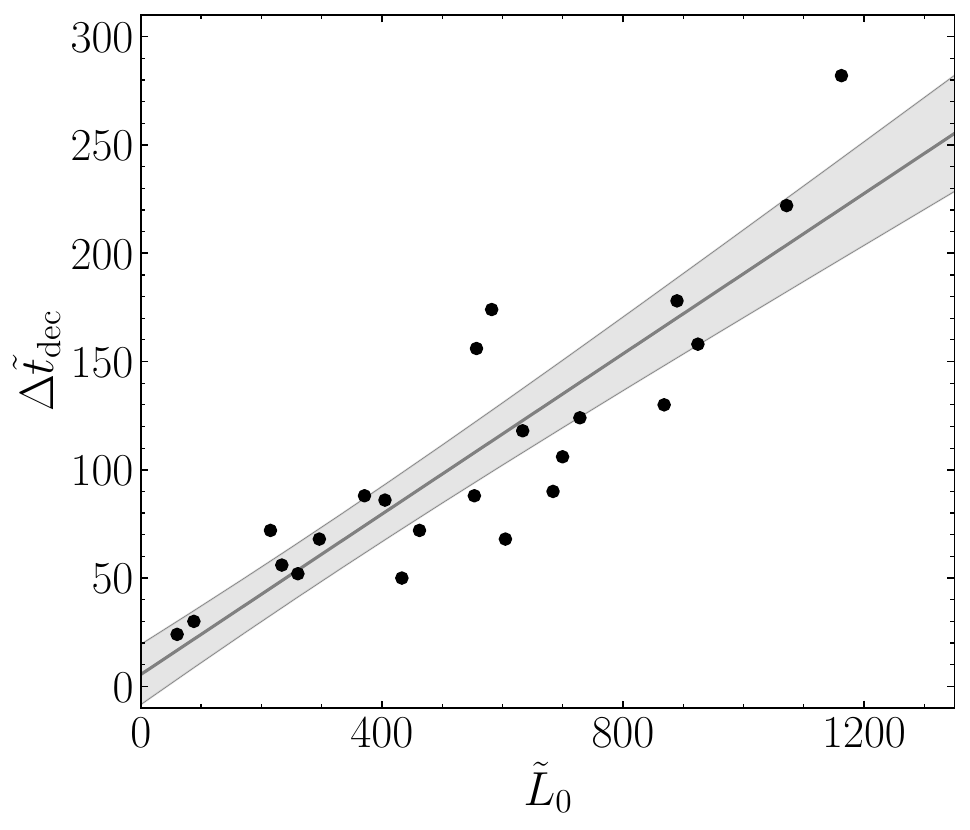}
        \caption{Network loops}
        \label{fig:global:lengthdecaynetwork}
    \end{subfigure}
    \begin{subfigure}{0.495\textwidth}
    \centering
       \includegraphics[width=\textwidth]{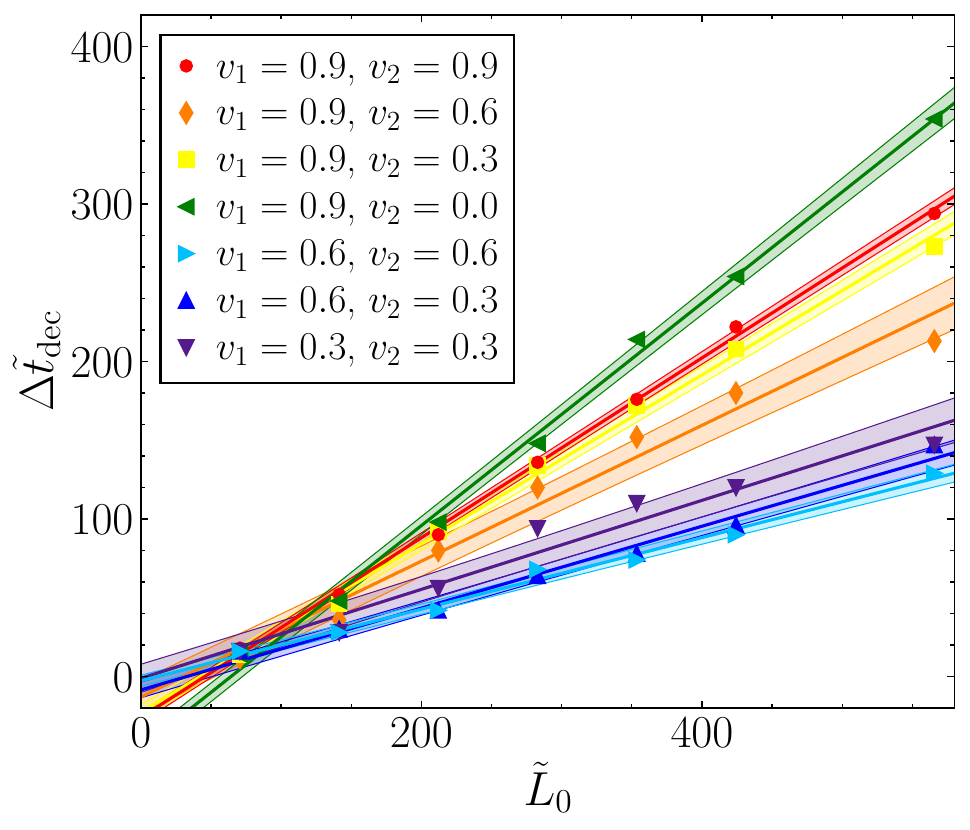}
        \caption{Artificial loops}
        \label{fig:global:lengthdecayartificial}
    \end{subfigure}
    \caption{Lifetime of the network (left) and artificial loops (right) as a function of their initial length. Lines and shaded regions are the results of a linear fit.
         }
    \label{fig:global:lengthdecay}\vspace{2cm}

\centering
\begin{tabular}{>{\centering\arraybackslash}p{4.5cm}>{\centering\arraybackslash}p{1.5cm}>{\centering\arraybackslash}p{1.5cm}>{\centering\arraybackslash}p{1.5cm}>{\centering\arraybackslash}p{1.5cm}}
\toprule
 Type of loop& $A\times 10^3$ & $B$ & $C\times 10^3$ & $D$   \\ \midrule 
 Network  & $185(22)$ & $5(14)$ & $89(13)$ & $-1(17)$ \\
 Artificial, $v_1=0.9, v_2=0.9$  & $571(10)$ & $-26(3)$ & $270(9)$ & $-22(6)$  \\ 
 Artificial, $v_1=0.9, v_2=0.6$ & $430(30)$ & $-13(11)$ & $223(14)$ & $-14(9)$   \\ 
 Artificial, $v_1=0.9, v_2=0.3$ & $534(14)$ & $-23(5)$ & $269(9)$ & $-20(6)$  \\ 
 Artificial, $v_1=0.9, v_2=0.0$ & $706(19)$ & $-45(6)$ &  $348(9)$ & $-40(6)$ \\ 
 Artificial, $v_1=0.6, v_2=0.6$ & $227(11)$ & $-3(4)$ & $125(5)$ & $-1(3)$ \\ 
 Artificial, $v_1=0.6, v_2=0.3$ & $260(15)$ & $-9(5)$ & $140(8)$ & $-6(5)$ \\ 
 Artificial, $v_1=0.3, v_2=0.3$ & $280(30)$ & $-1(9)$ & $154(15)$ & $2(9)$ \\ \bottomrule
\end{tabular}
\captionof{table}{Results of linear fits $\Delta\tilde{t}_{\rm dec} = A\tilde L_{0} + B$ and $\Delta\tilde{t}_{\rm dec} = C\tilde E_{\text{str,0}} + D$, for network and artificial loops. All artificial loops are simulated with $\sin\alpha=0.4$.
} 
\label{tab:global:decayfits}
\end{figure}

In \cref{fig:global:lengthdecaynetwork}, we show the decay time of network loops as a function of their initial lengths, the latter measured using the number of pierced plaquettes---see \cref{eq:global:lengthwindingdefinition}.   We observe that the results scale roughly linearly with $L_0$, indicating a scale invariant mechanism driving the decay of the loops. Similar behavior is also observed as a function of the initial string energy, $E_{\str,0}$, determined using \cref{eq:global:stringenergyweighteddefinition} at the time when the isolated loop is found. The results of linear fits of the form $\Delta\tilde{t}_\dec=A\tilde{L}_0+B$ and $\Delta\tilde{t}_\dec=C\tilde{E}_{\str,0}+D$ are presented in the first row of \cref{tab:global:decayfits}. The latter result allows us to estimate the particle-emission power,
\begin{equation}\label{eq:global:powernetworkresult}
\tilde{P}_\varphi=\frac{\d \tilde{E}_\str}{\d\tilde{\tau}}=\frac{1}{C}=11.2(1.6)\,,
\end{equation}
where $\tilde{P}_\varphi=\Pphi/v^2$. 

In \cref{fig:global:lengthdecayartificial}, we present the decay time of artificial loops as a function of $L_0$ for different choices of initial boost velocities. Here we approximate $L_0$ as $4\Lonefourth$, since the use of \cref{eq:global:lengthwindingdefinition} is not appropriate as the Manhattan effect underestimates the length on straight strings. We observe that artificial loops live up to three times more than network loops of the same length depending on the initial velocities, for the range of lengths that can be compared. As before, we observe that $\Delta\tilde{t}_\dec$ scales linearly with $L_0$ (and also $E_{\str,0}$), although the data shows a clear dependence on the velocity. The results of linear fits to each set of velocities are also presented in \cref{tab:global:decayfits}, from which the particle-emission power, as defined in \cref{eq:global:powernetworkresult}, can be obtained, $\tilde{P}_\varphi\approx 3-8$, depending on the initial velocities. Note that the $v_1=v_2=0.6$ case corresponds to the same setup used in \rcite{Saurabh:2020pqe}, which finds the  linear regression with coefficient $A$ to be $\sim 40\%$ bigger than ours.

\begin{figure}[!b]
    \centering
    \begin{subfigure}{0.7\textwidth} 
    \centering
        \includegraphics[width=1\textwidth]{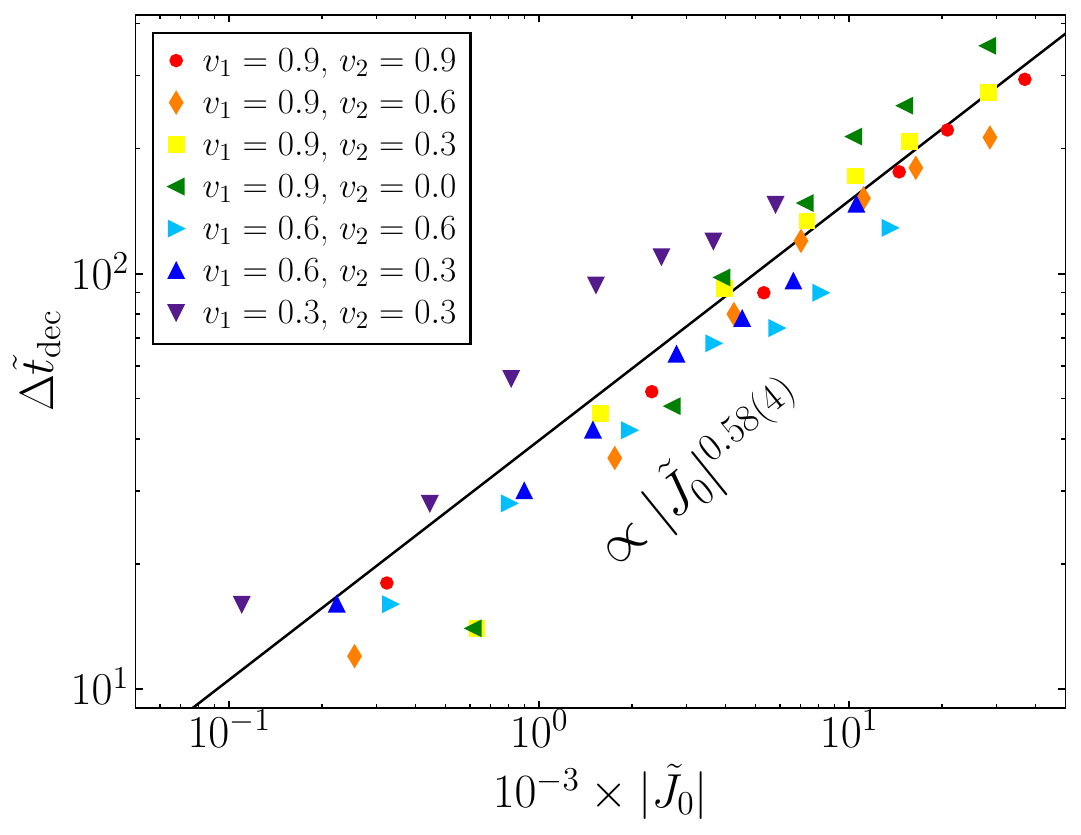}
    \end{subfigure}
    \caption{
         Lifetime of artificial loops with different initial velocities as a function of their initial angular momentum, measured using \cref{eq:global:stringJweighteddefinition}. The line is the result of a power-law fit to all families. }
    \label{fig:global:angularmomentum}
\end{figure}

In \cref{fig:global:angularmomentum}, we show $\Delta\tilde{t}_\dec$ as a function of ${J}_0$ for artificial loops. We observe all the loops follow a universal power-law dependence, that roughly scales as $\Delta\tau_\dec\propto J_0^{3/5}$. This highlights angular momentum as one of the key ingredients affecting global loop decay. Retrospectively, this also explains the observed dispersion of the points in \cref{fig:global:lengthdecaynetwork}, since we have no control over the angular momentum for network loops.

Finally, we have performed several consistency checks to ensure the robustness of our results. We observe very small variations of $\Delta \tilde{t}_\dec$ when changing $\delta x$, and for artificial loops we find changes smaller than $10\%$ when varying the ratio $L/\Lonefourth$. For artificial loops, we have also observed minimal dependence on $\Delta\tilde{t}_\dec$ on the boost direction, given by the choice of $\alpha$, which indicates that the long-range interactions rapidly take over any impact of this parameter, as it does not affect the total energy in the system.

\subsection{Particle emission}\label{sec:global:resultsspectra}

\begin{figure}[!b]
    \centering
    \begin{subfigure}{0.495\textwidth} 
    \centering
        \includegraphics[width=\textwidth]{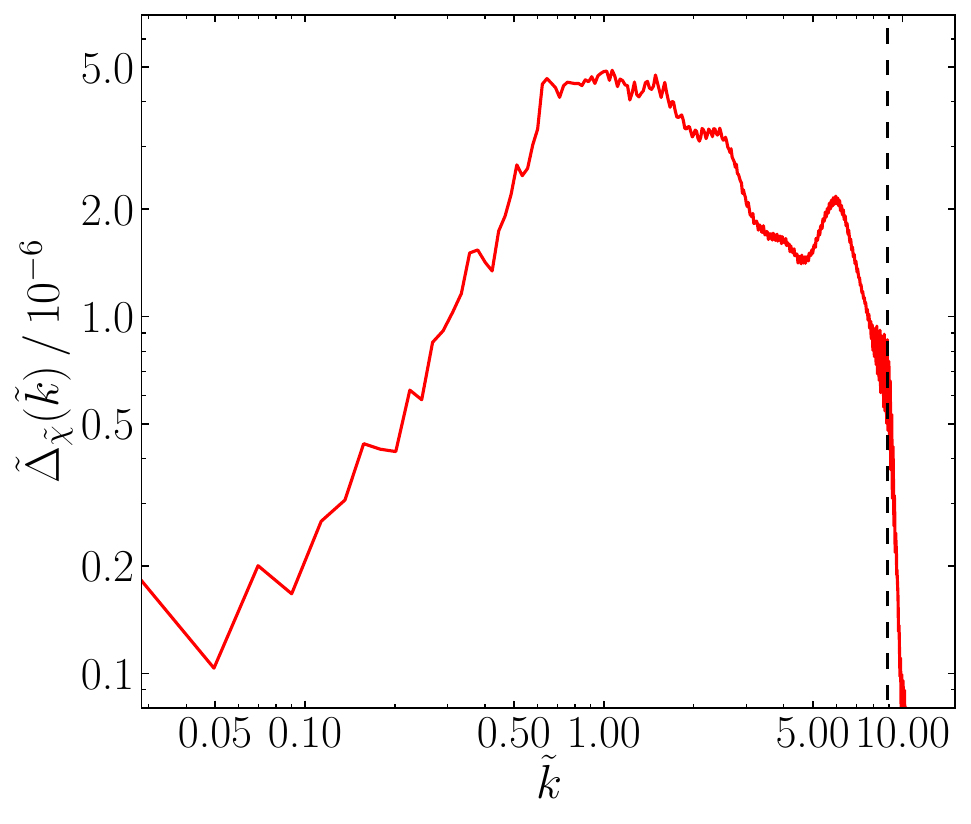}
        \caption{Massless modes}
        \label{fig:global:spectrachi}
    \end{subfigure}
    \begin{subfigure}{0.495\textwidth}
    \centering
       \includegraphics[width=\textwidth]{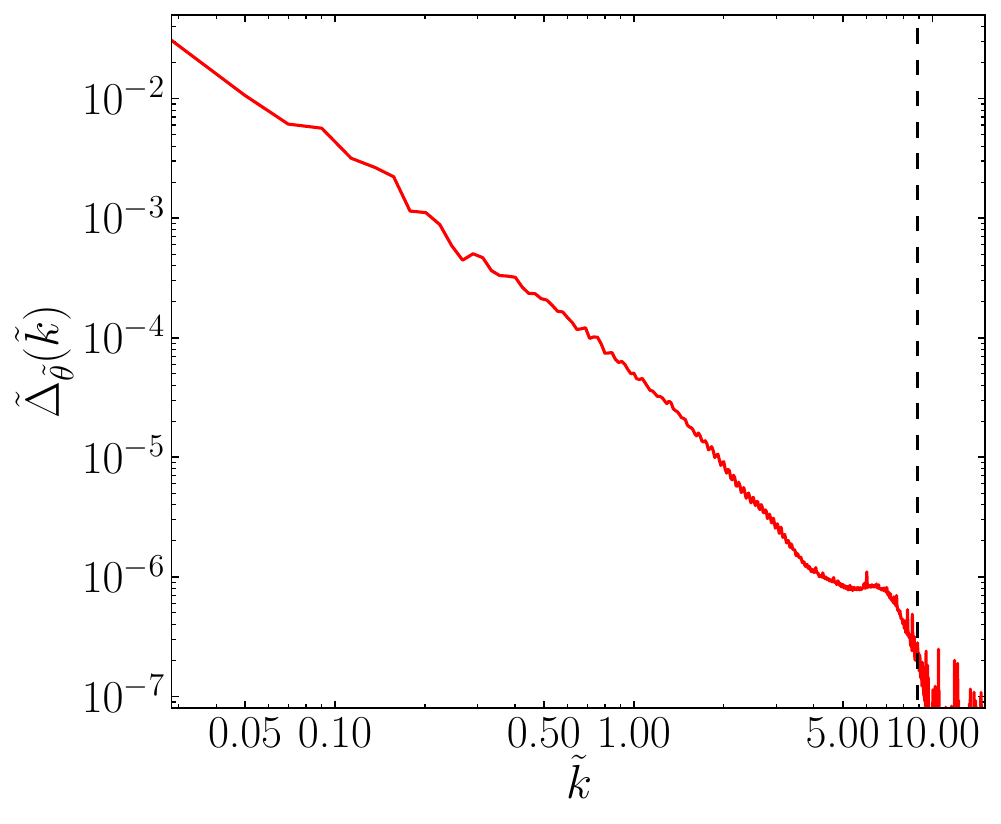}
        \caption{Massive modes}
        \label{fig:global:spectratheta}
    \end{subfigure}
    \caption{Power spectrum of massive (left) and massless (right) modes just after the collapse of an artificial loop generated with $v_1=v_2=0.6$ and $\sin\alpha=0.4$. Vertical dashed lines indicate the scale of the string core, $\tilde{k}_\text{c}=2\pi/\tilde{r}_\text{c}\approx9$.
         }
    \label{fig:global:spectrachitheta}
\end{figure}

We also study the distribution of massive, $\chi$, and massless, $\theta$, particles after the collapse of the loop, as defined in \cref{eq:global:fieldexcitationsdecomposition}. In \cref{fig:global:spectrachi,fig:global:spectratheta} we represent the power spectra of $\theta$ (left) and $\chi$ (right) at the end of the decay of an artificial loop generated with $v_1=v_2=0.6$, $\sin\alpha=0.4$ and $\tilde{L}_{1/4}=\tilde{L}/2\approx 141$. We observe the spectrum of massless models is power-law suppressed at high modes, while the spectrum of massive excitations peaks at $\tilde{k}\sim1$, albeit with an amplitude much smaller than that of the massless field spectrum at the same scale. Massless modes with $\tilde{k}\ll 1$ can easily be emitted, while the emission of massive particles, although also possible, is suppressed compared to the massless case.

\begin{figure}[!t]
    \centering
    \begin{subfigure}{0.7\textwidth} 
    \centering
        \includegraphics[width=1\textwidth]{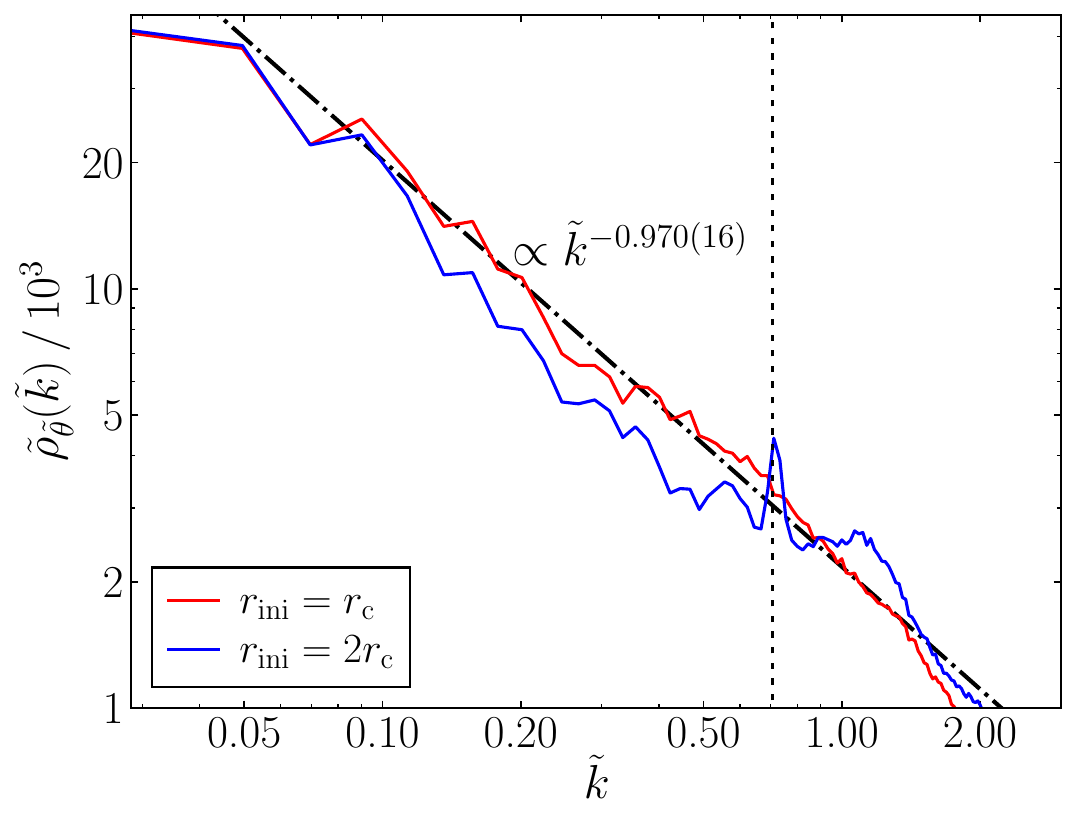}
    \end{subfigure}
    \caption{
         Lifetime of artificial loops with different initial velocities as a function of their initial angular momentum, measured using \cref{eq:global:stringJweighteddefinition}. The vertical dashed line indicates the scale of the core radius, $\tilde{k}_\text{c}=2\pi/\tilde{r}_\text{c}$, and the dot-dashed line is the result of a power-law fit to the $r_\text{ini}=r_\text{c}$ result. }
    \label{fig:global:spectrathetaenergy}
\end{figure}

We note that \rcite{Saurabh:2020pqe} presents the energy spectrum of massless modes, defined from \cref{eq:global:energymasslessmodespowerspectrum}, which shows a peak at a scale half that of the massive mode, $k=m_\chi/2$ (which corresponds to $\tilde{k}=\sqrt{2}/2$ in our program units). We present our results for this energy spectrum after the loop decays in red in \cref{fig:global:spectrathetaenergy}. From a power-law fit, we observe that the spectrum roughly scales as $\propto k^{-1}$, in agreement with \rcite{Saurabh:2020pqe}. However, we detect no presence of any such peak at $k=m_\chi/2$. Interestingly, if we set the initial radius of the string to be larger than its physical value by modifying the radial profile of the NO vortices used to generate the initial strings as $f(r)\rightarrow f(2r)$, we observe a peak appearing at the same scale as in \rcite{Saurabh:2020pqe}. We believe this peak is related to an excess energy in the radial mode of the string due to a excessively large core width set in the initial conditions.

\subsection{GW emission}\label{sec:global:resultsGWs}

We study the emission of GWs from decaying loops, following the procedure outlined in \cref{sec:Cosmo:GWsimulation}. In a Minkowski background, the equation of motion of the GWs becomes
\begin{equation}
\ddot{h}_{ij}-\partial_k\partial_k h_{ij}=\frac{4}{\mpl^2}\left[\Re(\partial_i\varphi\partial_j\varphi^*)\right]^\TT\,,
\end{equation}
where recall, $[...]^\TT$ refers to the transverse-traceless component, and we are neglecting backreaction of the GWs on the loop dynamics.  We define the fractional GW energy density normalizing with the total energy density of the complex scalar field, $\rho_\varphi$, which is conserved,
\begin{equation}\label{eq:global:GWfractionalenergystrings}
\Omega_\GW(k,t)=\frac{1}{\rho_\varphi}\frac{\d\rho_\GW(k,t)}{\d\log k}\,.
\end{equation}
The total GW energy emitted is computed by integration,

\noindent\begin{equation}\label{eq:global:GWenergystrings}
E_\GW=\rho_\varphi L^3\int\Omega_\GW(k,t)\d\log k\,.
\end{equation}
%In terms of program variables, $\tilde{E}_\GW=(\sqrt{\lambda}/v)E_\GW$.

We first analyze lattice discretization effects on the GW spectrum of a loop. For network loops, we run a high-resolution simulation with $\delta\tilde{x}=0.125$ and save the configuration when an isolated loop is found. This is used to create new coarser configurations with $\delta\tilde{x}^{(p)}=p\delta\tilde{x}$ ($p=2,3,4$), by eliminating $p-1$ sites of every $p$ consecutive points of the original lattice per dimension. For artificial loops, we run simulations with different $\delta\tilde{x}$, generated using equivalent initial conditions and parameters $\tilde{L}=192$, $\tilde{L}_{1/4}=96$, $v_1=0.6$, $v_2=0.7$ and $\sin\alpha=0.5$.

\begin{figure}[!p]
    \centering
    \begin{subfigure}{0.75\textwidth} 
    \centering
        \includegraphics[width=\textwidth]{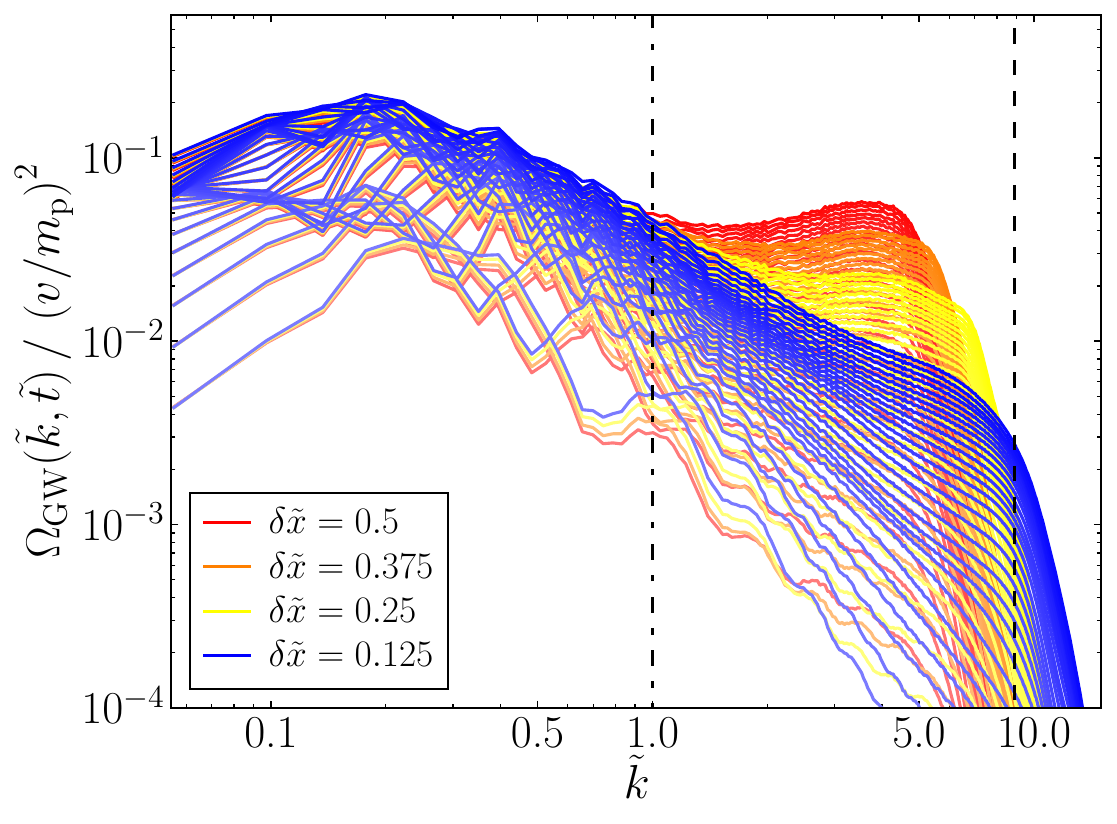}
        \caption{Network loop}
        \label{fig:global:GWUVnetwork}
    \end{subfigure}\vspace{0.8cm}
    \begin{subfigure}{0.75\textwidth}
    \centering
       \includegraphics[width=\textwidth]{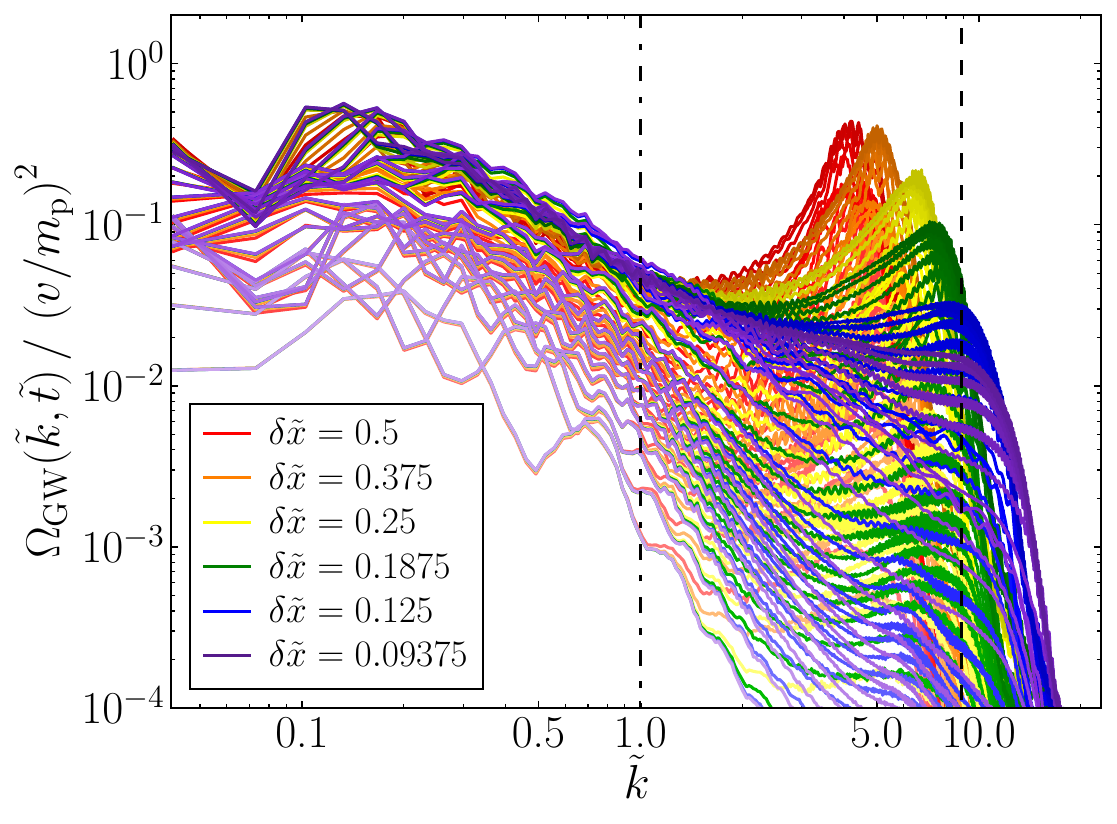}
        \caption{Artificial loop}
        \label{fig:global:GWUVartificial}
    \end{subfigure}
    \caption{Evolution of the GW energy density power spectrum with varying UV resolution and fixed lattice size. The vertical dashed line indicates the scale of the string core, $\tilde{k}_\text{c}=2\pi/\tilde{r}_\text{c}$, and the vertical dot-dashed line is the cutoff, $\tilde{k}_\text{cut}$, up to which we integrate the spectrum to compute the GW energy to prevent capturing UV artifacts. Spectra go from early to late times from bottom to top, with a separation of four and two units of program time between consecutive lines, for the top and bottom panels, respectively.
         }
    \label{fig:global:GWUV}
\end{figure}

The evolution of the resulting power spectra for network and artificial loops are shown in \cref{fig:global:GWUV}. Different lines correspond to different times, measured every four and two units of program time, respectively, with time going in general from bottom to top. In both cases, GW emission peaks at IR scales, $\tilde{k}\sim(2-6)\tilde{k}_0$, with $k_0=2\pi/L_0$ the scale of the initial string length, and there is good agreement between all resolutions up to $\tilde{k}\sim 0.1\tilde{k}_\text{c}$, where $k_\text{c}=2\pi/\rc$ is the scale of the core radius. A second peak emerges at higher modes, but it is suppressed as the UV resolution is improved, indicating it is indeed a lattice artifact arising when the string core is not well resolved. In view of this, we decide to compute the total GW energy emitted by loops integrating the spectrum only up to some cutoff scale, $\tilde{k}_\text{cut}\sim 0.1\tilde{k}_\text{c}$. This ensures the result does not capture the unphysical UV peak.%, while the neglected energy remains small since the true spectrum is suppressed in the UV.

\begin{figure}[!p]
    \centering
    \begin{subfigure}{0.75\textwidth} 
    \centering
        \includegraphics[width=1\textwidth]{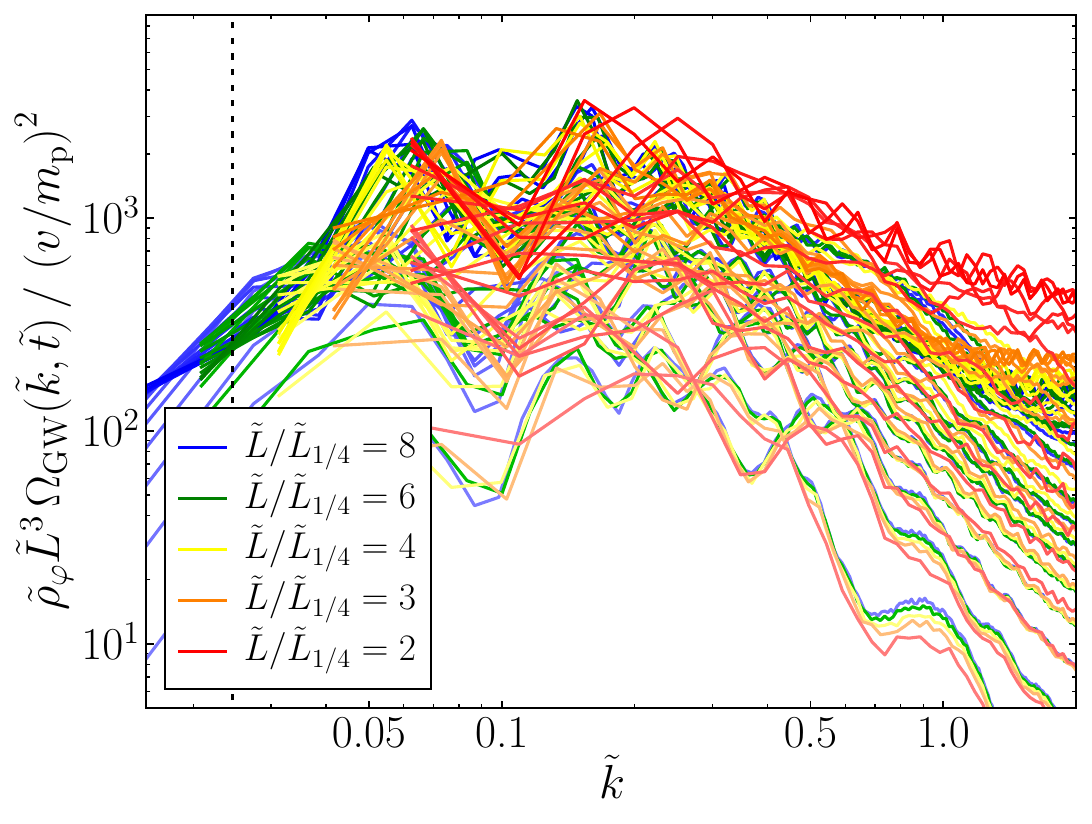}
    \end{subfigure}
    \caption{
         Evolution of the GW energy density power spectrum for artificial loops generated with $v_1=0.6$, $v_2=0.7$ and $\sin\alpha=0.5$, with fixed $\delta\tilde{x}=0.25$ $\tilde{L}_{1/4}=64$ and varying $L$. The vertical dotted line indicates the scale of the initial length of the string, $\tilde{k}_0=2\pi/\tilde{L}_0$. Spectra go from early to late times from bottom to top, with a separation of four units of program time between consecutive lines. }
    
    \label{fig:global:GWIR}\vspace{1cm}

    \centering
    \begin{subfigure}{0.75\textwidth} 
    \centering
        \includegraphics[width=1\textwidth]{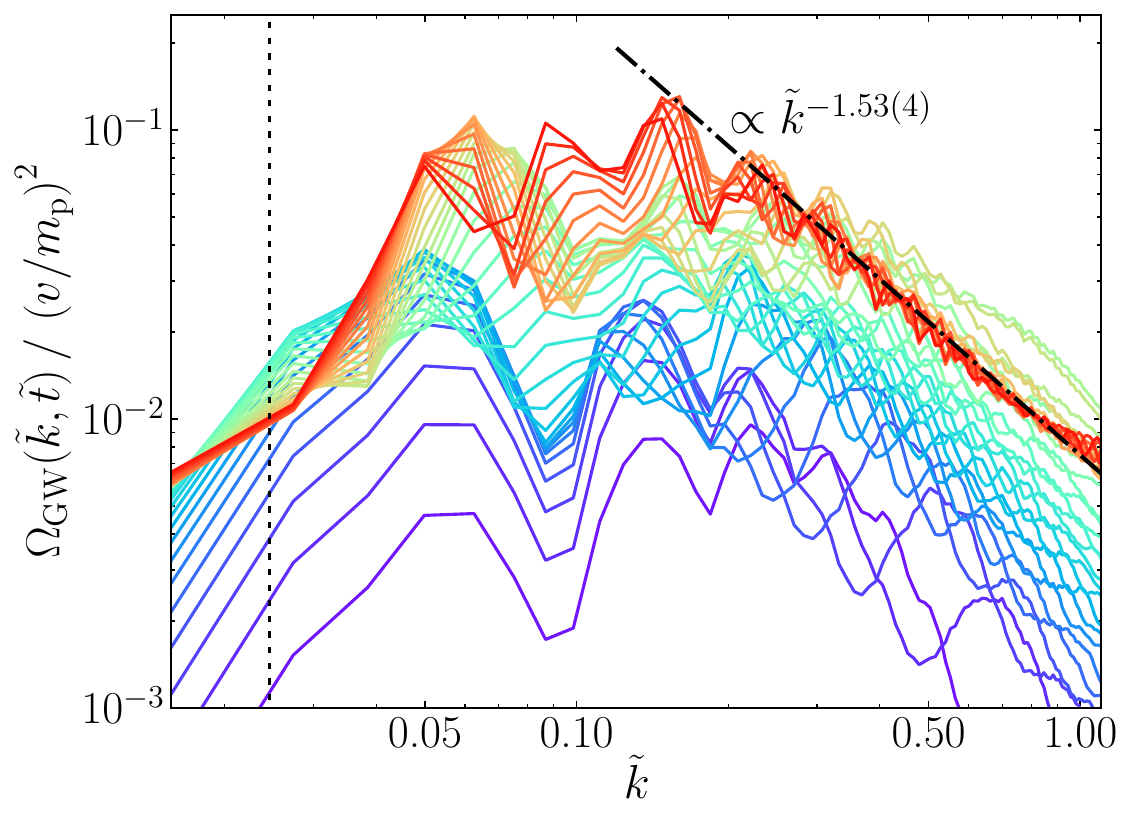}
    \end{subfigure}
   \caption{
         Evolution of the GW energy density power spectrum of an artificial loop with $\delta\tilde{x}=0.25$, $\tilde{L}=512$  and $\tilde{L}_{1/4}=64$, generated with $v_1=0.6$, $v_2=0.7$ and $\sin\alpha=0.5$  Each line corresponds to a different time, going from purple to red, with separation of two units of program time. The vertical dotted line indicates the scale of the initial length of the string, $\tilde{k}_0=2\pi/\tilde{L}_0$, and the dot-dashed line is a fit to the high-frequency tail of the final-time spectrum. } 
    \label{fig:global:GWIRzoom}\vspace{0.5cm}
\end{figure}

We also study the effect of varying the IR coverage for artificial loops. We set $\tilde{L}_{1/4}=64$, $\delta\tilde{x}=0.25$, $v_1=0.6$, $v_2=0.7$ and $\sin\alpha=0.5$, and vary the box size. The resulting spectra are shown in \cref{fig:global:GWIR}, multiplied by a factor that accounts for volume dependencies. The spectra are represented every four units of program time. We observe a noticeable discrepancy at intermediate scales for the smallest box, but the spectra converges rapidly as the box size increases. We also observe that the emission of GWs is suppressed for scales larger than the initial loop length, $k<k_\text{0}$, represented with a black dotted line.

A zoomed-in version of the spectra in the largest box is shown, every two units of program time, in \cref{fig:global:GWIRzoom}. Here we note the presence of various peaks in the spectrum. Although the peak structure resembles the harmonic pattern expected in NG strings---see \cref{sec:global:NG}---peak frequencies here are not in harmonic proportions, and their absolute and relative locations vary between early (purple) and late (red) times. From a fit to the UV tail of the spectrum, we observe it decays roughly as $k^{-3/2}$. This represents a steeper fall than standard NG predictions for a cusp-dominated emission, expected to be proportional to $ k^{-4/3}$.

\begin{figure}[!b]
    \centering
    \begin{subfigure}{1\textwidth} 
    \centering
        \includegraphics[width=1\textwidth]{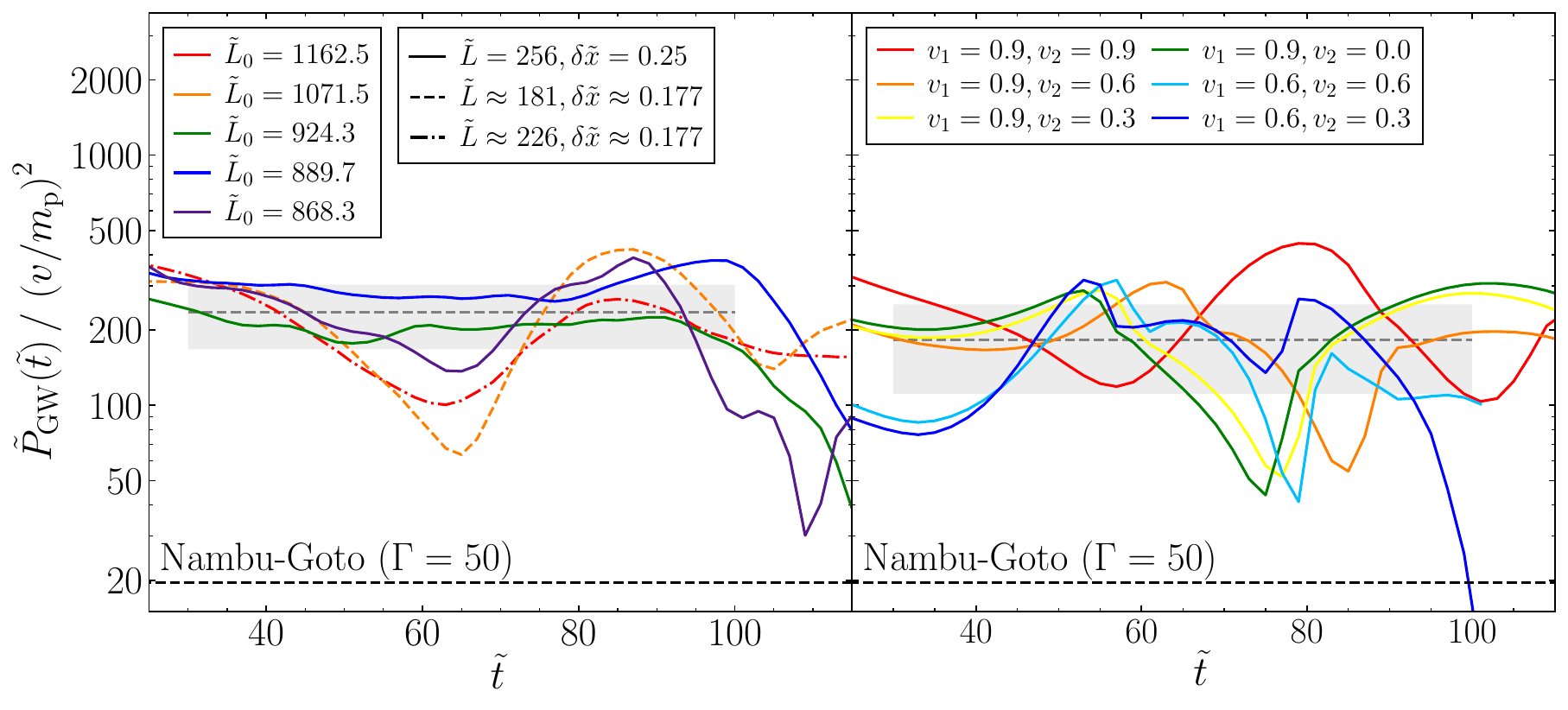}
    \end{subfigure}
   \caption{
         Rolling-averaged GW power emitted by network loops of several lengths (left) and artificial loops with different boost velocities (right), computed using \cref{eq:global:rollingaverage}. In the case of network loops, each color corresponds to a different initial length and the linestyle indicates the lattice parameters, as indicated in the plot legends. For artificial loops, each color refers to a different pair of boost velocities. For comparison, the typical NG result (for $\mu=\pi v^2$ and $\Gamma=50$) is shown as a horizontal dashed line. Finally, The grey bands represent the average value of the emission power in the range $\tilde{t}=30-100$ for each type of loop.} 
    \label{fig:global:GWpower}
\end{figure}

\newpage Having understood the main sources of systematic errors, we now turn our attention to the GW power emitted by a decaying loop. We define a rolling-average measurement of this quantity,
\begin{equation}\label{eq:global:rollingaverage}
P_{\rm GW}(t) \equiv %\langle E_\text{GW}'\rangle = 
\frac{L^3\rho_{\varphi}}{2T}\int^{t+T}_{t-T}\text{d}t'\int_0^{k_\text{cut}} \dot\Omega_\text{GW}(k,t')\,\text{d}\log k\,, 
\end{equation}
which, in terms of program variables, is $\tilde{P}_\GW=\PGW/v^2\times(v^2/\mpl^2)$. Our result for the GW power emitted is  shown in \cref{fig:global:GWpower} for a number of network and artificial loops, as a function of the lifetime of the loops. The rolling-averaged GW emission power is computed using $\tilde{T}=15$, which we observe is enough to remove rapid oscillations while still keeping features related to the string dynamics, and $k_\text{cut}=1$.

The left panel corresponds to network loops with different $L_0$, simulated for various $\delta x$ and $L$, as shown in the plot legend. In all cases, we observe the power emitted does not to depend on $L_0$ and is roughly constant in time, with fluctuations that depend on the specific details of the dynamics of each loop. At late time, the emission power drops down as the loop finally disappears. Remarkably, we do not observe evidence of any systematic variation of the GW power emission due to the shrinking of the loops. Between $\tilde{t}=30-100$, the average emission of all loops is $\tilde{P}_\GW=240\pm80$ (grey dahsed line and band).  Although  there is a priori no reason to expect this to be similar to the NG prediction, \cref{eq:global:NGpredictionGWs}, it is still instructive to make the comparison. Using $\mu=\pi v^2$ and $\Gamma=50$, one gets $\tilde{P}_\text{GW}\approx 20$, roughly an order of magnitude smaller than our result.

The right panel of \cref{fig:global:GWpower} shows our results for artificial loops with different velocities, produced in simulations with $\tilde{L}=450/\sqrt{2}\approx319$ and $\delta\tilde{x}=0.25/\sqrt{2}\approx0.177$, with initial string separation $\Lonefourth=L/3$, chosen to reduce IR effects. For all pairs $\{v_1,v_2\}$, the GW power emission is of a similar order as for network loops. It presents an amplitude that, as for network loops, shows no significant variation as the loops shrink. The power emission also shows some fluctuations that we believe are related to the very symmetric initial configuration. For the period $\tilde{t}=30-100$, the average emission is $\tilde{P}_\GW=190\pm80$.

Comparing the results for GW and particle-emission power, we obtain
\begin{equation}
{P_{\rm GW}\over P_{\varphi}} \approx \left\lbrace 
\begin{array}{ll}
\displaystyle {240 (80)\over 11.2(1.6)}\left({v\over m_\text{p}}\right)^2\,,\quad\quad&{\rm Network~loops,}\vspace*{1mm}\\[5pt]
\displaystyle{190(80)\over 5.2(2.5)}\left({v\over m_\text{p}}\right)^2\,,\quad\quad&{\rm Artificial~loops,}
\end{array}\right.
\end{equation}
where $P_\varphi$ for artificial loops is taken as an average over all families of initial velocities. %This is, in all cases, the emission of GWs compared to particle radiation is suppressed with a factor $\cO(10)(v/\mpl)^2$.
 As $v$ cannot be arbitrarily large (e.g. CMB constrains require $v/\mpl\leq 10^{-6}$~\cite{Lopez-Eiguren:2017dmc,Benabou:2023ghl}), we conclude $\PGW/\Pphi\ll 1$, indicating that the GW emission from decaying global string loops is completely subdominant compared to particle emission. Note this justifies a posteriori neglecting the backreaction of GWs on the matter fields.
 
\section{Conclusion}\label{sec:global:conclusions}

Studying the evolution and decay of cosmic strings requires the use of field-theory lattice simulations to fully capture the dynamics of the strings. In this chapter, we have reviewed the basics of cosmic strings in the early universe, and presented the results from \rcite{Baeza-Ballesteros:2023say}, where the simultaneously  emission of particles and GWs from string loops was studied using lattice simulations for the first time. Our results show that the particle emission completely dominates the decay of the loops. Indeed, we observe a universal result
\begin{equation}
\frac{\PGW}{\Pphi}\approx\cO(10)\left(\frac{v}{\mpl}\right)^2\ll 1\,,
\end{equation}
which holds with independence of the length, shape, energy and angular momentum of the loops.
This conclusion is robust for loops with length-to-width ratio $L_0/\rc\lesssim 1700$, with no indication this will change for longer loops. In particular, we observe no evidence of any logarithmic enhancement of the emission power of GWs with the length of the loops, which is sometimes assumed from combining field-theory results for the string tension---see \cref{eq:global:logdivergenttension}---with the NG prediction for the GW emission---see \cref{eq:global:NGpredictionGWs}. Note, however, that there is no reason why NG results must hold in the case of global strings.

Our result opens the door to a new technique to calculate the GWB from a network of strings. Current approaches are based on lattice simulations of networks~\cite{Figueroa:2012kw,Figueroa:2020lvo,Gorghetto:2021fsn}, which can only study a limited dynamical range, or the combination of field-theory and NG ingredients, usually assuming a logarithmic divergent string tension~\cite{Gouttenoire:2019kij,Chang:2019mza,Chang:2021afa,Gouttenoire:2021jhk}. We propose to obtain the GWB spectrum from combining our results for the  GW emission of isolated loops with predictions for the loop number density computed at cosmological scales~\cite{Ringeval:2005kr,Blanco-Pillado:2013qja,Klaer:2017qhr,Martins:2018dqg,Auclair:2019zoz,Auclair:2021jud}. We believe this will lead to a suppression of the GWB from global string networks compared to present predictions, as we find no evidence of a logarithmically dependent GW power.

\titleformat{\chapter}
{\thispagestyle{plain}\vspace{50pt}\normalfont\huge\bfseries}
{\filleft{\parbox[b][][b]{.26\textwidth}{\chapnumfont\color{chapnumcolor}1\hspace{-0.6cm}0}}}
{0pt}
{\Huge{\parbox[b][][t]{.74\textwidth}{\flushleft\hyphenpenalty=10000\chaptitlefont #1\vspace{0.0cm}}}}[\vspace{0.5ex}{\titlerule[0.7pt]}\vspace{-0.7cm}]

\chapter{GW emission from cosmic string loops: local case}
\label{sec:local}

%Cosmic strings are one dimensional topological defects predicted to originate in the early universe by many theories beyond the standard model. 
\Cref{sec:global} is focused on the study of global strings, which arise, for example, in axion models, and we represented in our simulations using a complex scalar field with a Mexican-hat potential. This type of string presents long-range forces, and closed loops were found to decay mainly through particle production, with a very suppressed gravitational wave (GW) emission power. Another type of cosmic strings are \textit{local strings}, which arise, for example, in grand unified theories~\cite{Copeland:2009ga,Copeland:2011dx}. These are characterized by short-range interactions and are typically represented using the Abelian-Higgs model. These strings are, a priori, expected to be well described by the Nambu-Goto (NG) approximation.

Recall that the NG model, introduced in \cref{sec:global:NG}, describes strings as infinitely thin objects that can only decay via the emission of GWs, while the production of massive particles is neglected. However, in a field-theory scenario both emission channels are available. In the work presented in this chapter we use field-theory lattice simulation to compare both decay routes simultaneously~\cite{Baeza-Ballesteros:stringsinprep}. We consider both network loops, generated in a similar manner to those in \rcite{Hindmarsh:2021mnl}, and two types of artificial loops, produced following \rrcite{Matsunami:2019fss,Hindmarsh:2021mnl}. We recall that we neglect backreaction of the GWs on the strings, and so the energy emitted in form of gravitational radiation is not subtracted from the energy budget of the strings. This is an assumption that we self-consistently check a posteriori.

The decay of local string loops into particles has already been explored using classical-field-theory simulations. In \rcite{Matsunami:2019fss}, artificial loops generated from the intersection of boosted infinite strings were studied, finding a lifetime proportional to the square of the string length, $\Delta t_\text{dec}^\text{part}\propto L^{2}$. Comparing these lattice results to the NG prediction for GW emission, $\Delta t_\text{dec}^\text{GW}\propto L$, this work concluded that above some critical length, $L_\text{crit}$, the decay of local string loops occurs predominantly via GW production. 

In \rcite{Hindmarsh:2021mnl},  loops originating from networks were investigated, finding their lifetime to be proportional to their length, $\Delta t_\text{dec}^\text{part}\propto L$. Ref.~\cite{Hindmarsh:2021mnl} also studied a new type of artificial loops originated from the intersection of static non-straight infinite strings. These were found to behave similarly to those in \rcite{Matsunami:2019fss}. This showed that the decay of both types of loops is dragged by fundamentally different mechanisms. These results point to the fact that, for loops originating after a phase transition in the early universe, the emission of GWs, based on NG predictions, would be suppressed compared to particle production.

In this chapter, we present lattice results on the simultaneous emission of particles and GWs from local string loops. In \cref{sec:local:model}, we introduce the Abelian-Higgs model and discuss some of the main properties of local strings. In \cref{sec:local:lattice}, we describe the observables we use to analyze the dynamics of the local strings, and the different types of initial conditions we consider. Results for particle and GW emission are presented in \cref{sec:local:results}, and we conclude in \cref{sec:local:conclusion} with a brief summary.

\section{The Abelian-Higgs model}\label{sec:local:model}

The Abelian-Higgs model contains a complex scalar field, $\varphi=(\phi_1+i\phi_2)/\sqrt{2}$, with $\phi_{1,2}$ real scalar fields, and an Abelian gauge field, $A_\mu$. It is characterized by the action,
\begin{equation}
S[\varphi, A]=\int\d^4 x\sqrt{-g}\left[(D_\mu\varphi)(D^\mu\varphi)^*-\frac{1}{4}F_{\mu\nu}F^{\mu\nu}-V(\varphi)\right]\,,
\end{equation}
where the scalar potential is the same as in the global case, 
\begin{equation}\label{eq:local:complexscalarpotential}
V(\varphi)=\lambda\left(|\varphi|^2-\frac{v^2}{2}\right)^2\,.
\end{equation}
Also, $D_\mu=\partial_\mu-ieA_\mu$ is the covariant derivative with $e$ the gauge coupling, and $F_{\mu\nu}=\partial_\mu A_\nu-\partial_\nu A_\mu$ is the field-strength tensor. This model is invariant under local gauge transformations, as given in \cref{eq:Cosmo:gaugetransformations}. 

%In conformal time and working in the temporal gauge, $A_0=0$, the dynamics of the fields are described by the equations of motion in \cref{eq:Cosmo:}, 
%\begin{equation}\label{eq:local:eomlocal}
%\begin{array}{rl}
%\varphi^{\prime\prime}+2\frac{a^\prime}{a}\varphi^\prime-D_iD_i\varphi=&-a^2\lambda(|\varphi|^2-v^2)\varphi\,,\\
%A_i^{\prime\prime}-\nabla^2 A_i+\partial_i(\nabla\cdot \bm{A})=&2a^2e\Im\left[\varphi^*D_i\varphi\right]\,.
%\end{array}
%\end{equation}

The time evolution of the fields is controlled by the equations of motion, given by \cref{eq:Cosmo:fieldseomcomplex,eq:Cosmo:fieldseomgauge}. We focus for our study in Minkowski background, where they take the form
\begin{equation}\label{eq:local:eomlocalMinkoski}
\begin{array}{rl}
\displaystyle\ddot{\varphi}-D_iD_i\varphi&\displaystyle=-2\lambda\left(|\varphi|^2-\frac{v^2}{2}\right)\varphi\,,\\[5pt]
\displaystyle\dot{F}_{0i}-\partial_jF_{ji}&\displaystyle=2e\Im\left[\varphi^*D_i\varphi\right]\,.
\end{array}
\end{equation}
A priori, these equations seem to depend on both $\lambda$ and $e$ separately. However, if we define dimensionless program variables using $\fstar=v$ and $\omegastar=\sqrt{\lambda} v$, and make the rescaling $A_\mu\rightarrow A_\mu/e$,  they take the form
\begin{equation}
\begin{array}{rl}
\displaystyle\ddot{\tilde{\varphi}}- \tilde{\partial}_i\tilde{\partial}_i\tilde{\varphi}&=\displaystyle-\left(|\tilde{\varphi}|^2-\frac{1}{2}\right)\tilde{\varphi}\,,\\
\displaystyle\dot{\tilde{F}}_{0i}-\tilde{\partial}_j\tilde{F}_{ji}&=\displaystyle2 \frac{e^2}{\lambda}\text{Im}\,[\tilde{\varphi}^*\tilde{D}_i\tilde{\varphi}]\,.
\end{array}
\end{equation}
This makes clear that the dynamics of the system only depends on the ratio $\beta=2\lambda/e^2$. In this work, we focus on the $\beta=1$ case.

As in the case of the model in used to represent global strings, given in \cref{eq:global:complexscalaraction}, the Abelian-Higgs model  is characterized by the presence of two phases. In a high-temperature thermal environment, the universe lies in a symmetric phase, $\langle\varphi\rangle=0$, in which the gauge symmetry is realized. At low temperatures, this local symmetry is spontaneously broken as the field falls into the true vacuum of the potential, $\langle \phi_1^2+\phi_2^2\rangle=v^2$. Similarly to the case of global strings, if the gauge symmetry is broken after the end of inflation, cosmic strings may form.

It is instructive to analyze the behaviour of the theory around the true vacuuum. We can expand the complex scalar field in terms of the radial and angular excitations, as we did in \cref{sec:global} for the global case,
\begin{equation}\label{eq:local:fieldexcitationsdecomposition}
\varphi(x)=\frac{v+\chi(x)}{\sqrt{2}}\text{e}^{i\theta(x)}\,.
\end{equation}
Using this expression, the action can be expanded as
\begin{multline}
S[\chi,\theta,A_\mu]=\int 	\d^4 x \sqrt{-g}\left[-\frac{1}{4}F_{\mu\nu}F^{\mu\nu}+\frac{1}{2}\partial_\mu \chi\partial^\mu \chi\right.\\ \left.-\lambda v^2\chi^2-\lambda v^3 \chi^3 - \frac{1}{4}\lambda \chi^4 +\frac{1}{2}e^2(v+\chi)^2\left(\frac{1}{e}\partial_\mu \theta-A_\mu\right)^2\right]\,.
\end{multline} 
Working in  the unitary gauge, $A_\mu'= A_\mu-\partial_\mu\theta / e$, makes evident that the massless mode $\theta$ is absorbed by the gauge field, which becomes massive. Thus, the low-energy effective theory contains two massive degrees of freedom: the vector field $A_\mu^\prime$, and the radial excitations, $\chi$. Their masses are, respectively,
\begin{equation}
m_A=ev\,,\quad\quad\quad m_\chi=\sqrt{2\lambda}v\,.
\end{equation}
In the $\beta=1$ case, in which we work, both masses are equal. Note that, while these conclusions have been deduced working in the unitary gauge, they hold in general, since the theory is gauge invariant.

The inverse of these masses are the Compton lengths of each of the excitations. In the context of cosmic strings, they characterize the rate at which the fields approach the vacuum and so the size of the string cores. This means that, in the $\beta=1$ case, local strings have a unique core width, $r_\text{c}\sim m_\chi^{-1}$.
The absence of massless degree of freedom also implies that interactions between strings are short-ranged.

Integrating out the massive modes, one can show that the dynamics of local strings can be effectively described by the NG action, discussed in \cref{sec:global:NG}. Therefore, one a priori expects that, according to the NG picture, local strings mainly decay via the emission of GWs. Testing this assumption using field-theory lattice simulations is one of the main goals of the work presented in this chapter.

\subsection{Infinite string solution}

\begin{figure}[!h]
    \centering
    \begin{subfigure}{0.7\textwidth} 
    \centering
        \includegraphics[width=\textwidth]{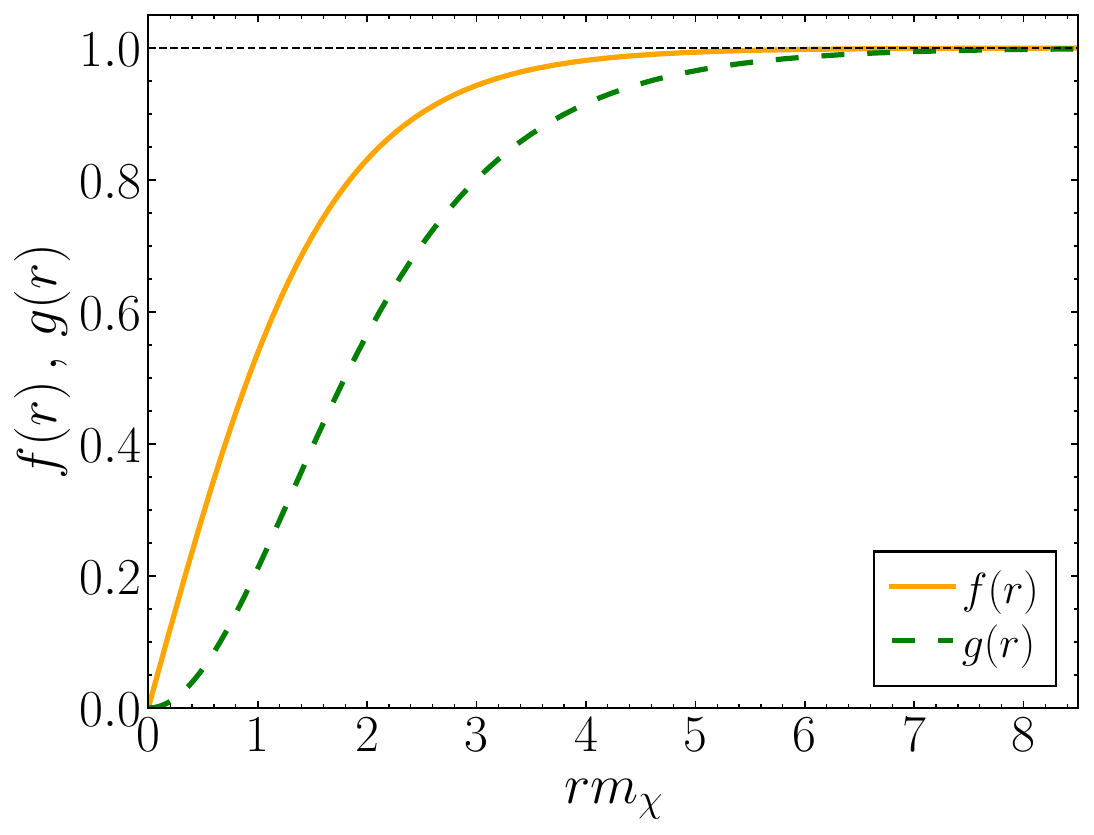}
    \end{subfigure}
    
    \caption{
        Radial profile of the NO vortex for $\beta=1$ and $k=1$, for the scalar and the gauge fields.  }
    \label{fig:local:NOprofile}
\end{figure}

Before presenting our lattice results, we consider the solution of an infinite straight string, the so-called Nielsen-Olsen vortex~\cite{NIELSEN197345}. Taking the core of string to lie on the $z$ axis, we can work with cylindrical coordinates, $(r, \theta, z)$. In the temporal gauge, $A_0=0$, we can use the following string ansatz,
\begin{equation}\label{eq:local:NOvortexansatzsolution}
\begin{array}{rl}
\displaystyle\varphi = f(r)\frac{v}{\sqrt{2}}e^{ik\theta}\,,\quad\quad\quad& \displaystyle{A}_0={A}_3=0\,.\\[10pt]
\displaystyle{A}_{1}=-\frac{g(r)}{e r}\sin\theta\,,\quad\quad\quad & \displaystyle{A}_{2}=\frac{g(r)}{e r}\cos\theta\,,
\end{array}
\end{equation}
where $k$ is the winding number of the string, and $f$ and $g$ are profile functions that can be determined numerically. Substituting this ansatz in \cref{eq:local:eomlocalMinkoski}, one finds the following equations for the profile functions,
\begin{equation}
\begin{array}{rl}
\displaystyle \frac{\partial^2 f}{\partial \tilde{r}^2}+\frac{1}{\tilde{r}}\frac{\partial f}{\partial\tilde{r}}-\frac{f^2k^2}{\tilde{r}^2}(1-g)^2-f(f^2-1)&=0\,,\\[10pt]
\displaystyle\frac{\partial^2 g}{\partial \tilde{r}^2} - \frac{1}{\tilde{r}}\frac{\partial g}{\partial\tilde{r}}+\frac{1}{2\beta}f^2(1-g)&=0\,,
\end{array}
\end{equation}
where $\tilde{r}=\sqrt{\lambda}v r$. This equations can be solved numerically using relaxation methods and imposing the boundary conditions $f(0)=g(0)=0$ and $f(\infty)=g(\infty)=1$. Our results for $\beta=1$ and $k=1$ are shown in \cref{fig:local:NOprofile}.

Analytically, it is possible to study the behavior of $f(r)$ and $g(r)$ at short and long distances from the string core. Close to the string, $r \ll m_\chi^2$, one finds
\begin{equation}
f(r)\propto r\,,\quad\quad\quad g(r)\propto r^2\,,
\end{equation}
and so $A_{1,2}\propto r$. At long distances, $r \gg m_\chi^2$, 
\begin{equation}
f(r)=1-FK_0(\sqrt{2} r)\,,\quad\quad\quad g(r) = 1 -G r K_1(\sqrt{2} r)\,,
\end{equation}
where $K_0$ and $K_1$ are are modified Bessel functions of the second kind, which obey the relation $K_1=-K_0'$, and $F$ and $G$ are positive constants that need to be determined from matching to the numerical solution. Asymptotically $K_0(x)\sim x^{-1/2}\text{e}^{-x}$, and so both $f$ and $g$ approach the vacuum exponentially fast. The full gauge field, on the other hand, only approaches zero at long distances as $A_{1,2}\propto r^{-1}$---see \cref{eq:local:NOvortexansatzsolution}.

From this analytic solution, one can extract conclusions for the interactions between strings. The   gradient, magnetic and potential energy density components of the infinite string are defined, respectively, as
\begin{equation}\label{eq:local:NOenergydensity}
\begin{array}{rl}
\rho_\text{G}(r)&\displaystyle=\frac{1}{2}\left(\frac{\d f}{\d r}\right)^2+\left(\frac{n f}{r}\right)^2\,,\\[10pt]
\rho_\text{B}(r)&\displaystyle=\frac{1}{4r^2}\left(\frac{\d g}{\d r}\right)^2\,,\\[10pt]
 V(r)&\displaystyle=\frac{\lambda}{4}(f^2-1)^2\,,
\end{array}
\end{equation}
and so the total energy density is $\rho=\rho_\text{G}+\rho_\text{B}+V$.  The tension of the string---energy per unit length---can be obtained from integrating this total energy density over the whole $(x,y)$-plane. While the separate integrals of gradient and the magnetic components do not converge at long distances, one can prove the combined result does. Indeed, the total energy density of the fields decays exponentially at long distances. The tension of a local string takes the simple form
\begin{equation}
\mu=\pi v^2B(\beta)\,,
\end{equation}
where $B(\beta)$ is a smooth function that takes the value $B(1)=1$~\cite{PhysRevD.37.263}. The exponential fall of the energy density at long distances means that, contrary to the global case, the interactions between strings are short-ranged, mediated by the two massive degrees of freedom.

\section{Lattice simulations of local strings}\label{sec:local:lattice}

Lattice simulations fully capture the field-theory structure of local strings, and  make it possible to investigate, on equal grounds, the emission of particles and GWs by the loops. We summarize the observables used to study the evolution and decay of string loops in \cref{sec:local:observables}. Then, \cref{sec:local:ICnetwork,sec:local:ICartificialI,sec:local:artificialtype2initialconditions} describe how the  loops we use in this study are generated. In particular, we consider three types of loops, which we call \textit{network loops} and \textit{artificial loops} of type \RNum{1} and \RNum{2}. As in the work presented in \cref{sec:global}, we study the decay of local string loops in Minkowski background, and also focus on the $\beta=1$ case. Program variables are defined as in the global case, using
\begin{equation}
\fstar=v\,,\quad\quad\quad\omegastar=\sqrt{\lambda}v\,,
\end{equation} 
and we recall that dimensionless variables are indicated with a tilde.
 %We also focus on the evolution and decay of loops in Minkowski background, as loop decay occurs at time scales much shorter than the expansion of the universe.

\subsection{Local-string observables}\label{sec:local:observables}

To study the dynamics and evolution of local strings, we use observables analogous to those presented in \cref{sec:global:observables} for the global case. We measure the location of the string cores by determining the pierced plaquettes, i.e., those with non-zero winding number. The winding number of a plaquette with lattice coordinate $\bm{n}$ on the $ij$-plane is computed similarly to the global case,
\begin{equation}\label{eq:local:windingplaquette}
W_{ij}(\bm{n})=\frac{1}{2\pi}\left[Y_i(\bm{n})+Y_j(\bm{n}+\hat{\bm{i}})-Y_i(\bm{n}+\hat{\bm{j}})-Y_j(\bm{n})\right]\,,
\end{equation}
where, in this case, the contribution from each link is defined in a gauge invariant way~\cite{Rajantie:1998vv,Hindmarsh:2017qff},
\begin{equation}\label{eq:local:windingnumberlink}
Y_{i}(\bm{n})=\left[e \delta x A_i(\bm{n}) +  \theta(\bm{n}) - \theta(\bm{n}+\hat{\bm{i}})\right]_\pi-e \delta x A_i(\bm{n})\,.
\end{equation}
Here, $\theta$ is the phase of the complex scalar field and $[\alpha]_\pi$ sets $-\pi<\alpha\leq \pi$. The location of the pierced plaquettes allows us to have real-time knowledge of the structure of the strings---for example, it allows us to determine if a single isolated loop if left in the simulation---, and to get an estimate of the string length in the lattice frame---see \cref{eq:global:lengthwindingdefinition}. %Also, algorithms exist to estimate the exact position of the string core within each plaquette, but we have not implemented them.

We also measure the string energy components. % and angular momentum. 
The kinetic, gradient, potential, electric and magnetic energy components of the strings are defined using a weight function, $W(\varphi)$. In the general case of an expanding background, working in conformal time, they take the values~\cite{Hindmarsh:2017qff}
\begin{equation}\label{eq:local:stringenergycomponentsweighteddefinition} 
\begin{array}{rlrl}
E_\text{K,str}&\displaystyle=a\displaystyle\int W(\varphi)|\varphi'|^2\d^3 x\,,\quad\quad\quad&
E_\text{G,str}&\displaystyle=a\displaystyle\sum_i\int W(\varphi)|D_i\varphi|^2\d^3 x\,,\\[10pt]
E_\text{V,str}&\displaystyle=a^3\displaystyle\int W(\varphi)V(\varphi)\d^3 x\,,\quad\quad\quad&
E_\text{E,str}&\displaystyle=\frac{a}{2}\displaystyle\int W(\varphi)\bm{E}^2\d^3 x\,,\\[10pt]
E_\text{B,str}&\displaystyle=\frac{a}{2}\displaystyle\int W(\varphi)\bm{B}^2\d^3 x\,,&&
\end{array}
\end{equation}
where $a$ is the scale factor, and the electric and magnetic fields are defined from the gauge field as $E_i=\dot{A_i}$ and $B_i=-\varepsilon_{ijk}F_{jk}/2$---see \cref{eq:Cosmo:electricmagneticfield}. 
From here, the total string energy is
\begin{equation}\label{eq:local:stringenergyweighteddefinition} 
E_\text{str}=E_\text{K,str}+E_\text{G,str}+E_\text{V,str}+E_\text{E,str}+E_\text{B,str}\,.
\end{equation}
%We  can also define the string angular momentum as~\cite{Matsunami:2019fss}
%\begin{equation}
%J_i=-\frac{1}{2}\varepsilon_{ijk}\int W(\varphi)\left\{x_j\left[\dot{\varphi}(D_k\varphi)^*+(\dot{\varphi})^*D_k \varphi\right]\varepsilon_{kml}E_lB_m\right\}\d^3x\,,
%\end{equation}
%where the electromagnetic contribution is related to the Poynting vector of the field, $\bm{E}\times\bm{B}$. %As in \cref{sec:global}, we use a weight function related to the potential of the fields, given in \cref{eq:global:weightfunction}.

A priori, there is no need to use a weight function in the case of local strings, as the total string energy is convergent. However, one wants to avoid including the energy of interstring radiation. In this work, we make use of the same weight function as used in the global case in \cref{sec:global},
\begin{equation}\label{eq:local:weightfunction}
W(\varphi)=\frac{4 V(\phi)}{\lambda v^4} \Theta\left(\frac{v^2}{2}-|\varphi|^2\right)\,,
\end{equation}
where $\Theta$ is the Heaviside function. A different weight function related to the Lagrangian has also been used in other works---see \rrcite{Hindmarsh:2017qff,Hindmarsh:2021mnl}. For such choice, however, there is no well-defined normalization of the weight function, and only the ratios between energy components have a physical meaning.

The measurements of the string energy components make it possible to obtain different estimates of the proper comoving length and mean-square velocity of the strings. For example, we can define~\cite{Hindmarsh:2017qff},
\begin{equation}
L_\text{str}=\frac{1}{a}\frac{E_\str-\Delta f E_\text{L,str}}{\mu(1+\Delta f)}\,,
\end{equation}
\begin{equation}
v^2=\frac{E_\str+E_\text{L,str}}{E_\str - \Delta f E_\text{L,str}}\,,
\end{equation}
where $E_\text{L,str}=E_\text{K,str}+E_\text{E,str}-E_\text{G,str}-E_\text{B,str}-E_\text{V,str}$ is the weighted Lagrangian energy.
Other estimators of the two quantities are also available---see \rcite{Hindmarsh:2017qff}. Here $\mu$ is the weighted tension of the static string and $\Delta f=f_B-f_V$ is the difference between the fraction of magnetic energy and the fraction of potential energy to the total energy of the static string. These are computed from the NO solution as
\begin{equation}
\begin{array}{rl}
\displaystyle\mu&\displaystyle=a^2\int W(\varphi)\left[\rho_\text{G}(r)+\rho_\text{B}(r)+V(r)\right]\d^2 x\,,\\[10pt]
\displaystyle f_\text{B} &\displaystyle=\frac{a^2}{\mu}\int W(\varphi)\rho_\text{B}(r)\d^2 x\,,\\[10pt]
\displaystyle f_\text{V} &\displaystyle=\frac{a^2}{\mu}\int W(\varphi)V(r)\d^2 x\,,
\end{array}
\end{equation}
where the components of the energy density of the NO vortex are defined in \cref{eq:local:NOenergydensity}. Using the weight function in \cref{eq:local:weightfunction}, we obtain $\mu=1.4415v^2$, $f_\text{B}=0.2047$ and $f_V=0.2056$.

\subsection{Generation of network loops}\label{sec:local:ICnetwork}

Network loops are generated from the decay of string networks in a similar fashion to the global case, outlined in \cref{sec:global:initialconditionnetworks}, and following the procedure in \rcite{Hindmarsh:2021mnl}. They are expected to have shapes and features similar to those loops that could form after a realistic phase transition in the early universe. 

Simulations are initialized with a Gaussian random realization of the scalar field in Fourier space, with the same power spectrum used for global network loops,
\begin{equation}
\Delta_{\phi_i}(k)=\frac{k^3 v^2 \ell_{\str}^3}{\sqrt{2\pi}}\exp\left(-\frac{1}{2}k^2\ell_{\str}^2\right)\,.
\end{equation}
Recall, this depends on a correlation length, $\ell_\str$, that controls the density of the resulting network. The gauge field and the time derivatives of both fields, on the other hand, are set to zero. 

The field configuration resulting from the previous step is too energetic and contains no gauge field. To get rid of the excess energy and allow the magnetic flux to form inside the strings, we evolve the configuration following diffusion equations of the form,
\begin{equation}\label{eq:local:eomfieldsdiffusion}
\begin{array}{rl}
\sqrt{\lambda} v \varphi^\prime-\partial_i\partial_i\varphi&=\displaystyle -2\lambda\left(|\varphi|^2-\frac{v^2}{2}\right)\varphi\,,\\[10pt]
\sqrt{\lambda} v F_{0i}-\partial_jF_{ji} & = 2e\Im\left[\varphi^*D_i\varphi\right]\,.
\end{array}
\end{equation}
The diffusive phase is applied for a total of $20$ units of program time, which we find to be enough for our purposes. %Note that Gauss' law is not strictly conserved during diffusion, but we empirically find that, by the end of this diffusive process, Gauss' law is still conserved almost at machine precision..

After the diffusion process, we let the network evolve in a radiation-dominated (RD) background, with scale factor $a(\tau)=\tau/\tau_0$, where $\tau$ indicates the conformal time and $\tau_0=70/\sqrt{\lambda} v$ in our simulations. %This ensures that the network comes close to the scaling regime.
 While it would be possible to obtain analogous results working in Minkowski background, as done in \rcite{Hindmarsh:2021mnl}, evolving the network in  RD dissipates some of the energy radiated from its decay. Moreover, we find the networks to decay slightly faster in an expanding background compared to a flat one.

As in the case of global strings, evolving local strings in an expanding background leads to a loss of resolution of the string core. To prevent this from happening, we perform an initial phase of extra-fattening, in which the fields are evolved with equations of motion,
\begin{equation}\label{eq:local:eomextrafattening}
\begin{array}{rl}
\displaystyle\varphi^{\prime\prime}+2\frac{a^\prime}{a}\varphi^\prime-D_i D_i\varphi & =\displaystyle -2a^{-2}\lambda\left(|\varphi|^2-\frac{v^2}{2}\right) \varphi\,,\\[10pt]
\displaystyle F_{0i}^\prime+4\frac{a^\prime}{a}F_{0i}-a^{-4}\partial_j F_{ji} & \displaystyle= 2a^{-2}e\Im\left[\varphi^*D_i\varphi\right]\,.
\end{array}
\end{equation}
This phase is set to last for a total of  $\Delta\tau_\ef=\sqrt{\tau_0(\Delta\tau_\HL+\tau_0)}$, where $\Delta\tau_\HL=L/2$ is the half-box-light-crossing time of the lattice. After this, the fields are evolved normally in a RD background, with equations of motion
\begin{equation}\label{eq:global:eomfields}
\begin{array}{rl}
\displaystyle\varphi^{\prime\prime}+2\frac{a^\prime}{a}\varphi^\prime-D_i D_i\varphi & \displaystyle= -2a^2\lambda\left(|\varphi|^2-\frac{v^2}{2}\right)\varphi\,,\\[10pt]
\displaystyle F_{0i}^\prime-\partial_j F_{ji} & \displaystyle= 2a^2e\Im[\varphi^*D_i\varphi]\,.
\end{array}
\end{equation}
for an additional time $\Delta \tau_\text{RD}=\Delta\tau_\HL-\Delta\tau_\ef$. The initial extra-fattening phase ensures that, after this additional time, the width of the string is equal to that at the end of diffusion.

We note that the extra-fattening evolution can be regarded, after redefining $A_\mu'= A_\mu/e$, as promoting the couplings to time-dependent variables, $\lambda\rightarrow a^{-4}\lambda$ and $e^2\rightarrow a^{-4} e^2$. This implies the mass of both massive modes decreases with time, so that string width grows. Naively substituting the couplings without redefining $\A_\mu$ would instead break gauge invariance of the action. Note that Gauss' law is conserved during the extra-fattening, but needs to be redefined accordingly,
\begin{equation}
\partial_i F_{0i} =2a^{-2}e\Im[\varphi^*\dot{\varphi}]\,.
\end{equation} 

\begin{figure}[!t]
    \centering
    \begin{minipage}{0.495\textwidth} 
    \centering
        \includegraphics[width=\textwidth]{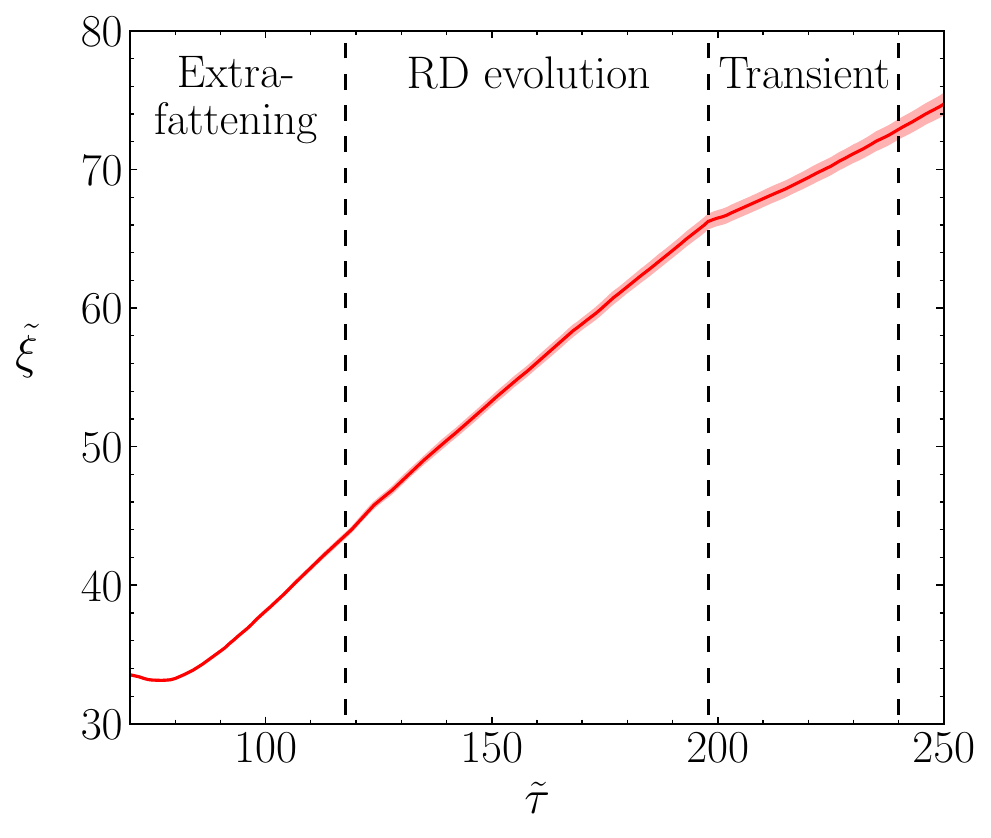}
    \end{minipage}
    \begin{minipage}{0.495\textwidth} 
    \centering
        \includegraphics[width=\textwidth]{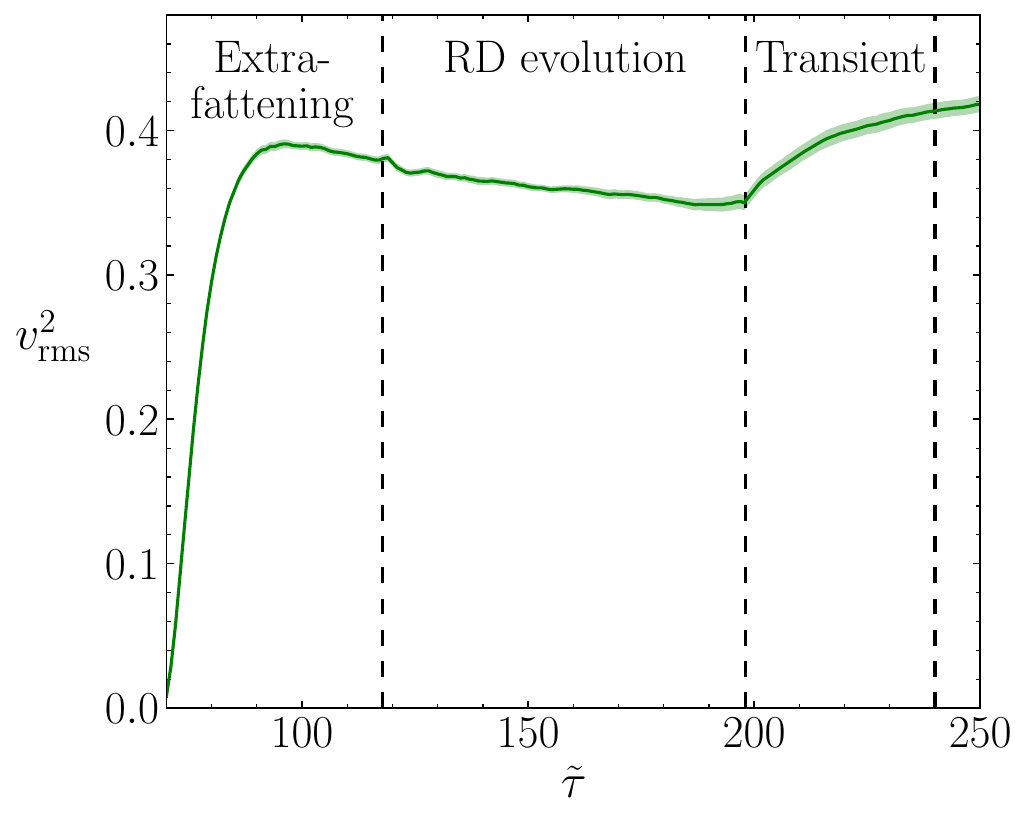}
    \end{minipage}
    
    \caption{
        Time evolution of the mean string separation and the mean-squared velocity of a network of local strings averaged over 20 independent realizations. Networks are generated with $\tilde{\ell}_\str=15$ and simulations are performed with $\tilde{L}=256$ and $\delta\tilde{x}=0.25$. The vertical lines corresponding to the end of the extra-fattening, RD evolution and the transient phases, as indicated. The time coordinate corresponds to comoving time during the first two phases, while it is the standard time coordinate in Minkowski spacetime. Bands indicate one standard deviation.}
    \label{fig:local:networkscaling}
\end{figure}

By the end of the evolution in a RD background, the string network is close to the scaling regime, with the comoving mean string separation, $\xi=(L^3/L_\text{str})^{1/2}$, linearly increasing, $\xi\propto\tau$, and an almost constant mean square velocity, $v_{\str}^2$. In \cref{fig:local:networkscaling} we represent the time evolution of these two quantities from the end of diffusion, averaged over 20 network realizations simulated with $\tilde{L}=256$, $\delta\tilde{x}=0.25$ and $\tilde{\ell}_{\str}=15$, for which the end of the evolution in RD corresponds to $\tilde{\tau}=198$.  %We believe this is related to the absence of long-range interactions. 

In most cases, however, the network has not yet decayed into a single loop at the end of the evolution in RD. We subsequently evolve the resulting network in a Minkowski background, for a maximum time of $2\Delta\tau_\HL$, which we found enough to typically have one isolated string remaining. Note this is much longer that the period $\Delta \tau_\HL/2$ used in the case of global networks, since we find local networks to decay slower.

Note that, when changing from RD to a Minkowski background, the network undergoes a transition period to adapt to the new background. This is clearly visible in \cref{fig:local:networkscaling}, where after $\tilde{\tau}=198$ the evolution of the mean square separation and the mean squared velocity changes from the previous regime. This is related to the fact that the characterisitcs of the scaling regime depend on the background metric.  After changing to Minkowski, we wait for $\Delta\tau_\HL/2$ after analyzing any loops, which we call a \textit{transient phase}. We believe this minimizes the impact of the sudden change of background. Even if a loop forms during this phase, we do not study its dynamics and GW emission until a time $\Delta\tau_\HL/2$ has passed from the end of RD evolution.  

After the transient period and if an isolated loop is found before a time $2\Delta\tau_\HL$ has passed since the end of the evolution in RD, we turn on the emission of GWs and study the evolution of the loop until it disappears. Approximately, $\sim 20\%$ of our simulations lead to isolated loops, of which $\sim 80\%$ can be used for our study. We discard those loops that self intersect forming several loops of similar size or infinite strings. Altogether, only $\sim16\%$ of the simulations are suitable for this study. %We find that the fraction of loops that self-intersect forming infinite string is much smaller than in the global case, possibly due to the lack of long-range attraction.

\subsection{Generation of artificial loops of type \RNum{1}}\label{sec:local:ICartificialI}

Local artificial loops of type $\RNum{1}$ are generated from the intersection of two pairs of parallel infinite boosted strings. We follow the procedure used in \rcite{Matsunami:2019fss}, similar in spirit to that presented for global artificial  loops in \cref{sec:global:initialconditionsartificial}. Note, however, that we do not isolate local strings loops, as was done in the global case. We have found that a naive generalization of the technique presented in \cref{sec:global:initialconditionsartificial} breaks Gauss' law. Instead, we choose the initial configuration so that one of the two loops resulting from the intersection of infinite strings is much larger than the other one, and wait for the smaller one to decay before starting our study of the longer loop.  We now describe the initialization procedure in detail. As in \cref{sec:global:initialconditionsartificial}, we consider one pair parallel to the $z$-axis and the other parallel to the $x$-axis, and we refer to each of them with subscripts ``1'' and ``2'', respectively, which should not be confused with the component index of the gauge field, $\mu=0,1,2,3$.

We first explain how the pair of strings parallel to the $z$-axis is generated. The starting point is the solution for the NO vortex in the temporal gauge, given in \cref{eq:local:NOvortexansatzsolution}. We refer to this solution as $\varphi^{(k)}_\NO$ and $A^{(k)}_{\NO,\mu}$, where $k=\pm 1$ indicates the winding number of the string. %We use $\mu\in\{t,x,y,z\}$ to denote the components of the gauge field thourgh this section to avoid confusion with the labels of each pair. 

The static NO configuration can be boosted in the $(x,y)$-plane with velocity $\bm{v}_1=v_1(\sin\alpha_1,\cos\alpha_1)$, resulting in
\begin{equation}
\begin{array}{rl}
\bar{\varphi}_{\bm{v}_1}^{(\pm)}(x,y;t)&=\varphi_\NO^{(\pm)}(x^\prime,y^\prime)\,,\\[5pt]
\bar{A}_{\bm{v}_1,0}^{(\pm)}(x,y;t)&=-\gamma_1 s_1 v_1 A_{\NO,1}^{(\pm)}(x^\prime,y^\prime)-\gamma_1 c_1 v_1 A_{\NO,2}^{(\pm)}(x^\prime,y^\prime)\,,\\[5pt]
\bar{A}_{\bm{v}_1,1}^{(\pm)}(x,y;t)&=[1+(\gamma_1-1)s_1^2] A_{\NO,1}^{(\pm)}(x^\prime,y^\prime)+(\gamma_1-1)s_1c_1 A_{\NO,2}^{(\pm)}(x^\prime,y^\prime)\,,\\[5pt]
\bar{A}_{\bm{v}_1,2}^{(\pm)}(x,y;t)&=(\gamma_1-1)s_1c_1 A_{\NO,1}^{(\pm)}(x^\prime,y^\prime)+[1+(\gamma_1-1)c_1^2] A_{\NO,2}^{(\pm)}(x^\prime,y^\prime)\,,\vspace{0.3cm}
\end{array}
\end{equation}
where we define $s_1=\sin\alpha_1$, $c_1=\cos\alpha_1$ and $\gamma_1=(1-v_1^2)^{1/2}$. Here $(x^\prime,y^\prime)$ are the coordinates in the rest frame of the string and $(t,x,y)$ are the coordinates in the boosted frame, related by

\noindent\begin{equation}\label{eq:local:relativisticboost}
\begin{array}{rl}
   x'&=-\gamma_1 v_1 s_1 t + [1+(\gamma_1-1)s_1^2] x+(\gamma_1-1)s_1c_1y\,,\\
   y'&= -\gamma_1 v_1 c_1 t + (\gamma_1-1)s_1c_1x+[1+(\gamma_1-1)c_1^2]y\,.
\end{array}
\end{equation}

The relativistic boost produces an undesired time component of the gauge field. To go back to the temporal gauge, we perform a gauge transformation, 
\begin{equation}
\varphi = \text{e}^{i\xi}\bar{\varphi}\,,\quad\quad\quad A_\mu=\bar{A}_\mu-\partial_\mu\xi\,,
\end{equation}
where $\xi$ is a function chosen so that $A_0=0$ in the boosted frame,
\begin{equation}
\dot{\xi} = \bar{A}_0\longrightarrow \xi=\int_0^t A_0\d t\,.
\end{equation}
As we evaluate the initial configuration at $t=0$, we can set $\xi=0$. However, $\dot{\xi}=\bar{A}_0$, which we need to take into account to compute the time derivatives of the fields,
\begin{equation}\label{eq:local:gaugetransformationderivative}
\begin{array}{rl}
\dot{\varphi}&=\dot{\bar{\varphi}}-ie\bar{A}_0\bar{\varphi}\,,\\
\dot{A}_i&=\dot{\bar{A}}_i-\partial_i\bar{A}_0\,.
\end{array}
\end{equation}

The product ansatz can then be used to generate a pair of parallel boosted strings. The complex fields for both strings, evaluated at $t=0$, are multiplied, while the gauge fields are summed,
\begin{multline}\label{eq:local:productansatzsinglepair}
    \varphi_1(x,y;t)=\displaystyle\frac{1}{v}\varphi_{{\bm v}_1}^{(+)}\left[x-\left(\frac{L}{2}+a_1\right),y-\left(\frac{L}{2}+b_1\right);t\right] \\
     \displaystyle \times \varphi_{-{\bm v}_1}^{(-)}\left[x-\left(\frac{L}{2}-a_1\right),y-\left(\frac{L}{2}-b_1\right);t\right]\,,
\end{multline}
\begin{multline}
    A_{1, \mu}(x,y;t) =\displaystyle A_{{\bm v}_1,\mu}^{(+)}\left[x-\left(\frac{L}{2}+a_1\right),y-\left(\frac{L}{2}+b_1\right);t\right] \\
     \displaystyle + A_{-{\bm v}_1,\mu}^{(-)}\left[x-\left(\frac{L}{2}-a_1\right),y-\left(\frac{L}{2}-b_1\right);t\right]\,,
\end{multline}
where $a_1$ and $b_1$ indicate the distance of each string to the center of the lattice, in analogy to the variables of the same name defined in the local case---see \cref{fig:global:initialpair}. 
The corresponding time derivatives are straightforward to evaluate. %In this work we restrict ourselves to symmetric configurations, where both strings in the pair have the same velocity and displacement in opposite directions.

The resulting configuration is then modified to fit in a periodic lattice, using a similar approach that that in \rcite{Matsunami:2019fss}. We do not modify the gauge field, as we observe its long-distance energy contribution to be negligible. The scalar field approaches the vacuum exponentially fast far from the string, and we only need to change its phase,
\begin{equation}\label{eq:local:hperiodicitymodification}
\varphi_1=|\varphi_1|\text{e}^{i\theta_1}\rightarrow \varphi_1^\per =|\varphi_1|\text{e}^{ih(x,y)\theta_1}\,,
\end{equation}
which also affects the time-derivative of the field. The filter function $h(x,y)$ is chosen so that the phase changes smoothly close to the boundary towards zero. We opt to use
\begin{equation}\label{eq:local:filterfunction}
h(x,y)=\left\{
\begin{array}{ll}
\displaystyle\frac{L/2-|x_L|}{L/2-L_h}\,, &\quad\quad\quad |x_L|> L_h\text{ and }|x_L|\geq|y_L|\,,\\[10pt]
\displaystyle\frac{L/2-|y_L|}{L/2-L_h}\,, &\quad\quad\quad |y_L|> L_h\text{ and }|x_L|<|y_L|\,,\\[5pt]
1\,, & \quad\quad\quad\text{otherwise}\,,
\end{array}
\right.
\end{equation} 
where we denote $x_L=x-L/2$ and $y_L=y-L/2$.
This differs from the choice in \rcite{Matsunami:2019fss}, which we find leaves some residual energy close to the $(x=0,L,y=L/2)$ boundaries that leads to instabilities at late times in the simulations. We use $\tilde{L}_h=\tilde{L}/2-16$ in our simulations, independently of the size of the lattice. We observe that varying $L_h$ up to a factor of four has a negligible effect on the final results. %Using the choice in \rcite{} leads to instabilities in the observables, but the effect on the lifetime of the loops is less that $10\%$.

Finally, we use again the product ansatz on two perpendicular string pairs and generate the initial conditions for our simulations,
\begin{equation}
\begin{array}{rl}
\varphi(x,y,z) &= \varphi_1^\per(x,y;t=0) \times \varphi_2^\per(z,y;t=0)\,,\\  
A_\mu(x,y,z) &= A_{1,\mu}(x,y;t=0) + A_{2,\mu}(z,y;t=0)\,.  
\end{array}
\end{equation}
The time derivatives of the fields are computed by successive differentiation, taking into account the gauge transformations in \cref{eq:local:gaugetransformationderivative} and the use of the filter function in \cref{eq:local:hperiodicitymodification}.

In this work, we consider different boost velocities for the two pairs, $v_1\neq v_2$ and set $\alpha_1=-\alpha_2=\alpha$, as we observe this leads to longer-lived strings. More concretely, we have found that other choices of the boost direction, such as $\alpha_1=\alpha_2$, lead rapidly to a \textit{double-line collapse} event, this is, a string configuration in which two antiparallel segments of the string approach each other, completely annihilating. We also consider $b_1,b_2\ll L$, so that the strings intersect rapidly after the start of the simulation. Finally, we set $a_1=a_2$ to be a small fraction of the box size, which we justify below. 

The initial configuration is evolved in a Minkoski background using \cref{eq:local:eomlocalMinkoski}, and the four strings soon intersect forming two loops. Taking advantage of the absence of long-range interactions between the strings, we opt to set $a_1=a_2$ small compared to the box size, so that the inner loop is much smaller than the outer one and collapses rapidly after the start of the simulation. After this happens, we start to measure the emission of particles and GWs from the loop.

The reason why we focus on the outer loop, rather than the inner one, is to keep the initial infinite strings in each pair far from the region modified by the filter in \cref{eq:local:filterfunction}. Note some parts of the strings of each pair still lie on top of the modified region of the opposite pair. The size of this region depends on the choice of $L_h$ and, as discussed above, we observe no effect on the dynamics of the loops from changing this parameter. Thus, we believe the effect of the chosen filter function on the loop to be negligible

\subsection{Generation of artificial loops of type \RNum{2}}\label{sec:local:artificialtype2initialconditions}

Artificial loops of type \RNum{2} are generated following the procedure introduced in \rrcite{Hindmarsh:2017qff,Hindmarsh:2021mnl}. This is based on initializing static strings of arbitrary shape by setting some magnetic flux on the plaquettes pierced by the initial strings. We note that this techniques relies on the use the hybrid or the compact formulation of the gauge theory, introduced in \cref{sec:Cosmo:latticemodel}. For our study, we use the hybrid formulation.

\begin{figure}[!b]
    \centering
    \begin{minipage}{\textwidth} 
    \centering
        \includegraphics[width=\textwidth]{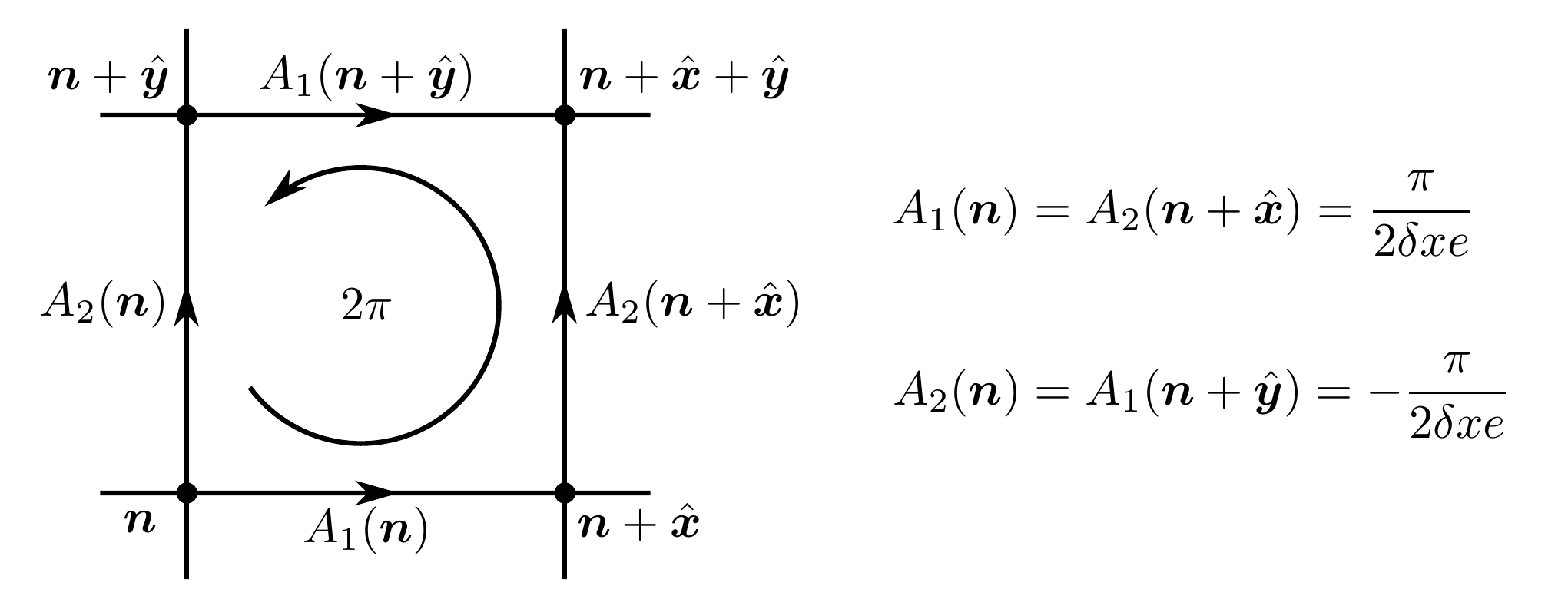}
    \end{minipage}
    
    \caption{
        Representation of a single plaquette in the $(x,y)$-plane pierced by a string going in the $z$ diraction (out of the paper). We indicate the initial values to which the different links are set. The scalar field is fixed to $\varphi=v$ everywhere. }
    \label{fig:local:singleplaquette}
\end{figure}

Given some shape of the desired one-dimensional strings, one can determine a set of plaquettes pierced by the string. The idea is to set the magnetic flux through these plaquettes to $\pm2\pi$, depending on the direction from which it is pierced by the string. This is achieved by setting the gauge variables on the links to
\begin{equation}
A_\mu(\bm{n})=\pm\frac{\pi}{2\delta x e}\,,
\end{equation}
where the sign depends on the orientation of link. The complex scalar field is set to be equal to the vacuum expectation value, $\varphi=v$, everywhere. A diagramatic representation for one plaquette is presented in \cref{fig:local:singleplaquette}, pierced by a string going out of the plane of the paper.

The initial configuration is then diffused for five units of program time, using \cref{eq:local:eomfieldsdiffusion}, which leads to the formation of strings with the expected radius, $r_\text{c}\sim m_\chi^{-1}$. Note that in the hybrid formulation the $\partial_j F_{ji}$ term in \cref{eq:local:eomfieldsdiffusion} needs to be discretized using the plaquette, 
\begin{equation}\label{eq:local:discretizationhybrid}
\left.\partial_j F_{ji}\right|_\text{hybrid}=\frac{1}{\delta x^2 e}\Delta_j^+ \sin\left[\delta x e \,\cC_{ji}(\bm{n})\right]\,,
\end{equation}
where $\Delta^+_j$ denotes the forward finite difference---see \cref{eq:Cosmo:finitedifferences}---and $\cC_{ij}(\bm{n})$ is the circulation of the gauge field on the plaquette,
\begin{equation}
\cC_{ij}(\bm{n})=\left[A_i(\bm{n})+A_j(\bm{n}+\hat{\bm{i}})-A_i(\bm{n}+\hat{\bm{j}})-A_j(\bm{n})\right]\,.
\end{equation}
The remaining terms in the diffusion equation can be naively discretized.

\begin{figure}[!h]
\centering
	\begin{subfigure}{0.45\textwidth} 
    \centering
        \includegraphics[width=1\textwidth]{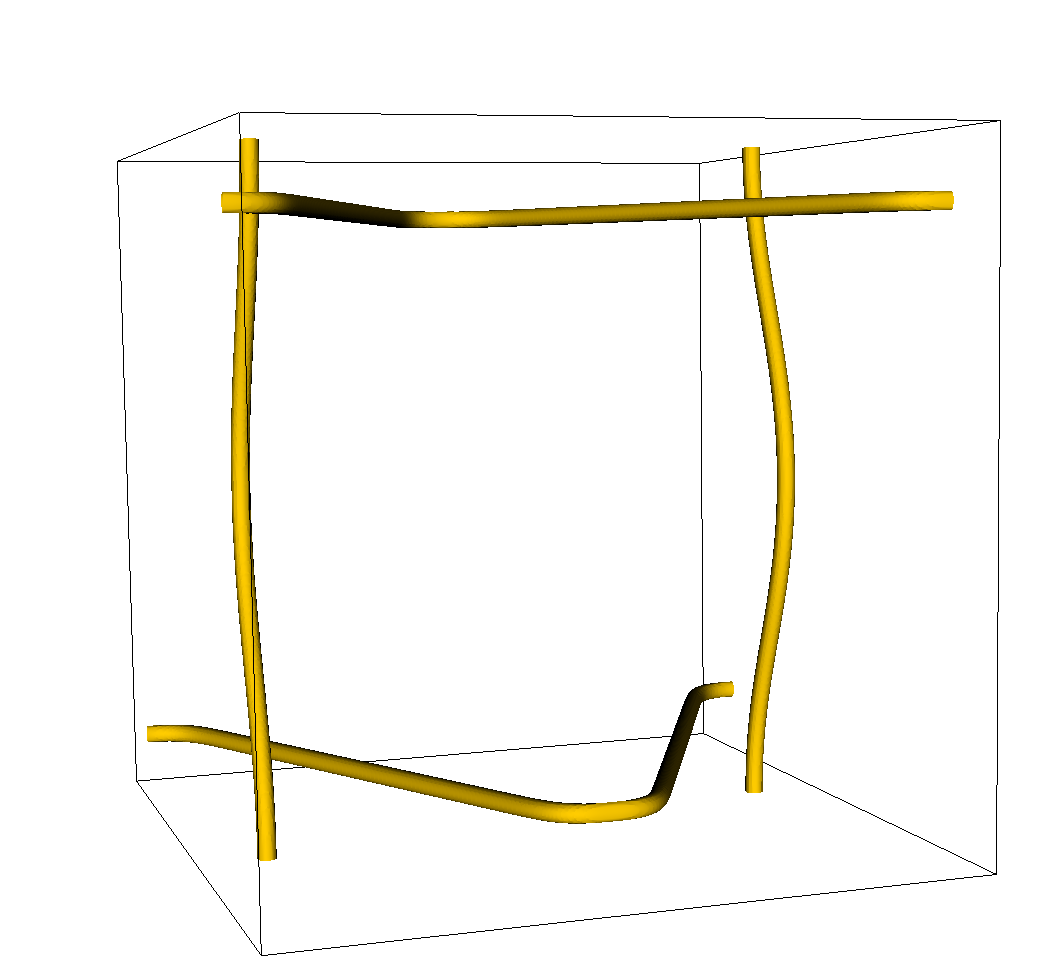}
        \caption{End of diffusion.}
        \label{fig:local:windingloopisolated1}
    \end{subfigure}\hspace{0.5cm}
    \begin{subfigure}{0.45\textwidth}
    \centering
       \includegraphics[width=1\textwidth]{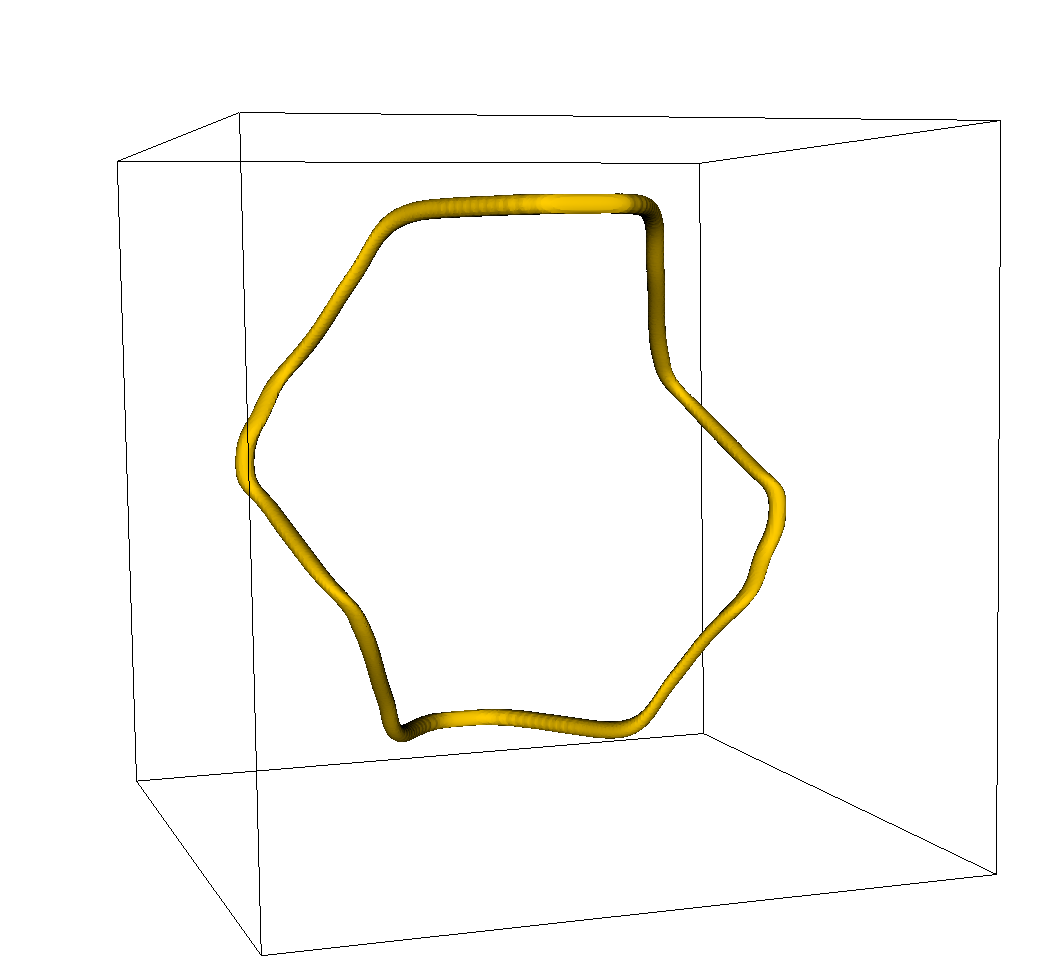}
        \caption{Instant after the outer loop disappears.}
        \label{fig:local:windingloopisolated2}
    \end{subfigure}

    \caption{
        Three-dimensional snapshots of $|\varphi|^2=0.2v^2$ surfaces of the simulation of an artificial loop of type \RNum{2}.}%, corresponding to the time at the end of diffusion (left) and the moment in which the outer loop completely disappears (right).}
    \label{fig:local:windingloopisolated}
\end{figure}

For our study, we use an initial configuration composed by four non-straight static strings, following \rcite{Hindmarsh:2021mnl}, which intersect soon after the start of the simulation forming two loops. We consider two strings at $y=L/10$ and $y=9L/10$ with sinusoidal form. The coordinates of the string core are given by
\begin{equation}
x=\pm A \cos(2\pi z/L)\,,
\end{equation}
with each sign corresponding to a different value of $y$. Also, we set $A=0.075L$. The other two strings have fixed $z$ with a sawtooth form,
\begin{equation}
x=\left\{\begin{array}{ll}
\displaystyle\pm B\left[\frac{y}{L/4}-1\right]\,,\quad\quad\quad & 0 \leq y \leq L/2\,,\\[10pt]
\displaystyle\mp B\left[\frac{y}{L/4}-3\right]\,,\quad\quad\quad & L/2 < y < L\,,
\end{array}\right.
\end{equation}
with $z=L/10$ and $z=9L/10$, for each string, and $B=L/2$. The signs, again, corresponds to each possible value of $z$. A representation of the resulting configuration, at the end the diffusion period,  is shown in  \cref{fig:local:windingloopisolated1}. 

After the diffusive phase, the strings are let to evolve in Minkoski spacetime. This is performed using a discretized version of  \cref{eq:local:eomlocalMinkoski} consistent with the hybrid formulation, using the discretization for the $\partial_jF_{ji}$ term in \cref{eq:local:discretizationhybrid}. The four infinite strings start to move as a result of their non-straight form, and they eventually intersect, forming two loops. Due to the initial position of the infinite strings, the outer one is much smaller than the inner one. Thus, we wait until the outer loop disappears and study the inner one afterwards. An example of the resulting isolated loop is presented in \cref{fig:local:windingloopisolated2}.

A drawback of this initialization method is that there remains a magnetic flux frozen on the  initialized plaquettes. When measuring the winding number on these plaquettes using \cref{eq:local:windingnumberlink}, one finds a non-zero results, which is  associated to a \textit{ghost} string~\cite{Hindmarsh:2017qff}. This is an unphysical consequence of this particular procedure, since these ghost strings neither contain energy, nor affect the dynamics of the physical strings. When measuring the length of artificial strings of type \RNum{2} from the number of pierced plaquettes, we subtract the number of initialized plaquettes. For our loops, the number of plaquettes belonging simultaneously  to both the real and the ghost string is negligible compared to the total number of plaquettes, and so this does not affect our ability to measure the length of the loops.

%The idea behind this procedure is the particular form of the discretized action used in the hybrid formulation. Setting the flux on some plaquettes to $2\pi$ discretization We note that this procedure only works if using the hybrid or compact formulation. We recall 

\section{Results}\label{sec:local:results}

We now present our results on the emission of particles and GWs from local string loops. In \cref{sec:local:particleproductionnetwork,sec:local:particleproductionartificial} we characterize the lifetime of  network and artificial loops, respectively, as a function of their length and energy, and determine the emission power of particles.  Then, \cref{sec:local:GWresults} presents results on the GW emission form network and artificial loops of type \RNum{1}. We recall that we are neglecting backreaction of the GWs on the matter fields, an assumption that we check self-consistently later.

\subsection{Particle emission from network loops}\label{sec:local:particleproductionnetwork}

We study the evolution of network loops and characterize their lifetime, $\Delta t_\dec$, as a function of their initial length, $L_0$, and energy, $E_\text{str,0}$. In total, we study 41 different network loops with length-to-width ratios up to $L_0/r_\text{c}\lesssim 3500$, using $\delta\tilde{x}=0.25$ in our simulations, as we justify later. %, as well as 14 artificial loops of type \RNum{1} and six of type \RNum{2}, with lengths $L_0/r_\text{c}\lesssim640$.

\begin{figure}[!b]
    \centering
    \begin{subfigure}{0.495\textwidth} 
    \centering
    
        \includegraphics[width=\textwidth]{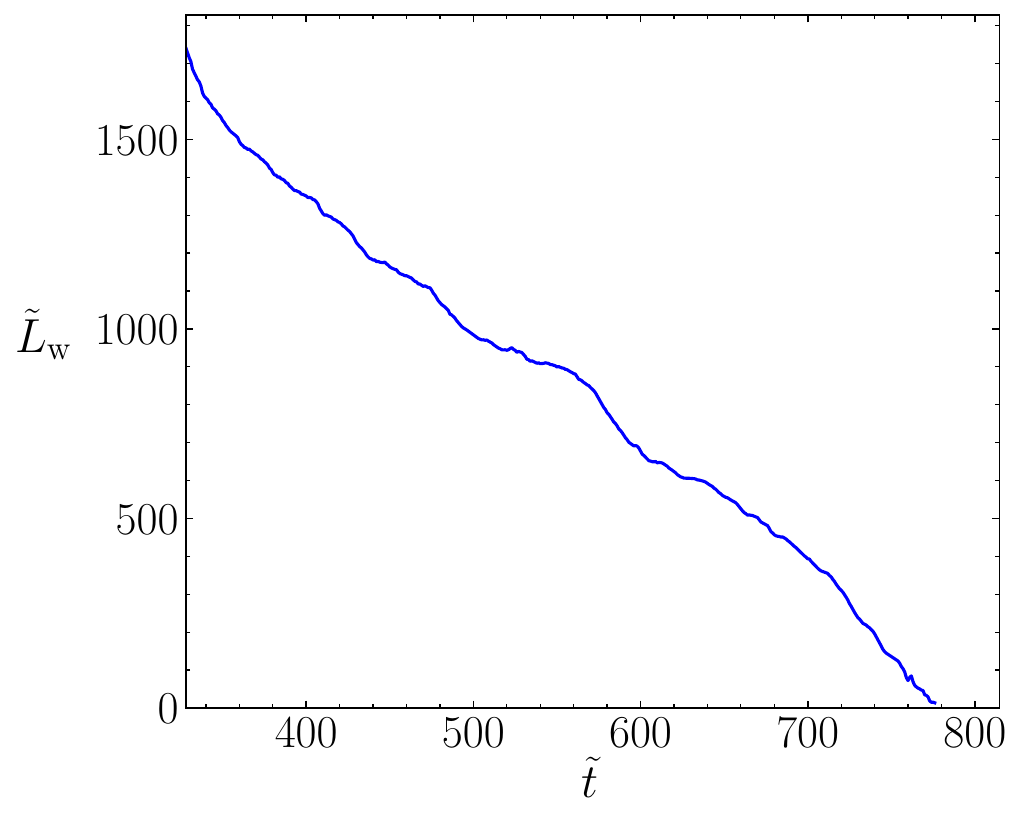}
        \caption{Network loop.}\label{fig:local:exampledecaynetworkloop}
    \end{subfigure}
    \begin{subfigure}{0.495\textwidth} 
    \centering
    \includegraphics[width=\textwidth]{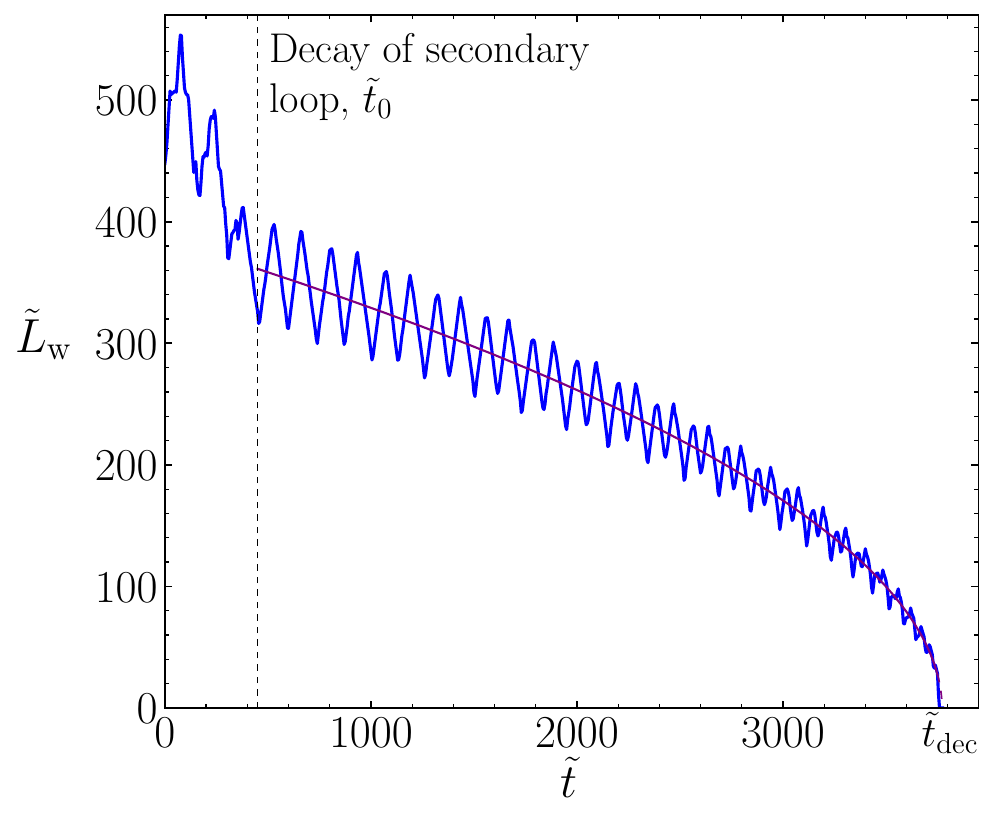}
        
        \begin{minipage}{0.9\textwidth}
    \caption{Artificial loop of type \RNum{1} for $v_1=v_2=0.6$.}\label{fig:local:exampledecayartificialloops1}
    \end{minipage}
    \end{subfigure}\vspace{0.5cm}
    
    \begin{subfigure}{0.495\textwidth} 
    \centering
    \includegraphics[width=\textwidth]{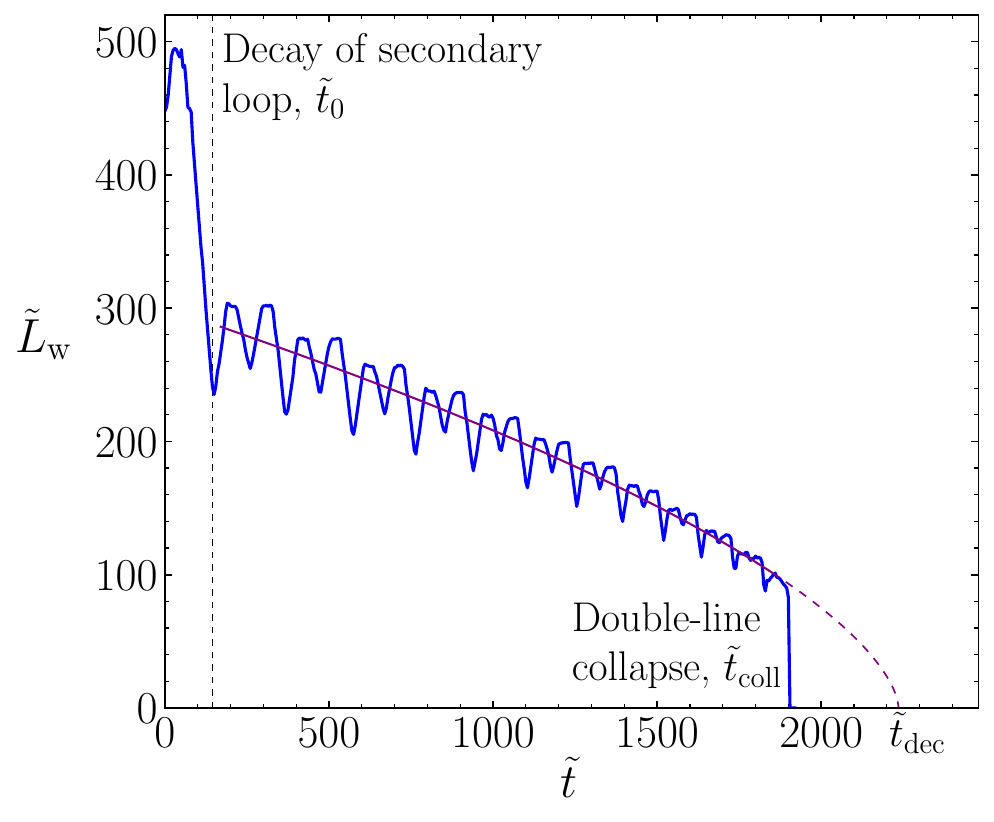}
        
        \begin{minipage}{0.9\textwidth}
    \caption{Artificial loop of type \RNum{1} for $v_1=0.3$ and $v_2=0.6$.}	\label{fig:local:exampledecayartificialloops3}
    \end{minipage}				
    
    \end{subfigure}
    \begin{subfigure}{0.495\textwidth} 
    \centering
    \includegraphics[width=\textwidth]{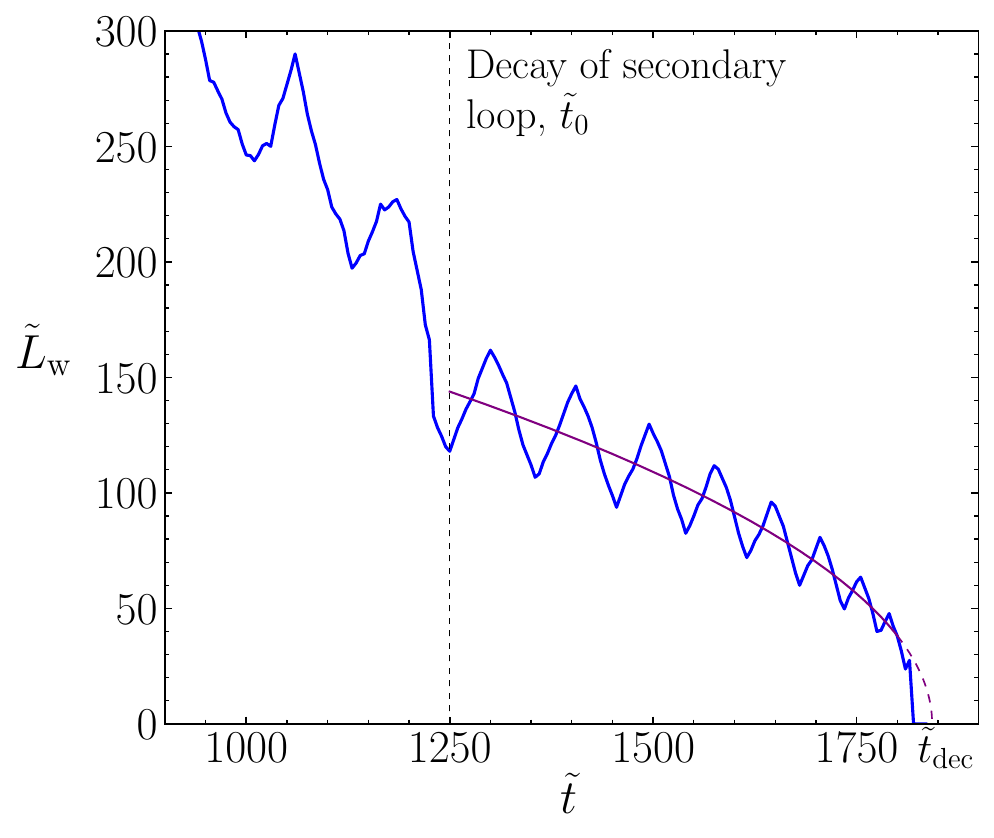}
    \begin{minipage}{0.9\textwidth}
    \caption{Artificial loop of type \RNum{2}.}\label{fig:local:exampledecayartificialloops2}
    \end{minipage}
        
    \end{subfigure}

    \caption{Example of the time evolution of the length of a network loop (top left) and artificial loops of type \RNum{1} (top right and bottom left) and type \RNum{2} (bottom right), measured from the number of pierced plaquettes. In the case of artificial loops, we indicate the time at which the secondary loop decays with a vertical dashed line. For these loops we also show the result from a fit to \cref{eq:local:artificiallengthfit}. This is used to estimate the decay time in the absence of double-line collapse, which is observed to occur in the last stages of the loop evolution for some initial configurations, as shown, for example, in the bottom left panel.
        }
    \label{fig:local:exampledecay}
\end{figure}

Network loops behave very similarly to their global counterparts, with their length decreasing almost linearly with time. An example of the time evolution of the length of a network loop, generated in a lattice with $\tilde{L}=256$ using $\tilde{\ell}_\varphi=15$, is shown in \cref{fig:local:exampledecaynetworkloop}. In general, network loops do not oscillate, having a lifetime smaller of their NG period, $T_{\NG}=L_\str/2$.

\begin{figure}[!t]
    \centering
    
    \begin{subfigure}{0.495\textwidth} 
    \centering
        \includegraphics[width=\textwidth]{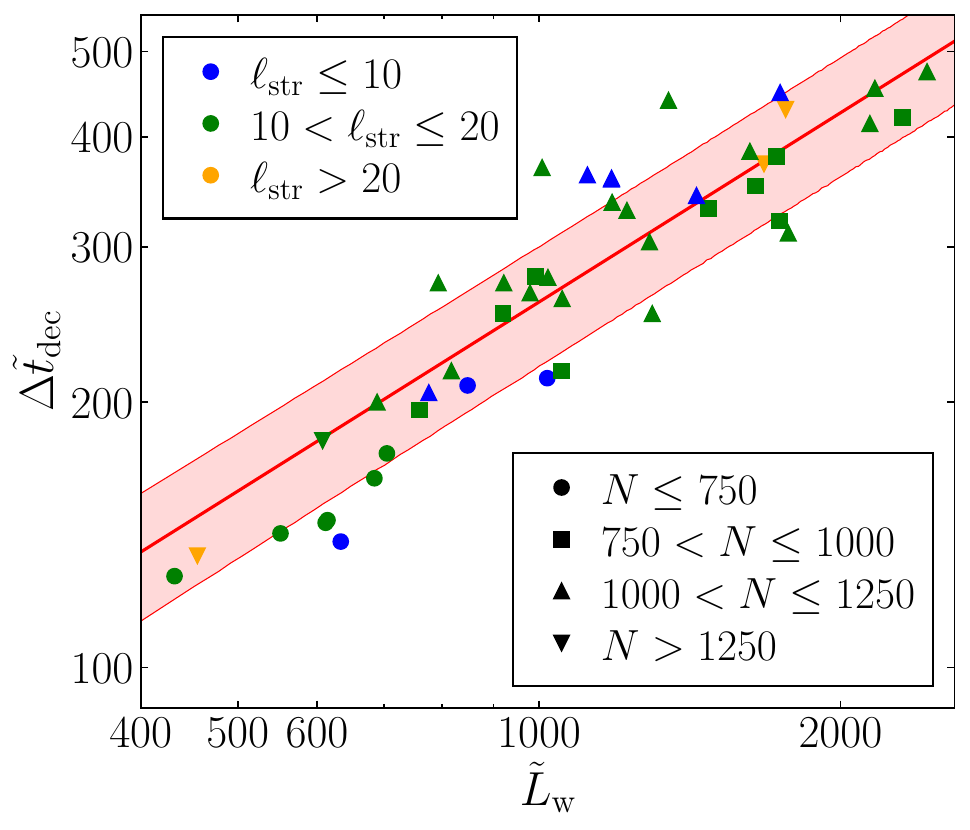}
        \caption{Power-law fit}\label{fig:local:decaynetwork1}
    \end{subfigure}
    \begin{subfigure}{0.495\textwidth} 
    \centering
        \includegraphics[width=\textwidth]{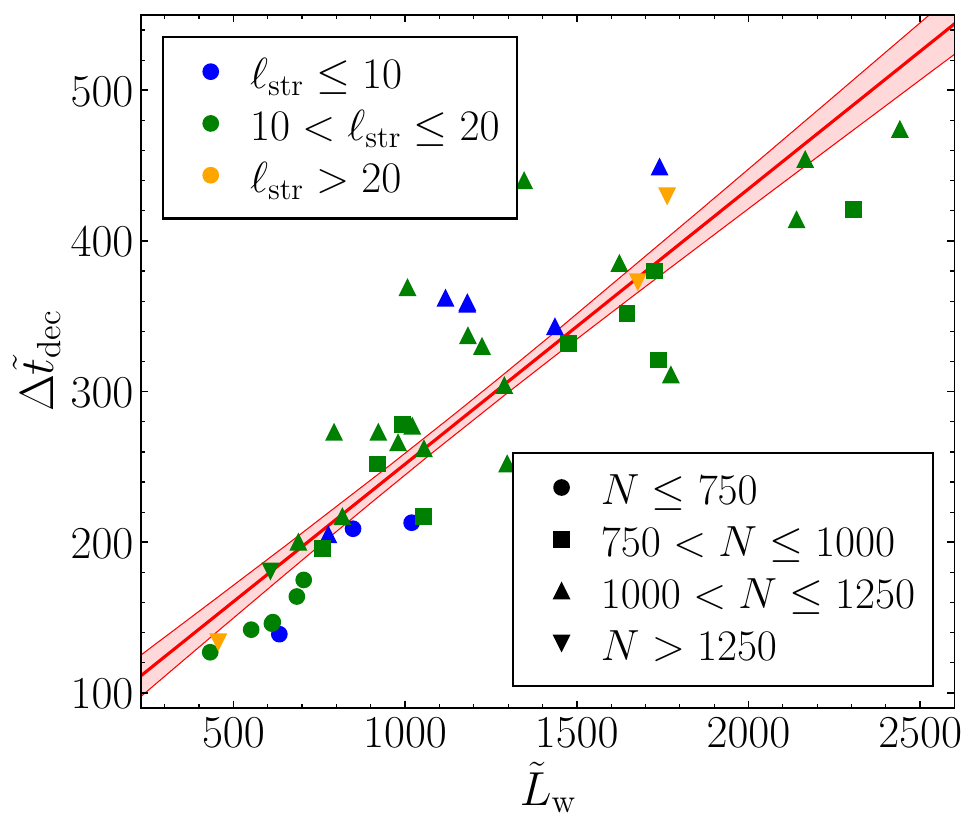}
        \caption{Linear fit}\label{fig:local:decaynetwork2}
    \end{subfigure}
    
    \caption{ Decay time of network loops as a function of their initial lengths. Lines and bands represent the best-fit result to a power law, $\Delta\tilde{t}_\dec=A\tilde{L}_0^\alpha$ (left, in logarithmic scale), and a linear dependence, $\Delta\tilde{t}_\text{dec}=c_1\tilde{L}_0+c_2$ (right, in linear scale). All simulations are performed with $\delta\tilde{x}=0.25$. }
    \label{fig:local:decaynetwork}
\end{figure}

In \cref{fig:local:decaynetwork1}, we show the decay time of all the studied network loops as a function of their initial lengths, measured from the number of pierced plaquettes---see \cref{eq:global:lengthwindingdefinition}. From a two-parameter fit of the form $\Delta\tilde{t}_\dec=A\tilde{L}_0^\alpha$,  we find that network loops decay slightly faster than linearly with their length, with $\alpha=0.71(5)$ and $A=1.9(0.7)$. The result of this fit is also presented in \cref{fig:local:decaynetwork}. % We observe that, as in the global case, the results scale roughly linearly with the initial length, $L_0$, indicating the decay of the loops is also driven by a scale invariant mechanism. We present in the figure the result of a linear fit of the form $\Delta\tilde{t}_\dec=A\tilde{L}_0$, from which we obtain $A=0.232(6)$. 
A similar relation is also obtained from the string energies, measured using \cref{eq:local:stringenergyweighteddefinition}. Fitting to $\Delta\tilde{t}_\dec=B\tilde{E}_{\str,0}^\beta$, we obtain $\beta=0.73(6)$ and $B=1.0(0.5)$.  

Note this has very important implications in the evolution of the strings. In particular, we can determine the particle emission power, 
\begin{equation}\label{eq:local:powerlocal}
\tilde{P}_\varphi=\frac{\d \tilde{E}_\text{str}}{\d \tilde{t}}\propto \tilde{L}^{\gamma}\,,
\end{equation}
where $\gamma=\alpha(1-\beta)/\beta$ and the power in program units is $\tilde{P}_\varphi=\Pphi/v^2$. From these results, we find that the emission power from network loops seems to increase slightly with the string length, with $\gamma=0.26(8)$. If GW emission power is independent of the length, as we show in \cref{sec:local:GWresults} and is predicted by NG, this implies that strings originating from phase transitions in the early universe would decay mainly via the emission of particles, with a very suppressed production of GWs.

It is worth commenting about systematic effects that could affect our results. 
We have studied the sensitivity of network loops to the UV resolution. We use a coarse graining procedure to generate low-resolution loop configurations from one with $\delta\tilde{x}=0.125$, finding that the dependence on $\delta\tilde{x}$ is be small for network loops. For the choice $\delta\tilde{x}=0.25$ used in this work,  we estimate the systematic error to remain only of a few percents. %This is in agreement with previous findings~\cite{Matsunami:2019fss,Hindmarsh:2021mnl}.

Finite-volume effects can also play a role. In some cases, we study loops that are several times longer than the box size. This could imply a higher curvature than typically expected in loops of the same size, if they were not constrained to fit in our lattices. Our loops may thus radiate faster than would be expected for loops of their size. However,  we do not observe evidence of this effect for the loop sizes studied, and so we expect that finite-volume effect on $\alpha$ and $\beta$ for network loops to be small. Even for a relative errors of $\sim30\%$ size, we would still obtain $\gamma\gtrsim 0$, meaning that the emission of particle radiation is, at most, independent of the size of the loops. 

Given this observation, it is instructive to also analyze our results assuming a linear dependence. The results to a fit of the form $\Delta\tilde{t}_\text{dec}=c_1\tilde{L}_0+c_2$ is presented in \cref{fig:local:decaynetwork2}, which is observed to be compatible with our data. From the fit we obtain $c_1=0.182(13)$ and $c_2=69(16)$. A similar fit to $\Delta\tilde{t}_\text{dec}=d_1\tilde{E}_{\text{str,0}}+d_2$ yields $d_1=0.097(7)$ and $d_2=63(18)$. If we assume this linear relation, we find that the emission power is independent of the length of the strings, with a power $\tilde{P}_\varphi^\text{linear}=10.3(7)$, which is almost equal to that obtained for global loops---see \cref{eq:global:powernetworkresult}. As we discuss in \cref{sec:local:GWresults}, even in this case, the decay of the loops would still be dominated by particle emission, with a suppression of GW production proportional to $v^2/\mpl^2$.

%From this, we can obtain an estimate for the particle emission power,
%\begin{equation}
%\tilde{P}_\varphi=\frac{\d \tilde{E}_\str}{\d\tilde{t}}=\frac{1}{C}=8.30(23)
%\end{equation}
%We observe how this result is only slightly smaller than that of global strings. This thus shows that the emission of particle radiation from local strings is not suppressed as expected from NG predictions, at least for the range of sizes considered in this work.

\subsection{Particle emission from artificial strings}\label{sec:local:particleproductionartificial}

\begin{figure}[!p]
    \centering
    \begin{minipage}{0.24\textwidth}
	\includegraphics[width=\textwidth]{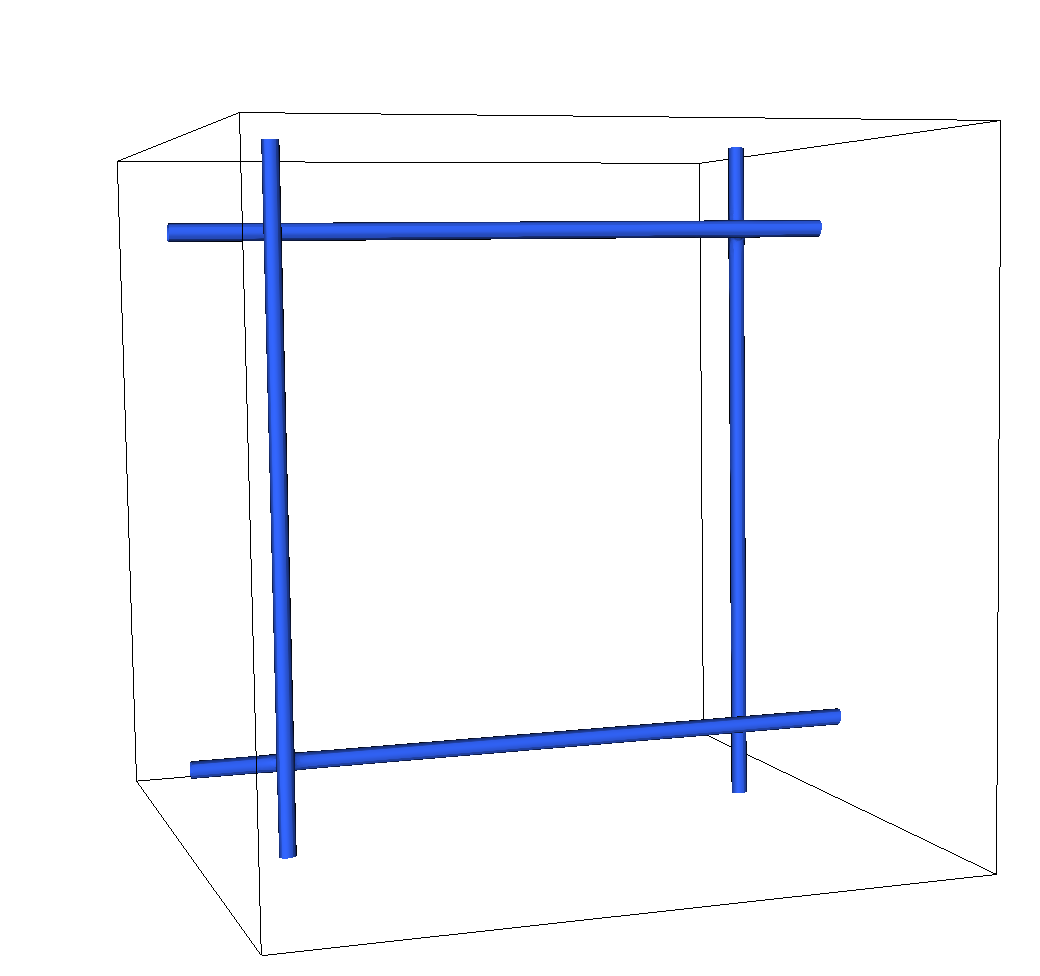}
	\centering $\tilde{t}=0$
\end{minipage}
	\begin{minipage}{0.24\textwidth}
	\includegraphics[width=\textwidth]{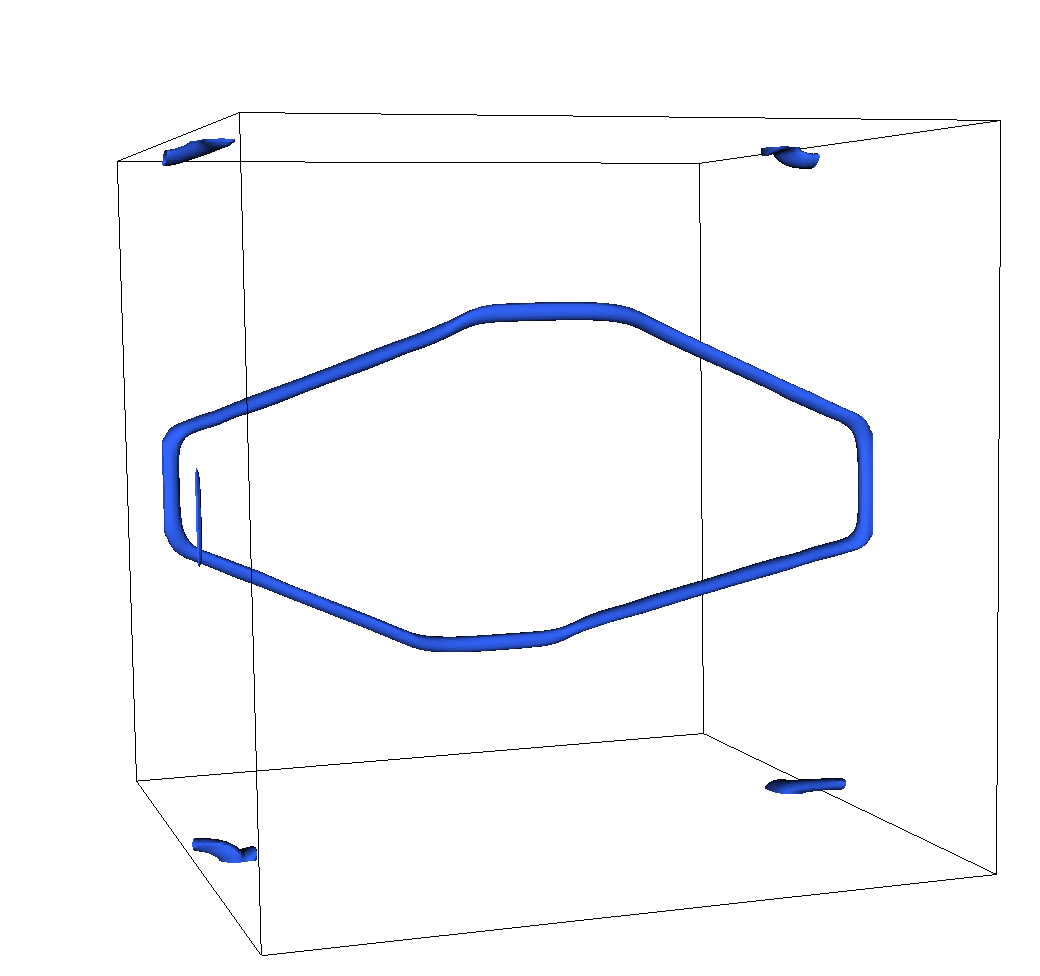}
	\centering $\tilde{t}=25$
\end{minipage}
	\begin{minipage}{0.24\textwidth}
	\includegraphics[width=\textwidth]{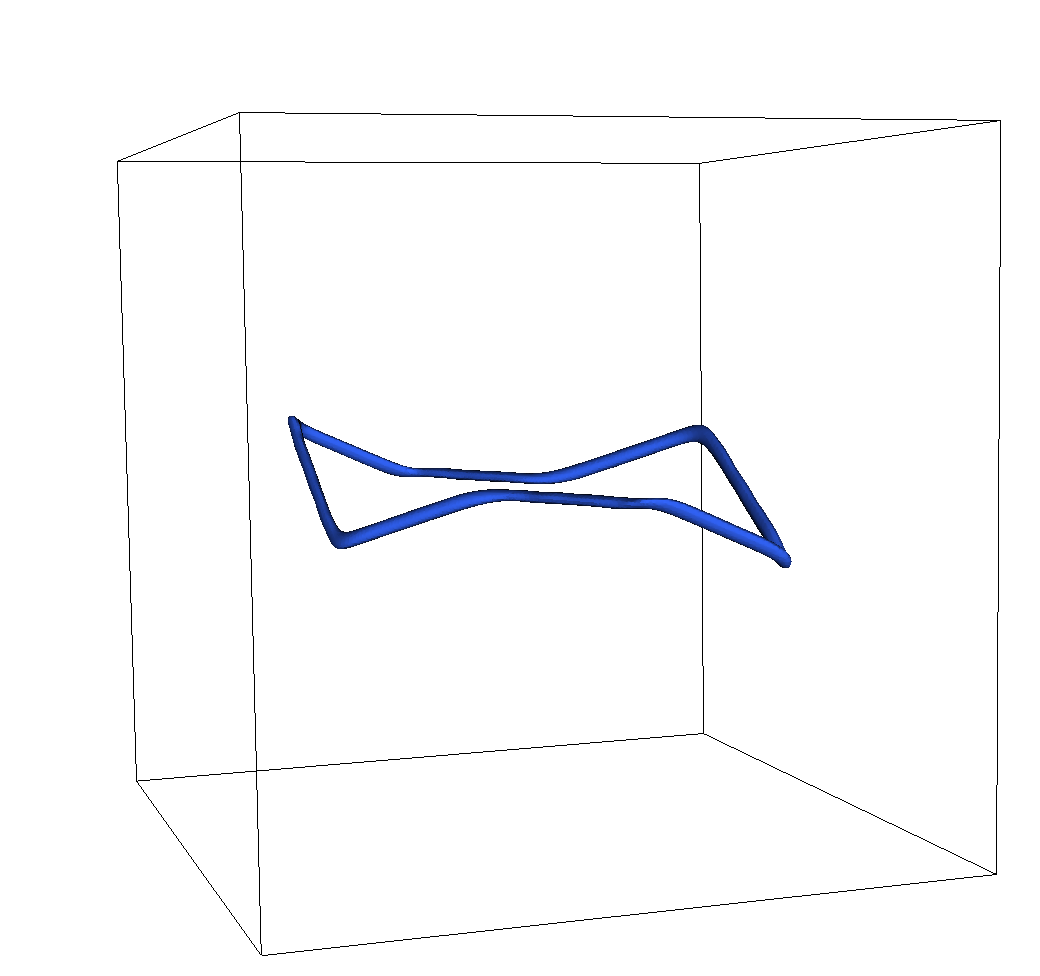}
	\centering $\tilde{t}=50$
\end{minipage}
	\begin{minipage}{0.24\textwidth}
	\includegraphics[width=\textwidth]{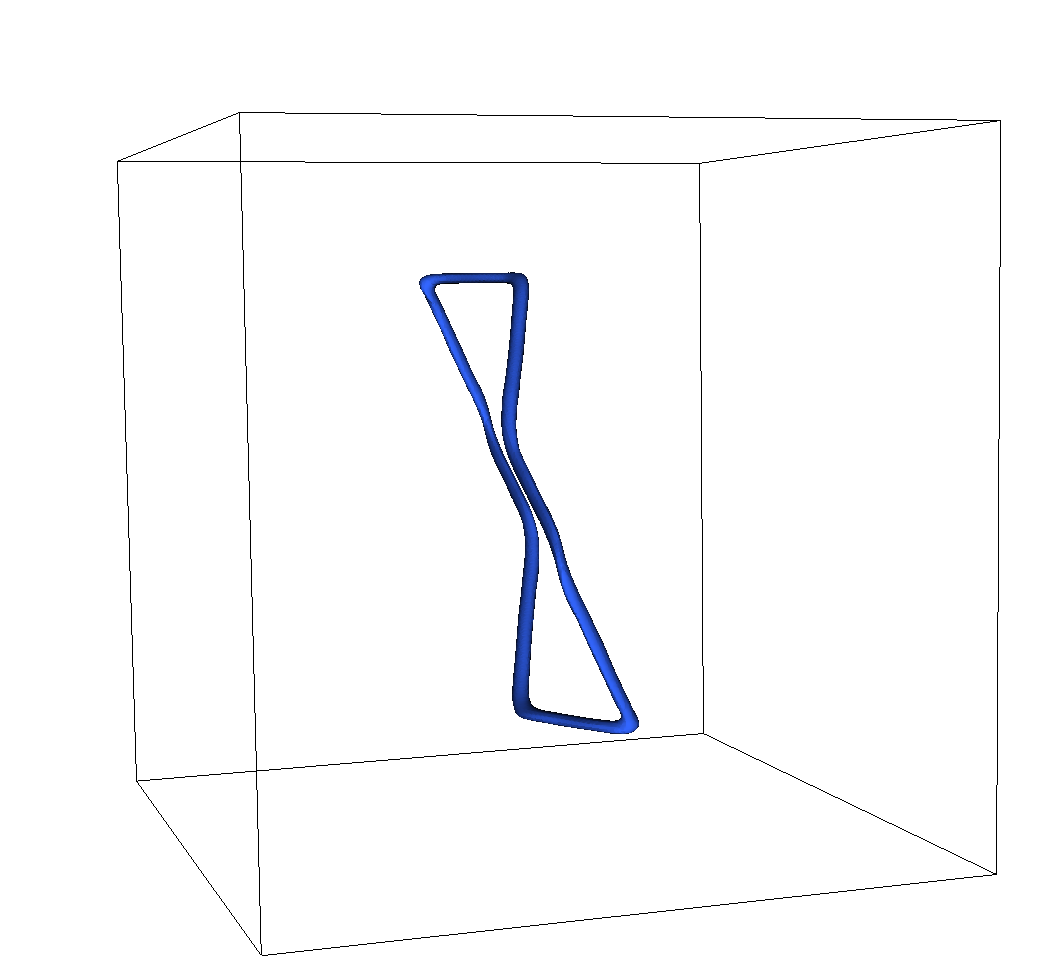}
	\centering $\tilde{t}=75$
\end{minipage}
\vspace{0.2cm}
	
	    \begin{minipage}{0.24\textwidth}
	\includegraphics[width=\textwidth]{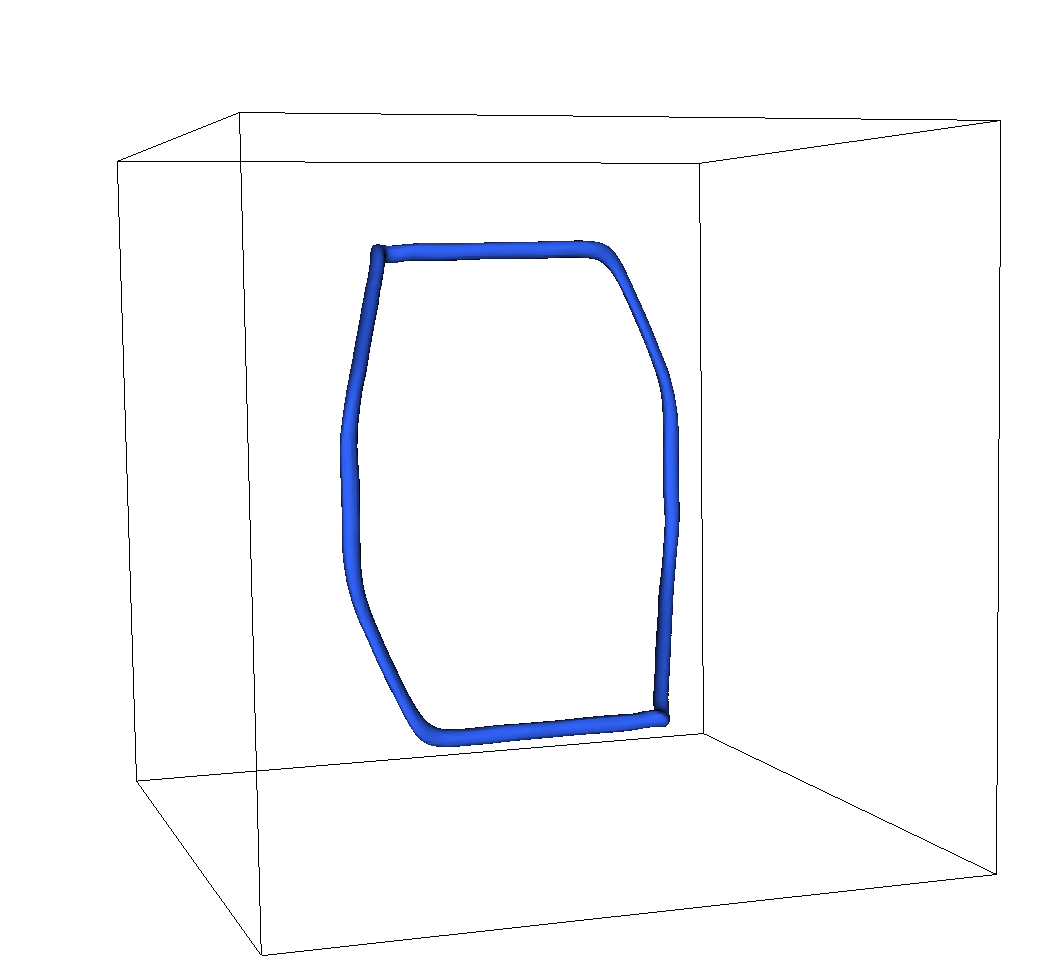}
	\centering $\tilde{t}=100$
\end{minipage}
	\begin{minipage}{0.24\textwidth}
	\includegraphics[width=\textwidth]{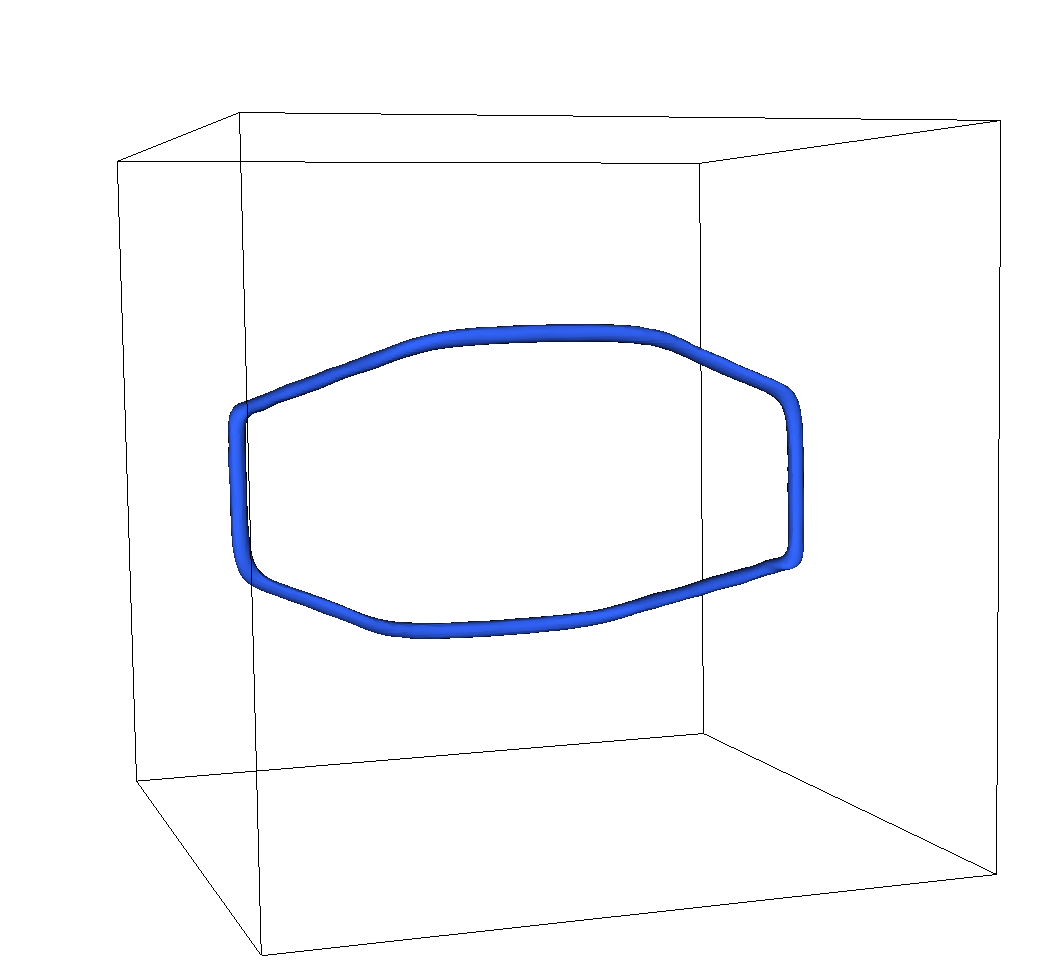}
	\centering $\tilde{t}=125$
\end{minipage}
	\begin{minipage}{0.24\textwidth}
	\includegraphics[width=\textwidth]{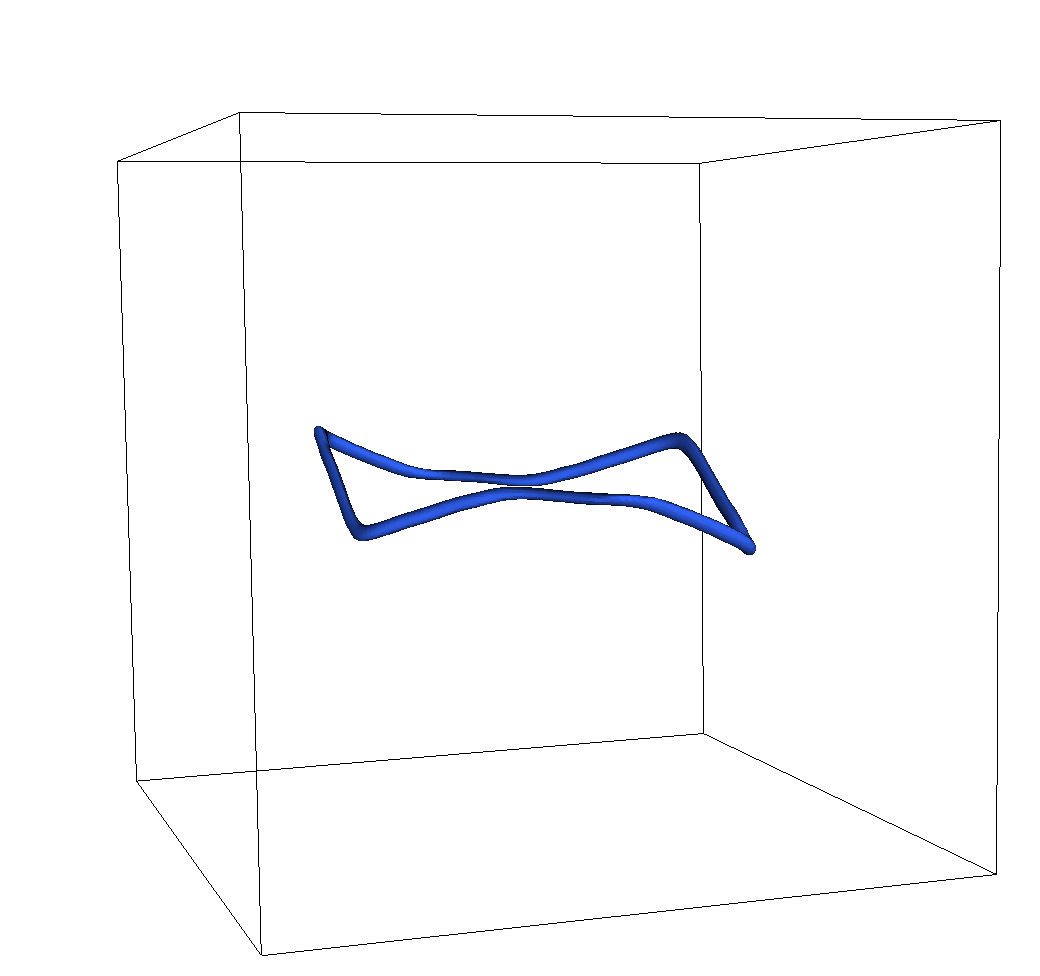}
	\centering $\tilde{t}=150$
\end{minipage}
	\begin{minipage}{0.24\textwidth}
	\includegraphics[width=\textwidth]{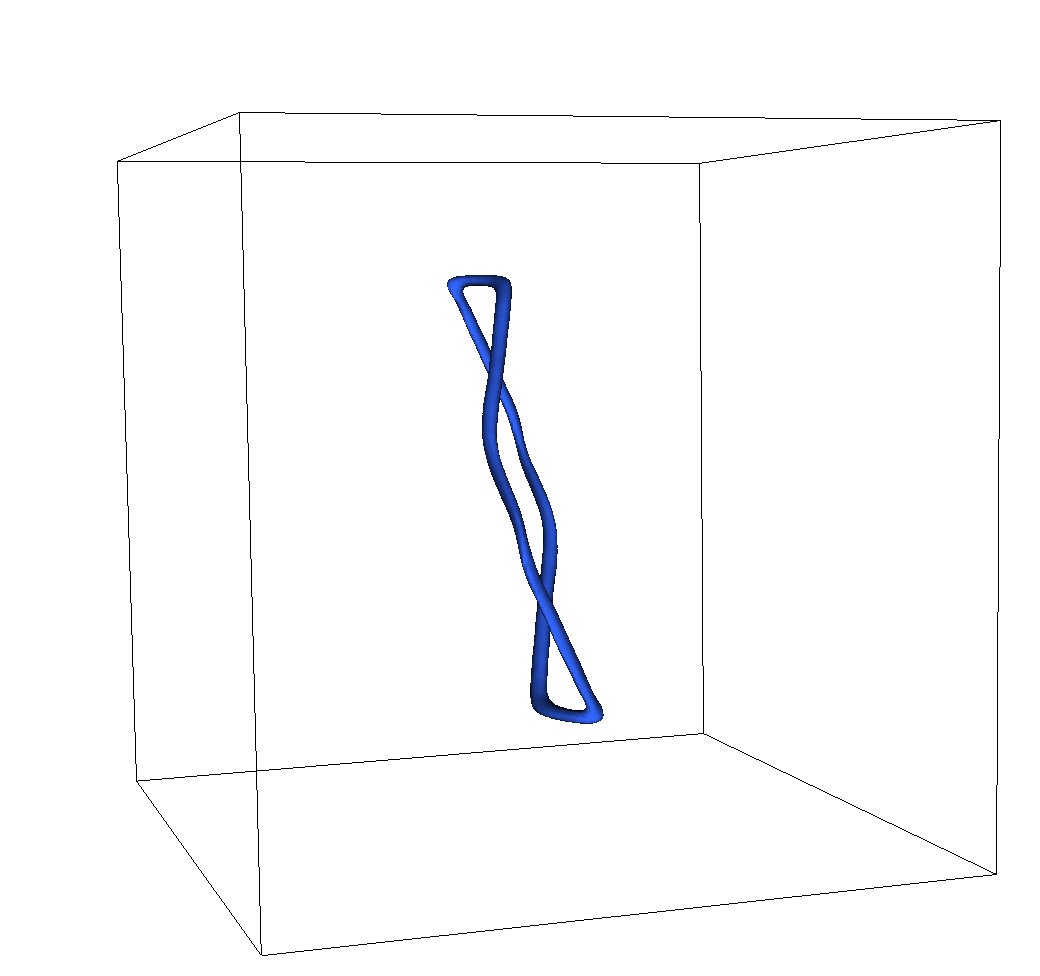}
	\centering $\tilde{t}=175$
\end{minipage}
\vspace{0.2cm}
	
	    \begin{minipage}{0.24\textwidth}
	\includegraphics[width=\textwidth]{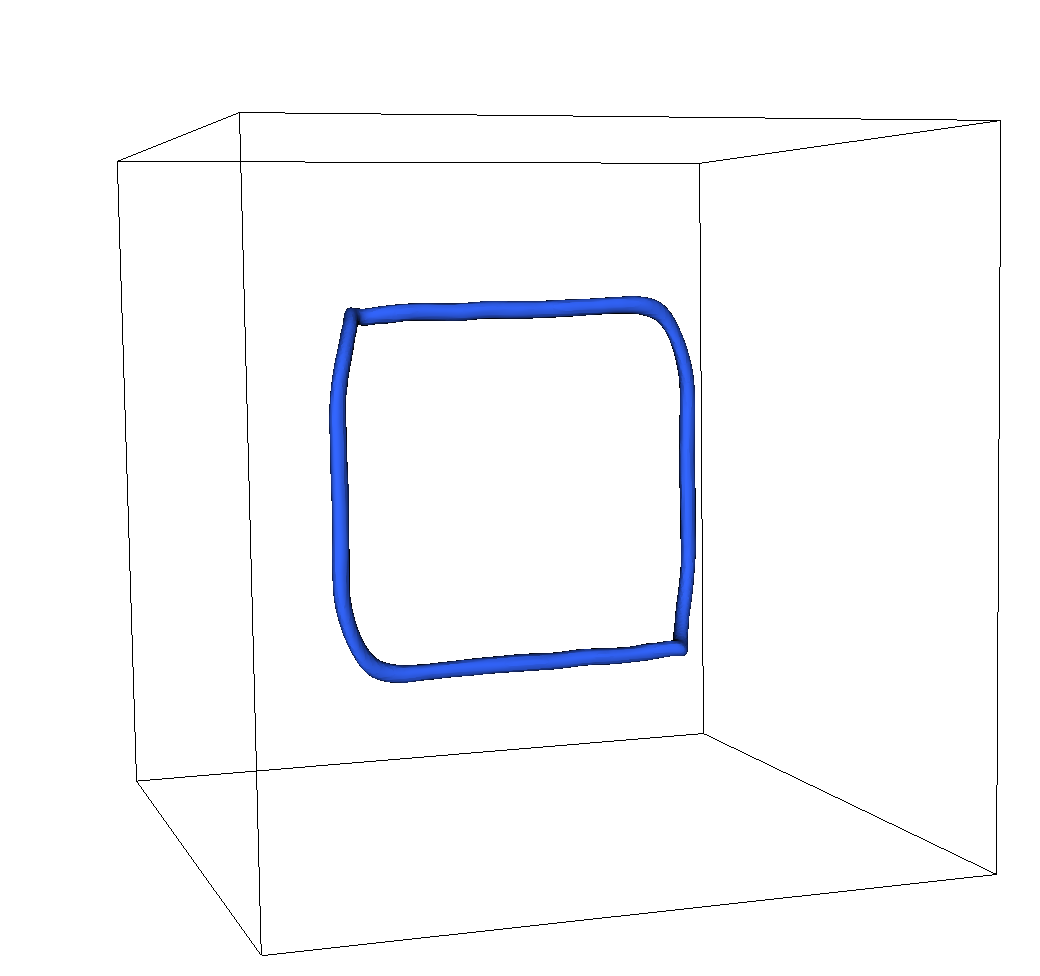}
	\centering $\tilde{t}=200$
\end{minipage}
	\begin{minipage}{0.24\textwidth}
	\includegraphics[width=\textwidth]{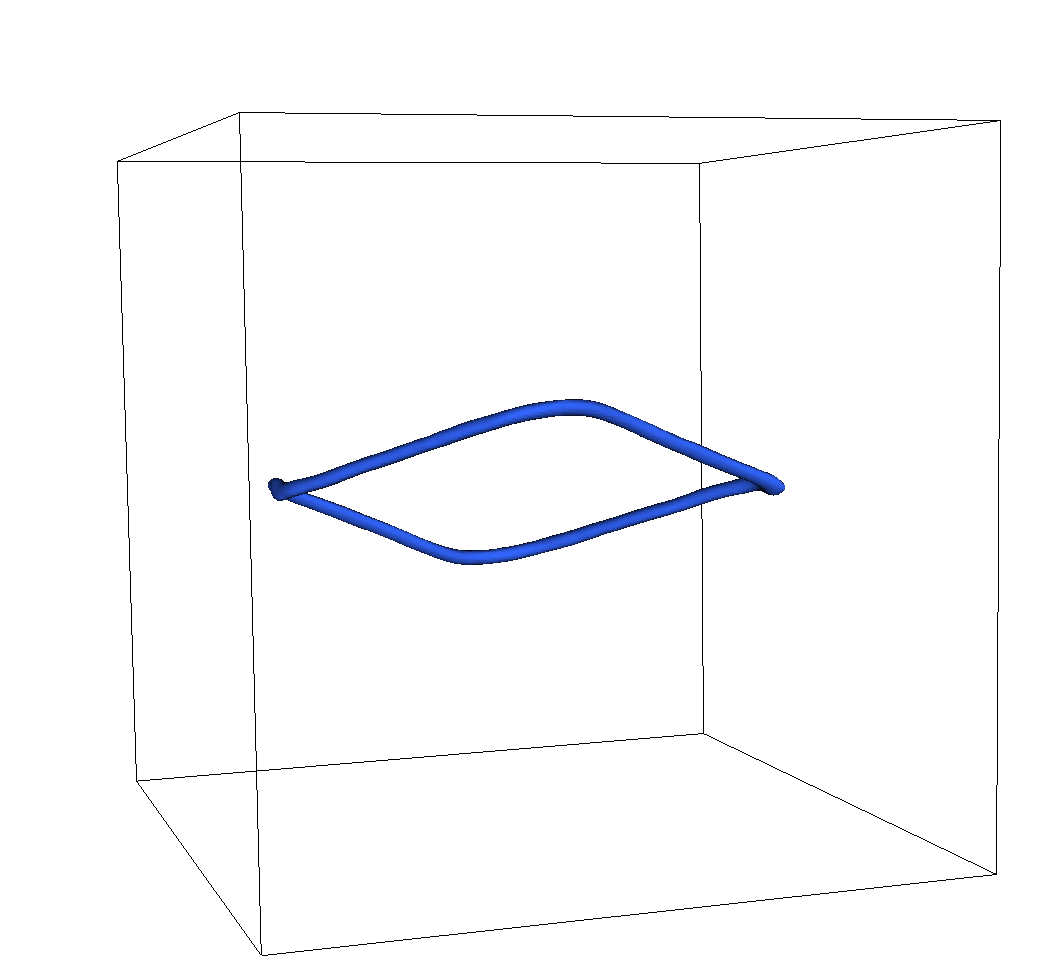}
	\centering $\tilde{t}=225$
\end{minipage}
	\begin{minipage}{0.24\textwidth}
	\includegraphics[width=\textwidth]{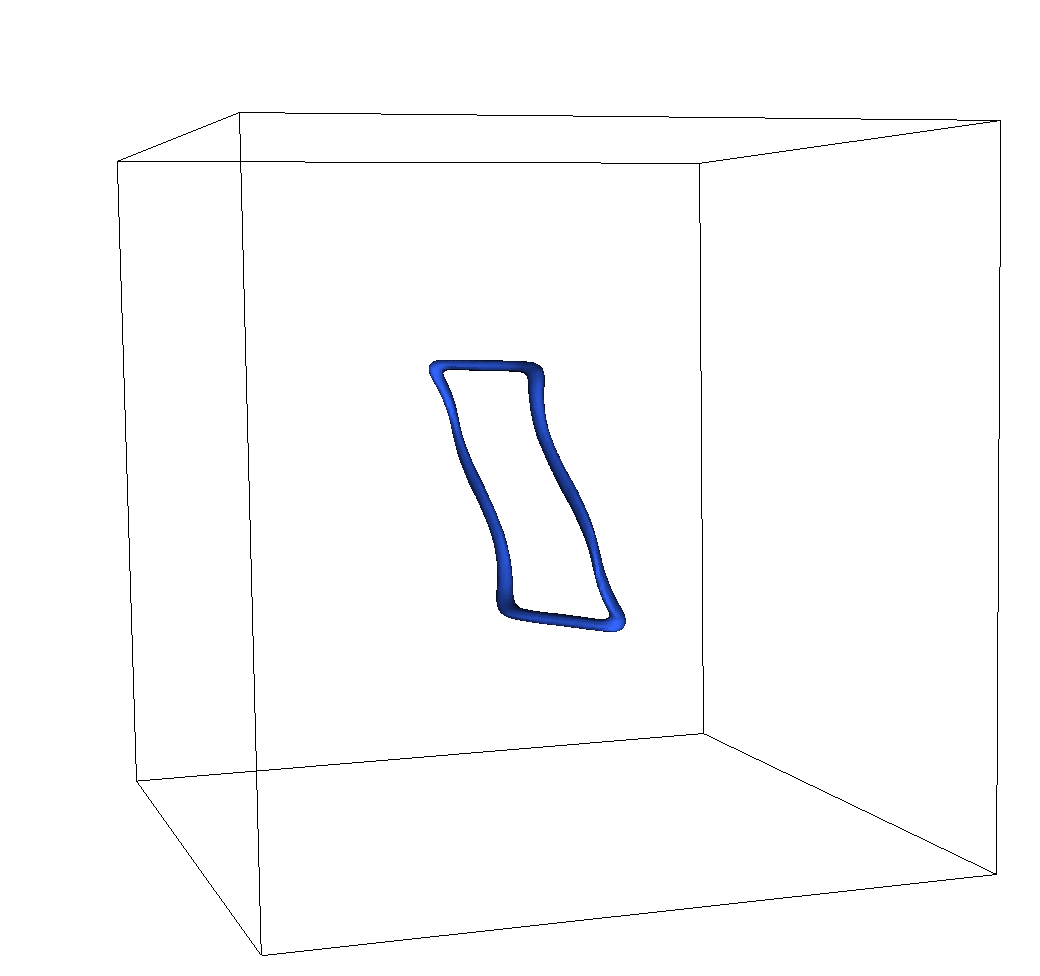}
	\centering $\tilde{t}=250$
\end{minipage}
	\begin{minipage}{0.24\textwidth}
	\includegraphics[width=\textwidth]{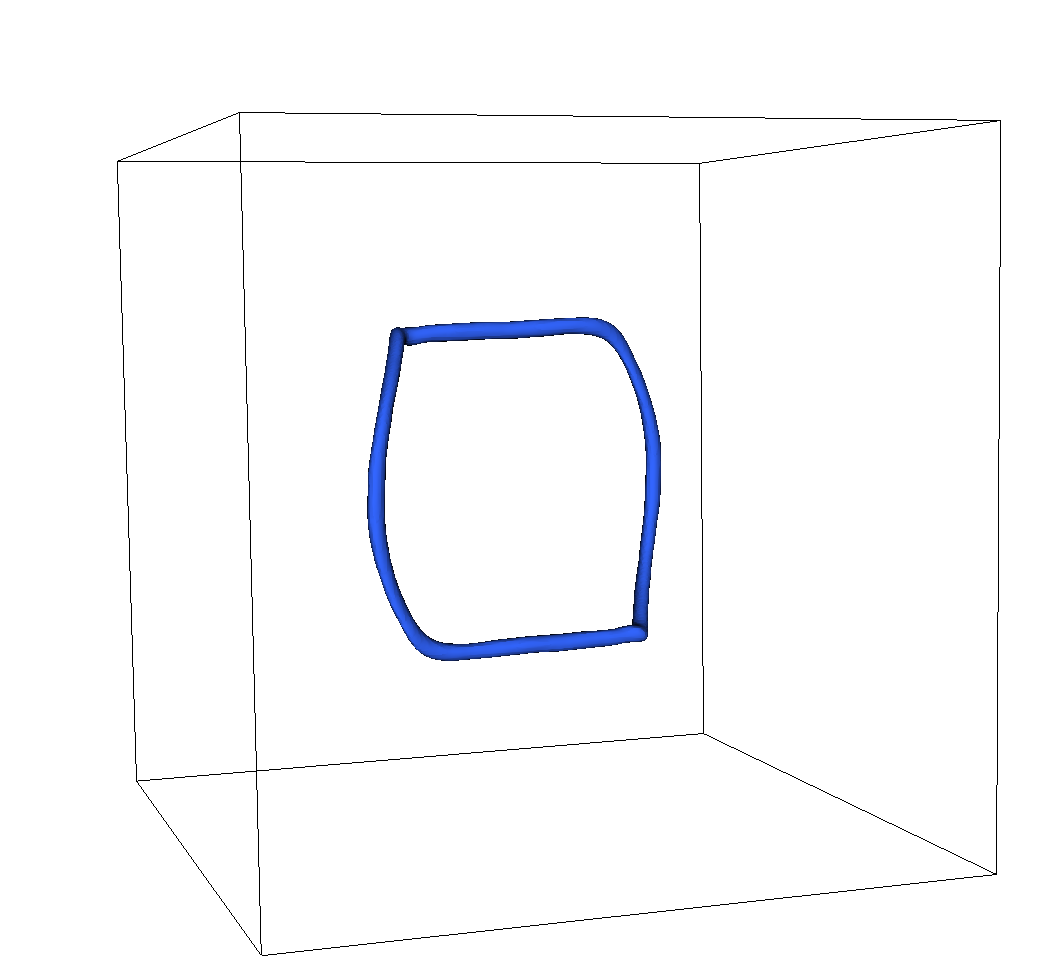}
	\centering $\tilde{t}=275$
\end{minipage}
\vspace{0.2cm}
	
	    \begin{minipage}{0.24\textwidth}
	\includegraphics[width=\textwidth]{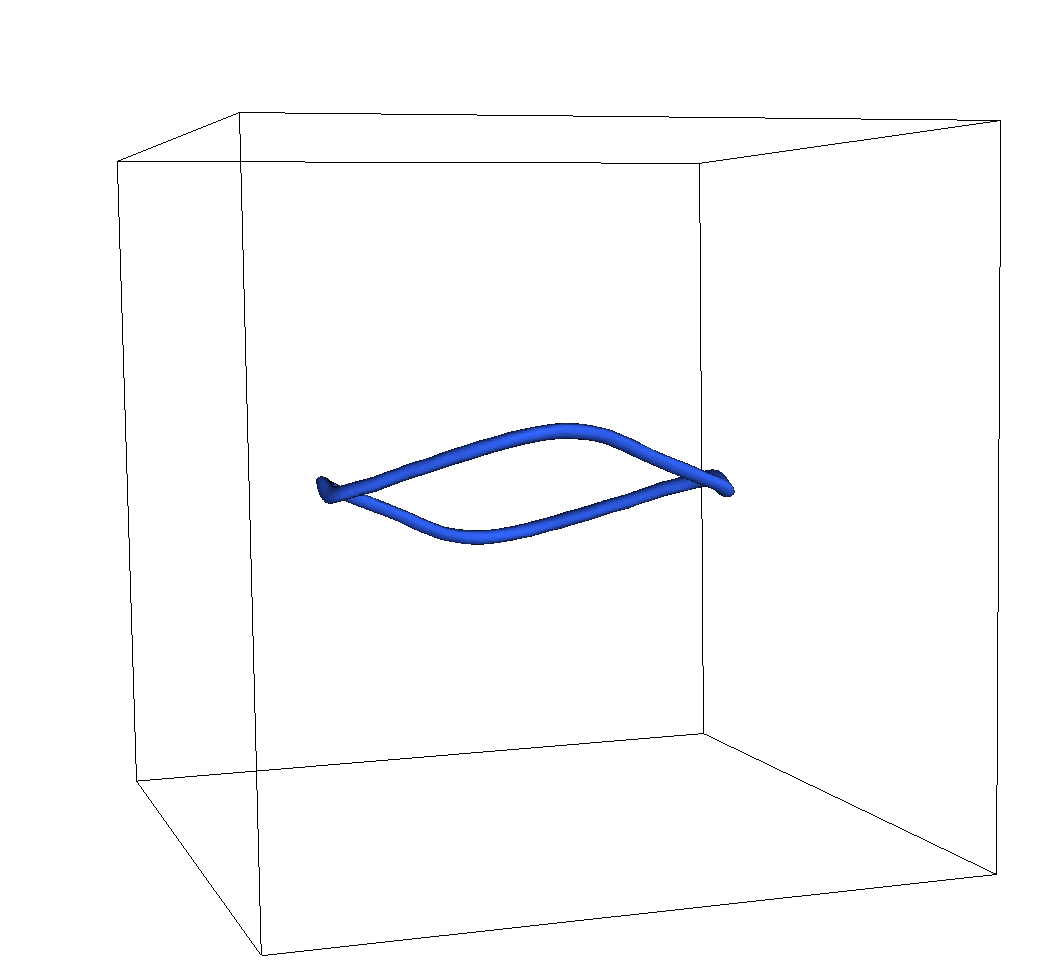}
	\centering $\tilde{t}=300$
\end{minipage}
	\begin{minipage}{0.24\textwidth}
	\includegraphics[width=\textwidth]{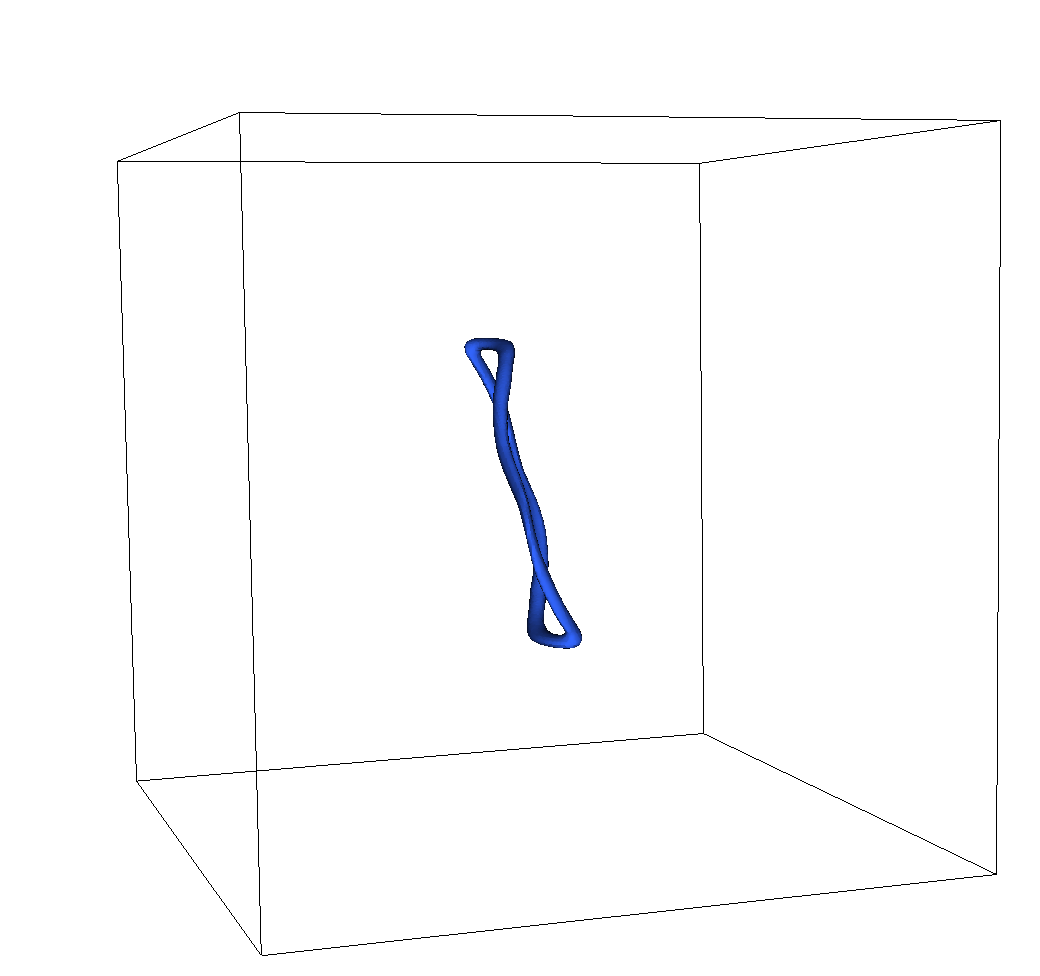}
	\centering $\tilde{t}=325$
\end{minipage}
	\begin{minipage}{0.24\textwidth}
	\includegraphics[width=\textwidth]{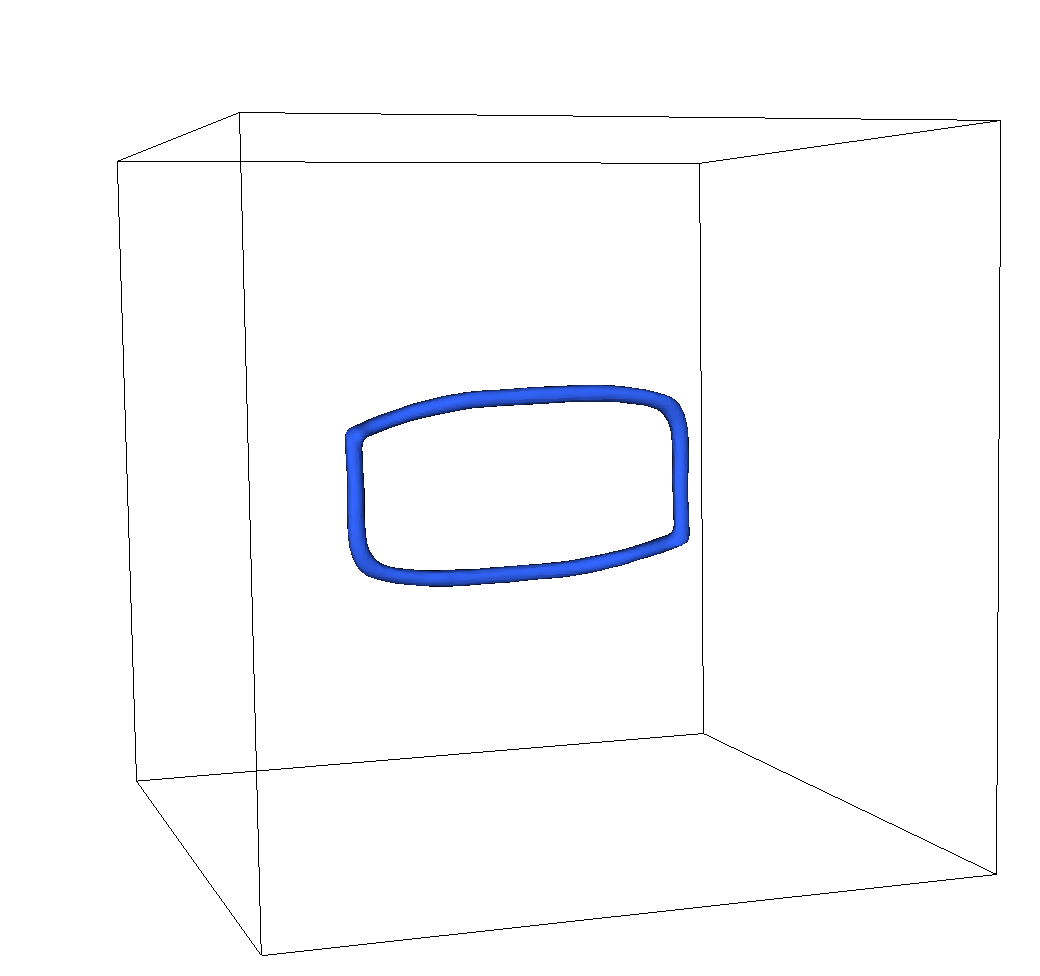}
	\centering $\tilde{t}=350$
\end{minipage}
	\begin{minipage}{0.24\textwidth}
	\includegraphics[width=\textwidth]{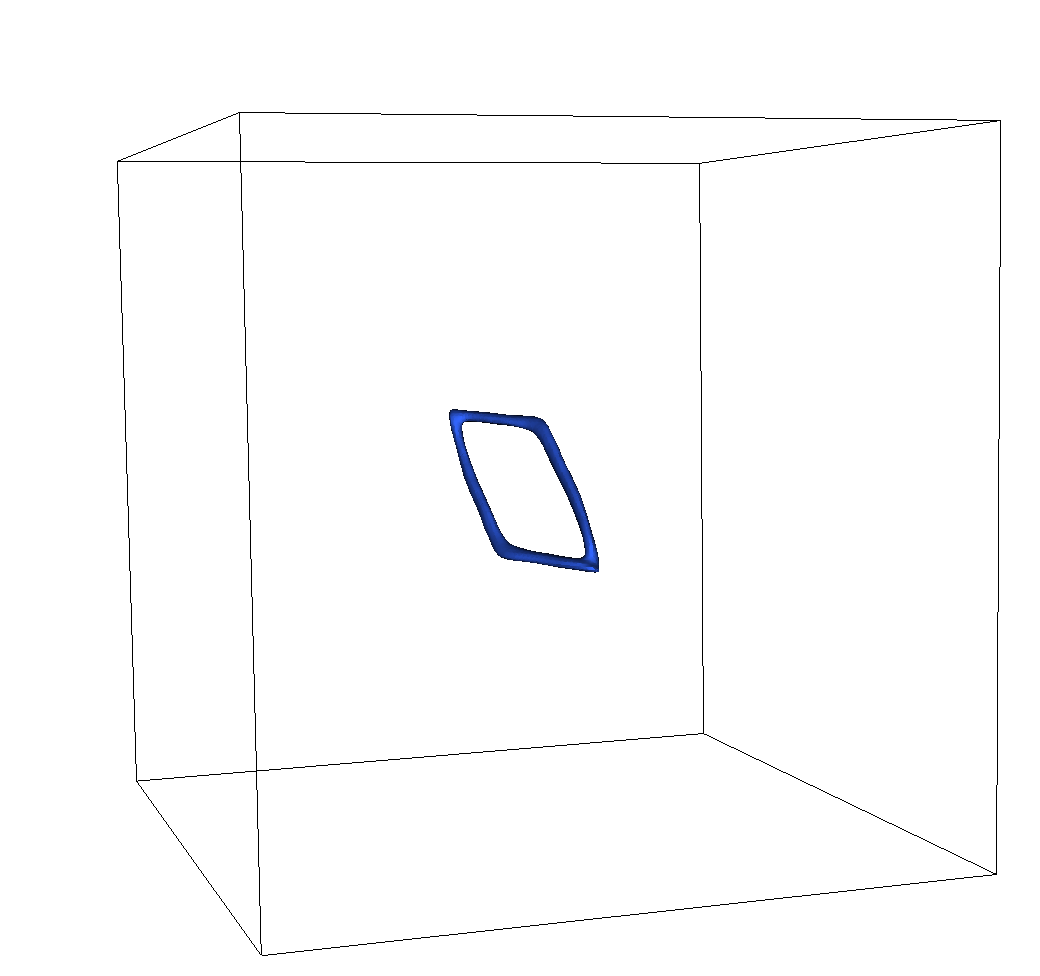}
	\centering $\tilde{t}=375$
\end{minipage}
\vspace{0.2cm}
	
	    \begin{minipage}{0.24\textwidth}
	\includegraphics[width=\textwidth]{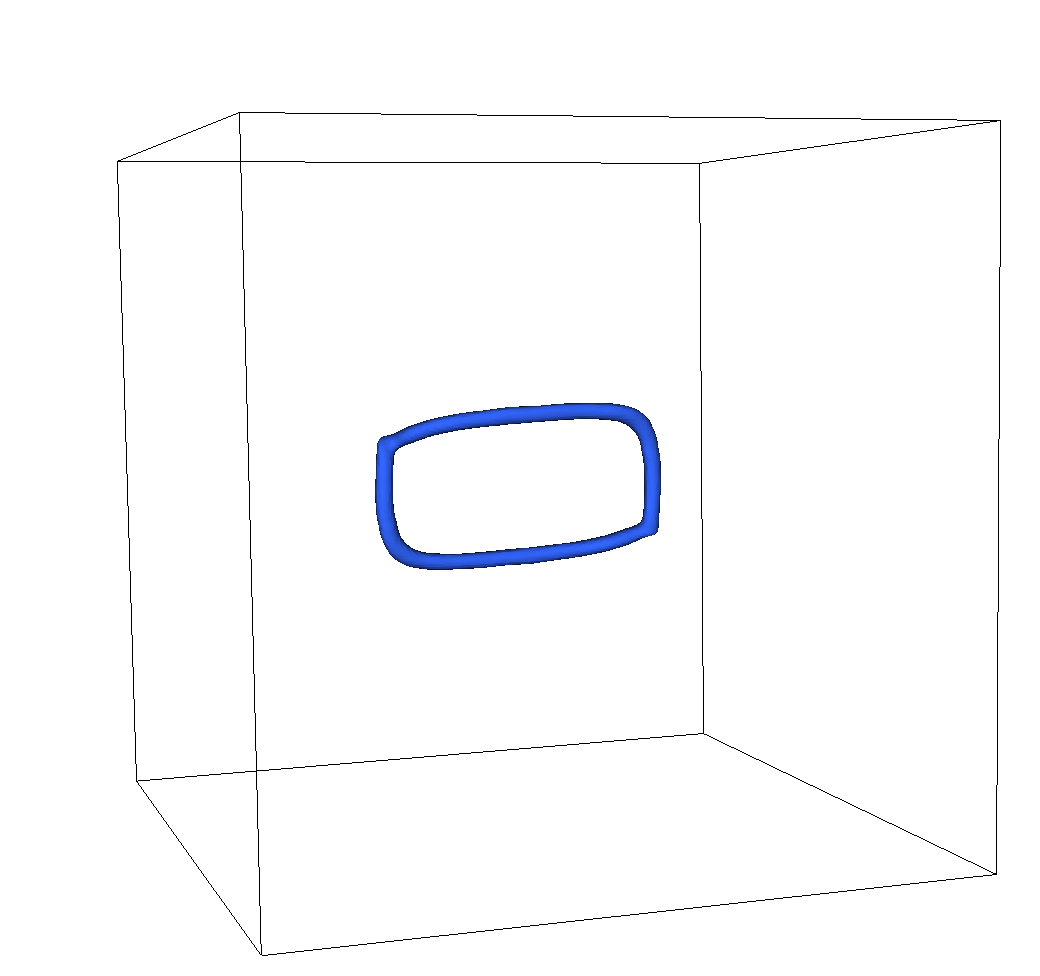}
	\centering $\tilde{t}=400$
\end{minipage}
	\begin{minipage}{0.24\textwidth}
	\includegraphics[width=\textwidth]{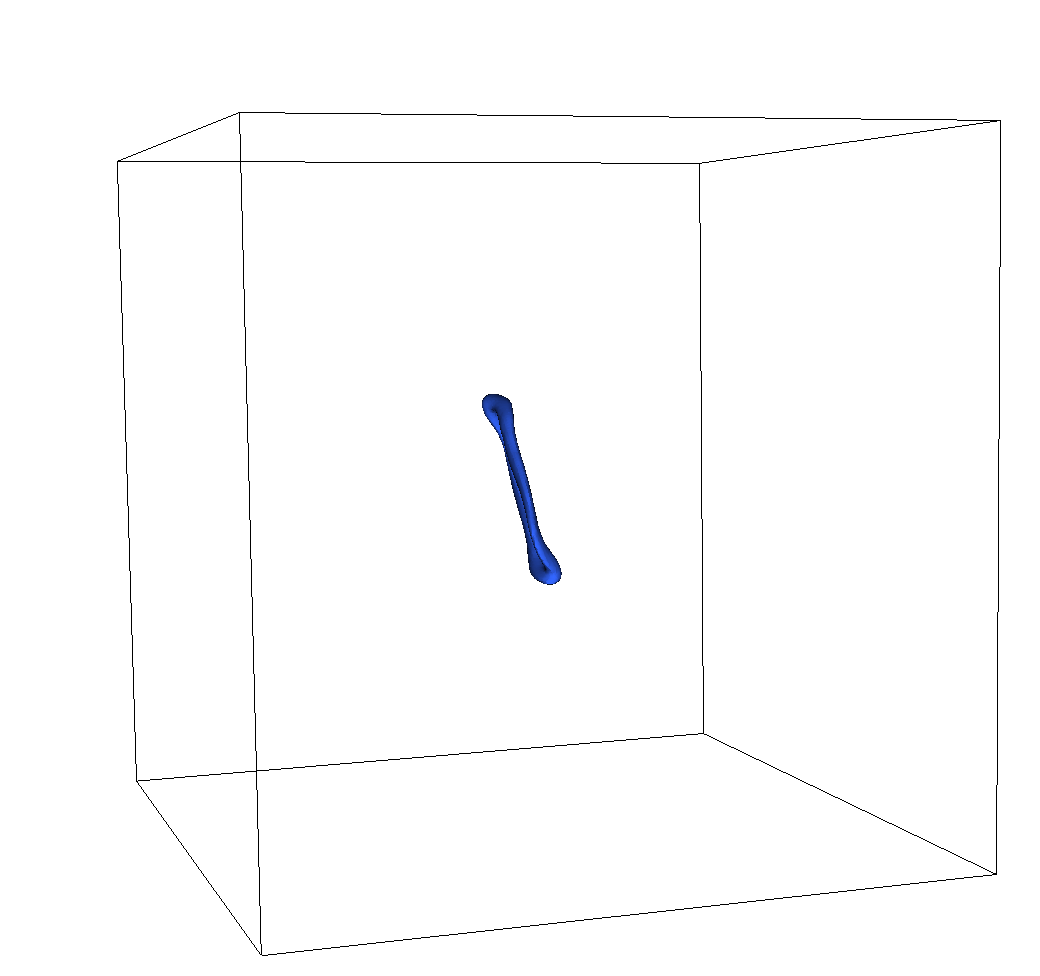}
	\centering $\tilde{t}=425$
\end{minipage}
	\begin{minipage}{0.24\textwidth}
	\includegraphics[width=\textwidth]{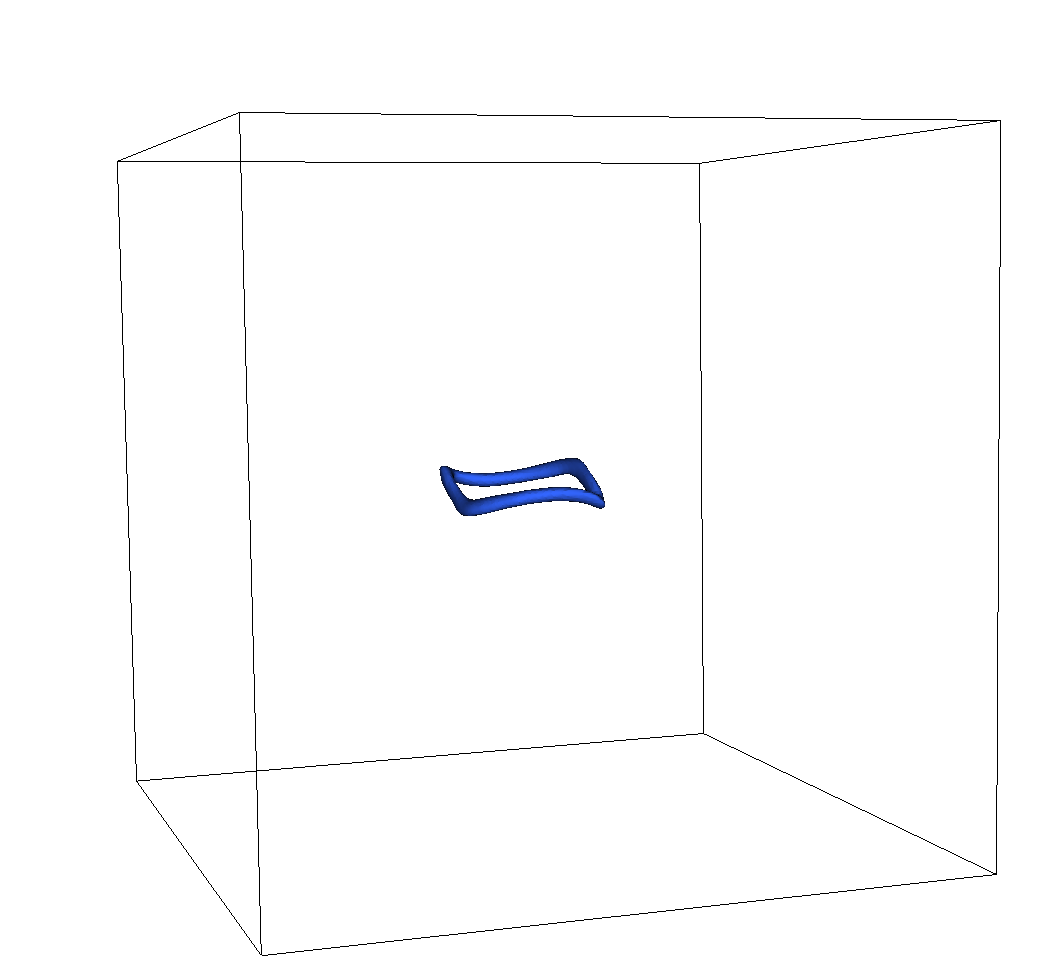}
	\centering $\tilde{t}=450$
\end{minipage}
	\begin{minipage}{0.24\textwidth}
	\includegraphics[width=\textwidth]{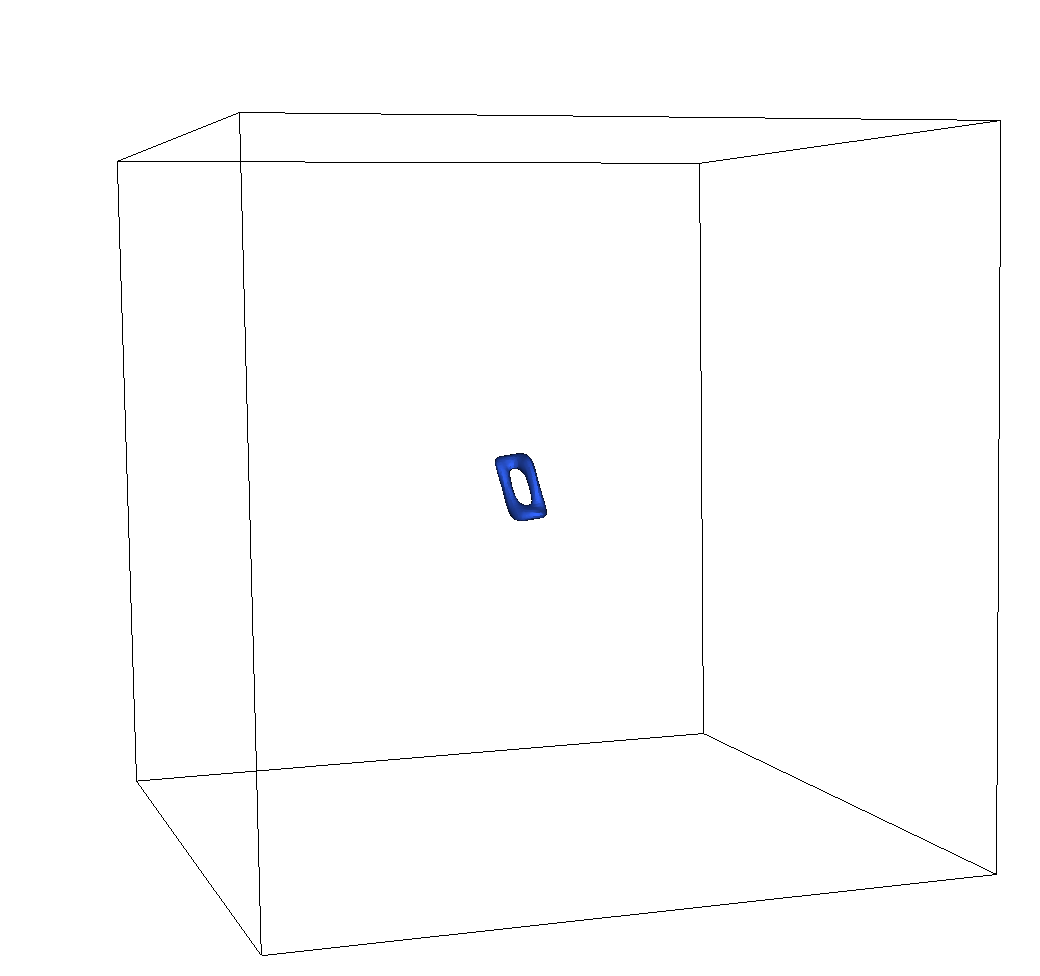}
	\centering $\tilde{t}=475$
\end{minipage}
\vspace{0.3cm}
   
    \caption{
        Three-dimensional snapshots of $|\varphi|^2=0.2v^2$ surfaces of different moments of the evolution of an artificial loop of type \RNum{1}, simulated with $\tilde{L}=64$, $\delta\tilde{x}=0.25$, $v_1=v_2=0.25$ and $\sin\alpha=0.4$. The secondary loop has already vanished at $\tilde{t}=50$, while the other one oscillates several times before decaying. The field has been periodically shifted by $L/2$ in the $x$ and $z$ directions so that the outer loop is centered in the figures.}
    \label{fig:local:stringsnapshotsartificial}
\end{figure}

We study the decay time of 14 artificial loops of type \RNum{1} and 6 loops of type \RNum{2}, with length to width ratios up to $L_0/r_\text{c}\lesssim 640$, as a function of their initial length and energy. We use simulations with $\delta\tilde{x}=0.1875$ and $\delta\tilde{x}=0.125$, which we justify below. Local artificial loops present a qualitatively very different dynamics than network loops, since they oscillate multiple times before they decay. Their lifetimes are thus much longer than that of network loops of the same size. As an example, we represent in \cref{fig:local:stringsnapshotsartificial} the evolution of a loop of type \RNum{1} generated with $v_1=v_2=0.6$ and $\sin\alpha=0.4$, simulated in a box with $\tilde{L}=64$ and $\delta\tilde{x}=0.25$.\footnote{We use a lattice of reduced size for the representation due to the size of the three-dimensional distribution savefiles, which becomes very large (tens of gigabytes) for bigger lattices.}  Notoriously, we find that even for a simulation of this reduced size, the loop oscillates multiple times before decaying.

Due to their long-lasting behaviour, the study of the decay time of artificial loops as a function of their initial length is more complicated than for network loops. Examples of the evolution of the strings length are presented in \cref{fig:local:exampledecay} for three artificial loops. In \cref{fig:local:exampledecayartificialloops1}, we represent the case of an artificial loop of type \RNum{1}, generated with $v_1=v_2=0.6$ and $\sin\alpha=0.4$, in a box with $\tilde{L}=168$ and $\delta\tilde{x}=0.1875$. In this case, we observe that after the inner loop disappears at $\tilde{t}_0\approx450$ (vertical dotted line), the length of the outer loop begins to decay slowly while oscillating, until it smoothly disappears at $\tilde{t}_\text{dec}\approx3800$. % The lifetime of the loop is thus much larger than the expected NG period, $T_\text{NG}\approx170$, which corresponds to the periodicity of the oscillations. 
We observe this behavior is very similar to that of an artificial loop of type \RNum{2}, presented in \cref{fig:local:exampledecayartificialloops2} for a simulation with $\tilde{L}=336$ and $\delta\tilde{x}=0.1875$.

A bit different is the evolution of the artificial loop of type \RNum{1} shown in \cref{fig:local:exampledecayartificialloops3}, generated with $v_1=0.3$, $v_2=0.6$ and $\sin\alpha=0.4$, in a box with $\tilde{L}=168$ and $\delta\tilde{x}=0.1875$. In this case, after oscillating a number of times, the loop disappears almost instantly. This is related to a double-line collapse event. Due to the initial square configuration, the loop eventually reaches a point in which two parallel segments approach each other, completely annihilating. 

Since the double-line collapse is a result of the artificial initial conditions, we opt to focus on the rate of decay while the loop oscillates. For all artificial loops of both types considered, we fit the time evolution of the length of the string to the following function
\begin{equation}\label{eq:local:artificiallengthfit}
\tilde{L}_\text{w}=A(\tilde{t}-\tilde{t}_\dec)^{p}\,,
\end{equation}
where $A$, $t_\dec$ and $p$ are the fit parameters. The fits are performed between the time the secondary loop disappears, $t_0$, and the time at which the loop collapses. The result of this fit is also presented in \cref{fig:local:exampledecayartificialloops1,fig:local:exampledecayartificialloops2,fig:local:exampledecayartificialloops3},  from which we obtain an estimate of the decay time of the loop, $t_\text{dec}$, if no double-line collapse happened. From this fit, we can get estimates of the initial length of the string, $L_0^\text{est}$, evaluating \cref{eq:local:artificiallengthfit} at $t_\text{0}$, which averages over the loop oscillation, and the lifetime of the loop if no double-line collapse  happened, $\Delta \tilde{t}_\dec^\text{est}=\tilde{t}_\text{min}-\tilde{t}_\dec$. We note that in the cases in which no double-line collapse occurs---see \cref{fig:local:exampledecayartificialloops1,fig:local:exampledecayartificialloops2}---this is very close to the measured lifetime of the loop.

\begin{figure}[!t]
    \centering
    
    \begin{subfigure}{0.495\textwidth} 
    \centering
        \includegraphics[width=\textwidth]{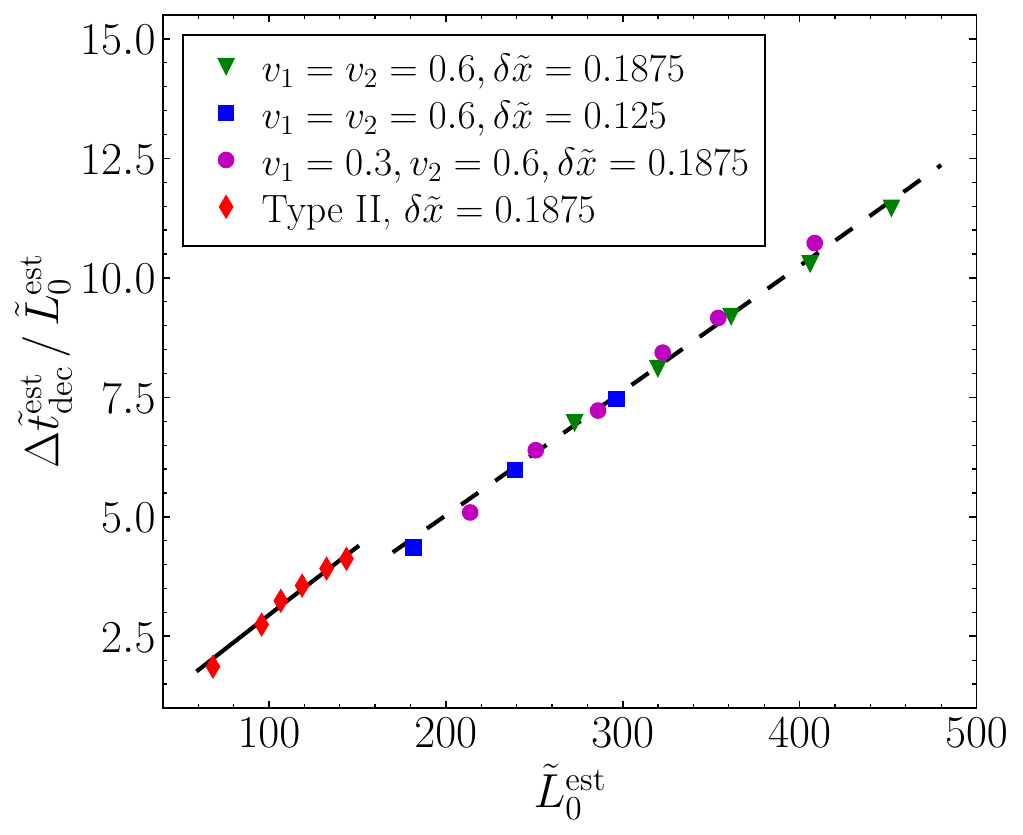}
        \caption{Estimated lifetime without double-line collapse}\label{fig:local:decayartificial1}
    \end{subfigure}
    \begin{subfigure}{0.495\textwidth} 
    \centering
        \includegraphics[width=\textwidth]{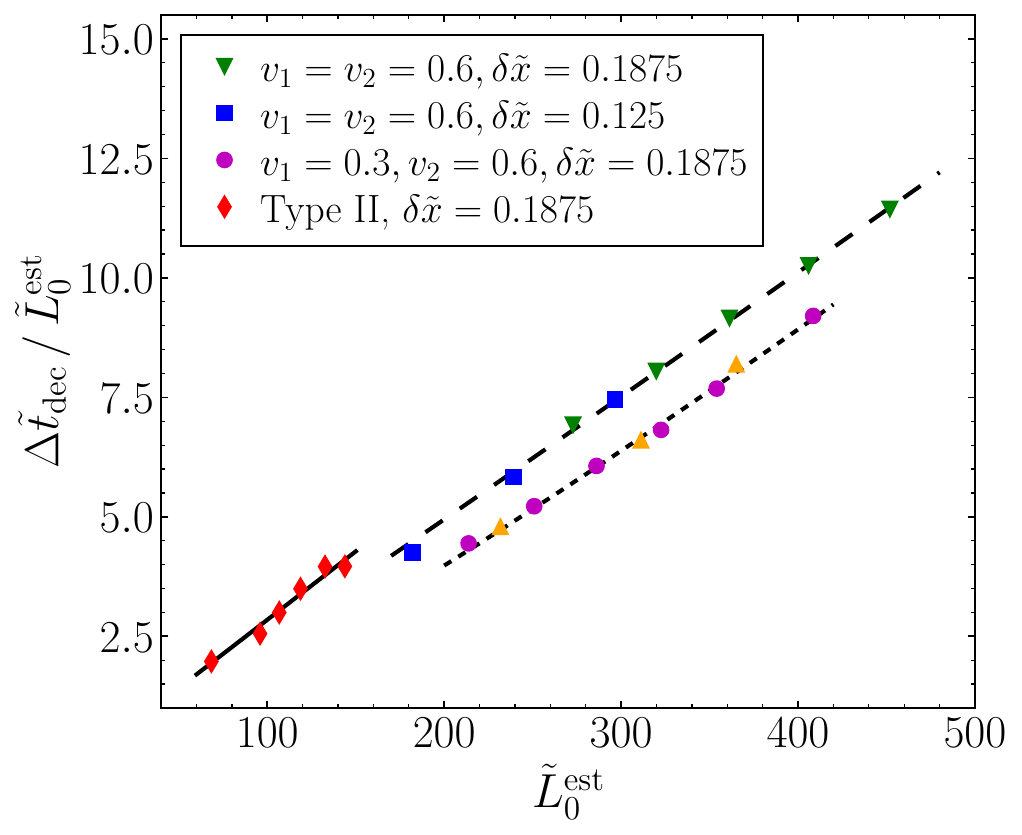}
        \caption{Measured lifetime with double-line collapse}\label{fig:local:decayartificial2}
    \end{subfigure}
    
    \caption{ Decay time of artificial loops, as a function of their initial lengths, with lifetime computed with (left) and without (right) extrapolating the decay time to remove the effect of a double-line collapse. All loops of type \RNum{1} are generated with $\sin\alpha=0.4$. Lines represent the best-fit result to a power law, and the initial length is estimated from fitting the evolution of the string length, as explained in the main text. We note that the effect of the double-line collapse is only notorious for those loops generated with $v_1=0.3$ and $v_2=0.6$ (magenta dots), which we fit separately from the other loops of type \RNum{1} in the right panel.}
    \label{fig:local:decayartificial}
\end{figure}

The results for the estimated initial length of the string as a function of the estimated decay time, are represented in  \cref{fig:local:decayartificial1} for several artificial loops of type \RNum{1} and \RNum{2}. We observe that the results for artificial loops of type \RNum{1} present a very similar behavior independently of the initial velocities, at least for the initial conditions considered in this work. The results can be simultaneously fitted to $\Delta\tilde{t}_\dec=A\left(\tilde{L}_0^{\text{est}}\right)^\alpha$, finding, $A=21(3)\times10^{-3}$ and $\alpha =2.027(25)$. A similar result is obtained for artificial loops of type~\RNum{2}, $A=33(12)\times10^{-3}$ and $\alpha =1.97(7)$. These results indicate that the mechanism underlying the decay of artificial loops is similar for both types, with $\alpha\approx 2$, and different from that of network loops, which have $\alpha\approx0.7$. 

Note that this conclusions are not affected if the measured collapse time is used to compute the lifetime of the loops instead of the extrapolated $\tilde{t}_\text{dec}$---see \cref{fig:local:decayartificial2}. In this case, we obtain significant differences for those loops generated with $v_1=0.3$ and $v_2=0.6$, which present a notorious double-line collapse event at the end of their decay. Although a universal behavior is no longer observed, loops with $v_1=0.3$ and $v_2=0.6$ still decay with an exponent $\alpha=2.16(3)$, close to that obtained in above.

%Note however that the actual rate of emission, related to the $A$ parameter, is very different between both types of artificial loops, 
%\begin{equation}
%\langle A^\RNum{1}\rangle =17(3)\times10^{-3}\,,\quad\quad\quad \langle A^\RNum{2}\rangle =8(5)\times10^{-3}\,.
%\end{equation}
%This implies that artificial loops of type \RNum{2} would live longer than loops of the other type for the same size. However, with our current technique we are not able to generate an isolated loop of this size.

A similar approach can be taken to study the lifetime of the loops as a function of their initial energy. We fit the evolution of the loop energy to
\begin{equation}
\tilde{E}_\str=B(\tilde{t}-\tilde{t}_\dec)^{q}\,,
\end{equation}
on the same range as for the previous analysis. Using the results, we can also determine an estimate of the initial string energy, $\tilde{E}_{\str,0}^{\text{est}}$. The results are fitted to $\Delta\tilde{t}_\dec=B\left(\tilde{E}_{\str,0}^{\text{est}}\right)^\beta$, obtaining in this case $B=5.4(1.0)\times10^{-3}$  and $\beta =1.988(26)$ for type \RNum{1}, and $B=3.4(2.5)\times10^{-3}$ and $\beta =2.08(13)$  for type \RNum{2}. Again, both exponents are compatible with $\beta\approx2$.

Using these results and \cref{eq:local:powerlocal}, we can compute an estimate for the particle emission power. Assuming $\alpha,\beta\approx 2$, we obtain that the power is inversely proportional to the length of the strings, \vspace{-\baselineskip}%$\Pphi\propto L^{-1}$, 

\indent\begin{equation}\label{eq:local:poweremissionparticlesartificial}
\Pphi=\frac{1}{2\sqrt{AB}}L^{-1}\,,
\end{equation}
in agreement with the results from~\rrcite{Matsunami:2019fss,Hindmarsh:2021mnl} for artificial loops. This implies that particle emission from these type of loops would be suppressed for longer loops. In particular, for a constant emission power of GWs, as predicted by NG and as we demonstrate in the next section, there would be some critical length, $L_\text{crit}$, above which the emission power of GWs will be higher to that of particles, in agreement to \rcite{Matsunami:2019fss}.

\begin{figure}[!b]
    \centering
    \begin{subfigure}{0.7\textwidth} 
    \centering
        \includegraphics[width=\textwidth]{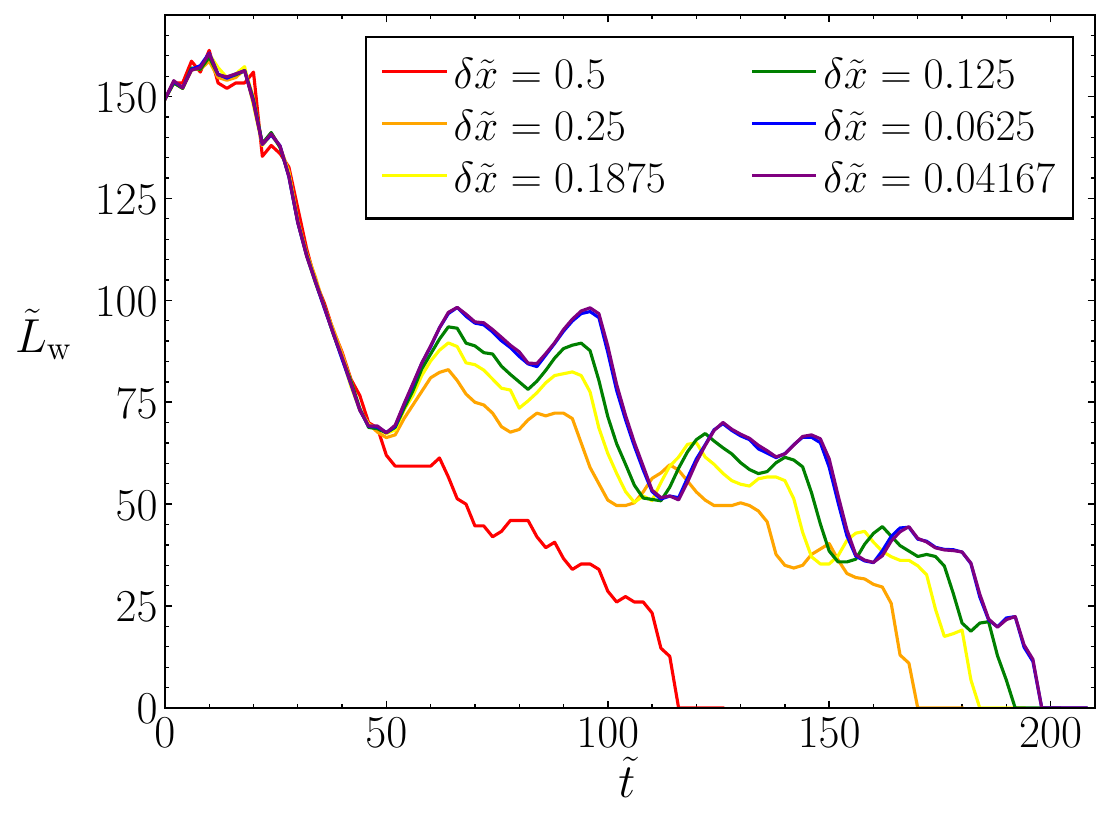}
    \end{subfigure}
    
    \caption{
        Time evolution of the length of an artificial loop of type \RNum{1}, generated with $v_1=0.3,v_2=0.6$ and $\sin\alpha=0.5$, in a lattice of size $\tilde{L}=56$, for varying UV resolution. The length is measured from the number of pierced plaquettes.%We observe how low resolution may lead to huge effects on the dynamics of the strings.  
        }
    \label{fig:local:decaylengthartificialUV}
\end{figure}

We also study possible discretization systematic effects on our results for artificial strings. We analyze the dynamics of the loops varying the UV resolution, by generating an equivalent initial configuration for several $\delta x$. The time evolution of the string length of artificial loops of type \RNum{1}, for different $\delta x$, is presented in \cref{fig:local:decaylengthartificialUV}. We observe a large dependence of the results on the resolution, with convergence only being reached for $\delta \tilde{x}\lesssim 0.1$. In particular, all simulations evolve similarly until the loop starts to oscillate, when it becomes clear that low resolutions vastly overshoots the decay rate. This justifies using $\delta\tilde{x}=0.1875$ in general for our study, as we observe systematic errors not bigger than $10$\%. This is also confirmed by our results in \cref{fig:local:decayartificial}, where no noticeable discrepancy can be found between loops with $\delta\tilde{x}=0.1875$ and $\delta\tilde{x}=0.125$. 

%Finally, we note that our results for network loops could be affected by finite-volume effects. In some cases, we study loops that are several times longer than the box size. This could imply a higher curvature of the loops to fit in the box, which may then radiate faster than would be expected for loops of their size. Note, however, that we do not observe evidence of this effect for the loop sizes studied, and so estimate that finite-volume effect on $\alpha$ and $\beta$ for network loops. Even if they were of a $\sim40\%$ size, we would still obtain $\gamma\lesssim 1$, meaning that the emission of particle radiatin is, at most, independent of the size of the loops.

\newpage\subsection{GW emission}\label{sec:local:GWresults}

We now study the emission of GWs from network and artificial loops of type \RNum{1}. The production of GWs is simulated following the procedure in \cref{sec:Cosmo:GWsimulation}. For the model under study in a flat background, the equations of motion of GWs are
\begin{equation}
\ddot{h}_{ij}-\partial_k\partial_k h_{ij}=\frac{2}{\mpl^2}\left\{2\Re[D_i\varphi (D_j\varphi)^*]-E_iE_j-B_i B_j\right\}^\TT\,.
\end{equation}
We define the fractional energy density of gravitational waves as
\begin{equation}\label{eq:local:GWfractionalenergystrings}
\Omega_\GW(k,t)=\frac{1}{\rho_\text{m}}\frac{\d\rho_\GW(k,t)}{\d\log k}\,,
\end{equation}
where we normalize with the total energy of the gauge and scalar fields, $\rho_\text{m}=\rho_\varphi+\rho_{\text{U}(1)}$. We also define the total energy of GWs by integrating over the whole spectrum, 
\begin{equation}\label{eq:local:GWenergystrings}
E_\GW=\rho_\text{m} L^3\int\Omega_\GW(k,t)\d\log k\,.
\end{equation}
We stress, again, that we are not considering backreaction of the GWs on the matter fields, which we justify below.

\begin{figure}[!p]
    \centering
    \begin{subfigure}{1\textwidth} 
    \centering
        \includegraphics[width=1\textwidth]{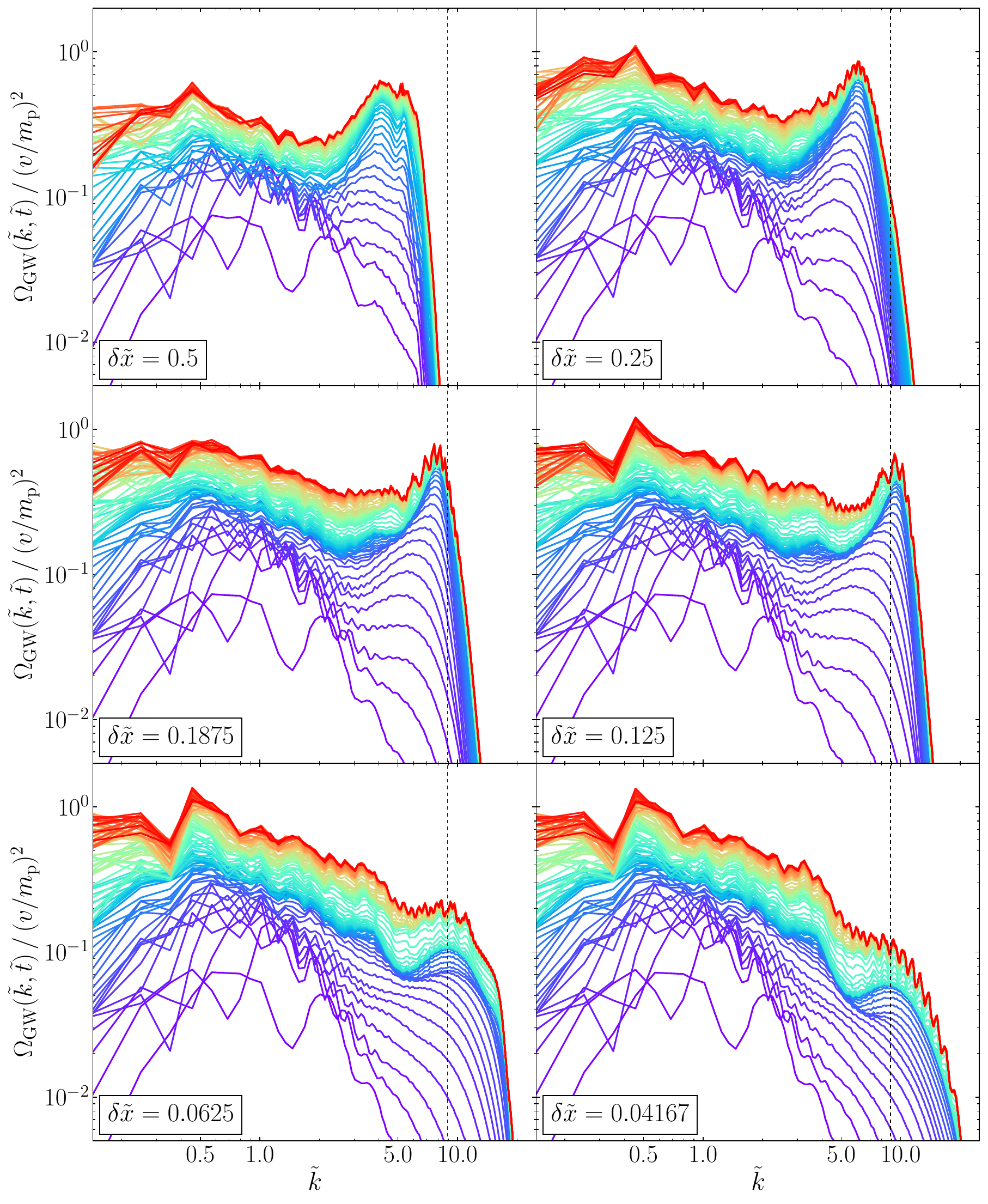}
    \end{subfigure}
   \caption{GW power spectra produced by an artificial loop of type \RNum{1}, generated with $v_1=0.3,v_2=0.6$ and $\sin\alpha=0.5$, in a lattice of size $\tilde{L}=56$ for varying UV resolution. The vertical dashed line indicates the scale of the string core, $\tilde{k}=2\pi/\tilde{r}_\text{c}$. Spectra are represented every unit of program time, going from early (purple) to late (red) times.} 
    \label{fig:local:GWUV}
\end{figure}

Given the UV dependence observed for the GW emission of global loops---see \cref{sec:global:resultsGWs}---and for the dynamics of local ones as discussed at the end of the previous section, we first study discretizations effects on the GW emission from local artificial loops. We make use of the same artificial loops presented in \cref{fig:local:decaylengthartificialUV}. The evolution of the power spectra is shown, for several values of $\delta\tilde{x}$, in \cref{fig:local:GWUV}, with early and late times represented in purple and red, respectively. We observe how the overall amplitude of the spectrum is suppressed for coarser lattices, as a result of the shorter lifetime of loops simulated with bad UV resolution. 

For all resolutions, furthermore, we observe the presence of a peak at IR scales, $\tilde{k}\sim0.4-0.5$. A second peak emerges in the UV for large $\delta x$ close to the scale of the core radius, $k_\text{c}=2\pi/r_\text{c}$, represented by a vertical dashed line. This peak, however, vanishes as the resolution is increased, indicating it is a lattice artifact. Deeper in the UV, the amplitude is exponentially suppressed. Overall, we observe that simulations with $\delta\tilde{x}\lesssim 0.1875$ agree up to scales $\tilde{k}_\text{cut}\sim2.5$. We perform our study with $\delta\tilde{x}= 0.1875$, and opt to compute the total energy density of GWs integrating up to this cutoff, as the contribution from the UV is very suppressed in this case, and even further suppressed for finer lattices. This cutoff will be used for both network and artificial loops. 

Regarding the IR coverage of the lattices, the spectra in \cref{fig:local:GWUV} seem to start decreasing for small scales, at $k\sim 0.3$. Ideally one would like to have better IR coverage. However, we are limited in this aspect by our simulation procedures. Network loops are typically of a length larger than the lattice size, and artificial loops are constructed to have a size similar to that of the lattice, since we want the secondary loop to vanish rapidly. Thus, our techniques do not allow us to easily increase the size of the lattice while keeping the strings of fixed length. We hope to come back to this point in the early future.

%Finally, we can use the results for the power spectrum of the finest resolution to determine the spectral index of the UV part of the spectrum. We present the result from a power-law fit in the bottom-left panel of \cref{fig:local:GWUV}, finding a result of the form $\propto k^{-0.56(2)}$. We note that this fall is much less steep than  NG predictions for cusp- or kink-dominated emission, expected to be proportional to $k^{-4/3}$ and $k^{-5/3}$, respectively.

\begin{figure}[!b]
    \centering
    \begin{subfigure}{1\textwidth} 
    \centering
        \includegraphics[width=1\textwidth]{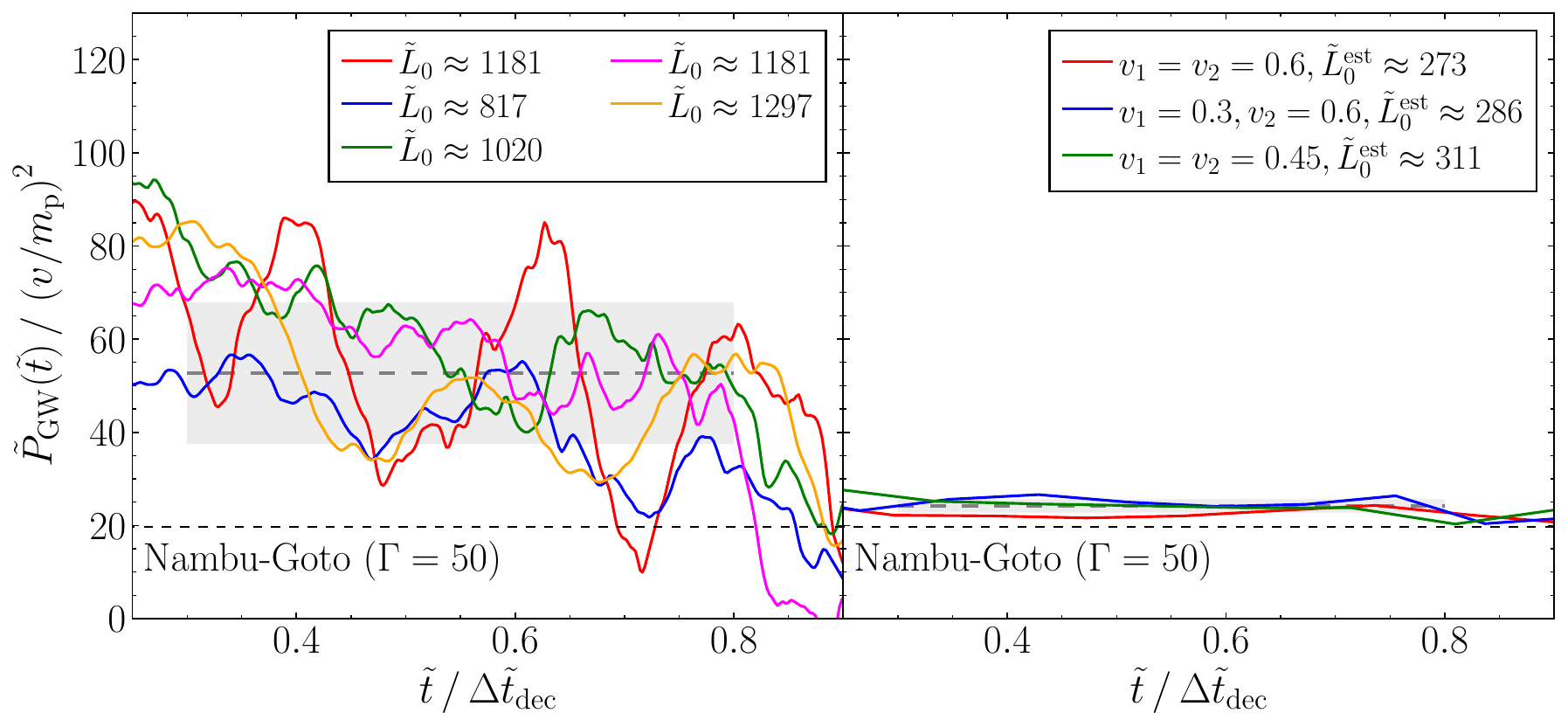}
    \end{subfigure}
   \caption{GW emission power for both network (left) and artificial loops of \mbox{type \RNum{1}} (right), computed using \cref{eq:global:rollingaverage} and \cref{eq:local:PowerAverageArtificial}, respectively. The grey band and line represents an average of the emission power, and we compare the NG predictions (horizontal dashed line), estimated with $\Gamma=50$ and $\mu=\pi v^2$. The power is represented as a function of the time since the emission of GWs started, normalized by the total decay time of each loop. Network and artificial loops are simulated with $\delta\tilde{x}=0.25$ and $\delta\tilde{x}=0.1875$, respectively.} 
    \label{fig:local:GWpower}
\end{figure}

We now present our result for the GW emission power of isolated loops. In the case of network loops, we follow the procedure used for global strings, and compute a rolling average of the total GW emission power, given in \cref{eq:global:rollingaverage}, using a window width $\tilde{T}=20$. We find this choice of $\tilde{T}$ to cancel fast oscillations, with very minor variations of the results if the window width is increased. Our results for network loops of different length are presented in the left panel of \cref{fig:local:GWpower} as a function of the time since we started measuring the GWs, normalized by the lifetime of each loop. In all cases, the emission power is roughly constant in time, with fluctuations that depend on the particular dynamics of the loops, and a minor decreasing trend during the life of the loop. At late times, the emission power rapidly drops down as the loop finally disappears.

From these results, we can obtain an average value of the emission power. In the range $\tilde{t}/\Delta\tilde{t}_\text{dec}\in[0.3,0.8]$, we obtain $\PGW=53(15)$ for network loops. This result is around five times smaller than the one obtained for global network loops. Comparing to NG expectations, $\tilde{P}_\text{GW}^\text{NG}\approx 20$, we observe that our result is still more than two times larger. %We note that performing the average including earlier times may need to incorrect results. The energy density of gravitational waves is only well defined if one takes into account all the  relevant frequencies of the system. T

The study of the GW emission power from artificial loops is more complicated. The main reason is the longer lifetimes of these loops, which can be dozens of times larger than  the half-box-light-crossing time of the lattice. For example, for the simulation presented in \cref{fig:local:decaylengthartificialUV,fig:local:GWUV}, $\Delta \tilde{t}_\HL=27$, while the string lives almost $500$ units of program time. This may have a severe impact in the study of GWs, since gravitational radiation propagates at the speed of light and can present interferences around the box. Indeed, we believe this to be the origin of the oscillations observed in the spectra in \cref{fig:local:GWUV} at late times.

To prevent said self-interferences, we take a different approach to obtain the GW emission power from artificial loops. We initialize the GWs to zero when the secondary loop disappears, and evolve them normally for a period of time $\Delta T_\text{GW}$, after which we measure the GW power spectrum. We  then reset the GWs to zero and repeat the procedure until the loop decays. The average emission power of GWs during each of these intervals is computed as
\begin{equation}\label{eq:local:PowerAverageArtificial}
P_\text{GW}=\frac{L^3 \rho_{\text{m}}}{\Delta T_\text{GW}}\int_0^{k_\text{cut}}\Omega_\text{GW}(k,t)\,\d\log k\,.
\end{equation}
The duration of the interval should be taken long enough to ensure all relevant frequencies of the system are captured, but small enough so that the emission power remains mostly constant and finite-volume effects are small. We find that using $\Delta \tilde{T}_\text{GW}=80-200$ leads to stable results, with discrepancies smaller than $10$\% from varying this quantity, and choose to use $\Delta \tilde{T}_\text{GW}=160$ in this work. We also focus only on artificial loops of \mbox{type \RNum{1}}, since those of type \RNum{2} have much shorter lifetimes. %, only a couple times longer that the chosen $\Delta \tilde{T}_\text{GW}$.

The results for the GW emission of artificial loops of type $\RNum{1}$ is presented in the right panel of \cref{fig:local:GWpower}. As in the case of network loops, the emission power remains mostly constant, with minor variations related to the oscillations of the loops. Averaging our results over the range $\tilde{t}/\Delta\tilde{t}_\text{dec}\in[0.3,0.8]$, we obtain $\tilde{P}_\text{GW}=24.2(1.4)$, which is very close to NG predictions.

We can finally compare these results to the particle emission power obtained in \cref{sec:local:particleproductionnetwork,sec:local:particleproductionartificial}. We find that GW are emitted with a constant power that is independent of the length of the strings, both for network and artificial loops, although with different emission power. Particle emission, on the other hand, shows a dependence on the length of the loops, which is very different between the two types of loops. 

For network loops, our results from a power-law fit indicate the the emission of particles increases slightly with the size of the loop, and so the emission of GWs is completely subdominant at large scales. Even if we assume a linear relation, which is also well reproduced by our data, we obtain, for reasonable values of $v$,

\begin{equation}
\frac{\PGW}{P_\varphi^\text{linear}}=\frac{53(15)}{10.3(7)}\left(\frac{v}{\mpl}\right)^2= 5.1(1.5)\left(\frac{v}{\mpl}\right)^2\,.
\end{equation}

\noindent Therefore, the emission of GWs is very suppressed compared to particle production for reasonable values of the vacuum expectation value, $v/\mpl\lesssim 3\times 10^{-5}$ (obtained using $\mu=\pi v^2$ combined with constrains of the string tension from the GWB\footnote{Contrains from the CMB require $v/\mpl\lesssim 10^{-3}$~\cite{Planck:2013mgr,Lazanu:2014eya,Lizarraga:2016onn}~\cite{Figueroa:2023zhu}.}). These results justify neglecting the backreaction of GWs on the strings.

 The opposite situation happens for artificial loops, for which particle emission depends roughly on the inverse of the length of the strings, $\Pphi\propto L^{-1}$, and so it is suppressed compared to GW radiation for long enough loops. Combining our results for the GW emission power and those for particle productions presented in \cref{sec:local:particleproductionartificial}---see \cref{eq:local:poweremissionparticlesartificial}---it is possible to obtain an estimate of the critical length at which the emission power of particles equals that of GWs, $\Pphi=\PGW$. We find
 
\begin{equation}
L_\text{crit}=2.8(3)\times\frac{\mpl^2}{\sqrt{\lambda} v^3}\sim\cO(1)r_\text{c}\left(\frac{v}{\mpl}\right)^{-2}\,.
\end{equation}

\noindent Given present constrains on the vacuum expectation values, $v/\mpl\lesssim 3\times 10^{-5}$~\cite{Figueroa:2023zhu}, the emission of GWs dominates for loops of a length-to-width ratio higher than $\sim 10^{9}$. 
For smaller loops, $L<L_\text{crit}$, particle emission is the main decay channel. For artificial loops with $L< L_\text{crit}$ neglecting the backreaction of the GWs is justified. %However, if one is interested in a model with large vacuum expectation value, $v\gtrsim \mpl^{2/3}$, this approximation is no longer valid. 

To summarize, we find that it is possible to create local loops (the artificial ones) for which the emission of GWs is the main decay route at cosmological scales. However, such loops are not observed to originate from a phase transition. For realistic loops forming in the early universe (network loops) our results indicate that the emission of GWs from string loops forming in the early universe is very suppressed compared to particle production at all scales. %that we have not found any evidence that such long-lived loops could originate from early-universe phase transitions, and so the emission of GWs from early universe local string networks is expected to be vastly suppressed. %Finally, if one would have liked to simulate artificial loops of a size at which GW emission overcame that of particles, one would have to take backreaction into account.

\newpage\section{Conclusions}\label{sec:local:conclusion}

In this chapter, we have presented results on a lattice study of the decay of local string loops into particles and GWs~\cite{Baeza-Ballesteros:stringsinprep}. We perform this study for two types of loops: network loops generated from the decay of string networks, with length-to-width ratios up to $L_0/r_\text{c}\lesssim 3500$, and artificial loops that form from the intersection of infinite strings, with $L_0/r_\text{c}\lesssim 640$. In both cases, we have found that the emission power of gravitational radiation is roughly constant in time, independently of the string length. This time-independent emission is in agreement with NG expectations, although with different magnitude, being around three times larger for network loops than for artificial ones.

 The emission of particles, on the other hand, shows two different dependencies on the string length for each type of loops. In the case of network loops, the results from a power-law fit indicate a residual positive length dependence, meaning that the emission of particles from network loops increases with their length. Our data, however, is also compatible with a linear fit, in which case the emission power of particles would be constant. In both cases, anyways, the production of GWs from loops is highly suppressed compared to particle emission at large scales. By contrast particle emission is inversely proportional to the length for artificial loops, $\Pphi\propto L^{-1}$, meaning that GW production dominates for loops longer than a critical length, $L_\text{crit}\sim \mathcal{O}(1) r_\text{c}(v/\mpl)^{-2}$. 

Our results have important consequences for the predicted GW background from a network of local strings. Studies of local loops are commonly based on the NG approximation, assuming that particle emission is absent at cosmological scales and the decay is dominated by GW emission. We have found that it is possible to fabricate loops for which this remains the case, as we have seen for artificial loops. However, this is not true for loops originating from phase transition. Our results for network loops clearly indicate that particle emission is not suppressed, but instead is the primary decay route for local string loops of cosmological size. 

As a next step, we plan to use the results presented in this chapter for the GW emission from local loops to obtain an estimate of the GW background generated from a network of local strings, by combining the results with predictions for the loop number density computed at cosmological scales from NG simulations~\cite{Ringeval:2005kr,Blanco-Pillado:2013qja,Klaer:2017qhr,Martins:2018dqg,Auclair:2019zoz,Auclair:2021jud}. We expect this to have an impact in suppressing the prediction for the GW background for local string networks, compared to estimations that ignore particle production, as we have found that the production of GWs is subdominant for realistic loops.

\clearpage
\makeatletter
\setlength{\@fptop}{0pt}
\setlength{\@fpbot}{0pt plus 1fil}
\makeatother

\backmatter
\bookmarksetup{startatroot}
\rhead[ \leftmark]{\bfseries \thepage}
\lhead[{\bfseries \thepage}]{ \leftmark}
\chapter*{Conclusions and outlook}
\thispagestyle{plain} 
\addcontentsline{toc}{chapter}{Conclusions and outlook}
\markboth{Conclusions and outlook}{}
\chaptermark{Conclusions and outlook}

The use of lattice techniques to study field-theory phenomena that cannot be captured by perturbative methods has become widespread during the last decades, tied to the development of large supercomputing facilities. Lattice computations have been successfully applied to study the strong interactions at low energies, for which perturbative approaches fail due to the phenomena of asymptotic freedom and confinement, and also to investigate non-linear processes taking place during the early universe, for which fluctuations of the fields cannot be treated as small perturbations. The work presented in this doctoral thesis is focused on the application of lattice techniques, complemented by other analytical approaches, to study two very different physical situations: hadron interactions in quantum chromodynamics (QCD), and the dynamics and emission of particles and gravitational waves (GWs) from cosmic string loops that could arise in the early universe.

\section*{Hadron interactions from lattice QCD}

Lattice QCD simulations allow one to study two- and three-particle scattering using the so-called quantization conditions, which relate the finite-volume energies of multiparticle states to infinite-volume scattering observables. Numerical computations can moreover be complemented by other analytical approaches, such as chiral perturbation theory (ChPT) or the limit of large number of colors, $\Nc$. \Cref{part:QCD} of this dissertation focuses on the study of hadron interactions, with special emphasis on the synergy between lattice QCD, ChPT and the large $\Nc$ limit.

\Cref{sec:largeNpions,sec:largeNmesons} are devoted to the study of the interactions between two mesons as a function of $\Nc$. \Cref{sec:largeNpions} presents the results from \rrcite{Baeza-Ballesteros:2022azb,Baeza-Ballesteros:2021nxu}, in which pion-pion interactions near threshold are investigated. We performed lattice simulations with $\Nc=3-6$ in a theory with four degenerate quark flavors, $\Nf=4$. We focused on two scattering channels: the $SS$ channel, which is analogous to the isospin-two channel of two-flavor QCD, and the $AA$ channel, which only exists for $\Nf\geq 4$. We used the two-particle quantization condition to match our results of the finite-volume energies to ChPT predictions including the $\eta'$ particle, and constrained the $\Nc$ scaling of the relevant low-energy constants. The results from this fit are presented in \cref{fig:largeNpions:finalfit}. We found that the leading $\Nc$ contribution is unnaturally small compared to subleading $\Nc$ corrections. %This work is a clear example of the existing synergy between lattice simulations, ChPT and the large $\Nc$ limit.

In addition, the $AA$ channel was found to have attractive interactions. This, together with its flavor quantum number makes it a candidate to contain a tetraquark resonance. This is supported by recent experimental discoveries at LHBb~\cite{LHCb:2020bls,LHCb:2020pxc,LHCb:2022sfr,LHCb:2022lzp}, which found tetraquark particles that would have the quantum numbers of the $AA$ channel in a theory with $\Nf=4$ degenerate flavors. All this motivated us to extend the study of meson-meson interactions to higher energies, with the main goal of shedding light on the nature of tetraquarks at large $\Nc$~\cite{Baeza-Ballesteros:largeNinprep,Baeza-Ballesteros:2024ogp}. In \cref{sec:largeNmesons}, we have present results of this ongoing work in which we have studied the scattering of two mesons as a function of $\Nc$. We have managed to characterize the scattering process of two pions as a function of $\Nc$ for both the $SS$ and the $AA$ channels, as shown in \cref{fig:largeNmesons:PSSS,fig:largeNmesons:PSAA}. While we have found no clear evidence of a tetraquark resonance in the $AA$ channel, the presence of a virtual bound state for $\Nc=3$ cannot be discarded and needs to be further investigated. We have also analyzed meson interactions in the $AS$ channel, which contains odd partial waves, finding that interactions are very weak in this channel.

Following this work, the study of the $\Nc$ dependence of other scattering processes is compelling. One possibility is the case of the interactions of two pions in the isospin-one channel, well-known to contain the $\rho(770)$ resonance. This investigation would allow us to constrain the $\Nc$ dependence of the decay width of this particle, $\Gamma$, expected to scale as $\Gamma\sim1/\Nc$. This work would open the door to use the large $\Nc$ dependence to characterize the nature of other resonances, such as the $\sigma(550)$, which has been proposed to be a tetraquark state~\cite{Maiani:2004uc}. %and discern between standard hadrons and other exotic states.

\Cref{sec:pipipiKmatrix,sec:isospinKmatrix} are devoted to the  study of three pion interactions in ChPT. Early lattice studies of three pions in the isospin-three channel found large discrepancies against leading-order (LO) ChPT predictions for the three-particle divergence-free $K$-matrix, $\Kdf$. This is an intermediate scheme-dependent quantity that appears in the relativistic-field-theory (RFT) three-particle quantization condition and is related to the scattering amplitude via some integral equations. In \cref{sec:pipipiKmatrix,sec:isospinKmatrix},  results for $\Kdf$ at next-to-leading order (NLO)  in ChPT are presented.

%, which motivated the investigation of the size of NLO corrections.  computation of the three-particle divergence-free $K$-matrix, $\Kdf$, for three pion systems at next-to-leading order (NLO) in ChPT. This is an intermediate scheme-dependent quantity that appears in the relativistic-field-theory (RFT) three-particle quantization condition, and is related to the scattering amplitude via some integral equations. 

\Cref{sec:pipipiKmatrix} presents the results from \rcite{Baeza-Ballesteros:2023ljl}, in which $\Kdf$ was determined at NLO in ChPT for the isospin-three channel.% in which $\Kdf$ for this channel was computed at NLO in ChPT, starting from the results for the scattering amplitude from \rcite{}. 
We found that, when working at NLO in ChPT, the relation between $\Kdf$ and the three-pion scattering amplitude, $\cM_3$, is reduced to a simple algebraic equation. We also developed a series of techniques to determine the various contributions to $\Kdf$ from different parts of amplitude and from subtractions. Our results showed that NLO corrections are indeed large, and help to reconcile ChPT predictions and lattice results---see \cref{fig:pipipiKmatrix:K0K1chpt}. The results, however, could also indicate that ChPT fails to converge in the range of pion masses studied. We also analyzed the convergence of the threshold expansion of $\Kdf$, which is typically the parametrization of $\Kdf$ used when fitting to lattice results. We found that, for energies in the three-pion elastic regime, the difference between the threshold-expanded and the exact result is smaller that 20\%.

Following this work, \cref{sec:isospinKmatrix} discusses the results from \rcite{Baeza-Ballesteros:2024mii}, in which we determined the three-pion $K$-matrix up to NLO in ChPT for all  isospin channels. As in the case of isospin three, we found that NLO corrections to LO are in general large. Our results for $\Kdf$ are depicted in \cref{fig:isospinKmatrix:results}, and provide insight on the convergence of the chiral expansion for three-particle quantities. They will be useful to compare against future lattice studies of three-pions at non-maximal isospin, and also to parametrize $\Kdf$ around threshold. Note, however, that the presence of resonances in these channels limits the range of validity of ChPT.

The work presented in \cref{sec:pipipiKmatrix,sec:isospinKmatrix} paves the path to computing $\Kdf$ in ChPT for  more complicated systems. A notorious example is the scattering of pions and kaons, or that of three kaons, which have already been studied on the lattice at maximal isospin~\cite{Blanton:2021llb,Draper:2023boj}. However, the determination of the corresponding scattering amplitudes at NLO  in ChPT is yet not available.

\Cref{part:QCD} of this dissertation is completed by \cref{sec:O3model}. This chapter presents ongoing work~\cite{Baeza-Ballesteros:O3inprep} on the study of two- and three-particle scattering in the (1+1)-dimensional O(3) non-linear sigma model. This model is commonly used as a toy model of QCD, as both theories share many qualitative features, such as asymptotic freedom or the existence of a low-lying spectrum of isospin-one particles. Moreover, the $S$-matrix of the O(3) model is known analytically, making the model ideal to test different formalisms. Our ultimate goal is to test the RFT three-particle formalism by performing a direct comparison between analytical predictions and lattice computations, to ensure that the different approximations made in the derivation of the formalism do not have a large numerical impact.

With the work presented in \cref{sec:O3model}, we take a first step in this direction. We have developed a three-particle generalization of the cluster algorithm, with which we have determined the finite-volume energy spectra of the isospin-two and -one channels for two particles, and the isospin-three, -two and -one channels for three particles. These energies have been extrapolated to the continuum using lattices with different lattice spacing and finely tuned physical volume. The results for two particles were compared against exact analytical predictions, finding good agreement to analytical results, raising the confidence in our simulating procedure. 

Results for three-particles, on the other hand, have been compared against analytical predictions made under the assumption that $\Kdf=0$,  computed using an adapted version of the the RFT formalism to 1+1 dimensions that we have developed. These predictions, together with our lattice results, are presented in \cref{fig:O3model:energiesthreeparticlesI3,fig:O3model:energiesthreeparticlesI2,fig:O3model:energiesthreeparticlesI0}. We have observed a discrepancy between the analytical predictions and the lattice results, indicating that $\Kdf$ is indeed non-zero. This work has also allowed us to gain insight on the intricacies of the RFT formalism, especially in relation to the role of the cutoff function and possible numerical instabilities. The future plan is to use our lattice results to constrain  $\Kdf$, and eventually compare them to an analytical prediction made from the factorizable $S$-matrix. This last point, however, would first require to adapt the integral equations that relate $\Kdf$ to $\cM_3$ to the (1+1)-dimensional world.

Beyond the O(3) model, the knowledge gained about the study of three-particle systems and the application of the  RFT formalism opens the door to investigate more complicated systems in QCD. Of particular interest are, for example, systems of three-pions at non-maximal isospin, which could allow one to characterize the nature of different resonances, such as the $\omega(782)$ resonance in the isospin-zero channel. Other states of three mesons are also relevant, such as the $DD\pi$ system~\cite{LHCb:2021auc}, expected to contain the $T_{cc}(3875)$ tetraquark. Finally, the application of lattice techniques to study light nuclei, such as the tritium of helium-3, is also very compelling.

\section*{Cosmic string loops from lattice simulations}

The second part of my doctoral research has been presented in \cref{part:strings}. This is focused on the application of classical-field-theory lattice techniques to study non-linear dynamics in an expanding background. In particular, this part of the thesis is devoted to analyzing the evolution of cosmic string loops using lattice simulations, capturing simultaneously the emission of particles and GWs. Cosmic strings are one-dimensional topological defects predicted to originate in the early universe by many theories beyond the Standard Model. For example, global strings that present long-range interactions are expected to form in axion and dark matter scenarios~\cite{Hui:2021tkt}, while local strings, having short-range interactions, are predicted by some grand unified theories~\cite{Copeland:2009ga,Copeland:2011dx}. Cosmic strings are expected to emit GWs which could potentially be detected by current and future GWs experiments. The numerical studies of string loops presented in this dissertation have been performed using the \CosmoLattice package. As part of my doctoral work, I have contributed to the development of this software~\cite{GWmodule:2022,GWmodule:2023}. %During my PhD, I also worked on the development of modules to simulate the emission of GWs from scalar and Abelian gauge theories, which can be found online. %Comentar sobre la relevance de GW

\Cref{sec:global} presents the results from \rcite{Baeza-Ballesteros:2023say}. This chapter focuses on cosmic string loops of global type, which are captured by a model containing a single complex scalar field with a Mexican-hat potential. Using field-theory lattice simulations, we studied the decay of this type of string loops into particles and GWs. Our results showed that, independently of the shape and length of the loops, and of the initial conditions used to generate the loops, the emission power of particles and GWs is approximately constant. This can be seen in \cref{fig:global:lengthdecay}, where we showed that the lifetime of the loops depends linearly on their size, and in \cref{fig:global:GWpower}. We found that the emission power of GWs is suppressed compared to particle radiation, $\PGW/\Pphi\approx\cO(10)(v/\mpl)^2$, with $\mpl\approx 2.44\times 10^18$ GeV the reduced Planck mass and $v$ the vacuum expectation value of the scalar field, constrained by experimental observations to values $v/\mpl\leq 10^{-6}$~\cite{Lopez-Eiguren:2017dmc,Benabou:2023ghl}. %we found no evidence of any logarithmic enhancement  of the GW radiation.

In \cref{sec:local} we present results  for local string loops~\cite{Baeza-Ballesteros:stringsinprep}, which we capture using the Abelian-Higgs model. Our conclusions are very different from the case of global strings. We found that the emission of particles from the strings depends on their shape and features. Loops originating from the decay of string networks, which are expected to possess features similar to loops arising in the early universe, have a particle emission power that seems to depend slightly on the size of the loops. This can be seen from \cref{fig:local:decaynetwork}, where we showed that the lifetime of the loops scales roughly as $\Delta t_\text{dec}\propto L^{-0.7}$ with their initial length, meaning that particle emission increases with the loop size. On the contrary, loops that are generated from the intersection of infinite strings have a lifetime that roughly scales quadratically with their initial length---see \cref{fig:local:decayartificial}---so that the emission of particles is suppressed for longer loops. In both cases we found that the emission of GWs is mostly independent of the size of the loops, as shown in \cref{fig:local:GWpower}. This means that it is possible to fabricate loops for which the emission of GWs is the main decay route at cosmological scales. However, these were not found to form from networks, and loops that originate naturally in the early universe after a phase transition decay mainly into particles, with a negligible emission of GWs.

The work presented in these two chapters represents a first step in the determination of the GW background (GWB) from a network of strings, which is of vital importance to properly interpret  future experimental detections. Combining our results with predictions for the loop number density in the early universe~\cite{Ringeval:2005kr,Blanco-Pillado:2013qja,Klaer:2017qhr,Martins:2018dqg,Auclair:2019zoz,Auclair:2021jud} will make it possible to determine the expected GWB emitted by a network of string loops, without relying on effective predictions of the GW emission. In addition, predictions for the GWB from a network of local strings could also be compared to the results from a full lattice simulation of the network dynamics that incorporates  the emission of GWs, which has not been performed before.%  local string could also be compared to %This technique would allow to overcome the limited separation of scales that can be achieved in field-theory simulations, and also avoid relying on approximations for the GW emission power from loops.% Another intetesting venue is the determination of the 

\section*{Final remarks}

Numerical lattice simulations are one of the main tools to address \mbox{problems} in high-energy physics from first principles. These include, for example, the use of lattice QCD to investigate the hadron spectrum, or the application of classical-field-theory lattice techniques to predict the GW emission from hypothetical early-universe phenomena. Throughout this dissertation, work in these two different topics has been presented. On the QCD side, two- and three-particle interactions have been investigated using numerical techniques complemented by other approaches such as ChPT and the large $N_\text{c}$ limit. The results presented contribute to the understanding of the non-perturbative regime of QCD, especially in relation to meson interactions and the possible existence of exotic resonances, such as tetraquarks. They also represent a starting point to study more complex systems, such as resonant three-meson systems and, eventually, light nuclei. The cosmology side of this dissertation, on the other hand, has been focused on analyzing the emission of particles and GWs from cosmic string loops using field-theory simulations. We have found that, for the type of the string loops that may form in the early universe, the production of GWs is very suppressed compared to particle emission. We leave for future work a prediction of the GWB, which we  believe  will be suppressed compared to present estimates.

%Whether for the study of the strong interactions or to investigate early universe phenomena, numerical lattice simulations have become a vital technique to answer the increasingly complicated questions in high energy physics and reach the precision level required by experiment, which cannot be reached with other approximate techniques. 

 %Lattice QCD 

\chapter*{Resumen}
\thispagestyle{plain} 
\addcontentsline{toc}{chapter}{Resumen}
\markboth{Resumen}{}
%\chaptermark{Conclusions and outlook}

La teoría de campos es el marco con que se describen la gran mayoría de fenómenos físicos a altas energías. Es el caso del modelo estándar de partículas, basado en la mecánica cuántica y la relatividad especial, y del modelo cosmológico estándar, que nace del principio cosmológico y la teoría de la relatividad general. En muchos casos, es posible realizar predicciones precisas mediante el uso de métodos perturbativos. Estas técnicas permiten, por ejemplo, estudiar en detalle procesos electromagnéticos en el modelo estándar, los cuales se pueden describir como una serie de potencias en la constante de estructura fina. Del mismo modo, ciertos fenómenos del universo primigenio, pueden analizarse mediante métodos perturbativos, tratando las fluctuaciones de los campos de materia y de la métrica como pequeños frente a un valor promedio homogéneo.

En otros contextos, sin embargo, no es posible aplicar métodos perturbativos. Uno de los ejemplos más conocidos es el estudio de las interacciones fuertes a baja energía mediante la cromodinámica cuántica (QCD, por sus siglas en inglés). A bajas escalas, debido al fenómeno de libertad asintótica, la constante de interacción fuerte crece, invalidando el uso de la aproximación basada en una serie de potencias de esta constante. En el universo primigenio, asimismo, muchos fenómenos están caracterizados por grandes fluctuaciones y no linealidades, que impiden el uso de métodos perturbativos. Algunos ejemplos son la producción de partículas durante y tras el fin de la época inflacionaria, o  la emisión  consiguiente de ondas gravitacionales (GW).

El uso de simulaciones numéricas en el retículo (o \textit{lattice}, en inglés), es una de las técnicas más extendidas para estudiar regímenes no perturbativos y/o no lineales. En el caso de QCD, la formulación en la \textit{lattice}, conocida como \textit{lattice QCD}, permite resolver la integral de camino mediante técnicas Monte Carlo.  Por otro lado, utilizando simulaciones de campos clásicos es posible estudiar la evolución temporal de los campos de materia y la producción de GW durante los primeros instantes del universo.

Esta tesis doctoral se centra en la aplicación de técnicas de \lattice al estudio de dos temas tan dispares como son la física hadrónica y el universo primigenio. En relación con el primero de estos temas, el trabajo presentado en la primera parte de la tesis se centra en el estudio de las interacciones entre dos y tres partículas, utilizando simulaciones de \lattice QCD junto a otros métodos como son las teorías efectivas---más concretamente la teoría quiral de perturbaciones---y el límite de gran número de colores. En la segunda parte de esta tesis, el trabajo presentado versa sobre la evolución y emisión de GW por parte de cuerdas cósmicas, estructuras unidimensionales que podrían formarse en el universo primigenio de acuerdo con diversas extensiones del modelo estándar.

\section*{Parte 1. Interacciones entre hadrones \mbox{mediante} técnicas de \textit{lattice}}

\subsection*{QCD a baja energía}

La cromodinámica cuántica es la teoría cuántica que describe las interacciones fuertes entre quarks y gluones~\cite{Fritzsch:1973pi}. Esta fuerza es responsable de mantener protones y neutrones unidos en núcleos atómicos, así como de la existencia de otros muchos hadrones insetables, con tiempos de vida media de tan solo una fracción de segundo. La teoría se basa en el grupo no abeliano SU($\Nc$), donde $\Nc$ es el número de colores, siendo $\Nc=3$ en el mundo real. La teoría contiene $\Nf$ sabores de quarks, representados por campos fermiónicos, y los mediadores de la fuerza fuerte, llamados gluones, dados por el campo \textit{gauge}. 

Una de las características distintivas de QCD es la propiedad de libertad asintótica~\cite{Gross:1973id,Politzer:1973fx}. A bajas energías, la constante de interacción fuerte crece sin límite. Esto impide la aplicación de métodos perturbativos a energías por debajo de la escala de QCD, $\Lambda_\text{QCD}$, pero tiene otras consecuencias de interés. La más importante es el fenómeno de confinamiento. A baja energía los quarks y gluones no se encuentran aislados en la naturaleza, sino que se hallan confinados formando hadrones. La mayor parte de los hadrones conocidos son mesones, formados por un quark y un antiquark, o bariones, formados por tres quarks~\cite{PDG:2020}. Durante las últimas décadas, no obstante, otros estados con una composición más exótica han sido descubiertos experimentalmente, como los llamados tetraquarks, formados por dos quarks y dos antiquarks~\cite{Belle:2003goh,Maiani:2004vq,LHCb:2020bls,LHCb:2020pxc,LHCb:2022sfr,LHCb:2022lzp,PDG:2020}.

Junto con la libertad asintótica y el confinamiento de quarks y gluones a bajas energías, otra de las propiedades características de QCD es la ruptura espontánea de la simetría quiral. En ausencia de masa, la acción de QCD es invariante bajo transformaciones globales del grupo de simetría,
\begin{equation*}
G=\text{SU}(\Nf)_\text{V}\times\text{SU}(\Nf)_\text{A}\times\text{U}(1)_\text{V}\times\text{U}(1)_\text{A}\,.
\end{equation*} 
A bajas energías, sin embargo, el vacío de QCD no respeta esta simetría, sino solo la parte vectorial, $H=\text{SU}(\Nf)_\text{V}\times \text{U}(1)_\text{V}$, que se corresponde con la simetría de isospín y el numéro bariónico. Los hadrones se pueden clasificar en representaciones irreducibles del grupo de isospín. El mejor ejemplo  es el conocido \textit{eightfold way}~\cite{Gell-Mann:1961omu,NEEMAN1961222}, que explica las similitudes observabas entre mesones pseudoescalares. Estos patrones se mantienen incluso en presencia de masa de los quarks, siempre que estas sean más pequeñas que $\Lambda_\text{QCD}$.

Al mismo tiempo, la simetría $\text{U}(1)_\text{A}$ está adicionalmente rota por efectos cuánticos, es decir, es anómala. Esto implica que el mesón singlete, la $\eta'$, tiene una masa mucho mayor que el resto de mesones pseudoescalares. La relación entre las masas viene dada por la fórmula de Witten-Veneziano~\cite{Witten:1979vv,Veneziano:1979ec}.

Debido a estas propiedades de QCD, su estudio analítico a bajas energías es especialmente complejo. Uno de los métodos más exitosos es la formulación de QCD en la \textit{lattice}~\cite{Wilson:1974sk,Wilson:1975id}, basada en el cálculo mediante técnicas de Monte Carlo de la integral de camino de QCD en un volumen finito y tiempo Euclídeo, utilizando una discretización del espacio-tiempo.

\textit{Lattice} QCD permite calcular funciones de correlación entre distintos operadores, a partir de las cuales se puede extraer información sobre observables de QCD.  En particular, las funciones de correlación a dos puntos permiten extraer las energías de estados de una o más partículas con los números cuánticos del operador en cuestión. Por ejemplo, las energías de los estados de dos o más partículas permiten extraer información sobre sus propiedades de dispersión.

Complementando las técnicas de \textit{lattice}, existen otras alternativas que permites estudiar las interacciones fuertes a baja energía. Una de ellas es el límite de gran número de colores, o límite the ‘t Hooft~\cite{tHooft:1973alw}, en el cual $\Nc$ se hace tender a infinito, manteniendo $\Nf$ constante. En este límite, QCD se simplifica manteniendo muchas de sus propiedades no perturbativas, como la libertad asintótica y la ruptura espontánea de la simetría quiral.  

Mediante el estudio de diagramas de Feynman en este límite, es posible caracterizar la dependencia con $\Nc$ y $\Nf$ de numerosos observables. Por ejemplo, se puede demostrar que las interacciones entre varios mesones están suprimidas a medida que $\Nc$ crece, y que QCD se convierte en una teoría de resonancias libres~\cite{tHooft:1973alw,Witten:1979kh,Coleman:1980nk}. Sin embargo, estimar las correcciones subdominantes al límite de gran $\Nc$ es muy complicado analíticamente. Las técnicas de \textit{lattice}, por otro lado, nos permiten cuantificarlas directamente, simulando a diferentes $\Nc$. Estas técnicas también nos permiten explorar otras preguntas abiertas en el límite de gran $\Nc$, como la posible existencia de hadrones exóticos, alrededor de la cual ha habido bastante polémica recientemente~\cite{Witten:1979kh,WeinbergTetra,Coleman:1980nk}.

Otra alternativa que permite el estudio analítico de QCD a baja energía es el uso de teorías efectivas~\cite{Weinberg:1978kz}, en particular la paradigmática teoría quiral de perturbaciones (ChPT)~\cite{Weinberg:1978kz,PhysRevLett.17.616,Gasser:1983yg,Gasser:1984gg}, que describe el régimen no perturbativo de QCD utilizando los mesones pseudoescalares como grados de libertad. La acción que describe las interacciones entre ellos contiene todos los términos compatibles con las simetrías de QCD, que dependen de ciertos parámetros, $L_i$, llamadas constantes de baja energía, cuyo valor ha de ser determinado experimentalmente o a partir de simulaciones \textit{lattice}. ChPT permite la realización de predicciones analíticas en potencias de $\Mpi^2/\Fpi^2$ o $p^2/\Fpi^2$, donde $\Mpi$ y $\Fpi$ son la masa y la constante de decaimiento del pion,  respectivamente, y $p$ indica su momento.

ChPT también permite estudiar el límite de gran $\Nc$~\cite{Gasser:1984gg,DiVecchia:1980yfw,Rosenzweig:1979ay,Witten:1980sp,Kawarabayashi:1980dp,Leutwyler:1996sa,Herrera-Siklody:1996tqr,Kaiser:2000gs}, aunque requiere de ciertas modificaciones. En particular, es necesario incluir el mesón singlete $\eta'$ junto al resto de mesones, ya que esta vuelve degenerada con el resto de mesones pseudoescalares a gran $\Nc$, e incluir el número de colores en el contaje de potencias de la expansion quiral.

\subsection*{Interacciones entre hadrones}

Los quarks y los gluones no se observan aislados en la naturaleza, sino formando estados compuestos llamados hadrones, cientos de los cuales han sido observados experimentalmente~\cite{PDG:2020}. La gran mayoría de estas partículas son inestables, con tiempos de vida medio tan pequeños como $10^{-22}$ segundos. Estos estados no pueden observarse directamente, sino que sus propiedades se infieren a partir de los productos de su decaimiento.

En general, se define un proceso de dispersión (o \scattering, en inglés), como aquel en que las partículas de un estado asintótico inicial, separadas inicialmente por distancias macroscópicas, se aproximan e interaccionan, dando lugar a un estado final generalmente diferente. La probabilidad de transición entre dos estados cualesquiera viene caracterizada por la amplitud de dispersión, $\cM$. 

En el caso de dos partíclas, dicha amplitud tiene una estructura analítica bien definida. En particular, contiene una serie de cortes para valores reales de la energía, relativos a los diferentes umbrales en los que diferentes estados finales de varias partículas aparecen. La dependencia de estos cortes con la energía viene caracterizada por las propiedades cinemáticas del sistema, y pueden sustraerse analíticamente. La parte restante de la amplitud se conoce como matriz $K$, $\cK_2$, la cual es una función meromorfa de las variables cinemáticas. La ubicación de polos en $\cK_2$ indica la presencia de estados intermedios en el proceso de dispersión. Polos en el eje real por debajo del umbral de dos partícular se corresponden con estados ligados de estas partículas, mientras que polos en el semiplano inferior en la segunda hoja de Riemann  indican la presencia de resonancias.

El caso de tres partículas es más complejo, ya que las interaciones están dadas no solo por interacciones de corto alcance de tres cuerpos, sino también por interacciones sucesivas de dos partículas. Al igual que en el caso anterior, es posible definir una matriz $K$ que contiene información únicamente sobre las interacciones de corto alcance de las tres partículas. Dadas $\cM_2$ y $\cM_3$, la matriz $K$ se obtiene resolviendo ciertas ecuaciones integrales, las cuales sustraen divergencias relacionadas con procesos intermedios de dos partículas. La matriz $K$ resultante, no obstante, depende de la regularización utilizada para tratar las dichas divergencias. El trabajo presentado en esta tesis utiliza una definición de la matriz $K$ propuesta en el contexto del formalismo de teoría de campos relativista (RFT) de tres partículas~\cite{Hansen:2014eka,Hansen:2015zga}, llamada matriz $K$ libre de divergencias o $\Kdf$.

En QCD, los procesos de interacción de dos y tres partículas pueden clasificarse en diferentes canales de isospín. En el caso de piones, partículas de isospín uno, las interacciones de dos partículas pueden ocurrir en tres canales de isospín ($\Ipp=2,1,0$) mientras que interacciones de tres piones tienen lugar en cuatro canales diferentes ($\Ippp=3,2,1,0$). En este último caso, algunos de los canales tienen multiplicidad mayor a uno, correspondiéndose con los posibles canales en que interacciones intermedias entre dos pions pueden ocurrir. Por ejemplo, en el canal de $\Ippp=2$, dos pions pueden interaccionar con $\Ipp=2$ ó $\Ipp=1$.

Una opción para describir las interacciones entre varios piones es el uso de ChPT. En el caso de dos piones con isospin dos, las predicciones para la amplitud de dispersión se conocen hasta segundo order (\textit{next-to-next-to-leading order}, NNLO)~\cite{Weinberg:1978kz,Gasser:1983yg,Bijnens:1995yn}. A orden cero (\textit{leading-order}, LO), estas reproducen los resultados experimentales de forma muy precisa, indicando un buen comportamiento de la expansión. En el caso de tres piones en isospín tres, la situación es muy diferente. La matriz $K$ de tres piones puede determinarse en ChPT a LO a partir de la amplitud de dispersión~\cite{Blanton:2019vdk,Bijnens:2021hpq,Bijnens:2022zsq} mediante relaciones algebraicas obtenidas a partir de las ecuaciones integrales. Sin embargo, se observan grandes discrepancias con resultados obtenidos en la \lattice~\cite{Blanton:2019vdk,Blanton:2021llb}, indicando posiblemente correcciones importantes procedentes de órdenes superiores en ChPT.

Otra posibilidad para analizar las interacciones de varias partículas en QCD es el uso de simulaciones \textit{lattice}. El estudio directo de procesos de dispersión no es posible en la \textit{lattice}, ya que estos ocurren en tiempo real y no se pueden definir estados asintóticos en volumen finito. Sin embargo, las propiedades de dispersión en volumen infinito se pueden determinar en la \textit{lattice} a partir del espectro de energía en volumen finito, usando las llamadas condiciones de cuantización.

En el caso de dos partículas, la condición de cuantización de Lüscher~\cite{Luscher:1986pf,Luscher:1990ux} relaciona $\cK_2$ con un factor geométrico que depende de las características de la \textit{lattice}, pero es independendiente de las propiedades de dispersión del proceso. En el caso de tres partículas se han propuesto varias alternativas~\cite{Hansen:2014eka,Hansen:2015zga,Mai:2017vot,Mai:2017bge,Hammer:2017uqm,Hammer:2017kms}. Esta tesis se centra en el  formalismo RFT~\cite{Hansen:2014eka,Hansen:2015zga}, el cual relaciona $\Kdf$ con un factor geométrico de tres partículas que depende de las características de la \textit{lattice}, y también de las interacciones entre pares de partículas. %Este formalismo, que ha experimentado un gran desarrollo en la última década, ha sido generalizado para i

\subsection*{Dispersión de mesones en el límite de gran $\Nc$}

Los capítulos~\ref{sec:largeNpions} y~\ref{sec:largeNmesons} de esta tesis se centran en el estudio de la interacciones de dos mesones como función del número de colores, utilizando simulaciones \textit{lattice} en una teoría con $\Nf=4$ quarks degenerados. En este modelo, las interacciones de dos mesones ocurren en siete canales diferentes, correspondientes con las representaciones irreducibles del grupo de isospín,  SU(4) en este caso.

En el capítulo~\ref{sec:largeNpions} se presentan los resultados de las \rrcite{Baeza-Ballesteros:2022azb,Baeza-Ballesteros:2021nxu}, centrados en el estudio de las interacciones entre dos mesones pseudoescalares cerca del umbral. En particular, se analizaron dos canales diferentes: el canal $SS$, equivalente al canal de isospín dos en QCD con dos sabores, y el canal $AA$, el cual solo existe en teorías con cuatro o más sabores degenerados. Las interacciones en estos canales se estudiaron utilizando el límite de gran $\Nc$ y la teoría quiral de perturbaciones incluyendo la $\eta'$. Se determinó por primera vez la amplitud para ambos canales en ChPT con la $\eta'$ a NNLO. Las amplitudes para cada canal dependen de combinaciones lineales de las LECs, llamadas $L_{SS}$ y $L_{AA}$. En el límite de gran $\Nc$, se espera que ambas tomen el mismo valor.

Para estudiar la dependencia de estas LECs con $\Nc$ se usaron simulaciones \lattice con $\Nc=3-6$ y diferentes valores de $\Mpi$, utilizando el software HiRep~\cite{DelDebbio:2008zf,DelDebbio:2009fd}. Las simulaciones permitieron determinar las energías en volumen finito de dos piones en el estado de mínima energía. Comparando dos discretizaciones diferentes de los fermiones de valencia y distintos espaciados de la \lattice se estudiaron también los efectos de discretización presentes en el canal $AA$. Finalmente, se usó el formalismo de Lüscher para determinar $L_{SS}$ y $L_{AA}$ a partir de las predicciones de ChPT. A partir de un ajuste simultáneo a ambos canales y todos los valores de $\Nc$ y $\Mpi$, se obtuvo,
\begin{equation*}
\frac{L_{SS,AA}}{\Nc}\times10^3 = - 0.02(8)-0.01(5)\frac{\Nf}{\Nc}\mp 1.76(20)\frac{1}{\Nc}+\cO(\Nc^{-2})\,.
\end{equation*}
donde el signo superior/inferior se corresponde con el canal $SS$/$AA$. 
Este resultado muestra que el valor de las variables en el límite de gran $\Nc$ es despreciable respecto a las correcciones subdominantes, que dominan a bajos valores de $\Nc$. Cabe notar que, si bien parece que una de las contribuciones subdominantes es mucho mayor que la otra, estas dependen de la escala de regularización (no así el término dominante).

Una de las observaciones más interesantes del trabajo expuesto en el capítulo~\ref{sec:largeNpions} es el hecho que el canal $AA$ presenta interacciones atractivas cerca del umbral. Esto, unido al hecho que contiene estados con cuatro sabores abiertos, hacen de este canal un candidato idóneo para explorar la existencia de tetraquarks y su dependencia con $\Nc$. La posible presencia de un tetraquark en el canal $AA$ podría estar relacionada con ciertos estados exóticos recientemente hallados en el LHCb~\cite{LHCb:2020bls,LHCb:2020pxc,LHCb:2022sfr,LHCb:2022lzp}, los cuales tendrían los números cuánticos del canal $AA$ en una teoría con cuatro sabores.

El capítulo~\ref{sec:largeNmesons} presenta resultados de un trabajo en progreso sobre las interacciones entre mesones como función de la energía a $\Nc$ variable~\cite{Baeza-Ballesteros:largeNinprep,Baeza-Ballesteros:2024ogp}, con el objetivo principal de estudiar la existencia de tetraquarks. En particular, se realizaron simulaciones \lattice con $\Nc=3-6$ para $\Mpi=590$ MeV y espaciado fijo, centradas en el estudio de los canales $SS$ y $AA$, así como del canal $AS$ que contiene ondas parciales impares. Para ello, se utilizó una base de operadores formada por operadores de dos partículas, tanto dos piones como dos mesones vectoriales, así como de operadores locales con forma de tetraquark.

Usando la condición de cuantización de Lüscher, se determinó la amplitud de dispersión para estos canales. En el caso de los canales $SS$ y $AA$, un ajuste de los resultados a $\Nc$ fijo permitió hallar la longitud de dispersión y el rango efectivo de ambos canales, los cuales se utilizaron para realizar una extrapolación a gran $\Nc$. Aunque no se ha observado la existencia de resonancias en el canal $AA$, los resultados para $\Nc=3$ sugieren la presencia de un estado ligado con una energía $E_\text{lig}/\Mpi=1.741(13)$.   Para el canal $AS$, se observaron interacciones muy suprimidas, en la línea de lo esperado de acuerdo con ChPT, que predice una amplitud de dispersión nula a LO. En el futuro, sería de interés estudiar más en detalle el estado observado en el canal $AA$ para $\Nc=3$, en particular su dependencia con la masa de los piones y su posible relación con las resonancias observadas.

Con este trabajo se da un paso más en la caracterización de la dependencia en $\Nc$ de observables de dispersión y se prepara el camino para el análisis de otros procesos de interés. Un ejemplo sería el estudio de resonancias, como la $\rho(770)$, que aparece en el canal de isospín uno, o la $\sigma(550)$, en el canal de isospín cero. En particular, el análisis de la dependencia de esta última con $\Nc$ podría proveer información sobre su naturaleza, y sobre si se trata de una resonancia mesónica o tiene una composición más exótica~\cite{Pelaez:2003ip,Pelaez:2005pi,Pelaez:2006nj}.

\subsection*{Matriz $K$ de tres piones a primer orden en ChPT}

Los capítulos~\ref{sec:pipipiKmatrix} y~\ref{sec:isospinKmatrix} presentan los resultados de las \rrcite{Baeza-Ballesteros:2023ljl,Baeza-Ballesteros:2024mii}, centrados en la determinación de $\Kdf$ para tres piones a primer orden en ChPT (\textit{next-to-leading-order}, NLO). Recientemente se observó que los resultados a LO en ChPT para la matriz $K$ de tres piones en el canal de isopín tres muestran una gran discrepancia con resultados \lattice para esta cantidad~\cite{Blanton:2019vdk,Blanton:2021llb}. El trabajo presentado en estos capítulos busca determinar si esta diferencia  se puede explicar debido a correcciones de gran tamaño a NLO.

En el capítulo~\ref{sec:pipipiKmatrix} se presentan los resultados de la \rcite{Baeza-Ballesteros:2023ljl}, centrados en el caso de isospín máximo. Al igual que a LO, las ecuaciones integrales que relacionan la amplitud de dispersión con $\Kdf$ se reducen a una simple ecuación algebraica a NLO en ChPT, $\Re\Mdf=\Kdf$, donde $\Mdf$ es la amplitud libre de divergencias, obtenida tras sustraer posibles divergencias relacionadas con interacciones sucesivas de dos partículas.

El cálculo de $\Kdf$ a este orden, partiendo de los resultados para $\cM_3$ de las \rrcite{Bijnens:2021hpq,Bijnens:2022zsq}, se puede dividir en tres partes, cada una de las cuales se trató con técnicas ligeramente distintas. La primera de ellas es la contribución proveniente de diagramas que contienen el intercambio de una partícula (OPE), justo con la sustracción de posibles divergencias que surgen cuando el cuadrimomento de dicha partícula obedece la condición de dispersión relativista, es decir, está \textit{on-shell}. Las otras dos contribuciones a $\Kdf$ son la llamada amplitud no OPE, asociada con el resto de la amplitud, y la sustracción correspondiente. En este caso, al contrario que para la parte OPE, ambas contribuciones pueden calcularse independientemente, ya que su parte real es siempre convergente. %En el caso de la sustracción, llamada comúnmente sustracción \textit{bull's head} (BH), el cálculo de una expansión alrededor del umbral necesita del usa de la prescripción de Hadamard para regular  

Los resultados de esta trabajo demuestran que las correcciones a NLO para los términos constante y lineal de una expansión de $\Kdf$ alrededor del umbral representan grandes correcciones respecto al resultado a LO. Aunque esto puede indicar una mala convergencia de la expansión quiral para observables de tres piones, también es  posible que la corrección a NLO sea anómalamente grande, y otras correcciones a órdenes superiores sean de menor tamaño. % que la aparición de nuevos diagramas a NLO dé lugar a grandes correcciones. %Este trabajo estudió también la convergencia de la expansión de $\Kdf$ alrededor del umbral.

En el capítulo \ref{sec:isospinKmatrix} se resumen los resultados de la \rcite{Baeza-Ballesteros:2024mii}, en la cual se extendió el trabajo realizado para el canal de isospín tres al resto de canales de tres piones, determinando $\Kdf$ a LO y NLO en ChPT. Este cálculo sigue las mismas líneas que el capítulo~\ref{sec:pipipiKmatrix}, con ciertas novedades. Estas son, entre otros, la presencia de una nueva contribución en el canal de isospín uno, relacionada con diagramas en los que una partícula se intercambia en el \mbox{canal $s$,} o la existencia de diferentes estructuras cinemáticas para los diferentes canales de isospín.

Al igual que el canal de isospín tres, el resto de canales también presentan grandes correcciones a NLO en ChPT. Esto proporciona valiosa información sobre la convergencia de la expansión quiral para observables de tres partículas. Además, los resultados obtenidos en este capítulo pueden ser de gran relevancia en el futuro, ya que podrán ser comparados con resultados \lattice o usados para parametrizar la dependencia quiral de $\Kdf$ cerca del umbral.

Siguiendo las líneas de este trabajo, el cálculo de $\Kdf$ para otros procesos puede ser de interés. En particular, sistemas de piones y kaones con isospín máximo ya han sido estudiados en la \lattice~\cite{Blanton:2021llb,Draper:2023boj}. Sin embargo, las correspondientes amplitudes de dispersión a NLO en ChPT, necesarias para el cálculo de $\Kdf$, aún no han sido calculadas.

\subsection*{Interacciones de dos y tres partículas en el modelo O(3)}

El capítulo~\ref{sec:O3model} cierra la primera parte de esta tesis doctoral. Está centrado en el estudio de la dispersión de dos y tres partículas en el modelo O(3) sigma no lineal en 1+1 dimensiones~\cite{Baeza-Ballesteros:O3inprep,Baeza-Ballesteros:2022bsn}. Este modelo es comúnmente usado para probar distintos formalismos~\cite{Luscher:1990ck,Bulava:2021fre}, antes de aplicarlos a QCD, ya que ambos comparten muchas propiedades, como la libertad asintótica o la existencia de un espectro de tres partículas con isospín uno. Además, el modelo O(3) es integrable, es decir, es posible obtener resultados exactos para su amplitud de dispersión~\cite{Zamolodchikov:1977nu,Zamolodchikov:1978xm}. El trabajo presentado en este capítulo tiene como fin último utilizar este modelo para estudiar el formalismo RFT de tres partículas, comparando resultados \lattice con predicciones analíticas obtenidas a partir de la amplitud de dispersión.

Como primer paso en esta dirección, se determinaron los espectros de energía de dos y tres partículas, y comparado con predicciones analíticas hechas bajo la suposición de una matrix $\Kdf$ nula. Se realizaron simulaciones \lattice para cuatro valores del volumen, con tres espaciados cada uno, usando una generalización a tres \textit{clusters} del algoritmo de actualización de \textit{cluster}~\cite{Wolff1}. Se trata de un algoritmo colectivo que permite reducir la correlación entre configuraciones y el ruido al medir funciones de correlación. Los resultados para las energías de volumen finito se extrapolaron directamente al continuo utilizando las simulaciones realizadas a igual volumen y diferente espaciado.

En el sector de dos partículas, se estudiaron los canales de isospín dos y uno. El espectro de energías se  comparó directamente con una predicción analítica, observando muy buena concordancia entre ambos. En el caso de tres partículas, se estudiaron los canales de isospín tres, dos y cero. Los resultados de las energías de volumen finito se compararon con predicciones obtenidas con una versión adaptada del formalismo RFT al caso de una dimensión espacial y asumiendo $\Kdf=0$. Se observaron discrepancias significativas, indicando que la matriz $K$ de tres partículas es en realidad no nula. Este estudio, además, permitió ganar conocimiento acerca de las complejidades de la condición de cuantización, como la posible existencia de inestabilidades numéricas o el rol de la función de \textit{cutoff} usada en el formalismo.

Siguiendo el camino marcado por este trabajo, el espectro de energías de tres partículas puede utilizarse para determinar $\Kdf$. Posteriormente, estos resultados pueden compararse con predicciones analíticas para esta cantidad. Dicha comparación requerirá reformular las ecuaciones integrales que relacionan la amplitud de dispersión con $\Kdf$ en una dimensión espacial. Más allá del modelo O(3), este trabajo prepara el camino para el estudio de sistemas de tres partículas en QCD más allá de sistemas de tres mesones con isospín máximo, en que se han centrado la gran mayoría de estudios \lattice hasta ahora~\cite{Mai:2018djl,Horz:2019rrn,Blanton:2019vdk,Mai:2019fba,Culver:2019vvu,Fischer:2020jzp,Hansen:2020otl,Alexandru:2020xqf,Brett:2021wyd,Blanton:2021llb,Draper:2023boj}. Por ejemplo, un caso interesante es el estudio de tres piones en isospín no máximo, ya que alguno canales de isospín continen resonancias de interés, como la $\omega(782)$ para el canal de isospín cero~\cite{PDG:2020}.

\section*{Parte 2. Cuerdas cósmicas mediante simulaciones de \lattice}

\subsection*{El universo primigenio}

El modelo cosmológico estándar está construido a partir de las bases de la relatividad general y el principio cosmológico. De acuerdo con este último, el universo es isótropo y homogéneo, es decir, idéntico en todas las direcciones y para cualquier observador independientemente de la posición en que se halle. Esto se traduce en la métrica de Friedman-Lemaître-Robertson-Walker (FLRW)~\cite{Robertson:1935zz}, caracterizada por el factor de escala, $a(t)$, que determina el ratio de expansión del universo. La evolución temporal del dicho factor de escala, a su vez, está determinada por la densidad de energía y la presión de la materia que llena el universo, de acuerdo con las ecuaciones de Friedman~\cite{Fridman2008}. 

Durante los primeros instantes del universo tras el fin de inflación, es probable que se alcanzaran energías muy por encima de las cotas alcanzables por aceleradores. Esto implica que, mediante el análisis experimental del universo primigenio, sería posible estudiar física más allá del modelo estándar. 

La descripción de ciertos fenómenos que tienen lugar durante el primer segundo del universo puede realizarse en el marco de la teoría de campos. Más concretamente, debido a los altos números de ocupación, es posible describir la dinámica del sistema utilizando métodos de teoría clásica de campos, basados en el estudio de la evolución de los campos de acuerdo con sus ecuaciones de movimiento.

En ciertas ocasiones esta evolución temporal se puede estudiar analíticamente, tratando las fluctuaciones como pequeñas perturbaciones. Sin embargo, en muchos casos el sistema puede desarrollar contribuciones no lineales de gran tamaño y la única opción es usar simulaciones numéricas en la \textit{lattice}. Las simulaciones  de campos clásicos se basan en la resolución numérica de las ecuaciones de movimiento de dichos campos, partiendo de unas condiciones iniciales dadas. Para ello se utiliza una versión discretizada de las ecuaciones de movimiento. Para la realización de simulaciones \mbox{\textit{lattice}} en los trabajos presentados en esta tesis doctoral se utilizó el software $\mathcal{C}$osmo$\mathcal{L}$attice~\cite{Figueroa:2020rrl,Figueroa:2021yhd}.

%La \lattice está caracterizada por un cierto tamaño, $L$, y un espaciado $\delta x$, que determinan los momentos que se pueden estudiar en la simulación. Para evitar posibles efectos sistemáticos, es necesario que la \lattice capture todos los modos físicos relevantes en el problema. 

El uso de técnicas de lattice permite estudiar también la emisión de GW en el universo primigenio~\cite{Maggiore:2007ulw,Caprini:2018mtu}. Una vez cesa la fuente que las genera, estas ondas viajan libremente hasta el día de hoy, y mediante una detección experimental sería posible determinar las características de los fenómenos que las producen. Sin embargo, una interpretación correcta de una potencial detección requiere de  predicciones teóricas precisas.

La simulación de GW en la lattice es especialmente compleja. La fuente de GW es la parte transversa y de traza nula (TT) del tensor anisótropo, y su cálculo en espacio real es una operación no local que requiere conocer el valor de los campos en todo el espacio. En espacio de Fourier, en cambio, la proyección al la componente TT es local, pero require transformar los campos a dicho espacio en cada paso temporal. Numéricamente, es más conveniente utilizar seis campos no físicos, $u_{ij}$, generados por un tensor anisótropo efectivo que no es TT~\cite{Garcia-Bellido:2007fiu}, evitando la necesidad de calcular transformadas de Fourier en cada paso de la evolución temporal. Solo cuando se desea medir el espectro de energía de las GW es necesario proyectar los campos $u_{ij}$ a su parte TT, que se corresponde con los campos de GW. La implementación de estas técnicas en $\mathcal{C}$osmo$\mathcal{L}$attice ha representado una parte del trabajo de esta tesis doctoral~\cite{GWmodule:2022,GWmodule:2023}.

\subsection*{Emisión de partículas y GWs por \loops cósmicos}

Las cuerdas cósmicas son defectos topológicos unidimensionales que podrían formarse en el universo primigenio tras una transición de fase, de acuerdo con ciertas extensiones del modelo estándar~\cite{Hindmarsh:1994re,Copeland:2009ga,Copeland:2011dx,Vachaspati:2015cma}. Se trata de defectos topológicos que se originan en teorías con un vacío no simplemente conexo. Un ejemplo son los modelos de gran unificación, que predicen la formación de cuerdas con interacciones de corto alcance, llamadas locales. Otro ejemplo de interés son los modelos de axiones~\cite{Peccei:1977hh,Peccei:1977ur}, los cuales incluyen una simetría U(1) anómala, la simetría de Peccei-Quin, que se rompe espontáneamente a bajas energías. Si esta ruptura de simetría ocurre tras el final de inflación, se podrían originar \textit{networks} de cuerdas cósmicas de tipo global, que interaccionan a larga distancia.

Las \textit{networks} de cuerdas están formadas por cuerdas infinitas, que se extienden a lo largo de todo el universo, y  cuerdas cerradas de tamaño subhorizonte, llamadas \textit{loops}. Típicamente, la evolución de las cuerdas se ha descrito utilizando la acción de Nambu-Goto (NG), en la cual se aproximan las cuerdas como objetos infinitamente finos~\cite{nambu1971lectures,Goto:1971ce}. En esta aproximación, las cuerdas infinitas decaen emitiendo \textit{loops}, mientras que estos últimos radían GW, dando lugar a un fondo estocástico de GW.

Una interpretación correcta de una hipotética detección del fondo de GW  depende de precciones teóricas precisas. En particular, una posibilidad es que los \loops decaigan mediante la emisión de partículas, la cual es despreciada por la aproximación de NG. Una comparativa entre ambas rutas de decaimiento requiere estudiar los dos canales simulaténeamente y un estudio en detalle que capture la estructura interna de las cuerdas solo es posible mediante el uso de simulaciones \textit{lattice}.

Los capítulos~\ref{sec:global} y~\ref{sec:local} de esta tesis se centran en el estudio de la emisión de partículas y GW por parte de cuerdas cósmicas. Cada capítulo se centra en una teoría diferente. El capítulo~\ref{sec:global} presenta los resultados de la \rcite{Baeza-Ballesteros:2023say}. Este trabajo trata el caso de cuerdas globales, las cuales se representan con una teoría cuyo contenido es únicamente un campo escalar complejo, con un potencial con forma $V(\varphi)=\lambda(|\varphi|^2-v^2/2)$ siendo $v$ el valor de expectación del vacío y $\lambda$ un parametro adimensional que regula las interacciones del campo escalar. Las cuerdas de este tipo decaen emitiendo radiación tanto masiva como sin masa. El capítulo~\ref{sec:local}, por otro lado, está centrado en el caso de cuerdas locales~\cite{Baeza-Ballesteros:stringsinprep}, las cuales se simulan utilizando el modelo Higgs abeliano, que contiene un campo \textit{gauge} U(1)  y un campo escalar complejo. Al contrario que las cuerdas globales, las cuerdas locales solo pueden emitir partículas masivas.

En sendos capítulos se presentan resultados relativos al estudio en la \textit{lattice} de la emisión de partículas y GW por parte de \loops globales y locales, respectivamente. En ambos casos, se analizan dos tipos de \textit{loops}. El primer tipo, referido como \network \textit{loops}, se genera tras el decaimiento de una \network de cuerdas, y se espera que presente características similares a aquellos \loops que se podrían formar en el universo primigenio~\cite{Hindmarsh:2019csc,Hindmarsh:2021vih,Hindmarsh:2021mnl}. El segundo tipo son lo llamados \loops artificiales, que se originan a partir de la intersección de cuerdas infinitas~\cite{Saurabh:2020pqe,Matsunami:2019fss} y permiten controlar las condiciones iniciales.

En el caso de las cuerdas globales, se encontró que ambos tipos de \loops se comportan de forma similar, emitiendo partículas con una potencia aproximadamente constante, con independencia de su longitud. Respecto a la emisión de GW, también se halló que ambos tipos de loops emiten con una potencia similar y constante, independiente de su longitud. Comparando los dos canales, se observó que potencia de la emisión de GW, $\PGW$, está muy suprimida comparado con la de partículas, $\Pphi$, 
\begin{equation*}
\frac{\PGW}{\Pphi}\approx\cO(10)\left(\frac{v}{\mpl}\right)^2\ll 1\,,
\end{equation*}
teniendo en cuenta que medidas experimentales del fondo cósmico de microondas restringen $v/\mpl\lesssim 10^{-6}$~\cite{Lopez-Eiguren:2017dmc,Benabou:2023ghl}. 

Para cuerdas locales, los resultados obtenidos fueron distintos. La \mbox{potencia} de emisión de partículas en este caso puede depender de la longitud de las cuerdas. Sin embargo, lo hace de forma muy distinta para cada tipo de \textit{loop}. Mientras que nuestros resultados indican que la  potencia de emisión de los \network \loops aumenta débilmente con su longitud (siendo los datos también consistentes con una emisión independiente de la longitud), los \loops artificiales emiten partículas más lentamente a mayor tamaño. Por otro lado, la emisión de GW parece ser independiente de la longitud para ambos tipos de \textit{loops}, aunque presenta mayor potencia para los \network \textit{loops}. Cabe notar que el resultado obtenido para \loops artificiales es muy próximo a la predicción de NG, siendo un factor tres veces más grande para los \network \textit{loops}.

Extrapolados a escalas cosmológicas, estos resultados tienen consecuencias relevantes. Si \loops locales con características similares a los artificiales se hubieran formado en el universo primigenio, decaerían principalmente emitiendo GW, dando lugar a un fondo de GW de gran amplitud. Sin embargo, no observamos que dicho tipo de \loops se forme a partir de \textit{networks} de cuerdas locales. Los \network \textit{loops}, que sí se esperan en el universo temprano, decaen principalmente emitiendo partículas, y la producción de GW está muy suprimida.

En el futuro, planeamos utilizar nuestros resultados sobre el decaimiento de \loops aislados, combinados con predicciones sobre la densidad de \loops en el universo primigenio, para determinar el fondo de GW emitido por una red de cuerdas cósmicas. Una predicción realizada de esta forma tendría la ventaja de no depender de predicciones de NG sobre la emisión de GW, ni de grandes extrapolaciones hechas a partir de resultados \lattice para redes de cuerdas cósmicas. Tanto para \loops globales como para locales, es de esperar que estos cálculos supriman la amplitud del fondo de GW respecto a predicciones actuales. %En particular, para las cuerdas globales, no observamos ninguna dependencia logaritmica con la longitud de las cuerdas, en contradicción con . %En el caso de \loops locales, por otro lado, hallamos que la emisión de partículas aumenta con la longitud, y por tan

\section*{Comentarios finales} 

Esta tesis doctoral está centrada en el uso de técnicas de \textit{lattice} para estudiar dos temas tan dispares como son las interacciones entre hadrones y el decaimiento y emisión de GW por cuerdas cósmicas. El trabajo presentado en la primera parte de esta tesis se centra en el estudio de interacciones entre dos y tres partículas, haciendo uso de simulaciones de \textit{lattice} QCD, junto con otros métodos como el límite de gran $\Nc$ y la teoría quiral de perturbaciones. Los resultados obtenidos representan un paso en la comprensión del régimen no perturbativo de QCD, en particular sobre la existencia de resonancias exóticas con cuatro quarks y las interacciones de tres mesones, y abren la puerta a estudiar sistemas más complejos.%, como resonancias de tres hadrones o núcleos ligeros.

La segunda parte de la tesis se centra en la evolución y decaimiento de \loops cósmicos, con especial énfasis en la emisión de partículas y GW. Los resultados presentados muestran que para \loops globales, la emisión de GW está universalmente suprimida comparada con la producción de partículas, con independecia del tamaño, forma o momento angular del \textit{loop}. Para \loops locales, por contra, es posible construir configuraciones para las cuales la emisión de GW podría dominar. Sin embargo, para \loops locales realistas que pudieran formarse en el universo primigenio, la ruta principal de decaimiento es la emisión de partículas, con una producción de GW muy suprimida. Combinando los resultados obtenidos con predicciones de la densidad de \textit{loops}, planeamos calcular una predicción para el fondo de GW. En vista de nuestro resultados, esperamos que dicho fondo esté muy suprimido, comparado con las predicciones actuales.

%%%%%%%%%%%%%%%%%%%%%%%%%%%%%%%%%%%%%%%%%%%%%%%%%%%%%%%%%%%%
% References
%%%%%%%%%%%%%%%%%%%%%%%%%%%%%%%%%%%%%%%%%%%%%%%%%%%%%%%%%%%%

\cleardoublepage

\phantomsection

%\addcontentsline{toc}{chapter}{{Bibliography}}

\renewcommand{\headrulewidth}{0.5pt}

\lhead[{\bfseries \thepage}]{The Bibliography}
\rhead[{The Bibliography}]{\bfseries \thepage}

%\bibliography{biblio}

\providecommand{\href}[2]{#2}\begingroup\raggedright\endgroup

%%%%%%%%%%%%%%%%%%%%%%%%%%%%%%%%%%%%%%%%%%%%%%%%%%%%%%%%%%%%
% Papers
%%%%%%%%%%%%%%%%%%%%%%%%%%%%%%%%%%%%%%%%%%%%%%%%%%%%%%%%%%%%

\end{document}